\documentclass[10pt,landscape,twocolumns,letterpaper]{article}
\usepackage[usenames, dvipsnames]{color}
\usepackage[greek,english]{babel}
\usepackage{amsmath}
\usepackage{amssymb}
\usepackage[usenames,dvipsnames]{color}
\usepackage{cancel}
\usepackage{graphicx}
\usepackage{mathrsfs}
\usepackage{bbm}
\usepackage[labelfont=bf,textfont=it]{caption}
\usepackage{IEEEtrantools}
\usepackage{makeidx}

\graphicspath{{./figs_compressed/}}

\usepackage[bordercolor=white,backgroundcolor=gray!30,linecolor=black,colorinlistoftodos]{todonotes}

\newcommand{\annote}[1]{}

\usepackage{wasysym} 

\def\Xint#1{\mathchoice
   {\XXint\displaystyle\textstyle{#1}}%
   {\XXint\textstyle\scriptstyle{#1}}%
   {\XXint\scriptstyle\scriptscriptstyle{#1}}%
   {\XXint\scriptscriptstyle\scriptscriptstyle{#1}}%
   \!\int}
\def\XXint#1#2#3{{\setbox0=\hbox{$#1{#2#3}{\int}$}
     \vcenter{\hbox{$#2#3$}}\kern-.5\wd0}}

\def\dashint{\Xint-}

\makeindex

\usepackage{geometry}
\usepackage{textcomp}

\usepackage{sidecap}
 
\usepackage{hyperref}

\newcommand{\doublestroke}{\mathbb}

\usepackage{bm}
\usepackage{yfonts}

\newcommand{\timout}[1]{{\textcolor{Purple}{\underline{#1}}}}

\hypersetup{
   colorlinks=true,         
   linkcolor=Red,          
   citecolor=red,        
   urlcolor=cyan            
}

\def\ts{\textstyle}

\def\sfrac{\textstyle \frac}

\newcommand{\ro}[1]{\text{#1}}

\newcommand{\Lie}{\pounds}

\newcommand{\st}[1]{\textrm{\tiny{#1}}}

\newcommand{\CL}{\ocircle}

\def\Q{{\sf Q}}
\def\shs{{\sf S}}

\def\pshs{{\sf PS}}

\def\Riem{{\sf Riem}}
\def\Superspace{{\sf Superspace}}

\def\Eucl{{\sf Eucl}}
\def\Sim{{\sf Sim}}

\def\e{\emph}

\def\laplacian{\Delta}

\def\thistext{Tutorial}

\newcommand{\doti}{\chi}		
\newcommand{\teph}{\mathrm{t}_\st{eph}}		
\newcommand{\dbm}{\mathfrak{D}}		
\newcommand{\almostHamiltonian}{\mathcal A}	

\newcommand{\mat}[1]{\mathrm{#1}}

\definecolor{gold}{RGB}{254,250,202}

\def\d{\mathrm{d}}

\def\A{A} 
\def\E{E} 

\usepackage{tocloft}

\begin{document}

\onecolumn

\thispagestyle{empty}

\pagecolor{black}

\color{gold}

\title{\vspace{1.5cm}\bf A Shape Dynamics Tutorial}

\author{\vspace{1.5cm}\\\Large\textbf{Flavio Mercati}\vspace{12pt}
\\\vspace{12pt}
\it  Perimeter Institute for Theoretical Physics,\\
\it  31 Caroline Street North,  Waterloo, ON, N2L 2Y5 Canada.
\vspace{4cm}\\
 email:~\href{mailto:fmercati@perimeterinstitute.ca}{fmercati@perimeterinstitute.ca}}
\date{}

\begin{center}

{\bf \huge A Shape Dynamics Tutorial (v2.0)}\\\vspace{12pt}
\includegraphics[height=0.65\textheight]{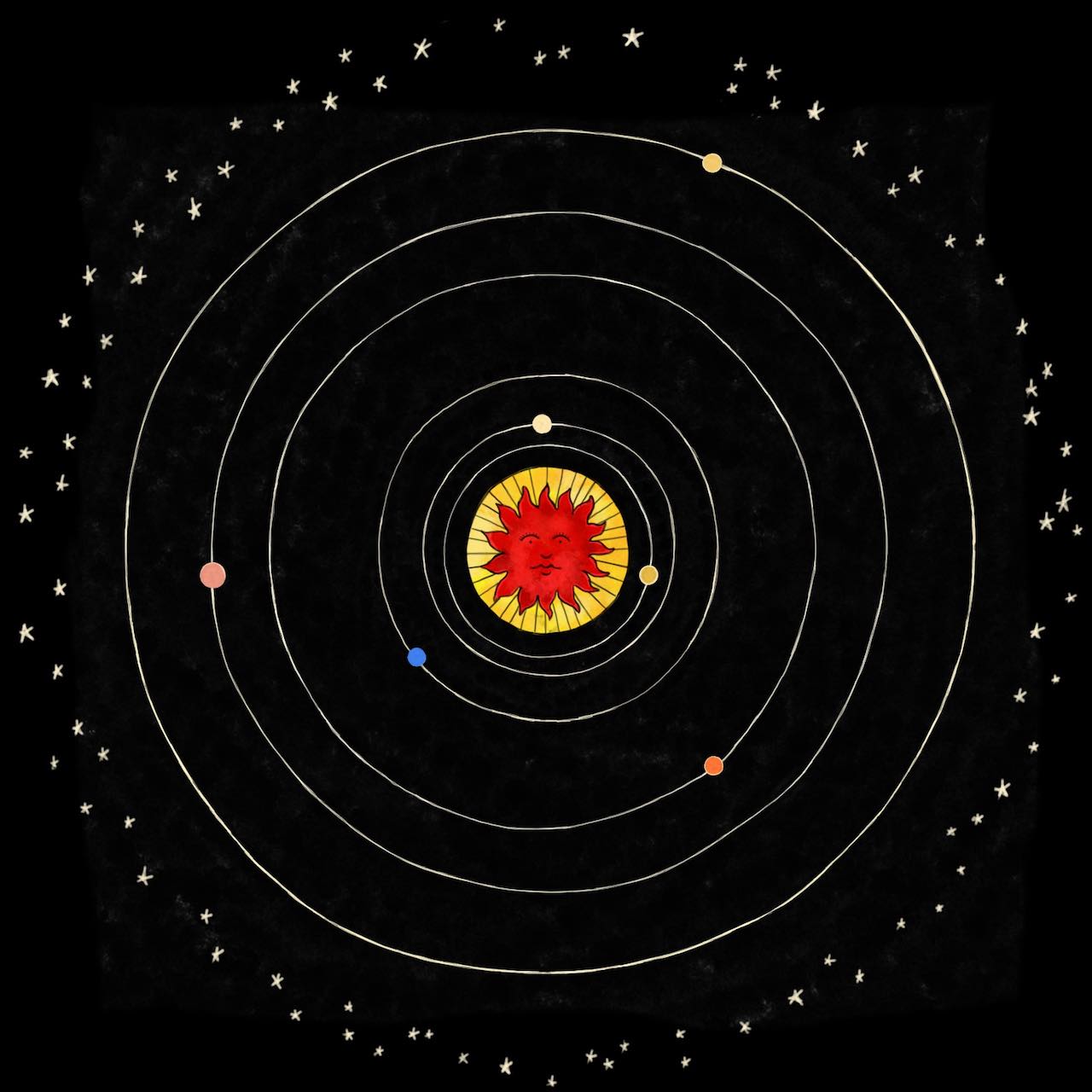}

\vspace{24pt}

{\Large\textbf{Flavio Mercati}}

\vspace{12pt}

\large
\it Dipartimento di Fisica, Universit\`a di Roma ``La Sapienza'',\\
\it P.le A. Moro 2, 00185 Roma, Italy;\\
\vspace{6pt}
\it  Perimeter Institute for Theoretical Physics,\\
\it  31 Caroline Street North,  Waterloo, ON, N2L 2Y5 Canada.\\
\vspace{6pt}
email:~\href{mailto:flavio.mercati@gmail.com}{flavio.mercati@gmail.com} ~~~ webpage:~\href{http://www.roma1.infn.it/~mercatif/}{www.roma1.infn.it/~mercatif/}
 
\end{center}

%
%
%
%
%
%


\twocolumn

\pagecolor{white}

\color{black}

\tableofcontents

\vfill

\begin{flushright}
Cover picture concept due to J. Barbour

\emph{``From henceforth the planets follow their paths through the ether like the birds in the air. We must therefore philosophize about these things differently.''}

Johannes Kepler (1571--1630)
\end{flushright}

\setcounter{page}{1}

\newpage

\begin{abstract}
Shape Dynamics (SD) is a new theory of gravity that is based on fewer and more fundamental first principles than General Relativity (GR). The most important feature of SD is the replacement of GR's relativity of simultaneity with a more tractable gauge symmetry, namely invariance under spatial conformal transformations.
This \thistext{} contains both a quick introduction for readers curious about SD and a detailed walk-through of
the historical and conceptual motivations for the theory, its logical development from first principles and an in-depth description of its present status. The \thistext{} is sufficiently self-contained for an undergrad student with some basic background in GR and Lagrangian/Hamiltonian mechanics. It is intended both as a reference text for students approaching the subject and as a review for researchers interested in the theory.
\annote{Change thistext to Book}
\end{abstract}

\section{Introduction}

\subsection{Forewords}

The main part of the introduction is Sec.~\ref{NutshellSubsec} (Shape Dynamics in a nutshell), where I attempted to offer a no-nonsense quick entry to the basic ideas of SD. This serves a dual purpose: on one hand, students interested in SD will have a brief overview of what the theory is about and what we hope to achieve with it; on the other, researchers curious about SD will find in Sec.~\ref{NutshellSubsec} a short description of the theory that is hopefully enough to decide whether these ideas are worth examining in depth. The minimum of notions needed to understand the core ideas of SD are outlined with the aim of making the Section as self-contained as possible. All the concepts are explained in detail in the rest of this \thistext{}, while taking an `historico-pedagogical' perspective and introducing them at the appropriate points of the story.
Sec.~\ref{NutshellSubsec} also includes a quick outline of basic concepts needed to understand SD, which are not part of normal undergrad curricula (like constrained Hamiltonian systems and gauge theories). However the Section is limited to just a few pages to be read quickly by experts, and the mentioned outline is by no means sufficient to understand properly those concepts. Its purpose is to give the undergrad readers a taste of the background knowledge that is necessary to understand SD and get the overall drift. Everything is exhaustively explained in the body of the text.

Part~\ref{PartHistorical} shows where SD comes from: we consider it as the most advanced stage of the
\emph{relational program,} which seeks to eliminate all absolute structures from physics. Absolute structures meaning anything that determines physical phenomena but is not determined by them. The chief example is Newton's  absolute space and time (or, in modern terms, inertial frames of reference).
The battlefield of Newton's absolutes has seen giants of science fighting the absolute-vs-relative debate: Galileo, Descartes, Newton himself, Leibniz, Mach, Poincar\'e and Einstein. Another example is scale, or size: SD tries to eliminate precisely this absolute structure from physics. One could imagine pushing this program further in the future: what determines the topology of space? Are the values of the physical constants a result of immutable laws or of a dynamical evolution? 

In Sec.~\ref{SectionNewtonsBucket} I explain in detail the fundamental problem of Newtonian dynamics: 
everything is based on the law of inertia, which in turn relies on the concepts of rest and uniform motion,
but these concepts are not defined by Newton.
Section \ref{SectionMachPoincarePrinciple} makes it clear what the problem with Newton's construction is. Stating in a mathematically precise way the defect of Newton's theory was an incredibly hard problem; Henri Poincar\'e solved it after more than two centuries. However even Poincar\'e's formulation (which we call the `Mach--Poincar\'e Principle') wasn't recognized for what it is until the work of Barbour and Bertotti in the 1960's.

Part~\ref{PartRPM} deals with relational dynamics in the simpler framework of systems of point particles. Relational dynamics is a reformulation of dynamics that satisfies the Mach--Poincar\'e Principle, as formulated by Barbour and Bertotti. It uses specific techniques that were invented on purpose, in particular that of `best matching'. These techniques turned out to be equivalent to the modern formulation, due to Dirac, of gauge theories as constrained Hamiltonian systems.
In Sec.~\ref{BBbestMatching} best matching is introduced at an intuitive level, while Sec.~\ref{SectionBestMatching} details it using the language of Principal Fibre Bundles, which are introduced to the reader. Sec.~\ref{SectionHamiltonianFormulation} describes the Hamiltonian formulation of best matching and links it to modern gauge theory. The techniques developed by Dirac for Hamiltonian constrained systems are needed in this Section and are therefore briefly explained.

Part~\ref{RelationalFieldTheory} deals with the more advanced framework of field theory.
Section~\ref{RWRsection} details (in a modern language) a series of results due to Barbour, O'Murchadha, Foster, Anderson, and Kelleher. These are striking results: they show that the principles of relational field theory alone are sufficient to derive GR, the general and special relativity principles, the universality of the light cone, Maxwell's electromagnetism, the gauge principle and Yang--Mills theory.
Section~\ref{YorksMethodSection} contains more background material: it presents York's method for the solution of the initial-value problem  in GR. This provides an important input for the formulation of SD. Section \ref{ExtendedRWR} deals with work I have done together with E.~Anderson,
and finally makes the connection from relational field theory to SD. The latter is shown to arise from the principles of relational field theory and the Mach--Poincar\'e Principle.
The final Section of this Part, number~\ref{CottonSquaredTheorySec}, describes the attempt to build a theory that incorporates the principles of relational field theory and assumes local scale invariance (also called \emph{conformal}, or \emph{Weyl} invariance) from the beginning (while in my derivation of SD local scale invariance emerges as a consistency requirements in the analysis of the constraints of the theory).
Such a theory would implement both a local notion of duration (and therefore invariance under local time reparametrization) and conformal invariance. Interestingly, this theory proves to be inconsistent, leaving us with Shape Dynamics as the only viable candidate for a theory of evolving conformal geometry. As SD only has global reparametrization invariance and implies a preferred notion of simultaneity, we have to conclude that refoliation invariance and conformal invariance are dual and alternative to each other: they cannot be kept simultaneously.

In Part~\ref{SDpart} SD is finally formulated in its current form. 
I begin with a brief account of the way the ideas at the basis of SD were developed in Sec.~\ref{HistoricalInterlude},
then in Sec.~\ref{SDLT} I proceed to derive the equations of SD from the point I left the theory in  Sec.~\ref{ExtendedRWR}. In Sec.~\ref{DOFsOfSD} I discuss the physical degrees of freedom of SD, which
are the conformally invariant properties of a 3-dimensional manifold, and their conjugate momenta.
In Sec.~\ref{SolutionOfProblemOfTimeinSD} I explain how SD represents a simple solution to the
problem of time of quantum gravity, and how one reconstructs the familiar 4-dimensional spacetime description of GR from a solution of SD.
Sec.~\ref{CouplingToMatterSec} deals with the coupling of Shape Dynamics to matter, which was analized by Gomes and Koslowski, who applied to SD previous results on the conformal method by Isenberg, Nester, \'O Murchadha and York.
In Sec.~\ref{ExperiencedSpacetimeSec} I summarize Koslowski's work on the emergence of the spacetime description. This work shows how the 4-dimensional, CMC-foliated line element that one deduces from a solution of SD is the spacetime that matter degrees of freedom experience in the limit in which backreaction can be ignored. In Sec.~\ref{SymmetryDoublingSec} I briefly describe the result by Koslowski and Gomes on the BRST formulation of SD, and finally in In Sec.~I summarize Gomes' work on a construction principle for SD along the line of rigidity theorems like that of Hojman, Kucha{\v{r}} and Teitelboim.
\\
The following Section~\ref{SecSolutionsOfSD} deals with the particular solutions of SD that have been studied so far.  In Sec.~\ref{SecBianchiIX} I study in detail the homogeneous solution with spherical topology (so-called `Bianchi IX' universes), and show what is perhaps the most striking consequence of SD: its solutions can be continued uniquely through the Big Bang singularity. In Sec.~\ref{SphericallySymmetricSec} I study spherically-symmetric solutions, which are the basis to discuss gravitational collapse and black holes, and present another striking result: the ADM-in-CMC-foliation description of a closed universe with collapsing matter fails at some point during the collapse (presumably when the system generates an event horizon), while the SD description seems well-defined at that point and after. In Sec.~\ref{AsymptoticallyFlatSD} I discuss in which sense one can talk about asymptotic flatness in SD (which is fundamentally a theory of compact universes), and I critically evaluate past results obtained in the asymptotically flat case.

The final Part of the \thistext{} contains the appendices, which are divided into a first,
major Appendix,~\ref{ADM-WDW}, with a brief account of the Hamiltonian formulation of
GR due to Arnowitt, Deser and Misner. This is the main tool of Canonical General
Relativity and is the theory we have to compare classical SD to. In this Appendix I give a 
standard derivation of this theory starting from GR and the Einstein--Hilbert action. The same
theory can be deduced from the axioms of relational field theory without presupposing
spacetime and without starting from the Einstein--Hilbert action, as was done in Sec.~\ref{RWRsection}.
This derivation assumes less and should be considered more fundamental than that
of  Arnowitt, Deser and Misner. However, I felt that the junior readers should be aware of the 
standard derivation. Finally, Appendix~\ref{OtherAppendices} contains a series of results
and derivations that are useful and referenced to throughout the text, but which are moved
to the end of the \thistext{} for the sake of clarity of exposition.

\noindent{\bf Acknowledgements}

A big thank to Julian Barbour and Henrique Gomes for extensive help at different stages of composition of this \thistext{}.
Thanks also to Lee Smolin, Tim Koslowski, David Wiltshire and Matteo Lostaglio for useful comments and
discussions. Special thanks to my students Andrea Napoletano, Monica Rincon Ramirez and
Mykola Murskyj for being my Guinea pigs and proofreading the drafts of this \thistext{}.  

%

\newpage

\subsection{Notation}
In the text we use a notation according to which the Greek indices $\mu,\nu,\dots$ go from
$0$ to $3$, while the lowercase Latin indices from the middle of the alphabet $i,j,k,\ell,m\dots$ are spatial and go from $1$ to $3$.
We assume a Lorenzian signature $(-,+,+,+)$. The lowercase Latin indices from the beginning of the
alphabet $a,b,c,$ refer to the particle number and go from $1$ to $N$.
Three-dimensional vectors will be indicated with Latin or Greek bold letters, $\mathbf{q},\mathbf{p},\bm{\theta},\bm{\omega},\dots$,
while three-dimensional matrices will be uppercase Roman or Greek $\Omega, \Theta,\mat{U},\mat{I},\dots$.
The spatial Laplacian $g_{ij} \nabla^i \nabla^j$ will be indicated with the symbol $\laplacian$, while for the d'Alembertian
$g_{\mu\nu} \nabla^\mu \nabla^\nu$ I'll use the symbol $\square$. The (spatial) conformal Laplacian $8 \Delta - R $ will
be indicated with the symbol $\CL$.

\subsection{Shape Dynamics in a nutshell} \label{NutshellSubsec}

Shape Dynamics (SD) is a field theory that describes gravity in a different way than General Relativity (GR). However the differences between the two theories are subtle: in most situations they are indistinguishable.

\subsubsection*{SD is a gauge theory of spatial conformal (Weyl) symmetry}

SD and GR are two different gauge theories defined in the same phase space, both of which admit a particular gauge fixing in which they coincide.
This does not guarantee complete equivalence between the two theories: a gauge fixing is in general not
compatible with every solution of a theory, in particular due to global issues. The equivalence between SD and GR
therefore fails in some situations.

What distinguishes SD from GR as a fundamental theory of gravity is its different \emph{ontology}. 

First, SD does without spacetime: the existence of a pseudo-Riemannian 4-dimensional manifold with Lorentzian signature is not assumed among the
axioms of the theory. Instead, the primary entities in SD are three-dimensional geometries that are fitted together by relational principles into a `stack' whose structural properties can be identified in some but not all cases with those of a four-dimensional spacetime which satisfies Einstein's field equations. The closest agreement with GR occurs if the three-geometries are spatially closed \index{closed spacelike hypersurfaces} when the relational principles of SD are fully implemented. However, there is also interest in partial implementation of SD's relational principles in the case in which the three-geometries are asymptotically flat.

\begin{figure}[t]
\begin{center}\includegraphics[width=0.4\textwidth]{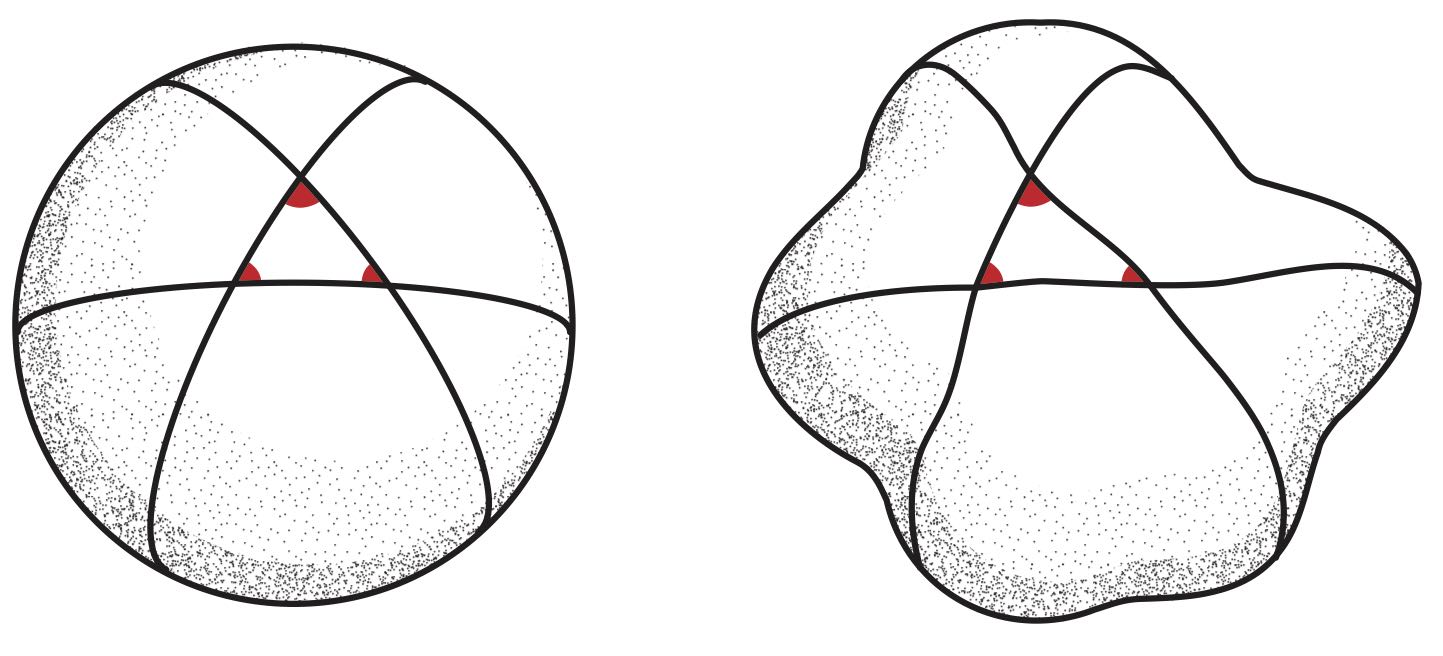}
\end{center}
\caption[What is a Conformal Transformation]{Conformal transformation of a 2-dimensional sphere. The triangle defined by the intersection
of three curves is transformed in such a way that its area and the lengths of its three edges are
changed, but the three internal angles (in red) are left invariant.}
\label{ConformalTransformation}
\end{figure}

Second, the spatial geometries which make the configuration space of SD are not Riemannian. They are \emph{conformal geometries,}
defined as equivalence classes of metrics under position-dependent conformal transformations (sometimes called `Weyl trasformations'; the fourth power of $\phi$ is chosen to simplify the transformation law of the scalar curvature $R$):
\begin{equation}\label{WeylTransformations}
\{ g_{ij} \sim g'_{ij} ~~ \text{\it if} ~~ g'_{ij} =\phi^4 ~ g_{ij} \,, ~~ \phi (x) > 0 ~ \forall x \} \,.
\end{equation}
Conformal transformations change lengths and preserve only angles (see Fig.~\ref{ConformalTransformation}).
Therefore a conformal geometry presupposes less than a Riemannian geometry, for which lengths determined by the metric are considered physical.  What is physical in SD is the \emph{conformal structure}, which is the \emph{angle-determining} part of the metric. 
Lengths can be changed arbitrarily and locally by a conformal transformation, which is a gauge transformation for SD.

So SD assumes less structure than GR, but it is in one sense a minimalistic lifting of assumptions: the next
thing in order of simplicity after Riemannian geometry is conformal geometry. Some other approaches to quantum gravity are decidedly more radical as regards the amount of structure they assume: either much more (\emph{e.g.} string theory) or much less (\emph{e.g.} causal sets). 

SD is based  on fewer and more basic kinematical first principles than GR:
\begin{quotation}
\emph{Spatial relationalism:} the positions and sizes of objects are defined relative to each other. This determines
what the physical configuration space is (see Sec.~\ref{MachsCritiqueOfNewtonSubsec}). In field theory this principle translates into conformal and diffeomorphism invariance, and the requirement of a spatially closed  \index{closed spacelike hypersurfaces} manifold. 

\emph{Temporal relationalism:} the flow of time is solely due to physical changes (see Sec.~\ref{SecTemporalRelationalism}).

The \emph{Mach--Poincar\'e Principle:} a point and a direction (or tangent vector, in its weak form) in the
physical configuration space are sufficient to uniquely specify the solution (see Sec.~\ref{PoincarePrincipleSubsec}).
\end{quotation}
There is no need for general covariance, the relativity principle, the existence
of spacetime, the existence of measuring rods and clocks. These concepts emerge from the solutions of
SD as characteristic behaviours or useful approximations. In this sense SD is more fundamental than GR
because it achieves the same with less. See Part~\ref{RelationalFieldTheory} for the full construction of
SD starting from its three first principles.

A common mistake is to regard SD just as a gauge-fixing of General Relativity. It is easy to see that this is not the case:
\emph{there are solutions of SD that are not solutions of GR, and vice-versa.} A satisfactory understanding of the GR solutions which SD excludes and of the SD solutions which GR excludes is still lacking. 

Let's now have a brief look at what exactly SD looks like.

\subsubsection*{Gauge theories are constrained Hamiltonian systems}

\begin{figure}[t]
\begin{center}\includegraphics[height=0.2\textheight]{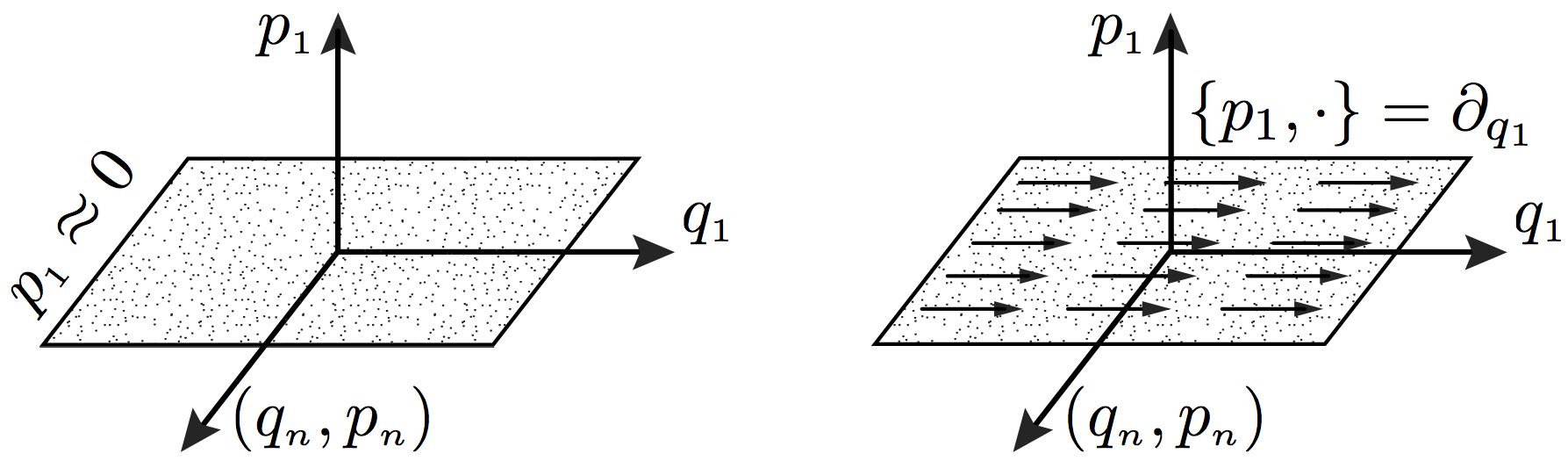}
\end{center}
\caption[The concept of a gauge constraint]{The constraint surface of a gauge constraint $\chi = p_1$ is represented in phase space,
where I put $p_1$ and $q_1$ on two axes, and all the other phase-space variables $(q_n,p_n)$,
$n=2,3\dots$ are represented collectively on the third axis.  On the right, I show the vector
field generated by $p_1$ through Poisson-brackets,  $\{  p_1,\cdot \} = \frac{\partial}{\partial q_1}$, 
which points towards the $q_1$ direction. The vector field is parallel to the constraint surface,
and its integral curves (the gauge orbits) lie on it.}
\label{GaugeConstraint}
\end{figure}

SD is more naturally formulated as a gauge theory in the Hamiltonian language. Gauge theories are theories with redundancies: one
uses more degrees of freedom than necessary in order to attain a simpler and \emph{local} description. 
In the Hamiltonian picture, this translates into \emph{nonholonomic constraints:} functions of the canonical variables $\chi = \chi(p,q)$ (with some dependency on the momenta everywhere on phase space) which need to vanish on the solutions of the theory $ \chi(p,q) \approx 0$.\footnote{With `$\approx$' we mean that the equation holds on the solutions of the constraint equations, following Dirac's notation.}  A single constraint identifies a codimension-1 hypersurface in phase 
space, the \emph{constraint surface}, on which the solutions of the theory are localized. For example, if a gauge constraint can be
written as $\chi = p_1$, where $p_1$ is one of the canonical momenta (as is always possible, thanks to Darboux's theorem~\cite{ArnoldBook}) \index{Darboux's theorem} the constraint surface is the hyperplane
$p_1 \approx 0$  shown in Fig.~\ref{GaugeConstraint}. But $p_1$ also plays  the role of the \emph{generator} of gauge transformations, which happen to be the translations in the $q_1$ direction:
through the Poisson bracket it defines a vector field on phase space $\{ p_1 , \cdot \} = \frac{\partial}{\partial q_1}$,  which is 
parallel to the $q_1$ axis (see Fig.~\ref{GaugeConstraint}). This vector field generates infinitesimal transformations on phase space (translations in the $q_1$ direction), and its integral curves are the \emph{gauge orbits} of the transformations. All the points on 
these curves are gauge-equivalent (they are related by gauge transformations: they have different representations but the same physical content). Moreover, the vector field $\frac{\partial}{\partial q_1}$
is parallel to the constraint surface $p_1 \approx 0$ by construction, and its integral curves lie on it. The physical meaning of 
a gauge constraint $\chi = p_1$ is that the $q_1$ coordinate is unphysical, like the non-gauge-invariant part of the electromagnetic
potentials ${\bf A}$ and $\varphi$, or like the coordinates  of the centre of mass of the whole Universe. 

\begin{SCfigure}
  \centering
\caption[The concept of a gauge fixing]{
The concept of gauge-fixing surface: the variable $q_1$ is unphysical and its value can be taken
arbitrarily, therefore we might choose a conventional value for $q_1$, to be determined by the value
of all the other phase-space variables, $q_1 = q_1(q_2,p_2,\dots)$. One way to obtain this is to intersect the
constraint surface $p_1 \approx 0$ with another surface, $\xi(q,p) \approx 0$, such that it is never parallel
to $p_1 \approx 0$ or, at the intersection, `runs along' the gauge orbits (represented by dashed lines on the constraint surface).
}
\label{GaugeFixing}
  \raisebox{0.025\textheight}{\includegraphics[height=0.3\textheight]{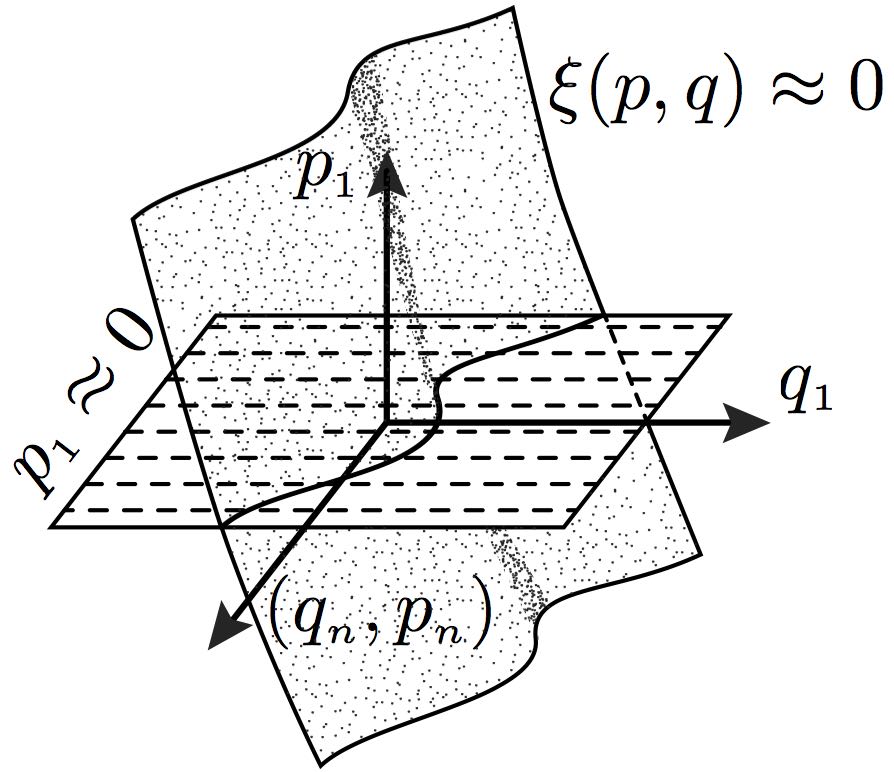}}
\end{SCfigure}

Since the $q_1$ coordinate is not physical, we can assign it any value along the solution without changing anything physical.
It is often useful (and necessary in quantum mechanics) to fix the value of $q_1$ by some convention. The standard
way of doing it is by choosing a \emph{gauge fixing:} we specify the value of $q_1$ as a function of the other variables, $q_1= q_1(q_2,p_2,\dots)$. This corresponds to intersecting the constraint surface $p_1 \approx0$ with another surface $\xi (p,q) \approx 0$ that specifies an intersection submanifold $\{ p,q \, ~\text{\it s.t.}~ \chi \approx 0 ,\xi \approx 0\}$ (see Fig.~\ref{GaugeFixing}). The gauge fixing should specify the gauge without ambiguity: it has to form a proper intersection with $p_1 \approx 0$, and therefore cannot be parallel to it where they intersect. Moreover, at its intersection with the constraint surface $\chi \approx 0$, the gauge-fixing surface $\xi \approx 0$ cannot `run along' (be tangent to) any of the gauge orbits: in that case there would be more than
one value of $q_1$ that corresponds to the same value of $q_2,p_2,\dots$. These two conditions define a good gauge-fixing surface.
For details on constrained Hamiltonian systems and gauge theories, see Sec.~\ref{CrashCourseInDiracsAnalysis}.

\subsubsection*{GR as a constrained Hamiltonian theory}

Arnowitt, Deser and Misner (ADM) formulated GR in the Hamiltonian language. They foliated spacetime into
a stack of spatial hypersurfaces and split the 4-metric $g_{\mu\nu}$ into a spatial part $g_{ij}$ and
four additional components $g_{0i}$ and $g_{00}$. The spatial metric components $g_{ij}$ represent
the canonical variables, and their momenta $p^{ij}$ are related to the extrinsic curvature of the 
spatial hypersurface with respect to its embedding in spacetime. The $g_{0i}$ and $g_{00}$ components (or better some combinations
thereof) enter the action without time derivatives, and are therefore Lagrange multipliers. They are associated with four \emph{local} constraints (meaning one constraint per spatial point). These constraints are
the so-called `superhamiltonian' $\mathcal H$ and `supermomentum' $\mathcal H^i$ constraint. 
Here I will call them the `Hamiltonian' and the `diffeomorphism' constraint.
The diffeomorphism constraint admits a simple geometrical interpretation: its vector flow sends configuration variables into themselves (one says it generates `point transformations' $\mathcal H^i  : g_{ij} \rightarrow g_{ij}$),
and there is no doubt about its being a gauge constraint. 

For the Hamiltonian constraint things aren't that simple:
it is quadratic in the momenta, and its vector flow does not admit the interpretation of a point transformation (it sends $g_{ij}$'s into both $g_{ij}$'s and $p^{ij}$'s). There is a large literature on the problem of interpreting $\mathcal H$. If it is interpreted as a gauge constraint, one would end up with the paradoxical conclusion that the dynamical evolution of GR is just a gauge transformation. There are also huge problems with the definition of what people call Dirac observables: quantities whose Poisson brackets with all the first-class constraints vanish on the constraint surface (meaning they must be invariant under the associated gauge transformations). In GR's case, that definition would lead to observables which are constants of motion  and don't evolve (`perennials', as Kucha\v{r}  \index{Kucha\v{r} perennial} called them~\cite{KucharReview}). Kucha\v{r} advocated a different notion of observables, namely ones which are only required to be invariant under diffeomorphisms. These would evolve, but they are too many: they would depend on \emph{three} polarizations of gravitational waves, while it is widely agreed that gravitational waves have \emph{two} physical polarizations.

The fact that $\mathcal H$ is quadratic in the momenta also causes major problems in its quantization. It leads to the notorious `Wheeler--DeWitt equation', for which there are many unsolved difficulties, above all its `timelsss' nature, but also ordering ambiguities and coincidence limits. The ADM formulation of GR is detailed in Sec.~\ref{ADM-WDW}, and the problems with this theory which lead to the introduction of SD
are explained at the end of Sec.~\ref{RWRsection} and in Sec.~\ref{YorksMethodSection}.

\begin{figure}[t!]
\begin{center}\includegraphics[width=0.3\textwidth]{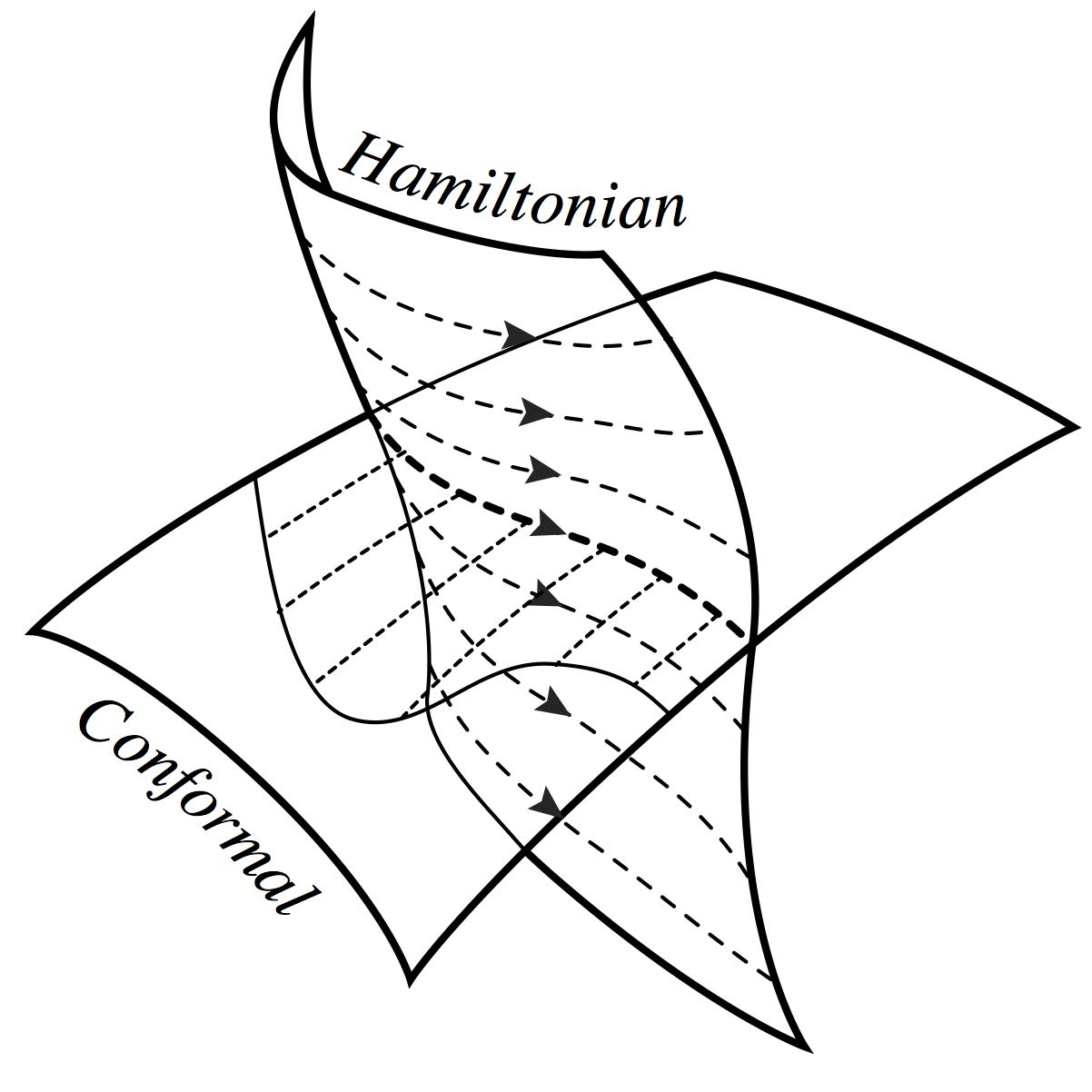}
\end{center}
\caption[The iconic diagram of Shape Dynamics]{
A schematic representation of the phase space of GR. In it, two constraints coexist, which
are good gauge-fixings for each other and are both first-class with respect to
the diffeomorphism constraint. One is the Hamiltonian constraint and the other is the conformal
(Weyl) constraint. The Hamiltonian constraint is completely gauge-fixed by the conformal constraint except for a single residual global constraint. It Poisson-commutes with the conformal constraint
and generates a vector flow on the Hamiltonian constraint surface  (represented in the figure), which is parallel to the conformal constraint surface. This vector flow generates the time evolution of the system in the
intersection between the two surfaces. Any solution can then be represented in an arbitrary
conformal gauge by lifting it from the intersection to an arbitrary curve on the conformal constraint
surface. All such lifted curves are gauge-equivalent solutions of a conformal gauge theory with conformally-invariant Hamiltonian.
}
\label{Surfaces}
\end{figure}

As illustrated in Fig.~\ref{Surfaces}, SD is based on the identification of the part of $\mathcal H$ which is not associated with a gauge redundancy and takes
it as the generator of the dynamics. The rest of $\mathcal H$ is interpreted as a gauge-fixing for another constraint $\mathcal C$. This constraint is linear in the momenta and generates genuine gauge transformations, constraining the physical degrees of freedom to be two per point.

\subsubsection*{Not every constraint corresponds to gauge redundancy}

That this is the case is pretty obvious: think about a particle constrained on a sphere or a plane, \emph{i.e.}, a holonomic constraint. Such a constraint obviously has nothing to do with gauge redundancy. However, there are constraints which Dirac \cite{DiracHamiltonianDynamics,dirac:lectures} argued can always be related to gauge symmetries: they are the so-called `first-class' constraints.
Being first-class means that they close an algebra under Poisson brackets with each other and with the Hamiltonian of the system. If that is the case, Dirac showed that one has freely specifiable variables in the system, one for each first-class constraint, and changing these variables does not change the solutions of the theory. But Barbour and Foster \cite{BF-DiraTheorem} have pointed out that the premises under which Dirac obtained his result do not hold in the important case in which the canonical Hamiltonian vanishes. In that case the Hamiltonian is just a linear combination of constraints, but that doesn't prevent the theory from having sensible solutions.
The solutions will be curves in phase space, and will still possess one freely specifiable variable for each constraint. But one of these redundancies will not change the curve in phase space: it will just change its \emph{parametrization}. Therefore one of the first-class constraints of the system will not be related to any gauge redundancy: there is not an associated unphysical `$q_1$' direction, like in the example above.
This counterexample to Dirac's statement is very important because it is realized in the theory we care about the most: General Relativity. One of the (many) constraints of GR \emph{should not} be associated with gauge redundancy. The Barbour--Foster argument is explained at the end of Sec.~\ref{CrashCourseInDiracsAnalysis}.

\subsubsection*{SD reinterprets  $\mathcal H$  as a gauge-fixing of conformal symmetry}

Shape Dynamics identifies another constraint surface $\mathcal C \approx 0$ in the phase space of GR, which is a good gauge-fixing for the Hamiltonian constraint. This gauge-fixing, though, happens to be also a gauge symmetry generator. It generates conformal transformations (\ref{WeylTransformations}) of the spatial metric, with the additional condition
that these transformations must preserve the total volume of space $V = \int \d^3 x \sqrt g$. The constraint $\mathcal C$, in addition, happens to close a first-class system with the diffeomorphism constraint $\mathcal H^i$, therefore  it is a matter of opinion whether it is $\mathcal C$ that gauge-fixes the system $(\mathcal{H},\mathcal{H}^i)$ or it is $\mathcal H$ which gauge-fixes $(\mathcal C,\mathcal H^i)$. If the real physics only lies in the intersection between  $\mathcal C \approx 0$ and  $\mathcal H \approx 0$ (which is the big assumption at the basis of SD, and doesn't hold if spacetime is assumed as an axiom), then the logic can be reversed and the Hamiltonian constraint can be interpreted as a special gauge-fixing for the conformal constraint. Then gravity can be reinterpreted as a gauge theory of conformal transformations, which admits a gauge-fixing that is singled out by some special properties. These properties, as I will show, have to do with the fact that it gives a `natural' notion of scale and proper time, which agree (most of the times) with those measured by physical rods and clocks.

\subsubsection*{SD's Hamiltonian constraint}

$\mathcal H$ and $\mathcal C$ do not entirely gauge-fix each other: there is a single linear combination of $\mathcal H(x)$ which
is first-class w.r.t. $\mathcal C$. This linear combination, $\mathcal H_\st{global} = \int d^3 x \, N_\st{CMC}(x) \, \mathcal H(x)$, is a single global constraint whose vector flow is parallel to both the $\mathcal C \approx 0$ and the $\mathcal H \approx 0$ surfaces on their intersection. This vector flow generates an evolution in the intersection: it has to be interpreted as the generator of time evolution. It is the part of our constraints which is not associated with a gauge redundancy and is instead associated with time reparametrizations of the solutions of the theory.

\begin{figure}\label{SDConstraintsScheme}
\begin{center}
\includegraphics[width=0.4\textwidth]{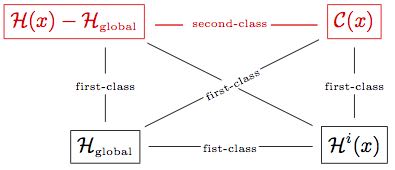}
\end{center}
\caption[Scheme of the constraints of SD and GR]{Scheme of the constraints of GR and of SD. GR's Hamiltonian constraint $\mathcal H$ has
been split into the global part $\mathcal H_\st{global}$ which is first-class w.r.t. the conformal constraint $\mathcal C$ and the part that is purely
second class, $\mathcal H (x)  - \mathcal H_\st{global}$. This second-class system admits two first-class subalgebras: the lower-left triangle, which
constitutes the constraint algebra of GR, and the lower-right triangle, making the constraint algebra of SD.}
\end{figure}

\subsubsection*{The `Linking Theory'}

SD pays a price for its conceptual simplicity: the generator of the evolution $\mathcal H_\st{global}$ contains the solution to a 
differential equation, $N_\st{CMC}$, and therefore is a nonlocal expression. But one can recover a local treatment by enlarging
the phase space. SD can in fact be considered as one of the possible gauge fixings of a first-class theory which is local (its constraints
are local) and lives in a larger phase space than that of GR. This phase space is obtained from that of GR by adjoining a scalar field $\phi$ and
its conjugate momentum $\pi$. The larger theory (called `Linking Theory') is defined by the constraints of GR, $\mathcal H$ and $\mathcal H^i$,
but expressed in terms of (volume-preserving-)conformally-transformed metrics $e^{4 \hat \phi} g_{ij}$ and momenta $e^{-4 \hat \phi} \left[ p^{ij}
- \frac 1 3(1- e^{6 \hat \phi})  \sqrt g g^{ij}\int p /V \right] $, where $\hat \phi = \phi - \frac 1 6 \log (\int \d^3 x \sqrt g \exp(6\phi) / V) $ and $V = \int \d^3x \sqrt g$. In addition, one has a modified conformal constraint which includes a term that transforms $\phi$. The new constraint is $\mathcal Q = 4 \,\pi - \mathcal C$ and generates simultaneous translations of $\phi$ and (volume-preserving) conformal transformations of $\pi$, so that the combination $e^{4 \hat \phi} g_{ij}$, is left invariant.
The constraint $\mathcal Q$ is now first-class with respect to $\mathcal H$ and $\mathcal H^i$. By completely gauge-fixing $\mathcal Q$, for
example with the condition $\phi \approx 0$, one obtains GR. On the other hand, one can use a different gauge-fixing, namely $\pi \approx 0$,
which is first-class with respect to $\mathcal Q$, but gauge-fixes $\mathcal H$ almost entirely, leaving only the global part $\mathcal H_\st{global}$ untouched.

One can then work with the Linking Theory, where all the equations of motion and constraints are local (apart from the dependence
on the total volume), and work out the solutions in this framework. As long as the solution is compatible with the gauge-fixing
$\pi \approx 0$, it's a legitimate SD solution.

All the details of the SD construction can be found in Sec.~\ref{ExtendedRWR} and in part~\ref{SDpart}.

\subsubsection*{The present status of SD}

Shape Dynamics' greatest hope is to provide a new approach to quantum gravity, based on a sum over histories of conformal 3-geometries. Such an approach is so far unexplored: 3D conformal symmetry plays a role in some  quantum gravity proposals, \emph{e.g.} Ho\v{r}ava--Lifshitz gravity or Causal Dynamical Triangulations, in which there is evidence that the theory admits conformally-invariant fixed points. However such symmetries are only asymptotic and do not characterize the physically-relevant regimes of these theories. In particular it seems that a common feature of quantum gravity models is the generation of additional dynamical degrees of freedom at the quantum level, \emph{e.g.} the scalar mode in  Ho\v{r}ava--Lifshitz gravity. A quantum gravity theory compatible with the principles of SD should, presumably, keep only two dynamical degrees of freedom all the way through its renormalization group orbit. Such a proposal seems in contradiction with the fact that quantum mechanics introduces a scale in physics through $\hbar$, and therefore break conformal invariance. This, however, shouldn't be taken as a sacred fact of nature: the fact that $\hbar$ is dimensionful can be a consequence of the fact that we measure its effects in a subsystem of the universe, and its scale might admit a relational expression in terms of the state of the matter in the universe. Indeed in~\cite{barbour2013gravitational} we formulated a toy-model of scale-invariant non-relativistic quantum mechanics, in which the fundamental ontology is that of a wavefunction evolving on shape space. The role of $\hbar$ is played by a dimensionless quantity which is related to the `spreading' of the wavefunction on shape space. Such quantity is intimately related to the particle equivalent of York time, \index{York time} the dilatational momentum $D$ (in appropriate units $\hbar$ and $D$ have the same dimensions). The  $\hbar$ which we use to describe phenomena we observe in the laboratory emerges, in such a model, when we concentrate on subsystems of the universe and model them semiclassically as quantum fluctuations around a classical solution. In this framework, doubling the size of the universe and simultaneously doubling $\hbar$ has no effect.

The chance of exploring an uncharted road to quantum gravity is one of the main motivations behind Shape Dynamics. However, so far, the work of the researchers in the field has been (mostly) limited to the classical theory. The reason for this is that before going quantum, we need to learn from scratch how to do physics without spacetime and relying only on the 3-dimensional conformal geometry of space.

The most important conceptual point that should be clarified at the classical level is whether SD is equivalent or not to GR. This should be investigated in those situations in which GR develops singularities, or when spacetime ceases to be CMC-foliable. The first evidence that SD can do better than GR in singular situation was~\cite{Budd:2011er}, in which Budd and Koslowski studied homogeneous cosmological solutions in 2 spatial dimensions with the topology of a torus. In this case the dynamics is that of the `Bianchi I' model, and inevitably it reaches `crushing' singularities in which $\det g \to 0$. In such situations the spacetime description ceases to make sense. However one can evolve the conformal geometry of space through those singularities, simply by requiring continuity of the shape degrees of freedom.
This result could have simply been a fluke of a lower-dimensional toy model, however in the recent~\cite{ThroughTheBigBang} we studied the much more complicated 3-dimensional `Bianchi IX' model of homogeneous cosmology (this model is described in detail in Sec.~\ref{SecBianchiIX}). In~\cite{ThroughTheBigBang}  we showed that at the singularity it is only the spatial volume and its conjugate momentum, the York time, which are singular. The conformal geometry at the singularity is degenerate, because it is flattened to a 2-dimensional object, but the shape degrees of freedom are not singular. The situation is analogue to that of a 3-body model in which the three particles go collinear: the triangle they describe is degenerate and its area is zero, but as shapes, the collinear configurations are perfectly regular. The shape of freedom can be continued through the singularity in a unique way just by requiring continuity, and on the other side the dynamics continues undisturbed, following the Bianchi IX equations of motion. In this way we end up joining two cosmological solutions of GR at the singularity: each half is an acceptable solution of GR and can be described as a spacetime, but the whole solution cannot. In this sense we proved that SD admits solutions which are not in GR, and it can do better than GR in dealing with singularities.

Another situation in which GR predicts singularities is black holes. Moreover, it is known that CMC foliations have a \emph{singularity-avoiding} property~\cite{Alcubierre,Gourg} in Schwarzschild's \index{Schwarzschild's spacetime} spacetime, so one could legitimately conjecture that the shape-dynamical description of black holes may be different from that of GR.\\
The first study on the subject was Gomes' paper~\cite{GomesBH} studying asymptotically flat, spherically symmetric vacuum solutions of ADM gravity in \emph{maximal slicing.}  This solution is derived and discussed in Sec.~\ref{AsymptoticallyFlatSD}. Interestingly, this solution does not have any singularity: it consists of a `wormhole' geometry with two asymptotically flat ends, and a `throat', that is, a sphere of minimal area. The geometry cannot support any concentric sphere (\emph{i.e.} spheres centred around the centre of symmetry) of area smaller than the throat. Of course I am now talking about the Riemannian geometry of spatial slices, which is not an observable property of Shape Dynamics: all of these geometries are conformally related to the 3-sphere with two piercings at the antipodes. However, this solution should be understood as a `background' carrying no dynamical degrees of freedom, over which matter perturbation can propagate. And the wormhole geometry should be what these matter perturbations experience in the limit in which backreaction can be ignored. After~\cite{GomesBH}, Gomes and Herczeg~\cite{Gomes:2013bbl} studied axisymmetric vacuum solutions corresponding to a maximal foliation of a portion of Kerr's spacetime which, again, does not cover the singularity. This solution too turned out to be `wormhole'-like. Recently Herczeg~\cite{Herczeg:2015mb} studied a few more solutions of ADM in maximal slicing, namely Rindler,  Reissner-N\"ordstrom and Van Stockum--Bonner solutions, and conjectured that the presence of a `throat', and the corresponding inversion symmetry around it (which exchanges the two asymptotic ends of the metric while leaving the throat invariant) are general properties of shape-dynamical black holes.

The series of works~\cite{GomesBH,Gomes:2013bbl,Herczeg:2015mb} on asymptotically flat maximally-sliced solutions did not attack the question whether the assumption of asymptotic flatness and maximal slicing is compatible with Shape Dynamics. A first problem was pointed out by me in~\cite{NoBirkhoff}, where I observed that in SD the assumption of spherical symmetry doesn't lead to a unique solution as in Birkhoff's theorem. There is a one-parameter family of solutions which depend on the expansion of matter at infinity. Only one of those solutions is Lorentz-invariant at the boundary, and it coincides with the `wormhole' found by Gomes in~\cite{GomesBH}. The others have different properties, and outside of an interval of the parameter, do not possess a `throat' anymore. I discuss this result in Sec.~\ref{AsymptoticallyFlatSD}.
More importantly, noncompact manifolds cannot be proper solutions of SD: they can at most be approximations to actual solutions. The solutions found in~\cite{GomesBH,Gomes:2013bbl,Herczeg:2015mb}
are acceptable approximations to physical situations in SD only if they can be shown to arise as the result of gravitational collapse of matter in a compact universe. For this reason in~\cite{ThinShellPaper1} we started the study of the gravitational collapse of the simplest form of matter: thin shells of dust. The solution of~\cite{GomesBH} can be shown to emerge as the result of the dynamical evolution of a shell if one assumes asymptotic flatness and zero expansion at the boundary. But these assumptions are still unphysical: one would like to study the actual case of a closed, \index{closed spacelike hypersurfaces} spherically symmetric universe with thin shells of dust. This is what I did, to the best of my present understanding, in~\cite{CompactThinShellsPaper}, which I reproduce (and expand) in Sec.~\ref{SphericallySymmetricSec}. The preliminary conclusion is that there is no evidence that the solution~\cite{GomesBH} forms. However I have the simplest dynamically closed model of gravitational collapse: a compact universe with some `spectator' matter (playing the role of `fixed stars') and a localized matter density which undergoes gravitational collapse. The solutions I find are exact and take into full account the back-reaction of matter on the metric degrees of freedom. My model shows, again, a departure of SD from GR: the ADM description in a CMC foliation breaks down at some point during the gravitational collapse (presumably when the system generates an event horizon), but the shape degrees of freedom seem unaffected by this breakdown, and can be continued smoothly through that point.

\newpage
~
\newpage

\part{Historical motivation} \label{PartHistorical}

\section{Newton's bucket} \label{SectionNewtonsBucket}

\subsection{The defects of the law of inertia}

Newton \index{Isaac Newton} based his  \emph{Principia} \cite{NewtonPrincipia} on the \emph{Law of inertia}\index{Law of inertia} (stated first by Galileo\index{Galileo Galilei}),
which he  made into the first of his three laws of motion:
\begin{quote} \it
A  body continues in its state of rest, or of uniform motion in a right line, unless it is compelled to change that state by forces impressed upon it.
\end{quote}
Assuming this law as a postulate, without first defining the notions of `rest', `uniform motion' and `right (or straight) line',  is inconsistent. 
In a Universe that is, in 
Barbour's words, like `bees swarming in nothing'~\cite{barbour:eot}, how is one to talk
about rest/uniform motion/straight lines? With respect to what?

The problem is that of establishing a notion of \emph{equilocality}\index{equilocality}: in an ever-changing Universe, what does it mean for an object to be 
at the same place at different times?

Newton anticipated these criticisms in the \emph{Scholium}\footnote{A `scholium' is an explanatory commentary.} \index{Newton's `Scholium'} at the beginning of the \emph{Principia}. \index{Newton's \emph{Principia}} He claims that rest/uniform motion/straight lines have to
be defined with respect to absolute space and  time:
\begin{quote} \it
I. Absolute, true, and mathematical time, of itself, and from its own nature, flows equably without relation to anything external, and by another name is called duration: relative, apparent, and common time, is some sensible and external (whether accurate or unequable) measure of duration by the means of motion, which is commonly used instead of true time; such as an hour, a day, a month, a year.
\vspace{6pt}

II. Absolute space, in its own nature, without relation to anything external, remains always similar and immovable. Relative space is some movable dimension or measure of the absolute spaces; which our senses determine by its position to bodies; and which is commonly taken for immovable space; [\dots] 
\end{quote}
These definitions make the \emph{Principia} a logically consistent system, which however relies on the scientifically
problematic concepts of absolute space and time. These affect the motion of material bodies in a spectacular
way - through the law of inertia - but aren't affected by them. Despite these shaky grounds, Newton's
Dynamics proved immensely successful over more than two centuries, and this tended to hide its foundational
problems. 

\subsection{Leibniz's relationalism}

The chief advocate for an alternative in Newton's time was Leibniz. In a correspondence with Clarke  \index{Gottfried Wilhelm Leibniz} \cite{LeibnizClarke} (writing basically on behalf of Newton)  he advocated
a \emph{relational understanding of space}, in which only the observable \emph{relative distances}  between bodies
play a role
\begin{quote} \it
I will show here how men come to form to themselves the notion of space. They consider that many things
exist at once and they observe in them a certain order of co-existence [\dots] This order is their \emph{situation}
or distance.
\end{quote} \index{Leibniz--Clarke correspondence}
We can say that Leibniz lost the argument with Clarke, mainly because he failed to provide a concrete, viable
way of implementing a relational mechanics. We'll see that he had no hope of doing that,
because much more sophisticated mathematics is  needed than was available at the time.

Leibniz's main argument against absolute space and time -- that they are not observable -- was actually anticipated
and countered by Newton in the \emph{Scholium}. He claimed one could prove the existence of absolute circular motion in his famous `bucket experiment' described as follows:
\begin{figure}[t]
\begin{center}\includegraphics[width=0.3\textwidth]{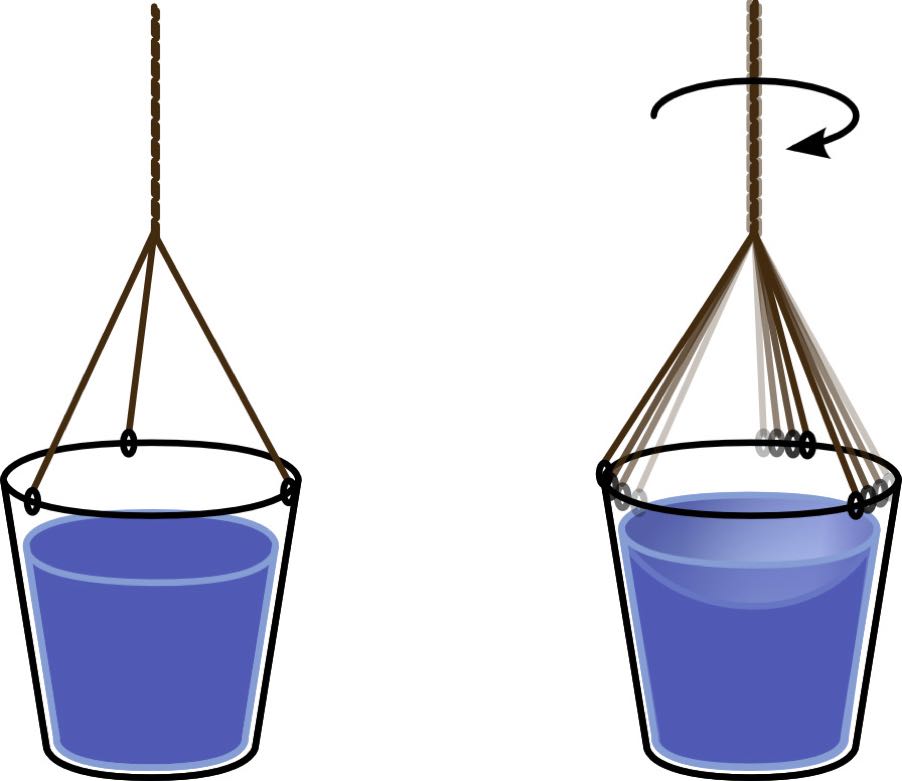}
\end{center}
\caption[Newton's bucket experiment]{Newton's bucket experiment: \index{Newton's bucket}
on the left, both the bucket and the water are at rest with respect to
the room. On the right, they are both rotating, and they have no motion relative
to each other. Then to explain the concave shape of the water in the picture
on the right, one has to invoke something external to the relative motions
of water and bucket.}
\label{NewtonsBucket}
\end{figure}
\begin{quote} \index{Newton's `Scholium'}
If a vessel, hung by a long cord, is so often turned about that the cord is strongly twisted, then filled with water, and held at rest together with the water; after, by the sudden action of another force, it is whirled about in the contrary way, and while the cord is untwisting itself, the vessel continues for some time this motion; the surface of the water will at first be plain, as before the vessel began to move; but the vessel by gradually communicating its motion to the water, will make it begin sensibly to revolve, and recede by little and little, and ascend to the sides of the vessel, forming itself into a concave figure [\dots] This ascent of the water shows its endeavour to recede from the axis of its motion; and the true and absolute circular motion of the water, which is here directly contrary to the relative, discovers itself, and may be measured by this endeavour. [\dots] And therefore, this endeavour does not depend upon any translation of the water in respect to ambient bodies, nor can true circular motion be defined by such translation. 
\end{quote}\index{Newton's `Scholium'}
In this passage Newton was, besides making a serious argument, covertly attacking the philosophy of Descartes \index{Ren\'e Descartes} \cite{JulianDiscoveryOfDynamics}, who had died in 1650 but whose ideas were still widely accepted. Descartes had declared position and motion to be relative and in particular had said that the `one true philosophical position' of a given body is defined by the matter immediately next to it. In the bucket experiment, it is obvious that the relative state of the water and the
sides of the bucket is \emph{not} the cause of the behaviour of the water: both at the beginning and at the end
they are at rest w.r.t. each other, but in one case the surface of the water is flat and in the other it's curved. Therefore
the curvature of the water must be caused by something else, which Newton identifies with the circular motion
w.r.t. absolute space. This argument largely settled the issue in the mind of most scientists until the end of the XIX
century.

\subsection{The \emph{Scholium} problem}

Newton was aware of the difficulties inherent in his tying the first law of motion to unobservable entities
like absolute space and time. Towards the end of the \emph{Scholium} he comments \index{Newton's `Scholium'} \index{Newton's \emph{Principia}}
\begin{quote} \it
It is indeed a matter of great difficulty to discover, and effectually to distinguish, the true motions of particular bodies from the apparent; because the parts of that immovable space, in which those motions are performed, do by no means come under the observation of our senses.
\end{quote}
But he believed that experiments like that of the bucket provided a handle on this problem:
\begin{quote} \it
Yet the thing is not altogether desperate; for we have some arguments to guide us, partly from the apparent motions, which are the differences of the true motions; partly from the forces, which are the causes and effects of the true motions.
\end{quote}
Newton then concludes the \emph{Scholium} with the grand words 
\begin{quote} \it
But how we are to obtain the true motions from their causes, effects, and apparent differences, and the converse, shall be explained more at large in the following treatise. For to this end it was that I composed it.
\end{quote}
Thus, he considers that deducing the motions in absolute space from the observable relative motions to be \emph{the fundamental
problem of Dynamics}, and claims that he composed the \emph{Principia} precisely to provide a solution to it.
Remarkably, he never mentions this \emph{Scholium problem} again in the \emph{Principia} and certainly doesn't solve it! What is more, nobody else attempted to for very nearly 200 years.

\vspace{12pt}
\noindent
\fbox{\parbox{0.98\linewidth}{
{\bf Further reading:} Newton's ``Principia'' \cite{NewtonPrincipia}, The Leibniz--Clarke correspondence \cite{LeibnizClarke}, Barbour's ``The Discovery of Dynamics'' \cite{JulianDiscoveryOfDynamics}.
}}\vspace{12pt}

\section{Origins of the Mach--Poincar\'e Principle}\label{SectionMachPoincarePrinciple}

\subsection{Tait's partial solution of the Scholium problem}
\label{TaitsScholiumSolution}

In 1884, Tait \cite{Tait} provided a solution to the Scholium problem in the simplest case of non-interacting, i.e., inertial, point masses. \index{Peter Guthrie Tait}
I give here my personal account of it.

Say we have $N$ point masses that don't exert any force upon each other (like a perfect gas), and we
are given only the succession of the relative distances\footnote{One could
include relativity of \emph{scale} in the picture.
If the whole universe consists only of those $N$ point particles, there is no external ruler with which
we can measure sizes, so there is absolutely no meaning in concepts like `the size of the universe'. Then, in this case,
the only truly  observable things in such a universe are $r_{ab}/r_{cd}$, the \emph{ratios} between $r_{ab}$'s.} between those particles $r_{ab}$, $a,b=1,\dots,N$, taken at
some unspecified intervals of time.
Those are $N  (N-1) / 2$ numbers, but they aren't all independent of each other. Newton would
say that the absolute space in which they move is three-dimensional, and this constrains the
$r_{ab}$'s to satisfy certain relationships. The simplest one is the triangle inequality between
triplets: $r_{13} \leq r_{12} + r_{23}$. Then, if $N\ge 5$, there are true equalities they have to satisfy, which
reduce their independent values to just $3N-6$.\footnote{$3N-7$ if we include relativity of size.} To convince yourselves about that just consider
that if the particles were represented as points in $\doublestroke{R}^3$ they would have $3N$ coordinates, but two
configurations that are related by a rigid translation of the whole system (3 degrees of freedom),
or a rigid rotation (3 further d.o.f.'s) would be equivalent, because
they would give the same $r_{ab}$'s. So we're down to $3N-6$ degrees of freedom.\footnote{With rescalings (1 d.o.f.) we go down by a further degree of freedom to $3N-7$.}
This is huge data compression, from something quadratic in $N$ to something linear.
M. Lostaglio \cite{MatteoPrivateCommunication} convinced me that this data compression should be taken as 
an \emph{experimental fact}, to which our senses are so used that it has become an intuitive
truth.  Geometry, in this case three-dimensional Euclidean geometry, is a synthesis of all those
relationships between observables.

To come back to the \emph{Scholium} problem, we have to determine the unobservable
positions in absolute space of the $N$ particles. These will be $N$ vectors ${\bf r}_a \in \doublestroke{R}^3$
($3N$ numbers), which must be determined from the $3N-6$ independent observables that can be extracted from $r_{ab}$
given as `snapshots' obtained at certain unspecified times. Following Tait, we assume for the moment an external scale is given. Tait's solution exploits the assumption that the particles are not interacting, 
and therefore according to Newton's first law they will move uniformly in a straight line in absolute space.
In a more modern language one is looking for the determination of an \index{inertial frame of reference} \emph{inertial frame of
reference},\footnote{In 1885, Lange \cite{Lange}, using a construction principle far more complicated than Tait's, coined the 
expressions `inertial system', in which bodies left to themselves move rectilinearly, and `inertial time scale', relative to which they also move uniformly. The two concepts were later fused into the notion of an inertial frame of reference.} in which the first law holds. By Galilean relativity, there will be infinitely many such inertial frames of reference all related to each other by Galilean transformations. \index{Galilean relativity} Tait's algorithm exploits this freedom and is as follows:

\begin{enumerate}
\item Fix the origin at the position of particle 1: then ${\bf r}_1 = (0,0,0)$.

\item Fix the origin of time $t=0$ at the instant when particle 2 is closest
to particle 1. At that instant, call $r_{12} = a$.

\item Orient the axes so that $y$ is parallel to the worldline of particle 2 and $x$ is orthogonal to it, so that $x_2 = a$ and $z_2 = 0$.

\item Set the unit of time with the motion of particle 2: ${\bf r}_2 = (a,t,0)$ (using the inertial motion of a particle as a clock is an idea due to Neumann \cite{Neumann}).

\item The motions of the remaining $N-2$ particles remain unspecified.
All one knows is that they will move along straight lines uniformly with respect to the time $t$ read by Neumann's inertial clock.
Their trajectories will therefore be ${\bf r}_a = (x_a,y_a,z_a)
+ (u_a,v_a,w_a) \, t$, where $(x_a,y_a,z_a)$ and $(u_a,v_a,w_a)$
together with $a$ will be $6N-11$ unspecified variables.
\index{inertial frame of reference}

\end{enumerate}

The conclusion we can draw is that one needs $6 N-11$ observable data to construct an inertial frame. Each `snapshot' we are given contains only $3N-7$ independent data ($3N-6$ independent 
relative data minus the time at which each snapshot has been taken, which is unknown). Therefore two snapshots aren't enough. 
They provide only $2(3N-7) = 6N -14$ data. We're short of three numbers in order to fix $a$, $(x_a,y_a,z_a)$ and $(u_a,v_a,w_a)$. We need a third snapshot\footnote{In fact a fourth as well if $N=3~\textrm{or}~4$. With only two particles, nothing can be done. Relational dynamics requires at least three particles. The Universe certainly meets that requirement!} to determine the
inertial frame. This is especially puzzling if one considers that $N$ can be as large as wanted, say a billion, but one would always need just three additional quantities.\footnote{If the scale is not given, then each snapshot carries only $3N-8$ independent
data. But the unspecified variables in this case aren't as many as
before: we are free to fix the scale so that $a=1$, and therefore
we only need to find $(x_a,y_a,z_a)$ and $(u_a,v_a,w_a)$ which are
$6N-12$. Two snapshots then fall short of $4$ data.}

The additional $3$ numbers that we need to specify through a
third snapshot are the direction of the rotation vector
of the system (which accounts for $2$ degrees of freedom)
and the ratio of the relative rotation to the expansion of the complete system as captured in the two snapshots. The point is that the values of $r_{ab}$ in the two snapshots are unaffected by a rotation of arbitrary magnitude of one snapshot relative to the other about an arbitrary direction. (Since the scale is assumed given and the centroid of the points can be determined in each snapshot, the overall expansion can be deduced. The difficulty is in the relative rotation.\footnote{In the scale-invariant case, we do not know the direction of the relative rotation vector, the rotation rate and the expansion rate. These are the four missing quantities. Note that the particle masses do not enter the law of inertia and can only be deduced in the presence of interactions.}

For what follows, the important thing that emerges from Tait's analysis is not so much that Newton's Scholium problem can be solved but the fact that two `snapshots' are never enough to do it. This, and the number of extra data needed due to the factors just identified, remains true in the much more realistic case of, say, $N$ point particles known to be interacting in accordance with Newtonian gravity.\footnote{In fact, one more datum is needed to determine the ratio of the kinetic to the potential energy in the system at the initial instant. Also, to determine the mass ratios of the particles, $N-1$ extra data will be needed.} Of course, the task is immensely more difficult, but in principle it is solvable.

\subsection{Mach's critique of Newton} \label{MachsCritiqueOfNewtonSubsec}

Ernst Mach \index{Ernst Mach} was a great experimental physicist who was also convinced one needed to know the history of science in order to make real progress.\footnote{This is part of the
reason why I'm putting so much emphasis on these historical notes} Being true to that belief,
in 1883 he  wrote a book on the history of mechanics \cite{mach:mechanics} which later proved immensely 
influential in the development of Einstein's General Relativity (and more recently in that of
Shape Dynamics). In this book Mach criticizes Newton's absolute space and time.

\subsubsection*{Mach's critique of absolute time}

\index{temporal relationalism}

\begin{quote} \it
It is utterly beyond our power to measure the changes of things by time. 
Quite the contrary, time is an abstraction at which we arrive through the changes of things.
\end{quote}

\index{Richard Feynman}
Richard Feynman quipped ``Time is what happens when nothing else does.'' Even if meant humourously, this does rather well reflect a deeply-rooted, fundamentally Newtonian concept of time that is still
widespread today. Mach would have answered to this: ``If nothing happens, how
can you say that time passed?''
Feynman's words express a view that is still unconsciously shared by theoretical physicists,
despite being seriously questioned by GR. According to this view, in the words of 
Barbour \cite{JuliansEssayOnTime} ``in some given interval of true time the Universe could do
infinitely many different things without in any way changing that interval of time.''
The Machian point of view is that this is correct only if one speaks about a \e{subsystem}
of the Universe, like the Earth or our Solar System. In that case it is true that all
the matter on Earth could do many different things without changing the interval of time.
But the whole Universe has to do something in order for that interval of time to be defined.
It is actually the other way around: an interval of time is defined by the amount of change 
that the state of the Universe undergoes.

One has to ask the following question: what do we mean when we say that one second passed?
Thinking about it, it becomes pretty clear that we always refer to physical things having changed.
Be it the hand of a watch that has ticked once, or the Earth rotated by $1/240$th of a degree,
we always mean that something has changed. The modern definition of a second is 
``9~192~631~770 periods of the radiation corresponding to the transition between the two hyperfine levels of the ground state of the caesium 133 atom'' \cite{WikipediaSecond}. These are a lot of oscillations.
The notion of a second (or, in general, duration) is useful, but it is obviously not fundamental:
one can always do without it and make direct reference to comparisons of changes in the Universe.
Instead of saying a car ``travels a quarter of a mile in 5.78 seconds'' one can equivalently say
the car ``travels a quarter of a mile as the Earth rotates through $0.024^\circ$''. The same holds
for any other possible measurements of duration. One can imagine representing the history of the Universe as a curve on some space, each point of which represents a unique configuration (\e{e.g.}, one point might represent the car on the start line and the Earth with the Sun on the Zenith above Indianapolis, IN,
and another point might correspond to the car on the end line, and the Sun at an angle of $0.024^\circ$
from the Zenith). The Universe passes through all the points of the curve, which contain information
about everything: from the position of the car to the psychological state of our brains. Then the speed
at which this curve is traversed doesn't make any difference: in this representation, if the
car covers its quarter mile track twice as fast, also the Earth would rotate and our brain states will evolve 
at double the speed. Nothing measurable will have changed. What counts is the sequence of
states the curve passes through, not the \e{parametrization} of the curve. I have here anticipated the precise mathematical realization of temporal relationalism that will be advocated in this
\thistext{}: \e{the history of the Universe can be represented by a curve in some configuration
space, independently of its parametrization.}

\index{unparametrized curve in configuration space}

We'll see later that the relationalist approach allows us to completely dispose of Newton's
absolute time, and to describe Newtonian Dynamics as a \emph{reparametrization-invariant
theory}, \index{reparametrization invariance} where there is no notion of time at all, there is just a succession of configurations
without any notion of duration. Then the requirement that the equations of motion
take the particularly simple form of Newton's  second law allows us to deduce a notion
of time, called \index{ephemeris time} \emph{ephemeris time}, which is a sort of average over all the changes 
in the positions of the particles in the Universe. This is the realization of Mach's ``abstraction
of time from change''.

\subsubsection*{Mach's critique of absolute space}

\index{Ernst Mach}
Mach, like Leibniz and the other advocates of relationalism, was opposed to visible effects
admitting an invisible cause. This is why he disagreed with Newton's interpretation of the
bucket experiment. Being a good experimentalist, Mach's intuition told him that the thin bucket wall couldn't possibly be responsible for the macroscopic concavity of the water's surface, it should admit a different cause. And here came an observation
which relied on the knowledge of the centuries-old practice of astronomers: Newton's
laws had not been verified relative to absolute space but to the \index{fixed stars} \emph{fixed stars},
with the rotation of the Earth as a clock (`sidereal' time).
Since antiquity astronomers noticed that the fixed stars (those that, unlike the wandering planets, do not change their observed relative positions on the sky)
provide a reference frame with respect to which all the motions are simpler. This practice had proved to be fruitful to such an extent that, when a discrepancy was observed, it was  attributed to non-uniformity of the rotation of the Earth or failure of 
Newton's law of gravity as happened in the \index{Moon crisis  of the 1890s} 1890s when astronomers observed an 
anomalous acceleration in the motion of the Moon.\footnote{One possibility was 
that the Earth's rotation speed was decreasing, giving rise to a spurious 
apparent acceleration of the Moon. However, it was also suggested that gravity 
could be absorbed by matter. Then during eclipses of the Moon the presence of 
the Earth between it and the Sun would reduce the gravitational force acting 
on the Moon and could explain the discrepancy. This possibility was only ruled 
out definitively by observations in 1939, which showed that the planets also 
exhibited the same anomalous acceleration as the Moon. Meanwhile, the creation 
of General Relativity in 1915 by Einstein had explained the longstanding 
anomalous advance of Mercury's perihelion as due to failure of Newton's 
gravitational law.} 
 The possibility that the fixed stars didn't identify an inertial frame of reference was never taken seriously. But this is actually the case, even if only to a microscopic degree, as we'll see in a moment.

So Mach, in \emph{The Science of Mechanics}, claimed that the cause of the concavity of the water's surface in Newton's bucket could be due to the distant stars.
This would have remained a rather bizarre claim had it not been for the incredibly suggestive insight
that followed:
\begin{quote} \it
No one is competent to say how the experiment would turn out if the sides of the vessel increased in thickness and mass till they were ultimately several leagues thick.
\end{quote}
It is then clear that Mach had in mind a sort of interaction between distant \emph{massive}
objects, and the local inertial frames. This observation made a great impression
on several people, most notably Einstein, for whom it represented a major
stimulus towards the formulation of General Relativity.

\subsection{Hoffman's experiment} 

\begin{figure}[t]
\begin{center}\includegraphics[width=0.4\textwidth]{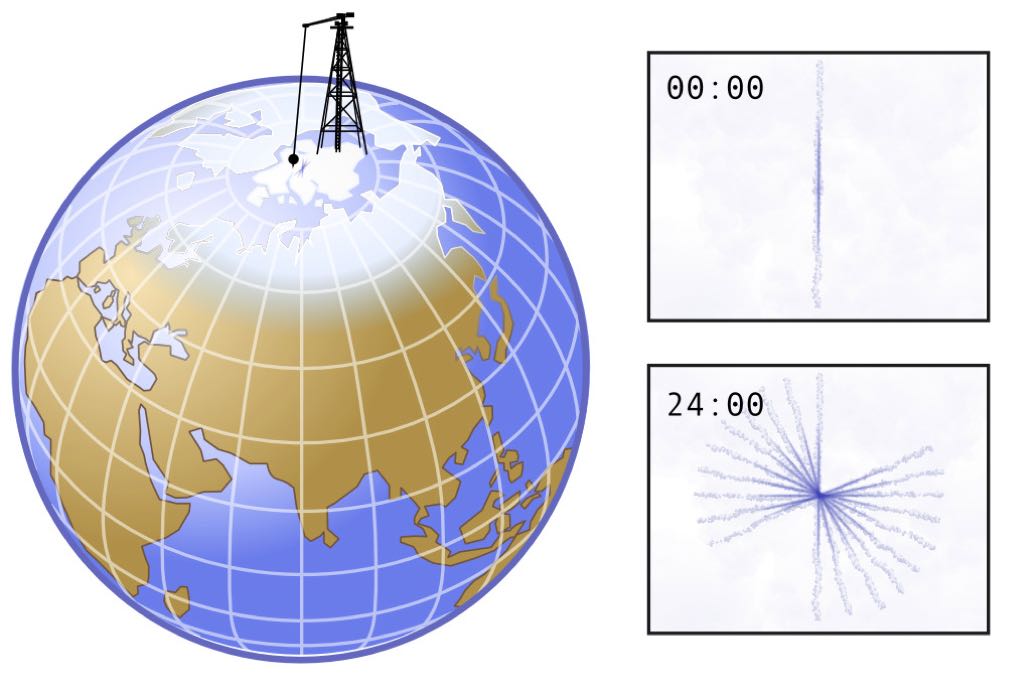}
\end{center}
\caption[Hoffman's experiment]{Hoffman's proposed experiment: according to Mach's conjecture, the plane of oscillation of a Foucault pendulum (for simplicity here at the North Pole) should be `dragged' by the Earth's rotation, and consequently should not, relative to the distant stars, complete a rotation in 24 hours, lagging
slightly behind.}
\label{Focault}
\end{figure}

In a 1904 book \index{Wenzel Hoffman} Wenzel Hoffman proposed a real experiment to test Mach's idea.\footnote{In fact, Hofmann's proposal had been anticipated in 1896 by the brothers \index{Benedict and Immanuel Friedlaender} Benedict and Immanuel Friedlaender (see \cite{NewtonsBucketConference}), one of whom actually did experiments with flywheels to text Mach's idea. I discuss Hoffmann's proposal because, more realistically, it uses rotation of the Earth.}
In the absence of buckets whose sides are ``several leagues thick'', he proposed to
use the Earth as the `bucket', and a Foucault pendulum as the water (let's put it
at one pole for simplicity (Fig.~\ref{Focault}).  If Newton is right, the rotation of the  Earth should
have no influence on the  plane of oscillation of the pendulum, which should
remain fixed with respect to absolute space. But if Mach is right, the large mass of the
Earth should `drag' the inertial frame of reference of the pendulum,
making it rotate with it very slowly. One would then see that, relative to the stars, the
pendulum would not complete a circle in 24 hours, but would take slightly longer.

\index{Focault's pendulum}

This experiment, as it was conceived, had no hope of succeeding. But it was actually
successfully performed in the early 2000's, in a slightly modified version. One just needs
the longest Earth-bound pendula that humans ever built: artificial satellites.
The \index{Lageos and Gravity probe B} Lageos and Gravity probe B satellites did the job, and they detected a rotation
of their orbital plane due to Earth's `frame-dragging' effect. After more than a century,
Mach has been proved right and Newton's absolute space ruled out.

\subsection{Poincar\'e's principle } \label{PoincarePrincipleSubsec}

The rather vague idea that Mach sketched in his \emph{Mechanics} was not
sufficiently precise to make an actual principle out of it. It has nonetheless been called
`Mach's principle' (Einstein coined the expression).
Einstein, despite being a strong advocate of Mach's principle for years, never found a satisfactory
formulation of it, and towards the end of his life even disowned it, claiming
that it had been made obsolete by the advent of field theory.\footnote{In his Autobiographical Notes \cite{EinsteinNote}, p.27, Einstein declares: ``Mach conjectures that in a truly rational theory inertia would have to depend upon 
the interaction of the masses, precisely as was true for Newton's other forces, a 
conception that for a long time I considered in principle the correct one. It presupposes 
implicitly, however, that the basic theory should be of the general type of Newton's 
mechanics: masses and their interaction as the original concepts. Such an attempt at a 
resolution does not fit into a consistent field theory, as will be immediately recognized.''}

In fact, a precise formulation of Mach's principle had been there in front of Einstein's
eyes all the time, but nobody recognized it for what it was until  Barbour
and Bertotti 
did in 1982 \cite{BarbourBertotti82}. This formulation is due to \index{Henri Poincar\'e} Henri Poincar\'e, in his
\emph{Science and Hypothesis} (1902). The delay in the identification of this important
contribution is due to the fact that Poincar\'e himself never thought of it as
a precise formulation of Mach's principle. In fact, although Poicar\'e can hardly have been unfamiliar with Mach's work, he did not cite it in \emph{Science and Hypothesis}. 

What Poincar\'e did ask was this: ``What precise defect, if any, arises in Newton's mechanics from his use of absolute space?''. The answer he gave can be understood in the light of 
our discussion of Tait's note: from \emph{observable} initial
configurations and their \emph{first derivatives} alone one cannot predict the future
evolution of the system.

The cause of this is, according to Poincar\'e, angular momentum. \index{angular momentum} There is no
way one can deduce the \e{total angular momentum of the system one is considering}
from the observable initial data $r_{ab}$ and their first derivatives alone.
This can be achieved by looking at the second derivatives, as was demonstrated
by Lagrange \index{Giuseppe Lodovico Lagrangia (Lagrange)} in 1772 \cite{Lagrange1772Paper} for the 3-body problem, but this remedy is 
unnatural, especially for the $N$-body problem when $N$ is large: one needs only 3 out of the $3N-6$ second derivatives.

Poincar\'e found this situation, in his words, ``repugnant'', but had to accept the
observed presence of a total angular momentum of the Solar System, and
renounce to further his critique. Interestingly, it didn't occur to Poincar\'e that
the Solar System isn't the whole Universe, it is instead a rather small part of
it as was already obvious in 1902.

Barbour and Bertotti therefore proposed what they called \e{Poincar\'e's principle}: The law of the Universe as a whole should be such that for it specification of initial inter-particle separations $r_{ab}$ and their rates of change should determine the evolution uniquely. There is a natural generalization of this law to dynamical geometry.

\vspace{12pt}
\noindent
\fbox{\parbox{0.98\linewidth}{
{\bf Further reading:} Mach's ``The Science of Mechanics'' \cite{mach:mechanics}, Barbour's essay ``The nature of time''  \cite{JuliansEssayOnTime}, Poincar\'e's ``Science and Hypothesis'' \cite{PoincareScienceAndHypothesis}. The great conference proceedings ``Mach's Principle: From Newton's bucket to quantum gravity'' \cite{NewtonsBucketConference}.
}}\vspace{12pt}

\newpage

\part{Relational Particle Dynamics} \label{PartRPM}

\section{Barbour--Bertotti best matching} \label{BBbestMatching}

As I said,  Julian Barbour and Bruno Bertotti 
in 1982 \cite{BarbourBertotti82} recognized 
that Poincar\'e had effectively given a mathematically precise formulation of Mach's principle and dubbed it the \index{Poincar\'e Principle} \e{Poincar\'e Principle} (Barbour now calls it the Mach--Poincar\'e Principle:) \index{Mach--Poincar\'e Principle}
\begin{quote} \it
Physical (or relational) initial configurations and their first derivatives alone should determine
uniquely the future evolution of the system.
\end{quote}
In the paper \cite{BarbourBertotti82}  Barbour and Bertotti  implemented this principle 
through what they called the intrinsic derivative and Barbour now calls \emph{best matching}, which allows one to establish a notion \index{equilocality}
of equilocality -- to say when two points are at the same position at different instants of time when only relational data are available.

\subsection{Best matching: intuitive approach}

\index{best matching}
The basic idea is this: say that you're an astronomer
who is given two pictures of three stars, taken some days apart (assume, for simplicity, that the stars
are fixed on a plane orthogonal to the line of sight; one could ascertain that, for example, by measuring their
redshifts). You're not given any information regarding the orientation of the camera at the time the two
pictures were taken. The task is to find an \emph{intrinsic} measure of change between the two pictures
which does not depend on the change in orientation of the camera. 

\begin{figure}[t]
\begin{center}\includegraphics[width=0.4\textwidth]{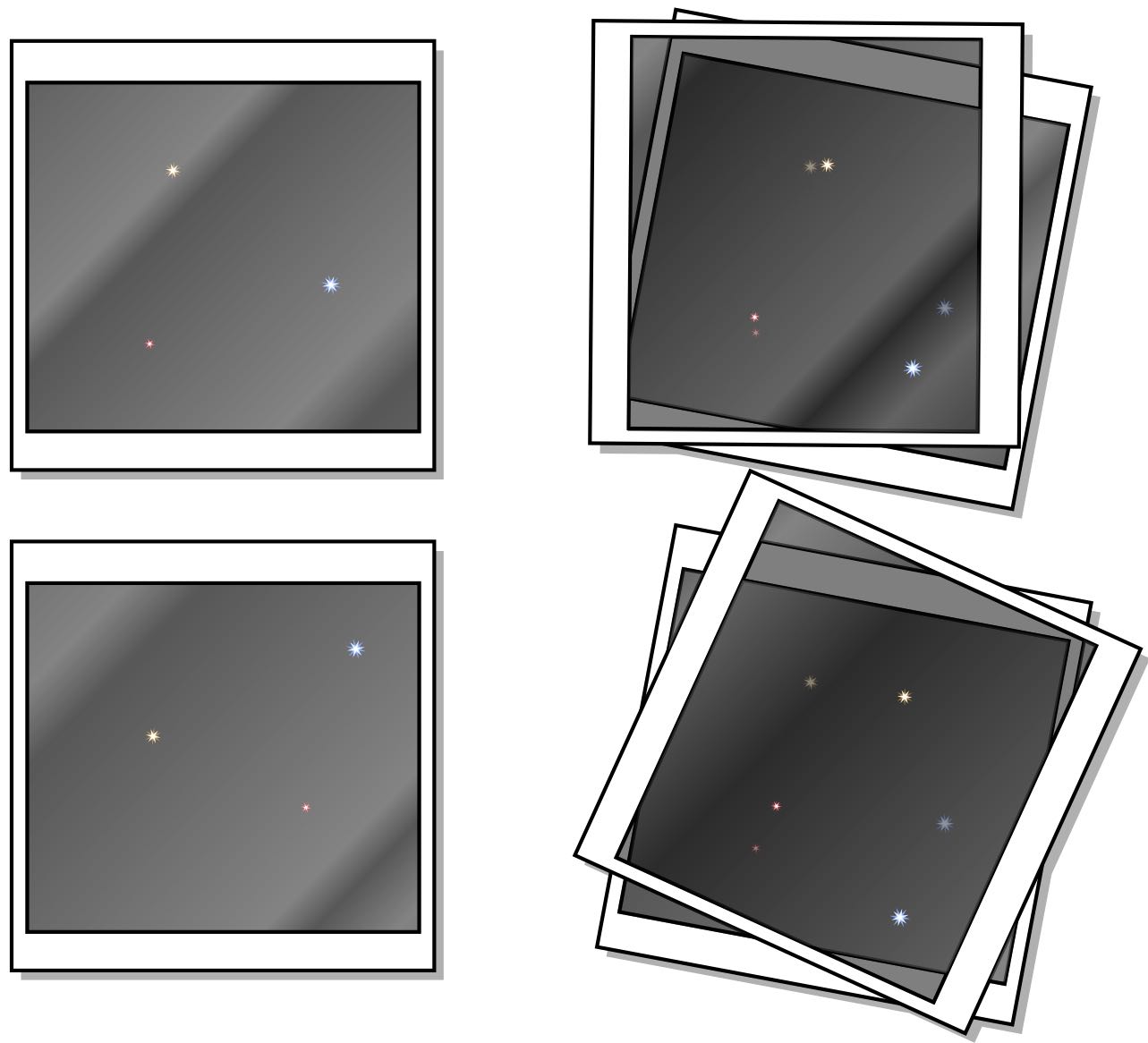}
\end{center}
\caption[Best matching illustrated with pictures of stars]{Different notions of \index{equilocality} equilocality: the two pictures on the left represent the same triple star system at different times. The relative positions of the stars have changed, and the astronomers are presented with the challenge of superposing picture 2 on picture 1 without having any reference background. On the right two such choices,
corresponding to different definitions of equilocality, \index{equilocality} are depicted.}
\label{PhotosOfStars}
\end{figure}

One is obviously only given the relative separations between the stars $r_{12}(t)$, $r_{23}(t)$, $r_{31}(t)$
at the two instants $t=t_i,t_f$. The task is to identify a Cartesian representation of the three particle positions ${\bf r}_a(t) = (x_a,y_a,z_a)(t) \in \doublestroke{R}^3$,
$a=1,2,3,$ at $t=t_i,t_f$  such that
\begin{equation}
\| {\bf r}_a(t_i) - {\bf r}_b(t_i)\| = r_{ab}(t_i)\,, ~~~ \| {\bf r}_a(t_f) - {\bf r}_b(t_f)\| = r_{ab}(t_f)\,,
~~~ \forall ~ a,b =1,2,3,
\end{equation}
where $\| {\bf r}_a \| = \| (x_a,y_a,z_a) \| =  |x_a|^2 +|y_a|^2 +|z_a|^2$.

Now make a tentative choice of Cartesian representation, ${\bf r}_a(t_i) = r^i_a$ and  ${\bf r}_a(t_f) = r^f_a$.
Notice that the Cartesian representation of the configuration at any instant consists of three arbitrary vectors of $\doublestroke{R}^3$, which can be repackaged into a single vector of $\doublestroke{R}^9$,
because of course $\doublestroke{R}^3\times \doublestroke{R}^3 \times \doublestroke{R}^3 = \doublestroke{R}^9$. So let's interpret the configurations at the initial and final instants as two vectors
of $\doublestroke{R}^9$: 
\begin{equation}
\begin{aligned}
q^i =& \oplus_{a=1}^3 {\bf r}_a^i = ({\bf r}_1^i,{\bf r}_2^i,{\bf r}_3^i) =(x_1^i,y_1^i,z_1^i,x _2^i,y_2^i,z_2^i,x_3^i,y_3^i,z_3^i)  \,,
\\
q^f =& \oplus_{a=1}^3 {\bf r}_a^i = ({\bf r}_1^f,{\bf r}_2^f,{\bf r}_3^f)=(x_1^f,y_1^f,z_1^f,x _2^f,y_2^f,z_2^f,x_3^f,y_3^f,z_3^f)  \,.
\end{aligned}
\end{equation}
We need a criterion to judge the `distance' between $q^i$ and $q^f$.  There is a
natural notion of distance on $\doublestroke{R}^9$ given by the Euclidean distance $\d : \doublestroke{R}^9 \times \doublestroke{R}^9 \to \doublestroke{R}$ which is just  the square root of the sum of the square of the difference between each component of the two $\doublestroke{R}^9$-vectors:
\begin{equation}\label{EuclDist}
\d \left(q^i_,q^f\right) = \left[ \sum_{\alpha=1}^9  (q^i_\alpha- q^f_\alpha)^2 \right]^{\frac 1 2} =\left( \sum_{a=1}^3 \| {\bf r}_a^i- {\bf r}_a^f \|^2 \right)^{\frac 1 2} \,.
\end{equation}
This expression depends both on the intrinsic change of configuration between the two triangles
and on the relative placement of picture 2 w.r.t. picture 1. We can remove the latter dependence
by trying all possible placements and finding the one that minimizes (\ref{EuclDist}). In other
words we have to find \index{best matching} $q^\st{BM} = \oplus_{a=1}^3{{\bf r}}_a^\st{BM}$ such that
\begin{equation}\label{BestMatchedDistanceFirstExpression}
\d \left(q^i, q^\st{BM} \right) =  \inf_{q'} ~ \d \left(q^i, q' \right)  \,, ~~~  \|  {\bf r}_a^\st{BM} -  {\bf r}_b^\st{BM}  \| = r_{ab}(t_f)\,,
\end{equation}
(I chose to move the second picture, but obviously I could equivalently have moved the first one). In other words, we minimize with respect to transformations of the second triangle that keep its
observable properties (the three $r_{ab}(t_f)$) unchanged. Those are the Euclidean transformations \index{Euclidean group}
$\Eucl(3) = ISO(3) = SO(3) \ltimes \doublestroke{R}^3$, which act on a single particle coordinate-vector as
\begin{equation}
{\bf r}_a  \to  \Omega \,{\bf r}_a  + \bm{\theta}  \,,
\end{equation}
if $q = {\bf r}_1 \oplus {\bf r}_2 \oplus {\bf r}_3$, a Euclidean transformation will act identically on all three particles:
\begin{equation}
q \to T [q] = \oplus_{a=1}^3 \left( \Omega \,{\bf r}_a  + \bm{\theta}\right)
\end{equation}
where $\bm{\theta} \in \doublestroke{R}^3$ and $\Omega \in SO(3)$. We can introduce the
auxiliary variables $\Omega$ and $\bm{\theta} $ in order to directly perform the constrained
variation to find $q^\st{BM}$ defined in (\ref{BestMatchedDistanceFirstExpression}):
\begin{equation}\label{BMDist}
\d_\st{BM} \left(q^i, q^f \right) =\inf_{T } \d (q^i , T[q^f])=  \inf_{\Omega , \bm{\theta} }  \left( \sum_{a=1}^3 \| {\bf r}_a^i - \Omega \,{\bf r}_a^f  - \bm{\theta}  \|^2 \right)^{\frac 1 2} \,.
\end{equation}
The result is a notion of distance, called \emph{best-matched distance}, that depends only on relational data. The quantity (\ref{BMDist}) is in fact
invariant under Euclidean transformations of either of the two vectors $q^i$ and $q^f$:
\begin{equation}\begin{aligned}
\d_\st{BM} \left(T'[q^i]\,,\, q^f \right) &= \inf_{\Omega\,,\, \bm{\theta} }  \left(  \sum_{a=1}^3 \| \Omega' \, {\bf r}_a^i + \bm{\theta}' - \Omega \,{\bf r}_a^f  - \bm{\theta} \|^2\right)^{\frac 1 2} \\
& = \inf_{\Omega\,,\, \bm{\theta} } \left(  \sum_{a=1}^3 \|  {\bf r}_a^i -  (\Omega')^{-1}  \Omega \,{\bf r}_a^f  - (\Omega')^{-1} (\bm{\theta} - \bm{\theta}') \|^2 \right)^{\frac 1 2}
 \\
& = \inf_{\Omega''\,,\, \bm{\theta}'' }  \left( \sum_{a=1}^3 \| {\bf r}_a^i -  \Omega'' \,{\bf r}_a^f  -\bm{\theta}''\|^2 \right)^{\frac 1 2} = 
\d_\st{BM}( q^i \,,\, q^f) \,,
\end{aligned}\end{equation}
and similarly $\d_\st{BM} \left(q^i\,,\, T'[q^f] \right) =\d_\st{BM} \left(q^i\,,\, q^f \right) $.

For the simple problem of three stars the best-matching condition can be solved explicitly as
a variational problem. Start with the translations, and consider a variation $\bm{\theta} \to \bm{\theta} + \delta \bm{\theta}$
that leaves the squared distance $\d^2 \left(q^i \, , \, T[q^f]  \right) $ stationary
(there's no need to vary the square-root, as the square root is a monotonic function)
\begin{equation}
\frac{\delta \, \d^2 \left(q^i \, , \, T[q^f] \right)}{\delta  \bm{\theta}} = 2\sum_{a = 1}^3 ({\bf r}_a^i  -  \Omega \,{\bf r}_a^f)  - 6 \, \bm{\theta} = 0 \,,
\end{equation}
which gives us the best-matching \index{best matching} condition
\begin{equation}
\bm{\theta}^\st{BM} =  \frac 1 3 \sum_{a = 1}^3 ({\bf r}_a^i  -  \Omega \,{\bf r}_a^f)  \,,
\end{equation}
this condition establishes that to best-match w.r.t. translations we just have to subtract the
barycentric  coordinates from both ${\bf r}_a^i$ and $\,{\bf r}_a^f$.
In other words, we have to make the barycentres of the two triangles coincide:
\begin{equation}\label{BMwithoutTranslations}
\d_\st{BM} \left(q^i \, , \, q^f  \right) = \inf_{\Omega }  \left( \sum_{a=1}^3  \| \Delta {\bf r}_a^i - \Omega \,\Delta{\bf r}_a^f   \|^2 \right)^{\frac 1 2} \,,
\end{equation}
where $\Delta {\bf r}_a = {\bf r}_a - \frac 1 3 \sum_a {\bf r}_a $ are the barycentric coordinates.

Now we need to vary with respect to rotations. Taking $\Omega \to \Omega + \delta \Omega$ (in the
sense of varying independently all the 9 components of $\Omega$) is wrong, because we have
to make sure the variation keeps the matrix an element of $SO(3)$. Imposition of this condition,
\begin{equation}
(\Omega + \delta \Omega)  (\Omega + \delta \Omega)^\st{T} = \mat{I} + \Omega \, \delta \Omega^\st{T}Ê+ \delta \Omega \, \Omega^\st{T} = \mat{I} \,,
\end{equation}
translates into the requirement of antisymmetry of the matrix $\delta \Omega \, \Omega^\st{T} $:
\begin{equation}
\delta \Omega \, \Omega^\st{T} = - \left(\delta \Omega \, \Omega^\st{T}  \right)^\st{T}  \,.
\end{equation}
In 3D any antisymmetric matrix can be written in terms of the `vector-product' operator 
with a vector  $\delta {\bm \omega}$
\begin{equation}\label{SO3AlgebraAction}
 \delta \Omega \, \Omega^\st{T}  = \delta {\bm \omega} \times
\,, ~~~ \Rightarrow ~~~   \delta \Omega  = \delta {\bm \omega} \times  \Omega \,,
\end{equation}
where ${\bm v} \times$ is to be understood as the matrix of components $\epsilon_{ijk} v_k$.
So our variation takes the form $\Omega \to \left(\mat{I} + \delta {\bm \omega} \times \right) \Omega$, 
where $\delta {\bm \omega}$ is an infinitesimal  vector that is parallel to the rotation axis. Imposing
stationarity of (\ref{BMwithoutTranslations}) w.r.t. $\delta {\bm \omega} $ variations, we get 
\begin{equation}\label{BMconditionForRotations}
\begin{aligned}
\sum_{a=1}^3  \| \Delta {\bf r}_a^i - (\mat{I} + \delta {\bm \omega} \times)\Omega \,\Delta{\bf r}_a^f   \|^2 =\sum_{a=1}^3  \| \Delta {\bf r}_a^i - \Omega \,\Delta{\bf r}_a^f   \|^2  \\
+ 2 \, \delta{\bm \omega} \cdot \sum_{a=1}^3 (\Omega \,\Delta{\bf r}_a^f )\times  \left( \Delta {\bf r}_a^i - \Omega \,\Delta{\bf r}_a^f   \right) =  2 \, \delta{\bm \omega} \cdot   \sum_{a=1}^3   (\Omega \,\Delta{\bf r}_a^f)  \times \Delta {\bf r}_a^i =0  \,.
\end{aligned}
\end{equation}
The above equation would be hard to solve were it not for a simplification: three particles always lie on
a plane. Therefore the infimum of $\d \left(q^i \, , \, T[q^f] \right)$ will
necessarily be found among those choices of $\Omega$ that make the two triangles coplanar. 
This is easily understood, because all the coplanar positionings of the two triangles are always
local minima of $\d \left(q^i \, , \, T[q^f]  \right)$ under variations
$\delta {\bm \omega}$ that break the coplanarity. Say that we choose the axes in such a way that  
$\Delta  {\bf r}_a^i $ lies on the  $x,y$ plane and start with an $\Omega $ that keeps
$\Omega \,\Delta {\bf r}_a^f$ on the same plane; then the quantity
(\ref{BMconditionForRotations}) is obviously parallel to the $z$ plane. Therefore the variations of
$\d \left(q^i \, , \, T[q^f] \right)$ in the $x$ and $y$ directions
vanish.
\index{best matching}

So let's assume that $\Delta  {\bf r}_a^i = (\Delta x_a^i , \Delta y_a^i , 0) $ and  $\Delta  {\bf r}_a^f = (\Delta x_a^f , \Delta y_a^f , 0) $, and take for $\Omega$ a rotation in the $x,y$ plane. Eq. (\ref{BMconditionForRotations})  becomes
\begin{equation}\begin{aligned}
 \sum_{a=1}^3 \left[ \Delta x_a^i  \left(\cos \phi \, \Delta y_a^f + \sin \phi \, \Delta x_a^f \right)  -
   \Delta y_a^i\left(\cos \phi \, \Delta x_a^f - \sin \phi \, \Delta y_a^f \right) \right] = 0\,,  
\end{aligned}\end{equation}
which simplifies to
\begin{equation}
\sum_{a=1}^3  \| \Delta {\bm r}_a^i \times \Delta {\bm r}_a^f  \|  \cos \phi + \sum_{a=1}^3  \Delta {\bm r}_a^i \cdot \Delta {\bm r}_a^f \,  \sin \phi =0 \,,
\end{equation}
which is easily solved
\begin{equation}
\phi_\st{sol} = \arctan \left( \frac{ \sum_{a=1}^3 \| \Delta {\bm r}_a^f \times \Delta {\bm r}_a^i \| }{ \sum_{b=1}^3  \Delta {\bm r}_b^i \cdot \Delta {\bm r}_b^f } \right) \,.
\end{equation}
The expression above transforms in a simple way under separate rotations of $\Delta {\bm r}_a^f $ and $ \Delta {\bm r}_a^i $: one can verify that (left as an exercise) under a rotation $\Omega(\alpha)$ in the $x-y$ plane:
\begin{equation}
\begin{aligned}
\Delta {\bm r}_a^i  \to \Omega(\alpha) \Delta {\bm r}_a^i  \,, \qquad \phi_\st{sol} \to \phi_\st{sol} - \alpha \,,
\\
\Delta {\bm r}_a^f  \to \Omega(\alpha) \Delta {\bm r}_a^f  \,, \qquad \phi_\st{sol} \to \phi_\st{sol} + \alpha \,,
\end{aligned}
\end{equation}
which implies that the best-matched distance 
\begin{equation}\label{BMfinal3stars}
\d_\st{BM} \left(q^i \, , \, q^f  \right) = \left( \sum_{a=1}^3  \| \Delta {\bf r}_a^i - \Omega(\phi_\st{sol}) \,\Delta{\bf r}_a^f   \|^2 \right)^{\frac 1 2} \,,
\end{equation}
is invariant under separate rotations and translations of $q^i$ or $q^f$.

We have obtained an expression that allows us to measure the amount of \emph{intrinsic} change
between the two configurations -- change that is not due to an overall translation or rotation.
This is the essence of best matching. And it is deeply connected to the theory of connections
on principal fibre bundles: it defines a \emph{horizontal derivative}. \index{horizontal derivative}

Now, the distance $\d_\st{BM} $ \index{best matching} we found is not very physical. First of all, it doesn't take into
account the masses of the particles. If we take two configurations and move around just
one particle, the $\d_\st{BM} $ between the two configurations changes independently
of the mass of the particle even if we move just an atom while the other particles have stellar masses.
This can be easily corrected by weighting the original Euclidean distances with the masses of the 
particles:
\begin{equation}\label{EuclideanDistanceMassWeighted}
\d \left(q^i ,q^f\right) =  \left( \sum_{a=1}^3 m_a \| {\bf r}_a^i - {\bf r}_a^f \|^2 \right)^{\frac 1 2} \,.
\end{equation}
Moreover, to introduce forces, as we shall see, we would like to weight different relative configurations differently, but
without giving different weights to configurations that are related by a global translation
or rotation. We can do that by multiplying by a rotation- and translation-invariant
function:
\begin{equation}\label{KineticDistance}
\d  \left(q^i, q^f \right) =   \left( U(r_{bc}) \,\sum_{a=1}^3 m_a \| {\bf r}_a^i - {\bf r}_a^f  \|^2 \right)^{\frac 1 2} \,.
\end{equation}
\begin{figure}[t]
\begin{center}\includegraphics[width=0.45\textwidth]{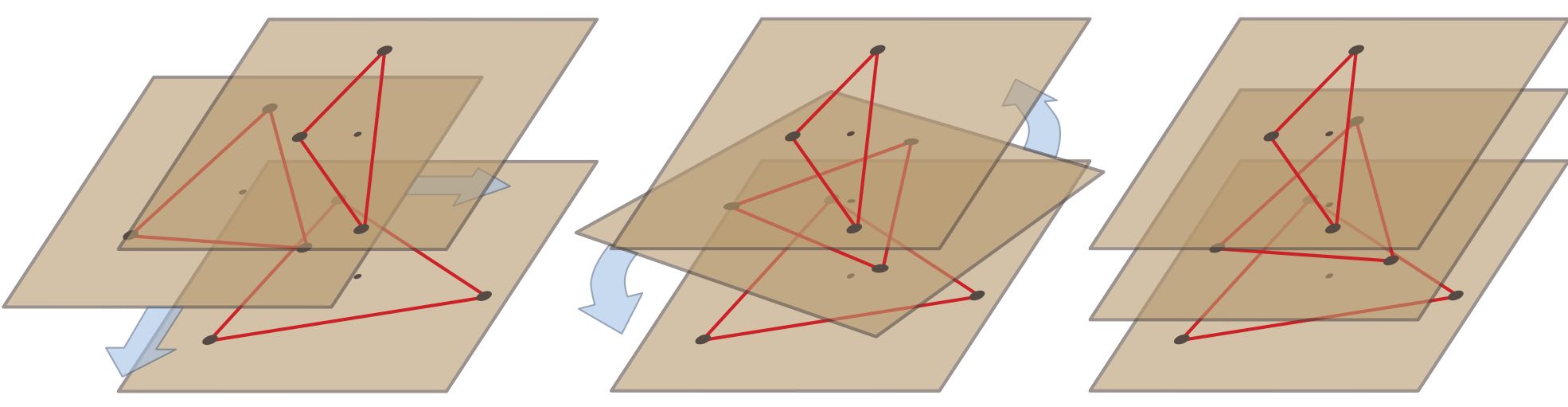}
\end{center}
\caption[Horizontal stacking]{Horizontal stacking: \index{horizontal stacking} in the picture a \emph{stacking} of three-body configurations is represented. The arbitrarily chosen stacking on the left is best matched (blue arrows) by translations so as to bring the barycentres  to coincidence, after which rotational best matching eliminates residual arbitrary relative rotation.}
\label{StackingPicture}
\end{figure}
The idea is now to perform this best-matching procedure for a complete history.  Take a series of 
snapshots of the kind depicted in Fig.~\ref{PhotosOfStars}, and represent them in an arbitrary
way in Euclidean space, $q^k = \oplus_{a=1}^3 {\bf r}_a^k$, $k = 1,2,\dots$. Then minimize
$\d \left(T_k [q^k] ,  T_{k+1} [q^{k+1}]\right)$ w.r.t. $\Omega_k$ and 
${\bm \theta}_k$ for all $k$. What we get is
a \emph{horizontal stacking} (by the blue arrows in Fig.~\ref{StackingPicture}) of all the configurations. 
In other words, a notion of \emph{equilocality}: the dots in the successive snapshots of the third part of  Fig.~\ref{StackingPicture} can all be said `to be in the same place'. This notion of position is not a pre-existing absolute structure but emerges out of the best matching.

\subsubsection*{Continuum limit}

Now consider the continuum limit, in which we parametrize the snapshots with a
continuous parameter $s$, with each successive snapshot separated by an infinitesimal $\d s$: 
\begin{equation}
v^k =v(s) \,, ~~~~ v^{k+1} = v(s +\d s) \,, ~~~~ v = {\bf r}_a \,, ~ \Omega \,,~ {\bm \theta}\,,
\end{equation}
Then the squared distance between two successive configurations is infinitesimal,
\begin{equation}\begin{aligned}
& \d^2 \left(T_k [q^k] , T_{k+1}[q^{k+1}] \right) =  \\
& U(r_{bc} ) \,\sum_{a=1}^3 m_a \|\Omega(s+\d s) \, {\bf r}_a (s + \d s)+ {\bm \theta}(s+\d s)
-\Omega(s) \, {\bf r}_a (s) - {\bm \theta}(s)  \|^2  \\
& \sim U(r_{bc} ) \,\sum_{a=1}^3 m_a \| \Omega  \, \d {\bf r}_a  + \d \Omega  \, {\bf r}_a   + \d {\bm \theta}  \|^2 \, \,,   \qquad  \d v = \frac{\partial v}{\partial s} \, \d s \,.   
\end{aligned}\end{equation}
For $\d \Omega$ too, we can repeat the argument for obtaining a variation of an $SO(3)$ matrix that remains
within the group, and we get
\begin{equation}\begin{aligned}
  U(r_{bc} ) \,\sum_{a=1}^3 m_a \| \Omega  \, \d {\bf r}_a  + \d {\bm \omega} \times  \Omega \, {\bf r}_a   + \d {\bm \theta}  \|^2  \,.     
\end{aligned}\end{equation}
\index{best matching}

We can get rid of the dependence on $\Omega$ by exploiting the invariance of the scalar
product under rotations:
\begin{equation}\begin{aligned}
\| \Omega  \, \d {\bf r}_a  + \d {\bm \omega} \times  \Omega \, {\bf r}_a   + \d {\bm \theta}  \|^2
= \|   \, \d {\bf r}_a  + \Omega^{-1} \d {\bm \omega} \times  \Omega \, {\bf r}_a   +\Omega^{-1} \d {\bm \theta}  \|^2
\end{aligned}\end{equation}

Here, $\Omega^{-1} \d {\bm \omega} \times  \Omega$ and $\Omega^{-1} \d {\bm \theta} $
are just the adjoint action of $SO(3)$ on $ISO(3)$. Minimizing w.r.t. $\d {\bm \omega}$
and $ \d {\bm \theta}$ makes this action irrelevant:
\begin{equation}\begin{aligned} 
\d_\st{BM} \mathcal L^2=   \inf_{ \d {\bm \omega} ,  \d {\bm \theta}} U(r_{bc} ) \,\sum_{a=1}^3 m_a \| \d {\bf r}_a  + \d {\bm \omega} \times  {\bf r}_a   + \d {\bm \theta}  \|^2 \,.     
\end{aligned}\end{equation}

The last expression can be integrated over a parametrized 
path ${\bf r}_a(s)$,
\begin{equation}\begin{aligned} \label{BestMatchedDistanceOverPath}
\int \d_\st{BM} \mathcal L  =   \inf_{ \d {\bm \omega} ,  \d {\bm \theta}} \int \d s \left( U(r_{bc} ) \,\sum_{a=1}^3 m_a \left \| \frac{\d {\bf r}_a}{\d s}  + \frac{\d {\bm \omega}}{\d s} \times  {\bf r}_a   + \frac{\d {\bm \theta}}{\d s}  \right\|^2\right)^{\frac 1 2} \,.     
\end{aligned}\end{equation}

We obtain a notion of length of the path that measures only the intrinsic, physical
change that occurs along the path. The above expression is in fact invariant
under \emph{local} ($s$-dependent) $ISO(3)$ transformations
\begin{equation}
{\bf r}_a(s) \to \Omega(s) \,  {\bf r}_a(s) + {\bm \theta}(s) \,.
\end{equation}
This kind of object is what is needed to define a variational principle for relational physics.
It realizes the kind of foundations for dynamics implied by Leibniz's criticism of Newton's concepts of absolute space and time.  
No wonder that, at the time, he was unable to make this precise.

\subsubsection*{Generating dynamics with best-matching}
\index{best matching}

One can use this measure of intrinsic change to assign a numerical value (weight) to any curve in the
`extended configuration space' (the space of Cartesian representations of $N$ particles), and thereby define an action to be minimized by the dynamical solutions. However, such an action principle is only capable of generating solutions with zero total momentum and angular momentum, as will be shown below. Therefore such a law cannot be used to describe a general $N$-particle system (like billiard balls on a table, or a gas in a box): it should rather be used to describe a complete universe.

It is in this sense that the novel foundation of dynamics I am describing satisfies Mach's principle and solves the puzzle of Newton's bucket (in the restricted case of a universe composed of point particles interacting with
instantaneous potentials): the dynamical law is such that it can only contemplate a universe with zero total angular momentum. In Newtonian dynamics, on the contrary, angular momentum is a constant of motion that is freely specifiable through the initial conditions of our solution. The law that the total angular momentum of the universe must be zero solves the problem of Newton's bucket in the following way: one makes a (small) error in assuming that the reference frame defined by the fixed stars is inertial. If the total angular momentum is zero,
and the Earth is rotating w.r.t. the fixed stars, that reference frame cannot be inertial. This is due to the fact that the total angular momentum of the rest of the universe must be equal and opposite to that of the Earth. The reference frame of the fixed stars must be rotating very slowly (because the stars greatly overweigh the Earth), and a truly inertial frame would be rotating both w.r.t. Earth and w.r.t. the fixed stars. Then one would, in principle, see exactly what Wenzel Hoffman proposed: the plane of a pendulum at the north pole would rotate around the axis of the Earth at a slightly slower speed than the Earth itself. This effect is, of course, impossible to detect
as it is too small. But this illustrates how a relational dynamics dissolves the puzzle of Newton's bucket. 
\index{Newton's bucket}

\vspace{12pt}
\noindent
\fbox{\parbox{0.98\linewidth}{
{\bf Further reading:} Barbour and Bertotti's 1982 seminal paper~\cite{BarbourBertotti82}.
}}

\newpage

\section{Best matching: technical details}\label{SectionBestMatching}
\index{best matching}\index{configuration space}\index{relative configuration space}\index{shape space}\index{pre-shape space}\index{similarity group}\index{quotient space}

I'll now make the ideas introduced above slightly more precise. First of all, we need to specify the various \emph{configuration
spaces} we're dealing with. The largest -- and simplest -- of them all is the \emph{extended configuration space}, or \emph{Cartesian space} $\Q^N = \doublestroke{R}^{3N}$.\footnote{For this and the other configuration spaces considered below, there is in fact a distinct space for each set of masses (or, better, mass ratios) of the particles.} It is the space of Cartesian
representations of $N$ bodies, and the mathematical embodiment of Newton's absolute space.
$\Q^N$ has a Euclidean metric on it, called the \emph{kinetic metric}:
\begin{equation}\label{KineticMetric}
\d \mathcal L^2_\st{kin} = \sum_{a=1}^N m_a \, \d {\bf r}_a  \cdot \d {\bf r}_a \,.
\end{equation}
Then there is the \emph{relative configuration space} $\Q_R^N = \Q^N / \Eucl(3)$, which is just
the quotient of $\Q^N$ by the Euclidean group of rigid translations and rotations.
Finally, if we insist that only ratios and angles have objective reality, we must
further quotient by scale transformations:
\begin{equation}
{\bf r}_a \to \phi \, {\bf r}_a \,~~~ \phi > 0 \,,
\end{equation}
which, together with the Euclidean group, make the \emph{similarity group} $\Sim(3) = \doublestroke{R}^+ \ltimes \Eucl(3)$. We will call this last quotient the \emph{shape space} of $N$ particles,
$\shs^N = \Q^N/\Sim(3)$.
However, since gauge-fixing or reducing w.r.t. rotations is hard except in some simple cases,
we also consider the configuration space obtained by quotienting just w.r.t. translations
and dilatations, $\pshs^N = \Q^N/\doublestroke{R}^+ \ltimes \doublestroke{R}^3$. We call
this \emph{pre-shape space}.

\begin{center}
\begin{tabular}{lcr}
\hline \hline
Reduced conf. space & $G$ & $B =\Q^N/G$ \\
\hline
Relative conf. space & $\Eucl(3) =ISO(3)$  & $\Q_R^N$ \\
Pre-shape space & $\doublestroke{R}^+ \ltimes \doublestroke{R}^3$ & $\pshs^N$\\
Shape space & $\Sim(3) = \doublestroke{R}^+ \ltimes ISO(3)$  & $\shs^N$ \\
\hline \hline
\vspace{-6pt}
\end{tabular}
\end{center}

Both mathematically and conceptually, it is important that the various reduced (quotient) spaces are not subspaces of the space from which they are obtained by reduction but distinct spaces.

\newpage

\subsubsection*{Principal fibre bundles}

\index{principal fibre bundle}
The groups acting on $\Q^N$ endow it with the structure of a \emph{principal $G$-bundle}\footnote{I believe the first to relate best matching with principal fibre bundles and connections on configuration space was H. Gomes in~\cite{Gomes:2008yq}} \cite{Frankelbook} (this actually holds only for the regular configurations in $\Q^N$, see below). Let's call the principal $G$-bundle $P$ and the group $G$. The reduced configuration space plays the role of the \e{base space}, which in a principal bundle is the quotient space $B = P/G$. The fibres are homeomorphic to the group $G$.\footnote{The fibres and the group are just \emph{homeomorphic} (meaning equivalent as topological spaces), not \emph{isomorphic}, because the fibres lack an identity element, which is an essential part of the structure of a group. D.~Wise alerted me that such a ``group that has forgotten its identity'' is called a \emph{torsor}. See J. Baez' description of \index{torsor} torsors \cite{BaezTorsors}.} As relationists, our prime interest is in the base space $B$, which we regard as the space of physically distinct configurations, but it is only defined through the quotienting process, and this poses problems: for example, $B$ inherits $P$'s structure of a smooth manifold
only if the group $G$ acts \emph{freely} (or \emph{transitively}) on $P$, \index{transitive action of a group} which means that there are no points in $P$ that 
are left invariant by any other transformation than the identity. But we know that this isn't the case
in $\Q^N$: there are symmetric configurations (e.g., collinear states or total collisions) for which the action is not free. These regions represent special parts of $B$, akin to corners or edges, where smoothness fails.\footnote{$B$ becomes a \emph{stratified manifold} \cite{StratifiedManifold}. \index{stratified manifold}}
Continuing a dynamical orbit after it crosses one of those points poses a challenge, but this is a technical, rather than conceptual, issue that I won't go into here.
\begin{figure}[ht]
\begin{center}\includegraphics[width=0.4\textwidth]{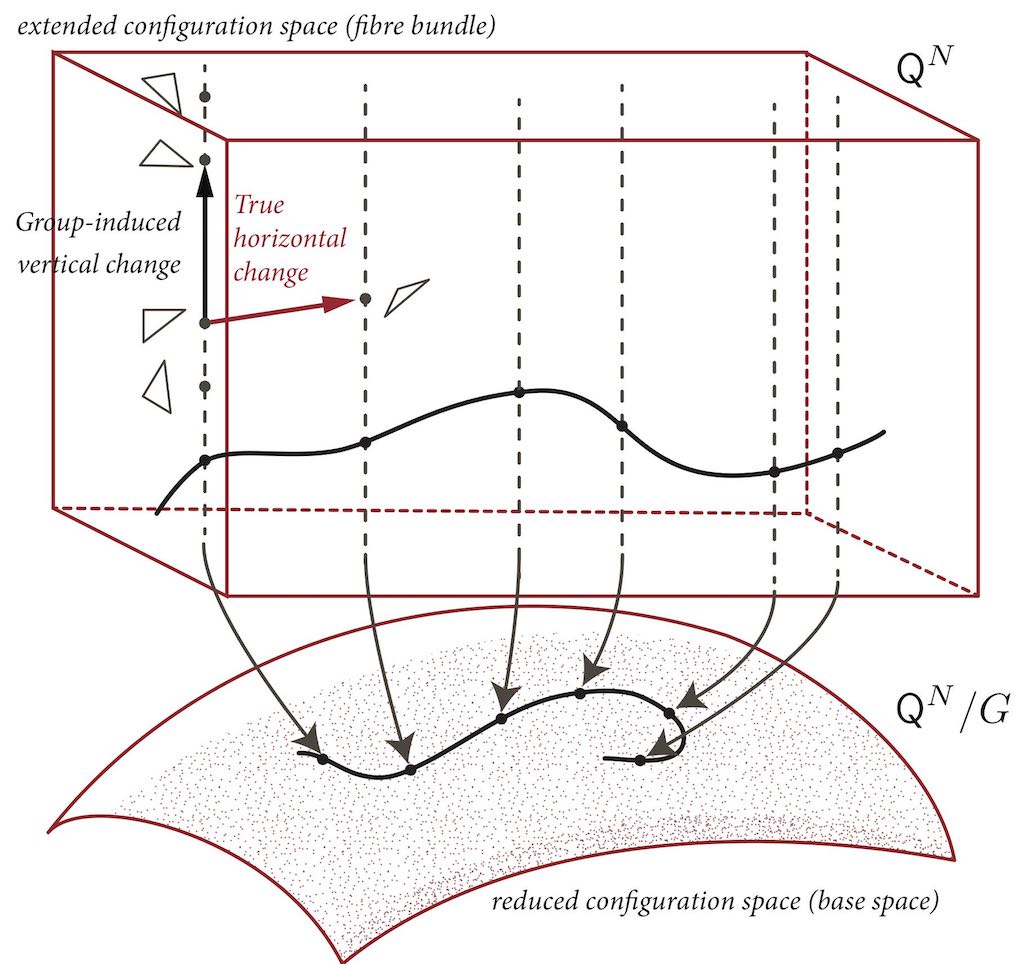}
\end{center}
\caption[3-body configuration space as a fibre bundle]{The fibre bundle structure of the three-body configuration space $\Q^3$,
with structure group $\Sim(3)$ \index{similarity group} and base manifold $\shs^3$ (the space of triangles).
Vertical motion changes the \emph{representation} of the triangle in Cartesian space,
while horizontal motion changes the shape of the triangle (\emph{e.g.,} its internal angles).}
\label{FibreBundleTriangle}
\end{figure}

Intuitively, a fibre bundle is a sort of generalization of a Cartesian product $ B \times G $ where one erects a tower of different representations in $\Q^N$ above every single relative configuration or shape
(see Fig. \ref{FibreBundleTriangle}). The difference with a true Cartesian product is
the lack of an origin. Obviously a point in, say, $\Q^3$ cannot
be uniquely represented as a point in $\shs^3$ (a triangle), together with a translation
vector ${\bm \theta}$, an $SO(3)$ matrix $\Omega$ and a dilatation constant $\phi$.
This doesn't make sense because  ${\bm \theta}$, $\Omega$ and $\phi$ are \emph{transformations}
that connect different Cartesian representations of the same triangle. This is where
local sections \index{local sections of a bundle} (and `trivializations') of the bundle enter: they provide a `conventional' choice of 
origin for each triangle. This means that a section associates with each triangle in a neighbourhood
of $\shs^3$ an oriented triangle with a definite size and position in $\doublestroke{R}^3$. This is purely 
conventional: for example I could decide that all triangles go onto the $x,y$ plane,
with their barycentre at the origin, and the most acute of their three vertices goes on the
$x$-axis at unit distance from the origin. Then I can represent any other element of 
$\Q^3$ through the transformation that is needed to bring the `reference' triangle
to congruence with the desired one.

One is forced to define those sections/trivializations \index{local sections of a bundle} \emph{locally},
that is only on a set of open neighbourhoods, because the sections
have to be smooth (continuous and infinitely differentiable), and unless
the bundle is `trivial' no section can be smooth everywhere.\footnote{For example, the section I 
defined in the above example fails to be continuous when the smallest-angle
vertex of the triangle changes: at that point I have an abrupt rotation of the
representative triangle in $\Q^3$. This might seem a quirk of the particular
section that I chose, but it is instead an obstruction of topological nature:
the topology of the Cartesian product $\shs^3 \times \Sim(3)$ is different
from that of $\Q^3$. This is easily seen: $\Q^3 \sim \doublestroke{R}^9$ is simply connected
while the rotation group $SO(3)$ is not, and therefore neither is  $\shs^3 \times \Sim(3)$.}

Physicists have a name of their own for local sections: \emph{gauges}. \index{gauge theory}

\subsubsection*{Connections}

\index{connection}
A principal $G$-bundle comes equipped with
a natural distinction between \emph{vertical} and \emph{horizontal} directions.
The first are defined as the subspace $V_p \subset T_pP$  of tangent vectors to $P$ that
are parallel to the orbits of $G$, and are related to unphysical, irrelevant or gauge motions. 
Horizontal directions represent physical, relevant, or true change (see Fig.~\ref{FibreBundleTriangle}),
and they are just defined as the vector-space complement to $V_p$, that is \index{tangent space (of a principal bundle)}
$T_pP = V_p \oplus H_p$. It's clear that this definition is ambiguous:
imagine that $T_pP = \doublestroke{R}^3$, and $V_p$ is one-dimensional.
Then any plane that is not parallel to $V_p$ is an equally good choice of
$H_p$. If one has a metric on $TP$, then $H_p$ can be defined
as the (unique) \emph{orthogonal} complement of $V_p$, but a principal bundle
does not always come equipped with a metric on $T_pP$. This is
what connections are introduced for. A connection on $P$ is a smooth
choice of horizontal subspaces in a neighbourhood of $P$.

\index{horizontal directions in a bundle}
\index{vertical directions in a bundle}

\begin{figure}[t]
\begin{center}\includegraphics[width=0.2\textwidth]{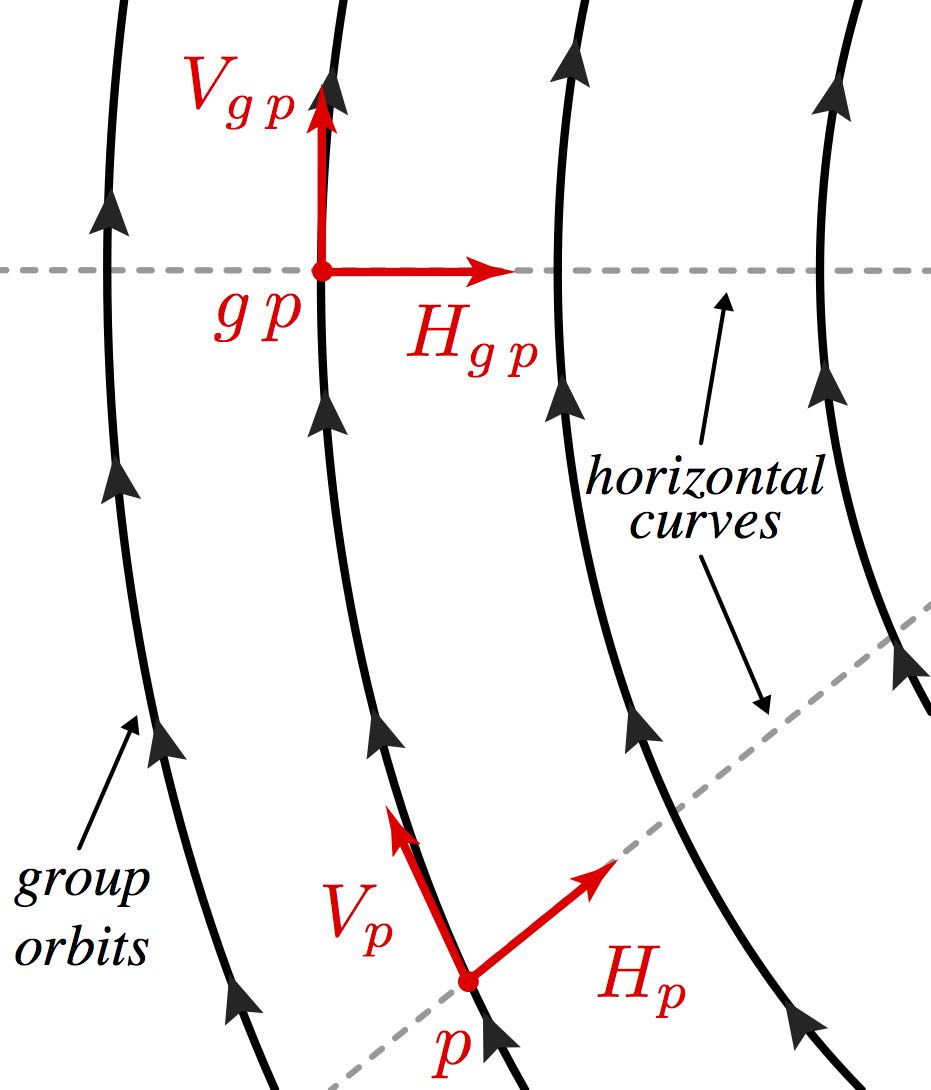}
\end{center}
\caption[$G$-invariance condition for a connection]{The $G$-invariance condition for a connection: if a horizontal curve through $p$
is transformed by $g\in G$, then the tangent vector at $g \, p$ to the transformed curve
must still be horizontal.}
\label{ConnectionGinvariance}
\end{figure}

A connection on $P$ defines a (conventional) notion of horizontality, and consequently
of \emph{horizontal curves}: those curves whose tangent vectors are horizontal. This
is the precise formalization of best-matched trajectories: they must be horizontal according
to some connection on $\Q^N$. Moreover a connection has to satisfy a compatibility
condition with the action of the group $G$, which basically states that the $G$-action 
sends horizontal curves to horizontal curves (see Fig. \ref{ConnectionGinvariance}).

As I said, if we have a metric on $TP$,  then that defines
a `natural' notion of horizontality: $H_p = V_p^\perp$. In order for this to define a
connection, the orthogonality condition defined by the metric has to be $G$-invariant,\footnote{A $G$-invariant metric associates the same scalar product to the tangent vectors of two curves
that intersect at a point and the corresponding tangent vectors
to the curves transformed under $G$.
\label{FootGinvariantMetric}} 
so that a $G$-transformation sends curves that are orthogonal to $V_p$ 
to curves orthogonal to $V_{g\,p}$.
A particularly simple situation is that of a $G$-invariant metric: in that case the
metric not only defines a connection, but all the horizontal curves are geodesics
of that metric. In \cite{BarbourBertotti82},  Barbour and Bertotti exploited the following
results to define a relational dynamical law, which I formulate in this way:

\begin{quote} \it
\textbf{Theorem:}  If a metric on $P$ is $G$-invariant,  given a \emph{sheet} in $P$ lifted above a single curve in $B$, all the horizontal curves on that sheet minimize the free-end point length
between the initial and final orbits. Moreover, all the horizontal curves on that sheet  have the same length according to the $G$-invariant metric.
\end{quote} 

The technical terms `sheet' and `free-end point' in this theorem will now be explained.

\subsubsection*{The two-stage variational procedure}

Now I will describe the variational principle that realizes best-matching.
The goal is to get an \emph{action}, \index{action principle} that is, a rule to associate a real
number to each path on the base manifold $P/G$, \e{i.e.,} the reduced configuration space, \index{reduced configuration space}
be it $\Q^N_R$ or $\shs^N$. It is pretty clear, at this point, that only the simplest
cases can be effectively worked out on $P/G$ itself \cite{EdwardBigReviewRelationalParticleMechanics}, 
since in general our only way to
represent that space is redundantly through $P$. So what we aim for is
an action principle on $P$ which is $G$-invariant, so that it associates
the same number to all paths that project to the same path in $P/G$
and correspond therefore to the same physical solution.\footnote{A note of
warning is in order here: I am not trivially talking about an action that is
invariant under a `global' translation or rotation of the system. For that
purpose the actions everybody is familiar with from basic physics courses
are perfectly good. Here I'm talking about an action which is invariant under
\emph{time-dependent} transformations. In this sense the step from 
elementary action principles is perfectly analogous to that from a global
to a local gauge symmetry in field theory. The only difference is that the
`locality' here is only in time, not in spacetime as in electromagnetism.}

\subsubsection*{Stage I: Free-end-point variation}\index{free-end-point variation}

\begin{figure}[t]
\begin{center}
\includegraphics[width=0.5\textwidth]{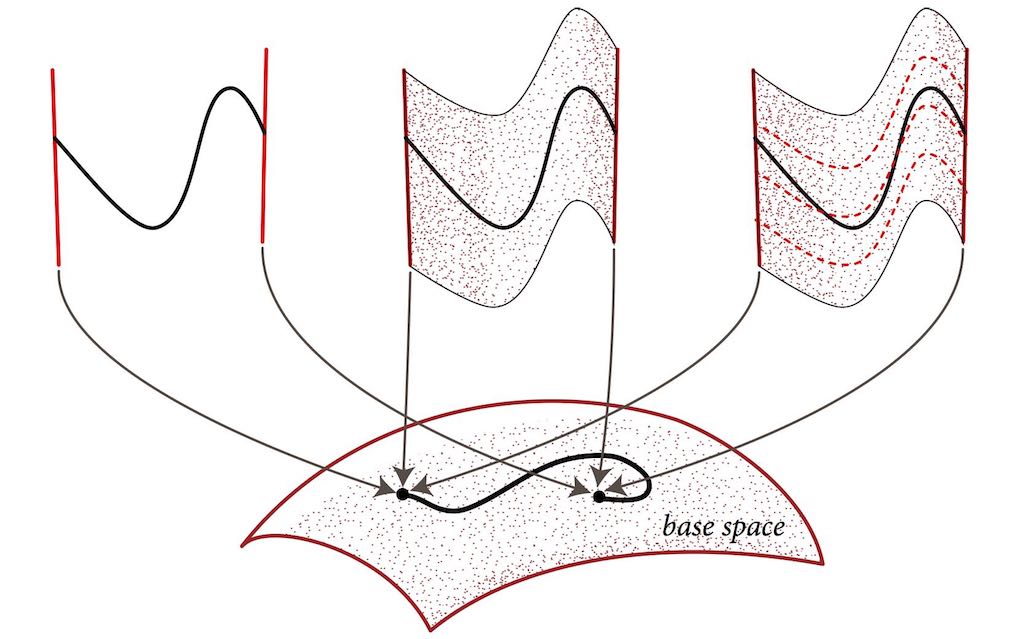}
\end{center}
\caption[Stage I of best-matching]{Stage I: Free-end-point variation on a sheet in $P$. Notice how this is all done
in the bundle, $P$, and the base space $B$ is abstractly defined as the collection of all gauge orbits (here represented as red lines).}
\label{2stageVariation1}
\end{figure}

The following pictures illustrate the first stage of variation: our end-points are two points in $P/G$, where the physics resides. In $P$ they map
to two fibres, the two red lines in Fig.~\ref{2stageVariation1}, which are two orbits of $G$.
\begin{enumerate}
\item 
First, take a trial curve in $P$ (in black) between the two (red) fibres ending anywhere on them. It projects to a trial curve in $B$.
\item Then \emph{lift} this curve  in $P$ to a sheet (in gray) in $P$ using the group action. All the
curves on this sheet correspond to the same \emph{physical} curve in $B$.
\item Consider all the horizontal curves (red dashed) in the sheet with endpoints \emph{anywhere} on the two red
fibres. If the metric in $P$ is $G$-invariant, these curves also minimize the arc-length
from the first fibre to the second. This is obtained through a \emph{free-end-point variation} (see below).
\item If the metric is $G$-invariant, all these horizontal curves will have the same length.
We'll define the value of the action on the curve in $B$ corresponding to the
considered sheet as the length of the curve. In this way the action is $G$-invariant.
\end{enumerate}\index{free-end-point variation}

\subsubsection*{Stage II: Physical variation}

Now we have an action associated with a \emph{sheet} in $P$ and consequently
with a single path in $P/G$. We can now evaluate it on every possible path in $P/G$ between the
two endpoints, which means on every possible sheet in $P$ between the two
red fibres: 
\begin{figure}[t]
\begin{center}
\includegraphics[width=0.5\textwidth]{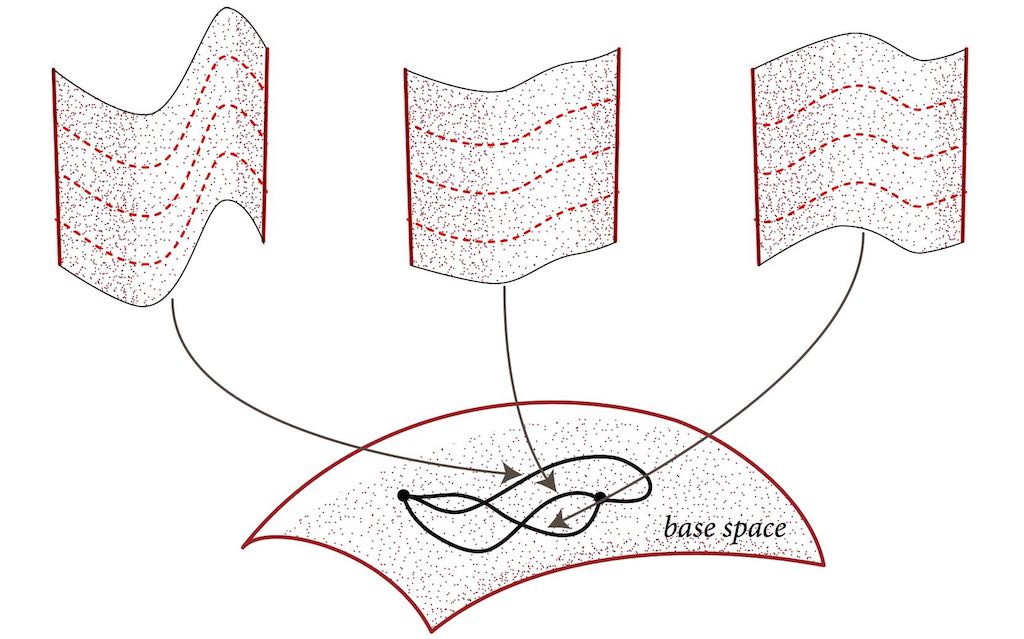}
\end{center}
\caption[Stage II of best-matching]{Stage II: Variation of the physical curve in $P/G$ to find the one (on the right) that realizes the extremum between the fixed end points in $P/G$.}
\label{2stageVariation2}
\end{figure}
\begin{enumerate}
\item[5] 
Consider all possible paths in $P/G$ joining the two red fibres.  Use the above rule to assign
a value of the action to each of them. If two of them lie on the same sheet, they have
the same action. The action is $G$-invariant.
\item[6] 
Minimizing with respect to this action will identify the physically realized sheet, and consequently
a unique curve in $P/G$.
\end{enumerate}
On this final sheet there is a $\text{dim}(G)$-parameter family of horizontal curves.
They are distinguished only by their initial positions on the fibre.
These curves are not more physical than the others on the same sheet, but they
realize a reference frame in which the equations of motion take the simplest 
form. In this sense they are special (Barbour and Bertotti introduced the notion of the \emph{distinguished representation}).
They give a preferred notion of equilocality (horizontal placement in Fig.~\ref{StackingPicture}). 

Before showing this technique in action, let me comment on temporal relationalism.

\subsection{Temporal relationalism: Jacobi's principle}
\label{SecTemporalRelationalism}
\index{temporal relationalism}
\index{Jacobi's principle}

So far, temporal relationalism found less space than spatial relationalism in this \thistext{}.
It is time to introduce it. I will show now how to realize Mach's 
aphorism \cite{mach:mechanics}
``time is an abstraction at which we arrive through the changes of things''
in a dynamical theory. 

The necessary mathematics had actually been created by Jacobi in 1837, \index{Carl Gustav Jacob Jacobi} nearly 50 years before Mach wrote that sentence. Jacobi was not thinking about the abstraction of time from 
change -- he seems to have been happy with Newton's concept of time. Rather, his aim was to give a mathematically correct formulation of Maupertuis's principle. \index{Maupertuis's principle} Ever since its original statement, it had been assumed, above all by Euler and Lagrange, \index{Leonhard Euler} \index{Giuseppe Lodovico Lagrangia (Lagrange)}  that all trial curves considered for comparison must correspond to the same total energy. The problem that Jacobi solved was the correct mathematical representation of such a condition. Euler and Lagrange had got the right answer using dubious mathematics.

Jacobi achieved his aim by reformulating Mapertuis's principle, for systems with a quadratic kinetic energy, in a `timeless' form. His principle determines the `true' (in the sense of
physically realized) trajectories of a dynamical system \e{with one fixed value $E$ of its total energy} \index{energy} as geodesics
in $\Q^N$, which are geometrical loci that do not depend on
any particular parametrization.\footnote{The same is true of a manifold, which doesn't depend
on any particular coordinate system.}

Jacobi's action is (integrating on a finite interval of the parameter $s\in [s_i,s_f]$) \index{action principle}
\begin{equation}
S_\st{J} = 2 \int_{s_i}^{s_f}  \d s \sqrt{\left( E - U \right) \, T_\st{kin}} \,, \qquad T_\st{kin} =  \sum_{a=1}^N {m_a\over 2} \, \frac{\d {\bf r}_a}{\d s} \cdot  \frac{\d {\bf r}_a}{\d s} \,,
\end{equation}
where $E$ is a constant (the total energy of the orbit we're interested in but here is to be regarded as a constant part of the potential, i.e., as part of the law that governs the system treated as an `island universe'), $U=U({\bf r}_a)$ is
the potential energy and $T_\st{kin}$ defined above is the kinetic energy. \index{kinetic energy} This expression
is reparametrization-invariant \index{reparametrizations}
\begin{equation}
\d s \to \frac{\partial s}{\partial s'} \, \d s' \,, \qquad  \sum_{a=1}^N m_a \, \frac{\d {\bf r}_a}{\d s} \cdot  \frac{\d {\bf r}_a}{\d s}
\to \left( \frac{\partial s'}{\partial s} \right)^2 \sum_{a=1}^N m_a \, \frac{\d {\bf r}_a}{\d s'} \cdot  \frac{\d {\bf r}_a}{\d s'} \,,
\end{equation}
(for reparametrizations $s'$ that preserve the end-point value of the parameter, $s'(s_i) = s_i$ and $s'(s_f)=s_f$)
thanks to the square-root form of the action. Jacobi's action is closely analogous to the expression for the arc-of-length in a Riemannian \index{arc-length}
manifold with metric \index{kinematic metric} $\d s^2 =  g^{ij}  \d x_i  \,\d x_j$:
\begin{equation}\label{GeodesicAction}
\int \d s \mathcal L = \int  \d s   \left( g^{ij} \frac{\d x_i}{\d s} \frac{\d x_j}{\d s}  \right)^{\frac 1 2} \,,
\end{equation}
and is precisely the same thing if our manifold is $\Q^N$ equipped with the
metric
\begin{equation}
\d \mathcal L^2 = 4 \, ( E - U )  \sum_{a=1}^N \frac{m_a}{2} \, \d {\bf r}_a  \cdot  \d {\bf r}_a  \,,
\end{equation}
which is conformally related to the kinetic metric by the factor $E - U$.

In the language of E. Anderson \cite{EdwardBigReviewRelationalParticleMechanics}  we can
rewrite the action in a manifestly \emph{parametrization-irrelevant}  form,
\begin{equation}
S_\st{J} = 2\int_{{\bf r}_a^i}^{{\bf r}_a^f}   \left( E - U \right)^{\frac 1 2} \left( \sum { m_a\over 2} \, \d {\bf r}_a  \cdot  \d {\bf r}_a \right)^{\frac 1 2}\,,\label{irrel}
\end{equation}
where now ${\bf r}_a^i ={\bf r}_a(s_i)$ and ${\bf r}_a^f ={\bf r}_a(s_f)$ are the end-points. 
This is in fact the form in which Jacobi originally formulated his principle.

\subsubsection*{The ephemeris time}

\begin{figure}[t]
\begin{center}
\includegraphics[width=0.4\textwidth]{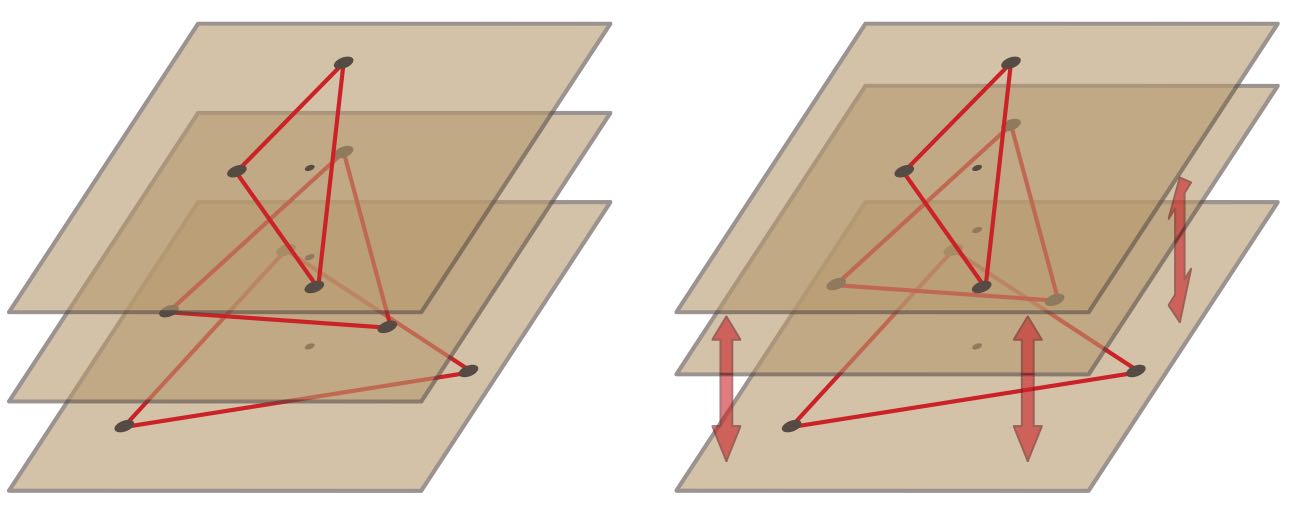}
\end{center}
\caption[Vertical stacking]{Vertical stacking:  \index{vertical stacking} the same stacking of three-body configurations as in Fig.~\ref{StackingPicture}
can be moved \emph{vertically} (red arrow) by a time reparametrization. While best matching fixes the horizontal stacking, ephemeris time fixes the vertical one.}
\label{VerticalStackingPicture}
\end{figure}

Jacobi's action gives rise to the following  Euler--Lagrange equations: \index{Euler--Lagrange equations}
\begin{equation}\label{RepInvariantEquationsOfMotion}
m_a \frac{\d}{\d s}\left( \left( \frac{E -U}{T_\st{kin}}\right)^{\frac 1 2}   \frac{\d {\bf r}_a}{\d s} \right) = -  \left( \frac{T_\st{kin}}{E - U}\right)^{\frac 1 2} \frac{\partial U }{ \partial {\bf r}_a} \,,
\end{equation}
which have a complicated form. But, as is always possible, now choose the parametrization
in which
\begin{equation} \label{paracond}
T_\st{kin}=E-U,
\end{equation}
with the increment of the new parameter given by
\begin{equation}\label{DifferentialOfTheInstant}
\d \teph = \d s \left( \frac{T_\st{kin}}{(E - U)}\right)^{\frac 1 2} =
\left( \frac{  \sum_{a=1}^N  m_a\d {\bf r}_a \cdot  \d {\bf r}_a }{2(E - U)}\right)^{\frac 1 2}
= \frac{\d \mathcal L}{2(E- U)} \,.
\end{equation}
Following the nomenclature of Edward Anderson, I call (\ref{DifferentialOfTheInstant}) \e{the differential of the instant}.  \index{differential of the instant} It measures the accumulation of a distinguished `duration' from one instant to another. 
The meaning of the subscript `eph' will be explained shortly. In the parametrization (\ref{DifferentialOfTheInstant}), the equations of motion take the  form
\begin{equation}\label{NewtonsEquations}
m_a \frac{\d^2 {\bf r}_a}{\d \teph ^2}  = - \frac{\partial U }{ \partial {\bf r}_a} \,,
\end{equation}
which is just Newton's equations  \index{Newton's equations} for $N$ particles interacting through the
potential $U$. Note that Eq.~(\ref{paracond}), rewritten as $E=T_\st{kin}+V$, would normally be interpreted as the expression of energy conservation. But, in a Machian approach $E$ for the Universe is to be interpreted, as I said, as a universal constant (like Einstein's cosmological constant), and then it, expressed explicitly through (\ref{DifferentialOfTheInstant}), becomes \e{the definition of time} or, better, \e{duration}.\footnote{We ask ``what's the time?'', if we want to know which \e{instant} of time is it. But we also ask ``How much time do we have?''.}
\index{duration}

Indeed, Eq.~(\ref{DifferentialOfTheInstant}) above is the closest thing to Mach's ideal we could imagine:
an increment of time which is a sum of all the  $\d {\bf r}_a$ in the particles' positions,
weighted by (twice) the difference between total and potential energy. Out of a timeless theory like
Jacobi's, a `natural'  parametrization emerges, which gives a notion of duration
as a \emph{distillation} of all the changes in the Universe \cite{JuliansReview}. 
In this time, the particles in the Universe, provided one identifies an inertial frame of
reference (I'll show how in a moment), will move according to Newton's laws. 

This is the place to explain the subscript `eph' \index{ephemeris time} in $\textrm dt_\st{eph}$ in (\ref{DifferentialOfTheInstant}). As I remarked above,  in the 1890s astronomers found an apparent deviation from Newton's laws
 in an anomalous acceleration in the motion of the Moon \cite{Clemence}. \index{Moon crisis of the 1890s} One possible explanation was that this was nothing to do with a failure of Newton's laws but arose from the use of the Earth's rotation to measure time. This was only definitively confirmed in 1939, when it was shown that the planets exhibited the same anomaly. This confirmed that 
 the rotating Earth doesn't tick a time in which Newton's laws are satisfied.
 The way out was to assume a more flexible notion of duration, in which 
 the second is defined by an `average' of the motions of the most prominent
 (and massive) objects in the solar system in such a way that Newton's laws are
 verified.  In practice, to match the increasing accuracy of observations it is necessary
 to include more and more objects in this definition of time, which is called `ephemeris time'.\footnote{From the Greek 
 {\greektext >efhmer\'is} 
  (eph\={e}meris) for ``diary''.
 Ephemerides are tables of the predicted positions of the celestial bodies over time.}
\index{ephemeris time}

\subsubsection*{What is a clock?}

In the physics community, there is a widespread misconception that to define clocks
one needs to make reference to periodic phenomena, which provide a time standard
because of their \emph{isochronicity}. \index{isochronicity}  This tradition finds its origin in the imitation of Einstein \cite{EinsteinPrincipedeRelativite}, \index{Albert Einstein}
with his clocks made with mirrors and light rays. 
But isocronicity of periodic phenomena \emph{per se} is a circular argument (if you define the second
as one period of a pendulum, then obviously the pendulum will always complete
a period in exactly one second!). A more refined version of this idea is based on the
hypothesis of the homogeneity (in time) of Nature: two phenomena that take place
under identical conditions should take the same time. But also this is a fallacy: there
are no two identical phenomena: ``You can never step into the same river twice'' as Heraclitus \index{Heraclitus} famously mused.
Indeed, if two phenomena had exactly the same attributes, they would have to
be identified (this Leibniz's principle of the \emph{identity of the indiscernibles}).

The relationalist point of view is that the main defining property of good clocks (both natural and 
artificial) is that they \emph{march in step}, and therefore they are useful for keeping
appointments \cite{JuliansEssayOnTime}. The fact that we (on Earth) can usefully make 
reference to an ever-flowing, ubiquitous notion of time is made possible by the way 
the objects in the world behave, with a lot of regularities. 
After Einstein and his relativity of simultaneity \index{relativity of simultaneity} \index{Albert Einstein} we cannot easily talk
about a universal notion of a present instant,\footnote{Here `easily' anticipates that there is a sense in which Shape Dynamics defines a universal present.} but the huge degree of regularity that our Universe
exhibits is by no means diminished: it just translates into the fact that we all can usefully
make reference to a unique spacetime, and all local measurements of clocks (and distances,
see below)  are mutually consistent with a notion of spacetime.
The simplest and prototypical example of a relationalist clock is Tait's \e{inertial clock},
which is the position of particle 2 in his solution to the Scholium problem (cfr. Sec.~\ref{TaitsScholiumSolution}).

\subsection{Best-matching `in action'}
\label{BestMatchingInAction}

\index{best matching}
\subsubsection*{Stage I: Free-end-point variation}
\index{free-end-point variation}

Assume that we have a $G$-invariant metric $( \d {\bf r}_a ,\d {\bf r}'_b )_\st{G} = \sum_{a,b} \mat{M}^{ab}  \, \d {\bf r}_a \cdot \d {\bf r}'_b$ on $TP$.\footnote{Here I mean $G$-invariant
in the sense described in footnote \ref{FootGinvariantMetric}.} In accordance with
Stage I of the best-matching procedure, we start with a generic path ${\bf r}_a (s) : [s_1 ,s_2] \to P $, then lift it \emph{locally} along the fibres with the group action
\begin{equation}\label{GroupActionOnr_a}
{\bf r}_a (s)  \to O(s) \, {\bf r}_a (s)  \,,
\end{equation}
and look for the horizontal paths. If the metric is $G$-invariant, those paths will
minimize the length defined by the metric \index{kinematic metric}
\begin{equation}
\d \mathcal L_\st{bare} = \sqrt{ \left( \d {\bf r}_a (s)  , \d {\bf r}_a (s) \right)_\st{G}}= \left\| \d{\bf r}_a (s) \right\|_\st{G}\,,
\end{equation}
so we \e{we only need to vary the group elements:}
\begin{equation}
 S_\st{BM} =  \inf_{O(s) } \int \d \mathcal L =  \inf_{O(s) } \int  \left\| \d \left(O(s) \, {\bf r}_a (s)\right) \right\|_\st{G}\,, 
\end{equation}
\emph{but keeping the end points free}. This last requirement is of paramount importance. One normally takes the variation while keeping the end points fixed, which makes it possible to discard some boundary
terms and obtain the Euler--Lagrange equations. Here variation of the end points along the fibres corresponds to unphysical motions, so nothing allows us to keep them fixed.
The $G$-invariance of the metric makes it possible to rewrite the
variational principle as
\begin{equation}
S_\st{BM}  = \inf_{O(s)} \int   \left\| \d  {\bf r}_a  +  O^{-1} \d O   \, {\bf r}_a \right\|_\st{G} \,.
\end{equation}
If $O$ is a  matrix representation of a Lie group, the expression $\d \epsilon = O^{-1} \, \d O $ is the differential of a representation of the corresponding Lie algebra $O(s) = \exp \, \epsilon(s)$. We can therefore replace $O(s)$ by an $s$-dependent 
Lie algebra element $\epsilon (s)$:
\begin{equation}\label{Best-matchedAction}
S_\st{BM}  = \inf_{\epsilon (s)} \int     \left\| \d  {\bf r}_a + \d \epsilon  \, {\bf r}_a \right\|_\st{G} \,.
\end{equation}
The last expression looks like a covariant differential, which we can call the \emph{best-matching differential}  \index{best-matching differential} \cite{EdwardBigReviewRelationalParticleMechanics}
\begin{equation}
\dbm {\bf r}_a =  \d  {\bf r}_a  + \d \epsilon  \, {\bf r}_a \,.
\end{equation}
The \emph{free-end-point} Euler--Lagrange equations (see Appendix \ref{AppendixFreeEndPointVariation} for their derivation) for (\ref{Best-matchedAction}) give \index{Euler--Lagrange equations}
\begin{equation}
\d \left( \frac{\delta \d \mathcal{L}}{\delta \d \epsilon} \right) =\frac{\delta \d \mathcal{L}}{\delta  \epsilon} \,, \qquad  \left.  \frac{\delta \d \mathcal{L}}{\delta \d \epsilon} \right|_{s={s_1}} = \left.  \frac{\delta \d \mathcal{L}}{\delta \d \epsilon} \right|_{s={s_2}} = 0\,,
\end{equation}
but thanks to the $G$-invariance of the metric the $\epsilon$ variable is cyclic, $\frac{\delta \mathcal{L}}{\delta  \epsilon} = 0$, and then the equations, together with the boundary conditions $ \left. \frac{\delta \d \mathcal{L}}{\delta \d \epsilon} \right|_{s={s_1}} = \left. \frac{\delta \d \mathcal{L}}{\delta \d \epsilon} \right|_{s={s_2}} = 0$, imply
\begin{equation}
\d \left(  \frac{\delta \d \mathcal{L}}{\delta \d \epsilon} \right) =  \left. \frac{\delta \d \mathcal{L}}{\delta  \epsilon} \right|_{s={s_1}} = \left. \frac{\delta \d \mathcal{L}}{\delta \d \epsilon} \right|_{s={s_2}} = 0  ~~~ \Rightarrow ~~~ \frac{\delta \d \mathcal{L}}{\delta \d \epsilon} = 0 \,. \label{InitialConditionsFromBestMatching}
\end{equation}
What we have here found are the conditions of horizontality. This procedure doesn't work if the metric is not $G$-invariant because then the `Lagrangian' depends on $\epsilon (s)$ as well as on $\d \epsilon (s)$.

Notice that the free-end-point variation of a cyclic coordinate is equivalent to the regular variation of a Lagrange multiplier:
if we define $\lambda = \d\epsilon$, then the Lagrangian depends on $\lambda$ but not on its derivative. Its fixed-end-point
variation gives $\frac{\delta \d \mathcal  L}{\delta \lambda} =0$, which is equivalent to (\ref{InitialConditionsFromBestMatching}).\footnote{In gauge theory, \index{gauge theory} there are many `multipliers' \index{Lagrange multipliers} (like the scalar potential $A_0$ in Maxwellian electrodynamics) that, dimensionally, are velocities (and hence cyclic coordinates because no quantities of which they are velocities appear in the Lagrangian). It is therefore strictly irregular to treat them as multipliers. Free-end-point variation explains why mathematics that strictly is incorrect gives the right answer.}

\subsubsection*{Stage II: Noether's theorem part I}
\index{best matching}

Now, if the metric is $G$-invariant, the quantities $\frac{\delta \d \mathcal{L}}{\delta \d \epsilon}$ are \emph{constants of motion} for the
paths that minimize the following uncorrected action:
\begin{equation}
S_\st{bare} = \int \d \mathcal{L}_\st{bare} = \int  \left\| \d {\bf r}_a \right\|_\st{G} \,.
\end{equation}
In fact, the action is invariant under \emph{global}, time-independent $G$ transformations
\begin{equation}
{\bf r}_a (s)  \to \exp  \epsilon \, {\bf r}_a (s)  \,,
\end{equation}
and Noether's theorem, part I \cite{NoethersPaper} establishes that 
\begin{equation}
\d \left( \left.\frac{\delta \d \mathcal{L}_\st{bare}(\epsilon \, {\bf r}_a (s) ) }{\delta \epsilon} \right|_{\epsilon = 0}\right) = 0 \,,
\end{equation}
which states that  the quantities $ \frac{\delta \d \mathcal{L}}{\delta \d \epsilon}$ of the preceding paragraph
are conserved along the solutions. This is true, in particular, if they have the value zero. \emph{Therefore the solutions of the Euler--Lagrange equations
for $S_\st{bare}$ are horizontal if they start horizontal: $\left. \frac{\delta \d \mathcal{L}}{\delta \d \epsilon}\right|_{s=s_1}=0$}.

Thanks to the $G$-invariance of our bare action, the best-matching condition
reduces to nothing more than an \emph{initial condition} on the data: it is sufficient to take
the `bare' action $S_\st{bare}$ and find the path that minimizes it with initial conditions
given by (\ref{InitialConditionsFromBestMatching}).

\subsubsection*{The Newtonian $N$-body problem}

\index{best matching}
\index{N-body problem}
Let's apply the technique of best-matching to Newtonian  gravity. I will use 
the reparametrization-irrelevant formulation (\ref{irrel}) with the variations expressed with respect to
ephemeris time. It is sufficient to take 
the following metric on $\Q^N$:
\begin{equation}
\d  \mathcal L^2_\st{New} = 4 \, (E - V_\st{New}) \sum_{a=1}^N  \frac{m_a}{2} \d {\bf r}_a \cdot  \d {\bf r}_a \,,
\end{equation}
which is conformally related to the kinetic metric through the positive-definite\footnote{$E-V_\st{New}$
is actually positive-definite `on-shell', that is, on the physical trajectories, where it is equal to the
positive-definite kinetic energy. Notice that if $E<0$, there are forbidden regions in configuration space: those where the potential energy is smaller than $E$.}
conformal factor $E-V_\st{New}$, where 
\begin{equation} \label{NewtonsPotential}
V_\st{New} = - \sum_{a<b}  \frac{m_a \, m_b}{\| {\bf r}_a - {\bf r}_b \|} \,.
\end{equation}
The conformal factor $E -V_\st{New}$ is translation- and rotation-invariant,
which allows us to find the horizontal curves by minimizing the action
\begin{equation}\label{NewtonianBestMatchingAction}
\int \d  \mathcal L_\st{New}  = 2 \int{   \bigg( (E- V_\st{New}) \sum_{a=1}^N \frac{m_a}{2}  \left\| \dbm  {\bf r}_a  \right\|^2 \bigg)^{\frac 1 2}} \,,
\end{equation}
where the best-matching differential here is
\begin{equation}
\dbm {\bf r}_a =  \d  {\bf r}_a    + \d  {\bm \omega}   \times {\bf r}_a + \d{\bm \theta}  \,.
\end{equation}
Varying w.r.t. $\d {\bm \omega}$ with free endpoints, I find
\begin{equation}\label{BMconditionRotationsNbodyProblem}
\frac{\delta \d \mathcal L_\st{New} }{\delta \d {\bm \omega}} = - \frac 1 {\d \doti}  \sum_{a=1}^N m_a  \, \dbm {\bf r}_a \times {\bf r}_a = 0 \,,
\end{equation}
where
\begin{equation}\label{NbodyDOTI}
\d \doti = (E-V_\st{New})^{-\frac 1 2} (\sum_a {\sfrac {m_a} 2} \|\dbm {\bf r}_a\|^2)^{\frac 1 2} \,.
\end{equation}
The expression here on the rhs is a sort of precursor of the ephemeris time, but it is not yet that since it depends on the auxiliary quantities
$\d {\bm \omega}$ and $\d {\bm \theta}$; I introduce $\d \doti$ merely to simplify the equations.\footnote{I use $\chi$ for
this differential quantity as the initial letter of the Greek word {\greektext qr\'onos} (chronos), referring
to a \emph{quantitative} notion of time (duration).} As I commented on Eq.~(\ref{DifferentialOfTheInstant}), E. Anderson calls it the \emph{differential of the instant} \index{differential of the instant}
 \cite{EdwardBigReviewRelationalParticleMechanics,EdFlavioPaper}.

 Varying w.r.t. $\d{\bm \theta}$, I get
\begin{equation}\label{BMconditionTranslationsNbodyProblem}
\frac{\delta\d  \mathcal L_\st{New}}{\delta \d {\bm \theta}} =   \frac 1 {\d \doti}  \sum_{a=1}^N m_a \, \dbm {\bf r}_a= 0 \,.
\end{equation}

If we now define the \emph{canonical momenta} \index{canonical momenta} \cite{Lanczos,Goldstein} ${\bf p}^a$ as
\begin{equation}
{\bf p}^a = \frac{\delta \d \mathcal L_\st{New}}{ \delta \d {\bf r}_a} = m_a \frac { \dbm {\bf r}_a}{\d \doti} \,,
\end{equation}
then the Euler--Lagrange equations read
\begin{equation}
\d {\bf p}^a = \frac{\delta \d \mathcal L_\st{New}}{ \delta {\bf r}_a} = - \d \doti  \frac{\partial V_\st{New}}{\partial {\bf r}_a} - \d  {\bm \omega}   \times{\bf p}^a \,,
\end{equation}
where the term $\d  {\bm \omega}   \times{\bf p}^a $ is due to the ${\bf r}_a$-dependence of  $\dbm {\bf r}_a $.
This term can be reabsorbed into a best-matched differential of the momentum:\footnote{The best-matched differential cannot act in the same way on  ${\bf r}_a$ and the momenta: for example the latter are translation-invariant. This will be made more precise in the Hamiltonian formulation.}
\begin{equation}
\dbm {\bf p}^a = \d {\bf p}^a + \d  {\bm \omega}   \times{\bf p}^a \,,
\end{equation}
and then the equations of motion take an almost-Newtonian form:
\begin{equation}
\frac{\dbm {\bf p}^a}{\d \doti} = 
m_a \frac{\dbm }{\d \doti} \left( \frac{\dbm {\bf r}_a}{\d \doti} \right) = -   \frac{\partial V_\st{New}}{\partial {\bf r}_a}  \,.
\end{equation}

As I said, $\d \doti$ is not an ephemeris time because of its dependence on the auxiliary
quantities $\d {\bm \omega}$ and $\d {\bm \theta}$. However, if we solve (\ref{BMconditionTranslationsNbodyProblem}) and (\ref{BMconditionRotationsNbodyProblem})
for $\d {\bm \omega}$ and $\d {\bm \theta}$, call the solutions 
 $\d  {\bm \omega}_\st{BM}$ and $ \d{\bm \theta}_\st{BM}$ and substitute them into  (\ref{NbodyDOTI}),
we do obtain the actual ephemeris time: \index{ephemeris time}
\begin{equation}\label{NewtonianNbodyEphemerisTime}
\d \teph = (E-V_\st{New})^{-\frac 1 2} \bigg( \sum_a m_a \| \d  {\bf r}_a    + \d  {\bm \omega}_\st{BM}   \times {\bf r}_a + \d{\bm \theta}_\st{BM} \|^2 \bigg)^{\frac 1 2}\,.
\end{equation}
We can call $\d  {\bf r}_a    + \d  {\bm \omega}_\st{BM}   \times {\bf r}_a + \d{\bm \theta}_\st{BM} $ the
`horizontal' differential because it measures the physical variation of ${\bf r}_a $ and contains no part due to a rigid
translation or rotation of the whole universe, and is therefore invariant under \Eucl. \index{Euclidean group} The ephemeris
time is invariant as well, and it measures a distillation of all the \emph{physical} change in the
universe.

The best-matching conditions (\ref{BMconditionRotationsNbodyProblem}), (\ref{BMconditionTranslationsNbodyProblem}) take a very intuitive form if expressed
in terms of the canonical momenta:
\begin{equation} \label{ConstraintsRotationTranslations}
{\bf  L} = \sum_{a=1}^N {\bf r}_a \times {\bf p}^a = 0\,, \qquad {\bf P} = \sum_{a=1}^N {\bf p}^a = 0\,.
\end{equation}

We can now exploit the $G$-invariance of the metric as a shortcut to stage II of
the best-matching procedure. The paths that minimize the bare action
\begin{equation}
S_\st{bare} =  2 \int{   \bigg( (E - V_\st{New}) \sum_{a=1}^N \frac{m_a}{2}  \left\| \d  {\bf r}_a   \right\|^2 \bigg)^{\frac 1 2}} 
\end{equation}
and start with zero total angular and linear momenta, \index{angular momentum}
\begin{equation}
\sum_{a=1}^N m_a \,  \d {\bf r}_a  = \sum_{a=1}^N m_a \, \d {\bf r}_a \times {\bf r}_a = 0\,, 
\end{equation}
project under $\Q^N \to \Q^N_R$ to the physical solutions of the best-matched theory in $\Q_R^N$. Minimizing $S_\st{bare}$ with the above initial conditions gives 
Euler--Lagrange equations (\ref{RepInvariantEquationsOfMotion}) that take a Newtonian form if expressed in terms of the ephemeris time for which $T_\st{kin} = E - V_\st{New}$.

If the energy is zero, $E=0$, we see here how this kind of dynamics satisfies the Mach--Poincar\'e Principle.  \index{Mach--Poincar\'e Principle}
The physical, observable initial data ($3N-6$ positions and $3N-6$ differentials) are alone enough to uniquely determine a physical trajectory. The total angular and
linear momenta are set to zero
by the constraints (\ref{ConstraintsRotationTranslations}) (the angular momentum represents the `missing' $3$ data in Poincar\'e's analysis). The dynamics
determines by itself an inertial frame of reference and an intrinsic notion
of duration (the ephemeris time) such that Newton's equations hold.
In this frame, the total angular momentum of the Universe must be zero, otherwise
the Mach--Poincar\'e Principle would be violated. We see that relational dynamics
not only provides a deeper and intrinsic foundation for Newton's dynamics
(finding the physical origin of reference frames and, ultimately, inertia), but it 
also imposes physical predictions that make it \emph{more restrictive} than Newton's theory.
If we could show that our Universe possesses angular momentum, that would rule
out Machian dynamics (linear momentum wouldn't be observable
anyway due to Galilean invariance). \index{energy}

A nonzero energy is an element of arbitrariness that would imply a (mild)
violation of the Mach--Poincar\'e Principle: observable initial data would
fail to be enough to determine the future evolution, by just a single datum,
and an observation of one single second derivative of the relational data
would be enough to fix the value of $E$. Therefore a nonzero energy,
despite being possible to describe in Machian terms, is disfavoured.

\subsubsection*{The scale-invariant $N$-body problem}

\index{scale-invariant N-body problem}
If we want to incorporate the relativity of scale in a simple way, we might choose
a different metric on $\Q^N$ which is manifestly scale-invariant:
\begin{equation}\label{ScaleInvarianMetric}
\d \mathcal L^2_\st{S} = - 4 \, V_\st{S} \sum_{a=1}^N  \frac{m_a}{2}  \d {\bf r}_a \cdot  \d {\bf r}_a  \,, \qquad V_\st{S} = \frac{V_\st{New}}{{I_\st{cm}^{\frac 1 2}}} \,,
\end{equation}
where $I_\st{cm}$ is what I call the `centre-of-mass moment of inertia':\footnote{That object is
actually half the trace of the centre-of-mass inertia tensor, defined as $\mat{I}=\sum_a m_a \, ({\bf r}_a - {\bf r}_\st{cm}) \otimes ({\bf r}_a- {\bf r}_\st{cm}) $.}
\begin{equation}
I_\st{cm} = \sum_{a=1}^N \, m_a \| {\bf r}_a - {\bf r}_\st{cm} \|^2  \,,
\end{equation}
or, written in a more relational way (due to Leibniz),
\begin{equation}\label{CenterOfMassMomentOfInertia}
I_\st{cm} = \sum_{a<b}  \frac{m_a\,m_b}{m_\st{tot}} \| {\bf r}_a - {\bf r}_b \|^2  \,.
\end{equation}
Notice that I can't put  values of the energy $E$ other than zero in (\ref{ScaleInvarianMetric})
because that would make the metric non-scale invariant.

The best-matching action is 
\begin{equation}
\int \d \mathcal L_\st{S} = 2 \int{\bigg( - V_\st{S} \sum_{a=1}^N \frac{m_a}{2}  \left\|  \dbm  {\bf r}_a  \right\|^2 \bigg)^{\frac 1 2}}  \,,
\end{equation}
where in this case
\begin{equation}
\dbm {\bf r}_a =  \d  {\bf r}_a +  \d  {\bm \omega}   \times {\bf r}_a +  \d \phi  \, {\bf r}_a +  \d{\bm \theta} \,.
\end{equation}
Note the scalar auxiliary variable $\phi \in \doublestroke{R}^+$, which corrects for dilatations.
The free-end-point variations of this action are identical in form to the ones of the previous 
paragraph. These now include the one related to scale transformations:
\begin{equation}\label{BMconditionDilatationsNbodyProblem}
\frac{\delta \d \mathcal L_\st{S}}{\delta \d \phi} =  \frac 1 {\d \doti}  \sum_{a=1}^N m_a  \, {\bf r}_a \cdot \left(  \dbm  {\bf r}_a  \right) = 0 \,,
\end{equation}
which expresses the vanishing of the total  \index{dilatational momentum} \emph{dilatational momentum}:\footnote{This quantity has the same dimensions as angular momentum, and the name for it was coined by analogy in~\cite{Anderson:2002ey}. It has not been given a name in the $N$-body literature and is usually denoted by $\mathcal J$, probably for Jacobi.}
\begin{equation}\label{DilatationalMomentumConstraint}
\ro{D} = \sum_{a=1}^N {\bf r}_a \cdot {\bf p}^a = 0\,.
\end{equation}
The equations of motion are
\begin{equation}
\frac{\dbm {\bf p}^a}{\d \doti} = - \frac{\partial V_\st{S}}{\partial {\bf r}_a} \,,
\end{equation}
where in this case the best-matching differential of the momentum gains a correction coming from dilatations:
\begin{equation}
\dbm {\bf p}^a = \d  {\bf p}^a + \d  {\bm \omega}   \times {\bf p}^a -  \d \phi  \, {\bf p}^a\,.
\end{equation}

In this case, the bare action,
\begin{equation}
S_\st{bare} =  2 \int{ \bigg( - V_\st{S} \sum_{a=1}^N \frac{m_a}{2}  \left\| \d  {\bf r}_a \right\|^2 \bigg)^{\frac 1 2}}  \,,
\end{equation}
also conserves the dilatational momentum, and therefore its Euler--Lagrange equations
give a representation in $\Q^N$ of the physical trajectories if one imposes the 
condition (\ref{DilatationalMomentumConstraint}) on the initial data.
The equations of motion for the bare action in ephemeris time are
\begin{equation}
\frac{\d {\bf p}^a}{\d t _\st{eph}}= m_a \frac{\d^2 {\bf r}_a}{\d t _\st{eph}^2}  =  - \frac{\partial V_\st{S} }{ \partial {\bf r}_a}
= - I_\st{cm}^{-\frac 1 2}  \frac{\partial V_\st{New} }{ \partial {\bf r}_a}
-  m_a \left( {\bf r}_a - {\bf r}_\st{cm} \right)  I_\st{cm}^{- \frac 3 2} \, V_\st{New} \,  \,.
\end{equation}
The dilatational momentum is zero, and  this implies that the centre-of-mass moment
of inertia is conserved, because
\begin{equation}
\frac{\d  I_\st{cm}}{\d t _\st{eph}} = 2 \, \ro{D} = 0 \,. 
\end{equation}
We therefore obtain a theory which is Newtonian gravity with $I_\st{cm}^{\frac 1 2}$ playing
the role of a gravitational constant, plus a `cosmic' force parallel to $ {\bf r}_a - {\bf r}_\st{cm}$,
and therefore pointing towards (or away from) the centre of mass, which keeps
$I_\st{cm}$ constant. In a universe made of at least $\sim 10^{80}$ particles, this 
`cosmic' force would be virtually undetectable through local observations (at the scale
of the solar system or even at the scale of galaxy clusters). In fact, the accelerations of localized systems due to this force would be almost identical, both in direction and magnitude, and thus undetectable by virtue of the equivalence principle.

\vspace{12pt}
\noindent
\fbox{\parbox{0.98\linewidth}{
{\bf Further reading:} Regarding principal fibre bundles, I suggest the book by G\"{o}ckler and
Schucker \cite{GockelerSchucker}, the review by Eguchi, Gilkey and Hanson \cite{EguchiHanson},
and Frankel's book \cite{Frankelbook}.
The masterpieces on the variational principles of mechanics
are Lanczos~\cite{Lanczos}, Goldstein~\cite{Goldstein} and Arnold~\cite{ArnoldBook}.
}}

\newpage

\section{Hamiltonian formulation}\label{SectionHamiltonianFormulation}

\index{Hamiltonian formulation}
The Hamiltonian formulation of the kind of systems we're interested in is nontrivial.
In fact the standard formulation fails to be predictive, precisely because of the 
relational nature of our dynamics. There are redundancies in the description, and
this means that the usual Legendre transform that is used to define the Hamiltonian
is singular, and the momenta are related to the velocities by one-to-many mappings. This situation is described through \emph{nonholonomic constraints} \index{nonholonomic constraints}
(meaning constraints that depend not only on the coordinates, but also on the momenta).

In fact, after the two-stage procedure described above, best matching \index{best matching}  leads to a 
simple set of constraints on the canonical momenta, namely Eqs.~(\ref{ConstraintsRotationTranslations}),
with the possible addition of (\ref{DilatationalMomentumConstraint}) if 
relativity of scale is assumed. There is also a constraint associated to temporal relationalism,
as I will show now.

\subsection{The Hamiltonian constraint}

Reparametrization-invariant theories are characterized by having a vanishing Hamiltonian.
We can see it in the prototype for these theories: the geodesic-generating action (\ref{GeodesicAction}),
whose Lagrangian is
\begin{equation}
 \mathcal L =  \left( g^{ij} \frac{\d x_i}{\d s} \frac{\d x_j}{\d s}  \right)^{\frac 1 2} \,.
\end{equation}
Its canonical momenta are \index{canonical momenta}
\begin{equation}
p^i = \frac{\delta \mathcal{L}}{\delta \dot x_i } =  \left( g^{k\ell} \frac{\d x_k}{\d s} \frac{\d x_\ell}{\d s}  \right)^{- \frac 1 2}  g^{ij} \frac{\d x_j}{\d s} \,,
\end{equation}
We see that the momenta, like the geodesic action $\int \d s \, \mathcal L $ \index{geodesic} from which they are derived, are themselves reparametrization-invariant: they can be written in the parametrization-independent form
\begin{equation}
p^i =  \left( g^{k\ell} \, \d x_k \,\d x_\ell   \right)^{- \frac 1 2}  g^{ij} \, \d x_j \,.
\end{equation}
The other thing one can observe is that the momenta have the form of $N$-dimensional
\emph{direction cosines}, because as a vector they have unit length, so the following
phase-space function vanishes \emph{weakly}:\footnote{Strong (=) and
weak ($\approx$) equations will be defined below. Here, the difference is immaterial. \index{weak equalities}}
\begin{equation}
 H = g_{ij} \, p^i \, p^j - 1 \approx 0 \,, \label{HamConstraint1}
\end{equation}
where $g^{ij}$ is the inverse metric. We see from (\ref{HamConstraint1}) that the velocities $\frac{\d x_i}{\d s} $
are under-determined by the momenta (their norm is not determined by the
momenta, which have unit norm). This is what Dirac \index{Paul Adrien Maurice Dirac} \cite{dirac:lectures} calls a 
\emph{primary constraint}: \index{primary constraint} an algebraic relation satisfied by the momenta
by virtue of their mere definition and not due to any variation.

By its definition, the canonical Hamiltonian vanishes, \index{canonical Hamiltonian}
\begin{equation}\label{CanonicalHam}
\mathcal H_\st{can} := \sum_i p^i \, \frac{\d x_i}{\d s} - \mathcal L ~{\equiv}~ 0 \,,
\end{equation}
but, as discussed below, the presence of the primary constraint 
(\ref{HamConstraint1}) implies that the true generator of the dynamics is not (\ref{CanonicalHam}) alone. 
It is instead (\ref{CanonicalHam}) 
plus a linear  combination of the primary constraints, which in our case is just (\ref{HamConstraint1}):
\begin{equation}\label{TotHamiltonian}
\mathcal H_\st{tot} = \mathcal H_\st{can} +  u \, \left( g_{ij} \, p^i \, p^j - 1  \right)  \,,
\end{equation}
where $u =u(s)$ is an arbitrary time-dependent function (whose conjugate momentum does not appear
in the total Hamiltonian and is assumed to vanish).

Dirac \cite{DiracHamiltonianDynamics} developed a general theory of constrained Hamiltonian systems and presented it in his beautiful lectures \cite{dirac:lectures}. I will now reformulate everything according to 
Dirac's theory. For the readers who are not familiar with the subject, I will start with a quick review of the technique.

\subsection{A crash course in Dirac's constraint analysis}\label{CrashCourseInDiracsAnalysis}

Consider a Hamiltonian system which is subject to a set of constraints like  (\ref{ConstraintsRotationTranslations}), (\ref{DilatationalMomentumConstraint}) and (\ref{HamConstraint1}). The constraints will be expressed through a set of  phase-space functions $\phi_a = \phi_a (p,q)$. When all of these functions vanish the constraints are satisfied. This identifies
implicitly a hypersurface in phase space (the \emph{constraint surface}). With the
notion of constraints comes that of \emph{weak equivalences}: two phase-space 
functions $f,g$ are weakly equivalent if their difference is a linear combination of the
constraints, $ f \approx g ~~ \Leftrightarrow f - g = \sum_a u_a \, \phi_a$. A function
that is equal to a linear combination of the constraints (which means it is zero on
the constraint surface) will be called \e{weakly vanishing}, \index{weak equalities}
and an equation that holds only on the constraint surface will be called a \e{weak equation},
as opposed to \emph{strong equations}, which hold everywhere in phase space.

Dirac \cite{dirac:lectures} starts by noticing that, in the presence of constraints, Hamilton's  \index{Hamilton's equations}
equations do not follow from the minimization of the canonical action $\delta \int \d s \left( p^i \, \dot q_i  - \mathcal H_\st{can} \right) $. In fact, when taking its variation,
\begin{equation} \label{CanonicalHamiltonianConstrained}
\delta \int \d s \left( p^i \, \dot q_i  - \mathcal H_\st{can} \right) 
= \int \d s  \left[ \left(\dot q_i - {\sfrac {\delta \mathcal H_\st{can}}{\delta p^i}} \right) \delta p^i  - \left(\dot p^i + {\sfrac{\delta \mathcal H_\st{can}}{\delta q_i}} \right) \delta q_i \right] = 0 \,, 
\end{equation}
one is not entitled to separately put to zero all the coefficients of $\delta p^i$ and $\delta q_i$. This because one cannot
take arbitrary variations $\delta p^i$, $\delta q_i$: they are constrained by the conditions $\phi_a \approx 0$.
The most generic variation one can take is one that keeps the phase-space vector $(\delta p^i , \delta q_i)$
tangent to the hypersurface $\phi_a = 0$ (a variation that keeps you on that hypersurface).

There is no metric on phase space, but there is enough to define orthogonality and parallelism: it's
the \emph{symplectic structure}. \index{symplectic structure} The reader can find more details in Arnold's book \cite{ArnoldBook}.
For our purposes, it is sufficient to say that, in the case of a variation constrained
to $\phi_a = 0$, Eq.~(\ref{CanonicalHamiltonianConstrained}) imposes a weaker set of conditions, namely
\begin{equation}
\frac{\delta \mathcal H_\st{can}}{\delta q_i} + \dot p^i =  \sum_a u^a  \, \frac{\partial \phi_a}{\partial q_i} \,, ~~~~ 
\frac{\delta \mathcal H_\st{can}}{\delta p_i} - \dot q_i =  \sum_a  u^a  \, \frac{\partial \phi_a}{\partial p^i} \,,
\end{equation}
for any choice of  $u_a$. The $u^a$'s purpose is to generate the whole tangent hyperplane to the  surface $\phi_a =0 $
at each point of it. The above equations are Hamilton's equation for a \emph{generalized Hamiltonian} \index{generalized Hamiltonian}
\begin{equation}\label{GeneralizedHamiltonian}
\mathcal H^* = \mathcal H_\st{can} + \sum_a u^a \,\phi_a \,.
\end{equation}
The evolution of a phase-space function $f$ under this generalized Hamiltonian
can be written in terms of Poisson brackets \index{Poisson brackets}
\begin{equation}\label{GeneralizedEOM}
\dot f = \{ f , \mathcal H^* \} =  \frac{\partial f}{\partial q_i} \frac{\partial \mathcal H^*}{\partial p^i} -
 \frac{\partial f}{\partial p^i} \frac{\partial \mathcal H^*}{\partial q_i}  \,.
\end{equation}
The equations of motion (\ref{GeneralizedEOM}) only make sense if the constraints $\phi_a$ are preserved by them, for otherwise
the evolution brings us out of the constraint surface $\phi_a = 0$. So, for consistency, we have to require that
\begin{equation} \label{ConsistencyCondition}
\dot \phi_a = \{ \phi_a , \mathcal H^* \} = \{\phi_a , \mathcal H_\st{can} \} + \sum_b u^b \{\phi_a , \phi_b \} \approx 0 \,.
\end{equation}
Recall that Dirac's weak equality `$\approx 0$' means that the result of the above calculation gives a combination of the constraints $\phi_a$ that vanishes when the on-shell condition $\phi_a \approx 0$ is imposed.

\index{weak equalities}

\index{primary constraint}
\index{secondary constraint}

Equations (\ref{ConsistencyCondition}) are the core of Dirac's analysis. There are 4 cases:
\begin{enumerate}
\item There are $a$'s for which Eqs.~(\ref{ConsistencyCondition}) have no solution. Then the system is not consistent
and the equations of motion admit no sensible solution, like the famous
Lagrangian $\mathcal L = q$, whose Euler--Lagrange equation is $1=0$.

\item Some of Eqs.~(\ref{ConsistencyCondition})  admit a nontrivial solution, and 
that solution does not depend on any $u_a$'s. Then all such solutions impose new constraints $\phi'_a$ 
on the $p^i$, $q_j$. Dirac calls the $\phi'_a$  \emph{secondary} constraints, 
and they need to be treated on the same footing as the $\phi_a$'s. Then one writes 
down a new modified Hamiltonian $\mathcal H^{**} = \mathcal H_\st{can} + \sum_a u^a \,\phi_a + \sum_b u'^b \,\phi'_b$
and applies the procedure again from the start. 

\item Some other equations might admit a nontrivial solution which depends on the $u^a$'s. Each such equation will fix one of the $u^a$'s as a function of $p,q$. Dirac calls these equations `specifiers'.
Each equation of this kind is associated with a \emph{second-class} constraint. Second-class constraints
are defined by the fact that their Poisson brackets with at least one other constraint in the
theory are not weakly vanishing.

\item All the equations that do not fall into cases 2 or 3 will  simplify to tautologies of the form $1=1$. 
There will be one such equation for each \e{first-class} constraint. First-class constraint are defined
by the fact that their Poisson brackets with every other constraint in the theory are weakly
vanishing.
\end{enumerate}
If we never stumble upon case 1, after a few iterations of the procedure all of the equations
will fall into cases 3 or 4 Then the system is well defined, and we can stop. 
The evolution will be finally generated by a \emph{total} Hamiltonian,
 \begin{equation}\label{TotalHamiltonian}
\mathcal H_\st{tot} = \mathcal H_\st{can} + \sum_{a \,\in \,\st{\it first-class}} \!\!\!\! u^a \,\phi_a
+  \sum_{b \,\in \,\st{\it second-class}} \!\!\!\!\!\! u^b(p,q) \,\phi_b \,,
\end{equation}
which is the canonical Hamiltonian plus a linear combination of all the leftover
first-class constraints (primary, secondary, \dots), and a linear combination of the second-class
constraints with the $u^a=u^a(p,q)$ which have been fixed by the specifier equations.

\index{first-class constraints} \index{second-class constraints}
\index{total Hamiltonian}

\subsubsection*{Dirac's theorem}

Whenever our algorithm stops (and the system turns out to be consistent), but some of the $u^a$'s are not specified by any 
`case 3' condition, then these $u^a$'s are \emph{gauge} degrees of freedom (like the position of the centre of mass in relational
systems).

As we said, the $u^a$'s that don't end up specified are those related to \emph{first-class} constraints, meaning that
they Poisson-commute with all the other constraints $\phi_a$ and with $\mathcal H_\st{can}$, so that for them
Eq.~(\ref{ConsistencyCondition}) falls into case 4.

\index{Paul Adrien Maurice Dirac}

The following is referred  to by some as `Dirac's theorem': 
\e{primary first-class constraints treated as generating functions of infinitesimal contact transformations lead to changes that do not affect the physical state}.

Consider the simplest case of a non-vanishing canonical Hamiltonian $\mathcal H_\st{can} $ and one single 
first-class constraint $\phi$: \index{canonical Hamiltonian}
$$
\dot f = \{f , \mathcal H_\st{can} \} + u \, \{ f , \phi\}.
$$
If $\phi$ is first-class (so that $\{ \phi , \mathcal H_\st{can} \} \approx 0$), then $u$ is not specified
by any condition and is left as arbitrary.  This arbitrariness is absent only if one considers phase-space
variables $y$ that are first-class with respect to $\phi$, $\{ y , \phi \} \approx 0$. Then $u$ does not
appear in the evolution of $y$. Variables like $y$ are gauge-invariant, and 
the absence of any arbitrariness in their evolution signals that they are physical
and are the only candidates for observables. Non-gauge invariant quantities  can be used in
the description of the system, but they depend on conventional choices, or gauges.

Let's show a simple example to make things concrete. Consider two free particles on a line,
with coordinates $q_1$ and $q_2$. The canonical Hamiltonian is
\begin{equation}
\mathcal H_\st{can} = \frac{p_1^2}{2m_1} + \frac{p_2^2}{2m_2}  \,,
\end{equation}
and say we have the constraint
\begin{equation}
\phi = p_1 + p_2\,,
\end{equation}
which is obviously first-class  as it is the one and only constraint in the model,
and a single constraint will always Poisson-commute with itself. 
The total Hamiltonian is $\mathcal H_\st{tot}  = \mathcal H_\st{can} + u \, \phi $,
and it generates the time evolution
\begin{equation}
\dot q_1 = \frac{p_1}{m_1} + u 
\,, \qquad 
\dot q_2 = \frac{p_2}{m_2} + u \,,
\,, \qquad 
\dot p_1 = \dot p_2 =0 \,.
\end{equation}
The solution to these equations is
\begin{equation}
q_i(s) =  \frac{p_i}{m_i} \, (s-s_1) +  \int_{s_1}^s \d s' \,  u(s') + q_i (s_1)
\,, \qquad 
p_i (s) = p_i (s_1)\,,
\end{equation}
and depends on four integration constants, $q_i (s_1)$, $p_i (s_1)$, and the arbitrary parameter $u(s)$.
The integration constants set the initial values of the phase-space variables, at $s=s_1$. 
The constraint $\phi \approx 0$ constrains these initial values to satisfy $p_1(s_1) = - p_2 (s_1)$.

We see that while the constraint $\phi \approx 0$ fixed the total canonical momentum $p_1 + p_2$  to be zero, the gauge choice
allows us to represent the system in a frame in which the total quantity $m_1 \, \dot q_1
+ m_2 \, \dot q_2 = m_\st{tot} \, u $ is nonzero.
The relative distance between the two particles $q_1 - q_2$ is gauge-invariant $\{ q_1 - q_2 , \phi \} = 0$, its equations of motion contain no arbitrariness,
\begin{equation}
\dot q_1 - \dot q_2 = \frac{p_1}{m_1} -
 \frac{p_2}{m_2}  \,,
\end{equation}
and it is the only gauge-invariant quantity in the system apart from the momenta $p_1$, $p_2$
(which however are constrained to sum to zero, and therefore have only one independent degree of freedom).
Therefore, there are two physical Hamiltonian degrees of freedom.
\index{gauge theory}
\index{physical degrees of freedom}

\subsubsection*{Barbour and Foster's exception to Dirac's theorem}\label{Barbour-FosterSubsec}

\index{first-class constraints} \index{second-class constraints}
\index{total Hamiltonian}
Julian Barbour and Brendan Foster found a bug in Dirac's theorem~\cite{BF-DiraTheorem}
that relates  to the case we're interested in: when the canonical Hamiltonian vanishes and the total Hamiltonian
is just a linear combination of constraints.

Consider again the case of a single first-class constraint $\phi$, but when the canonical Hamiltonian vanishes identically:\footnote{`Strongly' in Dirac's terminology.} {$\mathcal  H_\st{can}~\equiv~0$. Then
\begin{equation}
\dot f = \frac{\d f }{\d s} = u \, \{ f , \phi\} \,,
\end{equation}
and now it is true that $u$ is arbitrary, but the effect of changing $u$ is just
a reparametrization. The same equation written as
$$
\frac 1 {u} \frac{\d f }{\d s} =  \{ f , \phi\}
$$
is invariant under $s \to s'(s)$, $u \to u \, \frac{\partial s}{\partial s'}$.
Therefore this particular $\phi$ generates physical change.
If in the example above of the two particles on a line we had  $\mathcal  H_\st{can} =0$, then the equations of motion would reduce to
\begin{equation}
\dot q_1 = u 
\,, \qquad 
\dot q_2 = u \,,
\,, \qquad 
\dot p_1 = \dot p_2 =0 \,.
\end{equation}
with solution 
\begin{equation}
q_i = \int^s_{s_1} \d s' \, u(s') + q_i (s_1) \,, \qquad p_i = \text{\it const}\,.
\end{equation}
Now, we cannot say that the constraint  generates unphysical change, because it clearly moves the representative
point on a dynamical trajectory. Changing the function $u=u(s)$ merely amounts
to changing the parametrization of the same dynamical trajectory.

\subsection{Application to our systems}

\subsubsection*{Differential almost-Hamiltonian formulation}

We will here use the formulation proposed by Edward Anderson \cite{EdwardBigReviewRelationalParticleMechanics} as in Eq.~(\ref{irrel}).
In place of the Lagrange multipliers $u^a$, we use differentials $\d \xi^a$ of which we consider the  \index{free-end-point variation} 
free-end-point variation.\footnote{As we saw in Sec.~\ref{BestMatchingInAction} the free-end-point variation of a 
cyclic coordinate \index{cyclic coordinates} is equivalent to the regular variation of a Lagrange multiplier.} This enables us to
implement `parametrization irrelevance' at all stages. The Newtonian best-matching
action (\ref{NewtonianBestMatchingAction}) is extremalized by a path generated by the `differential-almost-Hamiltonian' object
\begin{equation}
\d \almostHamiltonian= \d \doti \, \ro{H} + \d {\bm \theta} \cdot {\bf P} + \d {\bm \omega} \cdot {\bf L} \,,
\end{equation}
which is a linear combination (through differentials) of the constraints \index{angular momentum} \index{canonical momenta}
\begin{equation} \label{ConstraintsRotationTranslationsHamiltonian}
{\bf  L} = \sum_{a=1}^N {\bf r}_a \times {\bf p}^a \,, \qquad {\bf P} = \sum_{a=1}^N {\bf p}^a \,, \qquad  \ro{H}  =  \sum_{a=1}^N \frac{ {\bf p}^a \cdot {\bf p}^a }{2 \, m_a} - U \,.
\end{equation}
The evolution of a phase-space function $f$ is generated by Poisson-commuting $f$ with $\d \almostHamiltonian$:
\begin{equation}
\d f = \{ \d \almostHamiltonian , f \} = \d \doti \,\{  \ro{H}  , f \} + \d {\bm \theta}  \cdot  \{{\bf P}  ,  f  \}  + \d {\bm \omega} \cdot \{ {\bf L} , , f \} \,,
\end{equation}
and therefore the best-matching differential of $f$ is generated by the Hamiltonian constraint $\ro{H}$ `smeared'  with $\d \doti$,\footnote{`Smearings' are defined below in the Relational Field Theory Part.}
\begin{equation}
\dbm f = \d \doti \,\{  \ro{H}  , f \}  = \d f - \d {\bm \theta}  \cdot  \{ {\bf P}  ,  f  \}  - \d {\bm \omega} \cdot \{ {\bf L}   , f \} \,.
\end{equation}
One sees immediately that this definition reproduces the correct action of the best-matching differential
on the coordinates and the momenta:
\begin{equation}\begin{aligned}
\d {\bf q}_a -\{  \d {\bm \theta}  \cdot   {\bf P}  , {\bf q}_a   \}  - \{ \d {\bm \omega} \cdot  {\bf L}   , {\bf q}_a  \} &= \d {\bf q}_a + \d {\bm \theta}   + \d {\bm \omega} \times {\bf q}_a  \,, \\
\d {\bf p}^a - \{ \d {\bm \theta}  \cdot  {\bf P}  , {\bf p}^a   \}  - \{ \d {\bm \omega} \cdot {\bf L}   , {\bf p}^a  \} &= \d {\bf p}^a + \d {\bm \omega} \times {\bf p}^a \,.  
\end{aligned}\end{equation}

The equations of motion are
\begin{equation}
\dbm   {\bf r}_a = \d \doti \,\{  \ro{H}  , {\bf r}_a \}   = \d \doti \,  \frac{ {\bf p}^a  }{ m_a}  \,,
\qquad 
\dbm  {\bf p}^a = \d \doti \,\{  \ro{H}  ,  {\bf p}^a \}  = \d \doti \, \frac{\partial U}{\partial  {\bf r}_a}   \,.
\end{equation}

The angular and linear momentum constraints close as a first-class system: 
\begin{equation}\begin{aligned}
\{ \ro{L}_i , \ro{P}_j \} = \epsilon_{ijk} \, \ro{P}_k \,, && \{ \ro{P}_i , \ro{P}_j  \} = 0 \,,
\end{aligned}\end{equation}
but their Poisson brackets with the Hamiltonian constraint depend on the potential $U$:
\begin{equation}\begin{aligned}
\{ {\bf  L}  , \ro{H}  \} =  \sum_{a=1}^N {\bf r}_a \times \frac{\partial U}{\partial  {\bf r}_a} \,, && \{  {\bf  P} ,  \ro{H}  \} =  \sum_{a=1}^N  \frac{\partial U}{\partial  {\bf r}_a}\,.
\end{aligned}\end{equation}

The only way to make the constraints propagate and obtain a consistent theory is to
have a potential that is invariant under global translations and rotations,
so that both of the above commutators vanish strongly.

If we insist on the relativity of scale, we have to add the dilatational momentum \index{dilatational momentum} constraint
\begin{equation} \label{ConstraintsDilatationHamiltonian}
\ro{D} = \sum_{a=1}^N {\bf r}_a \cdot {\bf p}^a \,,
\end{equation}
which is first-class with respect to the momentum constraints, but the same holds also for
the Hamiltonian constraint only for some choices of the potential,
\begin{equation}\begin{aligned}
\{ L_i , \text{D} \} = 0 \,, && \{ \text{D} , P_i  \} =  P_i \,, && \{ \text{H} , \text{D} \} = -  \sum_{a=1}^N \frac{ {\bf p}^a \cdot {\bf p}^a }{ m_a} +  \sum_{a=1}^N {\bf r}_a  \cdot \frac{\partial U}{\partial  {\bf r}_a}  \,,
\end{aligned}\end{equation}
namely, the energy has to be zero $E=0$  and the potential has to be \emph{homogeneous of degree -2} in order for $\{ \text{H} , \text{D} \} $ to weakly vanish. This is a consequence of Euler's homogeneous function theorem \cite{EulersHomgeneousFunctionTheorem}.

\subsection{A matter of units}

\index{best-matching}

Before moving to field theory, I want to make some remarks about dimensional analysis, \index{dimensional analysis}
which in a relational setting becomes a key -- and nontrivial -- point. There is much confusion
about the role of units in physics,\footnote{The reader might enjoy reading the interesting -- and inconclusive --  `trialogue'
between Duff, Okun and Veneziano \cite{duff2002trialogue} on the number of fundamental
constants in nature.} and the relational point of view highlights the issue and calls for clear thinking. I'll set the stage for my argument here, using only relational particle dynamics
as my prototype Machian theory.
The starting point is the best-matching action  for Newtonian gravity (\ref{NewtonianBestMatchingAction}).
The particle coordinates will be assumed of course to carry the dimensions of a length
$[{\bf r}_a] = \ell$. The masses are usually given dimension, but they are non-dynamical objects,
as they are constant in time and do not evolve. They are therefore `transparent' under 
all derivatives, and one can always remove the mass dimension from any equation. This can be easily achieved by dividing the action by the $3/2$th power of a reference mass, which is naturally assumed to be the total mass $m_\st{tot} = \sum_{a=1}^N m_a$.
Then the action only depends on dimensionless `geometric' masses $\mu_a = m_a / m_\st{tot}$
(notice how I rescaled and changed the units of the \index{energy} energy $E' =  E / m_\st{tot}^2$ which I can do because it is just
a constant):
$$
\int \d  \mathcal L_\st{New}  = 2 \int{   \bigg( (E' - V_\st{New}/ m_\st{tot}^2) \sum_{a=1}^N \frac{\mu_a}{2}  \left\| \dbm  {\bf r}_a  \right\|^2 \bigg)^{\frac 1 2}} \,.
$$
Let's now talk about Newton's constant \index{Newton's constant} $G$: in our expression for Newton's potential
(\ref{NewtonsPotential}) it didn't feature. In basic physics courses $G$ is introduced as
a conversion factor from mass$\times$length$^{-1}$ to accelerations. An acceleration
is length$\times$time$^{-2}$, but in our framework the independent variable is dimensionless: it is just a parameter on the evolution curve in configuration space, $[s] = 1$. It is ephemeris time \index{ephemeris time} $\d \teph$ (\ref{NewtonianNbodyEphemerisTime}) that plays the role of Newtonian absolute time.
Its dimensions can be read off its expression (in the mass-rescaled case):
$$
\d \teph = (E'-V_\st{New} /m_\st{tot}^2)^{-\frac 1 2} \bigg( \sum_a \mu_a \| \d  {\bf r}_a    + \d  {\bm \omega}_\st{BM}   \times {\bf r}_a + \d{\bm \theta}_\st{BM} \|^2 \bigg)^{\frac 1 2} \,.
$$
It is a length$^{3/2}$, $[\d \teph] = \ell^{3/2}$. Newton's law follows from our use of $\teph$ for
parametrization and of the best-matched coordinates ${\bf r}_a^\st{BM} ={\bf r}_a +  {\bm \omega}_\st{BM}   \times {\bf r}_a + {\bm \theta}_\st{BM} $,
\begin{equation}
\mu_a \, \frac{\d^2 {\bm r}_a^\st{BM}}{\d \teph^2}  =  \sum_{a<b} \frac{\mu_a \mu_b}{\| {\bf r}_a^\st{BM} - {\bf r}_b^\st{BM} \|^3} ( {\bf r}_a^\st{BM} - {\bf r}_b^\st{BM}) \,,
\end{equation}
(notice the appearance of rescaled masses only). The above equation is dimensionally consistent
without the need for any conversion factor -- $G$ simply doesn't appear. But what we call Newton constant and denote with $G$ was in effect measured by Henry Cavendish \index{Henry Cavendish} a bit over 200 years ago, so what is it?
\index{dimensional analysis}

Let's consider its definition~\cite{WikiNewtonConstant}: it is the gravitational force in Newtons exerted by one mass of 1~kg on another mass of 1~kg placed at a distance of 1~m. This definition is of course very unsatisfactory
from a relational point of view, and we would like to express it as a comparison. But there is 
another quantity in particle mechanics which has precisely the same definition, only with electric
charges in place of masses: it is Coulomb's constant. Its definition is~\cite{WikiCoulombsLaw}:
the electrostatic force in Newtons exerted by one charge of 1~C on another charge of 1~C placed at a distance of 1~m. This suggests a different understanding of Newton's constant: it is just the (dimensionless!) relative
magnitude between gravitational and other kind of interactions. If I were to include electrostatic interactions
in my relational particle model, I would use a potential of this form:
\begin{equation}
V_\st{N+C} =  - \sum_{a<b} \frac{\mu_a \mu_b}{\| {\bf r}_a - {\bf r}_b  \|} 
- \sum_{a<b} \frac{\epsilon_a \epsilon_b}{\| {\bf r}_a - {\bf r}_b \|}  \,,
\end{equation}
where $\epsilon_a$ are dimensionless electric charges. If the particles we are considering are
subatomic like the electron or the proton, the $\epsilon_a$'s are much larger than the $\mu_a$'s.
In fact, for a given particle, $\epsilon_a / \mu_a = (k_e/G)^{\frac 1 2} e_a / m_a$ where now
$k_e$ is Coulomb's constant and $G$ is Newton's constant if $e_a$ is expressed in Coulombs
and $m_a$ in kg. There's no such a thing as a universal gravitation constant or a
permittivity of vacuum: there are only (smaller or larger) dimensionless coupling constants.
The story changes in presence of `non-$1/r^2$' kinds of forces, like
harmonic oscillators or the Lennard--Jones potential.  If they are introduced in a na\"ive way, \emph{e.g.}:
\begin{equation}\label{LennardJonesPotential}
V_\st{N+C+L-J} =  - \sum_{a<b} \frac{\mu_a \mu_b + \epsilon_a \epsilon_b}{\| {\bf r}_a - {\bf r}_b \|} 
- \epsilon_0 \sum_{a<b} \left( \frac{r_m^{12}}{\| {\bf r}_a - {\bf r}_b \|^{12}} - \frac{2\,r_m^{6}}{\| {\bf r}_a - {\bf r}_b \|^{6}} \right)  \,,
\end{equation}
those forces clearly require the introduction of truly dimensionful constants, as the length 
$r_m$ of Eq.~(\ref{LennardJonesPotential}) where the Lennard-Jones potential has its minumum,
and $\epsilon_0$ with dimensions $\ell^{-1}$.
A dimensionful  constant like $r_m$ represents a conceptual challenge: it gives an absolute 
scale to the Universe, which doesn't make reference to any dynamical quantity in it.
In other words, a dimensionful constant is associated to a length, but what is it the length of?

As a first argument, we have to ask: how do we measure things like $r_m$ or $\epsilon_0$?
The answer is straightforward: we measure the equilibrium distance between two atoms which
interact through the Lennard-Jones potential. But that distance is always measured in relation
to something else: all measurements are, by their own nature, relational. If we doubled all
the distances in the universe there should be no visible effect: also $r_m$ and $\epsilon_0^{-1}$ should be doubled.
Therefore we should rather express $r_m$ and $\epsilon_0^{-1}$ as dimensionless constants times some other 
distance in the universe, but which one? The problem is that, according to laboratory
experience, $r_m$ and $\epsilon_0$ are, to a very good approximation, constant in time, while there are
no distances in the universe which are truly constant. If we want to give an answer that
satisfies the experimentalist, we can limit to a distance which is only approximately constant,
within the experimental error: the natural choice is $r_m , \epsilon_0^{-1} \propto \sqrt{I_\st{cm}}$, the square root of
the centre-of-mass moment of inertia of the whole universe defined in~(\ref{CenterOfMassMomentOfInertia}), essentially measuring the size of the universe. As we remarked in section~\ref{BestMatchingInAction}
the time dependence given by $r_m , \epsilon_0^{-1} \propto \sqrt{I_\st{cm}}$ would be very hard to detect
in the lab, within human time frames. 

\index{dimensional analysis}

I will now spell out a further argument in favour of  $r_m , \epsilon_0^{-1} \propto \sqrt{I_\st{cm}}$:
Van der Waals or harmonic forces can be considered as effective descriptions
of more fundamental physics which has just $1/r^2$-type forces, in which
one ignores or `coarse-grains' some degrees of freedom. Think about the electrical dipole force
generated by a pair of oppositely-charged particles: it falls off like $1/r^3$, but it is generated by
purely $1/r^2$-type forces. Therefore the dimensionful constant appearing in the
dipole potential must be related to some physical lengths: it is easy to convince oneself that
it is the size of the orbits of the pair of charges. This means that if every size in the universe was
scaled, the dimensionful constants would scale accordingly. Positing $r_m , \epsilon_0^{-1} \propto \sqrt{I_\st{cm}}$ finds then a justification in terms of effettive physics (or at least it appears 
more physical than having non-relational constants).

The two arguments above might seem convincing, but they ignore quantum mechanics. 
For instance, the argument about dipole forces doesn't make fully sense at a purely classical level: if we ignore quantum mechanics, the orbits of different pairs will have a continuum of different orbital element,
and these orbital elements
will change continuously due to interactions with the rest of the universe.
It would be unrealistic to assume something like $r_m , \epsilon_0^{-1} \propto \sqrt{I_\st{cm}}$
where the proportionality factors are (even approximately) constant in time: in a classical
world, these proportionality factors would change rapidly in time. Quantum mechanics
changes the picture: the orbitals of electrons in an atom are quantized, and they can only
jump by discrete quantities. At the deepest level, it is this discretization that allow us
to talk about `atoms', and to attribute to the dimensionless proportionality factors 
in  $r_m , \epsilon_0^{-1} \propto \sqrt{I_\st{cm}}$ a set of unchanging discrete values.
The issue of the origin of physical scales and units is a deep one, and requires a more careful
discussion, which would go beyond the purposes of this \thistext{}.
 
 \index{dimensional analysis}
 
\vspace{12pt}
\noindent
\fbox{\parbox{0.98\linewidth}{
{\bf Further reading:} For constrained systems I suggest Dirac's `Lectures'~\cite{dirac:lectures}, Henneaux-Teitelboim's book~\cite{HTbook}, and the book by Regge, Teitelboim ~\cite{HRTbook}.
Barbour and Foster's paper on Dirac's theorem~\cite{BF-DiraTheorem},
E. Anderson's review~\cite{EdwardBigReviewRelationalParticleMechanics}
for the `differential-almost-Hamiltonian' approach.
}}\vspace{12pt}

\newpage

\part{Relational Field Theory}
\label{RelationalFieldTheory}

\subsubsection*{The ontology of fields}

\index{Michael Faraday}
Faraday is credited with the introduction of the concept of field in physics. 
He found it extremely useful, in particular for the description of magnetic phenomena,
to use the concept of \emph{lines of force} (1830s) \cite{faraday1947experimental}:
\begin{quote}\it
3071. A line of magnetic force may be defined as that line which is described by a very small magnetic needle, when it is so moved
in either direction correspondent to its length, that the needle is constantly a tangent to the line of motion; [\dots]

3072. [\dots] they represent a determinate and unchanging amount of force. [\dots]  the sum of power contained in any one section of
a given portion of the lines is exactly equal to the sum of power in any other section of the same lines, however altered in form, or 
however convergent or divergent they may be at the second place. [\dots] 

3073.These lines have not merely a determinate direction,  [\dots] but because they are related to polar or antithetical power, have opposite
qualities in opposite directions; these qualities  [\dots] are manifest to us,  [\dots] by the position of the ends of the magnetic needle,  [\dots]
\end{quote}
\begin{figure}[b!]
\begin{center}\includegraphics[width=0.4\textwidth]{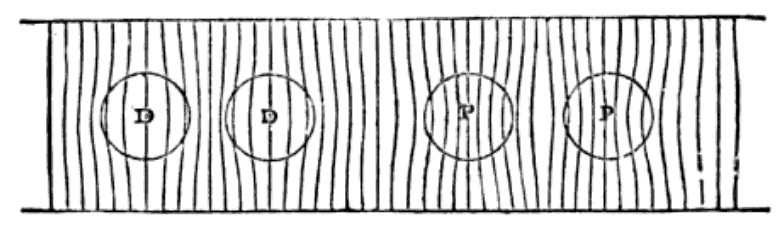}
\end{center}
\caption[Faraday's lines of force]{An illustration from Faraday's book  \cite{faraday1947experimental} showing his lines of magnetic force through diamagnetic (D) and paramagnetic (P) materials.}
\label{FaradaysLines}
\end{figure}
He speculated that the concept might be useful beyond magnetic phenomena,
\begin{quote}\it
3243. [\dots] The definition then given had no reference to the physical nature of the force at the place of action, and will apply
with equal accuracy whatever that may be; [\dots]

3251. Three great distinctions at least may be taken among these cases of the exertion of force at a distance; that
of gravitation, where propagation of the force by physical lines through the intermediate space is not supposed to
exist; that of radiation, where the propagation does exist, and where the propagating line or ray, once produced, has
existence independent either of its source, or termination; and that of electricity, where the propagating process has
intermediate existence, like a ray, but at the same time depends upon both extremities of the line
of force [\dots] Magnetic action at a distance has to be compared with these. It may be unlike any of them; for who
shall say we are aware of all the physical methods or forms under which force is communicated? It has been
assumed, however, by some, to be a pure case of force at a distance, and so like that of gravity; whilst others have
considered it as better represented by the idea of streams of power.
The question at presence appears to be, whether the lines of magnetic force have or have not a physical existence;
and if they have, whether such physical existence has a static or dynamic form [\dots].
\end{quote}
To Faraday it appeared clear that the issue of \emph{retardation} was key to determine whether physical
existence should be attributed to the lines of force of a certain interaction:
\begin{quote}\it
3246. There is one question in relation to gravity, which, if we could ascertain or touch it, would
greatly enlighten us. It is, whether gravitation requires \emph{time}. If it did, it would show undeniably
that a physical agency existed in the course of the line of force. [\dots]

3247. When we turn to radiation ph{\ae}nomena, then we obtain the highest proof, that though nothing ponderable passes,
yet the lines of force have a physical existence independent, in a manner, of the body radiating, or the body receiving
the rays. [\dots]
\end{quote}

\index{magnetic field}
\index{electric field}

\index{James Clerk Maxwell}

It was Maxwell who proved, with a monumental work, the superiority of the concept of fields for the description of electric and magnetic phenomena. In his 1855 seminal paper \emph{On Faraday's Lines of Force} \cite{clerk1864faraday} Maxwell modelled the field with an incompressible fluid whose velocity defined the field intensity and whose flux lines coincided with Faraday's lines of force. This analogy is particularly suited for forces that fall off as the square of the distance, like the electric and magnetic ones.
Maxwell's  \cite{clerk1864faraday} ends with the following lines:
\begin{quote}\it 
By referring everything to the purely geometrical idea of the motion of an imaginary fluid, I hope to attain generality and precision, and to avoid the dangers arising from a premature theory professing to explain the cause of the phenomena. If the results of mere speculation which I have collected are found to be of any use to experimental philosophers, in arranging and interpreting their results, they will have served their purpose, and a mature theory, in which physical facts will be physically explained, will be formed by those who by interrogating Nature herself can obtain the only true solution of the questions which the mathematical theory suggests.
\end{quote}
In his masterpiece, \emph{A Dynamical Theory of the Electromagnetic Field} \cite{maxwell1865dynamical}, Maxwell deduces the speed of light according to his model from the values of the electric permittivity and magnetic permeability of air measured by Weber and Kohlrausch. He then compares it to the direct measurement of the speed of light in air due to Fizeau and Foucault, finding good agreement:
\begin{quote}\it 
The agreement of the results seems to show that light and magnetism are affections of the same substance, and that light is an electromagnetic disturbance propagated through the field according to electromagnetic laws. [\dots]
Hence electromagnetic science leads to exactly the same conclusions as optical science with respect to the direction of the disturbances which can be propagated through the field; both affirm the propagation of transverse vibrations, and both give the same velocity of propagation.
\end{quote}
James Clerk Maxwell

\subsubsection*{The origins of Geometrodynamics}
\index{geometrodynamics}

\index{Carl Friedrich Gauss}
Carl Friedrich Gauss in his 1827 \emph{Disquisitiones generales circa superficies curvas (General investigations of curved surfaces)} \cite{gauss1902general} studied parametrized (coordinatized) 2d surfaces embedded in 3d Euclidean space. He was interested in those properties of the surface that are unaffected by a change of the way the surface is embedded in 3d space (as, for example, bending the surface without stretching it), or a change in the parametrization of the surface.
One natural such invariant quantity is the length of a curve drawn along the surface. Another is the angle between a pair of curves drawn along the surface and meeting at a common point, or between tangent vectors at the same point of the surface. A third such quantity is the area of a piece of the surface. The study of these invariants of a surface led Gauss to introduce the predecessor of the modern notion of metric tensor.
Among the important notions introduced by Gauss, there is the concept of \emph{intrinsic}, or \emph{Gaussian} curvature, for which the famous ``Theorema egregium'' holds  \cite{gauss1902general}:
\begin{quote} \it
If a curved surface is developed upon any other surface whatever, the measure of curvature in each point remains unchanged.
\end{quote}
Thus, the theorem states that the curvature of a surface can be determined entirely by measuring angles and distances on the surface, it does not depend on how the surface might be embedded in 3-dimensional space or on the parametrization of the surface.

\index{Georg Friedrich Bernard Riemann}

In 1854 Bernard Riemann had to give a habilitation lecture at the University of G\"ottingen and prepared three lectures, two on electricity and one on geometry. Gauss had to choose one of the three for Riemann to deliver and chose the lecture on geometry \cite{riemann8hypotheses}. This lecture revolutionized geometry, generalizing Gauss' results to any dimension, opening the possibility that the 3-dimensional space in which we live and do physics might not be Euclidean and could  possess intrinsic curvature \cite{riemann8hypotheses}. Riemann argued for an \emph{empirical} foundation for geometry:
\begin{quote}\it
Thus arises the problem, to discover the simplest matters of fact from which the measure-relations of space may be determined; a problem which from the nature of the case is not completely determinate, since there may be several systems of matters of fact which suffice to determine the measure-relations of space - the most important system for our present purpose being that which Euclid has laid down as a foundation. These matters of fact are - like all matters of fact - not necessary, but only of empirical certainty; they are hypotheses. We may therefore investigate their probability, which within the limits of observation is of course very great, and inquire about the justice of their extension beyond the limits of observation, on the side both of the infinitely great and of the infinitely small.
\end{quote}

Riemann's central concept was that of a metric, which characterizes the intrinsic geometry of a manifold. The importance of Riemann's work was so outstanding that now we talk about \emph{Riemannian geometry.} \index{Riemannian geometry}

Following that, around the end of the 19th century, \index{Gregorio Ricci-Curbastro} \index{Tullio Levi-Civita} Gregorio Ricci-Curbastro and Tullio Levi-Civita~\cite{ricci1900methodes} established the modern notions of tensors, as differential objects which are independent of  the coordinate system, laying the basis of modern differential geometry.
\index{tensor}

\index{Albert Einstein}
For our purposes, Einstein took the last step in this story by identifying spacetime with a 4-dimensional
manifold with Lorentzian signature, whose curvature was related to the energy--momentum tensor \index{energy--momentum tensor}
of matter through \index{Einstein's equations} Einstein's equations:
\begin{equation}
R_{\mu\nu} - {\sfrac 1 2} R \, g_{\mu\nu} =  8 \, \pi \, T_{\mu\nu} \,,
\end{equation}
in units where $G=c=1$. \index{Newton's constant} \index{speed of light} How this came about is an extremely interesting story that is recounted 
in great details elsewhere. My aim in this \thistext{} is to do something different.

\subsubsection*{An exercise in counterfactual history}

Einstein's discovery was strongly guided by the revolution brought about by Special Relativity,
after which the unity of space and time into a continuum appeared to be an inevitable property
of any future theory of physics. Had, as \timout{it} might have happened, the Machian ideas been made precise
earlier by someone like Poincar\'e, Barbour and Bertotti, the history of physics would probably have
taken a different course. I will now engage in this exercise in \emph{counterfactual history:}
imagine all I have explained about spatial and temporal relationalism in the previous Part had been understood in the 19th century, and add all the insights on the ontology of fields and geometry provided by Faraday, Maxwell, Gauss,
Riemann and Ricci--Levi-Civita. What could a relationalist physicist have done to extend Mach's ideas
to field theory?

She/he would have started by assuming a 3-dimensional perspective, in which space is a 3d Riemannian
manifold described by a metric tensor $g_{ab}$, on which other kind of fields live (scalar, vectors...).
Then she/he would have looked for a variational principle producing a dynamics which depended
on the fields and their first derivatives and with a Jacobi-type action. \index{Jacobi's principle} Any arbitrariness in the description,
like the coordinate system used to coordinatize the manifold and write tensors in components, should
be made redundant with best-matching, in analogy to the particle models.

In the following Sec.~\ref{RWRsection} and~\ref{ExtendedRWR} I will show how much could have been achieved with this handful of first principles. One can deduce the Special Relativity principle, the invariance of the speed of light and the universality of the light-cone. Moreover the whole of General Relativity can be derived, as well as the gauge principle, both in its abelian (electromagnetism) and non-abelian (Yang--Mills theory) form. In addition, the same principles allow for two different additional kinds of relativity (Galilean Relativity and Carrollian Relativity, or Strong Gravity) which are relevant in particular regimes of GR. Finally, by requiring the Mach--Poincar\'e Principle to be satisfied by those theories, one unambiguously obtains the theory which is the main subject of this \thistext{}: Shape Dynamics.

\index{Special Relativity}
\index{General Relativity}
\index{Carroll Relativity}
\index{light-cone}
\index{gauge theory}
\index{electromagnetism}
\index{Yang--Mills theory}
\index{Mach--Poincar\'e Principle}

Between Secs.~\ref{RWRsection} and~\ref{ExtendedRWR} I have inserted a Section, number~\ref{YorksMethodSection}, on York's method for solving the initial-value problem in General Relativity, which is necessary to understand the following,
where I show how implementing the Mach--Poincar\'e Principle leads to Shape Dynamics.

The following Sections will stress the fact that Shape Dynamics is logically independent of General Relativity.
It is in fact unnecessary to even know anything about GR to understand these Sections, and all
its main features will be derived independently, from more fundamental first principles. 
Needless to say, readers who are familiar with GR, and in particular its Hamiltonian formulation due
to Arnowitt--Deser--Misner will be able to appreciate the following Sections even more. I devote the whole Appendix~\ref{ADM-WDW} to reviewing the ADM formulation of GR. Readers unfamiliar with it should read 
 Appendix~\ref{ADM-WDW} either now, before the next Sections~\ref{RWRsection}-\ref{ExtendedRWR}, 
 or after them, in order to understand the connection with the spacetime picture. Also many readers
 who are familiar with ADM gravity might have never seen the Baierlein--Sharp--Wheeler action, \index{Baierlein--Sharp--Wheeler formulation}
 which is the key link between ADM and the formulations of the next sections, so they might
 consider reading  Appendix~\ref{ADM-WDW} now.

\subsubsection*{Nuggets of functional analysis}

\index{functional variation}
In this Part I will be working on a 3-dimensional Riemannian manifold $\Sigma$,
which is therefore endowed with a Riemannian 3-metric $g_{ab}$ and an associated integration 
measure $\int_\Sigma \d^3 x \sqrt g$. Throughout this Part I will assume the manifold to be
closed (compact without boundary). \index{closed spacelike hypersurfaces} This will be relaxed in Sec.~\ref{AsymptoticallyFlatSD} of the next Part,
and in Sec.~\ref{AsymptoticallyFlatADM} of Appendix~\ref{ADM-WDW} .

The fundamental objects of our study will be fields, understood as different multiplets  of 
functions on $\Sigma$. The prototype for our configuration spaces will be  $\mathscr{L}^2 (\Sigma)$,
the space of square-integrable functions on our spatial manifold $\Sigma$. This is also the
configuration space of scalar fields. More complicated fields like tensors will just be tensor
products of $\mathscr{L}^2 (\Sigma)$ (one for each component), with particular transformation
laws under diffeomorphisms and under the basic operations of tensor calculus (if the reader
needs an introduction to tensor calculus I suggest starting with Schutz \cite{SchutzBook},
continue with G\"ockeler--Sch\"ucker \cite{GockelerSchucker}, and ending up with Frankel \cite{Frankelbook}).

$\mathscr{L}^2 (\Sigma)$ can be made into an inner product space by defining
\begin{equation}
( \, f \, | \, h \, ) = \int \d^3 x \, \sqrt g \, f(x) \, h(x) \,,
\end{equation}
which satisfies all the axioms of an inner product. This inner product can be extended to dual tensors, e.g., 
\begin{equation}
( \, T^{ij}\, | \, S_{ij} \, ) = \int \d^3 x \, \sqrt g \, T^{ij} (x) S_{ij}(x) \,,
\end{equation}
and to tensor densities, e.g.,
\begin{equation}
( \, T^{ij}\, | \, S_{ij} \, ) = \int \d^3 x \, g^{\frac y 2} \, T^{ij} (x) S_{ij}(x)  \qquad y =1 -w-z \,,
\end{equation}
if $T^{ij}$ is a density of weight $w$ and $S_{ij}$ is of weight $z$.

\index{tensor densities}

The constraints in field theory are actually \emph{one constraint per space point} 
and \e{can contain spatial derivatives of the fields}. The proper way to understand
field theory constraints is through \e{smearings}: I can transform a constraint $\chi(x)$
into a real number by integrating it against a test function $( \, \chi \, | \, f \, )$.
Then $( \, \chi \, | \, f \, )$ becomes a \e{linear functional} of $f(x)$, of which one can take variations.
For example, if the metric $g_{ab}$ is the field we want to take variations of, we have
\begin{equation}
\delta ( \, \chi \, | \, f \, ) = \int_\Sigma \d^3 x \frac{\delta ( \, \chi \, | \, f \, )}{\delta g_{ij} (x)} \delta g_{ij} (x) \,,
\end{equation}
where the sum over repeated indices is understood. Here, $\frac{\delta ( \, \chi \, | \, f \, )}{\delta g_{ij} (x)} $
is a distribution called the \e{functional derivative} of $ ( \, \chi \, | \, f \, ) $ w.r.t. $g_{ij} (x)$.

\index{ultralocal}

If two fields are canonically conjugate, \e{e.g.}, $q(x)$ and $p(x)$, then the Poisson brackets
between two functionals $F[q,p]$, $G[q,p]$ on the phase space defined by $q$ and $p$ is
\begin{equation}
\{ F, G \} = \int_\Sigma \d^3 x \left(  \frac{\delta F}{\delta q(x)} \frac{\delta G}{\delta p(x)} - \frac{\delta F}{\delta p(x)} \frac{\delta G}{\delta q(x)} \right) \,.
\end{equation}
Notice that since one single field-theoretical constraint $\chi(x)$ is actually one constraint
\e{per spatial point}, the Poisson bracket of $\chi$ with itself can be nonzero. One has in
fact to smear it with two different smearing functions $f$ and $h$, and the Poisson bracket
will be
\begin{equation}
\{ ( \, \chi \, | \, f \, ) , ( \, \chi \, | \, h \, )  \} = \int_\Sigma \d^3 x \left(  \frac{\delta ( \, \chi \, | \, f \, ) }{\delta q(x)} \frac{\delta ( \, \chi \, | \, h \, ) }{\delta p(x)} - \frac{\delta ( \, \chi \, | \, f \, )}{\delta p(x)} \frac{\delta ( \, \chi \, | \, h \, ) }{\delta q(x)} \right) \,,
\end{equation}
which has no reason to vanish\timout{,} unless $\chi$ is \e{ultralocal} (it contains no derivatives of $q$, $p$), in
which case its variations will be linear in the smearings (with no derivatives acting on them), and
by antisymmetry in $f$ and $h$ the expression above will vanish. 

I will use the following notation for symmetrization and antisymmetrization of indices:
$$
T^{[ij]}= {\ts \frac 1 2} (T^{ij} - T^{ij}) \,, ~~~~~~~ T^{(ij)}= {\ts \frac 1 2} (T^{ij} + T^{ij}) \,.
$$

The fundamental extended configuration space we will start with is $\Riem$, the space of Riemannian
3-metrics. It is the geometrodynamical equivalent of Cartesian space $\Q^N =\doublestroke R^{3N}$ for particle dynamics.

\newpage

\section{Relativity Without Relativity}\label{RWRsection}

In a series of papers Julian Barbour, Brendan Foster, and Niall \'O Murchadha and Edward Anderson and I \cite{RWR-Ed-Strong-Coupled,RWR-Ed-Variations,RWR-VectorFields,RWR,EdFlavioPaper}, motivated by the desire to enforce 
reparametrization invariance and temporal relationalism, have shown how powerful
 the choice of a square-root form of the action is.
This approach makes it possible, without any prior spacetime assumptions, to arrive at General Relativity, Special Relativity and
gauge theory on the basis of relational first principles.   I'll give here first a simplified account 
of the original results in the `modern' language I used with E. Anderson in \cite{EdFlavioPaper},
 and in Section~\ref{ExtendedRWR} I'll get rid of all the simplifying assumptions 
 made in this Section and repeat the analysis in full generality (summarizing part of my recent contribution to this program with Edward Anderson \cite{EdFlavioPaper}).

Let's start from the following assumptions:
\begin{enumerate}
\item \emph{The action is a local functional of the 3-metric $g_{ab}$ and its first derivatives.}
\item \emph{The action is of Jacobi type, that is, the product of the square roots of a potential
and a kinetic term which is quadratic in the velocities $\d g_{ab}$.}
\item \emph{The theory must be free of any redundancy in the description of the fields,
and this independence must be realized locally through best-matching.}
\end{enumerate}\index{Jacobi's principle}\index{Jacobi-type action}
Then the simplest Lagrangian `line element' satisfying these assumptions that one can write is\footnote{Notice that the square root is \emph{inside} the integral. \index{local square root action} For this reason, the above action is not perfectly
analogous to a Jacobi action for particle mechanics: in particle models the analogue of the integral $\int \d^3 x$
is a sum over the particle index $a$ and the vector component $i$: $S_\st{Jacobi} = 2 \int [(E-U)\sum_{a,i} \frac{m_a}2 \, ( \d r^i_a)^2]^{1/2}$. This sum is inside the square root and makes the Jacobi action into a norm for the velocity vector $\d r^i_a $.  The action $\int \d \mathcal L_0$ is not a proper norm, but putting the integral inside the square root by analogy with $S_\st{Jacobi}$  would spoil the locality of the action. By locality I mean here that the action can be written as the integral of an expression that depends only on the values of the fields in a neighborhood of the point. As we shall see, this requirement of locality of the square root has very far-reaching consequences.} 
\begin{equation} \label{RWR-InitialLagrangian}
\d \mathcal L_0 =  \int \d^3 x \sqrt{g} \sqrt{R - 2 \, \Lambda} \sqrt{ \left(  g^{ik}g^{jl} -  \, g^{ij}g^{kl} \right)  \d  g_{ij} \, \d  g_{kl}}  \,,
\end{equation}
where $R$ is the 3-dimensional Ricci scalar and $\Lambda$ is a spatially constant parameter.
 The relative factor between  the two possible ways to contract the indices of the metric velocities $\d g_{ab}$ has
been put to $-1$. This is the only working hypothesis and will be relaxed in the next subsection.
It is important to notice that the above Lagrangian presupposes nothing about spacetime: it is merely the simplest
parametrization-irrelevant and local Jacobi-type expression one can form from a 3-metric. No spacetime
covariance (and no local Lorentz invariance) has been assumed. I'll show now how much can be deduced just
from this. Calling the local expression
\begin{equation}
\d \doti = \frac 1 2 \sqrt{ \frac{ \left(  g^{ik}g^{jl} - g^{ij}g^{kl} \right)  \d  g_{ij} \, \d  g_{kl}}{ R - 2 \, \Lambda}} \,,
\end{equation}
the field-theoretical version of the differential of the instant,\index{differential of the instant} the canonical momenta read
\begin{equation}
  p^{ij} = {\delta \d \mathcal L_0 \over \delta \d  g_{ij}} = \frac{\sqrt g }{2 \,\d \doti} (g^{ik}g^{jl} - g^{ij}g^{kl}) \,  \d  g_{kl}   \,.
\end{equation}
Due to the local square-root form of the action, the momenta satisfy at each space point the primary constraint
\begin{equation} \label{HamiltonianConstraintNo1}
\mathcal H =   {\ts\frac 1 {\sqrt g}} \left( p^{ij}p_{ij} -{\ts \frac 1 2} p^{2}\right) - \sqrt g \, (R - 2 \, \Lambda )  \approx 0\,,
\end{equation}
where $p = p^{ij} g_{ij}$. The Euler--Lagrange equations are
\begin{equation}\begin{aligned} \label{ELeqnRWR}
\d p^{ij}  = &  \sqrt g \left( R \, g^{ij} - R^{ij} + \nabla^i \nabla^j - g^{ij} \laplacian \right) \d \doti\\  
&-2 \, \Lambda \, \sqrt g \, g^{ij} \, \d \chi - \frac{2 \,\d \doti}{\sqrt g } \left( p^{ik} {p_k}^j -  {\ts \frac 1 2} p \, p^{ij} \right)\,.
\end{aligned}\end{equation}
In order for the theory to be consistent, the constraint (\ref{HamiltonianConstraintNo1}) must be propagated by the equations of motion,
\begin{equation}\begin{aligned}
\d \mathcal H &\approx  4\,\d \doti \sqrt g \, \nabla_i \d \doti \, \nabla_j p^{ij}  
+ 2 \, \d \doti \, \sqrt g \, \nabla_i \nabla_j p^{ij}  \\  
&= \frac{2}{\d \doti} \sqrt g \, \nabla_i \left( \d \doti^2 \, \nabla_j p^{ij} \right)  \,.
\end{aligned}\end{equation}
We see that the Hamiltonian constraint is not propagated unless $\nabla_j p^{ij} \approx 0$. This
is not guaranteed and has to be imposed on the initial conditions through the additional
constraint $\mathcal H_i = - 2 \, \nabla_j {p^j}_i$. However, doing so doesn't ensure that the
equations of motion will propagate the condition $\mathcal H_i \approx 0$: it has to be checked explicitly
\begin{equation}\begin{aligned}
\d \mathcal H_i & = - \nabla_i \d  \doti \, \mathcal H  \approx 0 \,.
\end{aligned}\end{equation}
We see that this additional constraint is preserved by the evolution, as its propagation gives a linear combination of Hamiltonian constraints, which therefore vanishes weakly.

The new constraint we had to introduce is, as I  illustrate in Appendix~\ref{ADM-WDW}, the diffeomorphism constraint, \index{diffeomorphism constraint} and
it acts on the metric as (\ref{3DiffeosGeneration}). Let's then realize diffeomorphism invariance at the level
of the action through best matching:
\begin{equation} \label{BestMatchedActionForGravity}
\d \mathcal L_\st{diff} =  \int \d^3 x \sqrt{g} \sqrt{R - 2 \, \Lambda} \sqrt{ \left(  g^{ik}g^{jl} -  \, g^{ij}g^{kl} \right)   \dbm  g_{ij} \, \dbm  g_{kl} }  \,,
\end{equation}
where the best-matching differential is 
\begin{equation}
\dbm g_{ij} =\d  g_{ij} + \Lie_{\d \xi} g_{ij}  =\d  g_{ij} + \nabla_i \d \xi_j + \nabla_j \d \xi_i \,.
\end{equation}\index{differential of the instant}
The local `differential of the instant' [cf.~(\ref{DifferentialOfTheInstant})] and the canonical momenta have the same structure with  the best-matching differential $\dbm$ in place of $\d$:
\begin{equation}
\d \doti = \frac 1 2 \sqrt{ \frac{ \left(  g^{ik} g^{jl} - g^{ij}g^{kl} \right)  \dbm  g_{ij} \, \dbm  g_{kl}}{ R - 2 \, \Lambda}} \,,
\end{equation} 
\begin{equation}
  p^{ij} = {\delta \d \mathcal L_\st{diff} \over \delta \d  g_{ij}} = \frac{\sqrt g }{2 \,\d \doti} (g^{ik} g^{jl} - g^{ij}g^{kl}) \,  \dbm  g_{kl}   \,.
\end{equation}
The local square-root form of the action still leads to  (\ref{HamiltonianConstraintNo1})
as a primary Hamiltonian constraint. The best-matching condition gives the diffeomorphism constraint
\begin{equation}\label{RWRdiffeoConstraint}
\frac{\delta \d \mathcal L_\st{diff}}{\delta \d \xi_i} = - 2 \, \nabla_j \left[\frac{\sqrt g }{2 \,\d \doti} (g^{ik} g^{jl} - g^{ij} g^{kl}) \,  \dbm  g_{kl} \right] =- 2 \, \nabla_j p^{ij} = \mathcal H^i \approx 0\,.
\end{equation}
The  Euler--Lagrange equations, and in particular the term $\frac{\delta \d \mathcal L_\st{diff}}{\delta g_{ij}}$,  are now changed, due to the dependence of the best-matching differential on the metric. This leads to the appearance of the term $ \Lie_{\d\xi} p^{ij}$ term in
\begin{equation}\begin{aligned}
\d p^{ij}  = & \sqrt g \left( {\sfrac 1 2} R \, g^{ij} - R^{ij} + \nabla^i \nabla^j - g^{ij} \laplacian \right) \d \doti\\  
&-2 \, \Lambda \, \sqrt g \, g^{ij} \, \d \chi - \frac{2 \,\d \doti}{\sqrt g } \left( p^{ik} {p_k}^j -  {\ts \frac 1 2} p \, p^{ij} \right) \\  
& + \Lie_{\d\xi} p^{ij} \,,
\end{aligned}\end{equation}
but this term, as we have already seen in particle models, can be brought to the left to form a best-matching differential \index{best-matching differential} of $p^{ij}$:
\begin{equation}\begin{aligned}
\dbm p^{ij}  = \d p^{ij} -  \Lie_{\d\xi} p^{ij} = & \sqrt g \left( {\sfrac 1 2}  R \, g^{ij} - R^{ij} + \nabla^i \nabla^j - g^{ij} \laplacian \right) \d \doti\\  
&-2 \, \Lambda \, \sqrt g \, g^{ij} \, \d \chi - \frac{2 \,\d \doti}{\sqrt g } \left( p^{ik} {p_k}^j -  {\ts \frac 1 2} p \, p^{ij} \right) \,.
\end{aligned}\end{equation}
The same thing happens with the propagation of the Hamiltonian and diffeomorphism constraint:
\begin{equation}\begin{aligned}
\dbm \mathcal H \approx \frac{2}{\d \doti} \sqrt g \, \nabla_a \left( \d \doti^2 \, \nabla_j p^{ij} \right)  \,,
~~~
\dbm \mathcal H_i  = \d  \doti \, \nabla_i \mathcal H - \nabla_i (\d  \doti \mathcal H ) \approx 0 \,,
\end{aligned}\end{equation}
The Hamiltonian constraint smeared with $\d \doti$ generates the same equations,
forming Anderson's `differential-almost-Hamiltonian':
\begin{equation}
\d \mathcal A = ( \d \doti | \mathcal H) \,, ~~~~ \dbm f = \{ \d \mathcal A , f \} \,,
\end{equation}
together with the definition of the momenta (which is one of the two Hamilton equations),
\begin{equation}
\dbm g_{ij} =  \{ \d \mathcal A , g_{ij} \}  = \frac{2 \, \d \doti}{\sqrt g } \left( g_{ik} g_{jl} - {\ts \frac 1 2} g_{ij} g_{kl} \right) \, p^{kl} \,.
\end{equation}

\noindent{\bf Rigidity of the choice of the Lagrangian}

In \cite{RWR} the authors tested a somewhat more general ansatz for the potential term.  They first considered an arbitrary power of the Ricci scalar, $R^\alpha$, and then a linear
combination of the terms $R^2$, $R^{ij}R_{ij}$ and $\laplacian R$, which are the only scalars
with dimensions $\ell^{-4}$ that can be built with the metric field alone ($R$ has dimensions 
$\ell^{-2}$ and no scalars with dimensions $\ell^{-3}$ exist). The propagation of the Hamiltonian constraint, in the words of the authors, ``leads to an explosion of unpleasant non-cancelling terms'', which rapidly end up trivializing the theory if included as new constraints (checking this explicitly is left as an exercise for the reader).
\index{supermetric}

The Lagrangian (\ref{RWR-InitialLagrangian}) is not the most general possibility also in other respects:
one could cancel the `$R$' term, leaving the potential as a constant. Or one could change
the relative factor between the two terms appearing in the kinetic term. These choices lead
to interesting \emph{and viable} alternatives. There was a preliminary discussion of them in \cite{RWR}
and in \cite{RWR-Ed-Strong-Coupled}, but a thorough analysis of these cases has only recently been 
completed by Anderson and myself in \cite{EdFlavioPaper}. I give a review of these results in
Section~\ref{ExtendedRWR}. For the moment, I'll just underline an important detail, which
at this level might seem unimportant and be ignored, but will become very relevant later.
The issue regards the relative factor in the `supermetric': if we generalize the Lagrangian (\ref{BestMatchedActionForGravity}) to
\begin{equation} 
\d \mathcal L_\st{diff-2} =  \int \d^3 x \sqrt{g} \sqrt{R - 2 \, \Lambda} \sqrt{ \left(  g^{ik}g^{jl} - \lambda \, g^{ij}g^{kl} \right)   \dbm  g_{ij} \, \dbm  g_{kl} }  \,
\end{equation}
by adding the coefficient $\lambda$ in the kinetic term, we get an additional term in the propagation of the Hamiltonian constraint:
\begin{equation}
\dbm \mathcal H = \frac{2 \,\sqrt g}{\d \doti} \, \nabla_i \left[  \d \doti^2 \, \nabla_j p^{ij}  - {\ts\frac{\lambda - 1}{ 3 \lambda -1}} \, \d \doti^2 \nabla^i p \right]  \,.
\end{equation}
which, if the diffeomorphism constraint $\nabla_j p^{ij} \approx 0$ is already implemented, introduces a new constraint $p \approx \rm{\it const}$. This is not propagated by the evolution, but it gives rise to a `specifier' equation
(case 3 of Dirac's analysis) which, in its turn, leads to a well-defined system with two propagating degrees of freedom (see Section~\ref{ExtendedRWR} for the details).\index{Weyl transformation}\index{conformal transformation}\index{conformal constraint}
The new constraint has a simple geometric interpretation as the generator of position-dependent  conformal transformations
\begin{equation}
g_{ab} \to \phi^4 \, g_{ab} \,, \qquad  \phi(x)>0\,,
\end{equation}
(also called Weyl transformations) of  the 3-metric. These transformations play an important role in \index{James W. York, jr.} York's solution of the initial-value problem of GR, 
which I review in Sec.~\ref{YorksMethodSection}, and, as I'll explain below, are needed to implement
the Mach--Poincar\'e Principle.\index{Mach--Poincar\'e Principle} It is striking that one ends up considering the same constraint $p \approx \rm{\it const}$ in the solution of the initial-value problem and by considering a generalized supermetric.

\noindent{\bf Inclusion of a scalar field: Special Relativity}

The ansatz for coupling of a scalar field to the metric field is
\begin{equation}\begin{aligned}
\d \mathcal L_\st{scalar} =  \int \d^3 x \sqrt{g} &\sqrt{ R - 2 \, \Lambda + k \, g^{ij} \, \nabla_i \varphi  \nabla_j \varphi + U(\varphi) }  \\  
\cdot  & \sqrt{ \left(  g^{ik}g^{jl} -  \, g^{ij}g^{kl} \right)   \dbm  g_{ij} \, \dbm  g_{kl}  + (\dbm \varphi)^2}  \,,
\end{aligned}\end{equation}\index{scalar field}

Here, $U(\varphi)$ can be any function of $\varphi$, and $\dbm \varphi = \d \varphi + \Lie_{\d \xi}  \varphi = \d \varphi + \d \xi^i \nabla_i \varphi$. A point to note here is that the form of the scalar kinetic term $(\dbm \varphi)^2$ is, if assumed quadratic, uniquely fixed by best matching, while its coefficient is free because it can be changed by rescaling of the field.  In contrast, the unknown constant $k$ will then appear multiplying the field propagation term 
$g^{ij} \, \nabla_i \varphi  \nabla_j \varphi$. This will have consequences, as we will soon see. Meanwhile, we see that the scalar field, together with the $g_{ij}$ field,  contributes directly to the local `differential of the instant',\index{differential of the instant}
\begin{equation}
\d \doti = \frac 1 2 \sqrt{ \frac{ \left(  g^{ik}g^{jl} - g^{ij}g^{kl} \right)  \dbm  g_{ij} \, \dbm  g_{kl} + \dbm \varphi^2 }{ R - 2 \, \Lambda + k \, g^{ij} \, \nabla_i \varphi  \nabla_j \varphi + U(\varphi) }} \,,
\end{equation}
but only indirectly (through $\d \doti$) to the metric momenta:
\begin{equation}\label{DefinitionOfTheMomenta3}
  p^{ij} = {\delta \d \mathcal L_\st{scalar} \over \delta \d  g_{ij}} = \frac{\sqrt g }{2 \,\d \doti} (g^{ik}g^{jl} - g^{ij}g^{kl}) \,  \dbm  g_{kl}   \,.
\end{equation}
The scalar field has its associated momentum
\begin{equation}
  \pi  = {\delta \d \mathcal L_\st{scalar} \over \delta \d  \varphi} = \frac{\sqrt g }{2 \,\d \doti}  \,  \dbm  \varphi   \,,
\end{equation}
and the  Hamiltonian constraint involves a quadratic combination of both the metric and the scalar-field momenta
\begin{equation} \label{HamiltonianConstraintNo3}
\mathcal H = \sqrt g \, (R - 2 \, \Lambda + k \, g^{ij} \, \nabla_i \varphi  \nabla_j \varphi + U(\varphi) ) -  {\ts\frac 1 {\sqrt g}} \left( p^{ij} p_{ij} -{\ts \frac 1 2} p^{2} + \pi^2 \right) \approx 0 \,.
\end{equation}
The equations of motion are
\begin{equation}\begin{aligned}
\dbm p^{ij}  = &   \sqrt g \left( {\sfrac 1 2}  R \, g^{ij} - R^{ij} + \nabla^i \nabla^j - g^{ij} \laplacian \right) \d \doti + k \, \nabla^i \varphi \nabla^j \varphi  \, \sqrt g \, \d \doti  \\  
&+ ( k \, \nabla^k \varphi \nabla_k \varphi  + U(\varphi) -2 \, \Lambda ) \sqrt g \, g^{ij} \, \d \chi - \frac{2 \,\d \doti}{\sqrt g } \left( p^{ik} {p_k}^j -  {\ts \frac 1 2} p \, p^{ij} \right) \,, \\
\dbm \pi = &  2 \, k \,  \sqrt g \, \nabla_i \left( \nabla^i \varphi \, \d \doti \right)  +   \frac{\delta U}{\delta \varphi} \, \d \doti \,, 
\end{aligned}\end{equation}
and best-matching  w.r.t.  diffeomorphisms gives
\begin{equation}
\frac{\delta \d \mathcal L_\st{scalar}}{\delta \d \xi_i} = - 2 \, \nabla_j p^{ij} + \pi \,  \nabla^i \varphi = \mathcal H^i \approx 0\,.
\end{equation}

Now I come to the key point: if we try to propagate the Hamiltonian constraint  (\ref{HamiltonianConstraintNo3}):
\begin{equation}\begin{aligned}
\dbm \mathcal H & = \frac{2}{\d \doti} \sqrt g \, \nabla_i \left( \d \doti^2 \, \nabla_j p^{ij} \right) 
+ (4 \, k+1) \frac{\sqrt g}{\d \doti} \nabla^i \left(    \pi \, \nabla_i \varphi \, \d \doti^2 \right)  \\  
& \approx (4 \, k+1) \frac{\sqrt g}{\d \doti} \nabla^i \left(    \pi \, \nabla_i \varphi \, \d \doti^2 \right) \,,
\end{aligned}\end{equation}
we find an obstruction. The option of introducing a new constraint $ \pi \, \nabla_i \varphi \approx 0 $
can be readily excluded, considering the fact that $ \pi \, \nabla_i \varphi $ is a vector constraint
which would kill 6 phase-space degrees of freedom whereas we introduced only two with the scalar field.
The only remaining possibility is to propagate $\mathcal H$ strongly by setting $k = - \frac 1 4$.
This result is very significant: we have found that the scalar field has to respect the same
light cone as the metric field, which in turn implies Special Relativity in small regions of space
for small intervals of time (local Lorentz invariance).\index{universal light cone}\index{Special Relativity}

To see this, we treat $\varphi$ as a test field that has no back reaction on the metric and consider a Euclidean patch, where the coordinates are chosen so that we
can write the metric as a small perturbation around a static Euclidean metric $g_{ij} = \delta_{ij} + h_{ab}$.
Since $\delta_{ij}$ is static, $\d \delta_{ij} = 0 $ and $\d g_{ij} =\d h_{ij}$. We can put $\d\xi$
to zero as it is used to fix the coordinate gauge, and $\d \chi = \d t$ can be used as definition of the unit of time.
 The variation of  Eq. (\ref{DefinitionOfTheMomenta3}) is
\begin{equation}\begin{aligned}
\frac{\d  p^{ij}}{\d t} = & {\ts \frac 1 2} \, \sqrt g  (g^{ik}g^{jl} - g^{ij}g^{kl}) \, \frac{\d^2 h_{kl}}{\d t^2} \\  
+ &  {\ts \frac 1 2} \, \sqrt g \left(  g^{ij} \frac{\d h_{ij}}{\d t} +  2 \, \frac{\d h^{ik}}{\d t} g^{jl} - 2\, \frac{\d h^{ij}}{\d t} g^{kl} \right)  \frac{\d h_{kl}}{\d t}  \,.
\end{aligned}\end{equation}
The only first-order term is the first one, so
\begin{equation}
\frac{\d  p^{ij}}{\d t} = {\ts \frac 1 2} (\delta^{ik}\delta^{jl} - \delta^{ij}\delta^{kl}) \, \frac{\d^2 h_{jl}}{\d t^2} + \mathcal O (h^2) \,.
\end{equation}
The other side of the Euler--Lagrange equations is (since we are treating $\varphi$ here as a test field with no backreaction on the metric, we ignore all the terms that depend on $\varphi$ in the following equation)
\begin{equation}\begin{aligned}
\frac{\d p^{ij}}{\d t}   = &   \left( \partial^k \partial^l h_{kl} \, \delta^{ij} - R^{ij}   \right)  - {\ts \frac 1 2}  (\delta^{ik}\delta^{jl} - \delta^{ij}\delta^{kl})  \frac{\d h_{ij}}{\d t} \frac{\d h_{kl}}{\d t}  \,,
\end{aligned}\end{equation}
where it's easy to show that
\begin{equation}
R_{ij}  ={\ts \frac 1 2 } \left( \partial_j \partial^k  h_{ik} + \partial_i \partial^k  h_{jk} - \partial_k \partial^k  h_{ij}-  \delta^{kl}  \partial_i \partial_j h_{kl} \right) \,.
\end{equation}
Taken together, the above equations give the equations of motion
of linearized gravity on a flat background. It is well known that these equations represent
spin-2 gravitons propagating, in the coordinates
we used, with a speed set to $1$ \cite{FeynmansLecturesOnGravitation}. \index{linearized gravity}

Now consider the scalar field: the two terms of its Euler--Lagrange equations are
\begin{equation}
\frac{\d \pi}{\d t}  = - 2 \, k \, \partial^i \partial_i \varphi   +   \frac{\delta U}{\delta \varphi} + \mathcal O (h^2) \,, ~~~~~~ \frac{\d \pi}{\d t} = {\ts \frac 1 2} \frac{\d^2 \varphi}{\d t^2} \,,
\end{equation}
which combined together give the following equation
\begin{equation}
\frac{\d^2 \varphi}{\d t^2} + 4 \, k \, \partial^i \partial_i \varphi   -   2 \, \frac{\delta U}{\delta \varphi}  =0\,,
\end{equation}
which is the equation of motion of a Klein--Gordon \index{Klein--Gordon equation} field with potential $2U$ \emph{and
propagation speed} $4 \,k$. The choice of $k$ that makes the theory consistent, $k=-\frac 1 4$, is 
therefore the one that fixes the propagation speed of the scalar field to $1$. \emph{Thus,
the scalar field respects the same light cone as the metric field.} Moreover, as is shown in \cite{RWR} and we shall shortly show for a one-form, the mechanism that underlies this result is \e{universal}: it works for all fields coupled to the metric field.

\index{universal light cone}\index{Special Relativity}
This result is very striking. It is an independent derivation, from Machian first
principles, \e{of the essence of Special Relativity}: constancy of the speed of light and a common light cone for all fields in nature. {What is more, it implies
local Lorentz invariance.  Let me stress that General Relativity assumes the universal light cone and Lorentz invariance among its founding principles, while this theory \emph{deduces them}
from a smaller set of first principles, which are therefore arguably \emph{more fundamental}.
\index{General Relativity}

\noindent{\bf Inclusion of a one-form field: Gauge Theory}

\index{one-form field}
The ansatz for inclusion of a one-form field is
\begin{equation}\begin{aligned}
\d \mathcal L_\st{1-form} =  \int \d^3 x \sqrt{g} &\sqrt{ R - 2 \, \Lambda +W(\A)+ V(g^{ij} \A_i \A_j) }  \\  
\cdot  & \sqrt{ \left(  g^{ik} g^{jl} -  \, g^{ij} g^{kl} \right)   \dbm  g_{ij} \, \dbm  g_{kl}  + g^{ij} \dbm \A_i  \dbm \A_j}  \,,
\end{aligned}\end{equation}
where $W(\A) =  \alpha  \,  \nabla_i \A_j  \nabla^i \A^j
+ \beta  \, \nabla_i \A_j  \nabla^j \A^i
+ \gamma \, \nabla^i \A_i  \nabla^j \A_j$,
$V$ can be any function of $g^{ij} \A_i \A_j$,
and $\dbm \A_i = \d \A_i + \Lie_{\d \xi} \A_i = \d \A_i + \d \xi^j \nabla_j \A_i + \nabla_i \d \xi^j \A_j $.
The local differential of the instant is
\begin{equation}
\d \doti = \frac 1 2 \sqrt{ \frac{ \left(  g^{ik}g^{jl} - g^{ij}g^{kl} \right)  \dbm  g_{ij} \, \dbm  g_{kl} +  g^{ij} \dbm \A_i  \dbm \A_j }{ R - 2 \, \Lambda +W+V  }} \,,
\end{equation}
the metric momenta are
\begin{equation}\label{DefinitionOfTheMomenta4}
  p^{ij} = {\delta \d \mathcal L_\st{1-form}  \over \delta \d  g_{ij}} = \frac{\sqrt g }{2 \,\d \doti} (g^{ik}g^{jl} - g^{ij} g^{kl}) \,  \dbm  g_{kl}   \,,
\end{equation}
the  momentum conjugate to $\A_a$ is 
\begin{equation}
  \E^i  = {\delta \d \mathcal L_\st{1-form}  \over \delta \d \A_i} = \frac{\sqrt g }{2 \,\d \doti}  \, g^{ij} \, \dbm \A_j   \,,
\end{equation}
and the  Hamiltonian constraint is
\begin{equation} \label{HamiltonianConstraintNo4}
\mathcal H = \sqrt g \, (R - 2 \, \Lambda + W + V ) -  {\ts\frac 1 {\sqrt g}} \left( p^{ij} p_{ij} -{\ts \frac 1 2} p^{2} - g_{ij} \E^i \, \E^j\right) \approx 0 \,.
\end{equation}
The equations of motion are
\begin{equation}\begin{aligned}
\dbm p^{ij}  = &   \sqrt g \left( {\sfrac 1 2}  R \, g^{ij} - R^{ij} + \nabla^i \nabla^j - g^{ij} \laplacian + U + W\right) \d \doti \\  
& + 2\, \sqrt g  \left( \alpha \, \nabla^k \A^i \nabla_k \A^j + \beta \, \nabla^k \A^i \nabla^j \A_k + \gamma \nabla^{(i} \A^{j)} \, \nabla^k \A_k\right) \d \doti
 \\  
& - \sqrt g \, V' \, \A^i \, \A^j \, \d \doti - \frac{\d \doti}{\sqrt g} \E^i\, \E^j- \frac{2 \,\d \doti}{\sqrt g } \left( p^{ik} {p_k}^j -  {\ts \frac 1 2} p \, p^{ij} \right) \,, \\
\dbm \E^i = &  - 2 \,  \sqrt g \,  \left[ \alpha \,  \nabla_j ( \nabla^j \A^i \, \d \doti)  + \beta \, \nabla_j ( \nabla^i \A^j\, \d \doti)+  \gamma \,  \nabla^i  \,(  \nabla^j \A_b\, \d \doti) \right] \\  
&+ 2 \, \sqrt g \, V' \, \A^i\,, 
\end{aligned}\end{equation}
while best-matching  w.r.t.  diffeomorphisms gives
\begin{equation}\label{DiffeoConstraint4}
\frac{\delta \d \mathcal L_\st{1-form} }{\delta \d \xi_i} = - 2 \, \nabla_j p^{ij} + \E^j \,  \nabla^i \A_j - \A^i \, \nabla_j \E^j = \mathcal H^i \approx 0\,.
\end{equation}
Propagation of the Hamiltonian constraint gives
\begin{equation}\begin{aligned}\label{PropagationHamConst4}
\dbm \mathcal H & = \frac{\sqrt g}{\d \doti} \, \nabla_i \left( 2 \, \d \doti^2 \, \nabla_j p^{ij} - \d \doti^2 \,  \E^j \,  \nabla^i \A_j + \d \doti^2 \, \A^i \, \nabla_j \E^j\right) \\  
& - \frac{2 \sqrt g}{\d \doti} \left[   (\alpha + {\ts \frac 1 4})  \nabla^k (\d \doti^2 \, \E^j \, \nabla_k \A_j) 
+ (\beta - {\ts \frac 1 4})  \nabla^k (\d \doti^2 \, \E^j \, \nabla_j \A_k)  \right] \\  
& - \gamma  \frac{2 \sqrt g}{\d \doti} \nabla^k (\d \doti^2 \, \E^k \, \nabla_j \A^j)   
-    \frac{\sqrt g}{\d \doti} \nabla^k \left( \d \doti^2 \, \nabla_j \E^j \, \A_k \right)  \,.
\end{aligned}\end{equation}
Here, the first line vanishes weakly due to the diffeomorphism constraint (\ref{DiffeoConstraint4}). 
In the second and third lines there are three terms that offer us the option of making them
vanish strongly through the choice of the parameters $\alpha$, $\beta$ and $\gamma$,
otherwise they would imply the additional constraints
\begin{equation}
  \E^j \, \nabla_k \A_j \approx {\rm \it const.} \,, ~~~
  \E^j \, \nabla_j \A_k \approx {\rm \it const.} \,, ~~~
  \E^k \, \nabla_j\A^j \approx {\rm \it const.} \,.
\end{equation}
None of these propagate, nor does any linear combination of them. Their propagation
would thus require the introduction of further tertiary constraints, the procedure
continuing (for any choice of $\alpha,\beta,\gamma$) until we find an inconsistency.
Therefore the only choice is to set $\alpha =  -\frac 1 4$,  $\beta =  \frac 1 4$ and $\gamma = 0$.
With this choice of parameters the $W$ term in the  potential  takes the form
\begin{equation}
W = {\ts \frac 1 4} (\nabla_i \A_j \nabla^j \A^i - \nabla_i \A_j \nabla^i \A^j)\,.
\end{equation}

In (\ref{PropagationHamConst4}) a further secondary constraint appears, this time without any option to kill it strongly:
\begin{equation}
\mathcal G = \nabla_i \E^i \approx 0  \,. 
\end{equation}
This is the `Gauss constraint' of electromagnetism. \index{Gauss constraint} Let's propagate it:
\begin{equation}
\dbm \mathcal G = 2 \, \sqrt g  \, \nabla_i (  \d \chi \, V' \, \A^i ) \,. 
\end{equation}
This shows that the only way to make it propagate is to take $V' = 0$, which means
no potential -- and, in particular, no mass term for the form field.

The secondary constraint we have found generates gauge transformations of $\A_i$.
Extending the definition of the Poisson brackets to include the form field and its
conjugate momentum $\{ F(E,A),G(E,A)\} = \int \d^3 x (  \frac{\delta F}{\delta \A_i } \frac{\delta F}{\delta E^i }
-  \frac{\delta F}{\delta E^i } \frac{\delta F}{\delta \A_i })$, we get 
\begin{equation}
\{ A_i , (\sigma|\mathcal G) \} =  - \nabla_i \sigma \,.
\end{equation}
We can easily implement this symmetry from the start,
through best-matching.

\noindent{\bf Best-matching gauge transformations} 

\index{gauge transformations}
The ansatz is
\begin{equation}\begin{aligned}
\d \mathcal L_\st{gauge} =  \int \d^3 x \sqrt{g} &\sqrt{ R - 2 \, \Lambda +{\ts \frac 1 4} (\nabla_i \A_j \nabla^j \A^i - \nabla_i \A_j \nabla^i \A^j)  }  \\  
\cdot  & \sqrt{ \left(  g^{ik}g^{jl} -  \, g^{ij}g^{kl} \right)   \dbm  g_{ij} \, \dbm  g_{kl}  + g^{ij} \dbm \A_i  \dbm \A_j}  \,,
\end{aligned}\end{equation}
with the modified best-matching differential
\begin{equation}
\dbm \A_i = \d \A_i + \Lie_{\d \xi} \A_i + \nabla_i \d \varPhi \,.
\end{equation}
With this choice everything is the same as above, except that now the momenta $\E^i$ have
the modified definition
\begin{equation}\label{Maxwell1}
  \E^i  =\frac{\sqrt g }{2 \,\d \doti}  \, g^{ij} \, ( \d \A_i + \Lie_{\d \xi} \A_i + \nabla_i \d \varPhi) \,,
\end{equation}
and we have to best-match with respect to $\d \varPhi$,
\begin{equation}\label{Maxwell2}
\frac{\delta \d \mathcal L_\st{gauge}}{\delta \d \varPhi} = -\nabla_i \E^i \,.
\end{equation}
We already know that the constraint algebra closes.
The equations of motion for the form-field are
\begin{equation}
\dbm \E^i = \d \E^i- \Lie_{\d\xi} \E^i -  \nabla_i \d \varPhi =  {\ts \frac 1 2}\, \sqrt g \,  \nabla_j (\nabla^j \A^i  \d \chi - \nabla^i \A^j \d \chi)\,.
\end{equation}
We see that the form field $\A_i$ enters the equations of motion only through the
combination $F^{ij} =\nabla^i \A^j - \nabla^j \A^i$. We can use the dual 
vector field density
\begin{equation}
B^i = {\ts \frac 1 2} \epsilon^{ijk} \nabla_j A_k \,,
\end{equation}
(notice that the contravariant form of the Levi-Civita symbol $ \epsilon^{ijk} $ is a density of weight 1,
while the covariant form  $ \epsilon_{ijk} $  has weight -1). Then the equations of motion take
the form
\begin{equation} \label{Maxwell3}
\d E^i - \Lie_{\d\xi} E^i -  \nabla_i \d \varPhi=   \sqrt g \, \epsilon_{ijk} \nabla^j B^k \,,
\end{equation}
but the vector density $B^i$ satisfies a transversality constraint
\begin{equation}\label{Maxwell4}
\nabla_i B^i = 0 \,.
\end{equation}
Let's now consider a flat background $g_{ij} = \delta_{ij}$ in Cartesian coordinates $\d \xi_i = 0$,
and use the ephemeris time $\d \chi = \d \teph$. Equations (\ref{Maxwell1}),
 (\ref{Maxwell2}), (\ref{Maxwell3}) and  (\ref{Maxwell4}) now take the form
 \begin{equation}\begin{aligned}
&{\bm \nabla} \cdot {\bf \E} = 0 \,, & &\frac{\!\!\!\! \d {\bf \E}}{\d \teph} =- {\bm \nabla }\times {\bf B} \,, \\
&{\bm \nabla} \cdot {\bf B} = 0 \,, & &\frac{\!\!\!\! \d {\bf B}}{\d \teph} =  {\bm \nabla }\times {\bf \E} \,,
 \end{aligned}\end{equation}
 These are obviously the source-free Maxwell equations. \index{Maxwell's equations} In addition we see that the definitions \index{magnetic field} \index{electric field}
 of $B^i$ and of $\E^i$ in terms of $\A^i$ and its space and time derivatives are nothing more
 than the standard expressions of the magnetic and electric fields in terms of the vector potential:
 \begin{equation}
 {\bf \E} = {\ts \frac 1 2}  \left( \frac{\!\!\!\! \d {\bf \A}}{\d \teph} +{\bm \nabla} \d \varPhi \right) \,, ~~~~
 {\bf B} ={\ts \frac 1 2} {\bm \nabla} \times {\bf \A}\,.
  \end{equation}
  
  \vspace{12pt}

\noindent{\bf Yang--Mills Theory}

\index{Yang--Mills field}
In \cite{RWR-VectorFields} Anderson and Barbour considered the case of $N$ 1-form fields coupled to 
each other in all possible ways compatible with a fairly general ansatz -- namely, that the potential
must be at most second order in the space derivatives and at most fourth order in the field variables.
For the kinetic term, the only freedom that was left was to have a symmetric matrix coupling the best-matched
velocities of the different 1-form fields.
If $A^\alpha_i$, $\alpha=1,\dots,N,$ are the various 1-form fields, propagation of the
Hamiltonian constraint requires the introduction of $N$ Gauss constraints:
\begin{equation}\label{GaugeTheory_GaussConstraints}
\mathcal G_\alpha = \nabla_i E^i_\alpha -  {C^\beta}_{\alpha\gamma} \, E^i_\beta \, A^\gamma_i \approx 0 \,,
\end{equation}
and these non-Abelian gauge transformations must be implemented in the best-matching differential as
\begin{equation}
\dbm A^\alpha_i = \d A^\alpha_i + \Lie_{\d \xi} A^\alpha_i + \nabla_i \d \varPhi^\alpha
+ {C^\alpha}_{\beta\gamma} \,A^\beta_i \, \d \varPhi^\gamma \,.
\end{equation}
The kinetic term is therefore
\begin{equation}
\d \mathcal L_\st{kin}^2 = \delta_{\alpha\beta} \, g^{ij} \, \dbm A^\alpha_i \, \dbm A^\beta_j \,,
\end{equation}
the potential term is constrained to be
\begin{equation}
W=  g^{ik} g^{jl} \, \delta_{\alpha\rho}
\left( \nabla_{[i} A^\alpha_{j]} + {\ts \frac 1 2} {C^\alpha}_{\beta\gamma} \, A^\beta_i \, A^\gamma_j \right)\left( \nabla_{[k} A^\rho_{l]} +  {\ts \frac 1 2}{C^\rho}_{\sigma\tau} \, A^\sigma_k \, A^\tau_l \right) \,,
\end{equation}
and the 3-index matrix ${C^\alpha}_{\beta\gamma}$ is constrained to satisfy the Jacobi rules
\begin{equation}
\delta_{\alpha\rho} \left(
{C^\alpha}_{\beta\gamma}  \,{C^\rho}_{\sigma\tau} +
{C^\alpha}_{\beta\tau}  \,{C^\rho}_{\sigma\gamma} +
{C^\alpha}_{\beta\gamma}  \,{C^\rho}_{\tau\sigma} 
\right) = 0\,,
\end{equation}
which implies, as Gell-Mann and Glashow \index{Murray Gell-Mann} \index{Sheldon Glashow} have shown \cite{Gell-Mann--Glashow}, that the matrix ${C^\alpha}_{\beta\gamma}$ 
represents the structure constants of a direct sum of compact simple and $U(1)$ Lie algebras. What we have
found is nothing less than Yang--Mills theory, thus covering the whole bosonic sector of the Standard Model.\index{Standard Model of Particle Physics}

\noindent{\bf Further generalizations}

In addition to the above, one could consider an antisymmetric tensor field $F^{ij}=-F^{ji}$ (the symmetric
case is already included in the treatment of the metric field), but this turns out to be completely equivalent to treating the dual 1-form field $B_i = \frac 1 2 \, \epsilon_{ijk} \, F^{jk}$, and similarly for the case of $N$ antisymmetric tensor fields. The proof of these statements is left as an exercise.

The question posed in \cite{RWR-VectorFields} of whether topological terms $\epsilon_{\mu\nu\rho\sigma} F^{\mu\nu} F^{\rho\sigma}$ as proposed by 't-Hooft \index{Gerard 't Hooft} can be accommodated in this approach remains to be answered. Terms of this sort appear to be needed to explain the low-energy spectrum of QCD \cite{tHooft1,tHooft2}, although they lead to the so-called
`strong CP problem' \cite{Peccei}.\index{strong CP problem}

In \cite{RWR-Ed-Strong-Coupled,RWR-Ed-Variations} Anderson extended the analysis to several new cases, for example Strong Gravity  and generalizations thereof, in particular a new class of theories that can be considered as strong-gravity limits of Brans--Dicke theory. \index{Brans--Dicke theory} \index{Strong Gravity} Particularly interesting (because they are  realized in Nature) are spin-1/2 fermions of the theories of Dirac, Maxwell--Dirac and Yang--Mills--Dirac. Fermions are introduced in a somewhat phenomenological fashion, with the introduction of a spin connection but without a first-order formulation of the gravitational degrees of freedom. The kinetic fermionic part of the action is therefore outside the local square root, and it does not contribute to the differential of the instant. The potential term, instead, does contribute. A more satisfying treatment involves Palatini's formulation of GR, which does not have a local square root at all. A discussion of first-order formulations of gravity in the relational setting is still missing. \index{Palatini formulation of General Relativity}

Finally, a paper \cite{EdFlavioPaper}  I recently wrote with E. Anderson completes the analysis of the metric field in full generality, moreover in a purely Hamiltonian setting, employing Dirac's method.  These results will be summarized in
Section~\ref{ExtendedRWR}, where they will be used to justify Shape Dynamics.

\vspace{12pt}
\noindent
\fbox{\parbox{0.98\linewidth}{
{\bf Further reading:} The complete literature on the `Relativity Without Relativity' approach
is \cite{RWR, RWR-VectorFields, RWR-Ed-Strong-Coupled, RWR-Ed-Variations,EdFlavioPaper}.
}}\vspace{12pt}

\noindent{\bf The problem of many-fingered time}

\index{many-fingered time}
The Machian first principles on which we have based our field theories up to now have proved to be quite powerful
in not only deriving the relativity principle from a simpler set of axioms, but also in identifying
all the kinds of (classical) fields that are presently thought to be fundamental in nature. 
However, these principles fail to realize, in the case of the gravitational field, what Barbour and Bertotti identified as the only precise formulation of Mach's principle: the Mach--Poincar\'e Principle. \index{Mach--Poincar\'e Principle}
The `culprit' is the local square-root form of the action, the very thing that plays such a key role in some of the most interesting results I have shown: the derivation of the universal light cone and gauge theory. It does this because it leads to \e{local}
Hamiltonian constraints, which constrain one degree of freedom at \e{each} space point.
This leads to a mismatch. For having identified the diffeomorphism symmetry and implemented it through best-matching, we attributed to the gravitational field 3 degrees of freedom per space point, which is
the dimensionality of $\Superspace$, \index{superspace} gravity's putative configuration space. But the  dynamical laws we found
do not realize the Mach--Poincar\'e Principle on $\Superspace$: given two points in it, there is a whole
`sheaf' of curves that extremalize our best-matching action (\ref{BestMatchedActionForGravity}).
The variational principle we have, in its most advanced form (the `differential-almost-Hamiltonian'
approach),  produces a curve in $\Superspace$ given initial data consisting of a point in $\Superspace$ and 
a transverse momentum, plus a  lapse that depends on both time and space (check appendix~\ref{ADM-WDW} for the definition of lapse).
But then any other lapse is equally good, and produces a different curve in $\Superspace$ which shares
only the endpoints with the original one (see Fig.~\ref{ManyFingeredTime}). The Einstein--Hilbert
action assigns the same value to both curves. The uniqueness required by the Mach--Poincar\'e
principle is absent.  In Fig.~\ref{ManyFingeredTime} I show graphically what this ambiguity means:
each curve in $\Superspace$ that can be generated by a different choice of lapse corresponds to a different foliation
of the same spacetime. \index{many-fingered time}

\begin{figure}[t!]
\begin{center}
\includegraphics[width=0.45\textwidth]{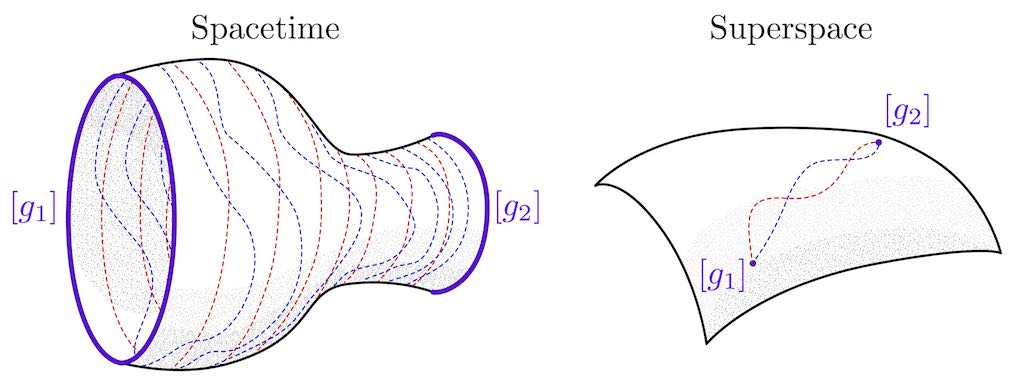}
\end{center}
\caption[The issue of Many-Fingered Time]{Many-fingered time issue: \index{many-fingered time}
a single solution to Einstein's equations (the slab of spacetime on the left) between two Cauchy hypersurfaces (in purple) where two initial and final 3-geometries $[g_1]$, $[g_2]$ (the $[\,\cdot\,]$ brackets stand for `equivalence class under diffeomorphisms') are specified, does not correspond to a single curve in superspace (on the right). Instead, for each choice of \emph{foliation} of the spacetime on the left that is compatible with the boundary conditions $[g_1]$, $[g_2]$ there is a different curve in superspace. Here I have represented two different choices of foliation in red and a blue.}
\label{ManyFingeredTime}
\end{figure}

From the Machian point of view, this situation is unacceptable, but the way out was actually anticipated  already in the 1970's.
As we can see from the counting of the degrees of freedom, the gravitational field has two
Lagrangian degrees of freedom per point ($\Riem$ has six per point, the diffeomorphisms reduce it to
$\Superspace$, which has three, and the Hamiltonian constraint further reduces them to two).
But $\Superspace$ has three degrees of freedom per space point, and therefore it cannot be the physical configuration space of GR. The Hamiltonian constraint, unlike the diffeomorphism constraint, does not
admit a simple geometrical interpretation in terms of transformations
on the configuration-space variables alone, and therefore cannot be used
to quotient $\Superspace$ into the physical configuration space of GR.
The orbits of the Hamiltonian constraint
do not lie on $\Riem$: they mix the metric and the momenta. In other words, 
the vector field corresponding to the Hamiltonian constraint is not a pointwise map of $\Riem$~onto $\Riem$: it lives inextricably in phase space, and the associated phase-space transformation
of the metric depends on the momenta.\footnote{The Hamiltonian constraint can map curves on $\Riem$ to curves on $\Riem$, what it cannot do is take points onto points without knowledge of the curve to which it belongs. Thanks to H. Gomes for the observation}
 
Moreover, the Barbour--Foster result \cite{BF-DiraTheorem} I sketched at the end of Sec.~\ref{CrashCourseInDiracsAnalysis} strongly suggests that the Hamiltonian constraint is not \e{exclusively} a generator of gauge transformations. There ought to be a single
linear combination of $\mathcal H(x)$, some $(f | \mathcal H) = \int \d^3 x \, f(x) \, \mathcal H(x)$
which is associated with genuine evolution. If not, one would reach the absurd conclusion that dynamical evolution
is a gauge transformation (look at Kucha\v{r}'s review \cite{KucharReview} for a criticism of such an idea).
The particular linear combination of $\mathcal H$ that generates reparametrizations actually depends 
on the gauge-fixing we choose for $\mathcal H$, which in turn is related to the way we foliate spacetime.
In a patch of spacetime, any positive function $f(x)$ can be used as a lapse defining a foliation,
and the corresponding linear combination $(f | \mathcal H)$ would then be our generator of true evolution. 

The Hamiltonian constraint must be an admixture of gauge and dynamics, and, at the first glance, there does not appear to be any obvious way to 
disentangle the two if one insists that refoliation invariance is sacrosanct. As I said, a way out of this puzzle was already proposed in the
70's, mainly thanks to J. W. York, who built on the work of Y. Choquet-Bruhat and A. Lichnerowicz.

\newpage

\section{York's solution to the initial-value problem}\label{YorksMethodSection}\index{initial-value problem}
\index{James W. York, jr.}\index{Yvonne Choquet-Bruhat}\index{Andr\'e Lichnerowicz}

I will now explain the core of York's work on the initial-value problem of GR.
What York did is highly relevant for both Shape Dynamics and the Machian program
in general: it indicated strongly that the physical degrees of freedom of GR are
conformally invariant, in addition to diffeomorphism invariant, and it identified a corresponding
 fully reduced phase space of GR. As I will show, it allows us to find a solution to the problem of many-fingered time and to formulate a theory 
of geometrodynamics that satisfies the Mach--Poincar\'e Principle and which we
call Shape Dynamics.

York found a general method for solving the `initial-value problem', which
means finding phase-space data $g_{ij}$, $p^{kl}$ on an initial Cauchy hypersurface that satisfy GR's Hamiltonian 
(\ref{QuadraticConstraint}) and diffeomorphism constraints (\ref{DiffeoConstraint}).

The diffeomorphism constraint merely requires the momentum $p^{ij}$
to be a tensor density whose covariant divergence w.r.t. its conjugate $g_{ij}$ vanishes. It is clearly closely analogous
to the Gauss constraint of electrodynamics, which can also be formulated on a Riemannian
manifold $(\Sigma,g)$ and fixes the divergence of a vector field to be zero. In electrodynamics
one can find an explicit solution of the Gauss constraint by identifying the longitudinal part of the vector
 field in terms of a scalar field that solves Poisson's equation. Subtraction of the longitudinal part then solves the problem. All of this is based on a very general
 result of differential geometry called the Helmholtz Decomposition Theorem,
 which allows one to identify the longitudinal and transverse components of a vector field
 on a general Riemannian manifold (see the first Section
 of Appendix~\ref{TTdecompositionAppendix}).
 
 \index{Helmholtz decomposition}
Berger and Ebin \cite{BergerEbin} found an analogue of Helmholtz's theorem for symmetric 2-tensors (later rediscovered by Deser \cite{deser1967decomposition}), which allows one to uniquely determine their `transverse' and `longitudinal'
parts (that is, the divergence-free part and the remainder) in terms of a vector field (analogous
to the scalar field mentioned above) which must satisfy a certain elliptic equation, for which existence
and uniqueness theorems hold. I won't reproduce this result here, since it is already contained
as a sub-case in York's treatment of the diffeomorphism constraint, which I describe in detail in
Appendix~\ref{TTdecompositionAppendix}. So the diffeomorphism constraint is no problem:
it always admits a unique solution, at least on a compact manifold.\footnote{In the asymptotically flat case, one can find
up to a 6-parameter family of solutions to the diffeomorphism constraint associated with the Killing 
vectors of flat space (see Appendix~\ref{TTdecompositionAppendix}).This is done by using special boundary conditions and is important for the covariant definition
of momentum and angular momentum of the gravitational field. However, it does not  need to bother
us at this stage.}

The problem is the Hamiltonian constraint. With the method sketched above, I can
build any solution I like of the diffeomorphism constraint by starting with a 3-metric  $g_{ij}$ and an arbitrary 
symmetric tensor $p^{ij}$ and extracting its transverse part w.r.t. $g_{ij}$ by solving an elliptic differential
equation. Now this transverse part will not in general satisfy the Hamiltonian constraint:
a quadratic combination of its components must be proportional to
the Ricci scalar of   $g_{ij}$ at each point. The constraint is scalar, so I could try to locally 
rescale the tensor by a scalar function $p^{ij} \to f(x) \, p^{ij}$, adjusting its magnitude
to solve the Hamiltonian constraint at each point, but then I would lose the transversality
with respect to $g_{ij}$. Alternatively, I could consider the Hamiltonian constraint as 
a (nonlinear and very complicated) differential equation for $g_{ij}$, but the
transverse part $p^{ij}_\st{T}$ of $p^{ij}$ is defined relative to $g_{ij}$ itself: $p^{ij}_\st{T}$
depends \emph{nonlocally} on $g_{ij}$, and the equation becomes an integro-differential
equation.  

\noindent{\bf Lichnerowicz's partial solution}

\index{Andr\'e Lichnerowicz}

In 1944 \cite{Lichnerowicz} Lichnerowicz had found a way to decouple the Hamiltonian and diffeomorphism constraint 
in the case of a \emph{maximal} Cauchy hypersurface, that is, one whose extrinsic curvature
has a vanishing trace $K = {K^i}_i=0$.
Consider the two constraints expressed in terms of the extrinsic curvature:
\begin{equation}\begin{aligned}
\nabla_b(K^{ij} - g^{ij} \, K ) = 0 \,,\qquad 
K^{ij} K_{ij} - K^2 - R =0\,,
\end{aligned}\end{equation}
Lichnerowicz's strategy to find a metric $g_{ij}$ and an extrinsic curvature $K^{ij}$ 
that solve the above equations was to start from an arbitrary pair $\bar g_{ij}$, $\bar K^{ij}$
that do not solve it and make a conformal transformation of the metric
\begin{equation}
g_{ij} = \phi^4  \bar g_{ij} \,, \qquad g^{ij} = \phi^{-4}  \bar g^{ij} \,, \qquad \phi  ~~ \text{\it smooth and positive}\,,
\end{equation}
and simultaneously of the extrinsic curvature (see Appendix \ref{TTdecompositionAppendix})
$K^{ij} = \phi^{-10}  \bar K^{ij} $ that transform $\bar g_{ij}$, $\bar K^{ij}$ into
$g_{ij}$, $K^{ij}$ which do satisfy the constraints. Then the problem of satisfying
the constraints become that of finding the correct $\phi$. This is possible
by virtue of the maximal slicing condition $K=0$ and the ellipticity
of the equation for $\phi$ that one finds. Let's see how this works.

As I show in Eq. (\ref{CovariantDerivativePhiTotheTenSymmetricTensor})
in  Appendix \ref{TTdecompositionAppendix} below, under the conformal transformation
\begin{equation}
 \nabla_j  W^{ij} =
 \nabla_j ( \phi^{-10}  \bar W^{ij} ) = \phi^{-10}  \bar  \nabla_j \bar W^{ij} 
- 2\, \phi^{-10} \, \bar W \, \partial^i \log \phi \,,
\end{equation}
so, if $W^{ij} =  K^{ij} - g^{ij} K$,  the diffeomorphism constraint transforms to 
\begin{equation}\begin{aligned}
\nabla_j( K^{ij} - g^{ij} \, K ) = \phi^{-10} \bar \nabla_j \bar K^{ij} + \phi^{10} \partial^i (\phi^{-10} \,\bar K)\,,
\end{aligned}\end{equation}
and therefore if $K=0$ the transformed diffeomorphism constraint does not depend on $\phi$.
It is this conformal covariance of the diffeomorphism constraint in the $K=0$ case that makes it possible
to decouple the two constraint equations and solve the Hamiltonian one separately
just for the `conformal factor' $\phi$.
The scalar curvature transforms according to the formula \cite{Lichnerowicz}
\begin{equation}\label{ConformalTransformationOfR}
 R = \phi^{-4} \, \bar R - 8 \, \phi^{-5} \bar \laplacian \phi \,.
\end{equation}
Therefore the two constraints take the following form in the barred variables
\begin{equation}\begin{aligned}  \label{LichnerowiczEquation}
\bar \nabla_j \, \bar K^{ij} = 0 \,,
\qquad
\phi^{-12} \bar K^{ij} \bar K_{ij}  - \phi^{-4} \, \bar R + 8 \, \phi^{-5} \bar \laplacian \phi  =0\,.
\end{aligned}\end{equation}
The last equation is strongly elliptic and quasilinear. For equations of this kind there are well-known theorems of existence
and uniqueness of the solutions. The initial-value problem \index{initial-value problem} can be therefore solved with the
following procedure:
\begin{enumerate}
\item Start with an arbitrary metric $\bar g_{ij}$ and an arbitrary  traceless  $\bar K^{ij}$.

\item Find the transverse part $\bar K^{ij}_\st{T}$ of $\bar K^{ij}$ with respect to $\bar g_{ij}$.

\item Solve the `Lichnerowicz equation' (\ref{LichnerowiczEquation}) for $\phi$.

\item Then the `physical' initial-value metric and extrinsic curvature, which satisfy the constraints, are $g_{ij} =
\phi^{4} \bar g_{ij}$ and $K^{ij} = \phi^{-10} \bar K^{ij}_\st{T}$. 
\end{enumerate}\index{Hidehiko Yamab}

Lichnerwicz's method has a serious limitation: on compact manifolds it works only for \emph{Yamabe-positive} metrics (see below).
The final physical data, in fact, satisfy the Hamiltonian constraint with $K=0$,
\begin{equation}\label{K=0HamiltonianConstraint}
K^{ij} K_{ij} = R\,,
\end{equation}
and the left-hand side is, by construction, non-negative. This implies that the techinque is 
only consistent if the scalar curvature on the right-hand side is non-negative \emph{at every
point} as well. This restricts the conformal equivalence class\footnote{The conformal equivalence class,
or conformal class for brevity, of a metric $g_{ij}$ is defined as $\{ \tilde g_{ij} / \exists~\phi >0, \phi~ \text{smooth}, \tilde g_{ij}=\phi^4 g_{ij}\}$. The space of conformal classes of metrics
is called \emph{conformal superspace}, which in this \thistext{} we also refer to as shape space $\shs$ when there is no risk of confusion with the shape space of the $N$-body problem. Shape Dynamics takes conformal superspace to be the reduced, physical configuration space of gravity.} of the metrics we can take
as initial data.

To justify the last statement, I need to introduce a result in pure mathematics: the Yamabe theorem \cite{YamabeConjecture,Schoen1984}, which Yamabe proposed as a conjecture and believed to have
found its proof in 1960. However Trudinger found an error in Yamabe's proof in 1968, and later Aubin and Schoen supplied the correct proof around 1984. The theorem states that
every Riemannian metric on a \index{closed spacelike hypersurfaces} \e{closed manifold}\footnote{The analogous statement on noncompact manifolds is 
\emph{wrong}, as proved by Jin \cite{Zhiren1988}.} (of dimension $\geq 3$) can be conformally transformed to a 
metric with \emph{constant scalar curvature}. This constant is obviously not uniquely determined
because  one can change the magnitude of the Ricci scalar with a rescaling (a constant conformal transformation), but its sign is a conformal invariant. In fact, if the manifold is compact there cannot exist a conformal
transformation mapping a metric $g_{ab}$ with $R>0$ everywhere to another metric $\bar g_{ij}$
with $\bar R<0$ everywhere. To see this, take Eq. (\ref{ConformalTransformationOfR}), multiply it by $\phi^{-1}$ and integrate it over all of $\Sigma$: 
\begin{equation}
\int_\Sigma \sqrt g \, R \, \phi^{-1} =   \int_\Sigma\sqrt{\bar g} \, (\bar R \, \phi - 8 \, \bar \laplacian \phi ) 
\equiv   \int_\Sigma\sqrt{\bar g} \, \bar R \, \phi \,.
\end{equation}
The final step in which the Laplacian is cancelled is legitimate only if $\Sigma$ is compact. The above equation
proves that if $R$ is definite everywhere it cannot change sign everywhere under a conformal transformation.\footnote{In the noncompact, asymptotically flat case one cannot discard the boundary value of  $\laplacian \phi$,
and it is these boundary conditions that allowed \'O Murchadha and York to define a mass-at-infinity for the gravitational
field in the asymptotically flat case \cite{Niall-York74}.}
The sign of the Ricci scalar in the representation in which $R$ is constant 
is therefore invariant under conformal transformations.

What is more, we can go further and introduce a real quantity, called the \emph{Yamabe constant},
which is a conformal invariant and represents precisely the value of the
Ricci scalar in the conformal gauge where it is a constant, rescaled by the volume
of the manifold so that it's invariant under constant rescalings as well:
\begin{equation}
y [\Sigma,g] = \inf_\phi \left\{ \frac {\int \d^3 x \sqrt g \,( \phi^2 \, R - 8 \, \phi \, \laplacian \phi )}{\int \d^3 x \sqrt g \, \phi^6} \right\} \,.
\end{equation}
A theorem~\cite{lee1987yamabe} by Yamabe, Trudinger, Aubin, and Schoen \index{Yamabe--Trudinger--Aubin--Schoen theorem} states that within a conformal equivalence class there exist
metrics which  realize the minimum, and they must have constant scalar curvature $R = \text{\it const.}$
Riemannian metrics can be classified according to the sign of the Yamabe constant, and divided into \emph{Yamabe positive, negative} and \emph{zero}, which means, according to the theorem, 
that they can be transformed into a metric with, respectively, positive, negative or zero constant
scalar curvature. 
In general a manifold can  be equipped with metrics belonging to different Yamabe classes. The following
quantity:
\begin{equation}
\mathcal Y [\Sigma] = \inf_g y [\Sigma,g]  \,,
\end{equation}
is a topological invariant of the manifold $\Sigma$, and is called the \emph{Yamabe invariant}. \index{Yamabe invariant}
The Yamabe--Trudinger--Aubin--Schoen theorem states also that no manifold
in dimension 3, 4 or 5 can have a Yamabe invariant larger than that of the 3-sphere:
\begin{equation}
\mathcal Y [\Sigma] \leq  \mathcal Y [S^n] \,, \qquad n=3,4,5\,.
\end{equation}
The manifolds themselves can then be classified according to the value of their Yamabe invariant:
Type -1 manifolds only admit negative-curvature constant-curvature metrics. Yamabe type 0 admit
constant-curvature metrics with zero or negative curvature. Type +1 manifolds admit at least one constant-curvature
metric with positive curvature.

We now observe that the Lichnerowicz equation (\ref{LichnerowiczEquation})
is conformally covariant in the sense that if $\phi$ is the solution of  (\ref{LichnerowiczEquation})
with data $\bar g_{ij}$, $\bar K^{ij}$, then for any smooth and positive $\theta$
one has that $\omega = \theta^{-1}\phi$ is the solution of  (\ref{LichnerowiczEquation}) for the 
data $\tilde g_{ij} = \theta^4\bar g_{ij}$, $\tilde K^{ij} = \theta^{-10}\bar K^{ij}$. Thus, one is
free to choose the $\theta$ such that $\theta^4\bar g_{ab}$ has Ricci scalar $R[\theta^4 \bar g_{ij};x) = {\rm \it const}$ with ${\rm \it const} $ either $= -1, =0$ or $=+1$ $\forall x$ (which exists by  Yamabe's theorem). 
It is now evident that Lichnerowicz's method cannot be used with Yamabe-negative metrics on compact manifolds. For in that case one would have a conformal transformation (generated by $\omega$)
that maps a metric $\tilde g_{ij}$ with $R=-1$ to a metric $g_{ab}$ which has an everywhere-positive $R$ [because by assumption $g_{ab}$ has to solve (\ref{K=0HamiltonianConstraint})].
The Yamabe zero case is distinguished: in that case Lichnerowicz's equation can be solved only
if $K^{ij} =0$ at every point, as can be seen from  (\ref{K=0HamiltonianConstraint}). This is a very
non-generic initial condition.

In the noncompact case, Eq.~(\ref{K=0HamiltonianConstraint})
admits a unique solution for any initial data if the manifold is asymptotically flat \cite{YorkTTdecomposition} (if the initial data include conditions at infinity).

An observation following the proof of the conformal covariance of Lichnerowicz's method:
for practical reasons, one starts by specifying a complete 3-metric $\bar g_{ij}$ (six components) 
and a complete symmetric 2-tensor $\bar K^{ij}$ (another six components), but then the 
end result, in the form of $g_{ab}$ and $K^{ij}$, depends only on the conformally-invariant and diffeo-invariant
part of $\bar g_{ij}$, that is, its conformal class (two degrees of freedom). Moreover,
it depends only on the TT-part of $\bar K^{ij}$, (which are another two degrees of freedom).
These would be physical data that solve the initial-value problem \index{initial-value problem} if Lichnerowicz's solution
were general, but it isn't. York succeeded in generalizing it to arbitrary Yamabe class, and can
consequently claim to have identified the physical degrees of freedom of the gravitational
field \cite{York:york_method_prl}.\footnote{\label{twocavs} Subject to a caveat, which we shall mention below.} Let's see what he did. 

\noindent{\bf York's general solution}

In a series of papers \cite{York:cotton_tensor,York:york_method_prl,YorkTTdecomposition,murchadha1974initial}, York 
and \'O Murchadha made  decisive progress  by letting the extrinsic curvature have a
spatially constant trace: 
\begin{equation}\label{YorkConditionK}
K^{ij} = K^{ij}_\st{TT} + {\ts \frac 1 3} \, \tau \, g^{ij}\,, \qquad \tau = \mathrm{\it const.}
\end{equation}
Here, $\tau$ is a spatial constant but it is time-dependent. In particular, it grows monotonically
whenever York's method can be applied, and can be used as a time parameter in its own right.
For this reason it is also referred to as the \emph{York time}. \index{York time}
A tensor such as (\ref{YorkConditionK}) is automatically transverse with respect to the metric $g_{ij}$ because
the covariant divergence of a constant times $g^{ij}$ is zero by the metric-compatibility condition.
A further assumption of York makes all the difference. The transformation law of 
$K^{ij}$ under conformal transformations is taken to be
\begin{equation}
K^{ij} = \phi^{-10} \, \bar K^{ij}_\st{TT} + {\ts \frac 1 3} \, \phi^{-4}  \tau \,\bar g^{ij}\,,
\end{equation}
so that after a conformal transformation the trace part is still
spatially constant:
\begin{equation}
\bar g_{ij} \, \bar K^{ij} =  g_{ij} \,  K^{ij}
\end{equation}
If that is the case, then Lichnerowicz's decoupling of the 
Hamiltonian and diffeomorphism constraints continues to hold: the TT-part is conformally
covariant, as proved in Appendix~\ref{TTdecompositionAppendix}, while the trace part is assumed to be invariant
and therefore transverse as well:
\begin{equation}
\nabla_j (\tau \, g^{ij}) = 0 \,, \qquad  \bar \nabla_j (\tau \, \bar g^{ij})
=  \nabla_j (\tau \, \bar g^{ij}) + \tau \, \bar g^{kj} \Delta \Gamma^i_{jk} + \tau \, \bar g^{ik} \Delta \Gamma^j_{jk} = 0\,,
\end{equation}
where $\Delta \Gamma^a_{jk} $ is defined below.
Thus when we pass from $g_{ij}$, $K^{ij}$ to $\bar g_{ij}$, $\bar K^{ij}$ no
term containing a derivative of $\phi$ appears in the diffeomorphism constraint to threaten the
transversality condition, which remains independent of $\phi$, so that the 
two constraints are still decoupled. But now the conformally transformed Hamiltonian constraint
gains a new term, quadratic in $\tau$:
\begin{equation}\label{LichnerowiczYorkEquation}
\phi^{-12} \, \bar g_{ik} \, \bar g_{jl} \, \bar K^{ij}_\st{TT} \bar K^{kl}_\st{TT} - {\ts \frac 2 3} \tau^2  - \phi^{-4} \, \bar R + 8 \, \phi^{-5} \bar \laplacian \phi  =0\,.
\end{equation}
Note the different powers of $\phi$ in front of the kinetic term and of $-\tau^2$, which 
is entirely due to York's assumption of the conformal invariance of $\tau$. 
Equation (\ref{LichnerowiczYorkEquation}) is called the Lichnerowicz--York equation. A solution of it exists and is unique on arbitrary compact or asymptotically flat manifolds,
regardless of the conformal class of the metric. Let's see how one studies the solutions
of Eq.~(\ref{LichnerowiczYorkEquation}). In accordance with the theory of quasilinear 
elliptic equations \cite{BookonSolutionOfEllipticQuasilinearEquations}, \index{quasilinear elliptic differential equation} Eq.~(\ref{LichnerowiczYorkEquation}) 
admits a unique solution \emph{iff} the polynomial 
\begin{equation}\label{LYPolynomial}
f(z)  = {\ts \frac 2 3} \tau^2 \, z^3 + R \, z^2 - {\rm \it KK} \,,
\end{equation}
(where I called $z = \phi^4$ and ${\rm \it KK} = \bar g_{ik} \, \bar g_{jl} \, \bar K^{ij}_\st{TT} \bar K^{kl}_\st{TT}$) admits a single positive root at every point.

The function $f(z)$ has its extrema at $z=0$ and $z= - R/\tau^2$, and its second derivatives at those
points are $f''(0) = 2 \, R$ and $f''( - R/\tau^2) = - 2\, R$. Thus if $R>0$, then $z=0$ is a local minumum
and  $z= - R/\tau^2$ a local maximum; if $R>0$, then vice versa. Moreover $f(0) = - {\rm \it KK}$ is always
negative, and changing the value of ${\rm \it KK}$ just shifts the whole function
downwards. Therefore we always fall into one of the two cases of Fig.~\ref{PolynomialFig}, 
where we either have a maximum at $z=0$ and a minimum at $z>0$, with a zero to its right (for $R<0$), 
or we have a maximum at $z<0$ and a minimum at $z=0$, with a zero to its right (for $R>0$), all of this regardless of the value of ${\rm\it KK}$.
If we have $R=0$, the zero is at  $z =({\ts \frac 2 3}{\rm\it KK}/\tau^2)^{1/3}$, which is
positive as long as ${\rm\it KK}\neq 0$, otherwise it's zero.
\begin{figure}[t]
\begin{center}
\includegraphics[width=0.45\textwidth]{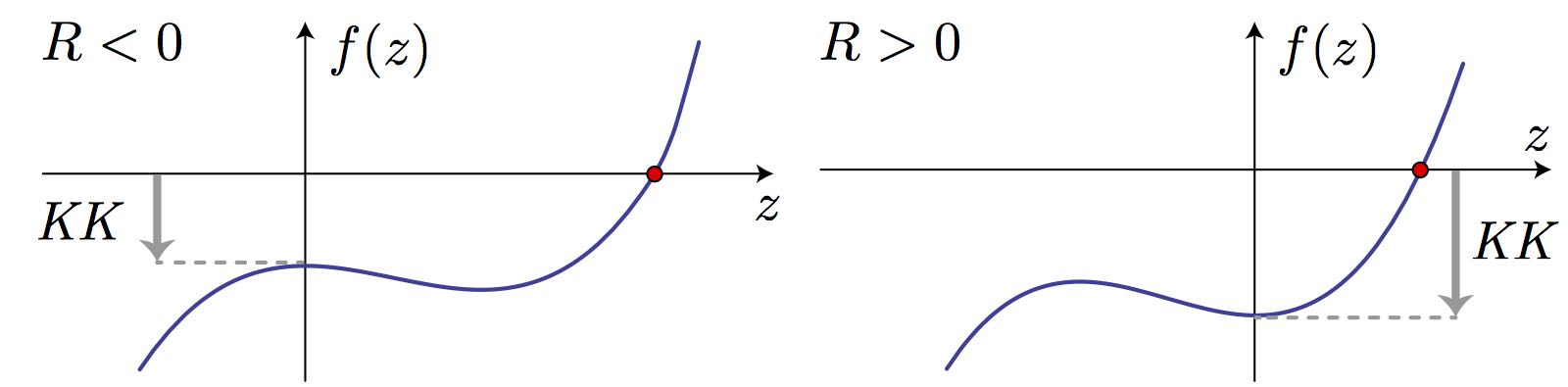}
\end{center}
\caption[Graphical identification of the root of the LY polynomial]{The polynomial (\ref{LYPolynomial}) admits a single
positive root for both signs of $R$. The term ${\rm \it KK}$ can
only push the curve down, moving the root but not changing its nature.\label{PolynomialFig}}
\end{figure}

York's general solution of the initial-value problem works in a spacetime neighbourhood of
a CMC (constant-mean-extrinsic curvature, meaning with $K ={\rm \it const.}$) Cauchy hypersurface. \index{CMC foliation} Therefore the caveat promised in footnote \ref{twocavs} is that York's method can only be applied to CMC-foliable spacetimes, which however are a large class \cite{IsenbergReview} and have nice singularity-avoidance properties.

\vspace{12pt}
\noindent
\fbox{\parbox{0.98\linewidth}{
{\bf Further reading:} Lichnerowicz' paper \cite{Lichnerowicz}, York's 1971 \cite{York:cotton_tensor}, 1972 \cite{York:york_method_prl} and 1973 \cite{YorkTTdecomposition} papers, \'O Murchadha's `readings of the LY equation' \cite{NiallVolumeYorkTime}, Misner--Thorne--Wheeler \cite{MTWbook}, Book on elliptic PDE's \cite{BookonSolutionOfEllipticQuasilinearEquations}. 
}}\vspace{12pt}

\newpage

\section{A derivation of Shape Dynamics} \label{ExtendedRWR}

I will now bring to an end this Part's journey from relational first principles to Shape Dynamics. I will use the analysis which E. Anderson and I made in \cite{EdFlavioPaper}, where we considered a pure-geometrodynamics theory in full generality, without any restriction on the potential term (apart from requiring a second-order potential  which is covariant under diffeomorphisms), and completing the Dirac analysis of the constraints. The analysis of \cite{EdFlavioPaper}  represents the `missing link' between the so-called `Relativity Without Relativity' approach and modern Shape Dynamics.  I will postpone to the beginning of the next Section a review of the actual work that led to the formulation of SD (in particular, the papers~\cite{Gomes:sd_paper,Gomes:linking_paper}).

The most general Jacobi-like, local square root Lagrangian we can take, with a lowest-order (dimension $\ell^{-2}$) potential term is 
\begin{equation}\label{AndersonMercatiInitialAction}
\d \mathcal L_\st{general} =  \int \d^3 x \sqrt{g} \sqrt{a\, R - 2 \, \Lambda} \sqrt{\left( \lambda_1 \, g^{ik}g^{jl} -  \lambda_2 \, g^{ij}g^{kl} \right)  \d  g_{ij} \, \d  g_{kl}}  \,,
\end{equation}
where $a$, $\Lambda$, $\lambda_1$ and $\lambda_2$ are spatially constant parameters.
Since a global rescaling of the action by a constant is irrelevant, it is only the ratio between the parameters $a$ and $\Lambda$ 
that has a physical meaning. For this reason we take $a \in \{ \pm1,0\}$, so that the parameter $a$ only determines
the sign of $R$ and whether the corresponding term is present or not.
The parameters $\lambda_1$, $\lambda_2$ present in the kinetic term parametrize the choice in the relative factor between  the two possible ways to contract the indices of the metric velocities ($\lambda_1=\lambda_2 =1$ is the `DeWitt value', which corresponds to General Relativity).

The `differential of the instant' is \index{differential of the instant}
\begin{equation}
\d \doti = \frac 1 2 \sqrt{ \frac{   \left( \lambda_1 g^{ik}g^{jl} - \lambda_2 \, g^{ij}g^{kl} \right)  \d  g_{ij} \, \d  g_{kl}}{a \, R - 2 \, \Lambda}} \,.
\end{equation}
The canonical momenta are
\begin{equation}
  p^{ij} = {\delta \d \mathcal L_\st{general} \over \delta \d  g_{ij}} = \frac{\sqrt g }{2\,\d \doti} ( \lambda_1 g^{ik}g^{jl} - \lambda_2  g^{ij}g^{kl} ) \,  \d  g_{kl}   \,.
\end{equation}
Due to the local-square-root form of the action, they satisfy at each space point the Hamiltonian constraint
\begin{equation}
\mathcal H = \sqrt g \, ( a \, R - 2 \, \Lambda ) -  {\ts\frac1 {\lambda_1 \, \sqrt g}} \left( p^{ij}p_{ij} - {\ts \frac{  \lambda_2 }{3 \lambda_2 -\lambda_1}} p^{2}\right) = 0\,.
\end{equation}
Note that $\frac{  \lambda_2 }{3 \lambda_2 -\lambda_1}$ diverges for $\lambda_2 = {\ts \frac 1 3} \, \lambda_1$. This singular case requires special care because for it the momenta satisfy a further primary constraint.



The Euler--Lagrange equations are
\begin{equation}\begin{aligned}
\d p^{ij}  = &  a\, \sqrt g \left( {\sfrac 1 2}  R \, g^{ij} - R^{ij} + \nabla^i \nabla^j - g^{ij} \laplacian \right) \d \doti\\  
&-2 \, \Lambda \, \sqrt g \, g^{ij} \, \d \chi - \frac{2 \, \,\d \doti}{\lambda_1 \,\sqrt g } \left( p^{ik} {p_k}^j -  {\ts \frac{  \lambda_2 }{3 \lambda_2 -\lambda_1}}  p \, p^{ij} \right) \,.
\end{aligned}\end{equation}
The same equations are generated by the total `differential-almost-Hamiltonian' object
\begin{equation}\label{DifferentialAlmostHamiltonianEvolutionEquations}
\d f = \{ \d \almostHamiltonian , f \}\,, \qquad \d \almostHamiltonian = ( \d \doti | \mathcal H ) \,.
\end{equation} 
In order for the theory to admit solutions, the Hamiltonian constraint must be first-class with respect to
itself, which, by virtue of (\ref{DifferentialAlmostHamiltonianEvolutionEquations}), implies $\d \mathcal H \approx 0$, i.e., that it is
propagated by the evolution. This is generated by the Poisson bracket
\begin{equation}
\{ (\d \chi | \mathcal H) , (\d \sigma | \mathcal H) \}
\approx  \frac a {\lambda_1} \,(  \d \chi \nabla^i \d \sigma - \d \sigma \nabla^i \d \chi \, |  \, -2 \nabla_j {p^j}_i + 2 {\ts\frac{\lambda_2 - \lambda_1}{ 3 \lambda_2 -\lambda_1}}
\, \nabla_i p )  \,.
\end{equation}
We have a first set of choices here: 
\begin{itemize}
\item[\bf I.] We can close the constraint algebra strongly by taking $a=0$.
\item[\bf II.] We can close the constraint algebra strongly by taking $\lambda_1\to \infty$ (with $\lambda_2$ fixed).
\item[\bf III.] We can introduce a new, secondary constraint
\begin{equation}
\mathcal Z_i = - 2 \, \nabla_j {p^j}_i  +2  \, \alpha \nabla_i p \,, ~~ \alpha ={\ts \frac{\lambda_2-\lambda_1}{3 \lambda_2-\lambda_1}}\,.
\end{equation}
\end{itemize}
The first two possibilities correspond respectively to the removal of $R$ (case {\bf I}) and of the
kinetic term (case {\bf II}) from the Hamiltonian constraint. 

{\bf  Case I} is known as \emph{strong gravity}, \index{strong gravity} \index{Carrollian Relativity} a theory in which the light cones collapse to lines: no signal can
be transmitted from one point to another, and each spatial point becomes causally disconnected from
the others. This was named `Carrollian relativity'\footnote{After Lewis Carroll, whose Red Queen character in \emph{Alice in Wonderland} says: ``Now, here, you see, it takes all the running you can do, to keep in the same place''.} by Levy-Leblond \cite{Levy-Leblond-Bacry,LevyLeblond}. Belinsky, Khalatnikov and Lifshitz conjectured that near a cosmological singularity the contribution of matter to gravity becomes negligible compared with the self-coupling of gravity, and the variation of the gravitational field from one point to another can be neglected \cite{HenneauxLRR}. The strong gravity regime would then describe physics near a singularity
according to this BKL conjecture. 

{\bf  Case II} corresponds to a non-dynamical metric which is constant in time. The addition of matter
gives rise to a dynamics in which signals are transmitted instantaneously across a fixed foliation.
This is just Galilean relativity, \index{Galilean Relativity} which is the limit in which the speed of light goes to infinity.

{\bf  Case III} is more complicated, and divides into several sub-cases. First of all, on introducing a 
secondary constraint we have to check whether it closes a first-class system with itself and with the
primary constraints.

The self-commutator of $\mathcal Z_i$ is
\begin{equation}
\{ ( \d \xi^i | \mathcal Z _i) , ( \d \chi^j | \mathcal Z_j) \} =   ( [\d \xi,\d \chi]^i | ~ \mathcal Z_i + (2 \, \alpha - 6 \, \alpha^2 ) \nabla^i p )  \,,
\end{equation}
where $[\d \xi,\d \chi]^a$ is the Lie bracket between the two vector fields $\d \xi$ and $\d \chi$.
So the  constraint closes on itself only if $6 \, \alpha^2 - 2 \alpha = 0$. This equation
admits two solutions: $\alpha = 0$ (corresponding  to the DeWitt value $\lambda_2 = - \lambda_1$) and $\alpha = \frac 1 3$.
Both of the choices  that make the constraint $\mathcal Z_i$ first-class w.r.t. itself have a clear geometrical meaning. The choice $\alpha = 1$ makes  $\mathcal Z_i \to \mathcal H_i =  - 2 \, \nabla_j p^{ij}$ into the generator of diffeomorphisms, while $\alpha =  \frac 1 3$ makes $\mathcal Z_i \to \mathcal S_i = -2 \left( \nabla_j p^{ij}  - \frac 1 3 \,\nabla_i p \right)$  into  the generator of
\emph{special diffeomorphisms}, \index{special diffeomorphisms} or unit-determinant diffeomorphisms, which are diffeomorphisms that leave
the local volume element $\sqrt g$ invariant.

%
%
%
%

Notice that special diffeomorphisms  (strongly) commute with conformal transformations (generated by $p^{ij}g_{ij}$)
\begin{equation}
\{ ( \d \xi^i | \mathcal S_i ) , ( \d \varphi | p ) \} = 0  \,,
\end{equation}
where $ \mathcal S_i $ corresponds to taking the traceless part of the momentum in the standard diffeomorphism constraint,
\begin{equation}
 ( \d \xi^i | \mathcal S_i )  = - 2 \int   \d\xi_i \nabla_j ( p^{ij}  - {\ts \frac 1 3} \, g^{ij} \, p ) \,,  
\end{equation}
or the Lie derivative of the unimodular metric $g^{-\frac 1 3} g_{ij}$,
\begin{equation}
 ( \d \xi^i | \mathcal S_i )  = \int  p^{ij}  \left( \Lie_{\d \xi} g_{ij} - {\ts \frac 2 3} \, g_{ij}\nabla_k \d \xi^k  \right) 
 = \int  p^{ij}  \, g^{\frac 1 3} \, \Lie_{\d \xi}( g^{-\frac 1 3} g_{ij}) \,.  
\end{equation}
Here $\Lie_{{\d\xi}}(g^{-\frac 1 3} g_{ij}) =  \nabla_i \d \xi_j + \nabla_j \d \xi_i - {\ts \frac 2 3} \, g_{ij}\nabla_k \d \xi^k $ is what York calls the \emph{conformal Killing form}
\cite{YorkTTdecomposition} associated with the vector $\d \xi$.

A related concept is that of \e{transverse} diffeomorphisms: \index{transverse diffeomorphisms} they are diffeomorphisms
generated by a transverse vector field $\nabla_i \xi^i = 0$.
It's easy to show that $ (\xi^i | \mathcal H_i )  =  (\xi^i | \mathcal S_i )- {\ts \frac 2 3}  ( \nabla_i \xi^i | p )$; therefore if a vector field is transverse the diffeomorphisms
it generates are special. The converse, however, is not true: there are special
diffeomorphisms that are generated by vector fields with a longitudinal component.

Making a coordinate change from $\lambda_1,\lambda_2$ to
$\alpha,\lambda_1$ the Hamiltonian constraint takes the form
\begin{equation}
\mathcal H = \sqrt g \, ( a \, R - 2 \, \Lambda ) -  {\ts\frac1 {\lambda_1 \sqrt g}} \left( p^{ij}p_{ij} -{\ts \frac 1 2} \left( 1 - \alpha \right) \, p^{2}\right)\,,
\end{equation}
Therefore the value $\alpha = 1/3$ corresponds to
\begin{equation}
\mathcal H = \sqrt g \, ( a \, R - 2 \, \Lambda ) -  {\ts\frac1 {\lambda_1 \sqrt g}} \left( p^{ij}p_{ij} -{\ts \frac 1 3} \, p^{2}\right)\,.
\end{equation}

\noindent{\bf Analysis of case III}

For any  value of $\alpha$ other than $0$ and $\frac 1 3$, we are forced to 
split $\mathcal Z_a$ into two secondary constraints: one is the diffeomorphism constraint $\mathcal H_i$, and the other should be a constraint implying $\nabla_i p = 0$.\footnote{Taking $\nabla_i p = 0$ itself as
a new constraint would be wrong, as it is highly reducible (meaning that its components
are not linearly independent) and we would be constraining too many (three) degrees of freedom. An equivalent scalar constraint is sufficient.} A moment's thought reveals
that the most generic possibility is to have $p ={\sfrac 3 2} \tau \, \sqrt g$, where $\tau$ is a spatial constant.
The constraint $p$ generates infinitesimal conformal transformations of the three-metric and its momenta:
\begin{equation}
\{ ( \varphi | p ),  g_{ij} \} =  \varphi \, g_{ji} \,, \qquad \{ ( \varphi | p ),  p^{ij} \} =  - \varphi \, p^{ij} \,.
\end{equation}
The addition of the constant term $\tau$ changes the generated transformations into 
\emph{volume-preserving conformal transformations} (VPCT's). \index{volume-preserving conformal transformations} In fact $\tau$ cannot be 
just any number: by consistency (if the base manifold $\Sigma$ is compact, as I am assuming in this Section) it is forced to be equal to (2 thirds of) the average of $p$ over the whole space:
\begin{equation}
\tau = {\sfrac 2 3}  \frac{\int \d^3 x \, p}{\int \d^3 x \sqrt g } \equiv {\sfrac 2 3}  \langle p \rangle \,,
\end{equation}
where the spatial average $\langle \, \cdot  \,\rangle$ is defined as
\begin{equation}
\langle \rho \rangle : = \frac{\int \d^3 x  \, \rho }{\int \d^3 x \sqrt g }  \,
\end{equation}
for a scalar density $\rho$. Rewriting the constraint as
\begin{equation}
\mathcal C = p - \langle p \rangle \, \sqrt g\,,
\end{equation}
we recognize its volume-preserving nature:
\begin{equation}\begin{aligned}
\{ g_{ab} , ( \varphi | \mathcal C ) \} &=  ( \varphi - \langle \varphi \rangle ) \, g_{ij} \,, \\    \{  p^{ij} , ( \varphi | \mathcal C ) \}  &=  -  ( \varphi - \langle \varphi \rangle ) ( p^{ij} +  {\ts\frac 1 2}  \langle p \rangle  \,\sqrt g  \, g^{ij})  \,.
\end{aligned}\end{equation}
The diffeomorphism and conformal constraints close as a first-class system among themselves:
\begin{equation}\begin{aligned}
\{ (\d \xi^i | \mathcal H_i) , (\d \chi^j | \mathcal H_j) \}  &= ([\d \xi,\d\chi]^i | \mathcal H_i) \,, \\  
\{ (\d \xi^i | \mathcal H_i) , (\d \varphi | \mathcal C) \}  &= (\pounds_{\d \xi} \d \varphi | \mathcal C) 
 \,, \\  
\{ (\d \varphi |  \mathcal C) , (\d \rho | \mathcal C ) \}  &= 0 \,.
\end{aligned}\end{equation}


Therefore case {\bf III} divides into three sub-cases:
\begin{itemize}
\item[\bf III.a] $\boxed{\alpha=0}$ Our secondary constraint is first-class w.r.t. itself and generates regular diffeomorphisms.
The diffeomorphism constraint $\mathcal H^i$ is propagated by $\mathcal H$:
\begin{equation}\begin{aligned}
\{ ( \d \doti| \mathcal H), ( \d \xi^i | \mathcal H_i) \}  = (\d  \doti| \pounds_{\d \xi} \mathcal H ) \,,
\end{aligned}\end{equation}
and therefore we end up with a first-class system, which is just ordinary GR in the ADM formulation.

At this point we discovered our symmetries, so we can encode them back into the action through best-matching, as we
did at the beginning of Sec.~\ref{RWRsection}:
\begin{equation}
\begin{aligned}
\d \mathcal L_\st{diff} =&  \int \d^3 x \sqrt{g} \sqrt{a \, R - 2 \, \Lambda} \sqrt{ \left(  g^{ik}g^{jl} -  \, g^{ij}g^{kl} \right)   \dbm  g_{ij} \, \dbm  g_{kl} }  \,,
\\
\dbm g_{ij} = & \d  g_{ij} + \Lie_{\d \xi} g_{ij}  =\d  g_{ij} + \nabla_i \d \xi_j + \nabla_j \d \xi_i \,,
\end{aligned}
\end{equation}
and the diffeomorphism constraint becomes primary, a consequence of the free-endpoint variation w.r.t. the best-matching field $\d \xi^i$.

\item[\bf III.b] $\boxed{\alpha = 1/3}$ Our secondary constraint is first-class  w.r.t. itself  and generates special diffeomorphisms. The propagation of the special diffeomorphism constraint $\mathcal S_i$ gives
\begin{equation}\begin{aligned}\label{PropagationOfS}
 \{ ( \d  \doti| \mathcal H), ( \d \xi^i | \mathcal S_i )  \} 
\approx  
- {\ts\frac 4 3} a \, ( \sqrt g ( \laplacian  - R - 3 \Lambda/a) \d \doti | \nabla^i\d \xi_i ) \,.
\end{aligned}\end{equation}
This falls into case 3 of Dirac's analysis (see Sec.~\ref{CrashCourseInDiracsAnalysis}): it is a `specifier' equation for the smearing  $\d  \doti$ of the Hamiltonian constraint. The system is second-class, and instead of each constraining two degrees of freedom per space point (giving a total of four physical dofs), the constraints  $\mathcal S_i$ and $\mathcal H$ gauge-fix each other and constrain the dofs only down to six per point in total. The specifier equation is
\begin{equation} \label{SpecifierVPdiffeos}
\nabla_i \left[   a ( \laplacian  - R- 3 \Lambda /a ) \d \doti \right] = 0\,,
~~ \Rightarrow ~~
 a \, ( \laplacian  - R- 3 \Lambda /a ) \d \doti = \text{\it const.} \,.
\end{equation}
The above equation is of the form  $\laplacian f(x) + g(x) \, f(x) + \text{\it const.}= 0$: it is an elliptic equation of the kind
we have encountered in the preceding Section, like the Lichnerowicz--York equation (\ref{LichnerowiczYorkEquation}). It admits a unique solution for each positive root of the equation $g(x) \, f(x) + \text{\it const.} = 0$ (considered as an equation for $f(x)$ at each point), and there is only a one-parameter family of zeroes: in fact the constant represents an integration
constant. One can write the equation as $( \laplacian  - R- 3 \Lambda /a ) \d \doti  = \langle ( \laplacian  - R- 3 \Lambda /a ) \d \doti \rangle$ or, since $\langle \laplacian \d \doti \rangle =0$, 
\begin{equation}
\laplacian \d \doti -( R + 3 \Lambda /a ) \d \doti  + \langle (R + 3 \Lambda /a ) \d \doti \rangle = 0 \,. \label{DOTIfixingEQspecialDiffeoTheory}
\end{equation}
The last equation is homogeneous in $\d \doti$ and therefore the overall normalization of $\d \doti$ is irrelevant. This last one-parameter ambiguity is related to the fact that we can reparametrize the solution
$\d \chi(x)$ with any time-dependent, spatially-constant function. It is the leftover reparametrization invariance
of the solutions one obtains once $\mathcal H$ has been gauge-fixed. If $\d \doti_\st{sol}$ is the solution of the 
equation for $\d \chi$, then the generator of this reparametrization symmetry is
\begin{equation}
\d \mathcal A_\st{global} = ( \d \doti_\st{sol} | \mathcal H ) \approx 0 \,,
\end{equation}
which is the last leftover of the Hamiltonian constraint that hasn't been gauge-fixed by $\mathcal S_a$.

We ended up with a theory with six Hamiltonian degrees of freedom per point, those constrained by $\mathcal S_a \approx 0$, and a single, global Hamiltonian constraint $\d \mathcal A_\st{global}$, which can only be calculated
by solving a differential equation (\ref{DOTIfixingEQspecialDiffeoTheory}). This equation is invariant under the transformations generated by $\mathcal S_a$,
\begin{equation}
\begin{aligned}
g_{ij} &\to  g_{ij} +  \Lie_{\zeta} g_{ij} - {\ts \frac 2 3} \, g_{ij}\nabla_k \zeta^k \\ 
p^{ij}  & \to  p^{ij} +  \Lie_{\zeta} p^{ij} - {\sfrac 2 3} p^{ij} \nabla_k \zeta^k - {\ts \frac 1 3} \, p \, \Lie_{\zeta} g^{ij} \,,
\end{aligned}
\end{equation}
(the proof of this is left as an exercise), and therefore the global Hamiltonian constraint is invariant as well.

Now, if we implement the special diffeomorphisms in the action through best-matching:
\begin{equation}
\begin{aligned}
\d \mathcal L_\st{Sdiff} =&  \int \d^3 x \sqrt{g} \sqrt{a \, R - 2 \, \Lambda} \sqrt{ \left(  g^{ik}g^{jl} - {\sfrac 1 3} g^{ij}g^{kl} \right)   \dbm  g_{ij} \, \dbm  g_{kl} }  \,,
\\
\dbm g_{ij} = & \d  g_{ij} +  \Lie_{\d \zeta} g_{ij} - {\ts \frac 2 3} \, g_{ij}\nabla_k \d \zeta^k \,,
\end{aligned}
\end{equation}
the above action can be rewritten as
\begin{equation}
\begin{aligned}
\d \mathcal L_\st{Sdiff} =& \int \d^3 x \sqrt{g} \sqrt{a \, R - 2 \, \Lambda} \sqrt{ g^{ik}g^{jl}  \dbm  g_{ij} \, \dbm  g_{kl} } \,,
\\
\dbm g_{ij} = & \d  g_{ij} - {\sfrac 1 3} g_{ij} g^{kl} \d g_{kl}+  \Lie_{\d \zeta} g_{ij} - {\ts \frac 2 3} \, g_{ij}\nabla_k \d \zeta^k \,,
\end{aligned}
\end{equation}
and the relation between momenta and velocities, $ p^{ij} =  \frac{\sqrt g }{2 \,\d \doti} (g^{ik}g^{jl} - {\sfrac 1 3} g^{ij}g^{kl}) (\d  g_{ij} +  \Lie_{\d \zeta} g_{ij})$, is not invertible. Therefore the momenta satisfy an additional primary constraint: $g_{ij} p^{ij} = 0$. We are forced anyway to include conformal transformations: this case therefore reduces to a sub-case of {\bf III.c}, which I now discuss.

\item[\bf III.c] $\boxed{\text{generic}~ \alpha}$ We have to introduce \emph{two} separate secondary constraints,
the diffeomorphism constraint $\mathcal H_i$ and the volume-preserving conformal constraint $\mathcal C = p - \langle p \rangle \, \sqrt g$. These constraints are first-class among themselves.
As in the last case, the propagation of $\mathcal C$,
\begin{equation}\begin{aligned}\label{PropagationOfC}
 \{ ( \d  \doti| \mathcal H), ( \d \sigma | \mathcal C)  \} \approx  2\, a ( & \sqrt g  (\laplacian  - R   - 3 \Lambda /a ) \d  \doti \\   &  - {\sfrac {1 - 3 \, \alpha } {2\lambda_1}}  \langle p \rangle \, p \, \d \doti  - \frac 3 2 \, \d \doti \mathcal H
\,   | \, \d \sigma - \langle \d \sigma \rangle  )\,. 
\end{aligned}\end{equation}
gives a `specifier' equation for  $\d \doti$,
\begin{equation}\begin{aligned}
2\, a \, (\laplacian  - R   - 3 \Lambda /a ) \d  \doti - {\sfrac {1 - 3 \, \alpha } {2\lambda_1}} \, \langle p \rangle^2 \, \d \doti= \text{\it const.} \,. \label{LFE}
\end{aligned}\end{equation}
This again is an elliptic equation and admits a unique solution.
This solution $\d \chi_\st{sol}$, if used to smear the Hamiltonian  constraint, $(\d \chi_\st{sol} | \mathcal H)$, gives the part of $\mathcal H$ that is first-class with respect to the corresponding conformal constraint. This first-class part is a single residual global Hamiltonian constraint
\begin{equation}
\d \mathcal A_\st{global} = ( \d \doti_\st{sol} | \mathcal H ) \approx 0 \,.
\end{equation}
This $\d \mathcal A_\st{global}$ is invariant under VPCT's (and diffeomorphisms), and therefore generates the dynamics in the reduced configuration space.

The theory we obtained satisfies the strong Mach--Poincar\'e Principle in the quotient of $\Superspace$
by VPCT's. The quotient of $\Superspace$ by ordinary conformal transformations is \e{conformal superspace}, \index{conformal superspace} \index{superspace} which we also
call \e{Shape Space} \index{Shape Space} $\shs$ in analogy to the particle models:
\begin{equation}
\shs := \Superspace /\Sim \,, \qquad  g_{ij}  \sim g'_{ij} ~~ \text{\emph iff} ~~ \exists ~~ \phi ~~~ \text{\it s.t.} ~~ g_{ij}  = e^{4\phi} g'_{ij} \,.  
\end{equation}
The quotient of $\Superspace$ by VPCT is just\footnote{Locally - one has to exclude degenerate metrics and have special care in the case of metrics with conformal isometries} the Cartesian product between
$\shs$ and the real positive line $\doublestroke{R}^+$, representing the dof associated with the total volume of space, $V = \int \d^3 x \sqrt g$.

These results do not depend on the value of $\alpha$, and therefore they hold also if $\alpha =0$ or $1/3$. The sub-cases {\bf III.a} and  {\bf III.b} are included in this one. The theory we are dealing with is (generalized) Shape Dynamics. I call it generalized because proper SD refers just to the case $\alpha=0$. It is not clear, at this point, whether the value of $\alpha$ has physical meaning. This is because in the shape-dynamical description its value can be reabsorbed into a rescaling of York time. Should it turn out that $\alpha$ has physical meaning, then it would have to be considered as a dimensionless coupling for Shape Dynamics. The chances then are that this coupling would run under renormalization group flow, and $\alpha =0$ would presumably be the IR value, where equivalence with General Relativity (and with it spacetime covariance) emerges as a sort of accidental IR symmetry. The $\alpha = 1/3$ value might then be the UV limit.
\end{itemize}

This table summarizes the different cases we have encountered:

\begin{center}
\begin{tabular}{lcccr}
\hline\hline
{\bf Case	}	& {\bf Parameter values	} 	& {\bf DOF's}	&	{\bf Theory}		&
\\
\hline
{\bf I}		& $a= 0$						&	4-12 p.p.				& Carroll Relativity 	&
\\
{\bf II}		&  $y=0$						&	0					& Galilean Relativity	&
\\
{\bf III.a}		& $\alpha = 0$				&	4 p.p.				& General Relativity	&
\\
{\bf III.b}		& $\alpha = {\ts\frac 1 3} $	&	4 p.p.+ 2 global 					&	?				&
\\
{\bf III.c}		& 	any						&	4 p.p. + 2 global		&	Shape Dynamics	&
\\
\hline\hline
\end{tabular}
\end{center}

I have reached the point at which I can finally introduce Shape Dynamics. As Henneaux and Teitelboim state
in their book~\cite{HTbook}, a second-class system like that of case {\bf III.c} can be seen as a gauge-fixing
of a first-class system. Doing this often requires enlargement of the phase space with further redundant (constrained)
degrees of freedom, and this is the case also in Shape Dynamics, where it is necessary to introduce a scalar field $\phi$ and its conjugate momentum $\pi$. Then one modifies the Hamiltonian, diffeomorphism and VPCT constraints:
\begin{equation*}
\begin{aligned}
\mathcal H  =& \frac {e^{-6 \hat \phi }} {\sqrt g} \left(p^{ab}-{\sfrac 1 3} p g^{ij}\right) \left(p_{ij}-{\sfrac 1 3} p g_{ij}\right)- (1-\alpha)\frac{e^{-6 \hat \phi}}{6 \sqrt g} \left(p-\sqrt{g} \left(1-e^{-6 \hat \phi }\right) \langle p\rangle \right)^2\\
&- \sqrt g \, R \,e^{2 \, \phi} - \sqrt g \, e^{\phi} \laplacian e^\phi  \approx 0 \,,\\
 \mathcal H^i  &= -2\, \nabla _j p^{ij} +  \pi \, \nabla ^i \phi  \approx 0\,,  \qquad    \mathcal Q = \pi -  4 \left(  p
-\langle p \rangle \, \sqrt g\right)  \approx 0 \,,
\end{aligned}
\end{equation*}
The above system is first-class. This theory is called the \emph{Linking Theory}, and it leads to SD as the following gauge-fixing: $\pi \approx 0$. That generates the set of constraints we found in case {\bf III.a} (for $\alpha = 0$).
A different gauge-fixing, namely $\phi \approx 0$, gives instead a first-class system (General Relativity) by killing just the modified VPCT constraint $\mathcal Q$. I'll define from scratch, and more carefully, the Linking Theory in Sec.~\ref{SDLT} after some historical background.

\vspace{12pt}
\noindent
\fbox{\parbox{0.98\linewidth}{
{\bf Further reading:} my paper with E.~Anderson \cite{EdFlavioPaper}.
}}

\newpage

\section{Cotton-squared theory}
\label{CottonSquaredTheorySec}

The \emph{Cotton tensor,} defined as \index{Cotton tensor}
\begin{equation}\label{CottonTensor}
C_{ijk} = \nabla_k R_{ij} - \nabla_j R_{ik} + {\sfrac 1 4} \left( \nabla_j R \, g_{ik} -\nabla_k R g_{ij}\right)  \,,
\end{equation}
is an object that characterizes the conformal geometry of the metric. It satisfies the identity $C_{[ijk]}=0$ and it is antisymmetric in the last two indices $j$ and $k$, and therefore can be considered a vector-valued 2-form. Moreover, all the contractions of two of its 3 indices with the metric are zero. 

It is convenient at this point to introduce  the covariant Levi-Civita density $\epsilon^{ijk}$, which is a tensor density of weight $+1$ of components \index{Levi-Civita tensor density}
\begin{equation}
\epsilon^{ijk} = \sqrt{g} \,  \delta^{[i}{}_1\,  \delta^j{}_1\,  \delta^{k]}{}_3 \,.
\end{equation}
The Levi-Civita density allows to introduce the `Hodge-$*$', \index{Hodge-$*$} a linear map from covariant antisymmetric $n$-tensors to contravariant $3-n$-tensor densities which does not erase any information about the components of the tensor (it is a isomorphism, or a pairing). This is because antisymmetric $n$-tensors have $\binom{3}{n}$ independent components, and antisymmetric $3-n$-tensors have the same number of independent components because $\binom{3}{3-n} = \binom{3}{n}$. Antisymmetric covariant $2$-tensors, \emph{e.g}. $A_{ij}=-A_{ji}$ are mapped to covariant vector densities: $V^i = {\sfrac 1 2} \epsilon^{ijk} A_{jk}$. It is easy to see that  $A_{ij}$ has 3 independent components just like $V^i$, so one loses nothing in the transformation. Antisymmetric covariant 3-tensors, \emph{e.g} $B_{ijk}$ only have 1 independent component (when all three indices are different), so they are mapped to scalar densities, like $S = {\sfrac 1 {3!}}\epsilon^{ijk} B_{ijk}$.
$C_{ijk}$ is antisymmetric only in the last two indices, so we may consider it a `vector-valued one-form', that is, a collection of three antisymmetric 2-tensors. We can then apply the Hodge-$*$ map to its last two component, 
\begin{equation}
C^i{}_j = {\frac 1 2}\epsilon^{i\ell m}C_{\ell m j} = \nabla_k \left( R_{j\ell} - {\sfrac 1 4} R \, g_{j \ell} \right) \epsilon^{k\ell i} \,,
\end{equation}
obtaining the \emph{Cotton--York density}, \index{Cotton--York density} first defined by York \index{James W. York, jr.} in~\cite{York:cotton_tensor}. As is easy to verify, this tensor density is symmetric in the sense that $C_{ij} = C_{ji}$, and it is trace-free $C^i{}_i =0$, a statement that coincides with the identity  $C_{[ijk]}=0$. Moreover, as one can verify explicitly, the Cotton--York density is \emph{transverse} by construction: $\nabla_i C^i{}_j =0$. In summary, $C_{ij}$ is a \emph{symmetric transverse-traceless} tensor density, which makes it a natural pure-spin-2, conformally invariant object.

The Cotton--York density is invariant under conformal transformations of the metric (the proof is left as an exercise):
\begin{equation}
C^i{}_j [ e^{4 \omega} g_{\ell m};x) = C^i{}_j  [g_{\ell m};x) \,.
\end{equation}
One can prove (another exercise for the reader, a full proof can be found in~\cite{EisenhartBook}) that \emph{local} conformal flatness and the vanishing of the Cotton--York density (or, equivalently, the Cotton tensor) are equivalent:
\begin{equation}
g_{ij} = e^{4 \, \omega(x)} \eta_{ij}  ~~ \Longleftrightarrow ~~ C^i{}_j  [g_{\ell m};x) = 0 \,,
\end{equation}
where $\eta_{ij}$ is the flat metric. \emph{Local} conformal flatness means that the statement above holds only in open neighbourhoods. With the Cotton--York density we can introduce an $L^1$ norm $\int_\Sigma \sqrt{C^i{}_j C^i{}_j} \d^3 x$, which is conformally-invariant and vanishes only if $C^i{}_j =0$ everywhere. So this norm might be considered as a distance of the conformal geometry from conformal flatness.

There is clearly an opportunity, here, to introduce a Weyl-invariant Jacobi-like action by using the Cotton tensor in the potential: 
\begin{equation}
\begin{aligned}
\d \mathcal L_\st{Weyl} =&  \int \d^3 x \sqrt{C^i{}_j  C^j{}_i } \sqrt{  g^{ik}g^{jl}  \dbm  g_{ij} \, \dbm  g_{kl} } 
\\
=&  \int \d^3 x \sqrt{\det g} \sqrt{C_{ijk} C^{ijk}} \sqrt{  g^{ik}g^{jl}  \dbm  g_{ij} \, \dbm  g_{kl} }  \,,
 \end{aligned}
\end{equation}
where
\begin{equation}
\dbm g_{ij} =  \d  g_{ij} + \Lie_{\d \xi} g_{ij}  + g_{ij}\,\d \rho  = \d  g_{ij} + \nabla_i \d \xi_j + \nabla_j \d \xi_i + g_{ij} \, \d \rho \,.
\end{equation}
Notice that there are no free parameters in this action: there is only one way of contracting the best-matching derivative $\dbm g_{ij}$ with itself, because its trace $g^{ij} \dbm g_{ij}$ does not contribute to the physical change and can be reabsorbed into the Weyl-compensating field $\rho$. The local `differential of the instant'  and the canonical momenta  in this case are:
\begin{equation}
\d \doti = \frac 1 2 \sqrt{ \frac{   g^{ik} g^{jl}  \dbm  g_{ij} \, \dbm  g_{kl}}{ C^i{}_j  C^j{}_i }} \,,
\qquad
  p^{ij} = {\delta \d \mathcal L_\st{Weyl} \over \delta \d  g_{ij}} = \frac{ g^{ik} g^{jl}  \dbm  g_{kl} }{2 \,\d \doti}  \,.
\end{equation}
The best-matching conditions give the diffeomorphism and Weyl constraints
\begin{equation}\label{CottonSquaredBestmatchingConstraints}
\begin{aligned}
\frac{\delta \d \mathcal L_\st{Weyl}}{\delta \d \xi_i} &= - 2 \, \nabla_j \left[\frac{g^{ik} g^{jl} \,  \dbm  g_{kl}}{2 \,\d \doti}  \right] =- 2 \, \nabla_j p^{ij} = \mathcal H^i \approx 0\,,
\\
\frac{\delta \d \mathcal L_\st{Weyl}}{\delta \d \rho} &= g_{ij} \left[\frac{g^{ik} g^{jl} \,  \dbm  g_{kl}}{2 \,\d \doti}  \right] = g_{ij} p^{ij} = \mathcal W\approx 0\,.
\end{aligned}
\end{equation}
Moreover,Êthe local square-root form of the action leads to the following primary Hamiltonian constraint:
\begin{equation}
\mathcal H = p^{ij} p_{ij} + C^i{}_j C^j{}_i  \approx 0 \,.
\end{equation}
In~\cite{York:cotton_tensor} York observed how the Cotton--York tensor $C^i{}_j$ and a momentum $p^i{}_j$ satisfying the constraints~(\ref{CottonSquaredBestmatchingConstraints}) together close a system of equations which are the `spin-2 analogue' of the scalar part of Maxwell's equations:
\begin{equation}
\begin{aligned}
&\nabla_i C^i{}_j = 0 \,, 
&&C^i{}_i = 0 \,,
&&\nabla_i p^i{}_j = 0 \,,
&&p^i{}_i = 0\,.
&
\end{aligned}
\end{equation}
The TT-momentum $p^i{}_j$ plays the role of the transverse electric field $\mathbf{\nabla} \cdot \mathbf{E} = 0$, while the Cotton--York tensor is analogue to the magnetic field $\mathbf{B}$, which is automatically transverse once one introduces the vector potential $\mathbf{A}$. For $C^i{}_j$ the metric $g_{ij}$ plays the role of potential, and Equation~(\ref{CottonTensor}) is analogue to $\mathbf{B} = \mathbf{\nabla} \times \mathbf{A}$, in that it makes $C^i{}_j$ automatically transverse-traceless. Just like $\mathbf{A}$ is not gauge-invariant, $g_{ij}$ is not diffeo- and Weyl-invariant. Just like $\mathbf{B}$ depends only on the transverse part of $\mathbf{A}$, $C^i{}_j$  depents only on the conformodiffeo-invariant part of $g_{ij}$. The potentials $\mathbf{A}$ and $g_{ij}$ introduce a representation of the respective fields which is redundant but local.

The analogy between Maxwell's equations and the equations of our Cotton-squared theory unfortunately ends here. The time-dependent Maxwell equations in vacuum relate the time derivatives of the electric and magnetic fields with the curl of each other. In the case of the Cotton-squared theory the Hamiltonian equations of motion are:
\begin{equation}
\dbm g_{ij} = 2 \d \chi \, p_{ij} \,, \qquad \dbm p^{ij}(x) = - 2 \int \d \chi(y) \, C^k{}_\ell(y)  \frac{\delta C^\ell{}_k(y)}{\delta g_{ij}(x)} \d^3y  \,,
\end{equation}
where the metric variation of the Cotton tensor is
\begin{equation} 
\delta C^i{}_j =   \epsilon^{i k \ell} \left[ \delta \left( \nabla_k R_{\ell j} \right) - {\sfrac 1 4} \left(\delta g_{\ell j} \nabla_k R + g_{\ell j} \nabla_k \delta R \right)\right]\,.
\end{equation}
This translates into the following  equations for $C^i{}_j$ and $p^i{}_j$:
\begin{equation}
\begin{aligned}
&\dbm C^i{}_j =2 \int \d \chi(y) \,  \frac{\delta C^\ell{}_k(y)}{\delta g_{ij}(x)}  p^k{}_\ell(y)\d^3y\,,
\\
&\dbm p^{ij} = - 2 \int \d \chi(y) \, C^k{}_\ell(y)  \frac{\delta C^\ell{}_k(y)}{\delta g_{ij}(x)} \d^3y \,.
\end{aligned}
\end{equation}
The above equations do look like Maxwell's vector equations with the curl replaced by the operator $\int \d \chi(y) \,  \frac{\delta C^\ell{}_k(y)}{\delta g_{ij}(x)} $. However this operator doesn't look at all like a tensor generalization of the curl.\footnote{There is a natural generalization of the curl for TT tensors: it is the linear differential operator $\mathcal O^{ij}  =   \sqrt{g} \epsilon^{kij} \nabla_k$. A nice analogue of Maxwell's vector equations could then be
$$
\d C^j{}_k =\sqrt{g} \epsilon^{ij\ell} \nabla_i p_{\ell k}
\,,
~~~~
\d p^j{}_k =\sqrt{g} \epsilon^{ij \ell} \nabla_i C_{\ell k} \,.
$$
These equations preserve the tracelessness of the tensors $p^i{}_j$ and $C^i{}_j$, because they are traceless themselves:
$$
\delta^k{}_j \d C^j{}_k =\sqrt{g} \epsilon^{ik\ell} \nabla_i p_{\ell k} =0
\,,
~~~~
\delta^k{}_j \d p^j{}_k =\sqrt{g} \epsilon^{ik\ell} \nabla_i C_{\ell k} =0
$$
However the operator $\mathcal O^{ij} $ does not preserve the transversality property, unlike Maxwell's vector equations which maintain this property because the divergence of the curl is zero:
$$
\nabla_j \d C^j{}_k =\sqrt{g} \epsilon^{ij\ell} \nabla_j \nabla_i p_{\ell k} \neq 0
\,,
~~~~
\nabla_j \d p^j{}_k =\sqrt{g} \epsilon^{ij \ell} \nabla_j \nabla_i C_{\ell k}\neq 0 \,.
$$
}

The Cotton-squared theory has worse problems than not being completely analogue to electromagnetism. If we study the constraint algebra, in fact, it turns out that it is not first-class, because the Hamiltonian constraint does not close with itself. The Poisson brackets of  $\mathcal H$  with itself give something of the form
\begin{equation}
\begin{aligned}
&\left\{ ( \mathcal H | \d \chi), (\mathcal H | \d \varphi) \right\} = 
\\
& \qquad\qquad \int \d \chi \left( a^{ijk} \nabla_i \nabla_j \nabla_k \d \varphi + b^{ij} \nabla_i \nabla_j \d \varphi + c^i \nabla_i \d \varphi +f \d \varphi \right)  \d^3 x \,,
\end{aligned}
\end{equation}
where  $a^{ijk}$, $b^{ij}$, $c^i$ and $f$ are complicated tensors (or scalars) depending on curvature invariants and (linearly) on the momentum $p^{ij}$. In the paper~\cite{CottonSquaredInconsistent}  this constraint algebra was studied, and it was found that the above Poisson bracket is zero if the Cotton--York tensor is zero. 
Adding $C^i{}_j = 0$ as secondary constraint would completely trivialize the dynamics. Moreover the operator acting on $\d \chi$ in the Poisson brackets is not invertible, therefore the above equation cannot  be interpreted as a `specifier' equation fixing the value of $\d \chi$. Therefore the authors of~\cite{CottonSquaredInconsistent} conclude that this system is inconsistent.\footnote{The system would not have trivial dynamics, however, if the Lagrangian was defined with a global square root, \emph{i.e.} $ \sqrt{\int \d^3 x \sqrt{C^i{}_j  C^j{}_i }  g^{ik}g^{jl}  \dbm  g_{ij} \, \dbm  g_{kl}  }$. Such a system has TT gravitational dofs but a single global Hamiltonian. It has been studied in detail in~\cite{A.-Gomes:2016fq}.}

\newpage

%
%
%
%
%
%
%
%

%
%
%
%
%

\newpage

\part{Shape Dynamics}\label{SDpart}

\section{Historical Interlude} \label{HistoricalInterlude}

In Sec.~\ref{SDLT}, I will present Shape Dynamics as the first-class extension in the manner of Henneaux and Teitelboim~\cite{HTbook}
of the theory we studied in case {\bf III.c} of last Section, when $\alpha = 0$. However, that's not how SD was originally discovered, which I now briefly recount before moving forward.

The stimulus came from Barbour's desire to create a scale-invariant Machian theory, first of particle dynamics and then dynamical geometry. The first step was the derivation in 1999 of a particle-dynamics model (published in 2003 \cite{Anderson:2002ey}) with dilatational best matching and the Newton gravitational potential $V_\textrm{New}$ (of degree $-1$) replaced by the potential $I_\textrm{cm}^{-1/2}V_\textrm{New}$, where $I_\textrm{cm}$ is the centre-of-mass moment of inertia. The dilatational constraint $D=\sum_a m_a{\bf r}_a\cdot{\bf r}_a$ resulting from the dilatational best matching commutes with this potential, and the resulting theory defines geodesics on shape space. The strong Mach--Poincar\'e Principle is therefore satisfied.}

Barbour and \'O Murchadha then tried to extend the underlying ideas of this particle model to dynamical geometry in~\cite{Barbour_Niall:first_cspv} in 1999. This is the first attempt  at a theory of gravity invariant under three-dimensional conformal transformations. This was expected to define a geodesic theory on conformal superspace that satisfies the strong Mach--Poincar\'e Principle and eliminates a perceived Machian defect of GR highlighted by York's work on its initial-value problem. York had shown that the initial-value problem \index{initial-value problem} could be solved by taking the following initial data:  a conformal equivalence class of 3-geometries, its variation,
and a single real quantity (the value of York time \index{York time} or, as \'O Murchadha showed \cite{NiallVolumeYorkTime}, the spatial volume). The necessity for this additional degree of freedom is quite puzzling. One could completely specify a solution of GR with initial data on conformal superspace $\shs$ (two degrees of freedom per point) if it were not for this additional single global degree of freedom, which is not conformally invariant. The theory of~\cite{Barbour_Niall:first_cspv} was a proposal to eliminate this puzzle, by `conformalizing' the Baierlein--Sharp--Wheeler action \index{Baierlein--Sharp--Wheeler formulation} [or rather the action~(\ref{AndersonMercatiInitialAction}) with $a=1$]. This was done by rescaling the metric by a scalar field $\phi$, $g_{ij} \to \phi^4 g_{ij}$, so that the action is invariant under a simultaneous conformal transformation of the metric and the scalar field:
\begin{equation}
g_{ij} \to \omega^4 g_{ij} \,, \qquad \phi \to \frac{\phi}{\omega} \,.
\end{equation}
This is the `St\"uckelberg trick' \index{St\"uckelberg trick}, which normally does not change a theory because it introduces one field while simultaneously introducing a gauge redundancy which makes the field unphysical.
The time derivative of the scalar field is eliminated from the action by best-matching with respect to conformal transformations and reabsorbing $\dot \phi$ into the best-matching field:
\begin{equation}
\d (\phi^4 g_{ij}) +g_{ij } \d \theta = \phi^4 \d g_{ij}  + g_{ij}( 4 \d \phi+ \d \theta) \,, \qquad   \d \theta' = 4 \d \phi+ \d \theta\,.
\end{equation}
Then the field $\phi$ is a Lagrange multiplier and extremizing the action w.r.t. $\phi$ gives the analogue of the lapse fixing equation (see below). The  local square root form of the action gives rise to a primary constraint which is the conformalized version of the Hamiltonian constraint of GR, that is, the Lichnerowicz equation. Finally, conformal  and diffeomorphism best matching imply that the momentum is transverse and traceless.  To implement the strong form of the Mach--Poincar\'e principle, and  eliminate the global `additional' degree of freedom from the initial-value problem, \index{initial-value problem}
  \'O Murchadha and Barbour divided the Lagrangian by the conformalized volume $\int \phi^6 \sqrt g \d^3x$. This was justified because it ensures that $\phi=0$ does not minimize the action. The resulting theory is  modification of GR which involves a nonlocal coupling between the volume $V$ of spatial slices and the local gravitational degrees of freedom. The theory has a major shortcoming, however: it lacks the York time \index{York time} term in the Hamiltonian constraint, and therefore is only valid when Lichnerowicz' method can be applied, \emph{i.e.} when the metric belongs to the Yamabe-negative class. Moreover, the global modifications to the dynamics of GR that this theory imply are hard to reconcile with cosmological observations.

The work on this theory was interrupted by the successes of the Relativity Without Relativity approach but then explored further by Barbour, Foster, \'O Murchadha, Anderson and Kelleher. The main breakthrough, in this phase, was represented by the observation, due to  B. Foster,  that volume-preserving conformal transformations can be implemented  in the following way:
\begin{equation} 
g_{ij} \to e^{4 \hat \phi} g_{ij} \,, \qquad \hat \phi(x):=\phi(x)- {\sfrac 1 6} \log \, \langle \sqrt g \, e^{6\phi}\rangle \,,\label{brendan}
\end{equation} 
in which $\phi$ is subject to no restriction except $\phi>0$ and would implement an unrestricted conformal transformation were it not for the correction $- {\sfrac 1 6} \log \, \langle \sqrt g \, e^{6\phi}\rangle$, which restores the total volume to the value it had before the transformation implemented by the unrestricted first term. This device led to the papers~\cite{Anderson:2004wr,Anderson:2004bq}, which are attempts at creating a VPCT-invariant theory of gravity. These attempts were flawed, because the authors did without the trick, used in~\cite{Barbour_Niall:first_cspv}, of disposing of the time derivative of $\phi$ by best-matching with respect to conformal transformations. They rather tried to use $\phi$ itself as a sort of best-matching field, which means that the resulting action depends on both $\phi$ and its time derivative, and therefore extremizing it w.r.t. $\phi$ leads to the Euler--Lagrange equations, which involve second time derivatives of $\phi$ and make it dynamical. In~\cite{Anderson:2004wr,Anderson:2004bq} the authors affirm that by considering  the \emph{free-end-point variation} of the action w.r.t. the field $\phi$ this problem is resolved, because this particular form of variation implies that the momentum conjugate to $\phi$ is bound to vanish throughout the solution. In Appendix~\ref{AppendixFreeEndPointVariation}  I show the correct way of treating free-end-point variations: the momentum conjugate to $\phi$ is bound to vanish \emph{only at the boundary of the integration interval:} in between it takes the values that are imposed by the Euler--Lagrange equations.
 The authors of~\cite{Anderson:2004wr,Anderson:2004bq} conclude that the whole package of York's method, namely, the Lichnerowicz--York equation with York time, \index{York time} the lapse-fixing equation, and the CMC constraint can be derived by a single variational principle from a BSW-type action which involves a single scalar field $\phi$. This conclusion is erroneous due to their misunderstanding of the meaning of free-end-point variation. \index{free-end-point variation}

Nevertheless,~\cite{Anderson:2004wr,Anderson:2004bq} represent an important step in understanding the underlying 3D conformally-invariant dynamics of GR, and setting the stage for the development of Shape Dynamics. In particular~\cite{Anderson:2004wr,Anderson:2004bq} patched a worrisome flaw in York's method: When attributing a constant trace (times $\sqrt g$) to the momentum:
\begin{equation}
g_{ij}  p^{ij} = \sqrt g \, \tau \,,
\end{equation}
York did it in a particular conformal frame, which was not the one in which the metric and the momenta satisfy the Hamiltonian constraint. But when making connection to that frame, York assumed that the trace of the momentum transforms like $\sqrt g$ under conformal transformations, that is, he assumed the York time $\tau$ \index{York time} to be invariant! York did not give a satisfactory justification of this assumption in his work, while~\cite{Anderson:2004wr}  clarifies what is behind what the authors call `York scaling': first, York's `CMC' condition should be understood as the nonlocal expression:
\begin{equation}
g_{ij}  p^{ij} (x) = \sqrt{g(x)} \, \langle  p \rangle = \sqrt{g(x)} \frac{\int d^3y \,p(y)}{\int d^3 z \sqrt{g(z)}}\,;
\end{equation}
second, the conformal transformations to the frame in which the scalar constraint is satisfied should only be \emph{volume-preserving} and implemented as in (\ref{brendan}). Those transformations then have a nontrivial action on the canonical momenta:
\begin{equation}
p^{ij} \to e^{-4\, \hat  \phi} \left[ p^{ij} 
- {\ts \frac 1 3} (  1 -e^{6 \hat \phi}   )  \langle  g_{kl} p^{kl}\rangle \sqrt g  \, g^{ij} \right] \,.
\end{equation}
It was several years before the next step was taken. To this I now turn.

\newpage

\section{Shape Dynamics and the Linking Theory}\label{SDLT} 

\index{Linking Theory}
In \cite{Gomes:sd_paper,Gomes:linking_paper}, Gomes, Gryb and Koslowski formulated Shape Dynamics the
way we understand it now, and they sharpened the idea of intersecting constraint surfaces with a global flow in the intersection which I illustrated in Fig.~\ref{Surfaces}. In the formulation of~\cite{Gomes:sd_paper,Gomes:linking_paper} Shape Dynamics is founded on GR, and on the realization that in the same phase space of ADM two first-class systems of constraints coexist which are dual to each other in the sense that they represent good gauge-fixings of each other. In the original papers the theory was defined as a `dual' formulation of GR, and is logically dependent on it. In this \thistext I would like to found SD on a set of first principles that make it logically independent of GR. My proposal is to tie it to the discussion of Sec.~\ref{ExtendedRWR}, where it is shown that the CMC constraint of York's method emerges as a secondary constraint by performing the Dirac analysis on the most general Jacobi-like local square root Lagrangian for geometrodynamics. The so-called `Linking Theory', introduced in~\cite{Gomes:sd_paper,Gomes:linking_paper}, is a first-class extension of the second-class system of case {\bf III.c}, when $\alpha = 0$.

\index{The Linking Theory}

The most elegant way to introduce the Linking Theory is to keep explicit equivalence with GR at every stage.
So we start with the ADM system:
\begin{equation}\label{ADMconstraintsSD}
\begin{aligned}
& \mathcal H  =  \frac 1 {\sqrt g} \, \left( p^{ij} p_{ij} - {\sfrac 1 2} p^2 \right)  - \sqrt g \, R  \,, \\
& \mathcal H^i  = - 2 \, \nabla_j \, p^{ij}  \approx 0 \,.
\end{aligned}
\end{equation}
Then we trivially extend the phase space (the cotangent bundle to $\Riem(\Sigma)$) with a scalar field $\phi$ and its conjugate momentum $\pi$. We also add a further constraint:
\begin{equation}
\mathcal Q = \pi \approx 0 \,,
\end{equation} \index{James W. York, jr.}
which makes $\phi$ into a gauge degree of freedom. This constraint is trivially first-class with respect to the
other ones (\ref{ADMconstraintsSD}). So we have a first-class system which has the same number of degrees
of freedom as ADM gravity and is trivially equivalent to it: ADM gravity can be recovered with the gauge fixing $\phi\approx 0$.

\index{canonical transformation} \index{generating functional}
Now we perform a canonical transformation with a type-2 generating functional:
\begin{equation}\label{CanonicalTransformationSD}
\begin{aligned}
&{\rm F} 
= \int \d^3 x \left( g_{ij}  \, P^{ij} + \phi \, \Pi +  g_{ij}  ({\ts e^{4\, \hat  \phi}} -1) P^{ij} \right) \,,\\
&\hat \phi(x):=\phi(x)- {\sfrac 1 6} \log \, \langle \sqrt g \, e^{6\phi}\rangle \,.
\end{aligned}
\end{equation}
The transformation rules are
\begin{equation}
\begin{aligned}
&  \frac{\delta {\rm F}}{\delta g_{ij}} = e^{4\, \hat  \phi} \, P^{ij} 
+{\ts \frac 1 3} (  e^{6 \hat \phi}  -1  )  \langle e^{4\, \hat  \phi} \, g_{kl} P^{kl}\rangle_g \sqrt g  \, g^{ij}
\,, &
& \frac{\delta {\rm F}}{\delta \Pi^{ij}} = e^{4\, \hat  \phi} \, g_{ij} \,, \\
& \frac{\delta {\rm F}}{\delta \phi} =  \Pi +  4 \left(  e^{4\, \hat  \phi} \, g_{ij}\,P^{ij} 
-\langle e^{4\, \hat  \phi} \, g_{ij}\,P^{ij} \rangle_g \, \sqrt g \, e^{6\, \hat  \phi} \right)  \,, &
& \frac{ \delta {\rm F}}{\delta \Pi} = \phi \,,
\end{aligned}
\end{equation}
which translate into $(g_{ij},\phi;p^{ij},\pi)\to (G_{ij},\Phi;P^{ij},\Pi) $, where
\begin{equation}\label{LinkingTheoryCanonicalTransform}
\begin{aligned}
&  P^{ij}= e^{-4\, \hat  \phi} \left[ p^{ij} 
- {\ts \frac 1 3} (  1 -e^{6 \hat \phi}   )  \langle  p \rangle \sqrt g  \, g^{ij} \right]
\,, &
&G_{ij} = e^{4\, \hat  \phi} \, g_{ij} \,, \\
& \Pi =  \pi -  4 \left(  p
-\langle p \rangle \, \sqrt g\right)  \,, &
&  \Phi = \phi \,.
\end{aligned}
\end{equation}
Now express the Hamiltonian, diffeomorphism and $\mathcal Q$ constraints in terms of the transformed variables
[this calculation is left as an exercise -- don't forget that the covariant derivative of a tensor density of weight $w=1$ is
$\nabla_k p^{ij} = \partial_k p^{ij} + \Gamma^i_{kl} p^{lj} + \Gamma^j_{kl} p^{il} -  \Gamma^l_{kl} p^{ij}$
and that the Christoffel symbols transform under conformal transformations $g_{ij} \to e^{4 \phi} g_{ij}$ as
$\Gamma^i_{jk} \to \Gamma^i_{jk} + 2 ( \delta^i{}_j \partial_k \phi + \delta^i{}_k \partial_j \phi - g_{jk} g^{il} \partial_l \phi)$], and obtain
\begin{equation}
\begin{aligned}
\mathcal H_{\hat \phi}  =& \frac{e^{-6 \hat \phi }}{\sqrt{g}}  \left(p^{ij} p_{ij}+\frac{1}{3} \sqrt{g}  \left(1-e^{6 \hat \phi }\right) \langle p\rangle \, p - \frac{1}{6} g \left(1-e^{6 \hat \phi }\right)^2 \langle p\rangle ^2-\frac{p^2}{2}\right) \\
&- \sqrt g \left( R \,e^{2 \, \hat \phi} - 8 \, e^{\hat \phi} \laplacian e^{\hat \phi} \right)   \approx 0 \,,\\
\mathcal H^i_{\hat \phi}  =& - 2\, e^{-4 \hat{\phi }}\left[\nabla _j p^{ij} - 2 \left(p-\sqrt{g} \, \langle p\rangle  \right) \nabla^i \phi  \right] \approx 0\,,  \\
  \mathcal Q_{\hat \phi} =& \pi -  4 \left(  p
-\langle p \rangle \, \sqrt g\right)  \approx 0 \,,
\end{aligned}
\end{equation}
which are equivalent to (another exercise for the reader)
\begin{equation}\label{FinalLinkingTheoryConstraints}
\begin{aligned}
\mathcal H_{\hat \phi}  =& \frac {e^{-6 \hat \phi }} {\sqrt g} \left(p^{ij}-{\sfrac 1 3} p g^{ij}\right) \left(p_{ij}-{\sfrac 1 3} p g_{ij}\right)-\frac{e^{-6 \hat \phi}}{6 \sqrt g} \left(\frac{\pi}{4}+ e^{6 \hat \phi } \sqrt{g} \, \langle p\rangle \right)^2\\
&- \sqrt g \left( R \,e^{2 \, \hat \phi} - 8 \, e^{\hat  \phi} \laplacian e^{\hat \phi} \right)  \approx 0 \,,\\
 \mathcal H^i_{\hat \phi}  &= -2\, \nabla _j p^{ij} +  \pi \, \nabla^i \phi  \approx 0\,,  \qquad�   \mathcal Q_{\hat \phi} = \pi -  4 \left(  p
-\langle p \rangle \, \sqrt g\right)  \approx 0 \,,
\end{aligned}
\end{equation}
where we recognize the Lie derivative of the scalar field in the smeared
version of the diffeomorphism constraint $(\mathcal H^i_{\hat \phi} | \xi_i) = \int \d^3 x \left( \pi^{ij} \Lie_\xi g_{ij} + \pi \Lie_\xi \phi \right)$.

\noindent{\bf The equations of motion of the Linking Theory}

Let's find the equations of motion generated by the total Hamiltonian of the Linking theory,
\begin{equation}
H_\st{tot} = ( \mathcal H | N ) + (\mathcal Q | \rho ) + (\mathcal H^i | \xi_i ) \,,
\end{equation}
they are, after application of the constraints,
\begin{align}
\dot g_{ij} =& 4 (\langle \rho \rangle -\rho ) \, g_{ij} + \Lie_\xi g_{ij}  + 2 \, \frac {e^{-6 \hat \phi } \, N } {\sqrt g} \left(p_{ij}-{\sfrac 1 3} p g_{ij}\right) 
\\
&- {\sfrac 1 3} \left\langle  N \, e^{6 \hat \phi } \sqrt{g} \right\rangle  \langle p\rangle \, g_{ij} \,,   
 \nonumber\\
\dot p^{ij} = &- 4 (\langle \rho \rangle - \rho  ) \left( \sigma^{ij} - {\sfrac 1 6}\sqrt g \langle p \rangle g^{ij} \right)   + \Lie_\xi p^{ij}
+ {\sfrac 1 {12}}  e^{6 \hat \phi } \sqrt{g} \, \langle p\rangle^2 \, N \, \, g^{ij}
\\
&- 2 \, N \, e^{-6 \hat \phi } \left( \frac 1  {\sqrt g} \sigma^i{}_k \sigma^{kj}  + {\sfrac 1 3} \langle p\rangle \, \sigma^{ij}\right) + e^{2 \, \hat \phi} \sqrt g \left( {\sfrac 1 2}  R \, N  - \Delta N  -  2 \, N \, \Delta \hat \phi\right) g^{ij} 
 \nonumber\\
& +e^{2 \, \hat \phi}  \, \sqrt g \left( \nabla^i \nabla^j N - R^{ij} \, N - 4 N \, \nabla^i \hat \phi \nabla^j \hat \phi - 4 \, \nabla^{(i} \hat \phi \, \nabla^{j)} N   + 2 N \, \nabla^i \nabla^j \hat \phi \right)  
\nonumber\\
&+{\sfrac 1 3} \langle p \rangle \left( p^{ij} - {\sfrac 1 2} \sqrt g \langle p \rangle g^{ij} \right)  \left\langle e^{6 \hat \phi} N \right\rangle + {\sfrac 1 {12 }} \left\langle \frac{\delta (\mathcal H_\phi | N)}{\delta \hat \phi} \right\rangle \left( e^{6 \hat \phi} -1\right) \sqrt g \, g^{ij}(x)  \,, 
\nonumber
\end{align}
where $\sigma^{ij} = p^{ij} - {\sfrac 1 3 } \, p \, g^{ij}$ is the traceless
part of the momenta. The derivation is again left as an exercise.

\noindent{\bf The GR gauge}

General Relativity is obtained by the gauge fixing $\phi \approx 0$, which is  second-class only with respect to $\mathcal Q_\phi[\phi,\pi;x)$:
\begin{equation}
\{  \phi (x) ,(\mathcal Q _{\hat \phi} | \rho) \} = \rho(x)  \,.
\end{equation}
This fixes $\rho \approx 0$, which in turn eliminates $\phi$ and $\pi$ from the theory, giving the ADM constraints.

\noindent{\bf The SD gauge}

Shape Dynamics follows from the gauge fixing $\pi \approx 0$, which has a single non-vanishing Poisson bracket\footnote{The Poisson bracket $\{ (\mathcal H^i_{\hat \phi} | \xi_i) , \pi(x) \} = 2 \, \nabla^i  \left[ ( p -\langle p \rangle \, \sqrt g ) \, \xi_i \right] $ vanishes on the gauge-fixing surface: if $\pi \approx 0$
then $\mathcal Q_{\hat \phi} \approx 0$ implies $p \approx  \langle p \rangle \, \sqrt g$.}
\begin{equation}\label{LapseFixingEq}
\{ (\mathcal H_{\hat \phi} | N) , \pi (x)\} = \frac{\delta  (\mathcal H_{\hat \phi} | N)}{\delta \hat \phi(x)} - \left\langle \frac{\delta  (\mathcal H_{\hat \phi} | N)}{\delta \hat \phi } \right\rangle ~ e^{6 \hat \phi(x)} \, \sqrt{g(x)} \,,
\end{equation}
where (after applying the constraint $\mathcal H_\phi \approx 0$)
\begin{equation} \label{LFEhomogeneous}
 \!\!\! \frac{\delta  (\mathcal H_{\hat \phi} | N)}{\delta \hat \phi(x)} \approx  \sqrt g \, e^{\hat \phi}  \left[ 56 \, N \,\laplacian  e^{\hat \phi} + 8  \,  \laplacian ( e^{\hat \phi} N ) - 2 \, N \left( 4  \, R  \, e^{\hat \phi} + e^{5 \hat \phi} \langle p \rangle^2 \right) \right]  \,,
\end{equation}
Define now the conformal Laplacian \index{conformal Laplacian} $\CL = 8 \, \laplacian -  R$, which is covariant under conformal
transformations, in the sense that if $g_{ab} = e^{4 \lambda} \bar g_{ab} $, $\CL f = e^{-4 \lambda}\bar \CL( e^{\lambda} f )$, so that when it is applied to $(f \, e^{\hat \phi})$ for any scalar $f$ it is invariant under the transformations generated by $\mathcal Q_{\hat \phi}$. We can then rewrite the expression 
above as follows
\begin{equation}
 \frac{\delta  (\mathcal H_{\hat \phi} | N)}{\delta \hat \phi(x)} \approx  \sqrt g \, e^{\hat \phi}  \left[ 7 \, N \,\CL e^{\hat \phi} +   \CL ( e^{\hat \phi} N )  - 2 \, N  \, e^{5 \hat \phi} \langle p \rangle^2   \right]  \,,
\end{equation}
which is an explicitly conformally-invariant expression.
%
%
%
%
%
%
%
This is the \emph{Lapse Fixing Equation}. \index{lapse-fixing equation} It can be solved for $N$ and admits a one-parameter set of solutions,
all related by a constant rescaling. Let's call a solution $N_\st{sol}$. Then $\mathcal H_\st{gl} = (\mathcal H_{\hat \phi} | N_\st{sol})$ is the part  of the
Hamiltonian constraint that is  first-class with respect to the gauge fixing $\pi$ and therefore survives it.

The general solution of Eq.~(\ref{LapseFixingEq}) consists of a linear combination of the two solutions $N_1$, $N_2$ of the homogeneous equation~(\ref{LFEhomogeneous}) plus a particular solution $N_0$ of the following nonhomogeneous equation:
\begin{equation}
\frac{\delta  (\mathcal H_{\hat \phi} | N)}{\delta \hat \phi(x)} = e^{6 \hat \phi(x)} \, \sqrt{g(x)} \,.
\end{equation}
The complete solution is then
\begin{equation}
N_\st{sol} = c_1 \, N_1+ c_2 \, N_2 + w \, N_0 \,,
\end{equation}
where $c_1$ and $c_2$ are spatial (but not necessarily temporal) constants, and $w$ is defined as
\begin{equation}
w = \left\langle \frac{\delta  (\mathcal H_{\hat \phi} | N)}{\delta \hat \phi } \right\rangle \,.
\end{equation}
Plugging the above solution into~(\ref{LapseFixingEq}), we get
\begin{equation}
\{ (\mathcal H_{\hat \phi} | N_\st{sol}) , \pi (x)\} = w \left[ 1  - \left\langle e^{6 \hat \phi} \, \sqrt{g} \right\rangle \right] e^{6 \hat \phi(x)} \, \sqrt{g(x)} \,.
\end{equation}
This is a secondary constraint, for which $w$ plays the role of a Lagrange multiplier. We can write this constraint as
\begin{equation}\label{GlobalConstraint}
\mathcal H_\st{gl} = \int \d^3 x \, \sqrt g \left( e^{6 \hat \phi} - 1 \right) \approx 0 \,.
\end{equation}
The above constraint is not trivial: in the constraints of the Linking Theory there is nothing ensuring that $\hat \phi$ is
actually volume-preserving. If one were to solve the constraint $\mathcal H_{\hat \phi} \approx 0$  for $\hat \phi$, the
solution $\hat \phi [ g,p;x)$  would not in general be volume-preserving. Since $\mathcal H_{\hat \phi} \approx 0$
completely fixes $\hat \phi$, the condition that $ \int \d^3 x \, \sqrt g ( e^{6 \hat \phi} - 1 )  = 0$ must be considered as an equation for $g_{ij}$, $p^{ij}$. We have therefore identified the leftover global constraint. In Eq.~(\ref{GlobalConstraint}), $\hat \phi$ must be considered as the solution $\hat \phi [ g,p;x)$ of the LY equation~(\ref{FinalLinkingTheoryConstraints}), and Eq.~(\ref{GlobalConstraint}) must be treated as a constraint
for the metric and metric momenta. It is obvious that $\mathcal H_\st{gl} $ commutes with the conformal constraint $\mathcal Q_{\hat\phi}$.

So, reducing phase space by integrating away $\phi$ and $\pi$, the final set of constraints we get is
\begin{equation}\label{FinalSDconstraints}
\begin{aligned}
&\mathcal H_\st{gl} = \int \d^3 x \, \sqrt g \left( e^{6 \hat \phi [ g,p;x)} - 1 \right) \,,  &
&\mathcal H^i = -2 \, \nabla_j p^{ij} \,, &
&\mathcal Q = 4( p - \langle p \rangle \, \sqrt g) \,,
\end{aligned}
\end{equation}
where $e^{6 \hat \phi [ g,p;x)}$ is the solution of the LY equation.
We recognise here the volume-preserving conformal constraint $\mathcal Q$ together with a conformally-invariant global Hamiltonian constraint $\mathcal H_\st{gl}$ that generates the evolution  and is
a nonlocal functional of the dynamical degrees of freedom $g_{ij}$ and $p^{ij}$.
The proof that  $\mathcal H_\st{gl}$ is both conformally and diffeo-invariant is left as an exercise, and it implies that the above system is
first-class.

\subsection{The degrees of freedom of Shape Dynamics}\label{DOFsOfSD}

As I said in Sec.~\ref{ExtendedRWR}, the  reduced configuration space of SD can be clearly identified: it is the quotient of $\Superspace$ by volume-preserving conformal transformations, that is $\shs \times \doublestroke{R}^+$, conformal superspace plus volume. \index{superspace} \index{conformal superspace} \index{Mach--Poincar\'e principle}
SD has the structure of a theory that satisfies the strong form of the  Mach--Poincar\'e principle on $\shs \times \doublestroke{R}^+$.
It satisfies the strong and not the weak form of this principle because of the reparametrization constraint $\mathcal H_\st{gl}$: one needs
just a point and a \emph{direction} in $\shs \times \doublestroke{R}^+$: we only need the increment in $\shs$, not the one in $\doublestroke{R}^+$
to determine the dynamical orbit -- thanks to  $\mathcal H_\st{gl}$.

This state of things might look a bit unnatural: there is just a single, global degree of freedom that does not belong to $\shs$ and yet
is necessary for the dynamics. York, \index{James W. York, jr.} noticing this fact, wrote~\cite{York:york_method_prl}:
\begin{quote}\it
The picture of dynamics that
emerges is of the time-dependent geometry of
shape (`transverse modes') interacting with the
changing scale of space (`longitudinal mode').
\end{quote}
With `transverse modes' he referred to the conformally-invariant degrees of freedom. Notice the singular in `scale of space' and `longitudinal mode': he didn't refer to the \emph{local} scales $\sqrt{g(x)}$ but to the single,
global one $V$.

But the reparametrization invariance of SD, expressed by the global constraint $\mathcal H_\st{gl}$, suggests that this
`heterogeneous' degree of freedom $V$ is not as physical as the conformally invariant ones living in $\shs$. The other ingredient
to understand what's going on here is due to N.~\'O Murchadha and J.~Barbour, who noticed \cite{Barbour:new_cspv} that in the initial-value problem, the choice of the initial volume $V$ is purely conventional: it is just a matter of units of lengths. There is an `accidental' symmetry in
the equations of Shape Dynamics which implies that $V$ and the York time $\tau$ are not genuine physical degrees of freedom.
We call this symmetry \emph{dynamical similarity}, \index{dynamical similarity} by analogy with a similar symmetry in Newtonian gravity~\cite{Barbour:2013goa,barbour2013gravitational}. This symmetry
can be expressed as $(p^{ij}_\st{TT}, \tau) \to (\alpha^4 p^{ij}_\st{TT}, \alpha^{-2}\tau)$, where $p^{ij}_\st{TT}$ is the transverse-traceless
part of the momenta, $\alpha$ is a constant (in space and time) and all the other dynamical degrees of freedom are left invariant. The transformation
I wrote connects a solution in $\shs \times \doublestroke{R}^+$ to another solution, but if we project these solutions down to $\shs$ they
both project to the same curve. The two solutions are just related by a change of length units, which cannot have physical significance.

So the punchline is: if, justified by the consideration that a single global unit of length is unphysical, we project  the solutions down to $\shs$,
we end up with a theory satisfying the \emph{weak} Poincar\'e principle in $\shs$.

\subsection{The solution to the Problem of Time in SD}\label{SolutionOfProblemOfTimeinSD}
\index{problem of time}

In~\cite{Barbour:2013goa} J. Barbour, T. Koslowski and I noticed how Shape Dynamics motivates a simple solution of the 
notorious \emph{problem of time} of quantum gravity, mentioned in Appendix~\ref{ADM-WDW}. The problem of time
in GR consists of two levels: first there is the problem of \emph{many-fingered time}, \index{many-fingered time} discussed at the end of Sec.~\ref{RWRsection}. For each choice of the lapse function $N(x,t)$ one obtains a solution of ADM gravity which is differently represented in $\Superspace$
but corresponds to the same spacetime, just foliated in another way. This is part of the reason why, upon na\^{i}ve quantization,
one obtains a Wheeler--DeWitt equation \index{Wheeler--DeWitt equation} that is time-independent (static) (see Sec.~\ref{WdWsection}). This problem is absent in SD because
the refoliation ambiguity is removed by the VPCT constraint: SD is compatible with only one particular foliation of spacetime,
that with constant trace of $p^{ij}$. However the theory is still reparametrization-invariant, even within that particular foliation.
Reparametrization-invariant theories have vanishing quantum Hamiltonians, and their quantization gives a static wavefunction(-al) which does not evolve.
\begin{figure}[t!]
\begin{center}\includegraphics[width=0.4\textwidth]{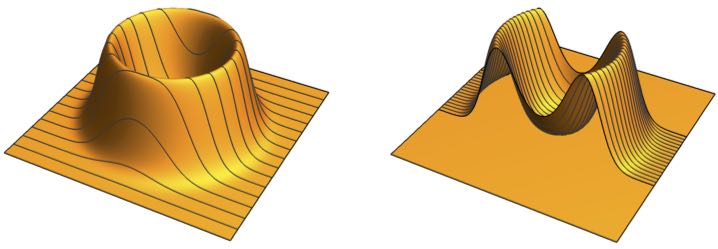}
\end{center}
\caption[2d quantum harmonic oscillator]{Probability density function for an eigenstate of the 2d quantum harmonic oscillator with high quantum number. The horizontal axes represent the two oscillator coordinates. One can `cut' the wavefunction along one axis, and obtain a marginal probability distribution for the other variable. Changing the location of the cut, one obtains a `time evolution' for the marginal distribution. However, the total probability is not conserved by this evolution. Moreover the evolution cannot be continued indefinitely because, sooner or later, near a `turning point' for the variable that is being used as internal clock,  this interpretation will break down altogether (figure concept: \cite{Bojowald:2010qw}).}
\label{2dOscillator}
\end{figure}
One strategy to circumvent this issue which has attracted interest in the literature is to accept that the Universe is described by a static wavefunctional and claim that our perception of time is the result of two factors: the wavefunctional being peaked around some semiclassical, high quantum number state, and us having access only to partial information about it~\cite{Bojowald:2010qw}.
Then the result of our measurements are comparisons of expectation values of \emph{partial observables}, and these evolve
only with respect to each other, in a limited sense. To understand this, imagine a 2d quantum harmonic oscillator described through
a Hamiltonian constraint like those we encountered in Sec.~\ref{SectionHamiltonianFormulation}. And imagine that the energy constant $E$ in
the Hamiltonian constraint $
\mathcal H = T - V -E \approx 0$ is equal to the energy of some high quantum number state, ${\sfrac 1 2} \hbar (n_1 + n_2)$ with
$n_1$ and $n_2$ large enough. Then the solution of the time-independent Schr\"odinger equation looks like Fig.~\ref{2dOscillator}: a volcano-shaped Probability Density Function.
One can, within a certain interval, use one of the two oscillators as an `internal clock' with respect to which the wavefunction
of the other oscillator is seen to evolve. This can be kept up as long as the chosen clock evolves monotonically: as soon as its value approaches a turning point where it inverts its motion, this description becomes untenable. Then a `grasshopper' strategy
is adopted, in which one jumps from one internal clock to the other, exploiting the intervals in which they evolve monotonically.
This strategy can be called \emph{Tempus Post Quantum}, \index{Tempus Post Quantum} in the sense that one seeks a physical definition of time in the
observables of the Universe \emph{after} quantizing. There are two problems with this: first, nothing ensures that the
wavefunction of the Universe will `oblige', and get into a semiclassical state. One would like to have a mechanism for
which this happens.\footnote{We think that our approach provides such a mechanism: read below and ref.~\cite{Koslowski:2014eu}.}  Second, the `grasshopper' strategy is problematic. As soon as one uses an internal clock, even far from the
`turning points', there will be violations of unitarity: one can easily see, for example, that in Fig.~\ref{2dOscillator} the area of the `cross-section' of the PDF is not conserved.

\index{Tempus Ante Quantum}
The strategy we adopted in~\cite{Barbour:2013goa} is the opposite: `\emph{Tempus Ante Quantum}'. If we identify our `internal clock' before quantizing, we will have a time-dependent Schr\"odinger equation. This is a desirable situation, but the problem with it is that it is not clear what should provide a good universal internal clock. It should be a quantity that grows monotonically in every
solution (no `turning points') and depends equally on every different part of space, in accordance with some measure). A priori, the search for such a quantity might look as hopeless as Kuchar's search for `unicorns'~\cite{KucharReview}. The advance that Shape Dynamics introduces is to single out such variable: I'm talking about $\tau = \frac 2 3 \,\langle p \rangle$, the York time. \index{York time} In addition to the good properties for a universal
internal time that I already mentioned, York time is geometrically distinguished: it's the only non-shape degree of freedom that plays a dynamical role in SD. $\tau$ is clearly  monotonic: this can be seen by writing its equations of motion in the Linking Theory and then applying all the constraints and the $\pi \approx 0$ gauge fixing~\cite{Barbour:2013goa}. To use it as an internal clock at the classical level,
one has to `deparametrize' the theory.

\noindent{\bf Deparametrization: non-autonomous description on $\shs$}
\index{deparametrization}

Deparametrization is a simple idea: take a reparametrization-invariant theory with a Hamiltonian constraint $\mathcal H(q_1,p^1,q_2,p^2,\dots) \approx 0$. Say you want to use the variable $q_1$, in the interval in which it's monotonic, as an internal clock.
Then $p^1$, which is conjugate to $q_1$ and generates $q_1$-translations, will play the role of a `Hamiltonian' that generates
evolution in the `time' $q_1$. Then we have to solve the Hamiltonian constraint w.r.t. $p_1$: $p_1= f(q_1,q_2,p^2,\dots)$ to obtain
$\mathcal H(q_1,f(q_1,q_2,p^2,\dots),q_2,p^2,\dots) \approx 0$. Then $f(q_1,q_2,p^2,\dots)$ will generate the evolution of all the other variables, $q_2,p^2,\dots$ with respect to $q_1$:
\begin{equation}
\frac{\d q_i}{ \d q_1} = \{ f(q_1,q_2,p^2,\dots)  , q_i \} \,, \qquad \frac{\d p^i}{ \d q_1} = \{ f(q_1,q_2,p^2,\dots)  , p^i \} \,,
\end{equation}
where $i = 2,3,\dots$. 

The  variable conjugate to York time \index{York time} is the volume $V= \int \d^3 x \sqrt g$,
\begin{equation}
\{\tau , V \} = {\sfrac 3 2} \{  \langle p \rangle, V \} =  \langle \sqrt g \rangle =  1  \,.
\end{equation}
Therefore to deparametrize SD with respect to $\tau$, we have to solve
the global Hamiltonian constraint for $V$.
The reparametrization
constraint of SD is $\mathcal H_\st{gl}$ in Eq.~(\ref{FinalSDconstraints}):
\begin{equation}
\mathcal H_\st{gl} = \int \d^3 x \, \sqrt g \, e^{6 \, \hat \phi [ g_{ij}, p^{kl} , \tau ; x) } - V \,,
\end{equation}
where $\hat \phi [ g_{ij}, p^{kl} , \tau ; x)$ is the solution of the LY equation~(\ref{FinalLinkingTheoryConstraints}), written in the form
{\thickmuskip=0mu
\medmuskip=0mu
\begin{equation}
\frac{e^{-6 \hat \phi}}{\sqrt g} \left(p^{ij}-{\sfrac 1 3} p g^{ij}\right) \left(p_{ij}-{\sfrac 1 3} p g_{ij}\right)-\frac{3}{8}  \sqrt g \, e^{6  \hat \phi} \tau^2- \sqrt g \left( R \,e^{2 \,  \hat \phi} - e^{ \hat \phi} \laplacian e^{ \hat \phi} \right)  = 0 \,.
\end{equation}}
Now notice that the LY equation is covariant under conformal transformations of the form $g_{ij} \to \lambda g_{ij}$, $p^{ij} \to \lambda \, p^{ij}$, $\hat \phi \to \hat \phi - \log \lambda $ and $\tau \to \tau$, where $\lambda= \lambda(x) >0$.
Therefore $\sqrt g \, e^{6 \, \hat \phi [ g_{ij}, p^{ij} , \tau ; x) } $ is fully conformally invariant: it cannot depend on the volume $V$. Deparametrizing with respect to $\tau$ is then immediate: the solution of $\mathcal H_\st{gl} \approx 0$ for $V$ is simply
\begin{equation}
V = \int \d^3 x \, \sqrt g \, e^{6 \, \hat \phi [ g_{ij}, p^{ij} , \tau ; x) } \,,
\end{equation}
and the Hamiltonian generating evolution in $\tau$-time is (see \cite{Koslowski:2013gua}, but essentially
the same Hamiltonian has been written by York)
\begin{equation}\label{SDHamiltonian}
H_\st{SD} = \int \d^3 x \sqrt g \, e^{6 \hat \phi [ g_{ij}, p^{kl} , \tau ; x) } \,.
\end{equation}
The above Hamiltonian depends on $\tau$, which we now take as the independent variable.
Hamilton's equations are consequently not autonomous (meaning that they depend
explicitly on the independent variable) and they are not invariant under $\tau$-translations.
This means that among the initial data needed to specify a solution we have to include a
value of $\tau$. The initial-value problem \index{initial-value problem} is completely specified by local shape initial data
(a conformal equivalence class and $TT$ momenta) plus $\tau$. However, one can 
rewrite this system as an equivalent one which is autonomous, at the cost of having
`friction' terms which make the equations of motion non-Hamiltonian~\cite{barbour2013gravitational}.
The key to do this is to do some dimensional analysis: initial data on shape space should be 
dimensionless, however the metric momenta are dimensionful. One can obtain dimensionless
momenta and equations of motion by multiplying $p^{ij}$ by an appropriate power of $\tau$, and reparametrize $\tau$  to $\log \tau$, but then the new momenta
won't satisfy Hamilton's equations (the difference will just be a dissipative term proportional to $p^{ij}$ in the equation for $\d p^{ij}/\d \log \tau$). Using $\log \tau$ as independent variables  allows us only to
describe half of each solution: the half in which $\tau$ is positive. The other half can be described
as a different solution of the same dissipative system. So each solution is split into two half at the
instant when $\tau =0$. 
This description is suggestive: one can do the same in the Newtonian N-body problem, where the role of $\tau$ is played by the dilatational momentum $\text{D}$. The dissipative nature of the equations of motion
imply an irreversible growth of a scale-invariant quantity (a function of shape space) which measures the degree of complexity of the shape of the universe. This defines a previously unnoticed arrow of time that points from the simplest and more homogeneous to the more complex and clustered states. In~\cite{barbour2013gravitational} we conjectured that an analogous arrow of time can be identified in
geometrodynamics, and may be a better way to think about the evolution of our Universe.
\index{deparametrization}\index{problem of time}

\subsubsection*{Construction of spacetime}

A solution of Shape Dynamics is a curve in conformal superspace, \index{conformal superspace} parametrized with York time $\tau$. A way to represent it is withÊa conformal gauge, for expample the unimodular gauge:
\begin{equation}
\tilde g_{ij}(x,t)= \frac{g_{ij}}{\det g^{1/3}} \,,
\end{equation}
from the tangent vector to the curve $\frac{\d {\tilde g}_{cd}}{\d \tau}$ we can build CMC momenta as
\begin{equation}
{\tilde p}^{ij} = \left( \tilde g^{ik}\tilde g^{jl} - {\sfrac 1 2} \tilde g^{ij}\tilde g^{kl} \right) \left( \frac{\d {\tilde g}_{kl}}{\d \tau} + \Lie_\xi \tilde g_{kl} \right) + {\sfrac 2 9} \, \tau \, \tilde g^{ij} \,,
\end{equation}
notice that, since $\tilde g_{ij}$ is unimodular,  $\left( \tilde g^{ik}\tilde g^{jl} - {\sfrac 1 2} \tilde g^{ij}\tilde g^{kl} \right) \left( \frac{\d {\tilde g}_{kl}}{\d \tau} + \Lie_\xi \tilde g_{kl} \right) $ is automatically zero-trace.
Then we can solve the diffeomorphism constraint for $\xi^i[\tilde g_{kl} , \tilde p^{kl} ; x)$ and make $\tilde p^{ij}$ transverse: $\tilde \nabla_j \tilde p^{ij} = 0$ (as remarked in York, such an equation for $\xi^i$  is elliptic
and admits a unique solution on a compact manifold).
 At this point we have everything that's necessary to solve the Lichnerowicz--York equation and get a 
scale factor $\phi[\tilde g_{ij} , \tilde p^{ij} ; \tau  , x)$ which can be
 used to define a proper Riemannian 3-metric as $g_{ij} = \phi^4  \, \tilde g_{ij}$  defining local scales (it is not unimodular). Finally, we can take the last step of solving the Lapse-fixing equation and get a lapse $N[ \phi, \tilde g_{ij},\tilde p^{ij};\tau,x)$, with which we can define a 4-dimensional Lorentzian metric:
\begin{equation}
g_{\mu\nu} = \left( 
\begin{array}{cc}
-N^2 + \phi^4 \, \tilde g_{ab}\, \xi^a \, \xi^b &   \phi^4 \tilde g_{ac} \, \xi^c\\   \phi^4 \tilde g_{bc} \, \xi^c &   \phi^4 \tilde g_{ab}
\end{array}
\right) \,.
\end{equation}
Notice that  nothing ensures that  $\det g_{\mu\nu} \neq 0$.  The above metric can be degenerate
and won't globally define, in general, a spacetime.

\subsubsection*{The emergence of rods and clocks}
\index{rods and clocks}

In \cite{EinsteinNote} Einstein remarked, about his theory of Relativity:\index{Albert Einstein}
\begin{quotation}
\it It is striking that the theory (except for four-dimensional space) introduces two kinds of physical things, i.e. (1) measuring rods and clocks, (2) all other things, e.g., the electromagnetic field, material point, etc. This, in a certain sense, is inconsistent; strictly speaking measuring rods and clocks would have to be represented as solutions of the basic equations... not, as it were, as theoretically self-sufficient entities. The procedure justifies itself, however,
because it was clear from the very beginning that the postulates of the theory are not strong enough to deduce from them equations for physical events
sufficiently  complete and sufficiently free from arbitrariness in order to base upon such a foundation a theory of measuring rods and clocks.
If one did not wish to forego a physical interpretation of the coordinates in general (something that, in itself, would be possible), it was better to
permit such inconsistency - with the obligation, however, of eliminating it at a later stage of the  theory.
\end{quotation}
The construction of a spacetime metric from a solution of SD I showed before is a first step in the
direction sought by Einstein:Êfrom scale-invariant and timeless shape-dynamic first principles, one \e{creates}, as opposed to simply postulating, a structure in which length and duration (proper time) are `there' to be measured.
\index{proper time}

But this is only the first step. We must show how the basic equations of the theory lead to the formation of structures which serve as rods and clocks that measure the dynamically created lengths and durations. We know that this happens in the actual Universe. Sufficiently isolated subsystems move along the
geodesics of this metric, and the proper time,
\begin{equation}
d s^2 =  \left( \phi^4 \tilde g_{ij} \, \xi^i \xi^j  - N^2 \right) d\tau^2 + 2 \, \xi_i \, dx^i \, d \tau + \phi^4 \tilde  g_{ij} \, dx^i \, dx^j \,,
\end{equation}
turns out to be the time ticked along their worldline by natural clocks belonging to sufficiently isolated and light subsystems, like for example the rotating Earth.
Isolated/light subsystems also provide natural rods, for example, rocks on the surface of the Earth, as their sizes can be compared with each other.

The challenge of the ``emergence of rods and clocks'' program  is to prove that, for suitable operational definitions of rods and clocks from material systemsÊ(\emph{e.g.} isolated enough galaxies, which have both a characteristic size and a rotation time), they will behaveÊ as is simply postulated in the spacetime description. Namely, rods will stay mutually  congruent when brought close to each other and compared (apart from Lorentz contraction if they are in relative motion), and clocks will all
approximate proper time along their trajectories (which will all be geodesics of the same spacetime
metric). In brief, all good rods and clocks will provide mutually consistent data that conspire to form
a unique spacetime manifold all observers will agree on. 

If this gets well understood, then the cases in which it fails (\emph{e.g.} black holes, the early universe) 
will become particularly interesting, and we will have a whole new perspective on them  that would not be available if the existence of spacetime is taken as the fundamental postulate. The point is that Shape Dynamics is capable of describing situations that cannot be described as a (single, smooth) spacetime manifold, and these are likely to be very relevant for cosmology and astrophysics (not to speak of quantum gravity, where the spacetime `prejudice' might have severely hampered progress).
\index{rods and clocks}

\subsection{Coupling to Matter and uniqueness}\label{CouplingToMatterSec}
\index{matter fields}

In~\cite{GomesCouplingToMatter} H.~Gomes and T.~Koslowski studied the coupling of SD to matter. A question immediately
arises: what is the transformation law of matter fields under conformal transformations?
In general, one could have
\begin{equation}
g_{ij} \to e^{4 \, \hat \phi} g_{ij}  \,, ~~~  \psi_\st{A}  \to e^{n_\st{A} \,\hat  \phi} \psi_\st{A}  \,, 
\end{equation}
for a generic field $\psi_\st{A}$, where  $n_\st{A}$ is a real constant. Or the transformation
law of $\psi_\st{A}$ could be even more complicated, for example involving derivatives of $\hat \phi$.

This issue has been discussed by J.~Isenberg, N.~\'O Murchadha and J.~York \cite{Isenberg:matter_coupling1}
in the context of York's method for solving the initial-value problem \index{initial-value problem} of GR, and in subsequent works by 
Isenberg and J.~Nester~\cite{Isenberg:matter_coupling2,Isenberg:matter_coupling3}. 
The result of these studies were that, at least in the case of physically-realized fields (Yang--Mills, scalars, Dirac fermions)
York's decomposition is still useful if we simply assume that all matter fields are conformally invariant, $n_\st{A} = 0$ ~$\forall~\rm{A}$.
\index{James W. York, jr}

In SD there is a simple argument for which gauge fields should be conformally invariant. If a Yang--Mills or Maxwell field transform under conformal transformation then the conformal constraint $\mathcal Q$ must be supplemented with a term proportional to $E^i_\alpha A^\alpha_i$ which is not gauge-invariant. The conformal constraint and the gauge constraint then would be second-class, ruining the structure of the theory.

In SD the matter Hamiltonian has to satisfy a bound. This bound is a consequence of the requirement that the conformal constraint $p$ completely gauge-fixes 
the  Hamiltonian constraint $\mathcal H$ (so that the CMC constraint $p -\langle p \rangle \sqrt g$ leaves just one single
degree of freedom unfixed). The linear differential operator $\Delta_\st{GF}$ obtained by taking the  Poisson bracket between the two constraints:
\begin{equation}\label{GaugeFixingOperator}
 \Delta_\st{GF} N  \ =  \{  p(x) , (\mathcal H  | N) \} 
\end{equation}
has to be invertible. If we assume that all the matter fields have zero conformal weight, and the Hamiltonian
constraint has the form
\begin{equation}
\frac 1 {\sqrt g} \, \left( p^{ij} p_{ij} - {\sfrac 1 2} p^2 \right)  - \sqrt g \, R + \mathcal H_\st{matter} \approx 0 \,,
\end{equation} 
then the operator (\ref{GaugeFixingOperator}) reads (on shell - that is, after applying the constraints):
\begin{equation}
\Delta_\st{GF}  \approx  2 \, \laplacian - {\sfrac 1 {6}} \langle p \rangle^2 - \frac{ \left( p - {\sfrac 1 3} g \, \text{tr} p\right)^2 }{g} + {\sfrac 1 {\sqrt g}} \left( \frac{\delta  \mathcal H_\st{matter} }{\delta g_{ij}} g_{ij} - {\sfrac 1 2} \,  \mathcal H_\st{matter}   \right) \,,
\end{equation}
where I used the index-free notation $\left( p - {\sfrac 1 3} g \, \text{tr} p\right)^2=\left( p_{ij} - {\sfrac 1 3} g_{ij} \, p\right)\left( p^{ij} - {\sfrac 1 3} g^{ij} \, p\right)$. The Laplacian is a positive operator, therefore $\Delta_\st{GF}$ has an empty kernel if
\begin{equation}
- {\sfrac 1 {6}} \langle p \rangle^2 - \frac{ \left( p - {\sfrac 1 3} g \, \text{tr} p\right)^2 }{g} + {\sfrac 1 {\sqrt g}} \left( \frac{\delta  \mathcal H_\st{matter} }{\delta g_{ij}}  g_{ij}  - {\sfrac 1 2} \,  \mathcal H_\st{matter}   \right) \leq 0 \,,
\end{equation}
which translates into a stronger bound on the matter Hamiltonian
\begin{equation}
 \frac{\delta  \mathcal H_\st{matter} }{\delta g_{ij}}  g_{ij} - {\sfrac 1 2} \,  \mathcal H_\st{matter}   \leq 0\,.
\end{equation}
The standard Hamiltonian for Yang--Mills and massless scalar fields satisfies the above bound.
Interestingly, a massive scalar field plus a cosmological constant:
\begin{equation}\label{MassiveScalarHamiltonian}
 \mathcal H_\st{matter} =  {\sfrac 1 2}  \frac{\pi^2}{\sqrt g} +  {\sfrac 1 2}  \sqrt{g} \left( \nabla_i  \varphi \nabla^i \varphi + m^2  \,  \varphi^2 \right) + 2 \Lambda  \sqrt g \,,
\end{equation}
respects the bound only if 
\begin{equation}
\frac{\delta  \mathcal H_\st{matter} }{\delta g_{ij}}  g_{ij} - {\sfrac 1 2} \,  \mathcal H_\st{matter}    = -  \frac{\pi^2}{\sqrt g} + \sqrt g \, \left(  {\sfrac 1 2} m^2  \varphi^2 + 2 \, \Lambda \right)   \leq 0  \,.
\end{equation}
The physical consequences of this limit have not been investigated in detail yet.

\subsection{Experienced spacetime}\label{ExperiencedSpacetimeSec}
\index{proper time}

With the possibility of coupling SD to matter field, we can now give an operational meaning to the 4-dimensional line element, following what has been done in~\cite{Tim_Proceedings_TheoryCanada9} and updating/simplifying the exposition.
Let's start with a background solution of SD, which we may represent in ADM gauge as a 3-metric $g_{ij}$ and CMC-momenta $p^{ij}$ that satisfy the Hamiltonian and diffeomorphism constraints of ADM gravity. We want to add to it some weak matter fields that we may use as probes. The simplest choice is a single component scalar field $\varphi$ with conjugate momentum $\pi$. We want the scalar field and its momentum to be small perturbations. So we introduce a formal infinitesimal parameter $\epsilon$ and we multiply each occurrence of $\varphi$ and $\pi$ by $\sqrt{\epsilon}$. So if the matter Hamiltonian is that of a free massive scalar as in Eq.~(\ref{MassiveScalarHamiltonian}) (without cosmological constant), it is quadratic in the field and its momentum and therefore it is linear in $\epsilon$.

We can calculate the first-order perturbation to the solution of the Lich\-ne\-ro\-wicz--York equation. To do that we perturb the conformal factor of the metric: $g_{ij} \to (1 + \epsilon \, \phi_1) g_{ij}$ and insert this ansatz into the Hamiltonian constraint with matter sources:
\begin{equation} \label{PerturbedLYeq}
{\sfrac 1 g} \left( p^{ij} p_{ij}  - {\sfrac 1 2} p^2 \right) - R
+ \epsilon \, \Delta_\st{GF} \phi_1 + \epsilon \,  \mathcal H_1 + \mathcal O(\epsilon^2) = 0 \,,
\end{equation}
where
\begin{equation}\label{FreeScalarPerturbationSDHamiltonian}
\mathcal H_1 = {\sfrac 1 2} \frac{\pi^2}{g} +  {\sfrac 1 2} \left(  g^{ij} \nabla_i \varphi \nabla_j \varphi + m^2 \varphi^2 \right)  \,,
\end{equation}
and 
\begin{equation}
\Delta_\st{GF} =  8 \Delta - R - {\sfrac 5 6}( \langle p \rangle^2 - 12 \Lambda)  - {\sfrac 7 g} \left( p^{ij}- {\sfrac 1 3}  g^{ij} \, \text{tr} p \right)\left( p_{ij}- {\sfrac 1 3}  g_{ij} \, \text{tr} p \right) \,,
\end{equation}
is an elliptic differential operator which depends only on the background metric.
The zeroth-order term of Eq.~(\ref{PerturbedLYeq}) is automatically zero (because $g_{ij}$ and $p^{ij}$ are a solution of the Hamiltonian constraint), so we can solve the first-order w.r.t. $\phi_1$:
\begin{equation}
\phi_1 = \int \d^3 y K(x,y) \mathcal H_1(y) \,,
\end{equation}
where $K(x,y)$ is the integral kernel $\Delta^x_\st{GF} K(x,y) = \delta^{(3)}(x-y)$.
Now we want to insert the first-order correction to the LY factor into the SD Hamiltonian~(\ref{SDHamiltonian}) 
\begin{equation}
\begin{aligned}
H_\st{SD} &= 6 \epsilon \int \d^3 x \sqrt{g} \, \phi_1 + \mathcal O(\epsilon^2)\\
&= 6 \epsilon \int \d^3 x \d^3 y\sqrt{g(x)} \, K(x,y) \mathcal H_1(y) + \mathcal O(\epsilon^2) \,.
\end{aligned}
\end{equation}
The above can be reorganized into the Hamiltonian~(\ref{FreeScalarPerturbationSDHamiltonian}) for a free scalar propagating on the background $g_{ij}$, smeared with an effective lapse function
\begin{equation}
N_\st{eff}  = 6 \epsilon \int \d^3 x \d^3 y\sqrt{g(x)} \, K(x,y)  \,.
\end{equation}
Now, the SD Hamiltonian will evolve the scalar field degrees of freedom, at first order in $\epsilon$, as
\begin{equation}
\dot \varphi = \left\{ \varphi , H_\st{SD} \right\} + \mathcal O(\epsilon) \,, \qquad
\dot \pi = \left\{ \pi , H_\st{SD}  \right\} + \mathcal O(\epsilon) \,.
\end{equation}
Once we integrate away the momenta $\pi$, turn into the following equations of motion for $\varphi$:
\begin{eqnarray}\label{ScalarPerturbationsEqOfMotion}
\ddot \varphi - N_\st{eff}^2 \Delta  \varphi + N_\st{eff}^2 m^2  \varphi + \dot  \varphi   \partial_\tau \left( \log {\sfrac {N_\st{eff}} {\sqrt{g}}} \right) + g^{ij}   N_\st{eff} \nabla_i N_\st{eff} \nabla_j \varphi = 0 \,.
\end{eqnarray}
T. Kolsowski in~\cite{Tim_Proceedings_TheoryCanada9}  now suggests a formal way of recovering the 4D line element that is experienced by the scalar field fluctuations:
\begin{enumerate}
\item observe that the first-order equations of motion~(\ref{ScalarPerturbationsEqOfMotion}) do not contain any mixed space-time derivative term, and therefore one does not need to implement a shift in the coordinate system when evolving in York time.\index{York time}

\index{light cone}
\item Deduce the light cone by studying the propagation of wave packets. There is a standard strategy to do this: one starts with a compact perturbation of an otherwise everywhere-zero $\varphi$, and observes how the support of the perturbation expands. There will be arbitrarily-high frequency modes (of arbitrarily-low amplitudes), which will travel at speeds that are arbitrarily close to the speed of light. These modes will only feel the principal symbol of the differential operator acting on $\varphi$ in~(\ref{ScalarPerturbationsEqOfMotion}), which is $\partial_\tau^2 - N_\st{eff}^2 g^{ij} \partial_i \partial_j$, because the non-quadratic terms will be suppressed by the frequency of the modes. Then we know that the experienced line element will be conformal to:
\begin{equation}\label{ConformallyRelated4Dmetric}
\d s^2_\st{conf} = - \d \tau^2 + N_\st{eff}^{-2} g_{ij} \d x^i \d x^j \,.
\end{equation}
This is a consequence of Malament's theorem~\cite{malament1977}, which states that the light-cone structure completely determines the conformal structure of spacetime.

\item To single out a particular spacetime metric out of the conformal class, we need to exploit the scale that is introduced by the nonzero mass $m$ of the scalar field (a massless field can only probe the conformal structure of spacetime). The strategy is to compare the propagation of waves of different wavelengths, which due to the mass have different local propagation speeds. Koslowski~\cite{Tim_Proceedings_TheoryCanada9} suggests to consider an infinitesimal spatial region, and define the wavelengths of the modes as integer fractions of the size of the region. Then, by observing the evolution of such a superposition of waves for an infinitesimal interval of York time, Eq.~(\ref{ScalarPerturbationsEqOfMotion}) implies that the interference patter of the field would change as if the mass was $N_\st{eff} m$. This implies that we have to reparametrize time using the lapse $N_\st{eff}$ in order to use everywhere the unit of time in which $c=1$. This implies that we have to multiply metric~(\ref{ConformallyRelated4Dmetric}) by the conformal factor $N_\st{eff}^2$. The experienced 4D line element is therefore:
\begin{equation}\label{Experienced4DMetric}
\d s^2_\st{exp} = -  N_\st{eff}^2 \d \tau^2 + g_{ij} \d x^i \d x^j \,.
\end{equation}

\end{enumerate}

\subsection{`Symmetry doubling': BRST formulation of SD}\label{SymmetryDoublingSec}
\index{Becchi--Rouet--Stora--Tyutin formalism}

In~\cite{Gomes:2012hh} Gomes and Koslowski studied the BRST (Becchi--Rouet--Stora--Tyutin) formulation of 
ADM gravity in maximal-slicing gauge.
 The BRST formalism is an approach to the quantization of gauge theories which makes mathematical sense of Faddeev--Popov ghosts and their exclusion from physical asymptotic states. Faddeev--Popov ghosts are unphysical fields with the statistics opposite to that predicted by the spin-statistics relation, which have to be introduced in perturbative calculations in quantum gauge theory in order to preserve gauge invariance at the perturbative level (in particular to ensure anomaly cancellation and unitarity). The path integral formulation of QFT in fact overcounts gauge-related configurations, which are physically equivalent, and it is therefore necessary to introduce a gauge-fixing that intersects gauge orbits only once in order to get finite results for scattering amplitudes. However the gauge-fixing procedure is not as straightforward as in classical gauge theory: in Feynman's path integral there is an integration measure over all of phase space, which gives a geometric factor when it is restricted over a gauge-fixing surface (\emph{i.e.} the determinant of the Jacobian of the transformation to coordinates on the gauge-fixing surface. This Jacobian is the same that one gets when integrating over a delta-function in curvilinear coordinates). The Jacobian determinant of Faddeev and Popov can be written as a functional Gaussian integrated over `dummy' fields, but these fields need to have wrong statistics (\emph{i.e.} they need to be Grassmannian even though they are scalars or vectors) in order to give the correct form of the determinant. One integrates over these so-called `ghost' fields in the path integral, but they should not appear in boundary states, because they are unphysical, they are just a calculational tool without which gauge invariant would be spoiled. In perturbation theory this means that ghost fields only appear in internal legs of Feynman diagrams, never in external legs. The role of these fields in perturbation theory is to compensate for unphysical/gauge-non-invariant modes, \emph{e.g.} the longitudinal mode of the gauge potential: each internal leg of a Feynman diagram  which involves a longitudinally-polarized gauge boson is cancelled by a leg of the corresponding ghost.
BRST symmetry is a kind of `supersymmetry' that characterizes the Lagrangian of a gauge-fixed gauge theory with Faddeev--Popov ghosts. This symmetry exchanges ghost fields with the other fields in the theory, which have opposite statistics. The operator generating this symmetry, $Q$, is nilpotent ($Q^2 =0$) and induces a grading on the extended Hilbert space of the theory, because its action on states raises the ghost number by one. Therefore $Q$ introduces what is known as a \emph{cochain complex} describing a \emph{cohomology}.
This allows to identify the physical Hilbert space (the space of asymptotic states) as the elements of its cohomology of the operator $Q$. The requirement that ghost do not enter external legs now has a precise topological expression. The existence of the operator $Q$ is sufficient to guarantee the consistency of the quantum gauge theory (at least if the gauge group is compact).

The crash-introduction to BRST given above is by no means supposed to be exhaustive, and I divert the reader to further readings (as~\cite{Becchi-BRST}) for more details on this method. My hope is that it is sufficient to get the gist of the results of~\cite{Gomes:2012hh}. In this paper the authors first observe that \emph{pure constraint} theories, \emph{i.e.} theories in which the Hamiltonian vanishes on-shell, like ADM gravity, have a gauge-fixed BRST Hamiltonian which takes the form of a pure bracket.
%
%
%
If $\chi_a$ are a set of first-class constraints which close a Poisson algebra $\{ \chi_a , \chi_b \} = U_{ab}^c \chi_c$ where $U_{ab}^c$ are, in general, structure \emph{functions} (functions of phase space),  
the BRST generator is defined as
\begin{equation}  
\Omega= \eta^a\chi _a-\frac{1}{2}\eta^b\eta^aU_{ab}^c P_c \,,
\end{equation}
where $\eta^a$ are the ghosts associated to the constraint transformations, and $P_b$ the canonically conjugate ghost momenta. 
The gauge-fixed Hamiltonian is constructed by choosing a ghost number $-1$ fermion, called the gauge-fixing fermion,
\begin{equation}
\tilde\Psi=\tilde\sigma^a P_a \,,
\end{equation}
where $\tilde\sigma^a$ is a set of \emph{proper} gauge fixing conditions (whose definition I won't get into). 
Denoting the BRST invariant extension of the on-shell Hamiltonian (where all constraints are set to vanish) by $H_o$, the general gauge fixed BRST-Hamiltonian is written as
\begin{equation}
 H_{\tilde\Psi}=H_o+\eta^a V_a{}^b P_b+\{\Omega,\tilde\Psi\} \,,
\end{equation}
where $\{H_o,\chi_a\}=V_a{}^b \chi_b$ and the bracket is extended to include the conjugate ghost variables. The gauge fixing  changes the dynamics of ghosts and other non-BRST invariant functions, but maintains evolution of all BRST-invariant functions. The crux of the BRST-formalism is that the gauge-fixed Hamiltonian $H_{\tilde\Psi}$ commutes strongly with the BRST generator $\Omega$. Although  gauge symmetry is completely encoded in the BRST transformation $s  := \{\Omega, . \}$, and we have fixed the gauge, the system retains a notion of gauge-invariance through BRST symmetry.

Applying this to a system with vanishing on-shell Hamiltonian $H_o \approx 0$, we find that the gauge-fixed BRST-Hamiltonian takes the form
\begin{equation}
 H_{\tilde\Psi}=\{\Omega,\tilde\Psi\} \,.
\end{equation}
Suppose now that $\sigma^a$ is both a classical gauge fixing for $\chi_\alpha$, and also a first class set of constraints: 
\begin{equation}
\{\sigma^a, \sigma^b\}= C^{ab}_c \sigma^c \,.
\end{equation}
We can then construct a nilpotent gauge-fixing $\Psi$ with the same form as the BRST charge related to the system $\sigma^a$,  the only difference being that ghosts and antighosts are swapped. This gauge-fixing fermion takes the form:
\begin{equation} 
\Psi=\sigma^a P_a -\frac{1}{2} P_b P_aC^{ab}_c \eta^c \,.
\end{equation}

Using this gauge-fixing fermion implies that the BRST extended gauge-fixed Hamiltonian would be invariant under two BRST transformations
\begin{equation}
   s_1  := \{ \Omega , \, .\, \} \,, \qquad
   s_2  := \{ \, .\, , \Psi \} \,,
\end{equation}
which follows  directly from the super-Jacobi identity and nilpotency of both $\Omega$ and $\Psi$.  
In~\cite{Gomes:2012hh} identified $\sigma^a$ with the maximal-slicing constraint $g_{ij} p^{ij} \approx 0$, together with a conformal-harmonic gauge $\left( g^{ij} \delta^k{}\ell + {\sfrac 1 3} g^{ik} \delta^j{}_\ell \right)\left(\Gamma^\ell_{ij} - \bar \Gamma^\ell_{ij} \right) \approx 0 $ (where $\Gamma^\ell_{ij}$ and $\bar \Gamma^\ell_{ij} $ are the Christhoffel symbols associated to, respectively, the metric $g_{ij}$ and a background metric $\bar g_{ij}$)  which gauge-fixes diffeomorphisms but is first-class with respect to $g_{ij} p^{ij}$. This proves the existence of the two BRST invariances in ADM gravity.

The consequences of this `doubling' of BRST symmetry for quantum gravity still has to be explored. Moreover the connection to Shape Dynamics has to be established by generalizing this result to CMC slicing $g_{ij} p^{ij} - \sqrt g \langle g_{ij} p^{ij} \rangle  \approx 0$.

\subsection{`Conformal geometrodynamics regained'}\label{HenriquesConstructionPrincipleSec}

The result on `symmetry doubling' described in the last Section inspired Gomes to conjecture that the existence of two symmetries which gauge-fix each other might be taken as a `construction principle' for gravity. This idea is closely related, in spirit, to the work of Hojman, Kucha{\v{r}} and Teitelboim~\cite{HKT}. \index{Hojman--Kucha{\v{r}}--Teitelboim derivation of GR} In that paper ADM gravity was uniquely derived by assuming the form of the `hypersurface deformation algebra', \index{hypersurface deformation algebra} which we built (piece by piece) in Sec.~\ref{ExtendedRWR}:
\begin{equation}\label{HypersurfaceDeformationAlgebra}
\begin{aligned}
&\{ (\d \xi^i | \mathcal H_i) , (\d \chi^j | \mathcal H_j) \}\approx ([\d \xi , \d \chi]^j | \mathcal H_j)  \,,
\\
&\{ (\d \xi^i | \mathcal H_i) , (\d \varphi | \mathcal H) \}  \approx (\pounds_{\d \xi} \d \varphi | \mathcal H)  \,,
\\
&\{ (\d \varphi | \mathcal H) , (\d \sigma | \mathcal H) \}
\approx  (  \d \varphi \nabla^i \d \sigma - \d \sigma \nabla^i \d \varphi \, |  \,  \mathcal H_i)  \,,
\end{aligned}
\end{equation}
these relations realize, in an algebraic way, the closure of a set of vector fields embedded in a foliated 4-dimensional spacetime. In particular, the first relation just means that the commutator of two spatial diffeomorphisms is identical to the diffeomorphism generated by the Lie bracket between the two smearing vector fields. The second relation means that the evolution of two infinitesimally-close points from one spatial hypersurface to the next one in the foliation, give two points which are on the same hypersurface. The last relation, similarly, implies that if we evolve the same point in two steps by the same two amounts, the final point will be on the same hypersurface (but not necessarily the same) no matter what the order in which we perform these steps. These relations are pictorically represented in Fig.~\ref{DiracAlgebroid}.
\begin{figure}[t!]
\begin{center}
\includegraphics[width=0.3\textwidth]{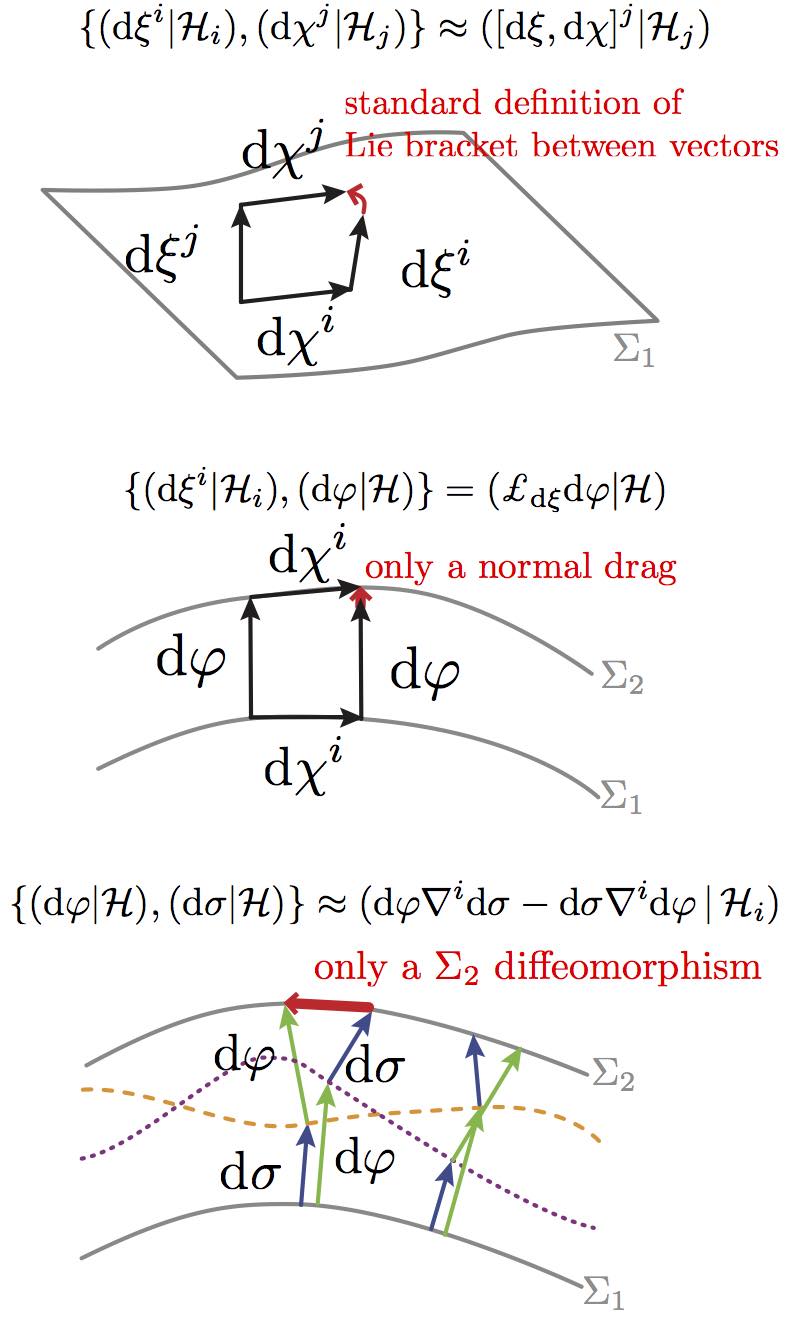}
\end{center}
\caption[The Dirac algebroid]{Pictorical representation of the hypersurface deformation algebra (originally appeared in my paper~\cite{EdFlavioPaper} with E. Anderson).\index{hypersurface deformation algebra}}\label{DiracAlgebroid}
\end{figure}
Hojman--Kucha{\v{r}}--Teitelboim assumed the relations~(\ref{HypersurfaceDeformationAlgebra}) to signify that the evolving 3-geometry described by the theory generates a consistent Lorentzian 4-geometry, and asked: `what are the possible forms that four constraints, functionals of $g_{ij}$ and $p^{ij}$, can take and close the Poisson algebra~(\ref{HypersurfaceDeformationAlgebra})?'. Their answer was that, with a few caveats, the most generic set of constraint on the ADM phase space which satisfy the relations~~(\ref{HypersurfaceDeformationAlgebra}) are the ADM Hamiltonian and diffeomorphism constraint. They titled their paper `Geometrodynamics Regained', clarifying their intention of finding a construction principle for ADM gravity which does not depend on General Relativity as its starting point. They write (p. 89):
\begin{quotation}
\it
We start by looking back instead of forward-looking back on how Einstein's law of gravitation was discovered and then placed on a pedestal of Òfirst principlesÓ (Section 1) and Iooking back on how geometrodynamics was derived from the Einstein law of gravitation in the late fifties (Section 2). After that we build a new pedestal. Its base is the set of deformations of a spacelike hypersurface embedded in an arbitrary Riemannian spacetime (Section 3). or rather a set of vector fields that generate those deformations. 
\end{quotation}

\index{hypersurface deformation algebra}
In~\cite{Gomes:2013naa}, eloquently titled `Conformal Geometrodynamics Regained', H. Gomes showed that the constraints of ADM gravity in maximal slicing can be uniquely derived (again, modulo a few caveats) from the assumption of `symmetry doubling'. He assumed spatial diffeomorphisms as the fundamental symmetry, and asked `what is the most general pair of constraints $\chi_1$, $\chi_2$ which separately close two first-class systems with the diffeomorphism constraint $\mathcal H_i$, and are proper gauge-fixing of each other (\emph{i.e.} they are completely second-class, up to a finite number of degrees of freedom)?'

It turns out the requirements of (i) being first class w.r.t. the diffeomorphism constraint and (ii) gauge-fix each other up to a finite-dimensional kernel are not sufficiently strong to uniquely fix the form of $\chi_1$ and $\chi_2$. But most of the constraint pairs that satisfy conditions (i) and (ii) do not have any dependence on the trace-free degrees of freedom of the momenta. This means that  the dynamics generated by an eventual  reparametrization constraint  (leftover from either element of the dual pair) will evolve \emph{spatial geometry  only through a changing volume form}. On the grounds that we are looking for a theory that has two \emph{dynamical}, propagating metric degrees of freedom,  one should  exclude pairs that have in them only the potential for a dynamical scale factor. We therefore add the requirement (iii) that the set of constraint represent a theory with the correct number of physical dynamical propagating degrees of freedom for gravity (transverse-traceless modes). With requirement (iii), modulo a few assumptions that are necessary in order to get a definite result, one obtains only two possible solutions.

The first simplifying assumption is that the constraints are local (finite number of derivatives), and they are polynomial in the metric $g_{ij}$, its inverse $g^{ij}$, the square root of its determinant  $\sqrt g$ and the momenta $p^{ij}$. This is a somewhat reasonable assumption that is often made in this sort of discussions, however I should point out that it excludes many non-analytic expressions which could represent perfectly legitimate constraints.

The second assumption is that the constraints contain spatial derivatives $\partial_j$ up to fourth order, and momenta $p^{ij}$ up to second order, and mixed terms up to second order in spatial derivatives and first order in momenta. In order to make the constraints first-class w.r.t. diffeomorphisms they will have to be polynomials in the Ricci tensor $R_{ij}$ (the Riemann tensor $R^i_{jk\ell}$ in 3D is just a function of the Ricci tensor\footnote{This is due to the \emph{Ricci decomposition:} $R_{ijk\ell} = - \frac R {(n-1)(n-2)} \left( g_{ik} g_{j\ell} - g_{i\ell} g_{jk} \right) + \frac 1 {n-2} \left( R_{ik} g_{j\ell} + R_{j\ell} g_{ik} - R_{i\ell} g_{jk} - R_{jk} g_{i\ell} \right) + W_{ijk\ell}$, where $n$ is the dimension and $W_{ijk\ell}$ is the Weyl tensor, which vanishes identically in 3D.}), the momenta and the metric and its inverse, and it will have to be homogeneous of degree one in the momenta and $\sqrt{g}$, so that it is a density of weight $1$. The most generic expression which satisfies all these conditions is:
\begin{equation}
\begin{aligned}
 \chi[\alpha,\beta,\vec \mu,\Lambda,j,k,l,a,b,c) := \left(\alpha ~ R^{ij}R_{ij}+\beta ~ \nabla^2R+ \sum_n\mu_n ~ R^n - 2 \Lambda \right)\sqrt g\\
+ j ~ \nabla^2 p + k ~ R^{ij} p_{ij} +l~R \, p +\frac{a \, p^{ij} p_{ij}+b \, p^2}{\sqrt g}+ c ~ p \,.
\end{aligned}
\end{equation}
 Note that the term $\nabla_i \nabla_j p^{ij}$ does not appear due to the diffeomorphism constraint, and the contracted derivatives of the Ricci tensor don't appear, since $\nabla_i\nabla_j R^{ij}=\frac{1}{2}\nabla^2R$ by the Bianchi identity.
The first main result of~\cite{Gomes:2013naa} then is thatÊ$\chi$ weakly commutes with itself for the following (non-exclusive) five families of coefficients: 

\begin{center}
\begin{tabular}{lccccccccccc}
\hline\hline
family  & $  \alpha $ & $  \beta $ & $  \mu_2 $ & $  \mu_1Ê$ & $  \Lambda $ & $  j $ & $  k $ & $  lÊ$ & $  a $ & $  b $ & $  c $
\\
\hline
 {\it I}  & $  \alpha $ & $  \beta $ & $  \mu_2 $ & $  \mu_1 $ & $  \Lambda $ & $  0 $ & $  0 $ & $  0 $ & $  0 $ & $  0 $ & $  0 $ 
\\
  {\it II} & $ 0 $ &  $  0 $ &Ê $  0 $ & $  0 $ & $  \Lambda $ & $  0 $ & $  0 $ & $  0 $ & $  a $ & $  b $ & $  c $
\\
 {\it III}Ê & $  \alpha $ & $  3\alpha+8\mu_2 $ & $  \mu_2 $ & $  \mu_1 $ & $  \Lambda $ & $  0 $ & $  0 $ & $  0 $ & $  0 $ & $  b $ & $  c $
\\
 {\it IV} & $  0 $ & $  0 $ & $  0 $ & $  \mu_1 $ & $  \Lambda $ & $  0 $ & $  0 $ & $  0 $ & $  a $ & $  -a/2 $ & $  c $
\\
 {\it V}  & $  0 $ &  $  0 $ & $  0 $ & $  0 $ & $  0 $ & $  j $ & $  0 $ & $  -8j $ & $  0 $ &  $  0 $  & $  0 $
\\
\hline\hline
\end{tabular}
\end{center}
\index{hypersurface deformation algebra}

The second, and central, result of~\cite{Gomes:2013naa} consists in the proof that there are only two possible \emph{pairs} of constraints $\chi_1$, $\chi_2$ belonging to the families above which gauge-fix each other. One is $\chi_1 \in$  {\it IV}  and $\chi_2 \in$  {\it II} (with $a=b=0$):
\begin{equation}\label{sol1_confgeomregained}
\left\{ 
\begin{aligned}
&\chi_1 =   (\mu_1 \, R - 2 \Lambda )\sqrt g  + {\sfrac{a}{\sqrt g}} \left(p^{ij} p_{ij}Ê- {\sfrac 1 2} \, p^2\right)+ c \, p
\\
&\chi_2 =   c' ~ p - 2 \Lambda'\sqrt g
\end{aligned}\right.\,,
\end{equation}
and the other is  $\chi_1 \in$  {\it IV}  and $\chi_2 \in$  {\it V}:
\begin{equation}
\left\{ 
\begin{aligned}\label{sol2_confgeomregained}
&\chi_1 =  (\mu_1 \, R - 2 \Lambda )\sqrt g  + {\sfrac{a}{\sqrt g}} \left(p^{ij} p_{ij}Ê- {\sfrac 1 2} \, p^2\right)+ c \, p
\\
&\chi_2 =  j' \left( \nabla^2 - 8  R \right) p
\end{aligned}\right.\,.
\end{equation}

The solution~(\ref{sol1_confgeomregained}) corresponds to ADM in CMC gauge with a possible additional term linear in the trace of the momentum (which we can exclude if we add the requirement of parity- or time-reversal-invariance of $\chi_1$. In fact the constraints can always be rescaled by a constant factor, which removes two of the parameters:
\begin{equation}
\left\{ 
\begin{aligned}
&\chi_1 =   ( R - 2 \Lambda )\sqrt g  + {\sfrac{x}{\sqrt g}} \left(p^{ij} p_{ij}Ê- {\sfrac 1 2} \, p^2\right)+ y \, p
\\
&\chi_2 =   p - \langle p\rangle \sqrt g
\end{aligned}\right.\,,
\end{equation}
and the parameter $x$ can be reabsorbed into a definition of the speed of light, which is a pure convention.

Solution~(\ref{sol1_confgeomregained}) corresponds to ADM in a gauge where the Hubble parameter is a conformal harmonic function. In many cases, such as if the metric is of positive Yamabe class, it reduces to maximal slicing gauge. In fact, due to the discrete distribution of eigenvalues of the usual Laplacian, this gauge is also just maximal slicing for a generic metric of Yamabe type strictly negative.
\index{Yamabe invariant}

\newpage

\section{Solutions of Shape Dynamics}\label{SecSolutionsOfSD}

I now turn to some examples of solutions of SD. For practical purposes, instead of working at the level of the Linking Theory, it is convenient to simply work in the conformal gauge in which SD is equivalent to GR in CMC gauge. We will therefore be studying solutions of Arnowitt--Deser--Misner gravity in CMC foliation. When such solutions exist, they are both solutions of GR and SD. However there are situations in which such solutions do not correspond to a well-defined solution of Einstein's equations, in particular at the Big-Bang singularity and around collapsed matter. However, by looking at the conformally-invariant degrees of freedom, one can check whether, as solutions of SD, 
they still make sense and can be continued past such breakdown points. My strategy is clear: I want to work with ADM gravity in CMC gauge as long as it is possible, and then focus on the shape degrees of freedom when my solutions evolve into something that cannot be described in GR.

The difficulty of solving the LY equation to define the SD Hamiltonian (\ref{FinalSDconstraints}) is in general enormous, and some simplifying assumptions are necessary. However, if the these assumptions are too strong one risks to completely trivialize the model into something with no genuine shape degrees of freedom. This is the case, for example, of a maximally symmetric spatial manifold, like the 3-sphere. The 3-sphere is but a point in shape space: the point corresponding to \emph{locally conformally flat} manifolds. The only degree of freedom that the 3-sphere has is its volume, which is not an SD observable, because it is not conformally invariant. This degree of freedom is however \emph{volume-preserving} conformally invariant, and it plays the role of SD Hamiltonian, generating evolution in York time. \index{York time} Such Hamiltonian is completely trivial, because there are no shape degrees of freedom to evolve.\footnote{However such a trivial solution makes sense as a background over which small fluctuations of geometric (gravitational waves) and matter degrees of freedom propagate, without disturbing too much the background geometry. This is analogous to homothetic collapse in the N-body problem: the shape of the $N$-body system never changes, so there is no objective change in the sense that the solution is a mere point in shape space. But we can use such a solution as a background solution to the $(N+m)$-body problem, in which $m$ particles are much lighter than the remaining $N$. The $N$ particles fall homothetically to the centre, and the other $m$ particles evolve over this background, perturbing it only slightly. Then there are  shape degrees of freedom that undergo genuine change.} 

The simplest generalizations of the maximally-symmetric case involve either reducing the degree of symmetry until some genuine shape degrees of freedom appear, or adding matter sources which contain some physical degrees of freedom.
I will examine the first case in the next Section, \ref{SecBianchiIX}. By assuming only homogeneity while relaxing the isotropy assumption, we obtain a two-parameter family of conformal geometries. These are known as `Bianchi IX universes', from the classification of homogeneous 3-geometries due to Luigi Bianchi~\cite{Bianchi1898}. In particular, Bianchi IX universes include all possible homogeneous geometries with $S^3$ topology, and for this reason they represent the most general background around which to expand cosmological solutions of SD with that topology (which is also the simplest closed topology in 3D). \index{closed spacelike hypersurfaces} This model is studied in detail in the next Sec.~\ref{SecBianchiIX}.

By renouncing homogeneity but not isotropy one gets a spherically-symmetric space, which is left invariant by an $SO(3)$ group of rotations, and has at least one pole that is stabilized by the group. If the space is compact, as we ought to assume, and with $S^3$ topology, the stabilizing poles are two.\footnote{Curiously such a geometry has been anticipated (or should we say \emph{invented?}) by Dante in his Divine Comedy. Dante describes a universe which has two poles, one centred on Jerusalem and the other on God, and any straight line emanating from one pole in any direction will necessarily end up at the other pole. The interpretation of Dante's cosmology in the Comedy as a hypersphere has originally been proposed By Andreas Speiser in 1925 (\cite{SpeiserBook}, pag. 53--54).}

Interestingly, renouncing homogeneity is not enough to introduce shape degrees of freedom: every spherically-symmetric 3-manifold is \emph{locally conformally flat}, which means that a conformal transformation can make the metric diffeomorphic to that of the 3-sphere (in the neighbourhood of any point). Therefore if the topology is assumed to be $S^3$, the conformal geometry is always conformally flat and equivalent to that of $S^3$ (this is proven in Sec.~\ref{SphericallySymmetricSec} below). We are then forced to introduce some matter degrees of freedom in order to obtain meaningful shapes. The simplest possibility is pressureless dust: a large number of vanishingly small particles, distributed in a spherically-symmetric way, all with equal radially-pointing momenta. One can choose the radial profile of such a distribution, and the simplest possibility is to have an infinitely thin `shell', which can be represented as distribution-valued sources for the Hamiltonian and diffeomorphism constraints. I will study this dynamical system to the extent that is possible analytically, without resorting to numerics, in Sec.~\ref{SphericallySymmetricSec}.

\subsection{Homogeneous solutions: Bianchi IX} \label{SecBianchiIX}

\index{homogeneous metrics} \index{Bianchi IX solutions}
The most generic homogeneous metric on the 3-sphere can be written as
\begin{equation}\label{BianchiIXTriads}
g_{ij} = \sum_{a,b=1}^3 q_{ab} \, \sigma^a_i \, \sigma^b_j \,,
\end{equation}
using the translation-invariant one forms~\cite{MTWbook}
\begin{equation}\label{Translation-invariant-one-forms-S3}
\begin{aligned}
\sigma^x &=  \sin r \, \d \theta - \cos r \, \sin \theta \, \d \phi \,,
\\
\sigma^y &=  \cos r \, \d \theta + \sin r \, \sin \theta \, \d \phi \,,
\\
\sigma^z &=  - \d r   - \cos   \theta \, \d \phi \,,
\end{aligned}
\end{equation}
which define an involutive distribution (an integrable system) because $\d \sigma^a = \frac 1 2 \epsilon^{abc} \sigma^b \wedge \sigma^c$. Notice that
\begin{equation}\label{DefOfDetG}
\det g = \sin^2 \theta ~ \det q \,.
\end{equation}
The vector fields dual to $\sigma^a_i$ are
\begin{equation}\label{Translation-invariant-vectors-S3}
\begin{aligned}
\chi_x &=  \cos r \, \cot \theta \, \frac{\partial}{\partial r} +\sin r \, \frac{\partial}{\partial \theta} - \cos r \, \csc \theta \, \frac{\partial}{\partial  \phi} \,,
\\
\chi_y &=  - \sin r \, \cot \theta \frac{\partial}{\partial r} +  \csc \theta  \left( \cos r \frac{\partial}{\partial \theta} + \sin r \, \frac{\partial}{\partial \phi } \right)\,,
\\
\chi_z &=  - \frac{\partial}{\partial r } \,.
\end{aligned}
\end{equation}
and can be used to define the momenta conjugate to $g_{ij}$:
\begin{equation}
p^{ij} = | \sigma^x \wedge \sigma^y \wedge \sigma^z | \sum_{a,b=1}^3 p^{ab} \, \chi^i_a \,\chi^j_b \,.
\end{equation}\index{symplectic structure} 
The pre-symplectic potential is canonical (except it gives double weight to off-diagonal terms),
\begin{equation}
\Theta = \int \d r \d \theta \d \phi ~ p^{ij} \, \delta g_{ij} = 4 \pi^2  \sum_{a=1}^3 p^{ab}  \, \delta q_{ab} \,,
\end{equation}
and the Hamiltonian constraint of the Arnowitt--Deser--Misner formulation of GR, smeared with a translationally-invariant (and therefore spatially constant) lapse $N$, 
reads\index{homogeneous metrics} \index{Bianchi IX solutions}
\begin{equation}
\begin{aligned}
(\mathcal H | N) = & N \, \int \d r \d \theta \d \phi \left[  \sqrt g \, R  -{\sfrac 1 {\sqrt g}} \left( p^{ij} p_{ij} - {\sfrac 1 2} p^2 
 \right)  \right]  \\
=& \frac {4 \pi^2 N} {\sqrt{|\det q|}}  \left[  p^{ab} q_{bc} p^{cd} q_{da} - {\sfrac 1 2} (p^{ab} q_{ab})^2  
+  \text{tr} q^2 - {\sfrac 1 2} (\text{tr} q)^2\right] \,.
\end{aligned}
\end{equation}
The diffeomorphism constraint needs to be smeared with a left-invariant shift,
which can be generically written as the linear combination $N^i = \xi^a \, \chi^i_a$, where $\xi^a$ are three real numbers. Then we have
\begin{equation}
(\mathcal H_i | N^i ) = -2 \int \d r \d \theta \d \phi \left(  p^{ij} \, \nabla_i \xi_j \right)
= 4 \pi^2 \epsilon_{ab}{}^c \xi^a p^{bd}q_{dc} \,,
\end{equation}
the three independent constraints implied by varying with respect to $\xi^a$ can be simply rewritten as
\begin{equation}\label{PandQcommute}
{\sfrac 1 2} \, [p,q]^a{}_b = 0 \,, 
\end{equation}
that is, the $p$ and $q$ matrices have to commute.
A gauge-fixing of~(\ref{PandQcommute}) is to impose that $q_{ab}$ is diagonal: $q_{xy} = q_{yz} = q_{xz} =0$: its Poisson brackets with (\ref{PandQcommute})  give
\begin{equation}
\begin{aligned}
&\{ [p,q]^y{}_z , q_{xy}  \} = - q_{xz} \approx 0 \,,&
&\{ [p,q]^x{}_z , q_{xy}  \} =  - q_{yz} \approx 0  \,,&
\\
&\{ [p,q]^x{}_y , q_{yz}  \} =  q_{xz} \approx 0 ,&
&\{ [p,q]^x{}_z , q_{yz}  \} =  - q_{xy} \approx 0  \,,&
\\
&\{ [p,q]^x{}_y , q_{xz}  \} = - q_{yz} \approx 0 \,,&
&\{ [p,q]^y{}_z , q_{xz}  \} =    q_{xy} \approx 0 \,,&
\end{aligned}
\end{equation}
and
\begin{equation}
\begin{aligned}
&\{ [p,q]^x{}_y , q_{xy}  \} =  (q_{xx} - q_{yy} ) \,,&
\\
&\{ [p,q]^y{}_z , q_{yz}  \} =  (q_{yy} - q_{zz} )  \,,&
\\
&\{ [p,q]^x{}_z , q_{xz}  \} =  (q_{xx} - q_{zz} ) \,,&
\end{aligned}
\end{equation}
we can see that this gauge-fixing is good only away from the hyperplanes $(q_{xx} = q_{yy})$, $(q_{xx} = q_{zz})$ and $(q_{yy} = q_{zz})$.

\subsubsection*{Diagonal model}

We can now introduce the Ashtekar--Henderson--Sloan variables~\cite{AHS} as
\begin{equation}\label{AHSvariables}
\begin{aligned}
&p^{ab} = \text{diag} ( P_x / C_x , P_y / C_y , P_z / C_z)  \,, &    & q_{ab} = \text{diag} ( C_x , C_y , C_z)   \,, & 
\end{aligned}
\end{equation}
then the symplectic form \index{symplectic structure}  and the Hamiltonian constraint look like (after a trivial rescaling, and introducing $P=\sum_a P_a$, $C = \sum_a C_a$):
\begin{equation}
\Theta = \sum_{a=1}^3\frac{ P_a  \,\delta C_a}{C_a} \,, \qquad 
 \sum_{a=1}^3 (P_a)^2 - {\sfrac 1 2} P^2 
 + \sum_{a=1}^3 C_a^2 - {\sfrac 1 2} C^2  = 0\,.
\end{equation}

\index{homogeneous metrics} \index{Bianchi IX solutions}
We can make a coordinate transformation that simultaneously diagonalizes the kinetic energy and symplectic form and separates scale and shape degrees of freedom. This is based on Jacobi coordinates for the $P_a$ variables 
and for the log of the $C_a$ variables (we can take their square roots and their logs because $C_a >0$ $\forall a$),
\begin{equation}
\begin{aligned}
& p_1 = \sqrt{\frac 2 3}  \left( {\ts P_z - \frac{P_y - P_x}{2} } \right) \,, &
& p_2 = {\ts \frac{P_y - P_x}{\sqrt 2}} \,, &
& D = {\ts \frac{P_x + P_y + P_z}{3}} \,, &
\\
& q^1 = \sqrt{\frac 2 3} \log  \left(  \frac{C_z}{\sqrt{C_y C_x}}\right) \,, &
& q^2 = \frac{1}{\sqrt 2}  \log \left(\frac{C_y}{C_x} \right)\,, &
& \alpha =  \left( C_x C_y C_z \right)^{\frac 1 3} \,, &
\end{aligned}
\end{equation}
we see that the $a$ and $b$ variables are scale-invariant, while $\alpha$ is not.
The inverses of the above relations are:
\begin{equation}\label{InvertedAHSvariables}
\begin{aligned}
& P_x= -\frac{p_2}{\sqrt{2}}-\frac{p_1}{\sqrt{6}} + D \,, &
&  P_y= \frac{p_2}{\sqrt{2}}-\frac{p_1}{\sqrt{6}}+ D \,, &
&  P_z= \sqrt{\frac{2}{3}} p_1 + D  \,, &\\
& C_x= \alpha \,  e^{- q^2/\sqrt{2} - q^1/\sqrt{6} }   \,, &
&  C_y= \alpha \, e^{ q^2/\sqrt{2} - q^1/\sqrt{6} }   \,, &
&  C_z= \alpha \, e^{ \sqrt{\frac{2}{3}} q^1 }  \,. &
\end{aligned}
\end{equation}
\index{homogeneous metrics} \index{Bianchi IX solutions}

\begin{figure}[t!]
\begin{center}
\includegraphics[width=0.35\textwidth]{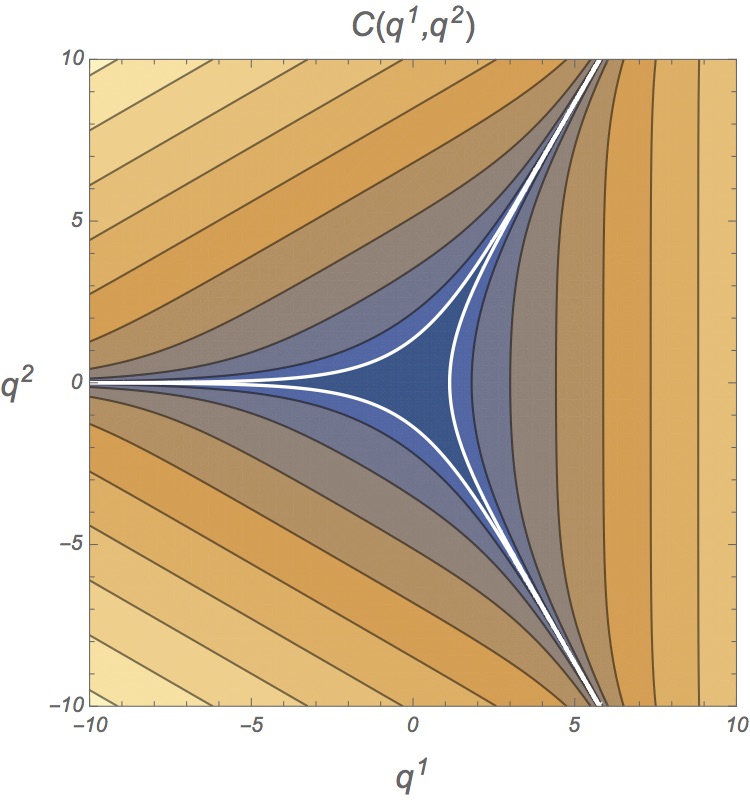}
\end{center}
\caption[Shape potential of Bianchi IX]{Shape potential of Bianchi IX. The curve in white is where the potential is zero. Inside it it is positive  (and it reaches its absolute maximum at $C(0,0) = 3/2$), and outside it is negative.}\label{ShapePotentialDiagram}
\end{figure}
The symplectic potential \index{symplectic structure}  now takes the following form:
\begin{equation}
\Theta = p_1 \, \delta q^1 + p_2 \, \delta q^2 + \frac{ 3 \, D \, \delta \alpha}{\alpha} \,,
\end{equation}
and the Hamiltonian constraint (choosing a lapse $N = \sqrt{| \det q |}/4 \pi^2$):
\begin{equation}\label{BIXHamiltonian1}
\mathcal H = {\sfrac 3 2} D ^2 - \left( p_1^2 + p_2^2 
\right) + \alpha^2 \, C(q^1,q^2) \,,
\end{equation}
where the `shape potential' $C(\vec q)$ (where $\vec q =(q^q,q^2)$) is
\begin{equation}\label{BIXpotential}
\begin{aligned}
C(\vec q) &:=  F (2 q^1) + F (  \sqrt{3} q^2 - q^1 ) + F ( -\sqrt{3} q^2-q^1 ) \,,
\\
F(x) &:= e^{- x/\sqrt{6}} - {\sfrac 1 2} e^{2x/\sqrt{6}}\,.
\end{aligned}
\end{equation}
In Fig.~\ref{ShapePotentialDiagram} we plot the potential, highlighting its triangular symmetry.
It is convenient to make the symplectic form diagonal, \index{symplectic structure}  by introducing the York time, \index{York time} which is conjugate to the volume $v = \alpha^{3/2}$, and it is therefore equal to $\tau = 2 \frac{D}{v}$, because $ 3 D \frac{\delta \alpha}{\alpha} = 2 \frac{D}{v} \delta v$.
Then one can rewrite everything as\footnote{Notice that one can include a cosmological constant by adding a term $-2 \, v^2 \, \Lambda$ to $\mathcal H$.}
\begin{equation}\label{YorkTimeBIXHamiltonian}
\Theta = p_1 \, \delta q^1 + p_2 \, \delta q^2 + \tau \, \delta v \,,
\qquad
\mathcal H = {\sfrac 3 8} v^2 \, \tau^2 - \left( p_1^2 + p_2^2 
 \right) + v^{4/3} \, C(\vec q) \,.
\end{equation}

The potential is exponentially steep. This means that if we start from a point in phase space where it is comparable to the kinetic energy $(p_1^2 + p_2^2 
)$ and move `downhill', it will become very quickly vanishingly small compared to the kinetic energy. This implies that the motion is, during most of the orbit, that of a free particle in the $(q^1,q^2)$-plane.  This free motion corresponds to a Kasner solution, in which, as the total volume shrinks, the metric is always growing in in one direction but contracting in the other two~\cite{AHS}. Only when the particle climbs up enough in the potential well, it will feel again the force and will bounce quite sharply against the potential walls~\cite{MTWbook}, and the Kasner exponents will be shuffled. Belinsky--Khalatnikov--Lifshitz~\cite{BKL} call these transitions \emph{Kasner epochs}. The resulting behaviour is similar to that of a billiard ball in a triangular pool table.
However, the potential is time-dependent, because its coupling is the mo\-no\-to\-ni\-cal\-ly-decreasing quantity $v^{4/3}$.  So the bounces of the particle against the potential walls are not elastic: the particle loses some kinetic energy with each bounce. 

I will now show what is known of the dynamics of this system in the usual language, which regards the volume and York time as dynamical variables on the same footing as the shape degrees of freedom. After that I will show how one can completely describe the system by referring only to shape space.

%

\subsubsection*{Kasner regime}
\index{homogeneous metrics} \index{Bianchi IX solutions}\index{Kasner epochs}\index{Bianchi I solutions}

Consider now the dynamics of the system far from the potential wells, in the Kasner regime. In this situation the term $v^{4/3} V_\st{S}$ in the Hamiltonian can be ignored, and, using the Ashtekar--Henderson--Sloan variables, we can write 
\begin{equation}
 \sum_{a=1}^3 (P_a)^2 - {\sfrac 1 2} P^2 
 \simeq 0 \,.
\end{equation}
The term $ \sum_a (P_a)^2 - {\sfrac 1 2} P^2$ is non-positive. Now, the region $\sum_a (P_a)^2 - {\sfrac 1 2} P^2 \leq 0$ is entirely contained in the two octants in which  all three $P_a$'s have the same sign, because it is bounded by a cone with axis of symmetry parallel to $(1,1,1)$, which is inscribed in the boundaries of the two octants. So either all the $P_a$'s are positive or they are all negative. The two cases are equivalent: one corresponds to an expanding universe and the other to a contracting one, and they are trivially related by time-reversal. So, without loss of generality, we can choose a sign for the $P_a$'s, for example the positive one.
\begin{figure}[t!]
\begin{center}
\includegraphics[width=0.3\textwidth]{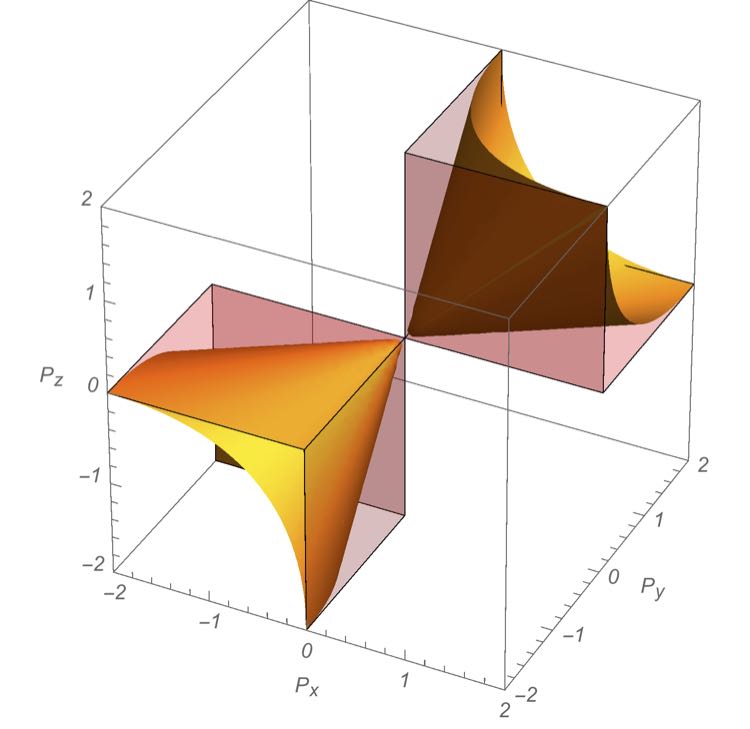}
\end{center}
\caption[On-shell surface in Kasner space]{Surface $\sum_a (P_a)^2 - {\sfrac 1 2} P^2=0$ in the $(P_x,P_y,P_z)$ space, in yellow. The surface is inscribed within the all-positive and the all-negative octants. The region $\sum_a (P_a)^2 - {\sfrac 1 2} P^2<0$ is bounded by the yellow surface.}\label{ConeFig}
\end{figure}
\index{homogeneous metrics} \index{Bianchi IX solutions}\index{Kasner epochs}\index{Bianchi I solutions}
The equations of motion, in Ashtekar--Henderson--Sloan variables, read
\begin{equation}\label{EOM-AHS}
\dot C_a = - C_a ( P -2 P_a) \,,
\qquad
\dot P_a =  C_a ( C -2 C_a) \,,
\end{equation}
and since we are far from the potential barriers, we can ignore the term $C_a ( C -2 C_a)$ so
 that  $\dot P_a  \simeq 0$. Now, calling: 
\begin{equation}
p_a = 1 - \frac{2 P_a}{P} \,,
\end{equation}
the solution of these equations are $p_a = \text{\it const}$, $C_a = e^{- P \, p_a (t - t_0)}$, so the metric  is
\begin{equation}\label{BIXmetricPsmallVariables}
g_{ij} = \sum_{a=1}^3 e^{- P \, p_a (t - t_0)} \, \sigma^a_i \, \sigma^a_j \,,
\end{equation}
and, after a choice of sign for the  $P_a$'s, as $P>0$, we see how the sign of $p_a$  determines whether the corresponding direction is contracting or expanding.
\index{homogeneous metrics} \index{Bianchi IX solutions}\index{Kasner epochs}\index{Bianchi I solutions}
\begin{figure}[t!]
\begin{center}
\includegraphics[width=0.3\textwidth]{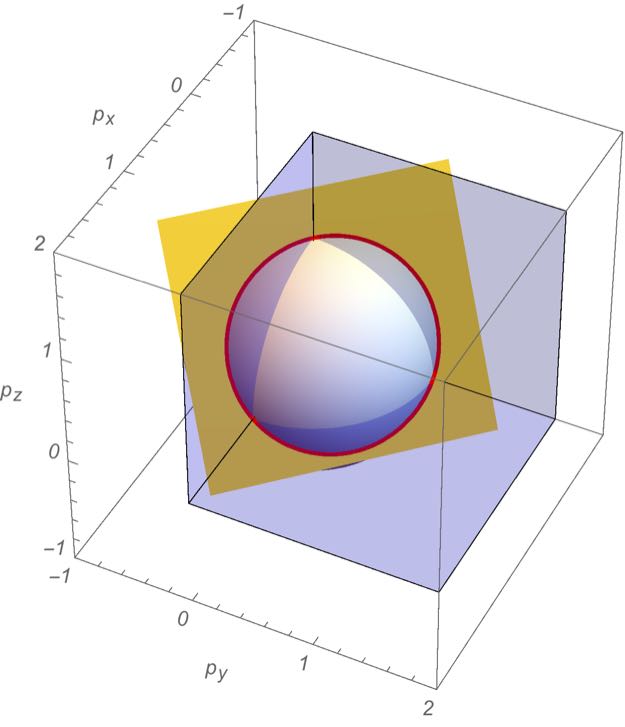}
\end{center}
\caption[Intersection between Kasner sphere and plane]{Intersection (in red) between the Kasner sphere $\sum_a  p_a^2 = 1 - p_\phi^2$ (white) and the Kasner plane $\sum_a  p_a = 1$ (yellow). The $p_x=0$, $p_y=0$ and $p_z=0$ planes are represented in transparent blue. The intersection is always outside of the region $p_x,p_y,p_z>0$, so one of the $p_a$'s has to be negative.}\label{KasnerIntersection}
\end{figure}
In the variables $p_a$ the Hamiltonian constraint becomes
\begin{equation}
\sum_{a=1}^3  p_a^2 \simeq 1  \,,
\end{equation}
which identifies a unit sphere in the space $(p_x,p_y,p_z)$, and the variables $p_a$ satisfy the further constraint
\begin{equation}
\sum_{a=1}^3  p_a = 1   \,.
\end{equation}
which identifies a plane.  The solution of the two constraints is therefore a circle, as can be seen in Fig.~\ref{KasnerIntersection}. As can be seen in the Figure, this circle lies in the three octants that share one face with the all-positive octant, and this implies that there is always one negative and two positive $p_a$'s. Therefore, inspection of the metric~(\ref{BIXmetricPsmallVariables}) reveals that there are always two contracting and one expanding directions. Similarly, in the  time-reversed case, one of the direction is always contracting while the other two are expanding.

\index{homogeneous metrics} \index{Bianchi IX solutions}\index{Kasner epochs}\index{Bianchi I solutions}

In the Kasner regime the motion in the shape plane $(a,b)$ is linear, like that of a free particle on the plane. During this phase, the dynamics of the volume too can be solved exactly:
\begin{equation}
\dot v = \frac{\partial \mathcal H}{\partial \tau } = {\sfrac 3 4} \, v^2 \, \tau \,,
\qquad
\dot \tau = - \frac{\partial \mathcal H}{\partial v } \simeq - {\sfrac 3 4} \, v \, \tau^2 \,,
\end{equation}
the quantity $D =  v \, \tau$ is conserved, $D(t) = D_0$, while the log-orthogonal variable $W =  \frac v \tau$ follows the equation
\begin{equation}
\dot W = {\sfrac 3 2} \, v^2 =   {\sfrac 3 2} \, D_0 \, W \,, ~~  \Rightarrow ~~ W = W_0 \, e^{ {\frac 3 2} \, D_0 \, t} \,,
\end{equation}
and inverting the definitions of $D$ and $W$ we get
\begin{equation}
v =  v_0 \,e^{ {\frac 3 4} \, D_0 \, (t-t_0)} \,, \qquad \tau = \tau_0 \,e^{ - {\frac 3 4} \, D_0 \, (t-t_0)} \,,
\end{equation}
so we see that the volume evolves exponentially in parameter time. Assuming, without loss of generality, that $D_0>0$, we notice that if the Kasner behaviour continued indefinitely as $t \to - \infty$ without any bounces, the volume would shrink to zero into a singularity. This would however take an infinite parameter time $t$. This is not the case, though, if we consider proper time. Recall that the lapse corresponding to parameter time is $N = \sqrt{| \det q |} /(4 \pi^2)$ (we used it to write the Bianchi IX Hamiltonian as in (\ref{BIXHamiltonian1}) and following equations), then $N = \frac{v}{4 \pi^2}$  and the 4-dimensional line element reads
\begin{equation}
\d s^2 =  -  \frac{v^2 \d t^2 }{16 \pi^4} + \sum_{a=1}^3 e^{- P \, p_a (t - t_0)} \, \sigma^a_i \, \sigma^a_j \,,
\end{equation}
now consider a timelike spacetime curve, e.g. $\psi =\theta=\phi =\text{\it const.}$, going from $t = -\infty$ to some finite $t = t_1$. The proper time measured on such curve is 
\begin{equation}
\Delta s = \int_{-\infty}^{t_1} \sqrt{\frac{v^2 \d t^2 }{16 \pi^4} } = \frac{v_0}{4 \pi^2} \int_{-\infty}^{t_1}  e^{ {\frac 3 4} \, D_0 \, (t-t_0) } \d t
=  \frac{e^{ {\frac 3 4} \, D_0 \, (t_1-t_0)}}{3 \, \tau_0 \, \pi^2}    < \infty \,.
\end{equation}
We conclude that, if the Kasner behaviour continued indefinitely without bounces up to the singularity,  it would take a finite proper time to reach the singularity $v \to 0$.
The actual solutions of Bianchi IX do not look like a single Kasner evolution up to the singularity, because of the bounces. I will consider, in the next part, the dynamics of a bounce, under some simplifying approximations which are very good in the vast majority of phase space (namely, when we are more than a few units away from the origin of shape space $q^1=q^2=0$). I will analyze a single bounce against one of the potential walls, and find an exact solution to the approximate equations of motion.  The solution tends to two different Kasner epochs in the two asymptotic ends, away from the bounce. After that I will study the dynamics of the volume, and find that the bounce does not change the fact that it takes a finite proper time to reach the singularity.
\index{homogeneous metrics} \index{Bianchi IX solutions}\index{Taub transitions}

\begin{figure}[t!]
\begin{center}
\includegraphics[width=0.35\textwidth]{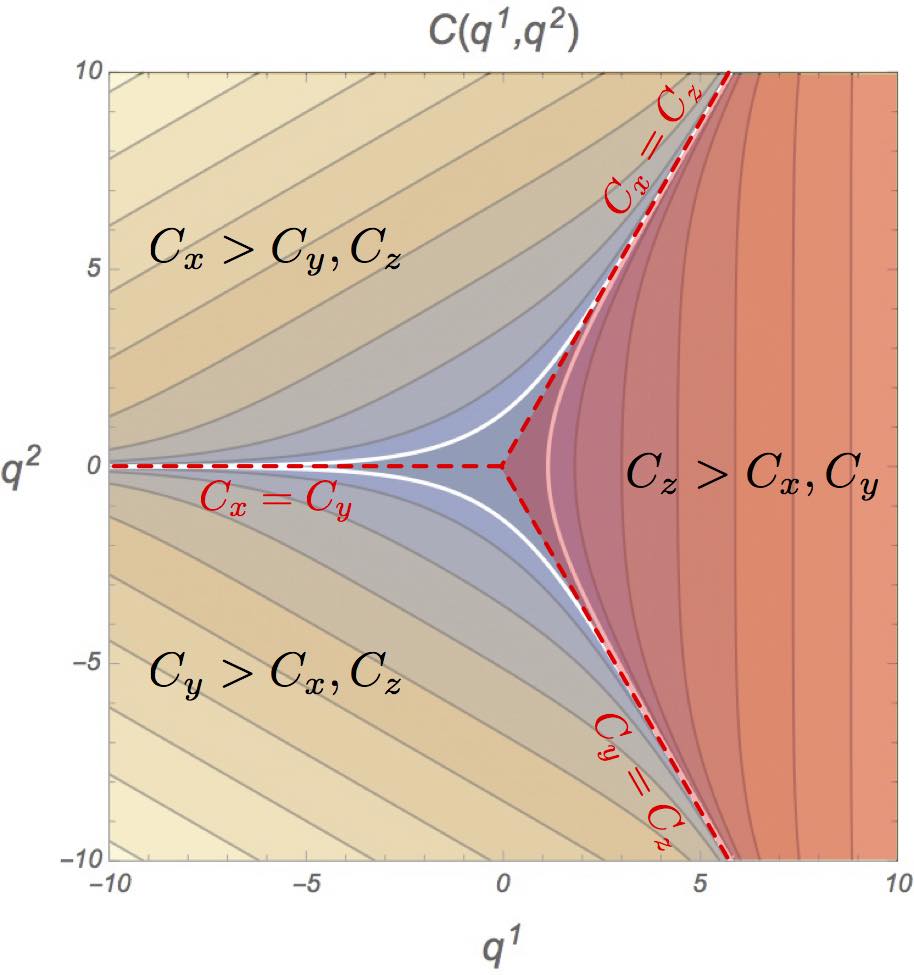}
\end{center}
\caption[Division of Bianchi IX shape space]{The Bianchi IX shape potential is symmetric under rotations of $120^\circ$ of the  $(q^1,q^2)$ space. These rotations correspond to discrete transformations of the manifold which exchange two of the axes. Shape space can thus be divided into three symmetric regions, and one can work in one of the three regions without loss of generality (except the particular case of the boundaries between the regions, indicated by the dashed red lines in the figure). Here I highlighted the region where $C_z>C_x,C_y$.}\label{BianchiIXShapeSpaceDivided}
\end{figure}

\index{homogeneous metrics} \index{Bianchi IX solutions}\index{Taub transitions}\index{Kasner epochs}\index{Bianchi I solutions}
\vspace{6pt}\subsubsection*{Taub transitions.} Assume now that $P_x > P_y >P_z \geq 0$ (which we can do without loss of generality) and that $C_z>C_x$, $C_z>C_y$ (that is, $\sqrt{2/3} \, q^1 > -  q^2/\sqrt{2}- q^1/{\sqrt{6}}$, $\sqrt{2/3} \, q^1  >  q^2/{\sqrt{2}}- q^1/{\sqrt{6}}$). The last condition just means that we are in one of the three regions in which shape space is divided by the three-fold symmetry of the shape potential (namely in the bottom one), see Fig.~\ref{BianchiIXShapeSpaceDivided}. Then, unless we are infinitesimally close to the boundaries $C_z=C_x$ or $C_z = C_y$, the variable $C_z$ is much larger than $C_x$ and $C_z$ -- because, as can be seen in the definitions~(\ref{InvertedAHSvariables}), they go like the exponentials of, respectively, $\sqrt{2/3} \, q^1$, $- q^2/{\sqrt{2}} - q^1/{\sqrt{6}}$ and $q^2/{\sqrt{2}}-q^1{\sqrt{6}}$.
Thus $C_x$ and $C_y$ can be neglected in the right-hand-side of the equations of motion~(\ref{EOM-AHS}), which reduce to
\begin{equation}
\begin{aligned}
&
\dot C_z = - C_z ( P -2 P_z) \,,
&
&
\dot P_z \sim  -C_z^2  \,,
&
&
\dot P_x \sim \dot P_y \sim 0  \,,
&
&
C_x \sim C_y \sim 0 \,.
&
\end{aligned}
\end{equation}
These equations can be solved exactly,
{\thinmuskip=0mu
\medmuskip=0mu
\begin{equation}
\begin{aligned}
P_z &= P_y + P_x + \sqrt{(P_y + P_x)^2+c} \left( \frac{(P_y + P_x)^2+c - e^{2(t-t_0) \sqrt{(P_y + P_x)^2+c}} }{(P_y + P_x)^2+c + e^{2(t-t_0) \sqrt{(P_y + P_x)^2+c}} } \right)\,,
\\
C_z &=  \frac{ 2 \left((P_y + P_x)^2+c\right) e^{(t-t_0) \sqrt{(P_y + P_x)^2+c}} }{e^{2(t-t_0) \sqrt{(P_y + P_x)^2+c}} + (P_y + P_x)^2+c} \,,
\end{aligned}
\end{equation}}
\begin{figure}[t!]
\begin{center}
\includegraphics[width=0.4\textwidth]{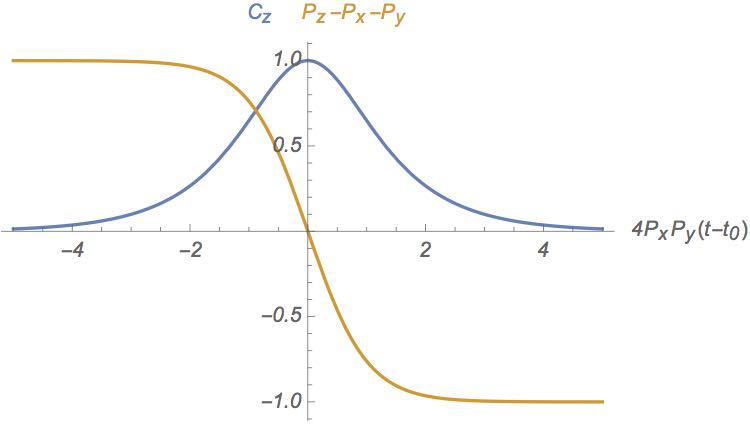}
\end{center}
\caption[Taub transition]{
Solution of the Bianchi IX dynamics in the Taub approximation $C_x \sim 0$, $C_y \sim 0$.
}\label{TaubDiagram}
\end{figure}
%
now, the integration constant $c$ can be fixed by imposing the constraint $\sum_a (P_a)^2 - {\sfrac 1 2} P^2 
\simeq 0$ asymptotically, for example when $t \to -\infty$. Then $c = - (P_y - P_x)^2 
$, and 
\begin{equation}\label{TaubSolution}
\begin{aligned}
P_z &= P_y + P_x  +  \sqrt{ 4 \,  P_y P_x 
} \left( \frac{4 \,  P_y P_x  
 - e^{2(t-t_0) \sqrt{4 \,  P_y P_x  
}} }{4 \,  P_y P_x  
+ e^{2(t-t_0) \sqrt{4 \,  P_y P_x  
}} } \right)\,,
\\
C_z &= \frac{2 \left(4 \,  P_y P_x  
\right)  e^{(t-t_0) \sqrt{4 \,  P_y P_x  
}} }{e^{2(t-t_0) \sqrt{4 \,  P_y P_x  
}} + 4 \,  P_y P_x  
} \,,
\end{aligned}
\end{equation}
so $C_z \to 0$ at both asymptotic ends, while the asymptotic values of $P_z$ are 
\begin{equation}
P_z (-\infty) = P_y + P_x +  \sqrt{ 4 \,  P_y P_x  
}  \,,
\qquad
P_z (+\infty) = P_y + P_x -  \sqrt{ 4 \,  P_y P_x 
}   \,.
\end{equation}
The two asymptotic values of $P_z$ correspond to the two values of $p_z$ on the intersection circle (red curve in Fig.~\ref{KasnerIntersection}) where it intersects with the plane $(1 - p_y)/P_y  = (1-p_x)/P_x$.
\index{homogeneous metrics} \index{Bianchi IX solutions}\index{Taub transitions}

We conclude that during a Taub transition $P_z$ switches from being the largest of the $P_a$'s to a smaller value, while $P_y$ and $P_x$ do not change significantly. The shape kinetic energy  then changes as
\begin{equation}
\Delta (p_1^2 +p_2^2)  =  -2 \left(\sqrt{P_x P_y} \left(P_x+P_y+P_z\right)+P_x P_y\right) \,,
\end{equation}
which is a negative quantity. As anticipated, the shape kinetic energy decreases after each bounce, because the potential is monotonically decreasing.

The $P_x$ and $P_y$ variables are conserved during the transition, but $C_x$ and $C_y$ are not. Their evolution however can be explicitly obtained by integrating their equations of motion:
\begin{equation}
\dot C_x = - C_x ( -P_x + P_y +P_z) \,, \qquad  \dot C_y = - C_y ( P_x - P_y +P_z) \,,
\end{equation}
where only $P_z$ has a time dependence given by Eq.~(\ref{TaubSolution}). The solution of the above equations is
\begin{equation}
\begin{aligned}
C_x = C^0_x \, e^{-2 \left(t-t_0\right) \left(\sqrt{P_x P_y}+P_y\right)} \left(e^{4 \left(t-t_0\right) \sqrt{P_x P_y}}+4 P_x P_y\right) \,,
\\
C_y = C^0_y \, e^{-2 \left(t-t_0\right) \left(\sqrt{P_x P_y}+P_x\right)} \left(e^{4 \left(t-t_0\right) \sqrt{P_x P_y}}+4 P_x P_y\right) \,.
\end{aligned}
\end{equation}
\index{homogeneous metrics} \index{Bianchi IX solutions}\index{Taub transitions}
We now have the dynamics of the volume during a Taub transition:
\begin{equation}
v = \sqrt{C_x C_y C_z}=e^{- \left( \sqrt{P_x P_y}+P_x+P_y \right)(t-t_0) } \sqrt{ e^{4 \sqrt{P_x P_y} (t-t_0)}+4 P_x P_y } \,,
\end{equation}
the above expression goes to zero as $t \to \infty$ and diverges as $t \to - \infty$. It is easy to see that $v$ is monotonic (its $t$-derivative is negative-definite). Moreover the proper time to reach the singularity is, again, finite, because $\int^\infty_{t_1} v \, \d t < \infty$. This proves that any solution of Bianchi IX will, at one time end, reach a singularity  in an infinite interval of parameter time $t$, but a finite interval of proper time. This  is often taken as an indicator that the singularity is physical: proper time normally coincides with the passage of time experienced by a comoving observer. However proper time is not fundamental in Shape Dynamics: time is always an `abstraction at which we arrive through the changes of things'. So, rather than proper time, what we should check is whether the singularity is approached after a finite amount of \emph{physical change,} that is, of change in shape space.
We can reformulate the question `is the singularity in the finite past?' as  `does it take a finite amount of change in shape to go back to the singularity?'. The answer is, as it turns out, no. In fact every Bianchi IX solution goes through an infinite amount of bounces before reaching the singularity. I will prove that in the following.
\index{homogeneous metrics} \index{Bianchi IX solutions}\index{Taub transitions}

\subsubsection*{Volume-time  description on shape space}

\index{homogeneous metrics} \index{Bianchi IX solutions}
The Hamiltonian~(\ref{YorkTimeBIXHamiltonian}) describes a point particle in the Euclidean plane under the influence of a potential, $v^{4/3} \, C$, whose `coupling constant' $v^{4/3}$ is dynamical. The equations of motion of the volume $v$ and its conjugate momentum, the York time $\tau$, \index{York time} are 
\begin{equation}
\begin{aligned}
\dot v &= \{ v , \mathcal H\} = {\sfrac 3 4} v^2 \, \tau \,,
\\
\dot \tau &= \{ \tau , \mathcal H\} = - {\sfrac 3 4} v \, \tau^2 - {\sfrac 4 3} v^{1/3} \, C
\approx - {\sfrac 1 4} v \, \tau^2  - {\sfrac 4 3} \left( p_1^2 + p_2^2 
\right) \, v^{-1} <0
\,,
\end{aligned}
\end{equation}
since the volume is non-negative $v \geq 0$, the right-hand-side of $\dot \tau$ is negative-definite, and therefore $\tau$ is monotonically decreasing. The volume, on the other hand, is piecewise monotonic whenever $\tau \neq 0$:
\begin{equation}
\dot{(v^{-1})} = - \frac{\dot v}{v^2}= - {\sfrac 3 4} \,\tau \,.
\end{equation}
The York time, being a monotonic function, can at most have one zero, which corresponds to maximal expansion/recollapse (because $v^{-1}$ has a minimum there: $\ddot{(v^{-1})} =- {\sfrac 3 4} \,\dot \tau >0$ and therefore $v$ has a maximum). So the volume is either always monotonic, or, in case there is a recollapse, it is monotonic before and after maximal expansion.
	
Determining the SD Hamiltonian is an algebraic problem: it is the volume $v = v(\tau,q^1,q^2,p_1,p_2)$ that solves the Hamiltonian constraint~(\ref{YorkTimeBIXHamiltonian}), which can be turned into a third-order equation in $v^{2/3}$.
The solution of a third-order differential equation can be written analytically, but it is a somewhat awkward expression, and moreover one has to take only the positive real roots.\footnote{By Descartes' rule of signs, if $C<0$ there is one and only one real positive root, while if $C>0$ there are either two or zero positive real roots.}
The SD Hamiltonian is a generator of evolution in York time, but we are not married to that particular parametrization: any dynamics that gives a curve in shape space is equally acceptable for SD, because the curve is all that is observable, regardless of the parametrization used. In this particular case we can take advantage of the fact that the volume is monotonic too (except for a single point of maximal expansion, possibly), and use $v$ as a parameter. The Hamiltonian that generates evolution in $v$ is the solution of Eq.~(\ref{YorkTimeBIXHamiltonian}) w.r.t. $\tau$:
\begin{equation}
H_\st{vol} =  \sqrt{{\sfrac 8 3}}\sqrt{ v^{-2} \left( p_1^2 + p_2^2  \right) - v^{-2/3} \, C} \,,
\end{equation}
where I chose the positive root, but any choice is equally valid as it corresponds to an arbitrary choice of time-direction. The above Hamiltonian generates a shape-space curve $q^1=q^1(v)$, $q^2=q^2(v)$, parametrized by $v$. The corresponding non-autonomous equations of motion are
\begin{equation}
\begin{aligned}
\frac{\d  \vec q }{\d v} & = \sqrt{{\sfrac 8 3}} \frac{v^{-2} \vec p}{\sqrt{ v^{-2} \left( p_1^2 + p_2^2  \right) - v^{-2/3} \, C}} \,,\\
\frac{\d \vec p}{\d v} & = - \sqrt{{\sfrac 2 3}} \frac{v^{-2/3} \, \vec \nabla C}{\sqrt{ v^{-2} \left( k_a^2 + k_b^2  \right) - v^{-2/3} \, C}} \,,
\end{aligned}
\end{equation}
\index{homogeneous metrics} \index{Bianchi IX solutions}
where $\vec \nabla C = (\partial_{q^1} C,\partial_{q^2} C)$.
The above description is still not entirely free of ambiguities: although the canonical variables $(q^1,q^2)$ are dimensionless, the conjugate momenta $(p_1,p_2)$ carry dimensions, as well as the parameter $v$. We can get rid of this by reparametrizing to a logarithmic time $\zeta = \log v/v_0$ (where we need to introduce a reference volume $v_0$, which makes the arbitrariness intrinsic in the dimensionful description explicit), and rescaling the momenta
by an appropriate factor of $v^{-2/3}$ which makes them dimensionless:~$\vec \pi = v^{-2/3} \vec p$. Any dimensionless quantity then nicely drops out of the equations of motion, which in turn become autonomous,
\begin{equation}\label{AutonomousVolumeTimeBIXeqs}
\begin{aligned}
\frac{\d  \vec q}{\d \zeta} & = \sqrt{{\sfrac 8 3}} \frac{\vec \pi }{\sqrt{ \pi_1^2 +  \pi_2^2  -  \, C}} \,,\\
\frac{\d \vec \pi}{\d  \zeta} & = - {\sfrac 2 3} \vec \pi -  \sqrt{{\sfrac 2 3}} \frac{\vec \nabla C}{\sqrt{\pi_1^2 + \pi_2^2  - C}} \,,
\end{aligned}
\end{equation}
and the above equations can be thought of as non-conservative equations
\begin{equation}\label{NonconservativeEOMBianchiIX}
\frac{\d \vec q }{\d \zeta}  = \frac{\partial H_\st{aut}}{\partial \vec \pi} \,,
\qquad
\frac{\d \vec \pi }{\d  \zeta}  = - {\sfrac 2 3} \vec \pi  -  \frac{\partial H_\st{aut}}{\partial \vec q}\,,
\end{equation}
whose Hamiltonian part is generated by
\begin{equation}
H_\st{aut} = \sqrt{\sfrac 8 3 } \sqrt{\pi_1^2 + \pi_2^2  -  C(\vec q)} \,.
\end{equation}
\index{homogeneous metrics} \index{Bianchi IX solutions}\index{Kasner epochs}

\subsubsection*{The bounces never end}

\begin{figure}[t!]
\begin{center}
\includegraphics[width=0.46\textwidth]{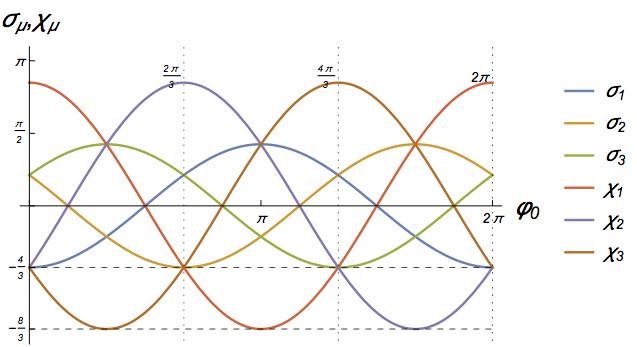}
\end{center}
\caption[Exponents in the Bianchi IX shape potential]{The exponents in the Bianchi IX shape potential during a Kasner epoch, as functions of the direction $\varphi_0$.}\label{BianchiIXexponentsPicture}
\end{figure}
Consider again the Kasner regime, in between two bounces: the shape potential energy is much smaller than the kinetic energy, $C \ll \pi_1^2 + \pi_2^2$, and the equations of motion~(\ref{NonconservativeEOMBianchiIX}) can be approximated as
\begin{equation}\label{BIXEOMnonquiescent}
\frac{\d \vec q}{\d \zeta}  \simeq \sqrt{{\sfrac 8 3}} \frac{\vec \pi}{\sqrt{ \pi_1^2 +  \pi_2^2 }} \,,
\qquad
\frac{\d \vec \pi}{\d  \zeta}  \simeq - {\sfrac 2 3} \vec \pi \,,
\end{equation}
and admit the following solution
\begin{equation}
\begin{aligned}
(\pi_1(\zeta),\pi_2(\zeta)) &= \pi_0  \, (\cos \varphi_0 , \sin \varphi_0)  \, e^{-\frac 2 3 \zeta} \,,
\\
(q^1(\zeta),q^2(\zeta)) &=  (q^1_0,q^2_0)  + \sqrt{{\sfrac 8 3}}  (\cos \varphi_0 , \sin \varphi_0) \, \zeta \,,
\end{aligned}
\end{equation}
where the integration constants $\pi_0$ and $\varphi_0$ determine, respectively, the initial magnitude of the shape momentum and its direction. $q^1_0 $ and $q^2_0$ determine the initial position of the point in shape space at $\zeta =0$. After sufficient volume-time $\zeta$ $q^1_0 $ and $q^2_0$  become irrelevant and can be ignored. \index{homogeneous metrics} \index{Bianchi IX solutions}
Now we can study the growth of the shape potential energy  $C[\vec q(\zeta)]$ (in which we ignore $q^1_0 $ and $q^2_0$) in comparison to the shape kinetic energy 
$(\pi_1^2 + \pi_2^2)$ during one Kasner epoch,
\begin{figure}[t!]
\begin{center}
\includegraphics[width=0.46\textwidth]{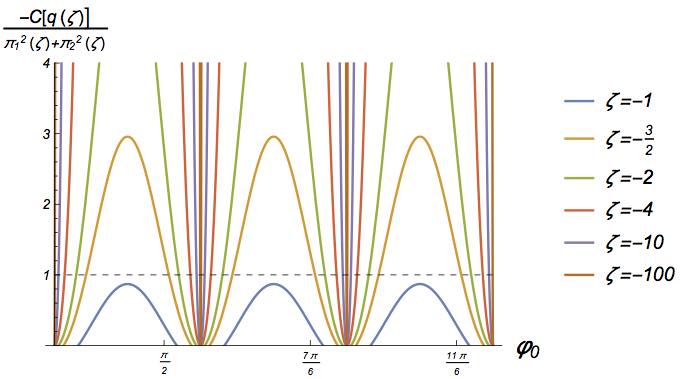}
\end{center}
\caption[Bianchi IX shape potential vs. shape kinetic energy]{The ratio between (minus) the shape potential energy $C(\vec q)$ and the kinetic energy $\pi_1^2 + \pi_2^2$ in a Kasner epoch, as a function of the direction of the motion of the particle $\varphi_0$, for different log-volume-times $\zeta$. As the singularity is approached $\zeta \to - \infty$, the potential is larger than the kinetic energy almost everywhere.}\label{BianchiIXPotVsKin}
\end{figure}
\begin{equation}\label{BIXKinAndPotEnergyDuringKasner}
\pi_1^2 + \pi_2^2  \simeq \pi_0^2 \, e^{-\frac 4 3 \zeta} \,, \qquad  C[\vec q (\zeta)]   \simeq \sum_{\mu=1}^3 \left(   e^{\sigma_\mu(\varphi_0) \, \zeta } - \frac 1 2  e^{\chi_\mu(\varphi_0) \, \zeta } \right) \,,
\end{equation}
where
\begin{equation}\label{BIXSigmaAndChiFunctions}
\begin{aligned}
&\sigma_1 = - {\sfrac 4 3}      \cos \varphi_0  ,  & &
\chi_1 = {\sfrac 8 3}      \cos \varphi_0  ,
\\
&  \sigma_2 = { \sfrac 2 3}   \left(  \cos \varphi_0 - {\sqrt 3} \sin \varphi_0 \right)  , & 
&\chi_2 =  {\sfrac 4 3}   \left(  \sqrt{3} \cos \varphi_0  - \sin \varphi_0 \right) , &
\\
& \sigma_3 =  { \sfrac 2 3}  \left(   {\sqrt 3}  \sin \varphi_0 + \cos \varphi_0\right)  , & &
\chi_3 =   - {\sfrac 4 3}   \left(\sqrt{3} \sin \varphi_0 +  \cos \varphi_0 \right) .&
\end{aligned}
\end{equation}
The six functions above are plotted in Fig.~\ref{BianchiIXexponentsPicture}. We can compare the potential and the kinetic energy at different values of $\zeta$, from $\zeta =0$ to $\zeta \to -\infty$, which corresponds to the singularity. In Fig.~\ref{BianchiIXPotVsKin} we see the ratio between $C(\vec q(\zeta))$ and $\pi_1^2 + \pi_2^2$ at different $\zeta$'s.  One can clearly see how the values of $\varphi_0$ for which this ratio is smaller than $1$ reduce to three point as $\zeta \to - \infty$, the points $\varphi_0 = {\frac 2 3} \pi  , {\frac 4 3}  \pi , 2 \pi$. 
We then proved that during a Kasner epoch, whatever the direction $\varphi_0$ of the motion\footnote{Except if the motion is along one of the three axes $\varphi_0 = {\frac 2 3} \pi  , {\frac 4 3}  \pi , 2 \pi$, but this is a measure-zero set of solutions.}, sooner or later the shape potential will grow larger than the shape kinetic energy, and a bounce will take place. We conclude that an infinite amount of bounces take place before the singularity is reached.
Even though it takes a finite amount of proper time to reach the singularity, any measure of time based on change of physical (shape) observables will be infinite, pushing the singularity into the infinite past (check~\cite{ThroughTheBigBang} for a discussion of this issue).

\subsubsection*{Quiescence}
\index{homogeneous metrics} \index{Bianchi IX solutions}\index{quiescence}

So far we considered an empty universe. As it turns out, there is a particular kind of matter which changes qualitatively the behaviour of the solutions: this is the case of a \emph{massless scalar field,} which modifies the Hamiltonian as
\begin{equation} \label{QuiescentHamiltonianConstraint}
\mathcal H = {\sfrac 3 8} v^2 \, \tau^2 - \left( p_1^2 + p_2^2 + {\sfrac 1 2} \pi_\phi^2
 \right) + v^{4/3} \, C - {\sfrac 1 2}  \sqrt g \, g^{ij} \nabla_i \phi \nabla_j \phi \,,
\end{equation} 
if the scalar field is homogeneous the derivative term $ {\sfrac 1 2}  \sqrt g \, g^{ij} \nabla_i \phi \nabla_j \phi$ is zero. One is then left just with a spatially constant kinetic term $\pi_\phi^2$, which by its equations of motion is also constant in time. In presence of the scalar field, the solution during a Kasner epoch of Eq.~(\ref{AutonomousVolumeTimeBIXeqs}) is
\begin{equation}
\begin{aligned}
(\pi_1,\pi_2) &= \pi_0  \, (\cos \varphi_0 , \sin \varphi_0)  \, e^{-\frac 2 3 \zeta} \,
\\
(q^1,q^2) &=  (q^1_0,q^2_0)  + \Xi \sqrt{{\sfrac 8 3}}  (\cos \varphi_0 , \sin \varphi_0) \, \zeta \,,
\end{aligned}
\end{equation}
where
\begin{equation}
\Xi = \sqrt{\frac{ \pi_0^2}{ \pi_0^2 + {\sfrac 1 2} \pi_\phi^2}}  \leq 1 \,.
\end{equation}
Now, during one Kasner epoch, all our functions $\sigma_\mu(\varphi_0)$ and $\chi_\mu (\varphi_0)$ in Eq.~(\ref{BIXSigmaAndChiFunctions}) will be multiplied by a factor $\Xi$. This factor changes the potential in Eq.~(\ref{BIXKinAndPotEnergyDuringKasner}) so that the minimum value (among the choices of $\varphi_0$) of the exponents in the terms $e^{\sigma_\mu(\varphi_0)\zeta}$ and $e^{\chi_\mu (\varphi_0) \zeta}$ is not $e^{-\frac 8 3 \zeta}$ anymore, but it is  $e^{-\frac 8 3 \Xi \,\zeta}$ instead, where $\Xi < 1$ unless $\pi_\phi = 0$.
This enlarges, at a given value of $\zeta$, the intervals in which the kinetic energy is larger than the potential energy. If $\Xi$ reaches the value $\frac 1 2$ or less, then the smallest exponent is smaller than $e^{-\frac 4 3 \zeta}$, and therefore it cannot compete with the kinetic energy: the ratio $C(\vec q) / (\pi_1^2 + \pi_2^2)$ always goes to zero as $\zeta \to - \infty$ for any value of $\varphi$.

Now, $\Xi$ is a constant of motion during the Kasner epochs, because so are $\pi_0^2 = p_1^2 + p_2^2$ and 
$\pi_\phi^2$. But during a Taub transition the momentum $\vec p$ change, while  $\pi_\phi$ is always conserved. As we saw above during the discussion of Taub transitions, the shape kinetic energy $p_1^2 + p_2^2$ always decreases after each transition. This means that the system will bounce against the potential only a finite number of times, until $\Xi$ reaches the value $\frac 1 2$. At this point the potential decreases too fast with respect to the kinetic energy, and the system stabilizes around one Kasner solution which goes on until the boundary of shape space $q_1^2 + q_2^2 \to \infty$ is reached. The singularity coincides with the instant the system hits the boundary of shape space, and it is now reached after a finite amount of change. 
This behaviour is known as \emph{quiescence:} the chaotic `mixmaster' behaviour of Bianchi IX is `tamed' by the presence of the massless scalar field.

\begin{figure}[t!]
\begin{center}
\includegraphics[width=0.46\textwidth]{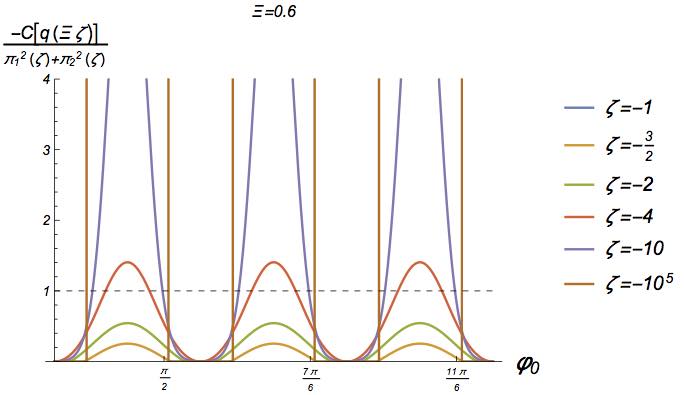}
\end{center}
\caption[Bianchi IX shape potential vs. shape kinetic energy, $\Xi<1$]{$-C(\vec q)/(\pi_a^2 + \pi_b^2)$ in a Kasner epoch, as a function of the direction of the motion of the particle $\varphi_0$, for different log-volume-times $\zeta$, and with a scalar field whose kinetic energy is such that $\Xi = \sqrt{ \pi_0^2/(\pi_0^2 + {\sfrac 1 2} \pi_\phi^2)} = 0.6$. Notice how three `gaps' open around the symmetry axes $\varphi_0 = {\frac 2 3} \pi , {\frac 4 3}  \pi , 2 \pi$.}\label{BianchiIXPotVsKin2}
\end{figure}\index{homogeneous metrics} \index{Bianchi IX solutions}\index{quiescence}

\subsubsection*{Orientation and the Shape Sphere}

The variables~(\ref{BianchiIXTriads}) we used to define the metric are a particular case of \emph{triads}, or \emph{vielbeins}. These are a way to represent a Riemannian metric in a redundant way, by using three 1-forms (covector fields) $e^a$, $a=1,2,3$,  with an internal $SO(3)$ gauge symmetry associated to the possibility of rotating the three forms into each other $e'^a = M^a{}_b e^b$.  The metric is written as
\begin{equation}
g_{ij} = \delta_{ab} \, e^a{}_i \, e^b{}_j \,.
\end{equation}
The variables~(\ref{BianchiIXTriads}) can be expressed in the basis of the $\sigma^a$ forms as $e^a{}_i = E^a_b \,\sigma^b{}_i $, and then the matrix $q_{ab} = \delta_{cd} E^c_a \, E^d_b$ is essentially the `square' of the matrix $E^a_b$. 
This representation is certainly redundant, as it introduces additional degrees of freedom and an associated gauge freedom, but it also contains more information than a Riemannian metric. In fact the determinant of $E^a_b$ is in principle allowed to take negative values, while the determinant of $g_{ij}$, just like that of $q_{ab}$, is proportional to $\det^2 E$. If we want to take into account both orientations, we end up with a double cover of shape space. 
We can choose an internal gauge in which $E^a_b$ is diagonal, and then the \emph{squares} of the diagonal components of $E^a_b$ can be coordinatized as
\begin{equation}
(E^x_x)^2= v^{\frac 2 3} e^{-\frac a {\sqrt 2} - \frac{b}{\sqrt 6}} \,, \qquad 
(E^y_y)^2= v^{\frac 2 3} e^{\frac a {\sqrt 2} - \frac{b}{\sqrt 6}} \,, \qquad
(E^z_z)^2= v^{\frac 2 3} e^{\sqrt{\frac 2 3} b } \,,
\end{equation}

\begin{figure}[h!]
\begin{center}
\includegraphics[width=0.3\textwidth]{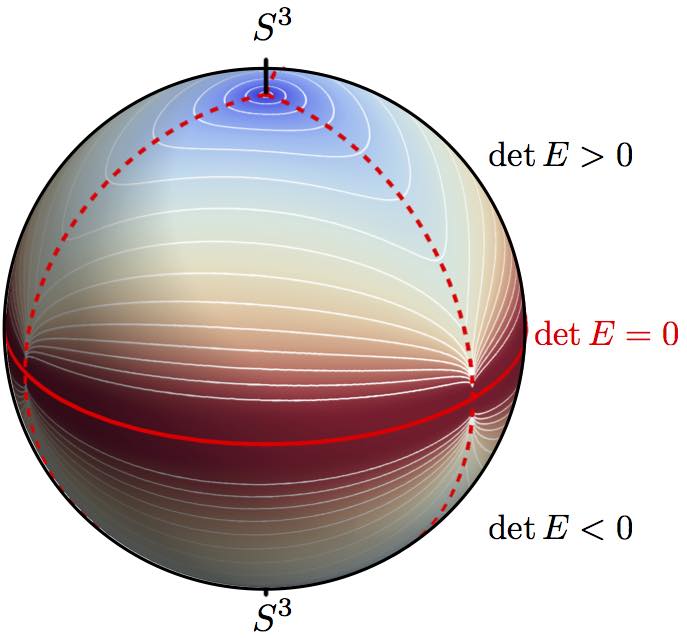}
\caption[Spherical Bianchi IX shape space]{Spherical representation of shape space, with round $S^3$ geometries on the poles and degenerate geometries at the equator. The dashed lines represent geometries with an additional symmetry between two directions, and correspond to $\alpha = \pi/6, \pi/2, 5\pi/6$.
The two hemisphere correspond to opposite orientations. The shape potential $C(\alpha,\beta)$ is represented as a colour plot on the sphere with equipotential lines in white.}\label{CompactifiedShapeSpace}
\end{center}
\end{figure}
\index{homogeneous metrics} \index{Bianchi IX solutions}\index{triads}\index{vielbeins}\index{orientation of space}

\index{homogeneous metrics} \index{Bianchi IX solutions}\index{triads}\index{vielbeins}\index{orientation of space}

It is convenient to represent this double cover with the single cover of a compact manifold. If we use the following  2-to-1 map:
\begin{equation}\label{SphericalShapeCoordinates}
q^1=|\tan\beta| \cos\alpha \,, ~~~ q^2=|\tan\beta| \sin \alpha \,, ~~~ \alpha \in [0,2\pi) \,, ~ \beta \in [0,\pi] \,,
\end{equation}
we are mapping the shape plane onto an hemisphere (half of $S^2$). Moreover if we add the condition
\begin{equation}
\det E = \frac{\tan \beta}{|\tan \beta|} \,,
\end{equation}
we map each hemisphere to one of the two leaves of the double cover of shape space. We just described
the \emph{shape sphere} (see Fig.~\ref{CompactifiedShapeSpace}): each hemisphere corresponds to all the shapes that our manifold can have for a given orientation. The equator $\beta = \pi/2$ is the border of the fixed-orientation shape space. There the shape coordinates $a$ and $b$ diverge, and corresponds to degenerate shapes in which one or two of the directions are infinitely larger than the other(s). We soldered the two hemispheres at the equator, and in doing so we chose a particular topology for shape space, in which two degenerate shapes with opposite orientation but same ratio $a/b$ coincide. Moreover this topology has a notion of continuity such that great circles in the shape sphere are continuous across the equator: this gives us a well-defined prescription to continue functions across the equator.

The shapes on the equator are much like collinear configurations in the 3-body problem: they are degenerate triangles of zero area, but as shapes they are perfectly acceptable. From what we discussed above, it is clear that the BIanchi IX system can only reach the Big Bang with a shape that belong to the equator. Then we have an intrinsic notion of big bang which is entirely expressed in shape space: the big bang singularity happens when the shape of the spatial slice is degenerate and therefore cannot support a nonzero volume.

\index{homogeneous metrics} \index{Bianchi IX solutions}\index{triads}\index{vielbeins}\index{orientation of space}

\subsection{Continuation through the singularity}

\index{continuation through the Big Bang singularity}

In~\cite{ThroughTheBigBang} we proved that the dynamics of the Bianchi IX system can be continued uniquely past the singularity. The equations of motion~(\ref{EOM-AHS}) with the addition of the scalar field $\phi$, $\pi_\phi$ take the form
\begin{equation}\label{BIXeqOfMotion}
\begin{aligned}
&\dot{\vec q} = 2 \, \vec p  \,,&  &\dot{\vec p} =  v^{4/3} \, \vec{\nabla} C(\vec q) \,,&  
\\
&\dot  v = - {\sfrac 3 4} v^2 \tau \,,&  &\dot  \tau = {\sfrac 4 3} v^{1/3} C(\vec q) + {\sfrac 3 4}  v \tau^2 \,,&
\\
& \dot \phi = \pi_\phi \,, & &\dot \pi_\phi = 0 \,,
\end{aligned}
\end{equation}
where $\vec{\nabla} C= \left( \frac{\partial  C}{\partial {q^1}} , \frac{\partial  C}{\partial {q^2}} \right)$.
As should be obvious after the previous discussion, Eqs.~(\ref{BIXeqOfMotion}) reach the singularity in an infinite parameter time $t$ (the one corresponding to the `` $\dot{}$ '' derivatives). If $\tau >0$, $v \to 0$ as $t \to +\infty$. I will show below that one can glue one such solution to a solution with $\tau <0$ that comes back from $v =0$ at $t = - \infty$.

We want to  decouple the evolution of the dimensionless relational degrees of freedom for the scale. To do that we need to find five dimensionless combinations of the variables $q^1$, $q^2$, $p_1$, $p_2$, $v$ and $\tau$ such that they are well-defined and they tend to a finite value at the singularity (which, in the compact polar representation, corresponds to $\beta = \frac \pi 2$). In the quiescent regime, in which the potential terms $\vec{\nabla} C(\vec q)$ and $C(\vec q)$ can be ignored, Eqs.~(\ref{BIXeqOfMotion}) admit the solutions
 \begin{equation}\label{BIsolution_1}
\left(\begin{array}{c}
q^1 (t)
\\
q^2 (t)
\end{array}\right)
=
\left(  2 \, p_0 \, t + a_0 \right) 
\left(\begin{array}{c}
\cos \varphi_0
\\
\sin \varphi_0
\end{array}\right)
+ b_0
\left(\begin{array}{c}
-\sin \varphi_0
\\
\cos \varphi_0
\end{array}\right) \,,
 \end{equation}
 \begin{equation}\label{BIsolution_2}
\left(\begin{array}{c}
p^1 (t)
\\
p^2 (t)
\end{array}\right)
=
 p_0
\left(\begin{array}{c}
\cos \varphi_0
\\
\sin \varphi_0
\end{array}\right) \,,
 \end{equation} 
 \begin{equation}\label{BIsolution_3}
  v(t) = v_0 \, e^{ - {\frac 3 4} v_0\, \tau_0 \, t} \,, \qquad
\tau(t) =  \tau_0 \, e^{ {\frac 3 4} v_0 \, \tau_0 \, t}  \,,
 \end{equation}
 \begin{equation}\label{BIsolution_4}
  \phi(t) = \pi_0 \, t + \phi_0 \,, \qquad
\pi_\phi (t) =  \pi_0  \,,
 \end{equation}
 where $p_0$, $a_0$, $\varphi_0$, $b_0$, $v_0$, $\tau_0$, $\pi_0$ and $\phi_0$ are integration constants.  The Hamiltonian constraint~(\ref{QuiescentHamiltonianConstraint}) imposes a relationship between these integration constants:
\begin{equation}
\frac{3}{8} v_0^2 \tau_0^2 - p_0^2 - {\sfrac 1 2} \pi_\phi^2 \approx 0 \,,
\end{equation}
where we ignored the potential term $v^{4/3} C(\vec q)$.

\index{homogeneous metrics} \index{Bianchi IX solutions}\index{continuation through the Big Bang singularity}

  \begin{figure}[t!]
\begin{center}
\includegraphics[width=0.23\textwidth]{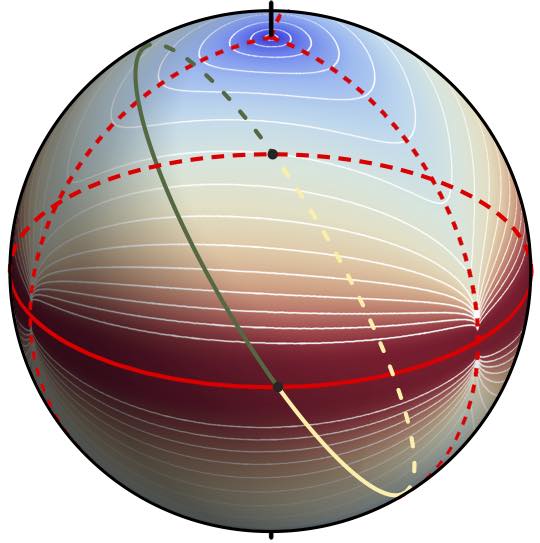}
\caption[Bianchi I solutions on Bianchi IX shape space]{Bianchi I-type solution (in green) on the Bianchi IX shape space. The curve in yellow is the natural continuation of the solution past the Big Bang (at the equator).}\label{BIsolutions_on_BIXshapespace}
\end{center}
\end{figure}

 The solution above traces a straight line in the $(q^1,q^2)$ plane as $t$ goes from $-\infty$ to $+\infty$.  In the `shape sphere' of Fig.~\ref{CompactifiedShapeSpace}  this shows as a great circle. In fact it is easy to prove that
 \begin{equation}\label{BIX_EquationOfGreatCircle}
 \sqrt{(q^1(t))^2 + (q^2(t))^2} \sin\left[ \arctan \left(q^2(t)/q^1(t)\right) - \varphi_0 \right] = b_0 \,,
 \end{equation}
 and since $(q^1)^2 + (q^2)^2 = \tan^2 \beta$, $\arctan \left(q^2/q^1\right)  = \alpha$, the equation above is that of a great circle on the upper hemisphere of the shape sphere.\footnote{This because the equation is $| \tan \beta |  \sin (\alpha -\varphi_0) = b_0$, with an absolute value around the $\tan \beta$, which is not differentiable at the equator $\beta = \frac \pi 2$ (and corresponds to two mirror images of the same semicircle on the two hemisphere, with a cusp at the equator). The equation for a complete, regular great circle is $\tan \beta   \sin (\alpha -\varphi_0) = b_0$.} In fact, embedding the unit sphere $S^2$ in $\mathbbm{R}^3$, the great circles are obtained by intersection with the planes through the origin:
 \begin{equation}\label{EquationOfGreatCircle}
 a \, x + b \, y + c \, z = 0 \,, \qquad  x^2 + y^2 + z^2 = 1 \,.
 \end{equation}
 The equation on the left has a 1-parameter redundance (it can be multiplied by a nonzero constant without changing the plane it identifies), and can therefore be parametrized by two parameters. Call $a = - \sin \varphi_0$, $b = \cos \varphi_0$ and $\frac{c}{\sqrt{a^2 + b^2}} = -b_0$, and the equations~(\ref{EquationOfGreatCircle}), written in spherical coordinates $x = r \, \sin \beta \, \cos \alpha$, $y = r \, \sin \beta \, \sin \alpha$, $x = r \, \cos \beta$, are equivalent to~(\ref{BIX_EquationOfGreatCircle}). 

There is one unique natural continuation of each great semicircle through the equator: it consists of completing the semicircle into a great circle.  We can do that monotonically by gluing a solution with $\tau_0>0$ (which reaches the singularity at $t = +\infty$) with another solution that has identical integration constants except that all momenta ($\vec p$, $\tau$, $\pi_\phi$) are inverted. In this way, as $t$ grows, the first solution flows towards the singularity and the second solution flows away from it. We can glue the two solutions into one by  extending the time parameter $t$ beyond $+\infty$ via an infinite reparametrization that compactifies it: $t = \tan \theta$. Then $t \xrightarrow[\theta \to (\frac \pi 2)^-]{} + \infty$,  $t \xrightarrow[\theta \to (\frac \pi 2)^+]{} - \infty$, and $\frac{\d t}{\d \theta} >0$  in a neighbourhood of $\theta = \frac \pi 2$.  The complete solution then can be written
\index{homogeneous metrics} \index{Bianchi IX solutions}\index{continuation through the Big Bang singularity}
 \begin{equation}\label{BIsolution_1_continued}
\left(\begin{array}{c}
q^1 (\theta)
\\
q^2 (\theta)
\end{array}\right)
=
\left(  2 \, s\, p_0 \, \tan \theta + a_0 \right) 
\left(\begin{array}{c}
\cos \varphi_0
\\
\sin \varphi_0
\end{array}\right)
+ s\, b_0
\left(\begin{array}{c}
-\sin \varphi_0
\\
\cos \varphi_0
\end{array}\right) \,,
 \end{equation}
 \begin{equation}\label{BIsolution_2_continued}
\left(\begin{array}{c}
p^1 (\theta)
\\
p^2 (\theta)
\end{array}\right)
=
 s \, p_0
\left(\begin{array}{c}
\cos \varphi_0
\\
\sin \varphi_0
\end{array}\right) \,,
 \end{equation} 
 \begin{equation}\label{BIsolution_3_continued}
  v(\theta) = v_0 \, e^{ - s \, {\frac 3 4} v_0\, \tau_0 \,\tan \theta } \,, \qquad
\tau(\theta) =  s \, \tau_0 \, e^{s \, {\frac 3 4} v_0 \, \tau_0 \, \tan \theta}  \,,
 \end{equation}
 \begin{equation}\label{BIsolution_4_continued}
  \phi(\theta) = s \, \pi_0 \, \tan \theta + \phi_0 \,, \qquad
\pi_\phi (\theta) = s\, \pi_0  \,,
 \end{equation}
 where $s = \text{sign}(\sin \theta) \equiv \text{sign}(\cos \beta) $. One might ask why we have to put the `$s$' in front of the term $ s\, b_0 \left(-\sin \varphi_0, 
\cos \varphi_0 \right)$ in $\left(q^1 (\theta), q^2 (\theta)\right)$. The reason is that this is the  only choice that guarantees the continuity of the curve $\alpha = \alpha(\beta)$ and all its derivatives $\frac{\d^n \alpha}{\d\beta^n}$. In fact notice that the equation for a great circle in spherical coordinates is
\begin{equation}
\tan \beta \, \sin ( \alpha - \varphi_0) = b_0 \,,
\end{equation}
and solving the above equation w.r.t. $\alpha$ we get:
\begin{equation}\label{GreatCircleEq_alpha}
\alpha(\beta) = \varphi_0 + \arcsin\left( \frac{b_0}{\tan \beta} \right) \,, 
\end{equation}
which is regular at the equator $\beta = \frac \pi 2$.  Now we can calculate the derivatives of $\alpha$ w.r.t. $\beta$ up to the fourth order, and show that
\index{homogeneous metrics} \index{Bianchi IX solutions}\index{continuation through the Big Bang singularity}
\begin{equation}\label{dnalphadnbeta_GreatCircle}
\begin{array}{rcl}
 \alpha(\beta)  \xrightarrow[\beta \to {\frac \pi 2}^-]{}  & \varphi_0  &
\xleftarrow[\beta \to {\frac \pi 2}^+]{} 
 \alpha(\beta)  \,,\\
\frac{\d \alpha(\beta) }{d \beta}  \xrightarrow[\beta \to {\frac \pi 2}^-]{}  & - b_0  &
\xleftarrow[\beta \to {\frac \pi 2}^+]{} 
\frac{\d \alpha(\beta) }{d \beta}  \,,\\
\frac{\d^2 \alpha(\beta) }{d \beta^2}  \xrightarrow[\beta \to {\frac \pi 2}^-]{}  &  0  &
\xleftarrow[\beta \to {\frac \pi 2}^+]{} 
\frac{\d^2 \alpha(\beta) }{d \beta^2}  \,,\\
\frac{\d^3 \alpha(\beta) }{d \beta^3}  \xrightarrow[\beta \to {\frac \pi 2}^-]{} &  -  b_0 (2 +b_0^2)  &
\xleftarrow[\beta \to {\frac \pi 2}^+]{}  
\frac{\d^3 \alpha(\beta) }{d \beta^3}  \,,\\
\frac{\d^4 \alpha(\beta) }{d \beta^4}  \xrightarrow[\beta \to {\frac \pi 2}^-]{} &   0  &
\xleftarrow[\beta \to {\frac \pi 2}^+]{} 
\frac{\d^4 \alpha(\beta) }{d \beta^4}  \,.
\end{array}
\end{equation}
We can do the same with Eqs. (\ref{BIsolution_1_continued}), inserted inside the expression of the  coordinates on the shape sphere as functions of $q^1$ and $q^2$:
\begin{equation}\label{BIX_DefBeta}
 \beta = (\sfrac{1 - s}{2}) \pi  + s  \arctan \sqrt{(q^1)^2 + (q^2)^2} \,,  ~~ \alpha = \arctan \left( {\sfrac{q^2}{q^1}} \right)  \,,
\end{equation}
where now $s = \text{sign} (\cos \beta) $.  By using the relation
\begin{equation}\label{dnalphadnbeta_BI}
\frac{\d^n \alpha  }{\d \beta^n} =  \frac{1}{\partial_\theta \beta } \frac{\partial}{\partial \theta} \left( \dots   \frac{1}{\partial_\theta \beta } \frac{\partial}{\partial \theta} \left(   \frac{\partial_\theta \alpha }{\partial_\theta \beta } \right) \dots \right) \,,
\end{equation}
where $\partial_\theta \alpha =\partial_\theta \alpha[q^1(\theta),q^2(\theta)]$ and  $\partial_\theta \beta =\partial_\theta \beta[q^1(\theta),q^2(\theta)]$. Calculating the left- and right-limit $\theta \to \frac \pi 2$ for the expressions~(\ref{dnalphadnbeta_BI})  we get the same results as in Eq.~(\ref{dnalphadnbeta_GreatCircle}), while if we didn't put the `$s$' in front of $b_0$ in Eq.~(\ref{BIsolution_1_continued}) we would have had a different sign for the two limits in each expression.

\index{homogeneous metrics} \index{Bianchi IX solutions}\index{continuation through the Big Bang singularity}
Now we would like to find a set of variables that are dimensionless and admit a well-defined value at the singularity. One can immediately verify that the definitions~(\ref{BIX_DefBeta}) have the same left- and right-limit:
\begin{equation}
\beta [q^1(\theta),q^2(\theta)]  \xrightarrow[\theta \to {\frac \pi 2}^-]{}   \frac \pi 2  \xleftarrow[\theta \to {\frac \pi 2}^+]{} \beta [q^1(\theta),q^2(\theta)] \,,
\end{equation}
and
\begin{equation}
\alpha [q^1(\theta),q^2(\theta)]  \xrightarrow[\theta \to {\frac \pi 2}^-]{}   \varphi_0 \xleftarrow[\theta \to {\frac \pi 2}^+]{} \alpha [q^1(\theta),q^2(\theta)] \,,
\end{equation}
so the spherical coordinates on shape space are two such variables.

As a further variable, the direction of the shape momentum $ \arctan \left( p_2/p_1 \right)$ is not good because it tends to $\varphi_0  + (s-1) {\frac \pi 2}$, and so it is not truly independent. A genuinely independent degree of freedom must measure the integration constant $b_0$, which is the impact parameter of the curve in shape space (the closest distance from the origin that the curve reaches). The impact parameter of an inertial particle in the plane is measured by the ratio between the angular momentum $ \vec q \times \vec p = (q^1 \, p_2 - q^2 p_1)$ and the norm of the momentum $|\vec p|$:
\begin{equation}
\gamma = \frac{\vec q \times \vec p }{p} \,,
\end{equation}
which tends to $-b_0$ from both directions:
\begin{equation}
\gamma [\vec q (\theta),\vec p (\theta)]  \xrightarrow[\theta \to {\frac \pi 2}^-]{}   - b_0 \xleftarrow[\theta \to {\frac \pi 2}^+]{} \gamma [\vec q (\theta),\vec p (\theta)]   \,.
\end{equation}
We can still form three independent dimensionless combinations out of the dimensionful quantities $v$, $\tau$, $p = \sqrt{(p_1)^2 + (p_2)^2}$ and $\pi_\phi$. The simplest which are also parity-invariant are:
\begin{equation}\label{BIX_DefXiSigmaKappa}
\xi = \frac{|\pi_\phi|}{p} \,,
\qquad
\sigma = \frac{v \, |\tau|}{p} \,,
\qquad
\kappa = \frac{v^{2/3}}{p} \,.
\end{equation}
\index{homogeneous metrics} \index{Bianchi IX solutions}\index{continuation through the Big Bang singularity}
All these quantities admit finite limits at the singularities:
\begin{equation}
\xi^2 [\pi_\phi (\theta),\vec p (\theta)]  
\xrightarrow[\theta \to {\frac \pi 2}^-]{}  
\frac{\pi_0^2}{p_0^2}
 \xleftarrow[\theta \to {\frac \pi 2}^+]{}
\xi^2 [\pi_\phi (\theta),\vec p (\theta)]  \,,
\end{equation}
\begin{equation}
\sigma [v (\theta) , \tau (\theta),\vec p (\theta)]  
\xrightarrow[\theta \to {\frac \pi 2}^-]{}  
\frac{v_0 \tau_0}{p_0^2}
 \xleftarrow[\theta \to {\frac \pi 2}^+]{}
\sigma [v (\theta) , \tau (\theta),\vec p (\theta)]  \,,
\end{equation}
\begin{equation}
\kappa^2 [v (\theta) , \vec p (\theta)]  
\xrightarrow[\theta \to {\frac \pi 2}^-]{}  
0
 \xleftarrow[\theta \to {\frac \pi 2}^+]{}
\kappa^2 [v (\theta) , \vec p (\theta)]  \,.
\end{equation}
The variable $\kappa$ does not tend to an arbitrary value: it is constrained to vanish at the singularity, when $v\to 0$. Now look at the Hamiltonian constraint~(\ref{QuiescentHamiltonianConstraint}) in the new variables:
\begin{equation}\label{Dimensionless_BIX_Ham_const}
\mathcal H = {\sfrac 3 8}  p^2 \, \sigma^2 - p^2  \left(1  + {\sfrac 1 2} \xi^2
 \right) + p^2 \, \kappa^2 \, C(\vec q) \approx 0 \,,
\end{equation}
$p^2$ factorizes, and what remains is a constraint between the variables $\sigma$, $\xi$ and $\kappa$. At the equator $\kappa^2 \, C(\vec q) $ vanishes (if the solution is quiescent, otherwise it can never reach the equator), and therefore $\sigma$ and $\xi$ are not independent:
\begin{equation}
1  + {\sfrac 1 2} \xi^2 - {\sfrac 3 8}  \sigma^2 \xrightarrow[\beta \to \frac \pi 2]{} 0 \,.
\end{equation}

\subsubsection*{Solution-determining data}

The dimensionless variables I considered so far ($\alpha$, $\beta$, $\lambda$, $\sigma$ and $\kappa$) are not enough to uniquely specify a curve in shape space. I would like to have five dimensionless variables whose asymptotic value at the equator completely fixes the initial-value problem (from now on I will call these `solution-determining data'). A Bianchi I curve on shape space like Eq.~(\ref{BIsolution_1_continued}) is uniquely fixed by specifying the integration constants $\varphi_0$ and $b_0$. The magnitude of the momentum $p_0$ and the integration constant $a_0$ only determine, respectively, the speed at which the same path is crossed by the representative point on shape space, and the position on the path of the point at $t=0$. Moreover the integration constants $v_0$ and $\tau_0$ don't affect the path on shape space at all, as the shape and volume/York time \index{York time} variables decouple in the Bianchi I limit. Therefore in this limit there are only two solution-determining data. However the full Bianchi IX solutions have more solution-determining data: the volume and York time couple to the shape degrees of freedom through the shape potential $C(\vec q)$, and the same asymptotic Bianchi I curve can have a different evolution away from the equator, if the value of the three independent dimensionless combinations of the integration constants $a_0$, $p_0$, $v_0$ and $\tau_0$ are different (only dimensionless combinations can affect the dimensionless curve on shape space. This is a consequence of the property of dynamical similarity, which I described in Sec.~\ref{DOFsOfSD} above). Three dimensionless combinations are, \emph{e.g.}, $a_0$,  $p_0 v_0^{- 3/2}$ and $\sqrt{p_0}  \tau_0$.
Of these, only two are truly independent, because the solution curve~(\ref{BIsolution_1_continued}) is invariant under time translation $ t \to t + c$ (or $\tan \theta \to \tan \theta + c$), which can be reabsorbed into the following transformation of the integration constants:
\begin{equation}\label{TimeTranslationTransformation}
\begin{aligned}
a_0 \to a_0 + 2 \, s \, p_0 \,c \,,
\\
v_0 \to e^{- \frac 3 4 s v_0 \tau_0 \tan \theta} v_0 \,,
\\
\tau_0 \to e^{\frac 3 4 s v_0 \tau_0 \tan \theta} \tau_0 \,.
\end{aligned}
\end{equation}
Any choice of integration constants $a_0$, $v_0$, $\tau_0$ that lie on the same orbit of the transformation~(\ref{TimeTranslationTransformation}) generates the same solution curve. Therefore our independent solution-determining data need to be invariants of the transformation~(\ref{TimeTranslationTransformation}). There are two such independent invariants which are also dimensionless:
\index{homogeneous metrics} \index{Bianchi IX solutions}\index{continuation through the Big Bang singularity}
\begin{equation}
\frac{v_0 \tau_0}{p_0} \,, \qquad a_0 - \frac{2 p_0 \log \left( v_0^2 \tau_0^6\right)}{3\,v_0 \tau_0} \,.
\end{equation}
The first invariant is obviously the value of  $\sigma = v \tau/p$ throughout a Bianchi I orbit, or the asymptotic value of $\sigma$ at the equator in a Bianchi IX orbit. The second invariant is the asymptotic value of a more complicated variable, which can be found by observing that the `dilatational shape momentum' $\vec q \cdot \vec p$ in a Bianchi I solution~(\ref{BIsolution_1_continued})  takes the value
\index{homogeneous metrics} \index{Bianchi IX solutions}\index{continuation through the Big Bang singularity}
\begin{equation}
\vec q \cdot \vec p = s p_0 \left(2 \, s \, p_0 \, \tan \theta + a_0 \right) \,,
\end{equation}
while the combination $\frac{2 p \log \left( v^2 \tau^6\right)}{3 \, v \tau}$ on the same solution takes the value
\begin{equation}
\frac{2 p \log \left( v^2 \tau^6\right)}{3\,v \tau} = 2 \,p_0 \, \tan \theta + s  \frac{2 p_0 \log \left( v_0^2 \tau_0^6\right)}{3\,v_0 \tau_0}\,,
\end{equation}
(notice the placement of the $s$ factors). So the difference between the first quantity divided by $p$ and the second quantity is time-independent:
\begin{equation}
\frac{\vec q \cdot \vec p}{p} - \frac{2 p \log \left( v^2 \tau^6\right)}{3\,v \tau} = s \left[ a_0 -  \frac{2 p_0 \log \left( v_0^2 \tau_0^6\right)}{3\,v_0 \tau_0} \right] \,.
\end{equation}
If we want to have a quantity that takes the same asymptotic value on both sides of the equator, we need to multiply the above by $s = \text{sign}(\tan \beta)$. This will be the definition of our new variable:
\index{homogeneous metrics} \index{Bianchi IX solutions}\index{continuation through the Big Bang singularity}
\begin{equation}
\omega = s \left[ \frac{\vec q \cdot \vec p}{p} - \frac{2 p \log \left( v^2 \tau^6\right)}{3\,v \tau} \right] \,,
\end{equation}
with this definition $\omega$ tends to the same value at the equator from both sides:
\begin{equation}
\omega  [\vec q (\theta),\vec p (\theta)]  \xrightarrow[\theta \to (\frac \pi 2)^-]{}   \left[ a_0 -  \frac{2 p_0 \log \left( v_0^2 \tau_0^6\right)}{3\,v_0 \tau_0}  \right] \xleftarrow[\theta \to (\frac \pi 2)^+]{} \omega [\vec q (\theta),\vec p (\theta)]   \,.
\end{equation}
From the asymptotic values of the dimensionless quantities $\alpha$, $\beta$, $\gamma$, $\omega$ and $\sigma$ we can identify uniquely a solution curve in shape space. A solution in the ADM phase space is uniquely determined once we specify the values of these five fields, plus a conventional dimensionful unit: the value of the magnitude of the momentum $p = |\vec p|$ at a point on the curve (for example, we could choose the asymptotic value $p_0$ at the equator, where $p$ is a conserved quantity). The ADM phase space is six-dimensional, so these six data are sufficient to determine a solution. Changing the value of $p_0$ changes the solution curve in the ADM phase space, but leaves its projection on shape space unchanged: the one-parameter family of solutions spanned by the values of $p_0$ is entirely redundant: all these solutions are physically equivalent. This will be made very explicit in the next Section (and it will be used to obtain the main result), where I will show that the equations of motion for $\alpha$, $\beta$, $\gamma$, $\omega$ and $\sigma$ completely decouple from $p$.

Our system has more degrees of freedom than the geometrodynamical ones: there is a scalar field $\phi$. Since $\phi$ is homogeneous and has no potential, it only interacts with the other fields through its kinetic energy $\frac 1 2 \pi_\phi^2$ which gravitates.  The scalar field kinetic energy influences the shape-dynamics of the system through the dimensionless quantity $\xi = |\pi_\phi|/p$, but not in such a way to change the counting of solution-determining data: this is due to the Hamiltonian constraint~(\ref{QuiescentHamiltonianConstraint}), which constrains the dimensionless quantities as in Eq.~(\ref{Dimensionless_BIX_Ham_const}): the constraint determines $\xi^2$ as a function of $\sigma$, $C(\vec q)$ and $\kappa^2 = \sigma^2 e^{-\frac{1}{2} \sigma\left( \frac{\vec q \cdot \vec p}{p}  - \omega \right)}$:
\begin{equation}\label{Dimensionless_BIX_Ham_const_solved_for_xi}
\xi^2  \approx  2 \, \sigma^2 \left[ {\sfrac 3 8}   +   e^{-\frac{1}{2} \sigma\left( \frac{\vec q \cdot \vec p}{p}  - \omega \right)} \, C(\vec q) \right]  - 2\approx 0 \,.
\end{equation}
$\xi$ changes radically the behaviour of the system by the fact that it is not zero: as I showed above (and will re-derive below in the new language I am using in this Section), if $\xi=0$ the solution never reaches the equator, while any nonzero value of $\xi$ induces quiescence. However, once we know that it is not zero, the particular value of $\xi$ does not enter the count of solution-determining data if we already specified $\alpha$, $\beta$, $\gamma$, $\omega$ and $\sigma$ , because it is dynamically locked to the other quantities through the constraint~(\ref{Dimensionless_BIX_Ham_const_solved_for_xi}). Alternatively, we could have taken $\xi$ as one of the solution-determining data, but one of the other ones, \emph{e.g.} $\sigma$, would then be determined by the constraint.

There is a last  degree of freedom that is left to discuss: the value $\phi$ of the homogeneous part of the scalar field. $\phi$  has no physically measurable consequences: it could be, for example, the phase of the Higgs field:\footnote{The phase of the Higgs field is the direction, in the internal field space, where the potential is flat, and therefore has identical equations of motion and an identical contribution to the Hamiltonian constraint as our field $\phi$.} a global change in direction of the Higgs expectation value cannot be measured. This is not the case for \emph{local} changes in direction: inhomogeneities of $\phi$ gravitate through the potential term ${\sfrac 1 2}  \sqrt g \, g^{ij} \nabla_i \phi \nabla_j \phi \,,$ in the ADM  Hamiltonian~(\ref{QuiescentHamiltonianConstraint}). Due to its equations of motion~(\ref{BIXeqOfMotion}),  $\dot \phi = \pi_\phi$, $\dot \pi_\phi = 0$, the field $\phi$ evolves by growing linearly in parameter time $t = \tan \theta$. If $\phi$ was the phase of the Higgs field, it would rotate an infinite amount of times before reaching the equator. It is clear then that $\phi$ does not admit a well-defined asymptotic value at the equator, just like the limit $\displaystyle \lim_{x \to 0} \sin(1/x)$ is not well-defined. If we had to include $\phi$ among the solution-determining data we would be in trouble: just like in the non-quiescent case, some of the physically-relevant degrees of freedom would not admit a well-defined value at the singularity, and therefore a continuation through the equator would not be meaningful.

The above discussion holds true for a completely homogeneous solution. If we were to include inhomogeneities in the picture, we would have to count the inhomogeneous modes of $\phi$ among the shape degrees of freedom (notice that in out convention $\phi$ is dimensionless). These, however, are suppressed near the singularity thanks to the BKL behaviour: as the volume approaches zero all spatial-derivative terms are suppressed w.r.t. the other degrees of freedom by some vanishing factor depending on the volume (or, rather, on $\kappa$). It remains to be seen what is the specific dynamics that these degrees of freedom tend to in the decoupling limit, and if one can find well-defined asymptotic values for them which are solution-determining data.

\subsubsection*{Asymptotic perennials}

\index{homogeneous metrics} \index{Bianchi IX solutions}\index{continuation through the Big Bang singularity}

Before moving on to the central proof, I would like to remark that the quantities $\gamma$, $\omega$, $\sigma$, $\xi$ and $p$ are \emph{Kucha\v{r} perennials} (see Sec.~\ref{NutshellSubsec}) \index{Kucha\v{r} perennial} for the Bianchi I system. In fact they are all conserved quantities, as one can explicitly check by showing that they Poisson-commute with the Bianchi I Hamiltonian $\mathcal H_\st{BI} = {\sfrac 3 8} v^2 \, \tau^2 - \left( p^2 + {\sfrac 1 2} \pi_\phi^2 \right)$. We can add one more conserved quantity which measures the angle $\alpha$ at the equator ($\varphi_0$ in the solution~\ref{BIsolution_1_continued}), for example:
\begin{equation}
\varphi = \arctan \left[ \frac{|\vec p|^2 \, q^2 + s  (q \times \vec p ) p_1}{|\vec p|^2 \, q^1 - s  (q \times \vec p ) p_2} \right] \,,
\end{equation}
where $s=\text{sign}(\tan \beta)$, and we would have a complete set of perennials that entirely determine the Bianchi I solution curve: $\varphi$, $\gamma$, $\omega$, $\sigma$, $\xi$ and $p$. The system admits a maximal set of conserved quantities: it is \emph{integrable}~\cite{ArnoldBook}.

Now, the Bianchi IX system is the exact opposite of integrable: due to its triangular pool table-shaped potential it is a chaotic system~\cite{Goldstein}. However, in presence of a massless scalar field, it exhibits quiescence, and its equations tend, near the equator, to those of Bianchi I. Therefore the quantities $\varphi$, $\gamma$, $\omega$, $\sigma$, $\xi$ and $p$, which are not conserved along a generic Bianchi IX solution, are \emph{asymptotically conserved}. We talk about \emph{asymptotic perennials:} quantities that are perennials only at one point on each solution curve. If the point where the asymptotic perennials are conserved is a well-defined submanifold of phase space, as is our case (it is simply the equator of the shape sphere), we can talk about a \index{Janus point} \emph{Janus surface}~\cite{barbour2013gravitational,Barbour:2014bga,Reply_to_Zeh,Entropy_and_Typicality_of_Universes}, if each solution is qualitatively symmetric around the point where it intersects the surface.  The quiescent Bianchi IX system is precisely an example of a \emph{Janus system} (just like the N-body problem~\cite{Entropy_and_Typicality_of_Universes,Reply_to_Zeh}), which allows to define an arrow of time based on the structure of the set of solution curves on shape space. The identification of asymptotic perennials is key to this construction, because it allows to prove that the space of physically-distinct solutions is isomorphic to the \emph{projective cotangent bundle} to shape space~\cite{barbour2013gravitational,Barbour:2014bga,Reply_to_Zeh,Entropy_and_Typicality_of_Universes}. In our case the situation is  more subtle: the space of physically-distinct solution is parametrized by a set of five asymptotic perennial, which only involve the specification of a point on a submanifold of shape space (the equator).

\subsubsection*{Intrinsic dynamics of dimensionless shape degrees of freedom}


It is time to express the dynamics of the system in terms of the variables $\alpha$, $\beta$, $\gamma$, $\omega$, $\sigma$ and $\xi$, relegating all the dimensions to the magnitude of the momentum, $p$. This will allow us to show that, once we choose an intrinsic parametrization on shape space (we will use first the arc-length parametrization $\d \ell = \sqrt{(\d q^1)^2+(\d q^2)^2}$ and, later, the $\beta$-parametrization), the variable $p$ completely decouples from the others, proving that it does not affect the solution curve on shape space.
\index{homogeneous metrics} \index{Bianchi IX solutions}\index{continuation through the Big Bang singularity}

Let's start by collecting the definitions of our variables:
\begin{equation}
\begin{aligned}
\alpha = \arctan(q^2/q^1) \,, \qquad \beta = (\sfrac{1 - s}{2}) \pi  + s  \arctan \sqrt{(q^1)^2 + (q^2)^2} \,,
\\
\gamma = \frac{\vec q \times \vec p }{p} \,,
~~
\xi = \frac{|\pi_\phi|}{p} \,,
~~
\sigma = \frac{v \, |\tau|}{p} \,,
~~
\omega = s \left[ \frac{\vec q \cdot \vec p}{p} - \frac{2 p \log \left( v^2 \tau^6\right)}{3\,v \tau} \right] \,.
\end{aligned}
\end{equation}
The above relations can be inverted,
\begin{equation}\label{BIX_definitions_of_dimensionless_variables}
\begin{aligned}
&q^1 = |\tan\beta| \cos\alpha \,, \qquad 
q^2 = |\tan\beta| \sin \alpha \,,
\\
&p_1 = s p  \cos\left[ \alpha + \arcsin\left( \frac{\gamma}{\tan \beta}\right) \right] ,
~~
p_2 = s p  \sin\left[ \alpha + \arcsin\left( \frac{\gamma}{\tan \beta}\right) \right] ,
\\
&\tau    = s  (p \, \sigma)^{-\frac 1 2}  e^{\frac{3}{8} \sigma  \left(\sqrt{\tan^2 \beta - \gamma^2} - \omega \right)}  \,,
~ v = (p \, \sigma)^{\frac 3 2} \, e^{-\frac{3}{8} \sigma  \left(\sqrt{\tan^2 \beta - \gamma^2} - \omega \right)} \,
\\
&\pi_\phi = s \, \xi \,  p\,.
\end{aligned}
\end{equation}
Now we can write the equations of motion, Eqs.~(\ref{BIXeqOfMotion}), which I reproduce here for convenience:
\begin{equation}\label{BIXeqOfMotion_copy}
\begin{aligned}
&\dot{\vec q} = 2 \, \vec p  \,,&  &\dot{\vec p} =  v^{4/3} \, \vec{\nabla} C(\vec q) \,,&  
\\
&\dot  v = - {\sfrac 3 4} v^2 \tau \,,&  &\dot  \tau = {\sfrac 4 3} v^{1/3} C(\vec q) + {\sfrac 3 4}  v \tau^2 \,,&
\\
& \dot \phi = \pi_\phi \,, & &\dot \pi_\phi = 0 \,,
\end{aligned}
\end{equation}
in terms of the variables $\alpha$, $\beta$, $\gamma$, $\omega$, $\sigma$, $\xi$ and $p$, simply by replacing the definitions~(\ref{BIX_definitions_of_dimensionless_variables}) into Eqs.~(\ref{BIXeqOfMotion_copy}), and solving for the first derivatives $\dot \alpha$, $\dot \beta$, $\dot \gamma$, $\dot \omega$, $\dot \sigma$, $\dot \xi$ and $\dot p$:
{\thinmuskip=0mu
\medmuskip=0mu
\thickmuskip=0mu
\begin{equation}\label{BIX_intrinsiceq_1}
\begin{aligned}
\dot \alpha ~=&~ 2 ~ p ~ \gamma ~ \cot^2 \beta \,,
\qquad
\dot \beta ~=~  2 ~ p ~ \cos^2 \beta ~ \sqrt{1 - \frac{\gamma^2}{\tan^2 \beta}} \,,
\\
\dot \gamma ~=&~ p ~ \sigma^2  \sqrt{1 -\frac{\gamma^2}{\tan^2 \beta}} \left( \sqrt{1 -  \frac{\gamma^2}{\tan^2 \beta} }  \, \frac{\partial C}{\partial \alpha} - \gamma \cos^2 \beta \,  \frac{\partial C}{\partial \beta}   \right) e^{{\frac \sigma 2} \left(\omega - \sqrt{\tan^2 \beta - \gamma^2}  \right) } \,,
\\
\dot \omega ~=&~\bigg{\{}  {\frac 4 3} s ~ p ~ \left[ \sigma \left(\sqrt{\tan^2 \beta - \gamma^2} - \omega \right) - 4 \right] C(\alpha,\beta) \\
&~~~+ p ~ \sigma^2 ~ |\cos \beta|^3 \left[    \left( 2  \gamma^2 - \tan^2 \beta \right) +  \omega \sqrt{\tan^2 \beta - \gamma^2  } \right]  \frac{\partial C}{\partial \beta}
\\
&~~~+ \frac{p ~ \gamma ~\sigma^2}{\tan^2 \beta} \left( 2 \sqrt{\tan^2 \beta - \gamma^2}  - \omega \right)  \frac{\partial C}{\partial \alpha}  \bigg{\}}e^{{\frac \sigma 2} \left(\omega - \sqrt{\tan^2 \beta - \gamma^2}  \right) }  \,,
\\
\dot \sigma ~=&~p~\sigma^2 \bigg{[}  -\frac{\sigma}{\tan\beta}  \left(  \frac{\gamma}{\tan\beta} ~ \frac{\partial C}{\partial \alpha} 
+  s ~ \cos^2\beta \sqrt{\tan^2 \beta - \gamma^2} \frac{\partial C}{\partial \beta}\right)
\\
&~~~~~~~~~ {\sfrac 4 3} s ~C(\alpha,\beta)\bigg{]} e^{{\frac \sigma 2} \left(\omega - \sqrt{\tan^2 \beta - \gamma^2}  \right) } \,,
\\
\dot p ~=&~\frac{p^2}{\tan^2 \beta} \left(  \gamma ~ \frac{\partial C}{\partial \alpha} 
 + s ~\sin (2 \beta) \sqrt{\tan^2 \beta - \gamma^2}  \frac{\partial C}{\partial \beta}  \right) e^{{\frac \sigma 2} \left(\omega - \sqrt{\tan^2 \beta - \gamma^2}  \right) }\,.
\end{aligned}
\end{equation}}
The equation for $\dot \xi$ is redundant: $\dot \xi = - \dot p \, \xi /p$, and in any case $\xi$ can be calculated as a function of the other dimensionless quantities using the Hamiltonian constraint, Eq.~(\ref{Dimensionless_BIX_Ham_const_solved_for_xi}), so we don't need to include it in the equations.
 \index{homogeneous metrics} \index{Bianchi IX solutions}\index{continuation through the Big Bang singularity}

The first thing that catches the eye of Eqs.~(\ref{BIX_intrinsiceq_1}) is that the rhs of all equations except $\dot p$ are homogeneous in $p$. This means that if we chose any intrinsic parametrization on shape space, the new time parameter will be homogeneous of degree one in $p$, and the latter will drop out of all equations except the one for $p$. For example, using 
the kinematic arclength on shape space $\d \ell = \sqrt{(\d q^1)^2+(\d q^2)^2}$, the equation of motion of $\ell$ in $t$ parametrization is $\dot \ell = 2 p$, and therefore, calling $f' := \d f /\d \ell$, the equations of motion take the form
{\thinmuskip=0mu
\medmuskip=0mu
\thickmuskip=0mu
\begin{equation}\label{BIX_intrinsiceq_2}
\begin{aligned}
\alpha' ~=& ~ \gamma ~ \cot^2 \beta \,,
\qquad
\beta' ~=~ \cos^2 \beta ~ \sqrt{1 - \frac{\gamma^2}{\tan^2 \beta}} \,,
\\
\gamma' ~=&~ \frac{\sigma^2}{2}  \sqrt{1 -\frac{\gamma^2}{\tan^2 \beta}} \left( \sqrt{1 -  \frac{\gamma^2}{\tan^2 \beta} }  \, \frac{\partial C}{\partial \alpha} - \gamma \cos^2 \beta \,  \frac{\partial C}{\partial \beta}   \right) e^{{\frac \sigma 2} \left(\omega - \sqrt{\tan^2 \beta - \gamma^2}  \right) } \,,
\\
\omega' ~=&~ {\frac 1 2}\bigg{\{}  {\frac 4 3} s ~ \left[ \sigma \left(\sqrt{\tan^2 \beta - \gamma^2} - \omega \right) - 4 \right] C(\alpha,\beta) \\
&~~~+\sigma^2 ~ |\cos \beta|^3 \left[    \left( 2  \gamma^2 - \tan^2 \beta \right) +  \omega \sqrt{\tan^2 \beta - \gamma^2  } \right]  \frac{\partial C}{\partial \beta}
\\
&~~~+ \frac{\gamma ~\sigma^2}{\tan^2 \beta} \left( 2 \sqrt{\tan^2 \beta - \gamma^2}  - \omega \right)  \frac{\partial C}{\partial \alpha}  \bigg{\}}e^{{\frac \sigma 2} \left(\omega - \sqrt{\tan^2 \beta - \gamma^2}  \right) }  \,,
\\
\sigma' ~=& \frac{\sigma^2}{2} \bigg{[}  -\frac{\sigma}{\tan\beta}  \left(  \frac{\gamma}{\tan\beta} ~ \frac{\partial C}{\partial \alpha} 
+  s ~ \cos^2\beta \sqrt{\tan^2 \beta - \gamma^2} \frac{\partial C}{\partial \beta}\right)
\\
&~~~~~~~~~ {\sfrac 4 3} s ~C(\alpha,\beta)\bigg{]} e^{{\frac \sigma 2} \left(\omega - \sqrt{\tan^2 \beta - \gamma^2}  \right) } \,.
\end{aligned}
\end{equation}}

\subsubsection*{Ephemeris time and scale}

\index{homogeneous metrics} \index{Bianchi IX solutions}\index{continuation through the Big Bang singularity}
\index{ephemeris time}\index{ephemeris scale}

Equations~(\ref{BIX_intrinsiceq_2}) do not contain any information regarding scale or duration. Units of scale and time need to be fixed once and for all at a point on a solution, and they are completely immaterial. What is not a matter of convention, however, is their subsequent evolution. This evolution is entirely determined by the shape degrees of freedom, and it can be calculated using two `ephemeris' equations determining the evolution of $p$ and the scale $v$. They are just the equations of motion of $p$ and $v$ in arclength parametrization:
\begin{equation}
\begin{aligned}
\frac{\d \log p}{\d \ell} =& \frac{e^{{\frac \sigma 2} \left(\omega - \sqrt{\tan^2 \beta - \gamma^2}  \right) }}{2\tan^2 \beta} \left(  \gamma ~ \frac{\partial C}{\partial \alpha} 
 + s ~\sin (2 \beta) \sqrt{\tan^2 \beta - \gamma^2}  \frac{\partial C}{\partial \beta}  \right) \,.
\\
\frac{\d \log v}{\d \ell} =& - \frac 3 8 ~ s ~ \sigma \,,
\end{aligned}
\end{equation}
The above equations have the same characteristics, which make them `ephemeris' equations: the unknown variables $p$ and $v$ do not appear on the right-hand-side, so they `parasite' on the evolution of the shape variables. Moreover the equations only determine the logarithms of $p$ and $v$, and therefore their solution are defined modulo a constant rescaling: this is the arbitrariness in fixing units at one point on the solution.

To reach the singularity $v=0$ from any finite point in shape space, either the rhs of the ephemeris scale equation diverges
 (which requires extra symmetry) or an infinite distance of kinematic arc-length is reached. The second condition is generic and states that the singularity is reached whenever the curve in shape space reaches the equator, since each point on the equator has infinite kinematic arc-length distance from any other point. We thus find the singularity condition $ \beta(\ell) =\frac{\pi}{2}$, where, in the spacetime description, the Big Bang occurs.

\index{homogeneous metrics} \index{Bianchi IX solutions}\index{continuation through the Big Bang singularity}
\index{ephemeris time}\index{ephemeris scale}

\subsubsection*{Crossing of the singularity} 

To study the crossing of $\beta=\frac \pi 2$ we parametrize (\ref{BIX_intrinsiceq_2}) by $\beta$, which is achieved by dividing the right-hand sides by $ \beta' = \cos^2 \beta \sqrt{1 - \frac{\gamma^2}{\tan^2 \beta}}$).
{\thinmuskip=0mu
\medmuskip=0mu
\thickmuskip=0mu
\begin{equation}\label{BIX_intrinsiceq_betaparametrization}
\begin{aligned}
\frac{\d \alpha}{\d \beta} ~=& ~ \frac{\gamma}{\sin^2 \beta ~ \sqrt{1 - \frac{\gamma^2}{\tan^2 \beta}}}  \,,
\\
\frac{\d \gamma}{\d \beta} ~=&~ \frac{\sigma^2}{2 \cos^2 \beta}    \left( \sqrt{1 -  \frac{\gamma^2}{\tan^2 \beta} }  \, \frac{\partial C}{\partial \alpha} - \gamma \cos^2 \beta \,  \frac{\partial C}{\partial \beta}   \right) e^{{\frac \sigma 2} \left(\omega - \sqrt{\tan^2 \beta - \gamma^2}  \right) } \,,
\\
\frac{\d \omega}{\d \beta} ~=&~ {\frac 1 2}\bigg{\{}  {\frac 4 3} s ~ \left[ \sigma \left(\sqrt{\tan^2 \beta - \gamma^2} - \omega \right) - 4 \right] \frac{C(\alpha,\beta)}{\cos^2\beta} \\
&~~~+\sigma^2 ~ |\cos \beta| \left[    \left( 2  \gamma^2 - \tan^2 \beta \right) +  \omega \sqrt{\tan^2 \beta - \gamma^2  } \right]  \frac{\partial C}{\partial \beta}
\\
&~~~+ \frac{\gamma ~\sigma^2}{\sin^2 \beta} \left( 2 \sqrt{\tan^2 \beta - \gamma^2}  - \omega \right)  \frac{\partial C}{\partial \alpha}  \bigg{\}} \frac{e^{{\frac \sigma 2} \left(\omega - \sqrt{\tan^2 \beta - \gamma^2}  \right) } }{\sqrt{1 - \frac{\gamma^2}{\tan^2 \beta}}}  \,,
\\
\frac{\d \sigma}{\d \beta} ~=& \frac{\sigma^2}{2} \bigg{[}  -\frac{\sigma}{\tan\beta}  \left(  \frac{\gamma}{\tan\beta} ~ \frac{\partial C}{\partial \alpha} 
+  s ~ \cos^2\beta \sqrt{\tan^2 \beta - \gamma^2} \frac{\partial C}{\partial \beta}\right)
\\
&~~~~~~~~~ {\sfrac 4 3} s ~C(\alpha,\beta)\bigg{]} \frac{e^{{\frac \sigma 2} \left(\omega - \sqrt{\tan^2 \beta - \gamma^2}  \right) } }{\cos^2 \beta ~ \sqrt{1 - \frac{\gamma^2}{\tan^2 \beta}}}\,.
\end{aligned}
\end{equation}}
This form of the equations holds whenever $\beta$ is monotonic in arc-length parametrization. This is in particular the case when the potential terms can be neglected. In this limit, the equations turn into Bianchi I equations (Kasner regime), whose solutions are straight lines in the $q^i$ plane, and the variables $\gamma$, $\omega$ and $\sigma$ become conserved:
\begin{equation}\label{BIeqs1}
\frac{\d \alpha}{\d \beta}  =  \frac{\gamma}{\sin^2 \beta \, \sqrt{1 - \frac{\gamma^2}{\tan^2 \beta}}}  \,, \qquad \frac{\d \gamma}{\d \beta} = \frac{\d \omega}{\d \beta} = \frac{\d \sigma}{\d \beta} = 0 \,.
\end{equation}
The general solution to the above equation is just Eq.~(\ref{GreatCircleEq_alpha}),
\begin{equation}
\alpha = \arcsin \left( \frac{- s \, \gamma}{\tan \beta} \right) + \text{\it{const.}}
\,,
~~ \gamma =  \text{\it{const.}} \,,
~~ \omega =  \text{\it{const.}} \,,
~~ \sigma =  \text{\it{const.}} 
\end{equation}
which represents a great circle on the shape sphere, which evolves smoothly through the equator. 

The ephemeris time/scale equations in this regime are,  in $\beta$-parametrization:
\begin{equation}
\frac{d \log p}{d \beta} = 0\,, ~~~ \frac{d \log v}{d \beta} = - \frac 3 8 \, s \,  \frac{\sigma}{\cos^2 \beta \sqrt{1 - \frac{\gamma^2}{\tan^2 \beta}}}   \,,
\end{equation}
whose solution is
\begin{equation}
p = 
\text{\it const.} \,,  ~~  v = v_\text{ref} \, e^{- \frac 3 8 \, s \, \sigma \sqrt{1 - \frac{\gamma^2}{\tan^2 \beta}} \tan \beta }  = v_\text{ref} \, e^{- \frac 3 8 \,  \sigma \sqrt{1 - \frac{\gamma^2}{\tan^2 \beta}} |\tan \beta| } \,,
\end{equation}
where  ${v}_\text{ref} $ is a reference scale chosen anywhere on the solution (except $\beta=\frac \pi 2$).
\index{homogeneous metrics} \index{Bianchi IX solutions}\index{continuation through the Big Bang singularity}

\subsubsection*{Continuing any quiescent Bianchi IX solution through the singularity}

So far I have shown that the equations evolve through $\beta=\frac \pi 2$ when the potential can be neglected (Bianchi I regime). To investigate the quiescent Bianchi IX regime  we need to find out what the conditions for the onset of `quiescence' are. The Hamiltonian constraint in the new variables is~(\ref{Dimensionless_BIX_Ham_const}), which I rewrite here  using the variable $\omega$ in place of $\kappa$:
\begin{equation}
1 + {\frac 1 2} \xi^2 -  \sigma^2 \left[ {\frac 3 8} +  e^{-\frac{\sigma}{2}   \left(\sqrt{\tan^2 \beta - \gamma^2} - \omega \right)}  C(\alpha,\beta) \right] \approx 0
\end{equation}
assuming quiescence, in the $\beta \to \frac \pi 2$ limit the constraint above becomes:
\begin{equation}
\sigma^2 \xrightarrow[\beta \to \frac \pi 2]{} {\frac 8 3}\left(  1 +{\frac 1 2} \xi^2\right) \,, 
\end{equation}
so if the scalar field is not zero $\sigma$ can take a continuous interval of values, while it is constrained if the scalar field is zero. In any case it has to respect the following bound (recall that $\sigma$ is positive definite):
\begin{equation}
\sigma \geq  \sqrt{\frac 8 3} \,.
\end{equation}
with the inequality only saturated when there is no scalar.

\subsubsection*{Bound on $\sigma$ from quiescence}

The `quiescence' bound on $\sigma$ comes from the requirement that the quantities
\begin{equation}\label{BIXpotentialTerms}
e^{ - {\frac \sigma 2} \sqrt{\tan^2 \beta - \gamma^2} } C(\alpha,\beta) \,, 
~~ 
e^{ - {\frac \sigma 2} \sqrt{\tan^2 \beta - \gamma^2} } \frac{\partial C}{\partial \alpha} \,,
~~
e^{ - {\frac \sigma 2} \sqrt{\tan^2 \beta - \gamma^2} } \frac{\partial C}{\partial \beta} \,,
\end{equation}¥
tend to zero as $\beta \to \frac \pi 2$, where the shape potential~(\ref{BIXpotential}) can be written as
\begin{equation}
\begin{aligned}
&C(\alpha,\beta) = \sum_{j=1}^6 c_j e^{\chi_j(\alpha) | \tan \beta |} \,,
\\
&c_j= \left(1,1,1,-{\sfrac 1 2},-{\sfrac 1 2},-{\sfrac 1 2} \right)\,,
\\
&\chi_j(\alpha) = \bigg{(} - {\sfrac{2\sin \alpha}{\sqrt 6}}, -{\sfrac{\sqrt{3} \cos \alpha - \sin \alpha}{\sqrt 6}},-{\sfrac{-\sqrt{3} \cos \alpha - \sin \alpha}{\sqrt 6}}, 
\\
& \qquad\qquad\qquad 2 {\sfrac{2\sin \alpha}{\sqrt 6}}, 2 {\sfrac{\sqrt{3} \cos \alpha - \sin \alpha}{\sqrt 6}}, 2{\sfrac{-\sqrt{3} \cos \alpha - \sin \alpha}{\sqrt 6}}\bigg{)} \,.
\end{aligned}
\end{equation}
We can extract the leading exponent from $C$ by taking
\begin{equation}
\chi_{\rm{max}} (\alpha)= \max \{ \chi_j(\alpha) , j=1,\dots 6 \}\,,
\end{equation}
and then rewrite $C$ as
\begin{equation}
C(\alpha,\beta) =e^{\chi_{\rm{max}} | \tan \beta |}  \left( c_{\rm{max}} + \sum_{j\neq \rm{max}}  c_j e^{\left[ \chi_j(\alpha)-\chi_{\rm{max}}(\alpha)\right]  | \tan \beta |} \right)  \,,
\end{equation}
then all the exponents $\chi_j(\alpha)-\chi_{\rm{max}}(\alpha)$ will be negative (this easily generalizes to the $\alpha$'s such that two exponents are identical).  Then 
\begin{equation}
\begin{aligned}
&e^{ - {\frac \sigma 2} \sqrt{\tan^2 \beta - \gamma^2} } C \\
&=e^{\chi_{\rm{max}} | \tan \beta | - {\frac \sigma 2} \sqrt{\tan^2 \beta - \gamma^2} }  \left( c_{\rm{max}} + \sum_{j\neq \rm{max}}  c_j e^{\left[\chi_j(\alpha)-\chi_{\rm{max}}(\alpha)\right] | \tan \beta |} \right)
\\
&\xrightarrow[\beta \to {\frac \pi 2}]{} e^{\left( \chi_{\rm{max}} - {\frac \sigma 2}\right) | \tan \beta |  }  \left( c_{\rm{max}} + \sum_{j\neq \rm{max}}  c_j e^{\left[\chi_j(\alpha)-\chi_{\rm{max}}(\alpha)\right] | \tan \beta |} \right)  \,,
\end{aligned}
\end{equation}\index{homogeneous metrics} \index{Bianchi IX solutions}\index{continuation through the Big Bang singularity}
and the condition for convergence of the above is simply  that the exponent of the prefactor is negative in the $\beta \to \frac \pi 2$ limit:
\begin{equation}\label{ConvergenceBound}
2 \, \chi_{\rm{max}}(\alpha) - \sigma < 0 \,.
\end{equation}
The above can be written as
{\thinmuskip=0mu
\thickmuskip=0mu
\thinmuskip=0mu
\begin{equation}\label{BoundOnSigma}
\begin{aligned}
\sigma > 2 \chi_{\rm{max}}(\alpha) \equiv  \sqrt{\sfrac 8 3} \max_{\alpha\in(0,2\pi]} \left( 2 \sin \alpha, \sqrt{3} \cos \alpha -  \sin \alpha , -\sqrt{3} \cos \alpha - \sin \alpha \right) \,.
\end{aligned}
\end{equation}}
in Fig.~\ref{BianchiIX_bound_in_new_variables_1} I show a plot of the bound.

\begin{figure}[t!]
\begin{center}
\includegraphics[width=0.375\textwidth]{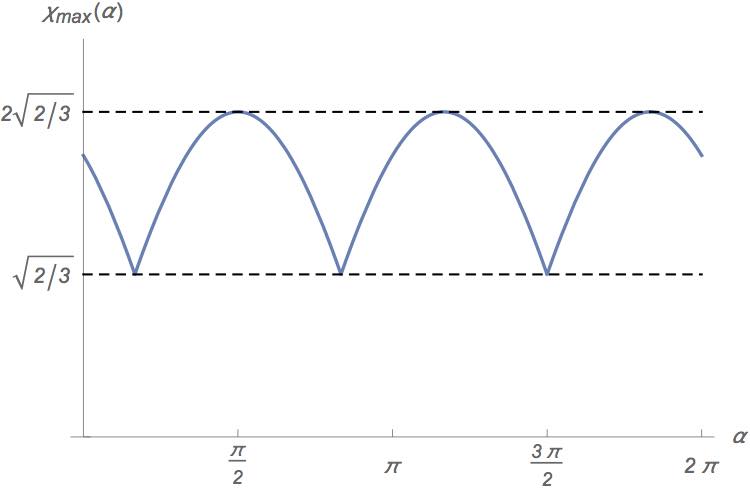}
\end{center}
\caption[Quiescence bound in dimensionless variables - 1]{Plot of the max of the coefficients $\chi_j(\alpha)$.}\label{BianchiIX_bound_in_new_variables_1}
\end{figure}

As a double-check, we can verify the following fact:
\begin{equation}
| \cos \beta | \, \log | C(\alpha,\beta) |  \xrightarrow[\beta \to \frac \pi 2]{}   \chi_{\rm{max}}(\alpha) \,,
\end{equation}
which of course implies the bound~(\ref{ConvergenceBound}). Plotting $|\cos \beta| \, \log | C(\alpha,\beta) |$ vs. $\alpha$ for a number of  values of $\beta$ that approach $\pi/2$ it can be seen that the function approaches $ \chi_{\rm{max}}(\alpha)$. This is shown in Fig.~\ref{BianchiIX_bound_in_new_variables_2}.  In the plot I highlighted (dashed line) the bound given by the Hamiltonian constraint $\sigma/2 \geq \sqrt{2/3}$. In absence of the scalar field $\sigma/2 = \sqrt{2/3}$ (dashed line), and there are only three points where the quiescent bound is respected. This proves the ``no quiescence without massless scalar field'' theorem: without scalar field you can achieve quiescence only if you stick on the three symmetry axes.
\index{homogeneous metrics} \index{Bianchi IX solutions}\index{continuation through the Big Bang singularity}

\begin{figure}[h!]
\begin{center}
\includegraphics[width=0.48\textwidth]{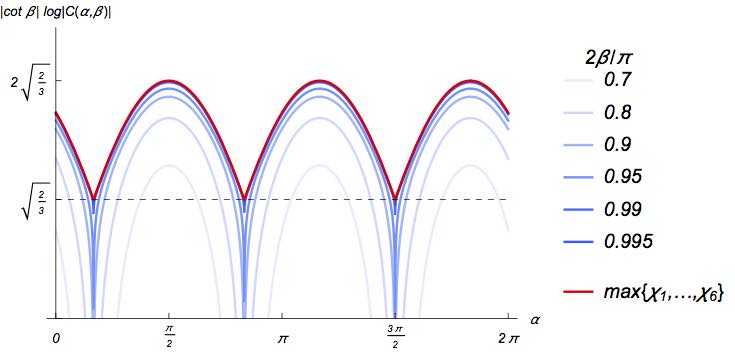}
\end{center}
\caption[Quiescence bound in dimensionless variables - 2]{Plot of $\cos \beta \, \log | C(\alpha,\beta) |$ vs. $\alpha$ for a series of values of $\beta$. It is apparent how the function approaches $\chi_{\rm{max}}(\alpha)$ as $\beta \to \frac \pi 2$.}\label{BianchiIX_bound_in_new_variables_2}
\end{figure}

\subsubsection*{Continuation through the Big Bang}

Now I am ready to prove the main thesis of this Section: Equations~(\ref{BIX_intrinsiceq_betaparametrization}) evolve uniquely through the equator $\beta = \pi/2$. To prove this I appeal to the Picard-Lindel\"of theorem~\cite{teschl2012ordinary}, which states that a system of ordinary differential equation like Eqs.~(\ref{BIX_intrinsiceq_betaparametrization}) admits a unique solution in an neughbourhood of a point if, in that neighborhood, the rhs is uniformly Lipschitz continuous in the dependent variables ($\alpha$, $\gamma$, $\omega$ and $\sigma$)  and continuous in the independent variable ($\beta$).\footnote{Similarly we could use the theorem to prove that the arc-length parametrized equations~(\ref{BIX_intrinsiceq_2}) admit a unique solution in a neighbourhood of a point on the equator by showing that  rhs is uniformly Lipschitz continuous in the dependent variables (in this case $\alpha$, $\beta$, $\gamma$, $\omega$ and $\sigma$) and continuous in the independent variable, which in this case is $\ell$.}
We can do more than that: we can prove that the rhs of Eqs.~(\ref{BIX_intrinsiceq_betaparametrization}) is differentiable in a neighbourhood of any point on the equator. To do this we first take the left- and right-limit $\beta \to \pi/2$ of the rhs, and show that they coincide:\index{homogeneous metrics} \index{Bianchi IX solutions}\index{continuation through the Big Bang singularity}
\begin{equation}
\begin{aligned}
\frac{\d \alpha}{\d \beta}
\xrightarrow[\beta \to {\frac \pi 2}^-]{}  
\gamma
 \xleftarrow[\beta \to {\frac \pi 2}^+]{}
 \frac{\d \alpha}{\d \beta} \,,
\\
\frac{\d \gamma}{\d \beta} 
\xrightarrow[\beta \to {\frac \pi 2}^-]{}  
0
 \xleftarrow[\beta \to {\frac \pi 2}^+]{}
\frac{\d \gamma}{\d \beta}  \,,
\\
\frac{\d \omega}{\d \beta} 
\xrightarrow[\beta \to {\frac \pi 2}^-]{}  
0
 \xleftarrow[\beta \to {\frac \pi 2}^+]{}
\frac{\d \omega}{\d \beta}  \,,
\\
\frac{\d \sigma}{\d \beta} 
\xrightarrow[\beta \to {\frac \pi 2}^-]{}  
0
 \xleftarrow[\beta \to {\frac \pi 2}^+]{}
\frac{\d \sigma}{\d \beta}  \,.
\end{aligned}
\end{equation}
Then we do the same for the left- and right- limit of the derivatives w.r.t. $\alpha$, $\beta$, $\gamma$, $\omega$ and $\sigma$:
{\thinmuskip=0mu
\medmuskip=0mu
\thickmuskip=0mu
\begin{equation}
\begin{aligned}
\frac{\partial}{\partial \alpha} \left( \frac{\d \alpha}{\d \beta} ,  \frac{\d \gamma}{\d \beta} ,  \frac{\d \omega }{\d \beta},  \frac{\d \sigma}{\d \beta}\right)
\xrightarrow[\beta \to {\frac \pi 2}^-]{}  
~(0,0,0,0)~
 \xleftarrow[\beta \to {\frac \pi 2}^+]{}
\frac{\partial}{\partial \alpha}\left( \frac{\d \alpha}{\d \beta} ,  \frac{\d \gamma}{\d \beta} ,  \frac{\d \omega }{\d \beta},  \frac{\d \sigma}{\d \beta}\right)\,,
\\
\frac{\partial}{\partial \beta} \left( \frac{\d \alpha}{\d \beta} ,  \frac{\d \gamma}{\d \beta} ,  \frac{\d \omega }{\d \beta},  \frac{\d \sigma}{\d \beta}\right)
\xrightarrow[\beta \to {\frac \pi 2}^-]{}  
~(0,0,0,0)~
 \xleftarrow[\beta \to {\frac \pi 2}^+]{}
\frac{\partial}{\partial \beta}\left( \frac{\d \alpha}{\d \beta} ,  \frac{\d \gamma}{\d \beta} ,  \frac{\d \omega }{\d \beta},  \frac{\d \sigma}{\d \beta}\right)\,,
\\
\frac{\partial}{\partial \gamma} \left( \frac{\d \alpha}{\d \beta} ,  \frac{\d \gamma}{\d \beta} ,  \frac{\d \omega }{\d \beta},  \frac{\d \sigma}{\d \beta}\right)
\xrightarrow[\beta \to {\frac \pi 2}^-]{}  
~(1,0,0,0)~
 \xleftarrow[\beta \to {\frac \pi 2}^+]{}
\frac{\partial}{\partial \gamma}\left( \frac{\d \alpha}{\d \beta} ,  \frac{\d \gamma}{\d \beta} ,  \frac{\d \omega }{\d \beta},  \frac{\d \sigma}{\d \beta}\right)\,,
\\
\frac{\partial}{\partial \omega} \left( \frac{\d \alpha}{\d \beta} ,  \frac{\d \gamma}{\d \beta} ,  \frac{\d \omega }{\d \beta},  \frac{\d \sigma}{\d \beta}\right)
\xrightarrow[\beta \to {\frac \pi 2}^-]{}  
~(0,0,0,0)~
 \xleftarrow[\beta \to {\frac \pi 2}^+]{}
\frac{\partial}{\partial \omega}\left( \frac{\d \alpha}{\d \beta} ,  \frac{\d \gamma}{\d \beta} ,  \frac{\d \omega }{\d \beta},  \frac{\d \sigma}{\d \beta}\right)\,,
\\
\frac{\partial}{\partial \sigma} \left( \frac{\d \alpha}{\d \beta} ,  \frac{\d \gamma}{\d \beta} ,  \frac{\d \omega }{\d \beta},  \frac{\d \sigma}{\d \beta}\right)
\xrightarrow[\beta \to {\frac \pi 2}^-]{}  
~(0,0,0,0)~
 \xleftarrow[\beta \to {\frac \pi 2}^+]{}
\frac{\partial}{\partial \sigma}\left( \frac{\d \alpha}{\d \beta} ,  \frac{\d \gamma}{\d \beta} ,  \frac{\d \omega }{\d \beta},  \frac{\d \sigma}{\d \beta}\right)\,.
\end{aligned}
\end{equation}}
The proof above can easily be completed by hand, by observing that the potential terms~(\ref{BIXpotentialTerms}) all go to zero (if quiescence is achieved, meaning that the bound~(\ref{ConvergenceBound}) is respected) like $\exp(- \text{\it{const.}} |\tan \beta|)$. Then all these terms, and  their derivatives w.r.t. $\alpha$, $\beta$, $\gamma$, $\omega$ and $\sigma$ all vanish at the equator. Then one is left with the Bianchi I terms, which are easily proved to be differentiable at the equator. Lipschitz continuity and continuity are weaker than differentiability, and therefore we have a proof of the main statement: each and every quiescent Bianchi IX solution can be continued uniquely through the singularity.
\index{homogeneous metrics} \index{Bianchi IX solutions}\index{continuation through the Big Bang singularity}

\newpage

\subsection{Spherically Symmetric Solutions}\label{SphericallySymmetricSec}
\index{spherically symmetric solutions}

As a warmup, let's start by writing the round metric on the 3-sphere $S^3$:
\begin{equation}\label{RoundMetricOnTheSphere}
\d s^2 = \d r^2 + \sin^2 r \left( \d \theta^2 + \sin^2 \theta \d \phi^2 \right) \,,
\end{equation}
where I used spherical coordinates $r \in [0,\pi]$, $\theta \in [0 , \pi]$, $\phi \in [0,2\pi)$.  Such a metric admits six killing vectors which close an $SO(4)$ algebra (verification left as an exercise). If we restrict to the killing vectors which leave invariant the two poles of the sphere, $r = 0$ and $r=\pi$, we find an $SO(3)$ subalgebra\footnote{This is called the \emph{isotropy subgroup}, or \emph{stabilizer} of $SO(4)$ with respect to the point $r=0$.}. These are the Killing vectors corresponding to 3D spherical symmetries. A spherically-symmetric tensor is defined as one whose Lie derivatives with respect to these three vectors vanish. As it turns out, the most general spherically-symmetric covariant 2-tensor which is symmetric in its two indices takes the form \index{Killing vector}
\begin{equation}
T_{ij} = \text{diag} \, \left\{  T_1(r) , T_2(r)  , T_2(r) \, \sin^2 \theta  \right\} \,,
\end{equation}
depending on only two arbitrary functions of the radial coordinate $r$. Similarly, a contravariant symmetric tensor takes the form
\begin{equation}
S^{ij} = \text{diag} \, \left\{  S_1(r) , S_2(r)  , S_2(r) \, \sin^{-2} \theta  \right\} \,,
\end{equation}
while a contravariant symmetric tensor \emph{density} gains an overall factor of $\sin \theta$, necessary to compensate the Jacobian of $SO(3)$ transformations when integrating over it:
\begin{equation}
Z^{ij} = \text{diag} \, \left\{  S_1(r) , S_2(r)  , S_2(r) \, \sin^{-2} \theta  \right\} \, \sin \theta \,.
\end{equation}
Finally, a contravariant vector field will only have one component, the radial one ~(verifying the last three statements is left as an exercise to the reader): 
\begin{equation}
X^i = \text{diag} \, \left\{  X_1(r) , 0,0  \right\} \,.
\end{equation}
For this reason we can take the following ansatz for the metric $g_{ij}$, the momenta $p^{ij}$ and the shift vector $\xi^i$:
\begin{equation} \label{SphericalSymmetry-g}
g_{ij} = \text{diag} \, \left\{  \mu^2 , \sigma  , \sigma \, \sin^2 \theta  \right\} \,,
\end{equation}
\begin{equation}\label{SphericalSymmetry-p}
p^{ij} =  \text{diag} \, \left\{  \frac{f}{\mu} , {\sfrac 1 2}  s   , {\sfrac 1 2}  s  \, \sin^{-2} \theta  \right\} \, \sin \theta \,,
\end{equation}
\begin{equation}\label{SphericalSymmetry-xi}
\xi^i = (\xi , 0 ,0 )\,,
\end{equation}
where $\mu$, $\sigma$, $f$, $s$ and $\xi$ are functions of $r$ and, possibly, time (and, obviously, $\mu^2>0$ and $\sigma>0$).

Notice that the above coordinate system can be defined on a compact or noncompact interval of $r$. The natural choice for the compact case is $r \in [0,\pi]$, such that the round metric on $S^3$ takes the form~(\ref{RoundMetricOnTheSphere}), while in the noncompact case $r \in [0, \infty)$, as one assumes in the case of the flat metric on $\mathbbm{R}^3$,
\begin{equation}
\d s^2 = \d r^2  + \d \theta^2 + \sin^2 \theta \d \phi^2  \,.
\end{equation}
Such choice is pure convenience and does not determine the metric topology (i.e. whether the metric makes the manifold compact). This is determined by the falloff conditions of the metric at the borders of the coordinate patch, which in turn are the borders of the interval on which $r$ is defined.

In order to have a compact metric topology, the metric has to satisfy the following regularity conditions at the two poles $r = r_a$, $a =\textrm{S},\textrm{N}$ (in addition to the ordinary ones $g_{ij} < \infty$, $g^{ij} < \infty$):
\begin{equation}\label{BoundaryConditionsMetric}
g_{rr} (r_a) \neq 0 \,, ~~~  g_{rr}'  , \,  g_{\theta\theta}  , \,  g_{\phi\phi} , \, g_{\theta\theta}' , \,  g_{\phi\phi}'  , \,   \xrightarrow[r \to r_a]{}  0 \,,
\end{equation}
where the symbol $'$ indicates the partial derivative with respect to $r$. The momentum tensor has to satisfy similar regularity conditions:
\begin{equation}\label{BoundaryConditionsMomenta}
p^{ij} < \infty , ~~~  (p^{rr})' , \,  p^{\theta\theta}  , \,  p^{\phi\phi}  , \,  (p^{\theta\theta})'  , \,  (p^{\phi\phi})' \xrightarrow[r \to r_a]{} 0 \,,
\end{equation}
the regularity conditions for a vector field are the easiest to understand:
\begin{equation}
\xi^i < \infty , ~~~ \xi^r (r_a) = (\xi^r)'(r_a) = 0 \,,
\end{equation}
these in fact ensure that the vector field is continuous and with continuous derivative in a neighbourhood of the pole.

\subsubsection*{Every spherically symmetric metric is locally conformally flat}
\index{spherically symmetric solutions}\index{conformal flatness}

The round metric on $S^3$~(\ref{RoundMetricOnTheSphere}) is locally conformally flat. In fact we can map it to the flat $\mathbbm{R}^3$ metric times a conformal factor with the following diffeomorphism:
\begin{equation}
\d r = \sin r \d y ~ \Rightarrow ~ y =  \log \tan {\sfrac r 2} \,, ~~~~  r = 2 \arctan e^y \,,
\end{equation}
then~(\ref{RoundMetricOnTheSphere})  transforms into
\begin{equation}
\d s^2 =  \frac{1}{\cosh^2 y} \left( \d y^2 + \d \theta^2 + \sin^2 \theta \d \phi^2 \right) \,.
\end{equation}
This proves that the $S^3$ metric is conformal to the $\mathbbm R^3$ flat metric, but this relation holds only locally. In particular it holds on any region that does not include the poles. In fact the diffeomorphism  $y =  \log \tan {\sfrac r 2} $ is not regular at the poles $r=0$ and $r=\pi$.

Similarly, we can find a diffeomorphism that puts a generic spherically-symmetric metric~(\ref{SphericalSymmetry-g}) into the form of a conformal factor times the round metric, by defining
\begin{equation}
\frac{\mu (r)}{\sqrt{\sigma(r)}} \d r = \frac{ \d y}{\sin y} \,,
\end{equation}
so that the metric reads
\begin{equation}
\d s^2 = \sin^2 y  ~\sigma(r(y)) ~ \left[ \d y^2 + \sin^2 y \left( \d \theta^2 + \sin^2 \theta \d \phi^2 \right)\right] \,.
\end{equation}
The coordinate transformation from $r$ to $y$ is not always well-defined (one says that this diffeomorphism gauge choice is not always \emph{attainable}), depending on the properties of the functions $\mu$ and $\sigma$.
It is obvious, then, that any spherically-symmetric metric is locally conformally flat (or \emph{conformally spherical}).
We reach the same conclusion if we calculate the Cotton tensor~(\ref{CottonTensor}) (introduced in Sec~\ref{CottonSquaredTheorySec}) of the metric~(\ref{SphericalSymmetry-g}): it is identically zero (another exercise for the reader).

A conformally flat geometry is but a point in shape space, so a vacuum spherically symmetric solution has no physical degrees of freedom. If we want a nontrivial dynamical model we have to add some form of matter. This is what I did in~\cite{CompactThinShellsPaper} (and its precursors~\cite{ThinShellPaper1,NoBirkhoff}), which I will explain below. To begin, I need to study the vacuum solutions of the ADM constraints in a CMC foliation.

\subsubsection*{Vacuum spherically symmetric solution of the ADM-CMC constraints}
\index{spherically symmetric solutions}

The vacuum\footnote{These constraints are not strictly the vacuum case: I included a cosmological constant, which doesn't complicate any calculation, and may be needed to ensure the existence and uniqueness of the solution to the constraints.} ADM constraints are
\begin{equation}
\begin{aligned}
\mathcal H  &=  \frac{1}{\sqrt g} \left( p^{ij} p_{ij} - {\frac 1 2} p^2 \right)
+ \sqrt g  (2\Lambda - R)   \,,
\\
\mathcal H_i  &= -2 \, \nabla_j p^j{}_i  \,,
\end{aligned}
\end{equation}
after replacing the spherically-symmetric ansatz, they turn into
\begin{equation}\label{SphericalADMconsts}
\begin{aligned}
&-\frac{1}{6 \sigma \mu ^2} \left[ \sigma ^2 \mu s^2 +  4 f^2 \mu ^3 -4 f \sigma \mu ^2 s + 12 \sigma \mu \sigma '' - 12 \sigma \sigma ' \mu ' - 3 \mu (\sigma')^2   \right. \\
& \left.- 12 \sigma \mu ^3 -( \langle p \rangle^2 - 12 \Lambda) \sigma ^2 \mu ^3  \right]   = 0\,,
\\
& \mu f' - {\sfrac 1 2} s \sigma' = 0 \,, 
\end{aligned}
\end{equation}
while the CMC constraint $p = \langle p \rangle \, \sqrt g$ reads
\begin{equation}
\mu f + s \sigma = \langle p \rangle \, \mu \, \sigma  \,.
\end{equation}
The last constraint can be solved algebraically, for example for $s$:
\begin{equation}\label{SolCMCconst}
 s  = \langle p \rangle \, \mu   - \frac{\mu}{\sigma} \, f  \,;
\end{equation}
and, after replacing~(\ref{SolCMCconst}) in~(\ref{SphericalADMconsts}), it is easy to see that the diffeomorphism constraint can be written as a total derivative:
\begin{equation}
\frac{\mu}{\sqrt \sigma} \left(  f \sqrt \sigma - {\sfrac 1 3} \langle p \rangle \, \sigma^{\frac 3 2} \right)'  = 0 \,.
\end{equation}
The solution of the last equation is
\begin{equation}\label{SolutionVacuumDiffConstraint}
 f = {\sfrac 1 3} \langle p \rangle \, \sigma  + \frac{A}{\sqrt \sigma}\,,
\end{equation}
where $A$ is an integration constant (meaning that it is spatially constant but can, in principle, still be a function of time. Finally, with a little work one can check that the Hamiltonian constraint can be rewritten as
\begin{equation}
\begin{aligned}
&- \frac{\sigma^{\frac 1 2} \mu}{\sigma'} \frac{\partial}{\partial r} \left[ \left(\frac{\sigma'}{\sigma^{\frac 1 4} \mu}\right)^2 - 4 \sqrt{\sigma} - \frac{f^2}{\sqrt \sigma} + {\sfrac 4 3} \Lambda\sigma^{\frac 3 2} \right] = \\
& \frac{2 \, f \,\mu}{\sigma'} \left( f' + {\sfrac 1 2} \, f \, \frac{\sigma'}{\sigma} - {\sfrac 1 2} \langle p \rangle \sigma' \right) \,,
\end{aligned}
\end{equation}\index{spherically symmetric solutions}
where the term $\left( f' + {\sfrac 1 2} \, f \, \frac{\sigma'}{\sigma} - {\sfrac 1 2} \langle p \rangle \sigma' \right)$ is identical to the diffeomorphism constraint and therefore vanishes on-shell. The remaining term is a total derivative, and we can solve the equation by introducing a new integration constant $m$,
\begin{equation}\label{DefinitionIntegrationConstant_m}
\frac{(\sigma')^2}{\sqrt{\sigma} \mu^2} - 4 \sqrt{\sigma} - \frac{f^2}{\sqrt \sigma} + {\sfrac 4 3} \Lambda\sigma^{\frac 3 2}  = - 8 \, m \,.
\end{equation}
Replacing the solution~(\ref{SolutionVacuumDiffConstraint}), and ordering terms by powers of $\sqrt{\sigma}$:
\begin{equation}
\frac{(\sigma')^2}{\sqrt{\sigma} \mu^2} - \frac{A^2}{\sigma^{\frac 3 2}}  + \left( 8 \, m - {\sfrac 2 3} \langle p \rangle  A \right) - 4 \sqrt{\sigma} + {\sfrac 1 9}  \left( 12 \, \Lambda - \langle p \rangle^2 \right) \sigma^{\frac 3 2}  = 0 \,.
\end{equation}
the above equation is a relation between $\sigma$ and $\mu$, and since the latter appears without derivatives, the easiest thing is to solve with respect to $\mu$:
\begin{equation}\label{VacuumSolutionForMu}
\mu^2 =  \frac{(\sigma')^2}{\frac{A^2}{\sigma }  + \left( {\sfrac 2 3} \langle p \rangle  A - 8 \, m  \right) \sqrt{\sigma}  + 4 \, \sigma - {\sfrac 1 9}  \left( 12 \, \Lambda - \langle p \rangle^2 \right) \sigma^2 } \,.
\end{equation}
We found a solution to all our constraints which apparently holds for any choice of the remaining free function $\sigma(r)$. This is a reflection of radial diffeomorphism invariance. In fact notice how $g_{rr} = \mu^2$ is homogeneous of degree two in $\sigma'$: the expression $\mu^2 \d r^2 \propto (\sigma' \d r)^2$ appearing in the metric is explicitly invariant under changes of radial coordinate.
However the choice of $\sigma(r)$ is not completely arbitrary. If we require regularity of the conformal geometry, there are obstructions to the values that $\sigma$ can take. In fact, by inspecting~(\ref{VacuumSolutionForMu}) we can see how the right-hand side is not guaranteed to be positive. All depends on the following sixth-order polynomial of $z = \sqrt{\sigma}/m$:
\begin{equation}\label{MordorPolynomial}
\mathscr P (z) = C^2 + \left( {\sfrac 1 3} \, \tau \, C - 2 \, \text{sign}(m) \right) \, z^3  + z^4 -  {\sfrac 1 {36}} \left(12 \, \lambda - \tau^2 \right) \, z^6  \,,
\end{equation}
where I made all parameters dimensionless by multiplying by the appropriate power of $m$:
\begin{equation}\label{DimensionlessVariablesMordor}
C = \frac{A}{2 \, m^2} \,, ~~~   \tau = m \, \langle p \rangle \,, ~~~ \lambda = m^2 \,\Lambda \,.
\end{equation}\index{spherically symmetric solutions}
Notice how the denominator of Eq.~(\ref{VacuumSolutionForMu}) coincides with $\frac{4 \, m^4}{\sigma}  \mathscr P \left(\frac{\sqrt{\sigma}}{m}\right)$. The polynomial~(\ref{MordorPolynomial}) may admit up to three real positive roots (because of Descartes' rule of signs: there are three possible sign-changes between coefficients). This means that, depending on the values of the coefficients $C^2$,  $ \left( {\sfrac 1 3} \, \tau \, C \pm 2\right)$ and ${\sfrac 1 {36}} \left(12 \, \lambda - \tau^2 \right) $ there may be intervals of values of $\sigma$ where the polynomial is negative. Such a situation is not problematic in GR: in those cases $\mu^2$ is negative and the 3-metric is Lorentzian. This just means that our CMC slices have turned timelike, and therefore the spatial metric changes its signature form Euclidean to Lorentzian somewhere. Here we see a core difference between SD and GR: in SD a signature change of the metric is a discontinuity of the conformal geometry, because the signature is a conformal invariant. Therefore in SD we have to assume that $\sigma$ is valued within the region(s) where the polynomial~(\ref{MordorPolynomial}) is positive. $\sigma$ might reach the boundary of such region(s), where the polynomial has a zero, but regularity of $\mu^2$ then demands that $\sigma$ reaches the boundary with zero derivative, in such a way that
\begin{equation}
\lim_{r \to \tilde r}  \mathscr P (\sqrt{\sigma}/m) = 0 
~~ \Rightarrow ~~
\lim_{r \to \tilde r}\left(  \frac{\sigma \, (\sigma')^2}{ \mathscr P (\sqrt{\sigma}/m) }  \right) < \infty  \,.
\end{equation}
\begin{figure}[t!]
\begin{center}
\includegraphics[width=0.46\textwidth]{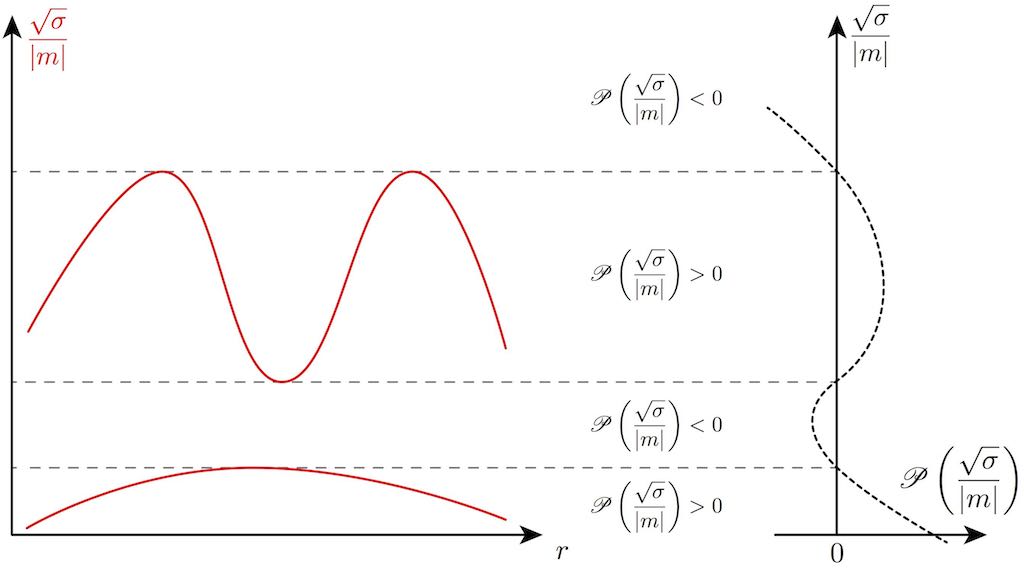}
\end{center}
\caption[Domain of areal radius]{On the right-hand side I show a possible  shape for the polynomial  $\mathscr P$ versus $\sqrt{\sigma}/m$ on the vertical axis. In parallel, on the left, I plotted two possible choices of $\sigma$ as a function of $r$. The intervals in which $\mathscr P>0$ from the left-hand side plot are projected onto the vertical axis of this last diagram, so that we one can see that $\sigma$ is confined within these intervals, and approaches their boundaries with zero derivative. Notice how in the upper interval, which is bounded from above and below, one can fit an arbitrary number of extrema of $\sigma(r)$, which can be made to `bounce' between these bounds an arbitrary number of times. In the case of the lower interval which is only bounded from above, $\sigma$ can have only one extremum: a maximum, and otherwise has to go to zero.\index{spherically symmetric solutions}}\label{DomainOfSigmaFig}
\end{figure}
In Fig.~\ref{DomainOfSigmaFig} I give an example of allowed choices of $\sigma(r)$ when $\mathscr P$ has three zeroes. Since $\sigma$ has to be continuous, and can only have extrema at the boundaries of the regions $\mathscr P >0$, it has to be monotonic in each interval between two consecutive extrema. This allows us to classify the possible behaviours of the metric at the poles.

Looking at the boundary conditions for the metric at the poles~(\ref{BoundaryConditionsMetric}) we see that we have to have $\sigma(r_a)  = 0$. Fortunately the polynomial is always positive at $\sigma = 0$ because the constant term $C^2>0$, so one of the positive intervals always include $\sigma =0$. A second condition from Eq.~(\ref{BoundaryConditionsMetric}) is that $\sigma' (r_a)=0$. But now look at Eq.~(\ref{VacuumSolutionForMu}): if $\sigma(r_a)  = \sigma' (r_a)=0$ then  the behaviour of $\mu$ at $r = r_a$ depends on the value of $A^2$. If $A\neq 0$, then $\mu(r_a) = 0$. This violates one of the conditions~(\ref{BoundaryConditionsMetric}), which states that $\mu$ has to be nonzero at the poles. So if we want a compact and regular metric around the poles  we need to have $A=0$. In that case 
\begin{equation}\label{DiscussionOfRegularityAtThePole}
\mu \simeq  \frac{(\sigma')^2}{- 8 m \sqrt{\sigma}} \,,
\end{equation}
and $\mu(r_a)$ might be made nonzero with an appropriate choice of $\sigma(r)$ (namely, that $\sigma'$ goes to zero as fast as $\sigma^{\frac 1 4}$), but its sign is decided exclusively by the sign of $m$. If $m>0$ then $\mu(r_a)<0$, which is unacceptable. So $m$ has to be negative. Then we can get a finite $\mu(r_a)$ if we assume $\sigma \sim a^2 \, (r-r_a)^{4/3} + \dots $ for small $(r-r_a)$. For $\mu^2$ we get
\begin{equation}
\mu^2  \sim - \frac 2 9 \left(  \frac{a^3}{m} + \frac{a^4 (r-r_a)^{\frac 2 3}}{2 \,m^2} + \dots \right) \,, 
\end{equation}
this implies
\begin{equation}
\mu^2 (r_a)  \sim  \frac{ 4 \, a^3}{ 9 \, m} \,, ~~~ \partial_r \mu^2 (r_a)  \to - \infty \,.
\end{equation}
Therefore $m<0$, $A=0$ is not acceptable either, because it violates the condition $(g_{rr})'(r_a) =0$. We conclude that in order to have regular poles, both integration constants $A$ and $m$ have to be zero. If we don't want to be confined to the completely trivial case of (Anti-)de Sitter spacetime, we have to couple the geometry to some matter degrees of freedom. Before doing that, I'll study the ADM equations of motion in vacuum, which will be necessary anyway also in the case with matter (because they will hold wherever the matter density is zero).

\subsubsection*{Equations of motion}

The ADM equations of motion require previous calculation of the CMC lapse, which is given by the following equation
\begin{equation}
\left( 8  \Delta -2 R +12 \Lambda  - \langle p\rangle ^2 \right) N 
- \frac{6 N}{g} \left(p-{ \sfrac 1 3} g \, \text{tr} p \right)^2
= \left\langle \sqrt g ~ \text{\it lhs} \right\rangle \,.
\end{equation}
Under the assumption of spherical symmetry  (which for a scalar function like the lapse is just $N=N(r)$), the lapse-fixing equation reduces to\index{spherically symmetric solutions}
\begin{equation}\label{SphSymmLFE1}
\begin{aligned}
\left(\frac{4 f     s }{\mu   \sigma  }  -\frac{4 f ^2   }{\sigma  ^2} -\frac{4    \mu'  \sigma' }{\mu^3 \sigma  }+\frac{4    \sigma'' }{\mu  ^2 \sigma  }-\frac{   (\sigma')^2}{\mu^2 \sigma^2}-\frac{4   }{\sigma}-\frac{   s^2}{\mu^2} \right)  N  +
\\
\left(12 \Lambda  - \langle p \rangle^2 \right) N  - \left( \frac{8 \mu'}{\mu  ^3}+\frac{8 \sigma' }{\mu^2 \sigma  }\right) N' + \frac{8 N'' }{\mu^2}
 =\left\langle \sqrt g ~ \text{\it lhs} \right\rangle  \,.
\end{aligned}
\end{equation}
The above equation is of the form $a(r)  N(r) + b(r) N'(r) + c(r) N''(r) = w$. The solution is a linear combination of the two linearly independent solutions of the homogeneous equation $a(r)  N(r) + b(r) N'(r) + c(r) N''(r) = 0$ plus a particular solution of the inhomogeneous one, which includes the constant term $w = \left\langle \sqrt g ~ \text{\it lhs} \right\rangle $.  We can look for a reformulation of Eq.~(\ref{SphSymmLFE1}) of the form
\begin{equation}\label{SphSymmLFE2}
F(r) \left( G(r) \left( H(r) \, N \right)' \right)' \approx w  \,,
\end{equation}
where the weak equality symbol $\approx$ indicates that the above equation is supposed to hold modulo the Hamiltonian, diffeomorphism and CMC constraints. 
As is easy to verify, the homogeneous part of Eq.~(\ref{SphSymmLFE2}) is formally solved by the linearly independent equations
\begin{equation}
N_1 =  \frac{1}{H(r)} \,, \qquad N_2 =  \frac{1}{H(r)} \int \frac{1}{G(r)} \d r \,,
\end{equation}
while for the particular solution of the inhomogeneous equation, we may use the ansatz
\begin{equation}
N_3 =  \frac{1}{H(r)} \int \frac{T(r)}{G(r)} \d r \,,
\end{equation}
which turns the equation into
\begin{equation}
F(r) \, T'(r)  \approx w  ~~~ \Rightarrow ~~~ T(r) = w \int \frac{1}{F(r)} \d r \,.
\end{equation}
Plugging the ansatz~(\ref{SphSymmLFE2}) into Eq.~(\ref{SphSymmLFE1}), we get that
\begin{equation}
H(r) = \frac{2 \mu \sqrt{\sigma}}{\sigma'}  \,, ~~~ G(r) = \frac{(\sigma')^2}{\mu^3} \,, ~~~ F(r) = \frac{4}{\sigma'\sigma^{1/2}}  \,.
\end{equation}
Then the formal solution can be written\index{spherically symmetric solutions}
\begin{equation}\label{SolutionLFEtwinshell}
N = \frac{\sigma'}{2 \mu \sqrt{\sigma}} \left( c_1 + c_2 \dashint \frac{\mu^3}{(\sigma')^2} \d r  + \frac w 6 \dashint \frac{\sigma^{3/2} \mu^3}{(\sigma')^2} \d r \right) \,,
\end{equation}
where $c_1$ and $c_2$ are the integration constants of the two solutions of the homogeneous equation. The symbol $\dashint$ refers to the principal value integral, which is needed because its arguments contain the term
\begin{equation}
\frac{\mu^3}{(\sigma')^2}  = \frac{|\sigma'|}{\left(\frac{A^2}{\sigma }  + \left( {\sfrac 2 3} \langle p \rangle  A - 8 \, m  \right) \sqrt{\sigma}  + 4 \, \sigma - {\sfrac 1 9}  \left( 12 \, \Lambda - \langle p \rangle^2 \right) \sigma^2\right)^{3/2}} \,,
\end{equation}
which diverges when $\sigma$ approaches a zero of $\mathscr P$ (which is an extremum of $\sigma$). This divergence has opposite sign on the two sides of the extremum (the left- and right- limite are opposite), and the degree of divergence is the same, so that the following quantity is finite:
\begin{equation}
\dashint_{r_1}^{r_2}\frac{\mu^3}{(\sigma')^2}   \d r = \lim_{\epsilon \to 0} \left( \int^{\tilde r - \epsilon}_{r_1} \frac{\mu^3}{(\sigma')^2} \d r  + \int_{\tilde r + \epsilon}^{r_2} \frac{\mu^3}{(\sigma')^2}   \d r \right) \,, ~~~ r_1 < \tilde r < r_2\,,
\end{equation}
where $\tilde r$ is the point where $\sigma$ has its extremum.

Once we have the lapse we can calculate the equations of motion for the metric
\begin{equation}\label{gdotEq}
\dot g_{ij} = \frac{2 N}{\sqrt g} \left( p_{ij}- {\sfrac 1 2} g_{ij} p \right) + \nabla_i \xi_j + \nabla_j \xi_i \,,
\end{equation}
using the spherical symmetry ansatz we get that the $\dot g_{\theta\theta}$ and  $\dot g_{\phi\phi}$  equations completely fix the shift vector:
\begin{equation}\label{SolEquation_gdotrr}
\xi_i = \delta^r{}_i  \left( f \, N + \dot \sigma \right) /\sigma' \,.
\end{equation}
Replacing the above solution of $\xi_i$ in the  $\dot g_{rr}$ equation (as well as the solutions of the ADM constraints), we find that it depend nontrivially on the lapse. Fortunately, replacing the solution of the lapse fixing equation~(\ref{SolutionLFEtwinshell}), we find that the two principal-value integrals never appear explicitly, they always have an $r$-derivative acting on them, and we can use the fact that
\begin{equation}
\partial_r \dashint \frac{\mu^3}{(\sigma')^2} \d r = \frac{\mu^3}{(\sigma')^2} \,, ~~~ \partial_r  \dashint \frac{\sigma^{3/2} \mu^3}{(\sigma')^2} \d r = \frac{\sigma^{3/2} \mu^3}{(\sigma')^2} \,,
\end{equation}
to simplify the expression. The equation then reduces to (with $\dot{\langle p \rangle}$ I mean $\partial_t \langle p \rangle$, that is, the derivative taken \emph{after} the spatial average)
\begin{equation}
\begin{aligned}
\left(\langle p \rangle (4\dot A  + 2 c_2) +A  (4 \dot{\langle p \rangle} + w)   - 96 \,\dot m \right) \sigma^{3/2} \\
+ \frac{\langle p \rangle}{3} \left(4 \dot{\langle p \rangle}+ w\right) \sigma^{3} + 6 A  \left(2 \dot A + c_2\right)= 0\,.
\end{aligned}
\end{equation}
In order for the above equation to hold for any choice of $\sigma(r)$ the only possibility is that\index{spherically symmetric solutions}
\begin{equation}\label{SolutionEqOfMotionThinShell}
\begin{aligned}
&c_2 = -2\,\dot A \,,
&&
w = - 4 \, \dot{\langle p \rangle} \,,
&&
\dot m  = 0 \,.
\end{aligned}
\end{equation}
We fixed two of the three integration constant present in the lapse, $c_2$ and $w$ (the third cannot be fixed because the system is reparametrization-invariant and one always has the freedom to specify the value of the lapse at a point). Moreover, we discovered a conserved quantity: the integration constant $m$.  This quantity has the significance of `mass-energy', and it is conserved because the system is spherically symmetric, so it cannot radiate its energy away in the form of gravitational waves, and a form of conservation of energy similar to what holds in field theory on Minkowsi spacetime holds. The quantity $m$ is actually what the \emph{Misner--Sharp mass}  \index{Misner--Sharp mass} reduces to in vacuum. This concept of quasi-local mass, introduced in 1964~\cite{MisnerSharpMass}, tries to capture this idea that in a spherically symmetric situation the only way mass-energy can escape from a sphere is by a physical flow of matter through the surface of the sphere. The Misner--Sharp mass is defined as
\begin{equation}
M_\st{MS} = \frac{\sqrt \sigma}{2} \left(  1 -  {^\st{(4)}g}^{\mu\nu}  \partial_\mu (\sqrt{\sigma}) \partial_\nu (\sqrt{\sigma}) \right) \,,
\end{equation}
where $\sqrt{\sigma} = g_{\theta\theta}$ is the \emph{aeral radius} coordinate of a spherically-symmetric metric. Using the definition of the ADM 4-metric, ${^\st{(4)}g}_{00} = - N^2 + g_{ij} \xi^i \xi_j$, ${^\st{(4)}g}_{0i} = g_{ij} \xi^j$, ${^\st{(4)}g}_{ij} = g_{ij}$, we get
\begin{equation}
M_\st{MS} 
=  \frac{\sqrt \sigma}{2} \left[ 1 -\frac{1}{4 \sigma }\left(\frac{(\sigma ')^2}{\mu ^2}-\frac{\left(\dot{\sigma }-\xi  \sigma '\right)^2}{N^2} \right) \right] \,,
\end{equation}
and, using Eq.~(\ref{SolEquation_gdotrr}) to replace $\xi$, and the maximal-slicing condition $s = - \mu f/\sigma$,
\begin{equation}
M_\st{MS} = \frac{\sqrt \sigma}{2}    -  \frac{(\sigma')^2}{8\sqrt{\sigma} \mu^2} + \frac{ f^2}{8\sqrt{\sigma}} = m \,,
\end{equation}
we get that $M_\st{MS}$ coincides with the expression~(\ref{DefinitionIntegrationConstant_m}) defining the integration constant $m$.

The equations of motion for the momenta are
\begin{equation}
\begin{aligned}\label{pdotEq}
\dot p^{ij} =& \frac{N}{2\sqrt g} g^{ij} \left( p^{k\ell} p_{k\ell} - {\sfrac 1 2} p^2\right)- \frac{2 N}{\sqrt g} \left( p^{ik} p_k{}^j - {\sfrac 1 2} p \, p^{ij} \right) 
\\
&  - N \sqrt g \left( R^{ij} - {\sfrac 1 2} g^{ij} R + \Lambda g^{ij} \right) 
\\
&+ \nabla_k (p^{ij} \xi^k) - p^{ik} \nabla_k \xi^j - p^{kj} \nabla_k \xi^i + \sqrt g \left( \nabla^i \nabla^j N - g^{ij} \Delta N \right)  \,,
\end{aligned}
\end{equation}
these equations are identically satisfied if one imposes the conditions (\ref{SolutionEqOfMotionThinShell}), and therefore add no further information. We have been able to solve exactly the spherically symmetric ADM-CMC equations in vacuum (plus, possibly, a cosmological constant). This is a significant result that can be exploited to build dynamically meaningful solutions of Shape Dynamics, for example by using some localized spherically symmetric distribution of matter, which leaves most of space empty.

\subsubsection*{Discussion of the roots of $\mathscr P$}
\index{spherically symmetric solutions}

I here reproduce Eq.~(\ref{MordorPolynomial}), rewriting it in a more convenient way:
\begin{equation}
\begin{aligned}
\mathscr P (z) &= C^2 + \left( {\sfrac 1 3} \, \tau \, C - 2 \, \text{sign}(m) \right) \, z^3  + z^4 - {\sfrac 1 {36}} \left(12 \, \lambda - \tau^2 \right) \, z^6 
\\
&=  {\sfrac{1}{36}}  \left(6 C+\tau  z^3\right)^2 - {\sfrac 1 3} \left(2 \, \text{sign}(m) \, z^3 +\lambda  \, z^6\right) + z^4 \,,
\end{aligned}
\end{equation}
this equation shows the polynomial $\mathscr P$ as a function of the dimensionless mass-weighted areal radius $z = \sqrt \sigma /m$, and of the dimensionless parameters $C = \frac{A}{2 \, m^2}$, $ \tau = m \, \langle p \rangle$, $\lambda = m^2 \,\Lambda$.

We can immediately see that there is a situation in which the polynomial is always positive: when both the cosmological constant and the Misner--Sharp mass are negative, $\lambda<0$ and $\text{sign}(m) = -1$ (of course $z$ is assumed positive). In this case the areal radius $\sqrt{\sigma}$ is allowed to take any value. In all the other cases, the condition $\mathscr P (z) >0$ gives the interval of values that $\sqrt{\sigma}$ is allowed to take, expressed in units of $m$. This condition defines a region in the space $z=\sqrt{\sigma} /m$, $C$ and $\tau$ which we can plot.

In Fig.~\ref{MordorFig1}, \ref{MordorFig2} and \ref{MordorFig3} I show the surface $\mathscr P (z) = 0$ for a set of values of $\lambda$ (positive, zero and negative), and for positive Misner--Sharp mass $m>0$. In Fig.~\ref{MordorFig3} I deal with the negative Misner--Sharp mass case, $m<0$. The axes of the three figures are compactified by taking the arctangent of the three dimensionless parameters, $z$, $\tau$ and $C$. 

Notice that the surface  $\mathscr P (z) = 0$  is not invariant under York-time reversal $\tau \to - \tau$, but instead it is invariant under simultaneous reversal of $\tau$ and $C$. In fact 
$C$ plays the role of a momentum, and momenta change sign upon time-reversal.\index{spherically symmetric solutions}
\begin{figure}[h!]
\framebox[0.49\textwidth]{\parbox{0.49\textwidth}{\center
$\bm \lambda < 0 \, , ~ \bm m>0$
\\
\includegraphics[width=0.23\textwidth]{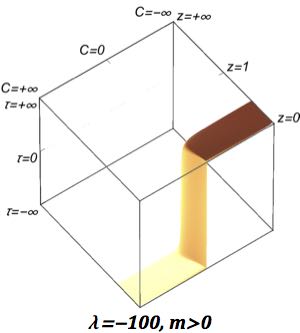}~\includegraphics[width=0.23\textwidth]{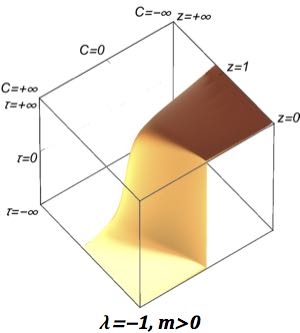}
}}
\caption[The surface $\mathscr P (z) =0$  for $m>0$, $\lambda < 0$]{The surface $\mathscr P (z) =0$ for \textbf{positive Misner--Sharp mass}  and two choices of \textbf{negative cosmological constant.}}\label{MordorFig1}
\end{figure}
\newpage
\begin{figure}[h!]
\framebox[0.49\textwidth]{\parbox{0.49\textwidth}{\center
$\bm \lambda \geq 0 \, , ~ \bm m>0$
\\
\includegraphics[width=0.23\textwidth]{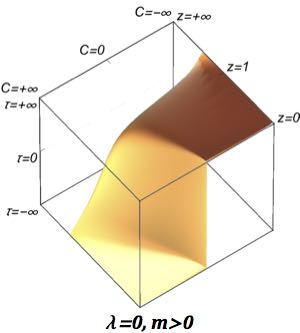}~\includegraphics[width=0.23\textwidth]{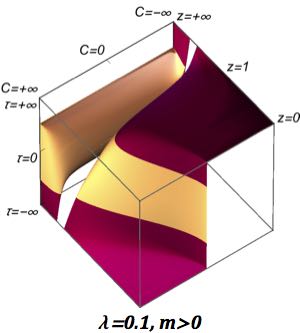}
\\
\includegraphics[width=0.23\textwidth]{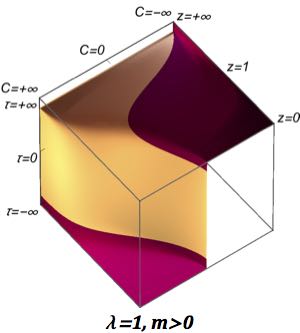}~\includegraphics[width=0.23\textwidth]{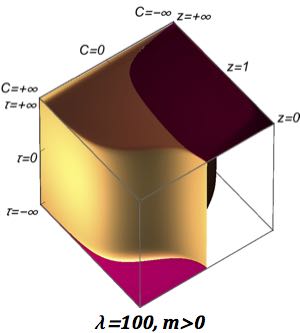}}}
\caption[The surface $\mathscr P (z) =0$  for $m>0$, $\lambda \geq 0$]{The surface $\mathscr P (z) =0$ for \textbf{positive Misner--Sharp mass}  and four choices of \textbf{zero or positive cosmological constant.}  The part of the surface where $\tau^2 < 12 \Lambda$ is in yellow, while the part $\tau^2 > 12 \Lambda$ is in red.}
\label{MordorFig2}
\end{figure}\index{spherically symmetric solutions}
\newpage
\begin{figure}[h!]
\framebox[0.49\textwidth]{\parbox{0.49\textwidth}{\center
$\bm \lambda > 0 \, , ~ \bm m<0$
\\
\includegraphics[width=0.23\textwidth]{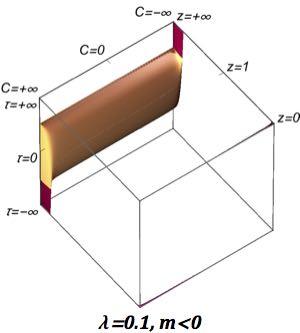}~\includegraphics[width=0.23\textwidth]{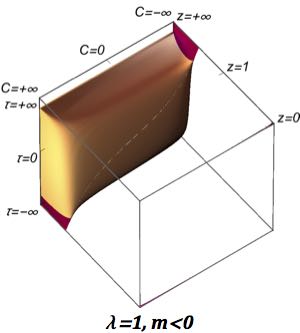}\\
\includegraphics[width=0.23\textwidth]{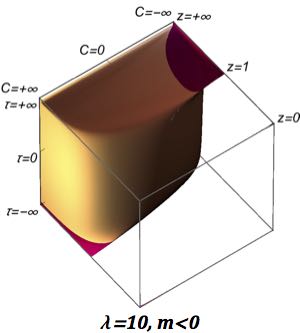}~\includegraphics[width=0.23\textwidth]{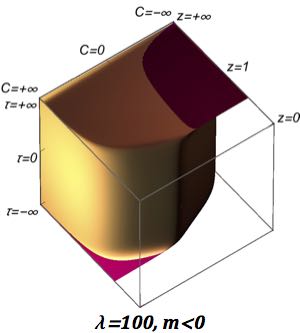}}}
\caption[The surface $\mathscr P (z) =0$  for $m<0$, $\lambda > 0$]{The surface $\mathscr P (z) =0$ for \textbf{negative Misner--Sharp mass}  and four choices of \textbf{positive cosmological constant.}  The cases of negative or zero cosmological constant are not included because the polynomial is always positive.}\label{MordorFig3}
\end{figure}\index{spherically symmetric solutions}
\newpage

The situation is different if the Misner--Sharp mass $m$ is zero, because we cannot use it to make everything dimensionless. In that case, however, we have a different quantity that is constant in time and we can use to rescale
everything: the cosmological constant $\Lambda$. We then introduce
$u = \sqrt{|\Lambda|} \sqrt{\sigma}$, $D = |\Lambda| \, A /2$, $t = \langle p \rangle/\sqrt{|\Lambda|}$
and the polynomial now takes the form
\begin{equation}
{\mathscr P}_0  (u) = D^2  +  {\sfrac 1 3} \, t \, D  \, u^3   + u^4 -   {\sfrac 1 {36}} \left(12 \, \text{sign} (\Lambda) - t^2 \right) \, u^6 \,,
\end{equation}
where again we see that if $\Lambda <0$ then ${\mathscr P}_0  (u)  >0$ for any value of $u$, $D$ and $t$. Therefore we are only left with the case $\text{sign} (\Lambda)  = +1$ to plot, in Fig.~\ref{MordorFig4}.
There would be a last case to consider, that of $\Lambda = m = 0$, but in this case the polynomial takes the form
\begin{equation}
m^4 \, \mathscr P \left({\sfrac{\sqrt{\sigma}}{m}}\right) =   {\sfrac{1}{36}}  \left(3 \, A + \langle p \rangle  \sigma^{\frac 3 2} \right)^2 + \sigma^2 \,,
\end{equation}
and is therefore always positive.\index{spherically symmetric solutions}
\begin{figure}[h!]
\begin{center}
\includegraphics[height=0.23\textwidth]{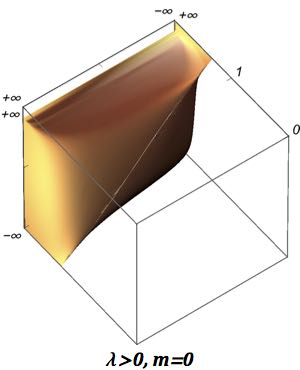}
\end{center}
\caption[The surface $\mathscr P (z) =0$  for $m=0$, $\lambda > 0$]{The surface $\mathscr P (z) =0$ for \textbf{zero Misner--Sharp mass}  and \textbf{positive cosmological constant.}}\label{MordorFig4}
\end{figure}
\newpage

\subsubsection*{Coupling a thin shell of dust}
\index{spherically symmetric solutions}\index{thin shell of dust}

Above I observed how, in order to have regular poles and a compact manifold, the integration constants $A$ and $m$ needed to be both zero, which completely trivializes the system. If we want a nontrivial dynamics we need to introduce some matter degrees of freedom. The simplest form of matter is a point particle, but a single one would break spherical symmetry, further complicating things. I will instead consider a collection of point particles, distributed on  the surface of a sphere, with momenta which are identical in norm and pointing radially.  In the limit of infinite, infinitesimally small-mass particles, one obtains a \emph{thin spherical shell of pressureless dust.} This is the system we are interested in.

We  are going to deduce the appropriate contribution of a thin shell of dust to the constraints of Shape Dynamics from that of a point particle. The Hamiltonian and diffeomorphism constraints of ADM gravity coupled to a massive point particle are
\begin{equation}
\begin{aligned}
\!\!\!\! &  \frac{1}{\sqrt g} \left( p^{ij} p_{ij} - {\frac 1 2} p^2 \right)
+ \sqrt g  (2\Lambda - R) + \delta^{(3)}(x^i- y^i) \sqrt{g^{ij} \, p_i \, p_j +m_0^2}  \,,
\\
\!\!\!\! & -2 \, \nabla_j p^j{}_i = \delta^{(3)}(x^i- y^i) \, p_i  \,,
\end{aligned}
\end{equation}
%
%
%
where $y^i$ are the coordinates of the particle, $p_i$ its momentum and $m_0$ its rest mass. The above constraints can be 
derived from the standard Einstein--Hilbert action coupled to a point particle, as shown in Appendix~\ref{AppendicePointParticle}. Note that $p_i$ is included as a cotangent vector, and this arises from minimal coupling.  It is not hard to show that the constraints above are first-class.

Now take a uniform distribution of point particles on the surface of a sphere, and take the continuum limit. The  constraints become
\begin{equation}
\begin{aligned}
\!\! &{\sfrac 1 {\sqrt g}} \left( p^{ij} p_{ij} - {\sfrac 1 2} p^2 \right) + \sqrt g  (2\Lambda - R) = - \sqrt h  \, \rho(R) \, \delta(r-R) \sqrt{g^{rr} \, p_r^2 +m_0^2 } \,, \\
\!\! & -2 \, \nabla_j p^j{}_i = \delta^r{}_i \, \sqrt h \, \rho(R) \, \delta (r-R) \, p_r \,,
\end{aligned}
\end{equation}
where $h_{ab}$ is the metric induced on the sphere by $g_{ij}$, and $\rho(R)$ is a scalar function to be determined [without weight: the additional weight $1/3$ is provided by the delta function $\delta (r-R)$].

To determine the function $\rho (R)$ we have to ask that changing the radius of the sphere 
does not change the number of particles $n$:
\begin{equation}
\int \, \d \theta \d \phi \d r \, \sqrt h  \, \rho(R) \, \delta(r- R)  = \int \d \theta \d \phi \, \sqrt{h(R,\theta,\phi)}  \, \rho(R) = 4 \pi \, n  \,,
\end{equation}
then we have
\begin{equation}
\rho(R) = \frac { 4 \pi \, n} { \int \d \theta \d \phi \, \sqrt{h(R,\theta,\phi)} } \,.
\end{equation}
now we can rescale the momentum $n \, p_r =  P$, and the mass $n \, m_0 = M$ of the single particle
into the momentum and the mass of the whole shell, so that $n$ drops out of the equations.
Now we can integrate over $\int d\theta d\phi$, 
\begin{equation}
\begin{aligned}
\!\!\!\!\!\!\!\!& \int \mathcal H \, d\theta d\phi   = - 4 \pi \, \delta(r-R) \sqrt{g^{rr} \, P^2 +M^2 }  \,, \\
\!\!\!\!\!\!\!\!& \int  \mathcal H_i \,  d\theta d \phi  =  4 \pi \, \delta^r{}_i \delta (r-R) \, P \,,
\end{aligned}
\end{equation}
and the three constraints, after assuming the spherically symmetric ansatz, look
\begin{eqnarray}\label{ConstraintsThinShell}
&&\begin{aligned}
&-\frac{1}{6 \sigma \mu ^2} \left[ \sigma ^2 \mu s^2 +  4 f^2 \mu ^3 -4 f \sigma \mu ^2 s + 12 \sigma \mu \sigma '' - 12 \sigma \sigma ' \mu ' - 3 \mu (\sigma')^2   \right. \\
& \left.- 12 \sigma \mu ^3 -( \langle p \rangle^2 - 12 \Lambda) \sigma ^2 \mu ^3  \right] =  \delta(r-R)  \sqrt{\frac{P^2}{\mu^2} +M^2}   \,,
\end{aligned}
\\
&& \mu f' - {\sfrac 1 2} s \sigma' = - {\frac {P} 2}  \delta(r-R)  \,, 
\\
&&\mu f + s \sigma = \langle p \rangle \, \mu \, \sigma  \,,
\end{eqnarray}
after solving the CMC constraint w.r.t. $s$, $ s  = \langle p \rangle \, \mu - \frac{\mu}{\sigma} f$ we can rewrite the  second constraint as
\begin{equation} \label{DiffeoConstraintWithShell}
\frac{\mu}{\sqrt \sigma} \left(  f \sqrt \sigma - {\sfrac 1 3} \langle p \rangle \, \sigma\right)'  = - {\sfrac 1 2}  P    \delta(r-R) \,.
\end{equation}\index{spherically symmetric solutions}\index{thin shell of dust}
The above equation has the form
\begin{equation}
F'(r) = G(r) \, \delta (r - r_0) \equiv G(r_0)  \, \delta (r - r_0) \,,
\end{equation}
in any open set which does not include $r_0$ the solution to such an equation is $F(r) = \text{\it const}$. But because of the delta function on the right-hand side we cannot assume the continuity of $F$. In fact, integrating the equation from $r=0$ to $r$ one gets
\begin{equation}
F(r) - F(0) = G(r_0) \, \Theta (r - r_0) + \text{\it const.} \,,
\end{equation}
where $ \Theta (x) = \left\{ \begin{aligned}
&0 \,, ~ x<0
\\
&1 \,, ~ x>0
\end{aligned}\right.$ is the Heaviside distribution. Alternatively we can write
\begin{equation}\label{FormaEqDiffconDelta}
F(r) = F_- \, \Theta (r_0-r) +F_+ \, \Theta (r - r_0) \,, \qquad  F_+ - F_- = G(r_0) \,.
\end{equation}
 So~(\ref{DiffeoConstraintWithShell}) is solved by
\begin{equation} \label{SolutionDiffeoForf}
  f (r) = {\sfrac 1 3} \langle p \rangle \, \sigma  + \frac{A_-}{\sqrt{\sigma}}  \Theta(R-r)+ \frac{A_+}{\sqrt{\sigma}}  \Theta(r-R) \,,
\end{equation}
with the `jump condition'
\begin{equation}\label{DeltaAjumpcondition}
 A_+ - A_- =  - \frac{\sigma^{\frac 1 2}(R) }{2\mu(R)} P  \,.\end{equation}
The Hamiltonian constraint can be written as
\begin{equation}
\begin{aligned}
&- \frac{\sigma^{\frac 1 2} \mu}{\sigma'} \frac{\partial}{\partial r} \left[ \left(\frac{\sigma'}{\sigma^{\frac 1 4} \mu}\right)^2 - 4 \sqrt{\sigma} - \frac{f^2}{\sqrt \sigma} + {\sfrac 4 3} \Lambda\sigma^{\frac 3 2} \right] = \\
& \frac{2 \, f \,\mu}{\sigma'} \left( f' + {\sfrac 1 2} \, f \, \frac{\sigma'}{\sigma} - {\sfrac 1 2} \langle p \rangle \sigma' \right) 
+ \delta(r-R)  \sqrt{\frac{P^2}{\mu^2} +M^2}\,,
\end{aligned}
\end{equation}
according to  the diffeomorphism constraint $\mu \left( f' +{\sfrac 1 2} f \frac{ \sigma' }{\sigma}- {\sfrac 1 2}  \sigma' \langle p \rangle \right) = - {\frac {P} 2}  \delta(r-R) $, so we may replace the second term above, getting 
\begin{equation}\label{fmultiplyingDelta}
\!\!\!\!\! \left[ \left(\frac{\sigma'}{\sigma^{\frac 1 4} \mu}\right)^2 - 4 \sqrt{\sigma} - \frac{f^2}{\sqrt \sigma} + {\sfrac 4 3} \Lambda\sigma^{\frac 3 2} \right]' =  \delta(r-R)  \left[ \frac{ f \,  P  - \sigma' \, \sqrt{\frac{P^2}{\mu^2} +M^2}}{{\sigma^{\frac 1 2} \mu}} \right]  ,
\end{equation}
which, again, is in the form~(\ref{FormaEqDiffconDelta}), and admits the solution (the $-8$ factor is there to identify $m_\pm$ with the Misner-Sharp masses of the respective regions)\index{spherically symmetric solutions}\index{thin shell of dust}
\begin{equation}
-  \left(\frac{\sigma'}{\sigma^{\frac 1 4} \mu}\right)^2 - 4 \sqrt{\sigma} - \frac{f^2}{\sqrt \sigma} + {\sfrac 4 3} \Lambda\sigma^{\frac 3 2}  =   - 8 \, m_-  \Theta(R-r) - 8 \, m_+  \Theta(r-R)  \,.
\end{equation}
The jump condition, however, in this case is less trivial to determine, because in the right-hand side of Eq.~(\ref{fmultiplyingDelta}) the discontinuous function $f$ is multiplying a Dirac delta, which is an expression we cannot make sense of (that equation should be taken as valid only outside of a neighbourhood of $r = R$). It is better to go back to the form~(\ref{ConstraintsThinShell}) of
the Hamiltonian constraint,
$$
\begin{aligned}
&-\frac{1}{6 \sigma \mu ^2} \left[ \sigma ^2 \mu s^2 +  4 f^2 \mu ^3 -4 f \sigma \mu ^2 s + 12 \sigma \mu \sigma '' - 12 \sigma \sigma ' \mu ' - 3 \mu (\sigma')^2   \right. \\
& \left.- 12 \sigma \mu ^3 -( \langle p \rangle^2 - 12 \Lambda) \sigma ^2 \mu ^3  \right] =  \delta(r-R)  \sqrt{\frac{P^2}{\mu^2} +M^2}   \,,
\end{aligned}
$$
and check which terms in the left-hand side can be divergent at $r = R$. $f$ and $s$ are not derived, and therefore they can contribute at most with a theta function, but not give any Dirac delta. $\mu$ and $\sigma$ have to be continuous because they are components of the metric.  Therefore their first derivatives $\mu'$ and $\sigma'$ can be at most discontinuous but not divergent, like $f$ and $s$. The second derivatives $\mu''$ and $\sigma''$, however, can be divergent, if the first derivatives are discontinuous. The only second derivative that appears is that of $\sigma$, so we can write
\begin{equation}\label{SingularPartOfSigma''}
\text{singular part of} \left(\frac{2 \, \sigma''}{\mu } \right)=  - \delta(r-R)   \sqrt{\frac{P^2}{\mu^2} +M^2}  \,.
\end{equation}

So we have to assume that
$\sigma$ is continuous, its first derivative has a jump, and its second derivative produces a
Dirac delta term. The prototype of one such  function is
\begin{equation}
y(r) = y_1(r) + (y_2(r) - y_2(R)) \Theta (r-R) \,,
\end{equation}
where $y_1(r)$ and $y_2(r)$ are continuous functions. Taking its second derivative:
\begin{equation}
y'' = y_1'' + y_2''  \Theta (r-R)+2 y_2'  \delta (r-R) +  (y_2 - y_2(R)) \delta' (r-R) \,,
\end{equation}
a distribution of the form $z(r) \delta' (r-R) $ is not simply  $z(R) \delta' (r-R) $:  
with a smearing it is easy to show that it is equivalent to the distribution $-z'(r) \delta (r-R) + z(R) \delta' (r-R)$, and in our case $z(R) = \lim_{r\to R}   (y_2(r) - y_2(R)) = 0$. Then the above equation reads
\begin{equation}
y''(r) = y_1''(r) + y_2''(r)  \Theta (r-R)+ y_2'(R)  \delta (r-R) \,.
\end{equation}
In terms of $y$, it is easy to see that the divergent term in $y''(r)$ \index{spherically symmetric solutions}\index{thin shell of dust}
can be written as $y_2'(R)  \delta (r-R) =  \left( \lim_{r\to R^+}  y'(r)  - \lim_{r\to R^-}  y'(r)  \right)  \delta (r-R) $. Then the jump condition for $\sigma''$ can be written as
\begin{equation} \label{lambdaJump}
\sigma '' = \left( \lim_{r\to R^+}  \sigma'(r)  - \lim_{r\to R^-}  \sigma'(r)  \right)   \delta  (r-R) + \text{\it regular part} \,.
\end{equation}
This (times $2/\mu$) is the only divergent part of Eq.~(\ref{SingularPartOfSigma''}), and therefore we can identify it with the right-hand side:
\begin{equation}
 \lim_{r\to R^+}  \sigma'(r)  - \lim_{r\to R^-}  \sigma'(r)   = - {\sfrac 1 2} \sqrt{P^2 + M^2 \mu^2(R)} \,,
\end{equation}
to produce our second jump condition.

It is convenient, at this point, to define some quantities which will appear in all the jump conditions below:
\begin{equation} \label{EqgttDotAtR}
 \lim_{r\to R^+}  \sigma'(r) = \gamma \,, \qquad   \lim_{r\to R^-}  \sigma'(r) = \kappa \,, \qquad \sigma(R) = \rho^2 \,,
\end{equation}
whatever diffeomorphism gauge we choose, around $r=R$,  $\sigma(r)$ can be written as
\begin{equation}
\sigma(r) \simeq \rho^2 + \gamma \, (r-R)   \Theta(r-R) +\kappa \, (r-R)   \Theta(R-r) + \mathcal O \left( (r-R)^2\right) \,,
\end{equation}
then, from the expression above, it is easy also to deduce that
\begin{equation}
 \lim_{r\to R^+} \dot  \sigma(r) = 2 \rho \dot \rho -  \gamma \dot R  \,, \qquad  \lim_{r\to R^-} \dot  \sigma(r) = 2 \rho \dot \rho -  \kappa \dot R  \,,
\end{equation}
this relation will be useful later.

We are now in position to demand the continuity of $\mu$. Its expressions inside and 
outside of the shell do not coincide:
\begin{equation}
\mu(r) = \left\{ \begin{array}{ll} 
 \frac{|\sigma'|}{\sqrt{\frac{A_-^2}{\sigma }  + \left( {\sfrac 2 3} \langle p \rangle  A_- - 8 \, m_-  \right) \sqrt{\sigma}  + 4 \, \sigma - {\sfrac 1 9}  \left( 12 \, \Lambda - \langle p \rangle^2 \right) \sigma^2 }} &  r<R
\\
 \frac{|\sigma'|}{\sqrt{\frac{A_+^2}{\sigma }  + \left( {\sfrac 2 3} \langle p \rangle  A_+ - 8 \, m_+  \right) \sqrt{\sigma}  + 4 \, \sigma - {\sfrac 1 9}  \left( 12 \, \Lambda - \langle p \rangle^2 \right) \sigma^2} }  & r>R
\end{array} \right.
\end{equation}
and we have to demand that the left and right limit of $\mu$ coincide:
\begin{equation}
 \lim_{r\to R^+} \mu(r) =  \lim_{r\to R^-} \mu(r) \,,
\end{equation}
that is,
{\thickmuskip=0mu
\thinmuskip=0mu
\medmuskip=0mu
\begin{equation}\label{ContinuityOfMu}
\!\!\!  \frac{|\kappa|}{\sqrt{ \left(\frac {A_- + \frac 1 3 \langle p \rangle \rho^3 } {2 \rho^{2}}\right)^2  - \frac{2 \, m_-}{\rho} +  1  - \frac{\Lambda  \rho^2}{3}     } }= \frac{|\gamma|}{\sqrt{ \left(\frac {A_+ + \frac 1 3 \langle p \rangle \rho^3 } {2 \rho^{2}}\right)^2  - \frac{2\,m_+}{\rho} +  1  - \frac{\Lambda  \rho^2}{3}     } } \,,
\end{equation}}
which is a new equation we have to take into account, together with the jump conditions above,
which, in the new notation, can be written as 
\begin{equation}\label{JumpGammaKappa}
\gamma  - \kappa   = - {\sfrac 1 2} \sqrt{P^2 + M^2  \mu^2(R) } \,.
\end{equation}\index{spherically symmetric solutions}\index{thin shell of dust}

\subsubsection*{Symplectic structure}
\index{spherically symmetric solutions}\index{thin shell of dust}\index{symplectic structure}

In order to discuss the dynamics of the system, we need to know which of the reduced-phase-space variables are canonically conjugate to each other. In other words, we need to calculate the symplectic form. By definition, the conjugate variables of the extended phase space are $g_{ij}$ and $p^{ij}$, as well as $R$ and $P$. Therefore the pre-symplectic potential is
\begin{equation}
\theta = \int_{S^3} \d r \d \theta \d \phi \, p^{ij} \, \delta g_{ij} + 4 \pi  \, P \delta R \,,
\end{equation}
restricting it through spherical symmetry and integrating in $\d\theta \d\phi$ we get
\begin{equation}
\theta = 4 \pi  \int_0^\pi \d r  \left( 2 f \, \delta \mu + s \, \delta \sigma\right)+ 4 \pi \, P \delta R \,.
\end{equation}
Now we may impose the CMC constraint~$\mu \, f = \mu \, \langle p \rangle \, \sigma - s \, \sigma$, and the solution to the diffeomorphism constraint~(\ref{SolutionDiffeoForf}),
\begin{equation}
\begin{aligned}
\theta &= 4 \pi  \int_0^\pi \d r  \left( 2 f \, \delta \mu -\frac{ \mu \, f}{\sigma} \delta \sigma + \langle p \rangle \, \mu \, \delta \sigma \right) + 4 \pi \, P \delta R  \\
&= - 4 \pi  \int_0^\pi \d r \frac{2 \mu}{\sqrt{\sigma}} \delta (f \sqrt \sigma)+ \langle p \rangle 4 \pi  \int_0^\pi \d r \mu \delta \sigma  + 4 \pi \, P \delta R  \\
&= - \frac{8 \pi}{3}  \int_0^\pi \d r \,\sigma \, \mu \, \delta \langle p \rangle  + 4 \pi  \, P \delta R \\
&-  8 \pi \int_0^\pi  \frac{\mu}{\sqrt \sigma}  \delta \left[ A_-  \, \Theta(R-r) +  A_+ \, \Theta(r-R)  \right]
\,.
\end{aligned}
\end{equation}
now, using Eq.~(\ref{DeltaAjumpcondition}) we observe that
\begin{equation}
\begin{aligned}
- 8 \pi  \int_0^\pi \frac{\mu}{\sqrt \sigma} \delta \left[  A_-  \, \Theta(R-r) +  A_+  \, \Theta(r-R)  \right] \\
= -  8 \pi  \left[ \delta  A_-  \int_0^{R} \d r  \frac{\mu}{\sqrt \sigma}  + \delta  A_+  \int_{R}^{\pi} \d r  \frac{\mu}{\sqrt \sigma}   \right]
\\ + 8 \pi \left[ (A_+  - A_-)  \frac{\mu(R)}{\sqrt{\sigma(R)}} \delta R\right] \\
= -  8 \pi  \left[ \delta  A_-  \int_0^{R} \d r  \frac{\mu}{\sqrt \sigma}  + \delta  A_+   \int_{R}^{\pi} \d r  \frac{\mu}{\sqrt \sigma}  \right]
- 4 \pi \, P \, \delta R   \,,
\end{aligned}
\end{equation}
and therefore the symplectic potential reduces to
\begin{equation}\label{SingleShellSymplectic1}
\begin{aligned}
\theta &=  - \frac{2}{3} V \delta \langle p \rangle -  8 \pi  \left[ \delta  A_-  \int_0^{R} \d r  \frac{\mu}{\sqrt \sigma}  + \delta  A_+  \int_{R}^{\pi} \d r  \frac{\mu}{\sqrt \sigma} \right]  \,,
\end{aligned}
\end{equation}
where $V = 4 \pi  {\displaystyle \int_0^\pi \d r} \, \sigma \, \mu $ is the on-shell volume.
\index{spherically symmetric solutions}\index{thin shell of dust}\index{symplectic structure}

Now, in isotropic gauge $\mu = \sqrt \sigma/\sin r$,~(\ref{SingleShellSymplectic1}) becomes
\begin{equation}\label{IsotropicSymplectic0}
\begin{aligned}
\theta &=  - \frac{2}{3} V \delta \langle p \rangle  +  8 \pi   ( \delta A_+ - \delta A_-) \log \left(\tan \frac {R} 2 \right)     \,,
\end{aligned}
\end{equation}
and, recalling Eq.~(\ref{DeltaAjumpcondition}), $ A_+ - A_- =  -  {\frac 1 2}  P \sin R   $, we get
\begin{equation}
\theta =  - \frac{2}{3} V \delta \langle p \rangle  - 4 \pi   \log \left(\tan \frac {R} 2 \right)   \delta \left(  P  \sin R  \right)  \,,\end{equation}
which, modulo an exact form, is identical to
\begin{equation}\label{IsotropicSymplectic}
\theta =  - \frac{2}{3} V \delta \langle p \rangle  - 4 \pi R \delta  P  \,.
\end{equation}
Everything I said so far applies identically to more than one shell. For example, the symplectic potential in isotropic gauge with many shells turns into $\theta =  - \frac{2}{3} V \delta \langle p \rangle  - 4 \pi  \sum_a R_a \delta P_a$.

At this point, we need to stop a moment to think about the consequences of having a single shell of dust in an otherwise empty universe. If we want a compact manifold, both $A$ and $m$ have to be zero at both poles, which with only one shell means $A_-=A_+ = m_- = m_+ =0$. Then from the jump condition associated to the diffeomorphism constraint~(\ref{DeltaAjumpcondition}) we deduce that $P = 0$: the shell cannot have a nonzero momentum. This in fact completely trivializes the dynamics of the shell, which has to be in static equilibrium throughout the whole solution. Such a solution, however trivial, is nevertheless a solution and should therefore be considered. For this reason, before generalizing the system, I will now analyze this case.

\index{spherically symmetric solutions}\index{thin shell of dust}\index{symplectic structure}
\subsubsection*{Single shell in static equilibrium}
\index{spherically symmetric solutions}\index{thin shell of dust}

\begin{figure}[t!]\begin{center}
\includegraphics[width=0.4\textwidth]{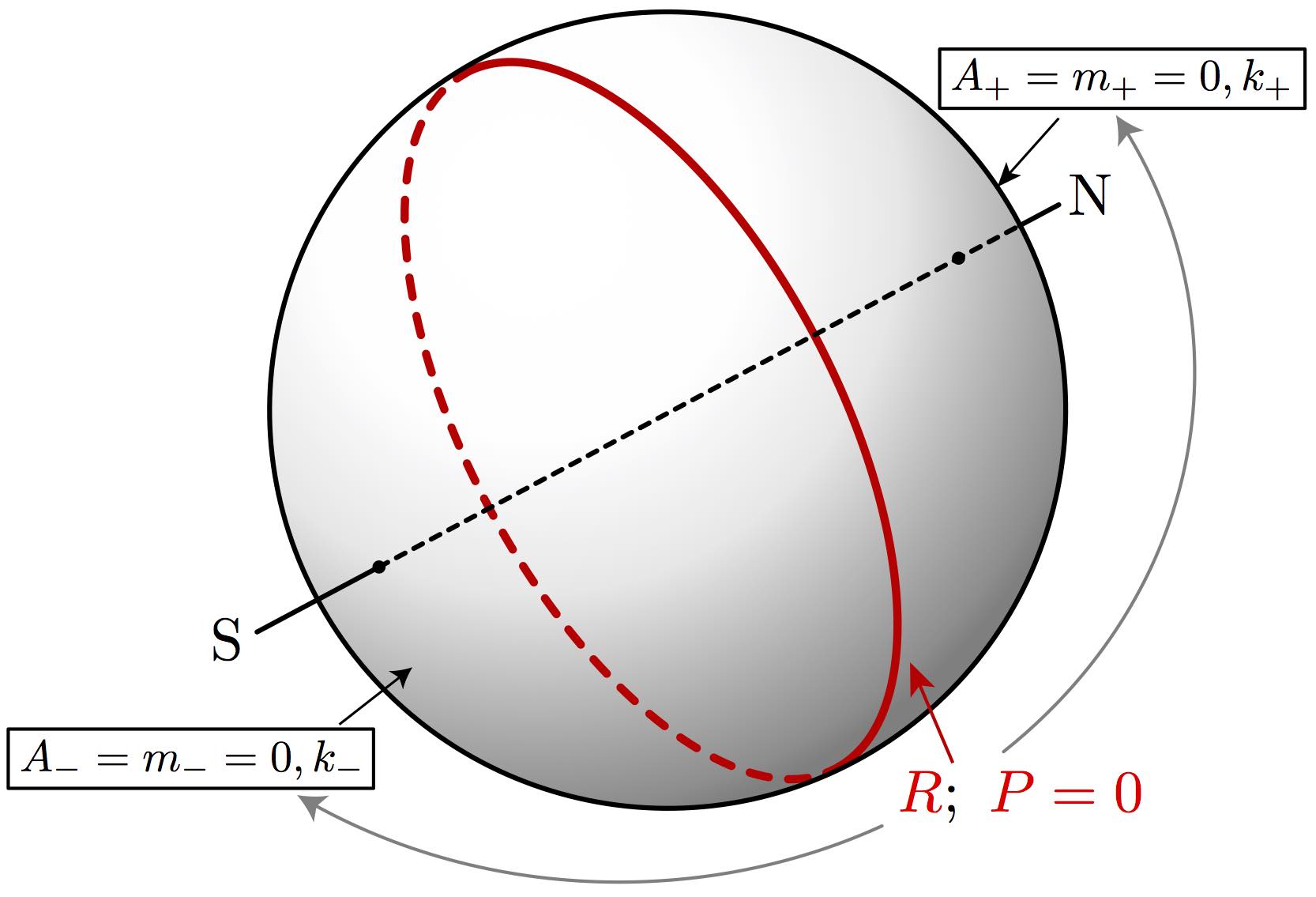}
\caption[Single shell universe diagram]{The `single shell' universe: the spatial manifold has the topology of the sphere $S^3$ and contains one thin shell  which divides the manifold into  the N and S polar regions. Both regions have to have $A_+=A_-=m_+=m_-=0$ in order for the geometry to be regular at the poles. The shell is characterized by a coordinate-radius degree of freedom $R$ and a radial-momentum degree of freedom $P$,   which are related to the jump in the integration constants $A$, $m$ and $k$. Since $A_+ = A_- = 0$, the momentum of the shell, $P$ is forced to vanish.}\label{SingleShell_SphereDiagram}
\end{center}
\end{figure}

If we include only one shell of dust, the manifold is divided into two regions (we will call them `$+$' and `$-$') which include a pole. Therefore the integration constants $A$ and $m$ are zero in both regions, $A_-=A_+ = m_- = m_+ =0$. Using this in the solution of the constraints, Eq.~(\ref{VacuumSolutionForMu}),
\begin{equation}
\mu^2(r) = \frac{(\sigma')}{4 \sigma - {\frac 1 9} (12 \Lambda - \langle p\rangle^2) \sigma^2} ~~~ \forall r < R\,, r > R \,.
\end{equation}
Then the continuity of $\mu$ across the shell imposes that $\displaystyle |\lim_{r \to R^+}\sigma'(r)| = |\lim_{r \to R^-} \sigma'(r)|$. Now, Eq.~(\ref{JumpGammaKappa}) imposes that
\begin{equation}
\lim_{r\to R ^+}  \sigma'(r)  - \lim_{r\to R ^-}  \sigma'(r)   = - {\sfrac 1 2} \sqrt{P ^2 + M ^2 \mu^2(R )} \,,
\end{equation}
while the first sets $P \approx 0$, because $A_+=A_-=0$. So, unless $M=0$, the left and right limits of $\sigma'$ must be equal in magnitude but opposite in sign:
\begin{equation}\label{SigmaDerivativeAtTheShell}
\lim_{r\to R ^+}  \sigma'(r)  = - \lim_{r\to R ^-}  \sigma'(r) =  -  {\sfrac 1 4} \sqrt{P^2 + M ^2  \mu^2(R ) } \approx  - {\sfrac 1 4}  M   \mu (R )  \,.
\end{equation}
Assuming $A_-=A_+ = m_- = m_+ =0$ we can calculate explicitly the metric in isotropic coordinates, such that $\d s^2 =  \mu^2(r) \left[ \d r^2 + \sin^2 r \left( \d \theta^2 + \sin^2 \theta \d \phi^2 \right) \right] $, which implies $\sigma = \sin^2 r \, \mu^2$. This last condition can be considered as a differential equation for $\sigma$:
\begin{equation}\label{SingleShell_IsotropicEquation}
\frac{(\sigma')^2}{4 \sigma - {\sfrac 1 9}  \left( 12 \, \Lambda - \langle p \rangle^2 \right) \sigma^2 } = \frac{\sigma}{\sin^2 r} \,,
\end{equation}
which is solved by
\begin{equation}\label{SingleShellSigma}
\sigma  = \frac{36}{12 \Lambda -\langle p \rangle^2} \left[ 1 - \left( \frac{1- k^2 \tan^2 \frac r 2 }{1 + k^2 \tan^2 \frac r 2} \right)^2 \right] \,.
\end{equation}\index{spherically symmetric solutions}\index{thin shell of dust}
In order for the solution above to be positive, we should assume $k$ real if $12 \Lambda -\langle p \rangle^2$, and imaginary otherwise. In other words, the quantity $k^2/(12 \Lambda -\langle p \rangle^2)$ is always positive.

\begin{figure}[t!]
\begin{center}
\includegraphics[width=0.4\textwidth]{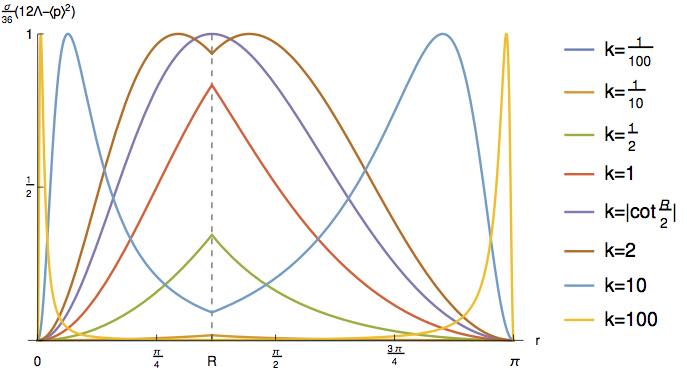}
\caption[Solutions of the constraints for the single shell universe]{Plot of the solution of Eq.~(\ref{SingleShellSigma2}) for a particular choice of $R $ for various values of $k$.}\label{SingleShellSigmaPlot2}
\end{center}
\end{figure}

As stated in Eq.~(\ref{SigmaDerivativeAtTheShell}) the function $\sigma(r)$ needs to flip sign of its derivative at  $r = R $.  If we call $k_-$$(k_+)$ the value of the integration constant $k$ at the left (right) of the shell, we can impose
\begin{equation}
\lim_{r\to R ^-} \sigma'(k_-,r) 
=
 - \lim_{r\to R ^+}\sigma'(k_+,r)  \,,
\end{equation}
such an equation admits the following solution:
\begin{equation}
k_+ = \frac{\cot^2 \frac {R } 2}{k_-}  \,,
\end{equation}
which implies that the full expression of $\sigma(r)$ at each side of the shell is
\begin{equation}\label{SingleShellSigma2}
\sigma =  {\sfrac{36}{12 \Lambda -\langle p \rangle^2}} \times   \left\{
\begin{array}{ll}
  1 - \left( \frac{1- k^2 \tan^2 \frac r 2 }{1 + k^2 \tan^2 \frac r 2} \right)^2   & \text{\it for } r<R  
\\
  1 - \left( \frac{k^2- \cot^4 \frac {R } 2 \,  \tan^2 \frac r 2 }{k^2  + \cot^4 \frac {R } 2 \, \tan^2 \frac r 2} \right)^2  
& \text{\it for } r>R 
\end{array}\right. \,,
\end{equation}
in Fig.~\ref{SingleShellSigmaPlot2} I plot the function $\sigma$ of Eq.~(\ref{SingleShellSigma2}), divided by $\frac{12 \Lambda -\langle p \rangle^2}{36}$ for a range of values of $k$ and for a particular choice of $R$.  
We then have a 1-parameter family of metrics which are exact solutions of the local parts of the constraints. All that is left over to solve are the jump conditions. The diffeomorphism one simply reduces to the constraint $P \approx 0$, while the Hamiltonian one takes a more complicated functional form:\index{spherically symmetric solutions}\index{thin shell of dust}
\begin{equation}
\begin{aligned}
h(k,R ,\langle p \rangle )  &= 
\lim_{r\to R ^+} \sigma'(r)  - \lim_{r\to R ^-} \sigma'(r)  + {\sfrac 1 2}    \sqrt{P ^2 + M ^2  \sigma(R ) \sin^{-2} R  }
\\
&=
\frac{8 \, k^2 \cos \left(\frac{R}{2}\right) \left[ k^2-\cot ^2\left(\frac{R}{2}\right) \right]}{\sin ^3\left(\frac{R}{2}\right) \left[ k^2+\cot ^2\left(\frac{R}{2}\right)\right]^3}  + {\sfrac 1 2}    \sqrt{P ^2 + M ^2  \sigma(R ) \sin^{-2} R  } \approx 0 \,.
\end{aligned}
\end{equation}
We are left with a 4-dimensional phase space, $(R,P,k,\langle p \rangle)$ and two constraints: $P\approx0$, $h \approx 0$. We need to perform a Dirac analysis and check whether the constraints are first- or second-class. To do so we need to calculate the symplectic form. Recall from the previous Section that the symplectic potential in isotropic gauge is $\theta =  - \frac{2}{3} V \delta \langle p \rangle  - 4 \pi R  \delta  P $. The volume $V$ is a function of $R $, $k$ and $\langle p \rangle$:
\begin{equation}\label{SingleShellExactVolume}
\begin{aligned}
&V= 4 \pi \int \d r \sigma \mu = 8 \pi \int_0^{\sigma(R )}  \frac{\sigma \d\sigma}{\sqrt{4 \sigma - {\sfrac 1 9}  \left( 12 \, \Lambda - \langle p \rangle^2 \right) \sigma^2}}\\
&=\textstyle
\frac{1728 \, \pi}{\left(12 \Lambda -\langle p \rangle^2\right)^{3/2}}  \left\{ \tan^{-1} \left[ k \tan \left(\frac{R }{2}\right)\right] - \frac{k \tan \left(\frac{R}{2}\right) \left(1-k^2 \tan ^2\left(\frac{R}{2}\right)\right)}{\left(k^2 \tan ^2\left(\frac{R}{2}\right)+1\right)^2}\right\} \,.
\end{aligned}
\end{equation}
The variation of the volume takes a simple form:
\begin{equation}
\delta V = \frac{8 k^2 \sin ^2 R (\sin R \, \delta k +k \, \delta R)}{\left(1 +k^2+(1 -k^2)  \cos R\right)^3} \,,
\end{equation}
so the symplectic potential is simply
\begin{equation}
\omega = \delta \theta =   - \frac{2}{3} w(R,k) \left( \sin R \, \delta k \wedge \delta \langle p \rangle +k \, \delta R \wedge \delta \langle p \rangle \right)  - 4 \pi \delta R \wedge  \delta  P \,,
\end{equation}\index{spherically symmetric solutions}\index{thin shell of dust}
where $w(R,k) = \frac{8 k^2 \sin ^2 R}{\left(1 +k^2+(1 -k^2)  \cos R\right)^3}$.
Now, the Poisson brackets between any two phase-space functions are given by the inverse of the symplectic form:
\begin{equation}
\{ f , g \} = \partial_i f \, (\omega^{-1})^{ij} \, \partial_j g \,, 
\end{equation}
and since

\begin{equation}\label{SingleShellSymplectic2-Form-Matrix}
\begin{aligned}
\omega_{ij} = \frac{1}{2} \left(
\begin{array}{cccc}
 0 & -\frac{2}{3} w \,  \sin R  & 0 & 0 \\
 \frac{2}{3} w \,  \sin R & 0 & \frac{2}{3} w \, k & 0 \\
 0 & - \frac{2}{3} w \, k & 0 & -4 \pi  \\
 0 & 0 & 4 \pi  & 0 \\
\end{array}
\right) \,,
\end{aligned}
\end{equation}
(where $i,j= \left(  k ,  \langle p \rangle ,  R  ,  P  \right)$, the inverse can be shown to be
\begin{equation}\label{SingleShellSymplectic2-Form-InverseMatrix}
\begin{aligned}
 (\omega^{-1})^{ij}  = \left(
\begin{array}{cccc}
 0  & \frac 3 {w \,  \sin R } & 0 & -  \frac 1 {2\pi} \frac{k}{\sin R} \\
 - \frac 3 {w \,  \sin R} & 0 & 0 & 0 \\
 0  & 0 & 0 &    \frac 1 {2\pi} \\
   \frac 1 {2\pi} \frac{k}{\sin R}   & 0 & -   \frac 1 {2\pi} & 0 
 \end{array}
\right) \,.
\end{aligned}
\end{equation}
Then the Poisson brackets take the following explicit form:
\begin{equation}
\begin{aligned}
\{ f , g \}  =&\textstyle \frac 1 {w \,  \sin R} \left[ 3 \left(\frac{\partial f}{\partial k}\frac{\partial g}{\partial \langle p\rangle }-\frac{\partial f}{\partial \langle p\rangle }\frac{\partial g}{\partial k}\right) 
-\frac{1}{2 \pi } w \, k
\left(\frac{\partial f}{\partial k}\frac{\partial g}{\partial P }-\frac{\partial f}{\partial P }\frac{\partial g}{\partial k}\right) \right] \\
&\textstyle+\frac{1}{2 \pi } \left(\frac{\partial f}{\partial R }\frac{\partial g}{\partial P }-\frac{\partial f}{\partial P }\frac{\partial g}{\partial R }\right) \,.
\end{aligned}
\end{equation}
The Poisson brackets between $h(k,R ,\langle p \rangle ) $ and $P $ then is:
\begin{equation}
\{ h , P  \} =
-\frac{1}{2 \pi } \frac{k}{\sin R} \, \frac{\partial h}{\partial k}   +\frac{1}{2 \pi } \, \frac{\partial h}{\partial R } \,,
\end{equation}
and an explicit calculation reveals that
\begin{equation}
\{ h , P  \} \approx - \frac{\cot R }{2 \pi} h \approx 0 \,,
\end{equation}
so the two constraints are first-class.

We have a four-dimensional phase space with two first-class constraints. One linear combination of the constraints can be interpreted as generating gauge transformations and indicating an unphysical degree of freedom, but the other linearly independent one cannot (see~Sec.~\ref{Barbour-FosterSubsec}), because it plays the role of Hamiltonian constraint generating the dynamics. It is convenient to take $P  \approx 0$ as the gauge constraint, which can be gauge-fixed with 
\begin{equation}
\chi = R  - \bar R \approx 0 \,,
\end{equation}
where $\bar R \in (0 , \pi)$ is any function of time (the simplest choice is a constant). $\chi$ is trivially first-class with respect to $h$ and second-class with respect to $P $. 
Replacing the gauge constraint $P  \approx 0$ and the gauge fixing $\chi  \approx 0 $ in the leftover Hamiltonian constraint $h\approx 0$ we get
\begin{equation}
h \propto \frac{M}{\sqrt{2 \cos^4 {\sfrac{\bar R}2}  \left(k^2 \tan ^2{\sfrac{\bar R}2}+1\right)^2}}
-
 \sqrt{\frac{36 \, k^2}{12 \Lambda - \langle p \rangle^2}}  \frac{  8 \, \cos {\sfrac{\bar R}2} \left( k^2-\cot ^2{\sfrac{\bar R}2}\right)}{\sin^3 {\sfrac{\bar R}2} \left( k^2+\cot ^2{\sfrac{\bar R}2}\right)^3}
\approx 0 \,,
\end{equation}\index{spherically symmetric solutions}\index{thin shell of dust}

Assuming that  $12 \Lambda > \langle p \rangle^2$, we can take $k$ real and positive, and the above equation  is equivalent to the following sixth-order polynomial in $k$:
\begin{equation}\label{k_polynomial}
\left(k^2+\cot ^2{\sfrac{\bar R}2}\right) \left(k^4 m-96 k^3 \cot {\sfrac{\bar R}2}+2 k^2 m \cot ^2{\sfrac{\bar R}2}+96 k \cot ^3{\sfrac{\bar R}2}+m \cot ^4{\sfrac{\bar R}2}\right)= 0 \,,
\end{equation}
where $m = M \sqrt{12 \Lambda - \langle p \rangle^2}$.

We can assume that $k^2 \neq - \cot ^2{\sfrac{\bar R}2}$ because $k \in \mathbbm{R}$ (otherwise $\sigma$ would be negative), so Eq.~(\ref{k_polynomial}) is equivalent to a fourth-order equation. Its discriminant is proportional to 
$$ m^2 \left(m^2-576\right)^2 \sin ^{20}(\bar R) \csc^8\left(\frac{\bar R}{2}\right) \,, 
$$
which is always positive. Therefore there are either four or zero real roots.
The former case holds only if both $\left(m^2-1728\right) \sin ^4\left(\frac{\bar R}{2}\right) \sin ^2(\bar R)  <0$ and $\left(m^2-576\right) \sin ^8\left(\frac{\bar R}{2}\right) \sin ^4(\bar R)<0$. So, in order for real roots to exist, we have to have $m^2 < 576$, that is,
\begin{equation}
M^2  < \frac{24^2}{12 \Lambda - \langle p \rangle^2} \,.
\end{equation}\index{spherically symmetric solutions}\index{thin shell of dust}
In summary, we found that the dynamics of the single-shell universe is completely trivial: the radial coordinate of the shell, $R$ is unphysical (even in isotropic gauge), because its conjugate momentum $P$ is a first-class constraint. All we can do is to impose $P\approx 0$ in $h(k,R,\langle p \rangle) \approx 0$ and we get a functional relation between $k$ and $\langle p \rangle$. For a given value of the rest mass $M$ this completely fixes the CMC metric as a function of the York time $\langle p \rangle$. \index{York time} In Fig.~\ref{SingleShellplots} I plot $\sigma(R)$ and the volume $V$ from Eq.~(\ref{SingleShellExactVolume}) as functions of York time $\langle p \rangle$, for a set of choices of $M$ between $0$ and the maximum $24/\sqrt{12 \Lambda - \langle p \rangle^2}$.

\begin{figure}[ht!]
\begin{center}
\includegraphics[height=0.28\textheight]{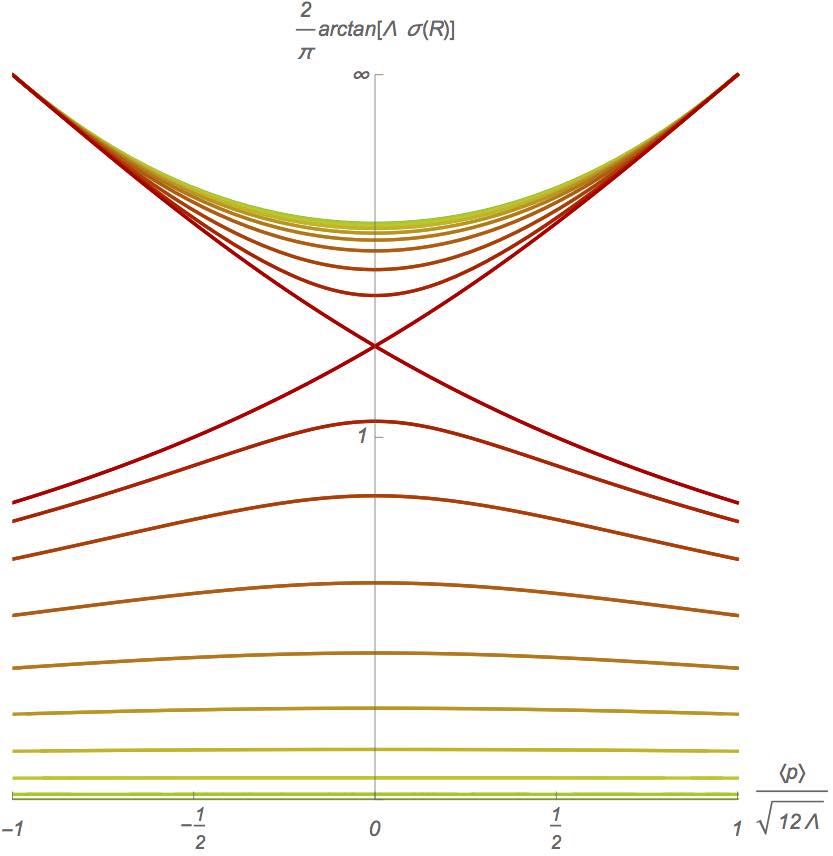}~~\includegraphics[height=0.28\textheight]{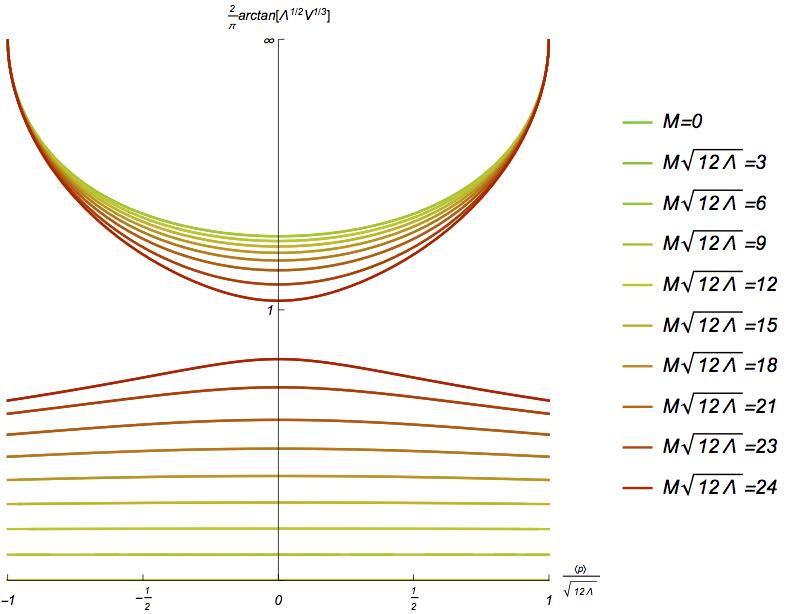}
\caption[$\sigma(R)$ and $V$ vs. $\langle p \rangle$ in single shell universe]{Plot of $\sigma(R)$ (left) and $V$ (right) as functions of York time $\langle p \rangle$ for a set of values of the rest mass. The vertical axis has been compactified by taking the $\arctan$.
As we can see, for each choice of the dimentionless parameter $M \sqrt{12 \Lambda}$ there are two conjugate solutions: one which goes from infinite volume/areal radius at the shell (at $\langle p \rangle = - \sqrt{12\Lambda}$) to a finite minimum (at $\langle p \rangle = 0$) and back to infinity (at $\langle p \rangle = +\sqrt{12\Lambda}$). Another one that goes from a finite volume/areal radius (at $\langle p \rangle = \pm \sqrt{12\Lambda}$) to a finite maximim (at $\langle p \rangle = 0$). In the zero-rest-mass limit the first kind of solutions tend to an acceptable result: two compact patches of de Sitter universe glued at their border, which is delimited by a lightlike shell. The second kind stops making sense as $M \to 0$: both the areal radius $\sigma(R)$ and the volume are zero throughout the  solution. In the opposite limit, $M \to \frac{24}{\sqrt{12 \Lambda}}$, the solution ceases to be smooth, because the first and second kind of solutions meet at a point at $\langle p \rangle =0$, and $d \sigma(R)/d\langle p\rangle$ has a discontinuity at that point. This is a signal that at that point the `cosmological horizon scale' associated to the cosmological constant and the `Schwarzschild horizon scale' \index{Schwarzschild's spacetime} associated to the mass-energy of the shell coincide. The physics of this family of solutions that I uncovered will be investigated in future works.
}\label{SingleShellplots}
\end{center}
\end{figure}\index{spherically symmetric solutions}\index{thin shell of dust}

We were able to solve analytically  the single-shell universe, because in this case the isotropic gauge condition leads to an equation for $\sigma$, (\ref{SingleShell_IsotropicEquation}), that can be solved exactly. However this will be a luxury that we cannot afford in the following section, and we need to be prepared to study the dynamics even when an explicit solution of the Hamiltonian constraint in isotropic gauge is not available. For this reason, in preparation to the next section, I here calculate again the symplectic potential without assuming any particular radial gauge. I will instead try to exploit as much as I can all the gauge-independent information we have about the form of the solution of the constraints.
To do so, I need to make only reference to the variables $\rho = \sqrt{\sigma (R)}$ (the areal radius of the shell), $A_+$ and $A_-$, which do not depend on the radial gauge (as opposed to $R$ and $P$, which, being related to a coordinate system, take a meaning only when a radial gauge is fixed).

Let's begin with the form~(\ref{IsotropicSymplectic0}) for the pre-symplectic potential:
\begin{equation}
\theta =  - \frac{2}{3} V \delta \langle p \rangle -  8 \pi  \left[ \delta  A_-  \int_0^{R } \d r  \frac{\mu}{\sqrt \sigma}  + \delta  A_+  \int_{R }^{\pi} \d r  \frac{\mu}{\sqrt \sigma} \right]  \,,
\end{equation}
introducing the theta functions
\begin{equation}
\Theta_+ (r) = \Theta(r-R) \,, \qquad  \Theta_- (r) = \Theta(R-r) \,, 
\end{equation}
we can write the potential as
\begin{equation}
\begin{aligned}
&     \sum_{\beta \in \{ +,-\}}  \int_{0}^{\pi} \frac{\left(\frac 1 3 \sigma^{3/2}(r) \, \delta \langle p \rangle  + \delta A_\beta  \right) \Theta_\beta(r) \, |\sigma'| \d r}{\sqrt{ A^2_\beta  + \left( {\sfrac 2 3} \langle p \rangle  A_\beta - 8 \, m_\beta  \right) \sigma^{3/2}  + 4 \, \sigma^2 - {\sfrac 1 9}  \left( 12 \, \Lambda - \langle p \rangle^2 \right) \sigma^3 }}
\\
=&  -   \int_0^{\sigma(R)} \left( \frac{\partial F_- [A_-,\langle p \rangle ,\sigma]}{\partial \langle p \rangle}  \delta \langle p \rangle  +  \frac{\partial F_- [A_-,\langle p \rangle ,\sigma]}{\partial A_-}  \delta A_- \right) d \sigma 
\\
&-    \int_0^{\sigma(R)} \left( \frac{\partial F_+ [A_+,\langle p \rangle ,\sigma]}{\partial \langle p \rangle}  \delta \langle p \rangle  +  \frac{\partial F_+ [A_+,\langle p \rangle ,\sigma]}{\partial A_+}  \delta A_+ \right) \d \sigma 
   \,,
\end{aligned}
\end{equation}
where
{\medmuskip=0mu
\thickmuskip=0mu
\thinmuskip=0mu
\begin{equation}
F_\beta  =  \log \left(A_\beta + {\sfrac 1 3} \langle p \rangle \sigma^{3/2}   \sqrt{ A^2_\beta  + \left( {\sfrac 2 3} \langle p \rangle  A_\beta - 8 \, m_\beta  \right) \sigma^{3/2}  + 4 \, \sigma^2 - {\sfrac 1 9}  \left( 12 \, \Lambda - \langle p \rangle^2 \right) \sigma^3 } \right) \,.
\end{equation}}
Then the symplectic form is
\begin{equation}
\begin{aligned}
\omega =& \delta \theta =  -  8 \pi    \bigg{(}
  \frac{\partial F_- [A_-,\langle p \rangle ,\sigma(R)]}{\partial \langle p \rangle}  \delta \langle p \rangle  +  \frac{\partial F_- [A_-,\langle p \rangle ,\sigma(R)]}{\partial A_-}  \delta A_-   \\
&+ \frac{\partial F_+ [A_+,\langle p \rangle ,\sigma(R)]}{\partial \langle p \rangle}  \delta \langle p \rangle  +  \frac{\partial F_+ [A_+,\langle p \rangle ,\sigma(R)]}{\partial A_+}  \delta A_+ \bigg{)} \wedge \delta \sigma(R)
   \,.
\end{aligned}
\end{equation}
applying $A_\pm = m_\pm =0$
\begin{equation}
\omega  =  -  \frac{16 \pi}{3}    \left( \frac{\sigma^{3/2}(R)}{\sqrt{ 4 \, \sigma^2(R) - {\sfrac 1 9}  \left( 12 \, \Lambda - \langle p \rangle^2 \right) \sigma^3(R)} } \right)   \delta \langle p \rangle  \wedge \delta \sigma(R)
   \,.
\end{equation}
The above equation is essentially stating that the variables $\sigma(R)$ and $\langle p \rangle$ are canonically conjugate.\index{spherically symmetric solutions}\index{thin shell of dust}

Let's now discuss the constraints imposed by the jump conditions. The diffeomorphism jump condition [Eq.~(\ref{DeltaAjumpcondition})] is now just a definition of $P$, which is not a dynamical variable anymore. The Hamiltonian jump condition  [Eq.~(\ref{JumpGammaKappa})] can be written in a way that depends only on $\rho$, $A_+$ and $A_-$. In fact, define $\kappa = \lim_{r\to R^-} \sigma'(r)$ and $\gamma = \lim_{r\to R^-} \sigma'(r)$. Using Eq.~(\ref{DeltaAjumpcondition}) into the second one:
\begin{equation}
\gamma - \kappa = - {\sfrac 1 2} \sqrt{4 {\sfrac{(A_+-A_-)^2}{\rho^2}} \mu^2(R) + M^2 \mu^2(R)} \,,
\end{equation}
and dividing by $|\mu(R)|$
\begin{equation}
\frac{\gamma}{|\mu(R)|} - \frac{\kappa}{|\mu(R)|} = - {\sfrac 1 2} \sqrt{4 {\sfrac{(A_+-A_-)^2}{\rho^2} } + M^2} \,,
\end{equation}
we can square the above equation and reorder
\begin{equation}
\frac{\gamma^2}{\mu^2(R)} + \frac{\kappa^2}{\mu^2(R)} - \frac{(A_+-A_-)^2}{\rho^2}  - {\sfrac 1 4} M^2  = 2 \frac{\gamma \, \kappa }{\mu^2(R)}    \,,
\end{equation}
and taking another square
\begin{equation}
\left( \frac{\gamma^2}{\mu^2(R)} + \frac{\kappa^2}{\mu^2(R)} - \frac{(A_+-A_-)^2}{\rho^2}  - {\sfrac 1 4} M^2 \right)^2  = 2 \frac{\gamma^2}{\mu^2(R)} \frac{\kappa^2}{\mu^2(R)}     \,,
\end{equation}
the equation only depends on  $\frac{\gamma^2}{\mu^2(R)} =\frac{\left(\lim_{r\to R^-} \sigma'(r)\right)^2}{\mu^2(R)}  $ and $\frac{\kappa^2}{\mu^2(R)} =\frac{\left(\lim_{r\to R^+} \sigma'(r)\right)^2}{\mu^2(R)}  $. Now we can use the left- and right- limits  of Eq.~(\ref{VacuumSolutionForMu}) to get rid of $\kappa/\mu^2(R)$ and $\gamma/\mu^2(R)$:
\begin{equation}
\begin{aligned}
\frac{\gamma^2}{\mu^2(R)} = \frac{A_-^2}{\rho} + \left( {\sfrac 2 3} \langle p \rangle A_- - 8 m_- \right) \rho + 4 \rho^2 - {\sfrac 1 9} \left( 12 \Lambda - \langle p \rangle^2 \right) \rho^4  \,,
\\
\frac{\kappa^2}{\mu^2(R)} =\frac{A_+^2}{\rho} + \left( {\sfrac 2 3} \langle p \rangle A_+ - 8 m_+ \right) \rho + 4 \rho^2 - {\sfrac 1 9} \left( 12 \Lambda - \langle p \rangle^2 \right) \rho^4  \,,
\end{aligned}
\end{equation}
and we have our contraint purely in terms of $\rho$, $A_+$ and $A_-$. Now we can use the boundary conditions at the poles, $A_\pm = m_\pm =0$, and the constraint simplifies to:
\begin{equation}\label{SingleShellOnshellConstraint}
\frac{M^2}{16} +   {\sfrac 1 9} \left( 12 \Lambda - \langle p \rangle^2 \right) \rho^4 - 4 \rho^2  =0 \,.
\end{equation}
The above constraint admits a real positive $\rho$ only when $M^2 (12 \Lambda - \langle p \rangle^2)<24^2$, which is the same upper bound on the mass that we found above.
Moreover, if we plot the solutions of~(\ref{SingleShellOnshellConstraint}) w.r.t. $\rho^2$ as functions of $\langle p \rangle$ we obtain the same diagram as the left one in Fig.~\ref{SingleShellplots}. We were therefore able to extract the same amount of information as before, but without having to fix the radial gauge.

We were able to solve every aspect of the `single-shell universe' analytically. The result is a system whose dynamics is completely trivial: the coordinate position of the shell $R$ is a gauge degree of freedom (the diffeomorphism constraint reduces to $P \approx 0$ which implies that the conjugate variable, $R$, is a gauge direction). The gauge-invariant degrees of freedom are all completely constrained: once we specify the rest mass of the shell $M$ in units of the cosmological constant $\Lambda$ the evolution is completely fixed and admits no integration constants: there are no adjustable parameters that we can choose to set initial data. The space of solutions is just a point.
This system is, therefore, too trivial for our purposes. We need to add degrees of freedom in order to have a nontrivial solution space.

The next thing in order of simplicity that we can do is to add a concentric shell, which adds a pair of canonical degrees of freedom and makes the system nontrivial.


\index{spherically symmetric solutions}\index{thin shell of dust}
\subsubsection*{Twin shell universe}\label{TwinShellSec}

\begin{figure}[t!]
\begin{center}
\includegraphics[width=0.4\textwidth]{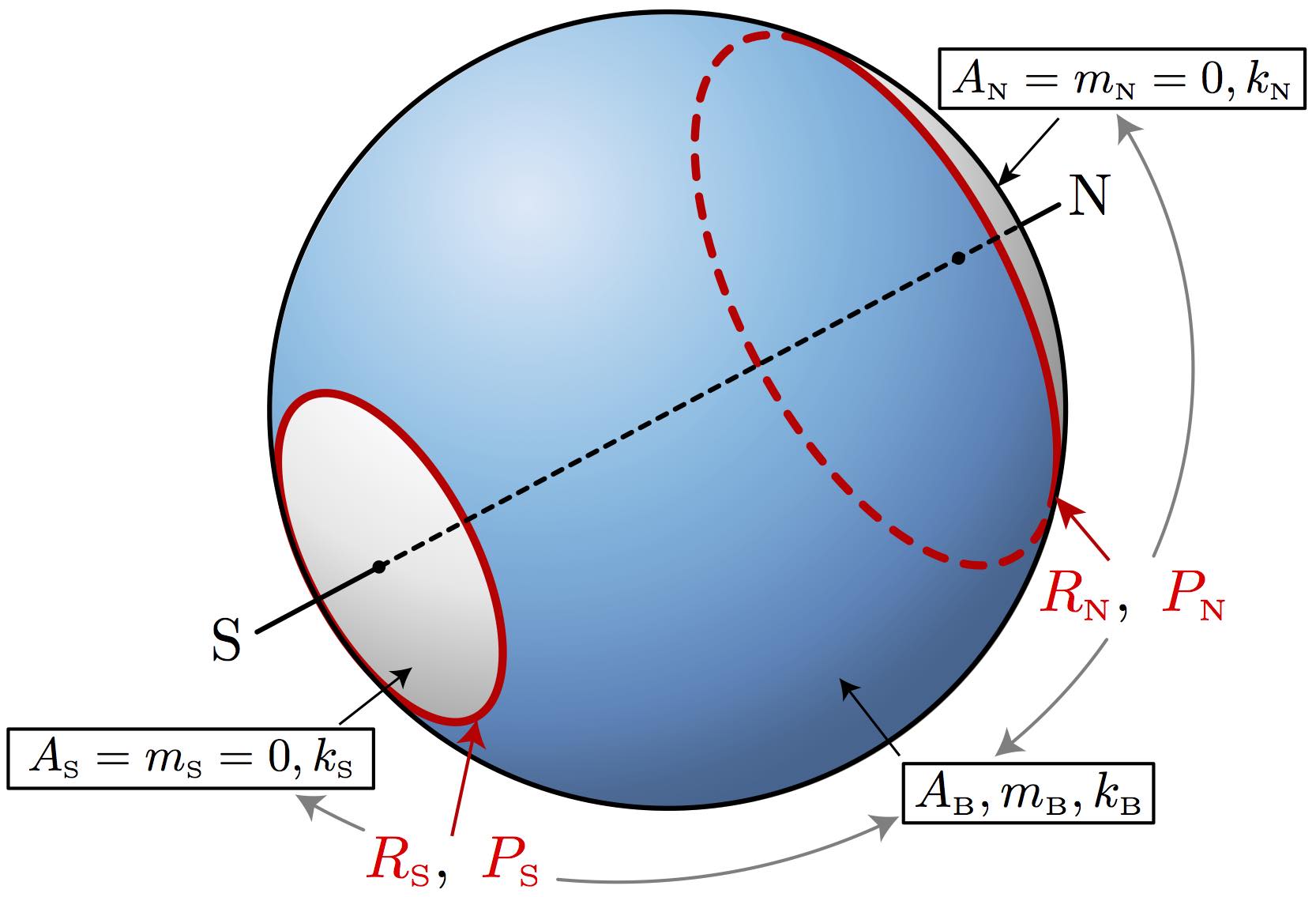}
\end{center}
\caption[Twin-shell universe diagram]{The `twin-shell' universe: the spatial manifold has the topology of the sphere $S^3$ and two concentric thin shells, which divide the manifold into three regions: the north and south polar regions indicated with N and S, and the `belt' region in between, indicated with a B. The shell closer to the north pole will be indicated as the N shell, and the other will be the S shell. Each region will have different values of the integration constants $A$ and $m$, but regularity demands that these constants both vanish in the two polar regions. Moreover the two shells will come equipped with a coordinate-radius degree of freedom $R_\st{S}$, $R_\st{N}$ and with a radial-momentum degree of freedom $P_\st{S}$, $P_\st{N}$. These degree of freedom will be related to the jump in the integration constants $A$ and $m$.
}\label{TwinShellPicture}
\end{figure}

\index{spherically symmetric solutions}\index{thin shell of dust}
Let's call the two shells N (`northern') and S (`southern'). Their canonical variables will be $R_\st{N}$, $R_\st{S}$, $P_\st{N}$ and $P_\st{S}$. The two shells divide the 3-sphere in three regions: N and S (the `north' and `south' polar caps), and B, the `belt' region between the two shells. The associated integration constants will be called $A_\st{N}$, $m_\st{N}$, $A_\st{B}$, $m_\st{B}$, $A_\st{S}$ and $m_\st{S}$.   

These integration constants will satisfy, at each shell, the same jump conditions we found above. The three $A$ integration constant will satisfy two conditions like Eq.~(\ref{DeltaAjumpcondition}):
\begin{equation}\label{Twinshell_DeltaAjumpcondition}
A_\st{B} - A_\st{S} = - \frac{\sqrt{\sigma(R_\st{S})}}{2\mu(R_\st{S})} P_\st{S} \,,
\qquad
A_\st{N} - A_\st{B} = - \frac{\sqrt{\sigma(R_\st{N})}}{2\mu(R_\st{N})} P_\st{N} \,.
\end{equation}
Now, calling  $\sigma(R_\st{S}) = \rho^2_\st{S}$, $\sigma(R_\st{N}) = \rho^2_\st{S}$, and 
\begin{equation}
\gamma_\st{S,N} = \lim_{r\to R_\st{S,N}^+} \sigma'(r) \,,
~~~~
\displaystyle \kappa_\st{S,N} = \lim_{r\to R_\st{S,N}^-} \sigma'(r) \,,
\end{equation}
the conditions that ensure the continuity of $\mu$ at the two shell are the same as Eq.~(\ref{ContinuityOfMu}),
{\medmuskip=0mu
\thinmuskip=0mu
\thickmuskip=0mu
\begin{equation}\label{Twinshell_ContinuityOfMu}
\!\!\!\!\!
\begin{aligned}\ts
\frac{|\kappa_\st{S}|}{\sqrt{ \left(\frac {A_\st{S} + \frac 1 3 \langle p \rangle \rho_\st{S}^3 } {2 \rho_\st{S}^{2}}\right)^2  - \frac{2 \, m_\st{S}}{\rho_\st{S}} +  1  - \frac{\Lambda  \rho_\st{S}^2}{3}     } } &= \frac{|\gamma_\st{S}|}{\sqrt{ \left(\frac {A_\st{B} + \frac 1 3 \langle p \rangle \rho_\st{B}^3 } {2 \rho_\st{B}^{2}}\right)^2  - \frac{2 \, m_\st{B}}{\rho_\st{B}} +  1  - \frac{\Lambda  \rho_\st{B}^2}{3}     } } \,,
\\
\ts
\frac{|\kappa_\st{N}|}{\sqrt{ \left(\frac {A_\st{B} + \frac 1 3 \langle p \rangle \rho_\st{B}^3 } {2 \rho_\st{B}^{2}}\right)^2  - \frac{2 \, m_\st{B}}{\rho_\st{B}} +  1  - \frac{\Lambda  \rho_\st{B}^2}{3}     } } &= \frac{|\gamma_\st{N}|}{\sqrt{ \left(\frac {A_\st{N} + \frac 1 3 \langle p \rangle \rho_\st{N}^3 } {2 \rho_\st{N}^{2}}\right)^2  - \frac{2\,m_\st{N}}{\rho_\st{N}} +  1  - \frac{\Lambda  \rho_\st{N}^2}{3}     } } \,.
\end{aligned}
\end{equation}}
Finally, the jump condition that ensures that the Hamiltonian constraint holds are two copies of Eq.~(\ref{JumpGammaKappa}):
\begin{equation}\label{Twinshell_JumpGammaKappa}
\gamma_\st{S}  - \kappa_\st{S}   = - {\sfrac 1 2} \sqrt{P_\st{S}^2 + M_\st{S}^2  \mu^2(R_\st{S}) } \,,
~~~
\gamma_\st{N}  - \kappa_\st{N}   = - {\sfrac 1 2} \sqrt{P_\st{N}^2 + M_\st{N}^2  \mu^2(R_\st{N}) } \,.
\end{equation}
Using Eq.~(\ref{Twinshell_DeltaAjumpcondition}) into Eqs.~(\ref{Twinshell_JumpGammaKappa}) 
\begin{equation}\label{Twinshell_JumpGammaKappa2}
\begin{aligned}\ts
\frac{\kappa_\st{S}}{| \mu(R_\st{S}) |}  - \frac{\gamma_\st{S}}{| \mu(R_\st{S}) |}  =  \sqrt{\frac{(A_\st{S} - A_\st{B})^2}{\rho_\st{S}^2} + {\sfrac 1 4} M_\st{S}^2 } \,,
\\
\ts
\frac{\kappa_\st{N}}{| \mu(R_\st{S}) |}   - \frac{\gamma_\st{N}}{| \mu(R_\st{N}) |}    = \sqrt{\frac{(A_\st{B} - A_\st{N})^2}{\rho_\st{N}^2} + {\sfrac 1 4} M_\st{N}^2 } \,.
\end{aligned}
\end{equation}
by taking twice the square of the above equations, we can make them independent of the signs of $\kappa_{S,N}$ and $\gamma_\st{S,N}$,
\begin{equation}\label{Twinshell_JumpGammaKappa3}
\begin{aligned}\ts
\left(\frac{\kappa_\st{S}^2}{\mu^2(R_\st{S})}  + \frac{\gamma_\st{S}^2}{\mu^2(R_\st{S})} -   \frac{(A_\st{S} - A_\st{B})^2}{\rho_\st{S}^2} -  {\sfrac 1 4} M_\st{S}^2 \right)^2 = 4 \frac{\kappa_\st{S}^2}{\mu^2(R_\st{S})} \frac{\gamma_\st{S}^2}{\mu^2(R_\st{S})}\,,
\\
\ts
\left( \frac{\kappa_\st{N}^2}{\mu^2(R_\st{S})}   + \frac{\gamma_\st{N}^2}{\mu^2(R_\st{N})}    - \frac{(A_\st{B} - A_\st{N})^2}{\rho_\st{N}^2} + {\sfrac 1 4} M_\st{N}^2 \right)^2 = 4  \frac{\kappa_\st{N}^2}{\mu^2(R_\st{S})}\frac{\gamma_\st{N}^2}{\mu^2(R_\st{N})} \,.
\end{aligned}
\end{equation}
now, using the definition of $\mu(r)$ at $r= R_\st{S}$ and $r = R_\st{S}$,
\begin{equation}
\begin{aligned}
&\ts
\frac{\gamma_\st{S}^2}{\mu^2(R_\st{S})}  = \left(\frac {A_\st{S} + \frac 1 3 \langle p \rangle \rho_\st{S}^3 }{\rho_\st{S}}\right)^2  - 8 \, m_\st{S} \, \rho_\st{S} + 4  \rho_\st{S}^2  - \frac{4\Lambda  \rho_\st{S}^4}{3} \,,
\\
&\ts
\frac{\kappa_\st{S}^2}{\mu^2(R_\st{S})} = \left(\frac {A_\st{B} + \frac 1 3 \langle p \rangle \rho_\st{S}^3 } {\rho_\st{S}}\right)^2  - 8 \, m_\st{B} \, \rho_\st{S} +4 \rho_\st{S}^2  - \frac{4\Lambda  \rho_\st{S}^4}{3} \,,
\\
&\ts
\frac{\gamma\st{N}^2}{\mu^2(R_\st{N})} = \left(\frac {A_\st{B} + \frac 1 3 \langle p \rangle \rho_\st{N}^3 } {\rho_\st{N}}\right)^2  - 8 \, m_\st{B} \, \rho_\st{N} +4 \rho_\st{N}^2  - \frac{4\Lambda  \rho_\st{N}^4}{3} \,,
\\
&\ts
\frac{\kappa_\st{N}^2}{\mu^2(R_\st{S})} = \left(\frac {A_\st{N} + \frac 1 3 \langle p \rangle \rho_\st{N}^3 } {\rho_\st{N}}\right)^2  - 8 \, m_\st{N} \, \rho_\st{N} + 4 \rho_\st{N}^{2}  - \frac{4 \Lambda  \rho_\st{N}^4}{3} \,,
\end{aligned}
\end{equation}
and recalling that, in order to keep the poles compact and smooth we need to have $A_\st{S} = A_\st{N}=0$ and $m_\st{S}=m_\st{N}=0$,\footnote{See the discussion on page \pageref{DiscussionOfRegularityAtThePole}.} we end up with the following two on-shell conditions:\index{spherically symmetric solutions}\index{thin shell of dust}
\begin{equation}\label{Twinshell_OnshellCondition}
\begin{aligned}
&\frac{M_\st{S}^4}{16}+4 A_\st{B}^2 \left(T \rho _\st{S}^2-4\right)+M_\st{S}^2 \rho _\st{S} \left(T \rho _\st{S}^3 - 4 \rho _\st{S}-2 X\right)+16 X^2 \rho _\st{S}^2 = 0\,,
\\
&\frac{M_\st{N}^4}{16} + 4 A_\st{B}^2 \left(T \rho _\st{N}^2-4\right)+M_\st{N}^2 \rho _\st{N} \left(T \rho _\st{N}^3 - 4 \rho _\st{N} -2 X\right)+16 X^2 \rho _\st{N}^2 =0 \,.
\end{aligned}
\end{equation}
where $T = \frac{1}{9} \left( 12 \Lambda - \langle p \rangle^2 \right)$ and $X = \frac 1 6 \langle p \rangle A_\st{B} - 2\, m_\st{B}$.
Conditions~(\ref{Twinshell_OnshellCondition}) are two identical equations 
involving the same $A_\st{B}$ and two different areal radii $\rho_a=(\rho_\st{S},\rho_\st{N})$ and rest-masses $M_a=(M_\st{S},M_\st{N})$. By rescaling both equations with appropriate powers of $m_\st{B}$ we can make them dimensionless. This requires introducing dimensionless variables analogue to those of Eq.~(\ref{DimensionlessVariablesMordor}):
\begin{equation}
C = \frac{A_\st{B}}{2 \, m_\st{B}^2} \,, ~~~   \tau = | m_\st{B} | \, \langle p \rangle \,, ~~~ \lambda = m_\st{B}^2 \,\Lambda \,, ~~~ z_a = \frac{\rho_a}{m_\st{B}} \,, ~~~ M_a = | m_\st{B} | \, \mu_a \,.
\end{equation}
Then the two Equations~(\ref{Twinshell_OnshellCondition}) can be written
{\medmuskip=0mu
\thinmuskip=0mu
\thickmuskip=0mu
\begin{equation}\label{Twinshell_OnshellCondition_dimensionless}
\!\!\!\!
\frac{\mu_a^4}{16} = \mu_a^2 z_a \left[   {\sfrac 2 3} C \tau + {\sfrac 1 9} z_a^3 \left(\tau^2-12 \lambda \right) + 4 z_a \mp 4\right] - {\sfrac{64}{3}} \left[ C^2 \left(\lambda  z_a^2-3\right) + 3 z_a^2 \mp C \tau  z_a^2\right] \,,
\end{equation}}
where in $\mp$ we choose the sign  $-$ if $m_\st{B}$ is positive, or $+$ if it is negative. The above equation gives, for each choice of constants $\lambda$ and $\mu_a$, and sign of $m_\st{B}$, a surface in the 3D space $z_a$, $\tau$, $C$, which is the same as that of Fig.~\ref{MordorFig1}, \ref{MordorFig2}, \ref{MordorFig3},  and \ref{MordorFig4}, which represented the regions which are excluded by the existence of a solution of the Hamiltonian constraint. We need to combine those diagrams with a plot of the on-shell surfaces~(\ref{Twinshell_OnshellCondition_dimensionless}) to check whether they intersect the forbidden region. To do this I will choose, for each choice of signs of $\lambda$ and $m_\st{B}$, one particular choice of $\lambda$, together with four on-shell surfaces corresponding to four representative choices of $\mu_a$. The following Figures~\ref{Mordor_Onshell_Fig_1}, \ref{Mordor_Onshell_Fig_2}, \ref{Mordor_Onshell_Fig_3} and \ref{Mordor_Onshell_Fig_4} show explicitly that the on-shell surface never crosses the forbidden region.
\index{spherically symmetric solutions}\index{thin shell of dust}

After that, in Fig.~\ref{Mordor_Onshell_Fig_5} and \ref{Mordor_Onshell_Fig_6} I will show two further cases, which were not included in Fig.~\ref{MordorFig1}, \ref{MordorFig2}, \ref{MordorFig3} and \ref{MordorFig4}, because in these cases there is no forbidden region. These are when the cosmological constant is negative and the Misner--Sharp mass is negative or zero. The forbidden region is absent, but the on-shell surface is still there.
\newpage
\begin{figure}[h!]
\framebox[0.49\textwidth]{\parbox{0.49\textwidth}{\center
$\bm \lambda < 0 \, , ~ \bm m>0$
\\
\includegraphics[width=0.23\textwidth]{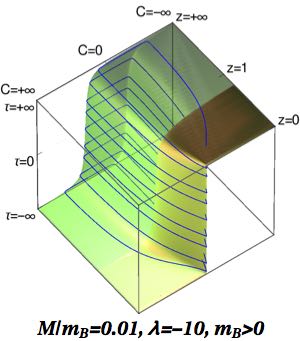}~\includegraphics[width=0.23\textwidth]{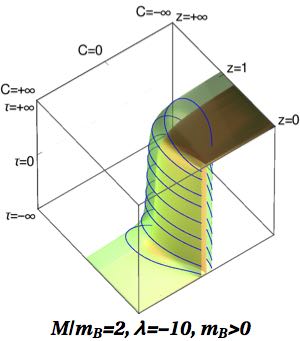}
\\
\includegraphics[width=0.23\textwidth]{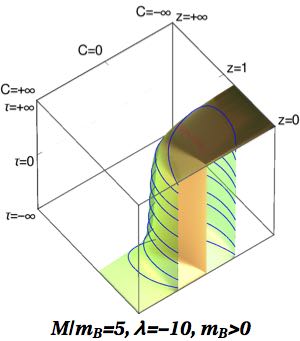}~\includegraphics[width=0.23\textwidth]{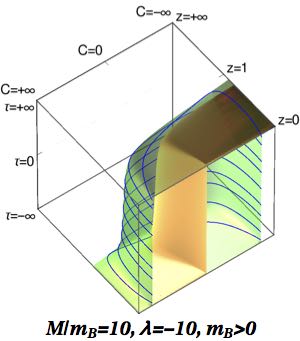}}}
\caption[On-shell surfaces for $\lambda<0$, $m_\st{B}>0$]{The surface $\mathscr P (z_a) =0$  of Fig.~\ref{MordorFig1}
 for $m_\st{B}>0$, $\lambda = -10$ (in yellow), together with the on-shell surface (in transparent green), for four choices of the ratio $M_a/m_\st{B}$. Some constant-$\tau$ curves on the on-shell surface are represented in blue to help localize the surface.}
\label{Mordor_Onshell_Fig_1}
\end{figure}\index{spherically symmetric solutions}\index{thin shell of dust}
\newpage
\begin{figure}[h!]
\framebox[0.49\textwidth]{\parbox{0.49\textwidth}{\center
$\bm \lambda > 0 \, , ~ \bm m>0$
\\
\includegraphics[width=0.23\textwidth]{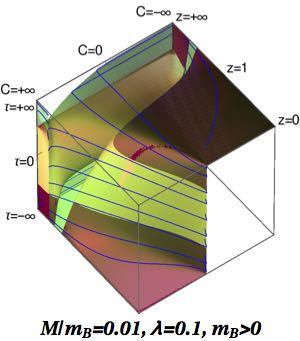}~\includegraphics[width=0.23\textwidth]{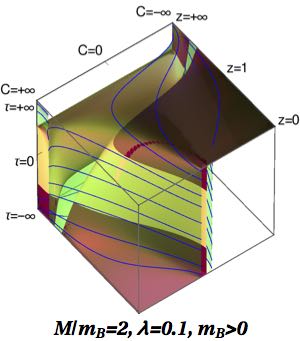}
\\
\includegraphics[width=0.23\textwidth]{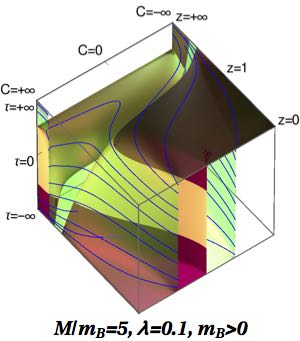}~\includegraphics[width=0.23\textwidth]{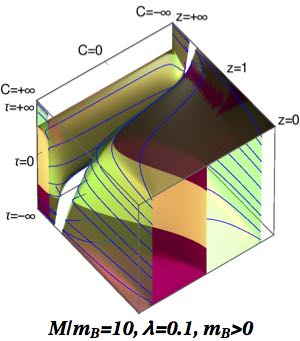}}}
\caption[On-shell surfaces for $\lambda>0$, $m_\st{B}>0$]{The surface $\mathscr P (z_a) =0$  of Fig.~\ref{MordorFig2}
 for $m_\st{B}>0$, $\lambda = 0.1 >0$ (in yellow/red), together with the on-shell surface (in transparent green), for four choices of the ratio $M_a/m_\st{B}$.}
\label{Mordor_Onshell_Fig_2}
\end{figure}\index{spherically symmetric solutions}\index{thin shell of dust}
\newpage
\begin{figure}[h!]
\framebox[0.49\textwidth]{\parbox{0.49\textwidth}{\center
$\bm \lambda > 0 \, , ~ \bm m<0$
\\
\includegraphics[width=0.23\textwidth]{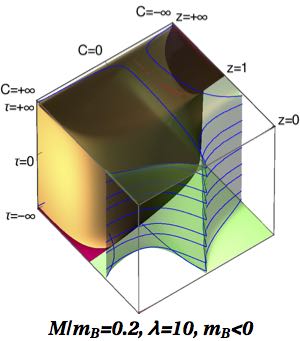}~\includegraphics[width=0.23\textwidth]{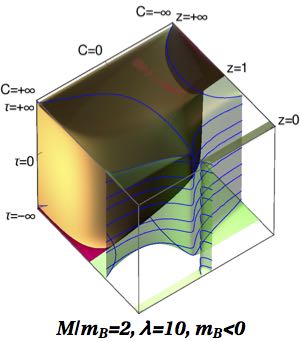}
\\
\includegraphics[width=0.23\textwidth]{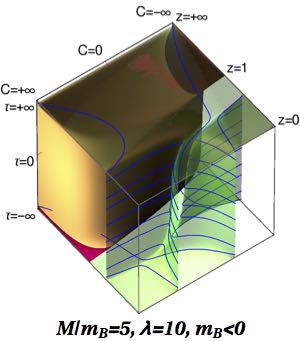}~\includegraphics[width=0.23\textwidth]{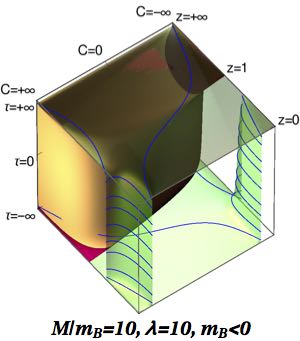}}}
\caption[On-shell surfaces for $\lambda>0$, $m_\st{B}<0$]{The surface $\mathscr P (z_a) =0$  of Fig.~\ref{MordorFig3}
 for $m_\st{B}<0$, $\lambda = 10 >0$ (in yellow/red), together with the on-shell surface (in transparent green), for four choices of the ratio $M_a/m_\st{B}$.}
\label{Mordor_Onshell_Fig_3}
\end{figure}\index{spherically symmetric solutions}\index{thin shell of dust}
\newpage
\begin{figure}[h!]
\framebox[0.49\textwidth]{\parbox{0.49\textwidth}{\center
$\bm \lambda > 0 \, , ~ \bm m=0$
\\
\includegraphics[width=0.23\textwidth]{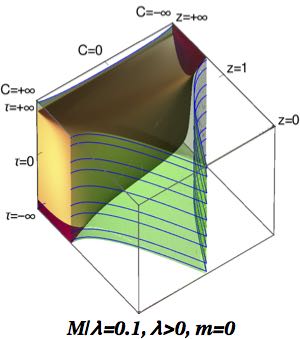}~\includegraphics[width=0.23\textwidth]{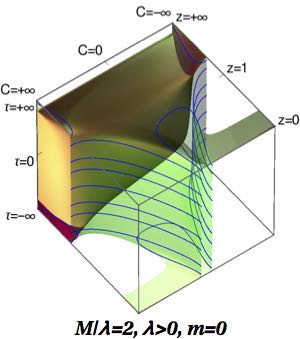}
\\
\includegraphics[width=0.23\textwidth]{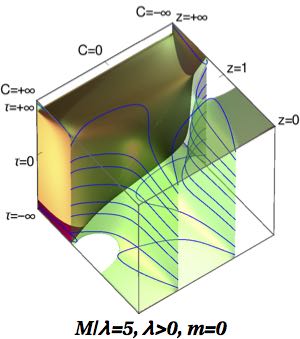}~\includegraphics[width=0.23\textwidth]{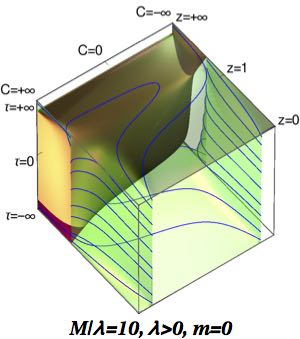}}}
\caption[On-shell surfaces for $\lambda>0$, $m_\st{B}=0$]{The surface $\mathscr P (z_a) =0$  of Fig.~\ref{MordorFig4}
 for $m_\st{B}=0$ (in yellow/red), together with the on-shell surface (in transparent green), for four choices of the ratio $M_a/\lambda$.}
\label{Mordor_Onshell_Fig_4}
\end{figure}\index{spherically symmetric solutions}\index{thin shell of dust}
\newpage
\begin{figure}[h!]
\framebox[0.49\textwidth]{\parbox{0.49\textwidth}{\center
$\bm \lambda < 0 \, , ~ \bm m<0$
\\
\includegraphics[width=0.23\textwidth]{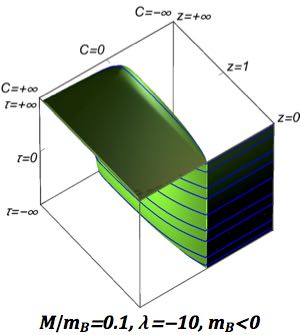}~\includegraphics[width=0.23\textwidth]{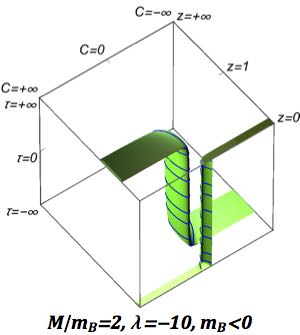}
\\
\includegraphics[width=0.23\textwidth]{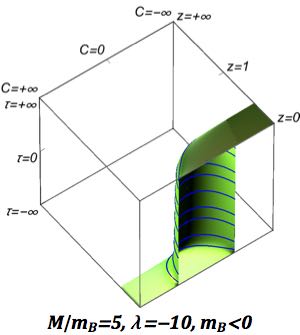}~\includegraphics[width=0.23\textwidth]{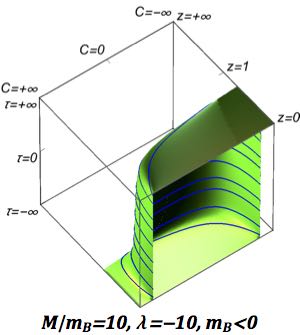}}}
\caption[On-shell surface  for $\lambda<0$, $m_\st{B}<0$]{On-shell surface for negative $\lambda$ and $m_\st{B}$, for four choices of the ratio $M_a/m_\st{B}$. In this case there is no excluded region because for this choice of signs of $\lambda$ and $m_\st{B}$ all values of the parameters are admissible.}
\label{Mordor_Onshell_Fig_5}
\end{figure}\index{spherically symmetric solutions}\index{thin shell of dust}
\newpage
\begin{figure}[h!]
\framebox[0.49\textwidth]{\parbox{0.49\textwidth}{\center
$\bm \lambda < 0 \, , ~ \bm m=0$
\\
\includegraphics[width=0.23\textwidth]{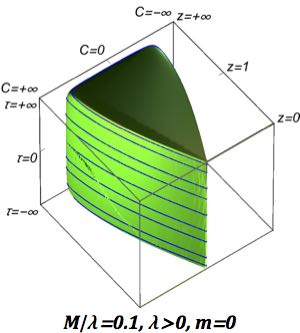}~\includegraphics[width=0.23\textwidth]{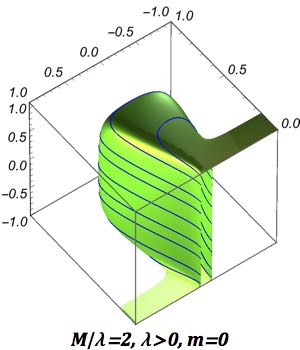}
\\
\includegraphics[width=0.23\textwidth]{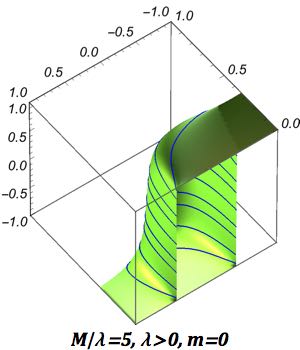}~\includegraphics[width=0.23\textwidth]{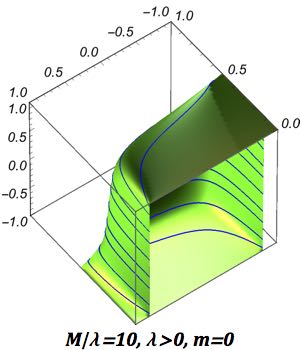}}}
\caption[On-shell surface  for $\lambda<0$, $m_\st{B}=0$]{On-shell surface for negative $\lambda$ and $m_\st{B}=0$, for four choices of the ratio $M_a/|\lambda|$. In this case too there is no excluded region because with $m_\st{B}=0$ and for this choice of signs of $\lambda$ all values of the parameters are admissible.}
\label{Mordor_Onshell_Fig_6}
\end{figure}\index{spherically symmetric solutions}\index{thin shell of dust}
\newpage

\index{spherically symmetric solutions}\index{thin shell of dust}

What one would like to do now is to solve all equations and identify the minimal core of dynamical variables that are needed for a description of the system, \emph{i.e.}, find the reduced phase space. This cannot be done in isotropic gauge as was done in the case of a single shell. However we can repeat what was done at the end of that section, and concentrate on gauge-independent variables ($\rho_\st{S}$, $\rho_\st{N}$, $A_\st{S}$,  $A_\st{B}$,  $A_\st{N}$) and try to calculate the symplectic form in terms of those variables alone. This turns out to be possible also in the `twin-shell' case.

\subsubsection*{Symplectic form}
\index{symplectic form}

The generalization of the pre-symplectic potential~(\ref{IsotropicSymplectic0}) to the case of two shells is:
\begin{equation}
\theta =  -  8\pi \sum_{\beta \in \{ \rm{S},\rm{B},\rm{N}\}} \int_{0}^{\pi}  \Theta_\beta(r)  \left[ \frac 1 3  \mu \,  \sigma \,  \delta \langle p \rangle + \frac{\mu}{\sqrt \sigma} \, \delta A_\beta \right] \d r  \,,
\end{equation}
where
\begin{equation}
\Theta_\beta(r)  = \left\{
\begin{array}{ll}
\Theta(R_\st{S}-r) \,  & \beta = \rm{S} \,,
\\
\Theta(r - R_\st{S})\Theta(R_\st{N}-r) \,  & \beta = \rm{B} \,,
\\
\Theta(r - R_\st{N}) \,  & \beta = \rm{N} \,.
\end{array}
\right.
\end{equation}
We can write
{\medmuskip=0mu
\thinmuskip=0mu
\thickmuskip=0mu\begin{equation}
\begin{aligned}
\theta 
&=  -  8 \pi    \sum_{\beta \in \{ \rm{S},\rm{B},\rm{N}\}}  \int_{\pi}^{0} \frac{\left(\frac 1 3 \sigma^{3/2}(r) \, \delta \langle p \rangle  + \delta A_\beta  \right) \Theta_\beta(r) \, |\sigma'| \d r}{\sqrt{ A^2_\beta  + \left( {\sfrac 2 3} \langle p \rangle  A_\beta - 8 \, m_\beta  \right) \sigma^{3/2}  + 4 \, \sigma^2 - {\sfrac 1 9}  \left( 12 \, \Lambda - \langle p \rangle^2 \right) \sigma^3 }} 
\\
&=  -  8 \pi  \sum_{\beta \in  \{ \rm{S},\rm{B},\rm{N}\}}   \int_0^\pi \left( \frac{\partial F_\beta [A_\beta,\langle p \rangle ,\sigma]}{\partial \langle p \rangle}  \delta \langle p \rangle  +  \frac{\partial F_\beta [A_\beta,\langle p \rangle ,\sigma]}{\partial A_\beta}  \delta A_\beta \right) | \sigma'| \d r    \,,
\end{aligned}
\end{equation}}
where
{\medmuskip=0mu
\thinmuskip=0mu
\thickmuskip=0mu
\begin{equation}
F_\beta ~=~   \log \left(\sqrt{ A_\beta^2   + \left( {\sfrac 2 3} \langle p \rangle  A_\beta  - 8 \, m_\beta   \right) \sigma^{3/2}  + 4 \, \sigma^2 - {\sfrac 1 9}  \left( 12 \, \Lambda - \langle p \rangle^2 \right) \sigma^3 }+A_\beta   + {\sfrac 1 3} \langle p \rangle    \sigma^{3/2}\right) \,.
\end{equation}}

The boundary conditions force $\sigma$ to be zero at the poles, and rise monotonically away from the poles up to the location of the two shells, $R_\st{S}$ and $R_\st{N}$. In the `belt' region, $\sigma$ has to be piecewise monotonic except when its value reaches a zero of the polynomial $ A^2_\st{B}  + \left( {\sfrac 2 3} \langle p \rangle  A_\st{B} - 8 \, m_\st{B}  \right) \sigma^{\frac 3 2}  + 4 \, \sigma^2 - {\sfrac 1 9}  \left( 12 \, \Lambda - \langle p \rangle^2 \right) \sigma^3$. A situation of particular interest is when $\Lambda >0$ and $\langle p \rangle^2 < 12 \,\Lambda$, so that there is a maximal positive root of the polynomial whose value is dominated by $\Lambda$ (a cosmological curvature scale). Then a consistent choice is to have $\sigma$  grow monotonically from $R_\st{S}$ to $r_\st{max}$, the location of its absolute maximum, and then decrease monotonically from $r_\st{max}$ to $R_\st{N}$ (see Fig.~\ref{TwinShell_DeterminingIntervalSigma}). This means that our pre-symplectic potential can be written
{\medmuskip=0mu
\thinmuskip=0mu
\thickmuskip=0mu\begin{equation}
\begin{aligned}
\theta =  -  8 \pi \bigg[  &  \int_0^{\rho^2_\st{S}}\left( \frac{\partial F_\st{S}}{\partial \langle p \rangle}  \delta \langle p \rangle  +  \frac{\partial F_\st{S}}{\partial A_\st{S}}  \delta A_\st{S} \right)  \, d \sigma 
+ \int_{\rho^2_\st{S}}^{\rho^2_\st{max}} \left( \frac{\partial F_\st{B}}{\partial \langle p \rangle}  \delta \langle p \rangle  +  \frac{\partial F_\st{B}}{\partial A_\st{B}}  \delta A_\st{B} \right)  \, d \sigma  +
\\
&
 \int_{\rho^2_\st{N}}^{\rho^2_\st{max}} \left( \frac{\partial F_\st{B}}{\partial \langle p \rangle}  \delta \langle p \rangle  +  \frac{\partial F_\st{B}}{\partial A_\st{B}}  \delta A_\st{B} \right) d \sigma 
+ \int_0^{\rho^2_\st{N}} \left( \frac{\partial F_\st{N}}{\partial \langle p \rangle}  \delta \langle p \rangle  +  \frac{\partial F_\st{N}}{\partial A_\st{N}}  \delta A_\st{N} \right)  d \sigma \bigg]      \,,
\end{aligned}
\end{equation}}
and since
\begin{equation}
 \frac{\partial F_\beta [A_\beta,\langle p \rangle , \sigma]}{\partial \langle p \rangle}  = \frac 1 3 \sigma^{3/2} \frac{\partial F_\beta [A_\beta,\langle p \rangle , \sigma]}{\partial A_\beta} \,,
 \end{equation}
 we can write
 \begin{equation}
\begin{aligned}
\theta =  -  8 \pi \bigg[  &  \int_0^{\rho^2_\st{S}} \frac{\partial F_\st{S}}{\partial A_\st{S}}   \left( {\sfrac 1 3} \sigma^{\frac 3 2} \delta \langle p \rangle  + \delta A_\st{S} \right)  \, d \sigma 
+ \int_{\rho^2_\st{S}}^{\rho^2_\st{max}}   \frac{\partial F_\st{B}}{\partial A_\st{B}} \left({\sfrac 1 3} \sigma^{\frac 3 2}  \delta \langle p \rangle  +  \delta A_\st{B} \right)  \, d \sigma  +
\\
&
 \int_{\rho^2_\st{N}}^{\rho^2_\st{max}}  \frac{\partial F_\st{B}}{\partial A_\st{B}}   \left( {\sfrac 1 3} \sigma^{\frac 3 2}  \delta \langle p \rangle  + \delta A_\st{B} \right) d \sigma 
+ \int_0^{\rho^2_\st{N}} \frac{\partial F_\st{N}}{\partial A_\st{N}}  \left( {\sfrac 1 3} \sigma^{\frac 3 2}  \delta \langle p \rangle  +   \delta A_\st{N} \right)  d \sigma \bigg]      \,.
\end{aligned}
\end{equation}
The symplectic form is then
{\medmuskip=0mu
\thinmuskip=0mu
\thickmuskip=0mu\begin{equation}
\begin{aligned}
\delta \theta =&  -  8 \pi     \left[  \frac{\partial F_\st{S} [\sigma = \rho^2_\st{S}]}{A_\st{S}}  \delta \rho^2_\st{S} \wedge \left( {\sfrac 1 3} \rho^3_\st{S} \delta \langle p \rangle  + \delta A_\st{S} \right)   -   \frac{\partial F_\st{B} [\sigma = \rho^2_\st{S}]}{A_\st{B}}  \delta \rho^2_\st{S} \wedge \left( {\sfrac 1 3} \rho^3_\st{S}   \delta \langle p \rangle  + \delta A_\st{B} \right)   \right] 
\\
&  -  16 \pi \,  \frac{\partial F_\st{B} [\sigma = \rho^2_\st{max} ]}
{A_\st{B}} \delta \rho^2_\st{max} \wedge  \left( {\sfrac 1 3} \rho_\st{max}^3 \delta \langle p \rangle  + \delta A_\st{B} \right) 
\\
& -  8 \pi  \left[ \frac{\partial F_\st{N} [\sigma = \rho^2_\st{N}]}{A_\st{N}}  \delta \rho^2_\st{N} \wedge  \left( {\sfrac 1 3} \rho^3_\st{N}  \delta \langle p \rangle  + \delta A_\st{N} \right)  - \frac{\partial F_\st{B} [\sigma = \rho^2_\st{N}]}{A_\st{B}} \delta \rho^2_\st{N} \wedge   \left( {\sfrac 1 3} \rho^3_\st{N} \delta \langle p \rangle  + \delta A_\st{B} \right)  \right]    \,.
\end{aligned}
\end{equation}}
We can prove that $\rho_\st{max}^2$ completely disappears from the symplectic form. In fact $\rho_\st{max}^2$ is a solution of the equation
\begin{equation}
A^2_\st{B}  + \left( {\sfrac 2 3} \langle p \rangle  A_\st{B} - 8 \, m_\st{B}  \right) \rho_\st{max}^3  + 4 \, \rho_\st{max}^4- {\sfrac 1 9}  \left( 12 \, \Lambda - \langle p \rangle^2 \right) \rho_\st{max}^6 = 0 \,.
\end{equation}
Varying the above equation w.r.t. $\rho_\st{max}$, $A_\st{B}$ and $\langle p \rangle$ we get an identity for $\delta \rho_\st{max}$:
\begin{equation}
\delta \rho_\st{max}^2 = f[\rho_\st{max},A_\st{B},\langle p \rangle] \left( {\sfrac 1 3} \rho_\st{max}^3 \delta \langle p \rangle + \delta A_\st{B} \right) \,,
\end{equation}
and therefore the only term containing $\rho_\st{max}$ vanishes:
{\medmuskip=0mu
\thinmuskip=0mu
\thickmuskip=0mu\begin{equation}
\begin{aligned}
 &-  16 \pi \,  \frac{\partial F_\st{B} [\sigma = \rho^2_\st{max} ]}
{A_\st{B}} \delta \rho^2_\st{max} \wedge  \left( {\sfrac 1 3} \rho_\st{max}^3 \delta \langle p \rangle  + \delta A_\st{B} \right) 
\\
&= -  16 \pi \,  \frac{\partial F_\st{B} [\sigma = \rho^2_\st{max} ]}
{A_\st{B}} f[\rho_\st{max},A_\st{B},\langle p \rangle] \left( {\sfrac 1 3} \rho_\st{max}^3 \delta \langle p \rangle + \delta A_\st{B} \right)\wedge  \left( {\sfrac 1 3} \rho_\st{max}^3 \delta \langle p \rangle  + \delta A_\st{B} \right) =0  \,.
\end{aligned}
\end{equation}}
We can finally use the boundary conditions at the poles, $A_\st{S} = m_\st{S} = A_\st{N} = m_\st{N} =0$, and we get the following nondegenerate 2-form ($\omega = \delta \theta$):
%
{\medmuskip=0mu
\thinmuskip=0mu
\thickmuskip=0mu
\begin{equation}\label{SymplecticFormTwinShell}
\begin{aligned}
&\omega \propto  \frac{ \frac{8 \pi}{3}    \rho^3_\st{S}   \delta \rho^2_\st{S} \wedge  \delta \langle p \rangle + 3 \, \delta \rho^2_\st{S} \wedge \delta A_\st{B}}{\sqrt{ A^2_\st{B}  + \left( {\sfrac 2 3} \langle p \rangle  A_\st{B} - 8 \, m_\st{B}  \right) \rho_\st{S}^3 + 4 \, \rho_\st{S}^4 - {\sfrac 1 9}  \left( 12 \, \Lambda - \langle p \rangle^2 \right) \rho_\st{S}^6 }}    ~-~ \frac{\frac{8 \pi}{3}   \rho^3_\st{S} \delta \rho^2_\st{S} \wedge \delta \langle p \rangle }{\sqrt{  4 \, \rho_\st{S}^4 - {\sfrac 1 9}  \left( 12 \, \Lambda - \langle p \rangle^2 \right) \rho_\st{S}^6 }} 
\\
& +  \frac{   \rho^3_\st{N}   \delta \rho^2_\st{N} \wedge  \delta \langle p \rangle + 3 \, \delta \rho^2_\st{N} \wedge \delta A_\st{B}}{\sqrt{ A^2_\st{B}  + \left( {\sfrac 2 3} \langle p \rangle  A_\st{B} - 8 \, m_\st{B}  \right) \rho_\st{N}^3 + 4 \, \rho_\st{N}^4 - {\sfrac 1 9}  \left( 12 \, \Lambda - \langle p \rangle^2 \right) \rho_\st{N}^6 }}  
~-~\frac{\frac{8 \pi}{3}    \rho^3_\st{N} \delta \rho^2_\st{N} \wedge \delta \langle p \rangle }{\sqrt{  4 \, \rho_\st{N}^4 - {\sfrac 1 9}  \left( 12 \, \Lambda - \langle p \rangle^2 \right) \rho_\st{N}^6 }}  \,,
\end{aligned}
\end{equation}}
The above 2-form is nondegenerate in the 4-dimensional phase space coordinatized by $\langle p \rangle$, $A_\st{B}$, $\rho_\st{S}$ and $\rho_\st{N}$. To reach the above expression we used every constraint that was at our disposal (the solution of the Hamiltonian, diffeomorphism and conformal constraint, and the two diffeomorphism jump conditions), except the two jump conditions associated to the Hamiltonian constraint. Notice that we didn't need to use a diffeomorphism gauge fixing to get a nondegenerate symplectic form, because we were able to recast the pre-symplectic form in a reparametrization-invariant form. In other terms, we avoided having to completely gauge fix our constraints by expressing the symplectic form in terms of a maximal system of gauge-invariant quantities.

\subsubsection*{Breakdown of the ADM description}\label{SecTheProblem}

\begin{figure}[b!]\center
~~~~~~~~~~~~~\includegraphics[width=0.5\textwidth]{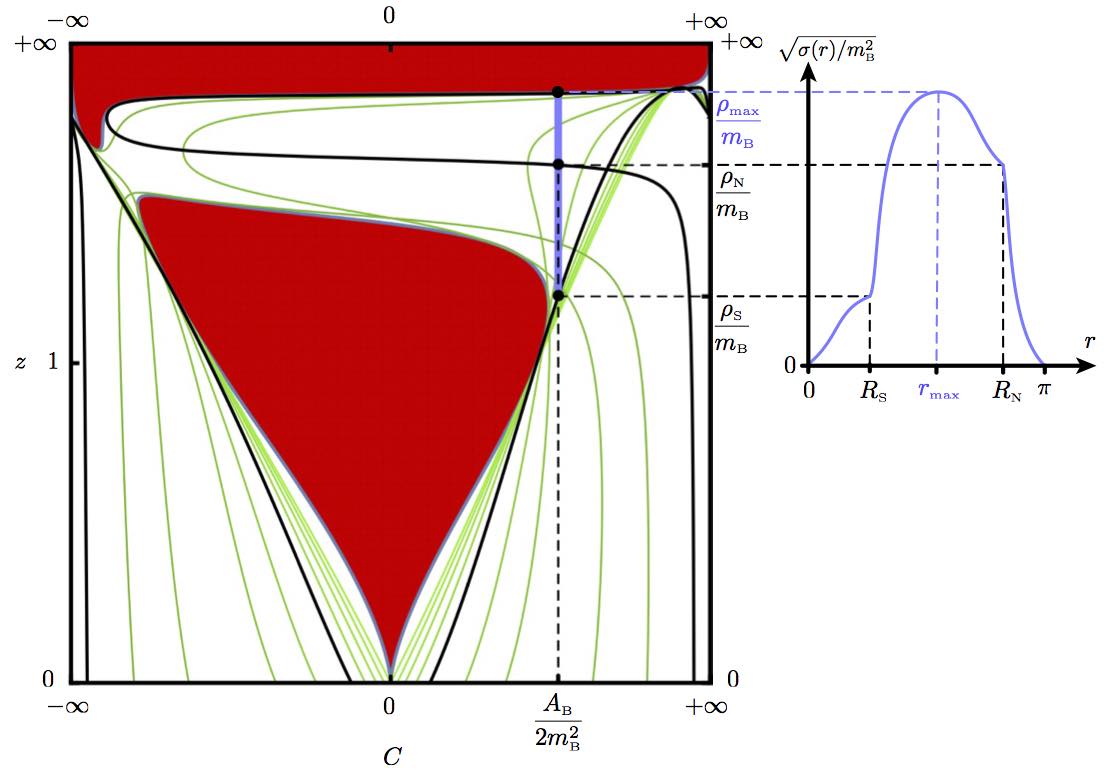}
\caption[Determining the codomain of $\sigma$ in the twin-shell universe]{A plot of the on-shell curves at a fixed York time $\tau =  0.46$  for a set of values of the rest mass $M_a/m_\st{B}$ (between 0 and 20), and for  $\lambda = 0.1$, $m_\st{B}>0$.
The excluded region $\mathscr P <0$ is in red.
Given the values of the rest masses of the two shells (in the figure $M_\st{S} = 2.5 \, m_\st{B}$ and  $M_\st{N} = 20 \, m_\st{B}$), specifying the value of the integration constant $A_\st{B}$ in the belt completely fixes  $\rho_\st{S}$, $\rho_\st{N}$ and $\rho_\st{max}$. Then the interval of values of the areal radius coordinate $\sigma(r)$ of the metric in the belt is fixed (light-blue strip). 
$\sigma$ will go from $\rho_\st{N}$ to a maximum given by the border of the excluded region (where $\sigma'$ is allowed to vanish), and then will go down until it reaches $\rho_\st{S}$. A  choice of $\sqrt{\sigma(r)}$ compatible with the boundary conditions imposed by the values of $\rho_\st{S}$, $\rho_\st{N}$ and $\rho_\st{max}$ is showed on the right.
}\label{TwinShell_DeterminingIntervalSigma}
\end{figure}

In this section I will discuss the conditions under which the ADM description of the system breaks down. To do this, I need first to show how the on-shell relations~(\ref{Twinshell_OnshellCondition_dimensionless}) are to be used to provide boundary conditions for the metric in a context with two shells. 
Consider a constant-York-time  slice $\tau = \text{\it const.}$. In Fig.~\ref{TwinShell_DeterminingIntervalSigma} I plot the `forbidden' region $\mathscr P <0$ in red in the plane $(C,z)$. In green I show a series of on-shell curves, solutions of~(\ref{Twinshell_OnshellCondition_dimensionless}) for different values of the rest-mass (normalized by $|m_\st{B}|$): $M/|m_\st{B}|$. Among these, two curves will correspond to the rest mass of the two shells, $M_\st{N}$ and $M_\st{S}$. I plot those in black. If we choose a value of $A_\st{B}$, through Eq.~(\ref{Twinshell_OnshellCondition_dimensionless}) we are also fixing the value of the areal radii of the two shells, $\rho_\st{S}$ and $\rho_\st{N}$, which can be read in the diagram as the ordinates of the corresponding points on the two black on-shell curves. This is like fixing the total energy of a one-dimensional system; the relation between position (areal radius) and momentum (given by $A$) is thereafter completely determined. The constraints of the system do not allow for independent behaviour of the two shells: they are `interlocked'.
Moreover, if $\lambda >0$, we also fix a \emph{maximum} areal radius $\rho_\st{max}$ that the metric can support, which is essentially determined by the cosmological constant (in Fig.~\ref{TwinShell_DeterminingIntervalSigma} it is the border of the top disconnected component of the red forbidden region).  Given all this data, we can determine an attainable form for the $\theta\theta$ component of the metric ($\sigma(r)$, the areal radius squared): it will monotonically interpolate $\sigma = 0$ with $\sigma = \rho_\st{S}^2$ (resp. $\rho_\st{N}^2$) from $r =0$ (resp. $\pi$) to $r= R_\st{S}$ (resp. $R_\st{N}$). Then its derivative will have, at $r= R_\st{S}$  (resp. $r= R_\st{N}$) a certain jump determined by Eqs.~(\ref{Twinshell_JumpGammaKappa}). In the region in between (the `belt' region) $\sigma$ will go from $\sigma(R_\st{S}) = \rho^2_\st{S}$ to a maximum $\sigma(r_\st{max}) = \rho^2_\st{max}$ and then down to $\sigma(R_\st{N}) = \rho^2_\st{N}$. Away from $r=r_\st{max}$, $\sigma$ will be monotonic. All of this is  illustrated by the Cartesian diagram on the right of Fig.~\ref{TwinShell_DeterminingIntervalSigma}.
Notice that, while interpolating in the belt region from one shell to the other, we could  have alternatively avoided having the areal radius reach the maximum value $\rho_\st{max}^2$ and bounce back. This is an acceptable choice if $\rho_\st{N} \neq \rho_\st{S}$, because the areal radius could monotonically interpolate between $\rho_\st{S}$ and $\rho_\st{N}$. But as we can see in the diagram above, the two black on-shell curves intersect at a point, which means that there exists a value of $A_\st{B}$ such that $\rho_\st{N} = \rho_\st{S}$ even though $M_\st{N} \neq M_\st{S}$. Then in this case we are forced to have the areal radius grow up to  $\rho_\st{max}^2$ and back, otherwise it could not possibly be interpolating between $\rho_\st{N}$ and $\rho_\st{S}$ while being monotonic. I conclude that the only consistent choice is that $\sigma$ \emph{always} bounces off the  value $\rho_\st{max}^2$, even when $\rho_\st{N} \neq \rho_\st{S}$.

\begin{figure}[b!]\center
~~~~~~~~~~~~~\includegraphics[width=0.5\textwidth]{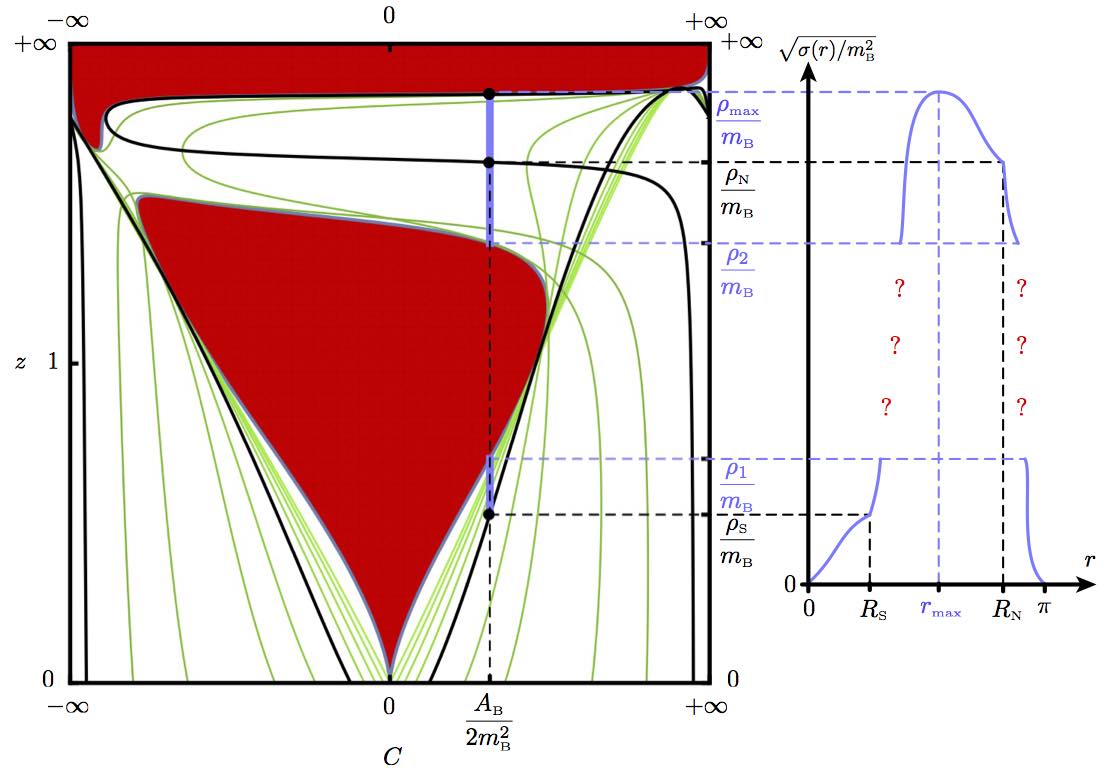}
\caption[Determining the codomain of $\sigma$ in the twin-shell universe]{Same plot as above, but with a value of $A_\st{B}$ such that the $C = \frac{A_\st{B}}{2 m_\st{B}}$ line crosses the bottom forbidden region (in red). The two points at which this crossing happens have $z = \frac{\rho_1}{m_\st{B}}$ and $z = \frac{\rho_2}{m_\st{B}}$.}\label{ProblemFig2}
\end{figure}

\index{spherically symmetric solutions}\index{thin shell of dust}

I am now ready to present the issue. Consider the diagram of Fig.~\ref{ProblemFig2}.
Now the chosen value of $A_\st{B}$ is such that the forbidden region crosses the line that connects $\rho_\st{S}$ with $\rho_\st{N}$. In this situation there is no acceptable solution to the constraint equations! In fact, the areal radius of the metric $\sqrt{\sigma}$ cannot  take all the values that are included in the interval $\left( \rho_\st{S},\rho_\st{N} \right)$, because a section of this interval is excluded.
There exists no metric that solves the Lichnerowicz--York equation in this situation. 
In the shape-dynamical interpretation of this system the spatial metric  is not itself physical, only its shape degrees of freedom are, and they live in a reduced shape-phase space, which is represented by the green on-shell surface, which never crosses the forbidden region and seems to be globally well-defined. On the other hand, in the ADM interpretation the spatial metric is the pull-back of the spacetime metric on a CMC hypersurface, and the fact that it is not well-defined implies that there is no spacetime metric, and the solution is not an acceptable solution of Einstein's equation. We identified a new point of departure between Shape Dynamics and GR: when the dynamical solution enters this region where the areal radius should interpolate between values that surround the forbidden region, the SD description is well-defined, while the GR one is not.

Notice that, as can be seen from the diagrams in this Section, the only case in which this departure is possible is that with positive Misner--Sharp mass $m_\st{B} >0$ and positive but small cosmological constant $\lambda >0$, $\lambda \ll  1$. The other choices of $m_\st{B}$ and $\lambda$ do not give rise to a `concave' allowed region where the on-shell surfaces of the two shells are separated by the forbidden region. Interestingly, the $m_\st{B} >0$, $0<\lambda \ll 1$ case is particularly physically relevant, as it seems to match our universe more closely.

\newpage

\subsection{Asymptotically flat Shape Dynamics} \label{AsymptoticallyFlatSD}
\index{asymptotic flatness}

Shape Dynamics makes sense primarily as a description of a spatially closed Universe. \index{closed spacelike hypersurfaces} However any successful theory of gravity also has to describe localized subsystems correctly. The simplest model is the asymptotically flat one, in which $\Sigma$ is open and the \emph{inferred 4-metric} tends to Minkowski's metric at infinity. This is supposed to model a small empty region of the Universe which is isolated by a vast empty region. In Fig.~\ref{NeighbouringStars} I schematically show the size of the empty space surrounding our Solar System, as opposed to the size of the orbits of the planets and of the Sun itself, which concentrates the vast majority of the mass present in that region. This is what I have in mind with a situation that is well-described by asymptotic flatness: an empty bubble, surrounded by a vast universe with a lot of faraway masses in it, at the centre of which there might be a concentration of mass (\emph{i.e.} the solar system). The dynamics of this concentration of mass is well described by an asymptotically flat background.

Now $\Sigma$ is open. A noncompact space does not go along very well with CMC foliations: the volume of space is not a well-defined concept (it is infinite), and therefore the meaning of  York time \index{York time}
 $\tau = {\sfrac 2 3} \langle p \rangle$, which would be its conjugate variable in the compact case, is not clear. A consistent choice is to put  $\langle p \rangle$ to zero. This foliation is called \emph{maximal slicing}. It makes sense as an approximation of a localized region of both space and time, in a time interval so short that the expansion of the Universe is negligible.

Moreover, stop for a second considering the spherically-symmetric solutions of ADM gravity in CMC foliation that we found in the previous section: Eq.~(\ref{VacuumSolutionForMu}) gives, for the radial component of the metric,
\begin{equation}
\mu^2 =  \frac{(\sigma')^2}{\frac{A^2}{\sigma }  + \left( {\sfrac 2 3} \langle p \rangle  A - 8 \, m  \right) \sqrt{\sigma}  + 4 \, \sigma - {\sfrac 1 9}  \left( 12 \, \Lambda - \langle p \rangle^2 \right) \sigma^2 } \,.
\end{equation}
unless $12 \, \Lambda - \langle p \rangle^2 =0$, as the areal radius $\sqrt{\sigma}$ goes to infinity, $\mu^2$ either goes to zero or becomes imaginary, signalling that there is a maximal radius. The only option to generically have a spatially asymptotically flat metric is that both $\Lambda =0$ and $\langle p \rangle =0$.
This should be physically understood in the following way: we are interested in scales which are much larger than the support of the matter fields we want to describe (\emph{e.g.} the solar system in Fig.~\ref{NeighbouringStars},  but are much smaller than the cosmological radius of curvature. So the interval of areal radii we are interested in is such that  the term $\langle p \rangle^2 \sigma^2$ is much smaller than $\frac{A^2}{\sigma } $Êand $ m \sqrt{\sigma}$.
\newpage

\begin{figure}[h!]
\begin{center}
\includegraphics[width=0.49\textwidth]{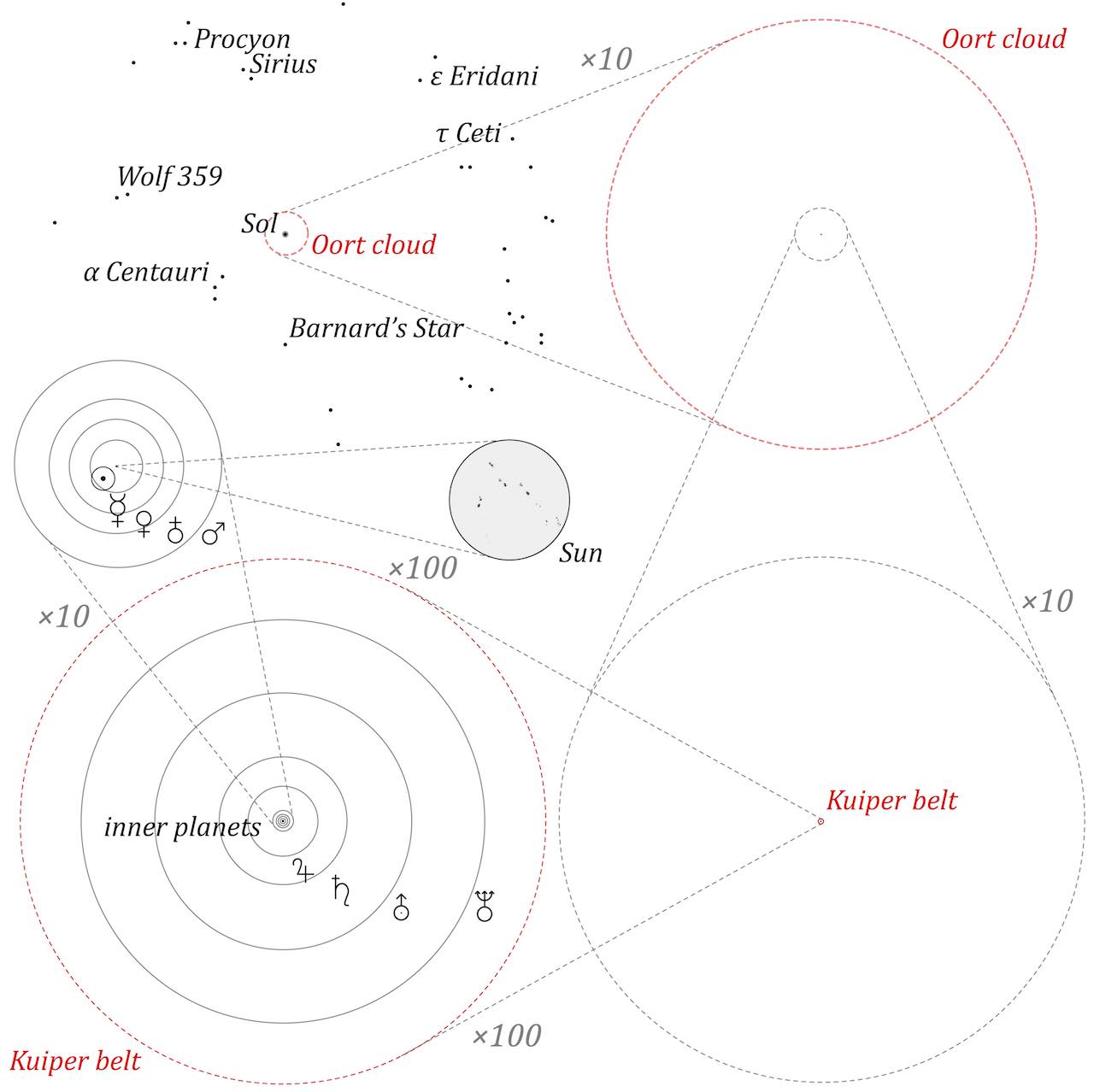}
\end{center}\vspace{18pt}
\caption[The neighbours of the Solar System]{The closest stars to the Solar System (within a radius of 13.1 light years). Magnifying 10 000 times the region of the Solar System, we can see the approximate size of the Kuiper Belt and the orbits of the outer planets. The solar system could be considered mostly empty itself, as more than 99.8\% of its mass is concentrated in the Sun, which is $\sim$6~000 times smaller than Neptune's orbit. The region of the Oort cloud is a very large void ($\sim 1$ ly $= 6~300$ AU) containing less than 5 earth masses of matter. The interstellar voids between the Oort cloud and the nearest neighbouring star systems are even more rarefied. 
}
\label{NeighbouringStars}
\end{figure}\index{asymptotic flatness}

\newpage

Having an open manifold, we cannot solve the constraints and equations of motion without boundary conditions for all of the fields. These, in the case of a manifold with an asymptotic infinity, are codified into \emph{falloff conditions} of the fields at infinity. In the GR literature, it is customary to assume the following falloff conditions for metric, the momentum, the lapse and the shift (see Appendix~\ref{AsymptoticallyFlatADM} for their standard derivation in GR):
\begin{equation}\label{AFfalloffConditions}
\begin{aligned}
&g_{ij} \to \delta_{ij} + \mathcal{O} (r^{-1}) \,, & p^{ij} \to  \mathcal{O} (r^{-2}) \,, \\
 & N \to 1 + \mathcal{O} (r^{-1}) \,, & \xi^i \to \mathcal{O} (1) \,.
\end{aligned}
\end{equation}
These conditions  should capture the effect of the rest of the Universe on the local system: they provide
a reference frame, scale, and time unit. Eqs.~(\ref{AFfalloffConditions}) are obtained from the Schwarzschild spacetime, \index{Schwarzschild's spacetime} which is the prototype of the asymptotically flat spacetime. Behind this assumption lies a very powerful result in GR: \emph{Birkhoff's theorem}, which states that any vacuum-spherically symmetric solution of Einstein's equations must be static and asymptotically flat. This in turn implies that if the all the matter in a spherically symmetric solution is concentrated inside a compact region, then the exterior solution is necessarily isometric to Schwarzschild's spacetime.

Now I reproduce the discussion of the state of Birkhoff's theorem in Shape Dynamics which I first published in~\cite{NoBirkhoff}.

\subsubsection*{Birkhoff's Theorem}
\index{asymptotic flatness}\index{Birkhoff's theorem}

In Sec.~\ref{SphericallySymmetricSec} I found the general solution of the ADM-CMC equations under the assumptions~(\ref{SphericalSymmetry-g}), (\ref{SphericalSymmetry-p}) and (\ref{SphericalSymmetry-xi}) of spherical symmetry:
\begin{equation}
\begin{gathered}\textstyle
g_{ij} = \text{diag} \, \left\{  \mu^2 , \sigma  , \sigma \, \sin^2 \theta  \right\} \,,
~~
p^{ij} =  \text{diag} \, \left\{  \frac{f}{\mu} , {\sfrac 1 2}  s   , {\sfrac 1 2}  s  \, \sin^{-2} \theta  \right\} \, \sin \theta \,,
\\
\xi^i = (\xi , 0 ,0 )\,.
\end{gathered}
\end{equation}
The solution can be found in Eq.~(\ref{SolCMCconst}), (\ref{SolutionVacuumDiffConstraint}), (\ref{VacuumSolutionForMu}), (\ref{SolutionEqOfMotionThinShell}), (\ref{SolutionLFEtwinshell}), which I reproduce here in the $\Lambda = \langle p \rangle =0$ case we're interested in now:
\begin{equation}\label{AFspherSymmSolution}
\begin{gathered}
 s  = 
    - \frac{\mu}{\sigma} \, f  \,, 
\qquad
     f = 
     \frac{A}{\sqrt \sigma}\,,
\\
\mu^2 =  \frac{(\sigma')^2}{\frac{A^2}{\sigma }  
- 8 \, m  \, 
 \sqrt{\sigma}  + 4 \, \sigma 
 } \,,
~~~~
N = \frac{\sigma'}{2 \mu \sqrt{\sigma}} \left( c_1 + c_2 \dashint \frac{\mu^3}{(\sigma')^2} \d r  
 \right) \,,
\end{gathered}
\end{equation}
while the equations of motion are identically solved if  $\xi = \left( f \, N + \dot \sigma \right) /\sigma' $,
$c_2 = -2\,\dot A$, 
 and $\dot m  = 0$ [from Eq.~(\ref{SolEquation_gdotrr}) and (\ref{SolutionEqOfMotionThinShell})].

It's easy to see that the metric $g_{ij}$ is asymptotically Euclidean: taking the areal radius as radial coordinate, $\sigma = r^2$, Eq.~(\ref{AFspherSymmSolution}) implies that $g_{rr} \xrightarrow[r\to \infty]{} 1$. Moreover, in this radial gauge, the falloff behaviour of the fields is
\begin{equation}
\begin{aligned}
&g_{ij} \to \delta_{ij} + \delta^r{}_i \delta^r{}_j {\frac {2 m} r} + \mathcal{O} (r^{-2})  \,, & p^{ij} \to \delta^i{}_r \delta^j{}_r {\frac A r} + \mathcal{O} (r^{-2}) \,, \\
 & N \to \text{\it const.} \left( 1 - \frac m {r} \right) + \mathcal{O} (r^{-2}) \,, & \xi^i \to \text{\it const.}  \frac{A}{2 r^2} + \mathcal{O} (r^{-3}) \,.
\end{aligned}
\end{equation}
we see that, if $A\neq 0$, the falloff conditions~(\ref{AFfalloffConditions}) are violated by $p^{rr}$.
This is a potentially alarming feature, but before setting blindly $A$ to zero to comply with~(\ref{AFfalloffConditions}), we should ask: what is the precise reason for conditions~(\ref{AFfalloffConditions})? The answer can be found in a 1987 paper by Beig and \'O Murchadha~\cite{beig1987poincare}, who studied the Poincar\'e invariance of the falloff conditions for asymptotically flat metrics.
\index{asymptotic flatness}\index{Birkhoff's theorem}\index{Poincar\'e transformations}\index{falloff conditions}

Beig and \'O Murchadha asked whether a given falloff behaviour for the metric and momenta $g_{ij}  \to r^{-n_{ij}}$, $p^{ij}  \to r^{-m_{ij}}$,  is preserved under infinitesimal Poincar\'e transformations, in the sense that the transformed metric too is such that $g'_{ij}  \to r^{-n_{ij}}$, $p'^{ij}  \to r^{-m_{ij}}$. To act with an infinitesimal Poincar\'e transformation on the canonical fields one needs to smear the total Hamiltonian  
\begin{equation}
H_\st{tot} = \int \d^3 x \left(N \, \mathcal H + \xi^i \, \mathcal H_i \right) \,,
\end{equation}
with appropriate lapse $N$ and shift $\xi^i$ fields that implement the wanted transformation.
It is easy to convince oneself that the following lapse:
\begin{equation}
N = \alpha^0 + \left( \beta^x \,  \sin \theta \cos  \phi +\beta^z \, \sin \theta  \sin \phi  + \beta^y  \cos  \phi \right) \,,
\end{equation}
implements, at the boundary, a time translation with parameter $\alpha^0$, and a boost with rapidity vector $  \beta^\st{A} = (\beta^x , \beta^y , \beta^z)$.\footnote{This lapse has been written in spherical coordinates. In cartesian coordinates it takes the more familiar form $N = \alpha^0 +  \beta^\st{A}  x_\st{A}/|x|$.}
The following shift:
\begin{equation}
\xi^i  =   \alpha^\st{A} \, \xi_\st{A}^i + \omega^\st{B} \, \chi_\st{A}^i \,,
\end{equation}
where
\begin{equation}\!\!\!\!\!\label{BoundaryKillingVectors}
\begin{aligned}
&\vec \xi_x = \left(\sin \theta \cos \phi, {\sfrac { \cos \theta \cos \phi} r} ,- {\sfrac { \csc \theta \sin \phi } r}\right)\,, & &
\vec \chi_x = \left( 0,\sin \phi ,\cot \theta \cos \phi  \right) \,,&
\\
&\vec \xi_y = \left(\sin \theta \sin \phi,{\sfrac { \cos \theta \sin \phi} r},{\sfrac { \csc \theta \cos \phi} r} \right) \,, & &
\vec \chi_y = \left( 0,-\cos \phi ,\cot \theta \sin \phi  \right) \,, &
\\
&\vec \xi_\st{z} = \left( \cos \theta,-{\sfrac { \sin \theta} r} ,0\right)\,,&
&
\vec \chi_\st{z} = \left( 0,0,1 \right) \,.&
\end{aligned}
\end{equation}
implements spatial translations with parameter $\alpha^\st{A} = ( \alpha^x, \alpha^y, \alpha^z)$ and rotations with parameter $\omega^\st{A} = ( \omega^x, \omega^y, \omega^z)$.\footnote{In cartesian coordinates these vector fields take the form $$\begin{aligned}
&\xi_x = \partial_x \,, & &
\xi_y = \partial_y \,, & &
 \xi_z = \partial_z \,,& \qquad
 & \text{(translations)}&
\\
&\chi_x = y \partial_z - z \partial_y \,,& &
\chi_y = z \partial_x - x \partial_z \,,& &
\chi_z = x \partial_y - y \partial_x \,.&
& \text{(rotations)} &
\end{aligned}$$}
These smearings generate every possible Poincar\'e transformation at infinity. The action of such transformations on the metric and momenta is given by the Poisson brackets with the total Hamiltonian,
smeared with the lapse and the shift introduced here. If these Poisson brackets do not generate a higher-order term in $r$, then the falloff conditions are preserved by Poincar\'e transformations. For this calculation, it is important to notice that the falloff conditions written above in cartesian coordinates are different in spherical coordinates:
\begin{equation}
\begin{aligned}
& g_{rr}  = 1 + \mathcal O ({\sfrac 1 r})  \,, & &  g_{r\theta} = {\sfrac{ g_{r\phi}}{\sin^2\theta}}  = r^2 + \mathcal O (1) \,, &
& g_{\theta\theta} =
 g_{\theta\phi}  =
 g_{\phi\phi} = \mathcal O (r)
\,, &\\
& p^{rr}  =  \mathcal O ({\sfrac 1 r})  \,, & &  p^{r\theta} =    p^{r\phi} =   \mathcal O ({\sfrac 1 r^2}) \,, &
& p^{\theta\theta} =
 p^{\theta\phi}  =
 p^{\phi\phi} = \mathcal O ({\sfrac 1 r^3})
\,. &
\end{aligned}
\end{equation}
The Poisson brackets with the momenta preserve the falloff conditions:
\begin{equation}
\begin{aligned}
& \{ p^{rr} ,   H_\st{tot} \} = \mathcal O (1/r)  \,, \qquad \{ p^{r\theta} ,   H_\st{tot} \}  =  \{ p^{r\phi} ,   H_\st{tot} \} =   \mathcal O (1/r^2)&
\\
&\{ p^{\theta\theta} ,   H_\st{tot} \} =  
 \{ p^{\theta\phi} ,   H_\st{tot} \} =  
\{ p^{\phi\phi} ,   H_\st{tot} \} =   \mathcal O (1/r^3) \,,&
\end{aligned}
\end{equation}
but the ones with the metric do not:
\begin{equation}
\begin{aligned}
& \{ g_{rr} ,   H_\st{tot} \} =  2 A \, \left( \beta^x \, \sin \theta \cos \phi + \beta^y \, \sin \theta \sin \phi + \beta^z \cos \theta  \right) + \mathcal O (1/r)  \,,\
\\
& \{ g_{r\theta} ,   H_\st{tot} \} =  \mathcal O (1/r) \,, ~~~~
 \{ g_{r\phi} ,   H_\st{tot} \} =   \mathcal O (1/r)&
  \\
&\{ g_{\theta\theta} ,   H_\st{tot} \} =   \mathcal O (r)  \,, ~~~~
 \{ g_{\theta\phi} ,   H_\st{tot} \} =   \mathcal O (r)  \,, ~~~~
\{ g_{\phi\phi} ,   H_\st{tot} \} =   \mathcal O (r) \,,&
\end{aligned}
\end{equation}
\index{asymptotic flatness}\index{Birkhoff's theorem}\index{Poincar\'e transformations}\index{falloff conditions}\index{Killing vector}

in particular, it is the $A$ integration constant that makes the system not asymptotically invariant under boost. Any boost in any direction will break the falloff conditions. If the $A$ integration constant is set to zero, then, as is shown in~\cite{beig1987poincare}, the falloff conditions are invariant.
To prove that the falloff conditions are invariant under spacetime translations and space rotations it is not sufficient to show that they are preserved by these transformations. One has also to show that the parity of the leading-order components $g_{ij} \sim \delta_{ij} + \frac 1 r \, \delta g_{ij}(\theta,\phi) + \mathcal O (r^{-2})$ and $p^{ij} \sim   \frac A r \, \delta p^{ij}(\theta,\phi) + \mathcal O (r^{-2})$ is preserved, in the sense that also $ \{ g_{ij}(\theta,\phi)  H_\st{tot} \} $ and $ \{ p^{ij}(\theta,\phi)  H_\st{tot} \} $ have the same parity at leading order. In particular, under a parity transformation $\theta \to \pi -\theta$, $\phi \to \phi +\pi$, $g_{r\theta}$, $p^{r\theta}$, $g_{\phi\theta}$ and $p^{\phi\theta}$ are odd, and all the other components are even. An explicit calculation confirms that the parity of both the metric and the momenta are preserved at leading order.

$A \neq 0 $ makes the boundary conditions not boost-invariant, but it is still invariant under Euclidean transformations (translations and rotations) and time translations. These are the symmetries of a universe which is homogeneous, isotropic and static. In fact, this is more than a coincidence: I intend to link the symmetry properties of the boundary conditions directly to those of the surrounding universe. The idea is the following: an asymptotically flat solution has to be interpreted as an approximation to a region of a spatially closed \index{closed spacelike hypersurfaces} cosmological solution which is mostly empty, and the geometry in this region tends to that of a Euclidean metric at the boundary in the limit in which the finite size of the region is ignored, if the rest of the universe is approximately homogeneous and isotropic.

The metric~(\ref{AFspherSymmSolution}) has six approximate Killing vectors, \index{Killing vector} which satisfy the Killing equation only at the border.\footnote{Only rotations are Killing vectors for the complete metric, translations satisfy the Killing equation only in the limit $r \to \infty$.} These vectors are those pf Eq.~(\ref{BoundaryKillingVectors}).
Each one of them is associated to a boundary term (a charge) that is needed to  make the diffeomorphism constraint differentiable (\emph{i.e.} well-defined under functional variation, see Appendix~\ref{AsymptoticallyFlatADM}).  The boundary charges associated to our Killing fields are the linear momentum is~\cite{beig1987poincare}:
\begin{equation}
\begin{aligned}
P_\st{A}  &=  2  \int_{\partial \Sigma}  \,\xi_\st{A}^i \, p_{ij}  \, \d S^i 
\\
&=  \lim_{r\to \infty} \frac{A}{r} \int \sin \theta \left (\sin  \theta   \cos  \phi  , \sin  \theta   \sin  \phi  ,\cos  \theta   \right)_\st{A} \d \theta \d \phi = 0,
\end{aligned}
\end{equation}
where $\d S^i = \delta^i_r \d \theta \d \phi$, and the angular momentum:
\begin{equation}
L_\st{A} = 2  \int_{\partial \Sigma}  \,\chi_\st{A}^i \, p_{ij}  \, \d S^i =2  \int_{\partial \Sigma}  \,\chi_\st{A}^r \, p_{rr}  \, \d \theta \d \phi = 0 \,.
\end{equation}
They are both zero.
\index{asymptotic flatness}\index{Birkhoff's theorem}\index{Poincar\'e transformations}\index{falloff conditions}\index{Killing vector}

One can also calculate the boundary charges associated to the \emph{conformal} Killing vectors, \index{conformal Killing vector} which together with the regular Killing vector close an $SO(4,1)$ algebra. These are the dilatation vector field $\varphi^i = x^i$,
and the three special conformal transformation vectors fields, $\kappa^i_\st{A} = 2 x^i x_\st{A} - \delta^i{}_\st{A}  x^i x_j$. Or, in spherical coordinates, $\vec \varphi = \left( r , 0 ,0\right)$, $\vec \kappa_x = \left( r^2 \cos \phi \sin \theta ,-r  \cos \theta  \cos \phi,r \csc \theta  \sin \phi  \right)$,
$\vec \kappa_y = \left( r^2 \sin \theta  \sin \phi ,-r  \cos \theta  \sin \phi ,-r \cos \phi \csc \theta \right)$ and $\vec \kappa_\st{z} = \left( r^2  \cos \theta ,r \sin \theta ,0\right)$,
\begin{equation}
D = 2  \int_{\partial \Sigma}  \,\varphi^i \, p_{ij}  \, \d S^i =4 \pi \, A  \,,
\qquad
K_\st{A} = 2  \int_{\partial \Sigma}  \,\kappa_\st{A}^i \, p_{ij}  \, \d S^i =  0\,.
\end{equation}
We see that the integration constant $A$ is proportional to the dilatation charge. This is the physical interpretation of $A$ that we have been looking for: it is associated to the dilatational momentum of the rest of the universe which lies outside of the boundary of our region. If the universe is compact, the total dilatational momentum has to be zero (we saw it in the previous section with the dilatational momenta of the two shells, which are equal and opposite), so this coincides with the momentum of our region. A nonzero $A$ signals that the region we are considering is collapsing or expanding with respect to the rest.

In conclusion, as I first observed in~\cite{NoBirkhoff}, in the asymptotically flat case one simply doesn't have enough informations to fix the boundary conditions with the assumption of asymptotic flatness. We need to know whether the region of the universe we are considering is expanding or collapsing. In General Relativity, instead, the additional condition we need is provided by the assumption of Lorentz invariance of the boundary.
I must observe, however, that if we construct a 4-metric with the solution~(\ref{AFspherSymmSolution}), then it is isometric to Schwarzschild, \index{Schwarzschild's spacetime} as can be proved by calculating all its Riemannian invariants (left to the reader as an exercise). This means that, whatever choice of $A$ we make, we are just considering different foliations of Schwarzschild's spacetime, at least in the vacuum region. A choice of foliation is unphysical in GR, but in SD it has a physical meaning, as it is a theory with a preferred foliation.

In the rest of this Section  I will describe solutions that have been studied in the past  in the case $A=0$.

\subsubsection*{The `wormhole' solution of Gomes}
\index{asymptotic flatness}\index{wormhole}

In~\cite{GomesBH} H. Gomes assumed the standard ADM asymptotically flat falloff conditions~(\ref{AFfalloffConditions}) in a spherically symmetric vacuum context. This corresponds to assuming $A =0$ in~(\ref{AFspherSymmSolution}). In this case we can solve explicitly all equations in a `isotropic' gauge $\sigma = \mu^2 r^2$, and we get the following 3-metric and momentum:
\begin{equation}\label{HenriquesWormholeMetric}
\d s^2 = \left( 1 + \sfrac m {4r} \right)^4\left[ \d r^2 + r^2 \left(\d \theta^2 + \sin^2 \theta \d \phi^2\right) \right] \,,  \qquad  p^{ij} = 0 \,,
\end{equation}
moreover the lapse and shift which solve the equations of motion are
\begin{equation}\label{HenriquesWormholeLapseShift}
N = \frac{4 r - m }{4 r + m} \,, \qquad \xi^i = 0 \,.
\end{equation}
Notice that the  metric~(\ref{HenriquesWormholeMetric})  is invariant under inversions $r \to \frac{m^2}{16 r}$ (check left to the reader). This implies that it has two asymptotically flat ends, one at $r =\infty$ and the other at $r=0$, and  a `throat', that is, it cannot support spheres of surface smaller than the one at $r = \frac m 4$ (the area of that sphere is $4 \pi \,  m^2$). The lapse function~(\ref{HenriquesWormholeLapseShift}) vanishes at the throat, and is negative in the interior region $r< \frac m 4$.
This `wormhole' solution is a maximal foliation of a region of the Kruskal extension of Schwarzschild's spacetime. 
The region is the left- and right-quadrants of the Kruskal extension, those which do not contain the singularity (see Fig.~\ref{FoliationOddlapseFig}). For this reason the paper~\cite{GomesBH}  caused some excitement, because it looked like the black hole solution of shape dynamics avoided the Schwarzschild singularity and predicted, instead, the formation of a wormhole.
However it is hard  to interpret this solution as a physical wormhole: if the trajectory of matter particles is traced ignoring backreaction, they will just reach the horizon in an infinite `maximal slicing' time (since the lapse vanishes at the throat, if we ignore backreaction we can only see test matter freeze at the throat). There is no reason why they should continue past the horizon, or even `jump' to the other quadrant after crossing the horizon, as was suggested in~\cite{Poplawski}. The litmus test for the wormhole nature of this solution would be the study of the motion of matter on this background while taking backreaction into account. 
However, before doing this, we need to convince ourselves that this solution is physical. The most important thing is to check whether this solution can arise from the gravitational collapse of matter.

\index{asymptotic flatness}\index{wormhole}
\subsubsection*{The `breather' solution}

In~\cite{ThinShellPaper1}, with H. Gomes, T. Koslowski and A. Napoletano, we studied an asymptotically flat model with matter, in order to check whether the `wormhole' solution is left as a result of gravitational collapse of matter. 
It is a less sophisticated version of the analysis of~\cite{CompactThinShellsPaper} (and the previous Section) with a single shell of matter, in which both the cosmological constant $\Lambda$ and the \index{York time} York time $\langle p \rangle$ are set to zero. Moreover $A$ was set to zero outside of the shell was put to zero as in~\cite{GomesBH}, so that the falloff conditions~(\ref{AFfalloffConditions}) would be respected. Of course the jump conditions~(\ref{DeltaAjumpcondition}) would then generically set the value of $A$ inside the shell to a nonzero value, and this is incompatible with compactness of the metric in that region. To make it compact we have to assume the existence of further matter surrounding the pole, which can `cap off' our manifold. For example, we could be dealing with two shells, like in an asymptotically flat version of the `twin shell universe' studied above, but we are focusing on the dynamics of one of the two shells and the region of space that surrounds it.

In the present case we have to set $A_+=0$ (the dilatational momentum outside of the shell), but $m_+\neq 0$, because the Misner--Sharp mass in the asymptotically flat case coincides with the ADM mass \index{ADM mass} and should not be set to zero.
Then Eq.~(\ref{VacuumSolutionForMu}) in isotropic gauge $\mu^2=r^{-2}\sigma$ becomes
\begin{equation}\label{AFSingleShell_IsotropicEquation}
\frac{(\sigma')^2}{\frac{A_+^2}{\sigma} - 8 \, m_+ \sqrt{\sigma} +4 \sigma} = \frac{\sigma}{r^2} \,,
\end{equation}
whose solution is
\begin{equation}\label{AFSingleShellSigma}
\sigma  = \frac{m_+^2}{4}  \left[ \left( \frac {k_+\, r} {2 \, m_+} \right)^{\frac 1 2}  +  \left( \frac {2 \, m_+ }{k_+ \, r} \right)^{\frac 1 2}  \right]^4\,.
\end{equation}
The areal radius has a minimum at $r = 2 m_+ /k_+$ where its value is $\sigma = 4 m_+^2$, which is where $\sqrt{\sigma}/m_+$ reaches the only zero of the polynomial $\mathscr P(\sqrt\sigma /m_+)$. This minimum plays the same role as the `throat' of the `wormhole' solution. 
The on-shell condition~(\ref{SingleShellOnshellConstraint}) is, in isotropic gauge,
{\thinmuskip=0mu
\thickmuskip=0mu
\begin{equation}\label{AFOnshellCondition2}\textstyle
\left[ 4 (m_+ + m_-) \, \rho - 4 \, \rho^2  +  {\sfrac 1 8} M ^2 \right]^2 =  \left(4  \rho^2 - 8 \, m_+ \, \rho   \right) \left(  \frac{R ^2 P ^2} {4 \rho^2}  - 8 \, m_- \, \rho +4 \rho^2   \right) \,.
\end{equation}}
The equation above depends on $R $ through $\rho = \sqrt{\sigma(R )}$, and has also a dependence on $m_-$, the Misner--Sharp mass inside the shell, which we don't know anything about (because we decided to remain agnostic regarding the matter content inside the shell), except that it's dynamically conserved. Expressing everything in units of $m_+$:
\begin{equation}
R  = m_+ \, Y \,, ~~ P  = m_+ \, X  \,, 
~~
m_- = m_+  \, \alpha \,, ~~ M  =  m_+ \, Z \,,
\end{equation}
then Eq.~(\ref{AFOnshellCondition2}) becomes
\begin{equation}\label{OnShell_dimensionless}
\textstyle  \frac{Z^4}{64}  + \frac{(2 Y+1)^4}{r^2} \left( (\alpha -1)^2 - Z^2 \frac{\left(4 Y^2-4 \alpha  Y+1\right)}{16 (2 Y+1)^2}\right) =   \frac{(1-2 Y)^2 Y^2}{(2 Y+1)^2}  Y^2\,,
\end{equation}
In Fig.~\ref{OnShellPlots} I show the on-shell curves $Y$ vs. $X$, for any possible choice of rest-mass $Z$, and for a set of choices of $\alpha$. Notice that the constant $\alpha$, on physical grounds, should be smaller than one (and larger than zero), as the ADM mass \index{ADM mass} inside the shell should be smaller than outside.

\begin{figure}[h!]
\includegraphics[width=0.15\textwidth]{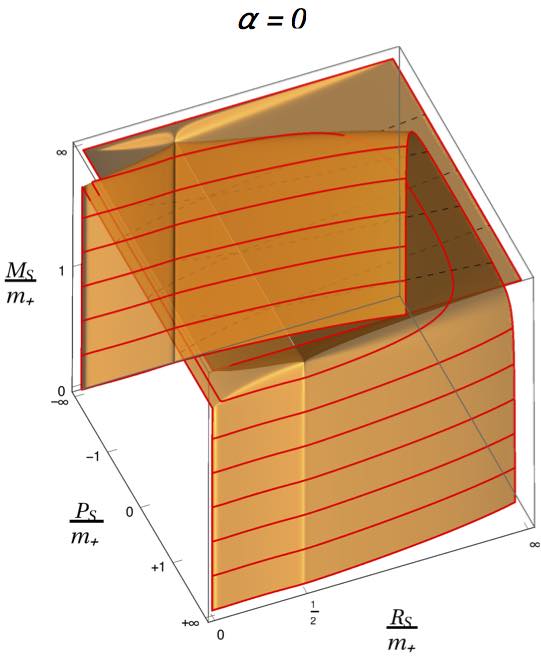}
~
\includegraphics[width=0.15\textwidth]{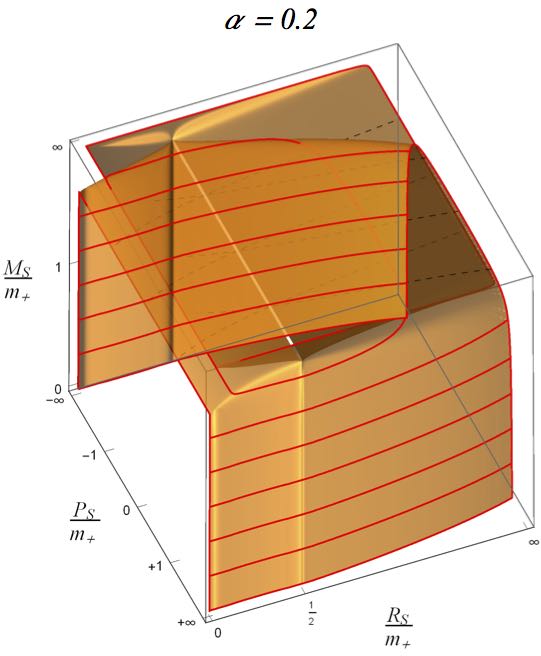}
~
\includegraphics[width=0.15\textwidth]{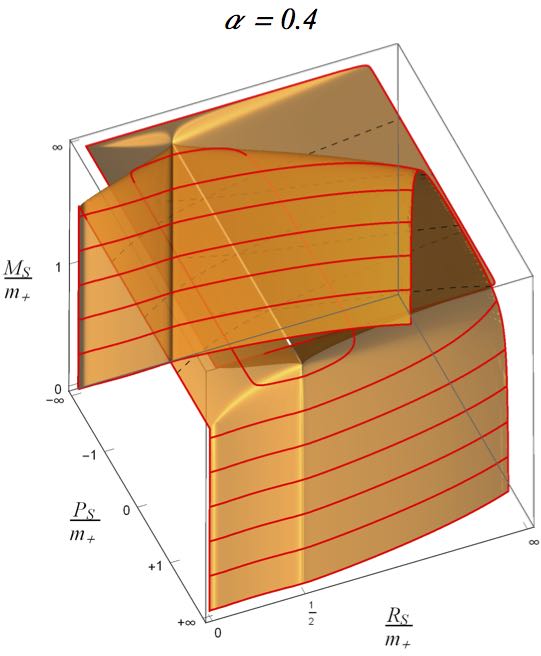}
\\
\includegraphics[width=0.15\textwidth]{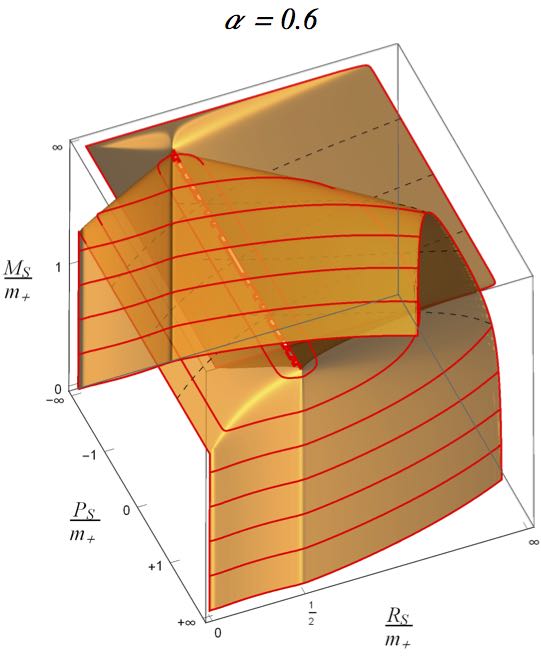}
~
\includegraphics[width=0.15\textwidth]{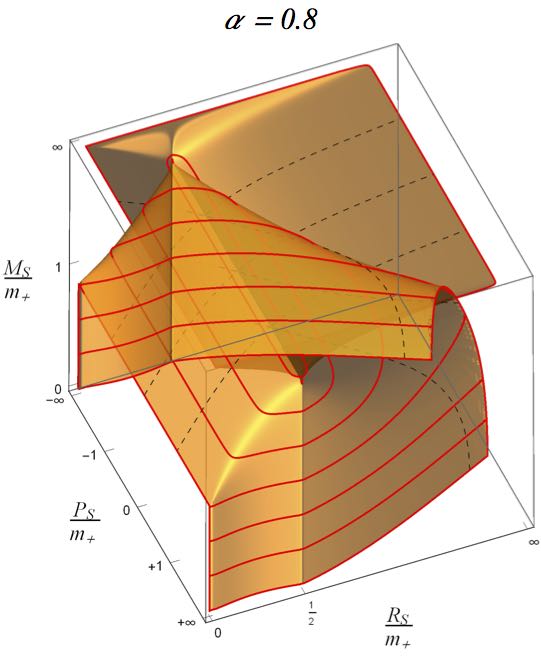}
~
\includegraphics[width=0.15\textwidth]{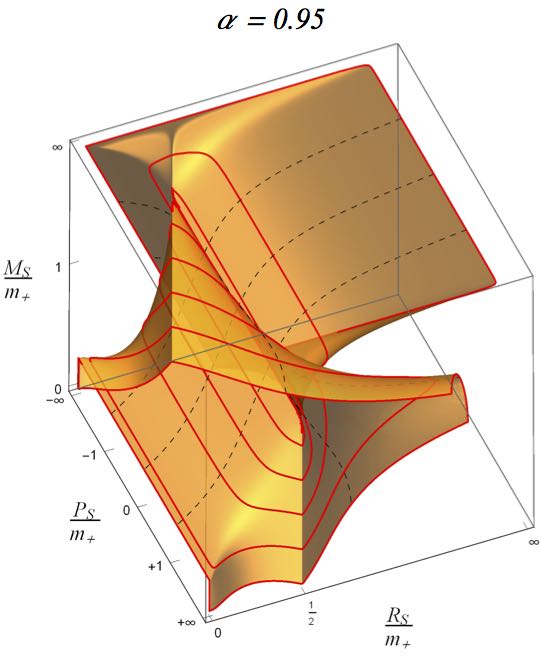}
\caption[AF single-shell on-shell surfaces]{Plots of the on-shell surface~(\ref{OnShell_dimensionless}) in the space $X \in (-\infty,\infty)$, $Y \in [0,\infty)$,  $Z \in [0,\infty)$, and for certain fixed values of $\alpha = m_- / m_+ \in  [0,1)$.  The three variables $X$, $Y$ and $Z$ have been compactified by taking their $\arctan$. The red curves represent the constant-$Z$ cross-sections, which are on-shell curves in the phase space $X$, $Y$. Notice how all the curves `bounce' on the $X = \pm \infty$ boundary of phase space at $Y = 1/2$.}\label{OnShellPlots}
\end{figure}

\index{asymptotic flatness}\index{wormhole}
A study of the Hamiltonian vector flow in the reduced phase-space $(R ,P )$ can tell us the speed at which the dynamics in maximal slicing time crosses the on-shell curves.  Consider Eq.~(\ref{AFOnshellCondition2}) as a condition on $m_+$: if we take into account the definition of $\rho = \sqrt{\sigma(R )}$, it turns into an eight-order equation for $m_+$. Let us write it as $F(m_+,R ,P ;M,m_-) = 0$. Its solution gives the ADM energy  $m_+$ as a function of the dynamical variables $R $ and $P $ (as well as the constant parameters $m_-$ and $M$). This is the Hamiltonian generator of evolution in maximal-slicing time. If all we are interested in are the equations of motion of $R $ and $P $ in this time variable, we can avoid having to explicitly solve $F=0$.  We can instead differentiate $F$ w.r.t.  all of the dynamical variables: $\frac{\partial F}{\partial m_+} d m_+ + \frac{\partial F}{\partial R } d  R  + \frac{\partial F}{\partial  P } d P  = 0 $, which implies that $\frac{\partial m_+ }{\partial R } = -\left.  \frac{\partial F}{\partial R } / \frac{\partial F}{\partial m_+} \right|_{F =0} $ and $\frac{\partial m_+ }{\partial P } = - \left.  \frac{\partial F}{\partial P }/ \frac{\partial F}{\partial P_+} \right|_{F =0}$.
Then the Hamiltonian equations of motion generated by $m_+$ are
\begin{equation}\label{HamiltonianEquationsOfMotion} \textstyle
\dot R  =  - \left. \frac{\partial F}{\partial P } \left(\frac{\partial F}{\partial P_+} \right)^{-1}\right|_{F =0} \,,
\qquad
\dot P  = \left. \frac{\partial F}{\partial R } \left( \frac{\partial F}{\partial m_+} \right)^{-1} \right|_{F =0}\,,
\end{equation}
which, before replacing the solution $F=0$, are two perfectly tractable functions of $R $, $P $ and $m_+$.

Eq.~(\ref{HamiltonianEquationsOfMotion}) allows us to  study the Hamiltonian vector flow in phase space. In particular we can check its behaviour at the `throat' $R  \to m_+/2$ without having to solve $F=0$. It turns out that the vector flow vanishes at the throat:
\begin{equation}
\dot R   \xrightarrow[R  \to m_+/2]{} 0 \,,
\qquad
\dot P    \xrightarrow[R  \to m_+/2]{} 0\,.
\end{equation}
The limits before are the same irrespective of the direction they are taken from.
so, as expected, in maximal-slicing time the shell `freezes' at the throat. One can also prove that it takes an infinite amount of maximal-slicing time for the shell to reach the throat, by explicitly integrating the vector flow. Keep in mind that Maximal-slicing time has no intrinsic physical meaning: one of the fundamental relational underpinnings of SD is that time should be abstracted from the change of physical (\emph{i.e.} shape) degrees of freedom. In this sense maximal slicing time is a useful approximation to the change in the DOFs of a clock far away from the origin.
\index{asymptotic flatness}\index{wormhole}

We can now  give at least partial answers to the question we set forth at the beginning: does the `wormhole'-like line element found in~\cite{GomesBH} emerge from the gravitational collapse of spherically symmetric matter? Under the same assumptions of asymptotically flat boundary conditions (\emph{i.e.} $p^{ij} \to r^{-2} ~\Rightarrow ~ A_+ =0$) at infinity the answer is clearly positive. The line element given by the areal radius~(\ref{AFSingleShellSigma}) outside of the shell when $k_+ = 4$ is identical to that of ref.~\cite{GomesBH}, so, as it collapses, the shell leaves in its wake the `wormhole' line element.

The `on-shell' relation~(\ref{OnShell_dimensionless}) produces, for any value of $\mu = \frac{M }{m_+}$ and  $\alpha = \frac{m_-}{m_+}$ a curve in the $P$-$R$ space, which reaches the boundary of phase space $P 
\to \pm \infty$ at $R = \frac 1 2$, that is, $R  = \frac {m_+} 2$. This value of $R $ coincides with the throat of the wormhole line element with mass $m_+$. This result implies that the collapsing shell does not reach the throat in a finite maximal-slicing time. This time parameter coincides with the experienced reading of a clock of an inertial observer at infinity. The preliminary conclusion is that the shell `freezes' at the throat and cannot be observed to cross it. However, as we know, maximal-slicing time can at best be an infinitely-thin layer of \index{York time} York time (the time parameter of CMC slicings). 

If we focus on the intrinsic features the on-shell curves of~(\ref{OnShell_dimensionless}), we observe that they continue past the point $R = \frac 1 2$ where they reach the boundary of phase space.  The solution curves fall  into two topologically-distinct kinds: the closed and the open ones. The former are closed loops which touch the boundary of phase space at two points. They correspond to the cases in which the shell does not have enough kinetic energy to reach infinity, and recollapses back. Interestingly, this behaviour is observed on both sides of the throat  $R = \frac 1 2$, so the shell recollapses also when it is in the region beyond the throat.
The other kind of curves are the open ones, which reach the asymptotic boundary $R \to \infty$, and the other asymptotic infinity at $R \to 0$.

\index{asymptotic flatness}\index{wormhole}
\subsubsection*{Back to the `wormhole' solution}

Once we established that the `wormhole' solution emerges as the result of the gravitational collapse of ordinary matter, it remains to be established whether it can be considered an approximation to a proper solution of Shape Dynamics, \emph{i.e.}, a compact solution. In fact it is clear that one cannot simply put $A_\st{B}=0$ in the `twin-shell' models studied in the previous Section, because the integration constant $A_\st{B}$ is a fundamental dynamical part of the system which cannot be ignored: it is associated to the radial momentum of matter, and there can be no gravitational collapse without radial motion. Looking at the diagrams in Fig.~\ref{Mordor_Onshell_Fig_1}, \ref{Mordor_Onshell_Fig_2}, \ref{Mordor_Onshell_Fig_3}, \ref{Mordor_Onshell_Fig_4}, \ref{Mordor_Onshell_Fig_5} and \ref{Mordor_Onshell_Fig_6}, we can see that there are some intervals of values of $A_\st{B}$ and  $\langle p \rangle$ in which the `forbidden region' prevents the existence of spheres of surface smaller than a minimum value (\emph{e.g.} Fig.~\ref{Mordor_Onshell_Fig_2}). This is exactly the same mechanism that is responsible for the existence of a throat in~\cite{GomesBH}. However this is not a generic feature of the system considered there, and in particular one can even see how most of the on-shell surfaces (representing the kinematically allowed values of the dynamical variables of the shells) reach the $\rho =0$ plane, which represents a zero-area shell. This puts the conjecture of the formation of a wormhole in serious doubt.

\begin{figure}[t!]\center
\includegraphics[width=0.4\textwidth]{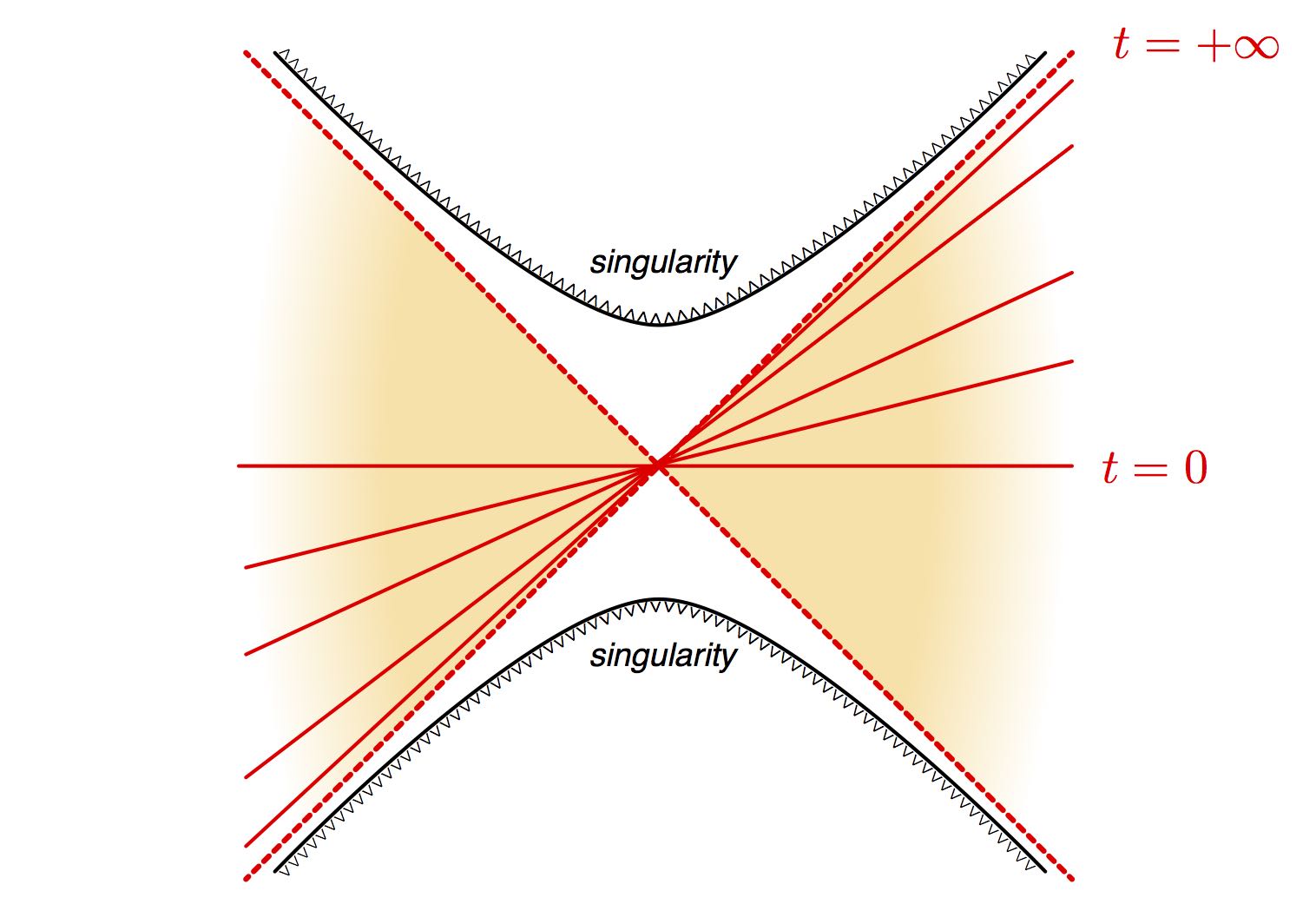}
\caption[Regions of Kruskal's extension occupied by the `wormhole' solution]{
Regions of Kruskal's extension of Schwarzschild \index{Schwarzschild's spacetime} which is occupied by the `wormhole' solution, together with a representation of the maximal slices.}
\vspace{-12pt}
\label{FoliationOddlapseFig}
\end{figure}
In~\cite{Gomes:2013bbl} Gomes and Herczeg studied an axially-symmetric solution which appears to be the rotating equivalent of the solution studied in~\cite{GomesBH}. Here, too, one only covers with maximal slices a part of the Kruskal extension of Kerr spacetime. 

\index{asymptotic flatness}\index{wormhole}\index{boundary charges}
\subsubsection*{Well-posedness of the asymptotically-flat variational principle}

In view of the problems of the `wormhole' solution(s),  in~\cite{NoBirkhoff} I critically reassessed the boundary conditions assumed in~\cite{GomesBH}, which are the common assumption of~\cite{GomesBH,Gomes:2013bbl,ThinShellPaper1}, and arguably their most determining one. Their use in GR is justified by the requirement of Poincar\'e invariance of the falloff conditions~\cite{beig1987poincare}, but in SD one is only authorized to assume symmetries  of the spatial slices at infinity, not of the spacetime metric. As I showed at the beginning of this Section,  the integration constant $A$, which is set to zero by the boundary conditions assumed in~\cite{GomesBH},  breaks  the invariance under asymptotic Lorentz transformations of the falloff conditions. The asymptotic invariance under spatial rotations and translations, and under time translations is still respected when $A \neq 0$. Lorentz invariance of the boundary is not a legitimate request for an asymptotically flat solution of Shape Dynamics: it is only the spatial slices which have to develop the isometries of Euclidean space at the boundary.  So the parameter $A$ cannot be put to zero in the same way as in GR,  and has  to be taken to remain as an arbitrary spatially-constant function of time. However, as can be deduced from the analysis of~\cite{GomesPoincareInvariance} by H. Gomes, the condition $A=0$  seems to be required in order that more general \textit{spatial} asymptotic rotations be associated with finite charges. In~\cite{GomesPoincareInvariance} it was shown that the standard asymptotically flat falloff conditions of GR (which imply $A=0$) are necessary in order to ensure the well-posedness of the variational problem. In other words, if we relax the assumption of spherical symmetry, falloff conditions that allow $A \neq 0$ will attribute infinite values to some of the boundary charges (like angular momentum), which means that one cannot define counterterms that make the action differentiable. As soon as we depart from perfect spherical symmetry, asymptotically flat SD with $A\neq 0$ is not a well-defined dynamical system. This seems to be a powerful argument in favour of fixing $A=0$, however the analysis of the previous Section proves, in a closed universe \index{closed spacelike hypersurfaces} this argument doesn't hold (there are no boundary charges and the variational problem is always well-posed), and the integration constant $A$ may admit values other than zero (it is determined by the state of motion of matter and setting it to zero `by hand' is inconsistent). Therefore, if asymptotically flat SD with $A\neq 0$ turns out to be inconsistent, it cannot be a good approximate description of a nearly-empty region in a larger closed universe. The issue of what is the right noncompact model of such a situation remains to be studied.

\newpage
~

\newpage

\part{Appendices}

\appendix

\section{Arnowitt--Deser--Misner Gravity}  \label{ADM-WDW}
\index{Arnowitt--Deser--Misner formalism}

\subsection{The Arnowitt--Deser--Misner (ADM) formalism}

To write GR in Hamiltonian form, we need a notion of time. Therefore assume that spacetime $\mathcal M$ is globally hyperbolic, then choose a foliation by spacelike hypersurfaces $\Sigma_t$, where $t$ will be a monotonic label for the leaves.

Now consider  on $\mathcal M$ a system of coordinates \emph{adapted to the foliation}: on each hypersurface 
$\Sigma_t$,  introduce some coordinate system $(x_1,x_2,x_3)$. If it varies smoothly between neighboring hypersurfaces, then $(x_1,x_2,x_3,t)$ constitutes a well-behaved coordinate system on $\mathcal M$. 
The theory of foliations tells us that in such a coordinate system the 4-metric ${^{(4)}g}_{\mu\nu}(x,t)$ can be 
decomposed into the induced metric on the leaves $g_{ij}(x,t)$ plus a scalar $N(x,t)$, called the \emph{lapse},
and a three-vector $N_a: \mathcal M \to T(\Sigma_t)$, called the \emph{shift}. Then
\begin{equation}\label{Covariant4Metric}
{^{(4)}g}_{00} =  - N^2 + g_{ij} \, N^i \, N^j \,, \qquad {^{(4)}g}_{0i} = g_{ij} \, N^j \,,
\qquad {^{(4)}g}_{ij} = g_{ij} \,,  
\end{equation}
and the inverse metric  ${^{(4)}g}^{\mu\nu}(x,t)$ is
\begin{equation}
{^{(4)}g}^{00} =  - 1/N^2  \qquad {^{(4)}g}^{0i} = N^i / N^2 \,,
\qquad {^{(4)}g}^{ij} = g^{ij} - N^i \, N^j / N^2 \,. \label{Contravariant4metric}
\end{equation}

Let $n^\mu(x,t)$ be a unit timelike 4-vector field, $g_{\mu\nu} n^\mu n^\nu =-1$, normal to the three-dimensional hypersurfaces. Its components are
\begin{equation}
n^\mu = \left( 1/N , - N^i / N \right) \,.
\end{equation}
This equation clarifies the meaning of the lapse $N$ and the shift $N^i$.
The lapse $N(x,t)$ expresses the proper time elapsed between the point 
$(x,t)\in \Sigma_t$ and the point $(x',t+\delta t)$ on the following infinitesimally close hypersurface 
$ \Sigma_{t+\delta t}$ towards which $n^\mu$ points. Starting from the point $(x',t+\delta t)$, one has to
move `horizontally', on the spatial hypersurface, by an amount $N^i(x',t+\delta t)$ 
to reach the point with coordinates $(x,t+\delta t)$ (see Fig.~\ref{LapseShiftFigure}).
     
\begin{figure}[t]
\begin{center}\includegraphics[width=0.3\textwidth]{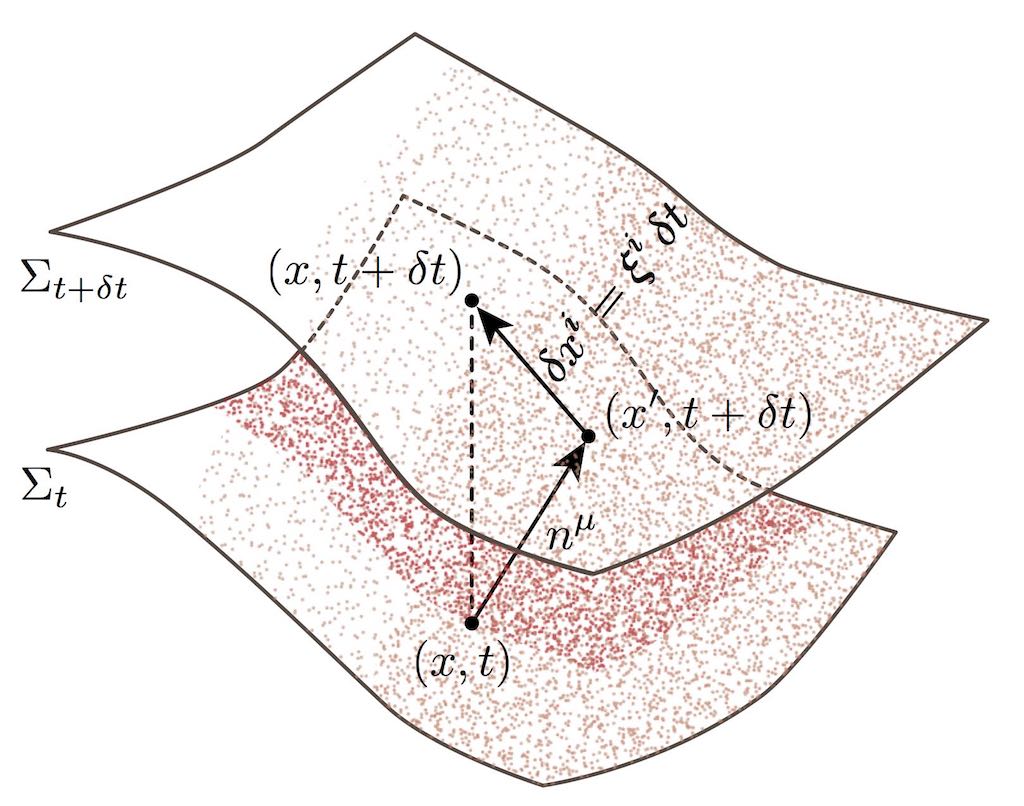}
\end{center}
\caption[Lapse and Shift]{Graphical representation (in 2+1 dimensions) of the meaning 
of the lapse and the shift.}
\label{LapseShiftFigure}
\end{figure}

To decompose the Einstein--Hilbert action (plus cosmological constant $\Lambda$), 
\begin{equation}
S_{\st{EH}} = \int \d^4 x \, \sqrt{- {^{(4)}g}} \left( {^{(4)}R} -2 \, \Lambda \right) \,, \label{EinsteinHilbertAction}
\end{equation}
into its spatial components according to the chosen foliation, we need an expression
for the determinant of the 4-metric, which is readily obtained from Eq.~(\ref{Contravariant4metric}),\index{Arnowitt--Deser--Misner formalism}
\begin{equation}
\sqrt{- {^{(4)}g}} =  N \, \sqrt g \,.
\end{equation}
We also need the decomposition, named after Gauss and Codazzi (see \cite{Frankelbook} sec. 8.5a pag. 229), of the 4D Ricci scalar ${^{(4)}R}$ into the 3D intrinsic scalar curvature
$R$ and the \emph{extrinsic curvature},
\begin{equation}
K_{ij} = \frac{1}{2 N} \left(\text{\textsterling}_{\vec N} g_{ij} -  \frac{d g_{ij}}{d t}  \right) 
= \frac{1}{2 N} \left( \nabla_i N_j + \nabla_j N_i -  \frac{\d g_{ij}}{\d t}  \right) \,, \label{ExtrinsicCurvature}
\end{equation}
of the leaves ($ \text{\textsterling}_{\vec N} $ is the Lie derivative w.r.t. the 3-vector field $N^i$). We get
\begin{equation}
 {^{(4)} R} =  R  + K_{ij} \, K^{ij} -  K^2 - 2 \, \nabla_\mu \left( K n^\mu \right) - \frac{2}{N} \nabla_i \nabla^i N \,.
\end{equation}
The Einstein--Hilbert action (\ref{EinsteinHilbertAction}) then reads
\begin{equation}\begin{aligned}
S_{\st{EH}} = \int \d^4 x \, \sqrt g  & \left\{  N \left( R - 2 \, \Lambda  + K_{ij} \, K^{ij} -  K^2   \right)  \right.\\  
& \left. - 2 \, N \,\nabla_\mu \left( K n^\mu \right) - 2 \,\nabla_i \nabla^i N \right\} \,,
\end{aligned}\end{equation}
where the next-to-last term is obviously a 4-divergence,
\begin{equation}\label{4-divergenceTerm}
-2 \int \d^4 x \, \sqrt g  \, N \,\nabla_\mu \left( K n^\mu \right)  
= - 2 \int \d^4 x \, \sqrt{- {^{(4)}g}} \nabla_\mu  \left(  K n^\mu \right) = 0 \,,
\end{equation}
while the last term is a 3-divergence:
\begin{equation}\label{3-divergenceTerm}
- 2 \int \d^4 x \, \sqrt g \, \nabla_i \nabla^i N  = - 2 \int d t \, \int \d^3 x \,\nabla_i \left(  \nabla^i N \right)  = 0 \,.
\end{equation}
Thus the final form of the 3+1 decomposition of the Einstein--Hilbert action is
\begin{equation}
S_{\st{EH}} = \int \d^4 x \, \sqrt g \,  N \left( R - 2 \, \Lambda  + K_{ij} \, K^{ij} -  K^2   \right)\,.
\end{equation}

Let's write this in the Hamiltonian language. The coordinates are the 3-metric $g_{ij}$, the lapse $N$
and the shift $N_i$ (the same number of degrees of freedom, $10$, as there were in the original 4-metric).
The only time derivatives that appear are those of $g_{ij}$, through the extrinsic curvature (\ref{ExtrinsicCurvature}), and in particular we have no time derivatives of $N$ and $N_i$, which
therefore are just Lagrange multipliers.\index{Arnowitt--Deser--Misner formalism}

The momenta conjugate to $g_{ij}$ are
\begin{equation}
p^{ij} = \frac{ \delta \mathcal{L}_{\st{EH}}}{\delta \dot g_{ij}} = \sqrt g \left( K \, g^{ij} - K^{ij} \right) \,, \label{MomentaK}
\end{equation}
(notice the $\sqrt g$ factor: $p^{ij}$ is a symmetric tensor \emph{density} of weight 1).

The canonical Hamiltonian is given by the Legendre transform $\mathcal H_{\st{ADM}}= \int \d^3 x \left( p^{ij} \, \dot g_{ij} - \mathcal{L}_{\st{EH}} \right)$ and (with some boundary terms discarded) is equivalent to
\begin{equation}\label{ADMhamiltonian}
\mathcal H_{\st{ADM}} = \int \d^3 x \left( N \, \mathcal H +  N_i \, \mathcal H^i \right) \,,
\end{equation}
where (with the trace $p=g_{ij}p^{ij}$)
\begin{equation}
\mathcal H = \frac{1}{\sqrt g} \left( p_{ij} p^{ij} - \frac{1}{2} p^2 \right) + \sqrt g \left(   2 \, \Lambda  - R \right) \approx 0 \,,
\label{QuadraticConstraint}
\end{equation}
is called the \emph{Hamiltonian}, or \emph{quadratic} constraint, and
\begin{equation}
\mathcal H^i = - 2 \, \nabla_j p^{ij} \approx 0  \,, \label{DiffeoConstraint}
\end{equation}
is the \emph{diffeomorphism}, or \e{momentum} constraint.
The ADM Hamiltonian is a linear combination of constraints with $N$ and $N_i$ playing  the role of Lagrange
multipliers. It therefore vanishes, and there is no preferred notion of time.
This is an expression of the
reparametrization invariance of GR, and leads to the `problem of time'.

Notice the minus sign of $p^2/2$ in the kinetic term $p_{ij} p^{ij} - \frac{1}{2} p^2$ of  (\ref{QuadraticConstraint}).
It is related to the Gauss--Codazzi equations, and has nothing to do with the Lorentzian signature of spacetime.
The Lorentzian signature can be read off the sign in front of $R$, which is negative, and would take the
opposite sign if spacetime were Euclidean.

A comment on the interpretation of the constraint (\ref{DiffeoConstraint}). 
I said it  generates 3-diffeomorphisms. In fact, if it is smeared with a vector field $\xi_i$,
$(\xi_i | \mathcal H^i) = \int \d^3 x \, \xi_i (x) \, \mathcal H^i (x) $ and
the Poisson brackets with the metric are taken,
\begin{equation}
\left\{ g_{ij}(x), (\xi_k |  \mathcal H^k) \right\} = \nabla_i \xi_j + \nabla_j \xi_i = \text{\textsterling}_\xi g_{ij} \,.\label{3DiffeosGeneration}
\end{equation}
then the metric transforms as $g_{ij} \to g_{ij} + \text{\textsterling}_\xi g_{ij}$. The Lie derivative \cite{Frankelbook,StewartBook}
$\text{\textsterling}_\xi$ is the way an infinitesimal diffeomorphism like $x'_i = x_i + \xi_i $ acts on tensor
fields of any kind. See appendix~\ref{LieDerivativeAppendix} for details.
\index{Arnowitt--Deser--Misner formalism}

\vspace{12pt}
\noindent
\fbox{\parbox{0.98\linewidth}{
{\bf Further reading:} ADM's review of their original papers \cite{ADM}, and the more recent review \cite{DeserScholarpedia}. Misner, Thorne, and Wheeler's \e{Gravitation} \cite{MTWbook}, Frankel's book on the geometry of physics \cite{Frankelbook} and Schutz's \e{Geometrical Methods of Mathematical Physics} \cite{SchutzBook}. 
}}\vspace{12pt}

\subsection{The Wheeler--DeWitt equation}
\label{WdWsection}
\index{Wheeler--DeWitt equation}

The quantization of the ADM representation of Einstein's theory can be understood -- only formally -- 
in the language of the Schr\"odinger functional $\Psi : \Riem \to \doublestroke{C}$, where
$\Riem$~ is the space of Riemannian three-metrics. The ADM constraints  (\ref{QuadraticConstraint}) and (\ref{DiffeoConstraint})  translate into operator equations on the wave functional $\Psi[g]$. 
The one associated with the quadratic constraint is called the Wheeler--DeWitt equation and, ignoring operator ordering issues, is
\begin{equation}
\hat{\mathcal H} \, \Psi = 0 \,, \qquad \hat{\mathcal H} = \frac{1}{g} \left( g_{ik} \, g_{jl} - \frac{1}{2}  g_{ij} \, g_{kl} \right)
\frac{\delta}{\delta g_{ij}} \frac{\delta}{\delta g_{kl}} - R +2 \, \Lambda \,. \label{WdWEquation}
\end{equation}
This equation is the functional analogue of a time-independent
Schr\"odinger equation\footnote{As a differential equation, it is closer to a Klein--Gordon
equation, being a hyperbolic functional differential equation (due to the minus sign in the
kinetic term), whereas in non-relativistic quantum mechanics the time-independent Schr\"odinger equation
is elliptic.} for the Hamiltonian $\hat{\mathcal H}$ with eigenvalue
$2 \Lambda$.
The other equation is:
\begin{equation}
\hat{\mathcal H}^i \, \Psi = 0 \,, \qquad \hat{\mathcal H}^i = \nabla_j \frac{\delta}{\delta g_{ij}} \,, \label{QuantumDiffeoConstraint}
\end{equation}
which enforces invariance of the wave functional $\Psi[g]$ under diffeomorphisms.

The puzzling feature of the Wheeler--DeWitt equation is that it is completely static: its solutions, whatever they are, will be `frozen' wave-functionals on $\Riem$, analogous to the solutions of the time-independent Schr\"odinger equation, which only represents the eigenstates of a Hamiltonian but no dynamics. This issue is usually referred to as the \emph{problem of time.} 

\subsection{The Baierlein--Sharp--Wheeler action}\label{BSWsubsec}
\index{Baierlein--Sharp--Wheeler formulation}

In 1962 Baierlein, Sharp and Wheeler \cite{BSW} found an action for GR which is of 
Jacobi type, explicitly enforcing reparametrization invariance. Consider the Einstein--Hilbert
action in ADM Lagrangian variables (with the explicit metric velocities $\dot g_{ij}$ in place of
the extrinsic curvature). The action has an interesting dependence on the lapse $N$:
\begin{equation}\begin{aligned}
S_{\st{EH}} =  \int \d^3 x \, \d t \, \sqrt g \,& \left[ N (R - 2 \, \Lambda) + {\ts\frac 1 4}  N^{-1} \, T \right] \,,  
\end{aligned}\end{equation}

where the kinetic term $T$ is
\begin{equation}
T = (g^{ik}g^{jl} - g^{ij}g^{kl}) \left[{d  g_{ij} \over
d  t} -  \pounds_{\vec N} g_{ij}\right]\left[{d  g_{kl} \over
d  t} -  \pounds_{\vec N} g_{kl} \right] \,.
\end{equation}
Varying the action w.r.t. $N$,
\begin{equation}
- {\ts\frac 1 4} N^{-2}  \, T + R - 2 \, \Lambda = 0 \,,
\end{equation}
we can solve the resulting equation for $N$,
\begin{equation}
N =  {\ts\frac 1 2}\sqrt{\frac T {R - 2 \, \Lambda}} \,,
\end{equation}
and, substituting last expression in the action, eliminate the lapse from it:
\begin{equation}
S_{\st{BSW}} = \int \d^3 x \, dt \,  \sqrt{g} \sqrt{R-2\, \Lambda} \sqrt{{T}}
 \,. \label{BSWaction}
\end{equation}
This is the BSW action. We got this action from ADM, let's show that we can do the converse.
The canonical  momenta are
\begin{equation}
    p^{ij} = {\delta {\cal L} \over \delta \dot g_{ij}} = \sqrt{g ( R  - 2 \, \Lambda) \over {T}}(g^{ik}g^{jl} - {\sfrac 1 2}
g^{ij}g^{kl}) \left({d  g_{kl} \over d  t} -
 \pounds_{\vec N} g_{kl}\right) \,.
\end{equation}
As usual in Jacobi-type actions, there is a primary constraint involving the momenta
that comes from the square-root form of the action. In this case, the constraint
is just the ADM quadratic constraint (\ref{QuadraticConstraint}).

The vector field $N_i$ appears without any time derivative, and is therefore a Lagrange
multiplier. This implies a primary constraint stating that the momentum conjugate
to $N_i$ vanishes:
\begin{equation}
    p^i_N = {\delta {\cal L} \over \delta \dot N_i} = 0\,.
\end{equation}
From the Euler--Lagrange equations for $N_a$, we get a secondary constraint, saying
that the variation of the action with respect to $N_a$ vanishes. This is the ADM diffeomorphism
constraint:
\begin{equation}
   {\delta {\cal L} \over \delta  N_i} = 2 \, \nabla_j p^{ij} = 0 \,.
\end{equation}

If we calculate the Hamiltonian, through a Legendre transformation, we get 
\begin{equation}\begin{aligned}
H_{\st{BSW}} = & \int \d^3 x \left(  p^{ij} \, \dot g_{ij}  - \mathcal L_{\st{BSW}} \right) \\
=&  \int \d^3 x \sqrt g  \, \sqrt T \left( \frac{ p_{ij} \, p^{ij} - p^2 /2}{\sqrt{R-2\,\Lambda}} +  p^{ij} \, \pounds_{\vec N} g_{ij} - \sqrt{(R-2\,\Lambda) \, T} \right) \,.  
\end{aligned}\end{equation}
If now we define $N = \sqrt{\frac{T}{4(R-2\,\Lambda)}}$    and integrate by parts $ p^{ij} \, \pounds_{\vec N} g_{ij}$:
\begin{equation}
H_{\st{BSW}} = H_{\st{ADM}} = \int \d^3 x \sqrt g \left( N \, \mathcal H + N_i \, \mathcal H^i \right) \,,
\end{equation}
we see that the theory is equivalent to GR in the ADM formulation.
\index{Baierlein--Sharp--Wheeler formulation}

\subsection{Asymptotically flat ADM} \label{AsymptoticallyFlatADM}
\index{asymptotic flatness}\index{boundary charges}

The ADM hamiltonian (\ref{ADMhamiltonian}), which I reproduce here:
$$
 H_{\st{ADM}} = \int_\Sigma \d^3 x \left( N \, \mathcal H +  N_i \, \mathcal H^i \right) \,,
$$
does not generate Einstein's equations if $\Sigma$ is not compact. In fact one has to take into account the boundary
terms I discarded in (\ref{4-divergenceTerm}) and (\ref{3-divergenceTerm}). To calculate the necessary modifications
to the Hamiltonian, we can vary (\ref{ADMhamiltonian}) and pay attention to each integration by parts
\begin{equation}\label{BoundaryTermsVariationADM}
\begin{aligned}
\delta  H_{\st{ADM}} =& \int_\Sigma \d^3 x \left\{  A^{ij} \, \delta g_{ij} + B_{ij} \, \delta p^{ij} \right\} \\
&+ \int_\Sigma \d^3 x \, \sqrt g \, \nabla_k \left( N^k \, p^{ij} \, \delta g_{ij} - 2 \, N^i \, p^{jk} \, \delta g_{ij} - 2 \, N_j \, \delta p^{jk}  \right) \\
&+ \int_\Sigma \d^3 x \, \sqrt g \, \nabla^j \left(  \nabla^i N \, \delta g_{ij}  - N \,  \nabla^i \delta g_{ij} \right) \\
&+ \int_\Sigma \d^3 x \, \sqrt g \, \nabla^j \left(  N \, g^{kl} \,   \nabla_i \delta g_{kl} -  \nabla_i N \, g^{kl} \, \delta g_{kl} \right)  \,.
\end{aligned}
\end{equation}
Instead of considering this problem in full generality, I will specialize to the asymptotically flat case.
The boundary conditions are these: the 4-metric has to reduce to Schwarzschild \index{Schwarzschild's spacetime} at spatial infinity,
which in Cartesian coordinates reads
\begin{equation}
ds^2   \xrightarrow[r \to \infty]{} - \left( 1 - \frac{m}{8 \, \pi \, r} \right) \, dt^2 + \left( \delta_{ij}  + \frac{m}{8 \pi} \frac{x_i x_j}{r^3} \right) d x^i  d x^j + O (r^{-2}) \,.
\end{equation}
Therefore the spatial metric in a generic spacelike hypersurface goes to the Euclidean one like $g_{ij} - \delta_{ij} \sim r^{-1}$,
and its derivatives $g_{ij,k}  \sim r^{-2}$. The lapse and the shift can be read off (\ref{Covariant4Metric}) and go like $N-1 \sim r^{-1}$
and $N^i \sim r^{-1}$. Their derivatives will go like $N_{,i} \sim r^{-2}$ and $N^i{}_{,j} \sim r^{-2}$. The momenta, defined as (\ref{MomentaK}) and (\ref{ExtrinsicCurvature}), have to go like $p^{ij} \sim r^{-2}$. Now, considering the boundary terms in (\ref{BoundaryTermsVariationADM}),
the only ones that contribute are those that go like $r^{-2}$, because the surface integrals go like $r^2$. The only compatible terms belong to the last two lines, which contribute with the following leading order:
\begin{equation}
\begin{aligned}
 \int_{\partial \Sigma} d^2 \sigma^i \left(g^{kl} \, \partial_i  \delta g_{kl}  - \partial^j \delta g_{ij} \right) =
-  \delta E[g_{ij} ]   \,, \\
E[g_{ij} ] = \int_{\partial \Sigma} d^2 \sigma^j \left(g_0^{ik} \partial_k g_{ij} - g_0^{kl} \partial_j  g_{kl}   \right) \,,
\end{aligned}
\end{equation}
where $g_0^{ij}$ is the flat metric on the boundary ($\delta^{ij}$ in Cartesian coordinates). We have found a local boundary integral that can be added to the ADM Hamiltonian to give a well-defined variational principle:
\begin{equation}
\delta \left(H_{\st{ADM}} + E[g_{ij} ] \right) = \int_\Sigma \d^3 x \left\{  A^{ij} \, \delta g_{ij} + B_{ij} \, \delta p^{ij} \right\} \,.
\end{equation}\index{asymptotic flatness}\index{boundary charges}\index{ADM mass}
Remarkably, we end up with a generator of the dynamics which is not pure constraint: it is a \emph{true Hamiltonian},
which doesn't vanish on the solutions of the equations of motion. Rather it takes the value $E[g_{ij}]$, which depends on
the leading order of the metric at infinity. Moreover, $E[g_{ij}]$ is a conserved quantity: in fact the equations of motion
are exactly identical to the compact case where $E[g_{ij}] = 0 $ (a boundary term does not affect the equations of motion),
and therefore $\mathcal H_\st{ADM}$ alone is conserved and always identical to zero on any solution. Then, since the 
total Hamiltonian is conserved by definition (because it is time-independent), the boundary term is conserved.

The boundary conditions we have considered for the shift ($N^i \sim r^{-1}$) are too restrictive. In fact  the most generic ones are $N^i \sim \xi^i + r^{-1}$ where $\xi^i$ is a vector which is tangential to the boundary and $\xi^i \sim r$ because of Killing vectors \index{Killing vector} at infinity. The new contribution to the variation is
\begin{equation}
 \int_{\partial \Sigma} d^2 \sigma_k \left(\cancel{\xi^k   p^{ij}   \delta g_{ij}} -2   \xi^i    p^{jk}  \delta g_{ij}  -2   \xi_j \delta p^{jk} \right)
\end{equation}
where the first term vanishes because $\xi^i$ is parallel to the boundary and therefore $ d^2 \sigma_k \xi^k =0$. The remaining 
terms can be written as a total variation
\begin{equation}
-2  \int_{\partial \Sigma} d^2 \sigma_k \left(\xi^i  \,  \delta(p^{jk} \,   g_{ij}) \right)
=
-2 ~ \delta  \int_{\partial \Sigma} d^2 \sigma_k \left(\xi_j \, p^{jk} \right) \,.
\end{equation}
So we get the boundary terms
\begin{equation}
B[g_{ij},\xi_k] = 2 ~ \int_{\partial \Sigma} d^2 \sigma_k \left(\xi_j \, p^{jk} \right)  \,,
\end{equation}
which are such that the variational problem is well-posed:
\begin{equation}
\delta \left( \mathcal H_{\st{ADM}} + E[g_{ij} ] + B[g_{ij},\xi_k] \right) = \int_\Sigma \d^3 x \left\{  A^{ij} \, \delta g_{ij} + B_{ij} \, \delta p^{ij} \right\} \,.
\end{equation}

In the pure Schwarzschild \index{Schwarzschild's spacetime} case we have $g_{ij} = \delta_{ij}  + \frac m {8 \pi}  \frac{x_i x_j}{r^3}$,
and therefore $\partial_k g_{ij} =  \frac{m}{8 \, \pi \, r^3} \left( \delta_{ik} \, x_j + \delta_{jk} \, x_i - 3 \,  \frac{x_i x_j x_k}{r^2} \right)$,
$\delta^{ij} \partial_k g_{ij} = - \frac{m \, x_k}{8 \pi \, r^3}$ and therefore (since $d \sigma^i$ is parallel to $x^i$ and is normalized
so that its integral on a sphere of radius $R$ is $4 \, \pi \,R^2$,
\begin{equation}
E_\st{Schw} = \frac{m}{4 \pi} \, \int_{S^2}  d \cos \theta \, d \phi = m \,,
\end{equation}
the boundary energy coincides with the Schwarzschild mass.
\index{asymptotic flatness}\index{boundary charges}\index{ADM mass}

\newpage

\section{Other Appendices}\label{OtherAppendices}

\subsection{The case for closed spacelike hypersurfaces}
\index{Gauss constraint}\index{closed spacelike hypersurfaces}

It is well known that on a compact space the total electric charge must be zero. This is
a consequence of the Gauss constraint: ${\bm \nabla} \cdot {\bm E} = \rho$. Defining a 
region $\Omega \in \Sigma$ with a well-defined boundary $\partial \Omega$, and calling 
the total charge inside that region $Q(\Omega) = \int_\Omega \d^3 x \sqrt g \, \rho$, we can prove
that $Q(\Omega) = - Q(\Sigma \setminus \Omega)$ which implies that the total charge in
$\Sigma$, $Q(\Sigma) =Q(\Omega) + Q(\Sigma \setminus \Omega) = 0$. The proof 
 makes use of the Gauss law:
\begin{equation}
\left.
\begin{aligned}
Q(\Omega) =  \int_\Omega \d ^3 x  \sqrt g \, {\bm \nabla} \cdot {\bm E} = \int_{\partial \Omega} \d {\bm \sigma} \cdot {\bm E}\\
Q(\Sigma \setminus \Omega)  = \int_{\Sigma \setminus \Omega} \d ^3 x \sqrt g  \, {\bm \nabla} \cdot {\bm E} = - \int_{\partial \Omega} \d {\bm \sigma} \cdot {\bm E}
\end{aligned}\right\} \Rightarrow Q(\Sigma) = 0 \,.
\end{equation}

In the case of a 3-metric $g_{ab}$ with a Killing vector \index{Killing vector} $\xi_a$, defined by the Killing equation
$\Lie_\xi g_{ab} = \nabla_a \xi_b + \nabla_b \xi_a = 0$, we can prove that an analogous result holds
as a consequence of the diffeomorphism constraint:
\begin{equation}
-2 \nabla_j p^{ij} = j^i \,,
\end{equation}
where $j^a$ is the contribution to the diffeomorphism constraint due to matter fields. For example
$j^i = - \pi \nabla^i \varphi$ for a scalar field and $j^i = E^j \nabla^i A_j - A^i \nabla_j E^j$ for
an electromagnetic field. Now, the Killing equation and the diffeomorphism constraint imply that the projection of $j^i$
along the $\xi_i$ direction, $j^i \, \xi_i$, is a divergence:
\begin{equation}
-2 \nabla_j ( p^{ij} \xi_i)  = j^i \, \xi_i \,.
\end{equation}
Therefore, by the same argument for electric charge, $\int_\Sigma \d^3 x \, j^i \, \xi_i = 0$ on a 
closed $\Sigma$. \index{closed spacelike hypersurfaces} Depending on the isometries of $g_{ij}$, the quantity $j^i \, \xi_i$ might either
represent some components of the linear or angular momentum of matter. Therefore, if the metric
has isometries, one finds that the Machian constraints (vanishing total linear and angular momentum)
arise just as a consequence of the closedness of space.

This argument makes use of Killing vectors, \index{Killing vector} but the generic solution is not guaranteed to possess isometries. 
This is related to the fact that in the generic case  the gravitational field will carry  some angular and linear momentum 
 (for example in the form of gravitational waves), and one cannot limit consideration to the
matter contribution. The way to properly take into account the gravitational contribution requires
a much more subtle treatment which I will not go into here.

Assuming we have an analogous result for the generic case, the consequences of the observation reported
here are clear: a closed spatial manifold is Machian, while manifolds with a boundary or open
manifolds are subject to boundary conditions that spoil the self-contained nature of the theory. They
cannot be descriptions of the whole Universe, because for example they admit the presence of
a nonzero angular momentum. They might be at most descriptions of subsystems of the Universe,
which do not take into account what is going on outside. In the case of a compact manifold with a boundary
this seems pretty obvious, but in the noncompact case it is not. In fact the most popular choice
that theoreticians make of the spatial manifold and related boundary conditions is asymptotically flat. This
choice is particularly non-Machian, as it requires one to specify the value of an external angular momentum
and energy at infinity, which are by necessity externally given and not fixed by the dynamical degrees of
freedom inside the Universe. Asymptotically flat spaces are still very useful to describe isolated regions
of space, but they shouldn't be used as models for the whole Universe!
\index{closed spacelike hypersurfaces}

\subsection{Free-end-point variation}

\label{AppendixFreeEndPointVariation}
\index{free-end-point variation}

Let $S$ be an action I wish to extremalize over a principal $G$-bundle, $q$ be the canonical coordinates  
(the particle coordinates, or the metric and matter fields) 
and $\phi$ the compensating coordinates that move us on the fibre,
\begin{equation}\label{ActionFreeEndpointVariation}
S  = \int_{s_1}^{s_2}\d s\, \mathcal L (q,\dot q, \phi , \dot \phi) \,.
\end{equation}
My aim is to extremalize the action given boundary values for the fields $q$ at the endpoints of
the trial curve, $q(s_1)$ and $q(s_2)$, but I want to leave the endpoint values of the
compensating fields free, so that I actually just specify an initial and final \emph{gauge orbit}.
Taking the variation of the action w.r.t. $\phi$ and $\dot \phi$, I am led to the condition
\begin{equation}
\delta S =  \int_{s_1}^{s_2} \d s  \left\{ \frac{\delta \mathcal L}{\delta \phi}  - \frac \d {\d s} \left( \frac{\delta \mathcal L}{\delta \dot \phi} \right) \right\} \delta \phi
+  \left. \frac{\delta \mathcal L}{\delta \dot \phi}  \delta \phi \right|_{s=s_1}^{s_2} \,.
\end{equation}
Now, the action has to be stationary with respect to \emph{all} variations $\delta \phi$ around
the extremalizing trajectory $\phi(t)$. So it has to be stationary also under  fixed-endpoint
variations $\{ \delta \phi ~~ \text{s.t.} ~~ \delta \phi(s_1) = \delta \phi (s_2) = 0 \}$. This implies that the extremalizing trajectory has to satisfy the Euler--Lagrange equations
\begin{equation}
\frac{\delta \mathcal L}{\delta \phi}  - \frac \d {\d s} \left( \frac{\delta \mathcal L}{\delta \dot \phi} \right)  = 0 \,,
\end{equation}
but the extremalizing trajectory must also make the action stationary, $\delta S =0$, and therefore the
only possibility is that it is at the same time such that
\begin{equation}
 \left. \frac{\delta \mathcal L}{\delta \dot \phi} \right|_{s=s_1} =    \left. \frac{\delta \mathcal L}{\delta \dot \phi} \right|_{s=s_2} = 0 \,.
\end{equation}
Both boundary terms must vanish because the action has to be stationary 
with respect to variations with one fixed endpoint and one free endpoint as well.
%

\subsection{Lie derivative}
\index{Lie derivative}

\label{LieDerivativeAppendix}

The Lie derivative is a map from any kind of tensor and a vector field to a tensor of the same kind.
Here I will only give the definition and some useful properties. 

The Lie derivative of a tensor with respect to a vector field $\xi^c$ is
the directional derivative in the direction of $\xi^c$. 
Associated with $\xi$ there is a vector flow that induces a one-parameter
family of diffeomorphisms $\phi_s$:
$$
\frac{\d \phi_s(x)^i}{\d s} =\xi^i \,.
$$
Calling  $\phi_t^*$ the \emph{pullback} \cite{Frankelbook}
of the diffeomorphism on tensor fields, we have that
$$
\phi^*_s T^{i_1 \dots i_m}_{j_1 \dots j_n} (\phi_s (x)) 
$$
is the value of $T^{i_1 \dots i_m}_{j_1 \dots j_n}$ at  $\phi_s(x)$ pulled
back to  the point $x$. Then the Lie derivative of $T^{i_1 \dots i_m}_{j_1 \dots j_n}$ w.r.t. $\xi$ at $x$ is
\begin{equation}\label{DefLieDerivative}
\Lie_\xi T^{i_1 \dots i_m}_{j_1 \dots j_n} 
= \lim_{s \to 0} \left( \frac{\phi^*_s T^{i_1 \dots i_m}_{j_1 \dots j_n} (\phi_s (x))  - T^{i_1 \dots i_m}_{j_1 \dots j_n} (x) }{s} \right)
\,. 
\end{equation}
It is clear then how the Lie derivative represents the action of
an infinitesimal diffeomorphism generated by the vector field $\xi$ on
an arbitrary tensor field.

\subsubsection*{Coordinate expression}

If $T^{i_1 \dots i_m}_{j_1 \dots j_n}$
is a $m,n$-tensor density of weight $w$, its Lie derivative with respect to a vector field $\xi_i$ is
\begin{equation}\begin{aligned}\label{CoordinateExpressionLieDerivative}
\Lie_\xi T^{i_1 \dots i_m}_{j_1 \dots j_n}  &=  \xi^k  \partial_k  T^{i_1 \dots i_m}_{j_1 \dots j_n}  + w \, \partial_k \xi^k   \,T^{i_1 \dots i_m}_{j_1 \dots j_n} \\  
& -  \partial_k \xi^{i_1}  \, T^{k \dots i_m}_{j_1 \dots j_n}  \dots - \partial_k \xi^{i_m} \, T^{i_1 \dots k}_{j_1 \dots j_n}  \\  
& +\partial_{j_1} \xi^{k} \, T^{i_1 \dots i_m}_{k \dots j_n}  + \dots + \partial_{j_n} \xi^{k} \, T^{i_1 \dots i_m}_{j_1 \dots k}  \,.\end{aligned}\end{equation}

Notice that, although not obvious at first sight, the expression on the right-hand side is covariant; in fact we can
replace all the partial derivatives $\partial_c $ with covariant derivatives $\nabla_c$, and the Christoffel symbols
cancel:
\begin{equation}\begin{aligned}
\Lie_\xi T^{i_1 \dots i_m}_{j_1 \dots j_n}  &=  \xi^k \nabla_k  T^{i_1 \dots i_m}_{j_1 \dots j_n}  + w \, \nabla_k \xi^k   \,T^{i_1 \dots i_m}_{j_1 \dots j_n} \\  
& -  \nabla_k \xi^{i_1}  \, T^{k \dots i_m}_{j_1 \dots j_n}  \dots - \nabla_k \xi^{i_m} \, T^{i_1 \dots k}_{j_1 \dots j_n}  \\  
& +\nabla_{j_1} \xi^{k} \, T^{i_1 \dots i_m}_{k \dots j_n}  + \dots + \nabla_{j_n} \xi^{k} \, T^{i_1 \dots i_m}_{j_1 \dots k}  \,.\end{aligned}\end{equation}

The Lie derivative obeys the Leibniz rule with respect to both the inner and the tensor product, meaning that
\begin{equation}\begin{aligned}
\Lie_\xi \left( \delta^{\dots}_{\dots} \, T^{i_1 \dots i_m}_{j_1 \dots j_n} \, S^{k_1 \dots k_r}_{l_1 \dots l_s} \right) &=
\delta^{\dots}_{\dots}  ~  \Lie_\xi T^{i_1 \dots i_m}_{j_1 \dots j_n} \, S^{k_1 \dots k_r}_{l_1 \dots l_s} \\
& +\delta^{\dots}_{\dots}  ~ T^{i_1 \dots i_m}_{j_1 \dots j_n} \, \Lie_\xi  S^{k_1 \dots k_r}_{l_1 \dots l_s}  \,,  
\end{aligned}\end{equation}
where $\delta^{\dots}_{\dots}$ denotes any combination of Kronecker deltas,
as for example ${\delta^{k_6}}_{i_3} {\delta^{j_3}}_{k_2}  {\delta^{j_7}}_{k_1}$.

Unlike the covariant derivative, the Lie derivative does not depend on the metric. For this reason,
when one takes metric variations of expressions containing the Lie derivative, the variation commutes with it:
$$
\frac \delta {\delta g_{ij}} \left(T^{i_1 \dots i_m}_{j_1 \dots j_n} \, \Lie_\xi  S^{k_1 \dots k_r}_{l_1 \dots l_s} \right)
= \frac \delta {\delta g_{ij}} T^{i_1 \dots i_m}_{j_1 \dots j_n} \, \Lie_\xi  S^{k_1 \dots k_r}_{l_1 \dots l_s} 
+ T^{i_1 \dots i_m}_{j_1 \dots j_n} \, \Lie_\xi  \left(  \frac \delta {\delta g_{ij}}  S^{k_1 \dots k_r}_{l_1 \dots l_s} \right) \,.
$$

\index{Lie derivative}
\subsubsection*{Examples}

The Lie derivative of the metric field is
\begin{equation}
\Lie_\xi g_{ij} = \nabla_i \xi_j + \nabla_j \xi_i \,,
\end{equation}
(note that the right-hand side depends on the metric,
but only because the metric appears in the first place
on the left-hand side).
As can be deduced from the expression (\ref{DefLieDerivative}), the Lie derivative respects also the symmetry of the
tensors it acts on. For example, its action on a covariant (meaning with downstairs indices) symmetric tensor like the extrinsic curvature is
\begin{equation}
\Lie_\xi K_{ij}  =  \xi^k  \nabla_k  K_{ij} + \nabla_{i} \xi^{k} \, K_{kj} \,.
\end{equation}
If the tensor is contravariant (meaning with upstairs indices) and a tensor density of weight $w=1$ like the metric momentum,
then its Lie derivative is
\begin{equation}
\Lie_\xi p^{ij}  =  \xi^k \nabla_k p^{ij}   + \nabla_k \xi^k  \, p^{ij} 
-  \nabla_k \xi^i  \, p^{kj}  - \nabla_k \xi^j \, p^{ik}  \,.
\end{equation}

\subsection{TT decomposition of tensors}

\label{TTdecompositionAppendix}

\noindent{\bf Helmholtz Decomposition Theorem}
\index{York's decomposition} \index{Helmholtz decomposition}

Any vector field ${\bf F}$ on a $3$-space $\Sigma$ (which we assume endowed with a Riemannian
metric $g_{ab}$, even though it's not necessary for the theorem) can be written as the sum of a transverse ${\bf F}_\st{T}$ and 
longitudinal ${\bf F}_\st{L}$,
\begin{equation}
{\bf F} = {\bf F}_\st{T} + {\bf F}_\st{L} \,,
\end{equation}
where ${\bm \nabla} \cdot {\bf F}_\st{T} = 0$ and  ${\bm \nabla} \times {\bf F}_\st{L} = 0$.
The two parts can be written respectively as the curl  of a vector field ${\bm \theta}$ and
the divergence of a scalar field $\phi$:
\begin{equation}
 {\bf F}_\st{T}  = {\bm \nabla }\times {\bm \theta} \,, \qquad {\bf F}_\st{L} = {\bm \nabla } \phi \,.
\end{equation}
If $\Sigma$ is compact, or if it is noncompact but all the fields have appropriate fall-off
conditions at infinity, the two parts are mutually orthogonal with respect to the natural global inner product
between vector fields (this is where the metric plays a role), as can be proved with an integration by parts:
\begin{equation}\begin{aligned}
({\bf F}_\st{T} | \, {\bf F}_\st{L} ) & = \int \d^3 x \, \sqrt g ~ {\bf F}_\st{T}   \cdot {\bf F}_\st{L} = \int \d^3 x \, \sqrt g ~   {\bm \nabla }\times {\bm \theta}  \cdot  {\bm \nabla } \phi  \\  
& = - \int \d^3 x \, \sqrt g ~  {\bm \nabla } \cdot ({\bm \nabla }\times {\bm \theta}) \, \phi 
= 0 \,.
\end{aligned}\end{equation}

The decomposition can be made by solving the equation ${\bm \nabla} \cdot {\bf F} = {\bm \nabla} \cdot {\bf F}_\st{L}$ and ${\bm \nabla} \times {\bf F} = {\bm \nabla} \times {\bf F}_\st{T}$
for $\phi$,
\begin{equation}
\laplacian \phi = {\bm \nabla} \cdot {\bf F} \,.
\end{equation}
This is Poisson's equation for $\phi$ with source ${\bm \nabla} \cdot {\bf F}$, which is well-known to admit a unique solution. Once the solution has been found, let's call it $\nabla^{-2} {\bm \nabla} \cdot {\bf F}$,
the transverse part can be readily defined as $ {\bf F}_\st{T}  = {\bf F} -{\bm \nabla} (\nabla^{-2} {\bm \nabla} \cdot {\bf F})$.

This decomposition is unique modulo a harmonic part, i.e., a field which solves the Laplace equation
\begin{equation}
{\bf F}_\st{L}  = {\bm \nabla} \Lambda \,, \qquad \laplacian \Lambda = 0 \,.
\end{equation}
This equation admits only the zero solution on closed spaces, and  the same holds for
simply connected noncompact spaces if vanishing boundary conditions are chosen.

\noindent{\bf York's conformally covariant decomposition of  symmetric tensors}

In close analogy to the Helmholtz decomposition theorem, symmetric tensors admit a decomposition into a transverse-traceless (spin-2) part, a longitudinal (spin-1) part, and a pure trace, scalar part. 

\begin{equation}\label{TTdecomposition2-tensor}
X^{ij} = X^{ij}_\st{TT}  + X^{ij}_\st{L} + X^{ij}_\st{tr} = X^{ij}_\st{TT} + (LY)^{ij} + {\ts \frac 1 3} \, X \, g^{ij} \,, 
\end{equation}
where $X = g_{ij} X^{ij}$, and 
\begin{equation}
(LY)^{ij} = \nabla^i Y^j + \nabla^j Y^i - {\ts \frac 2 3} \, g^{ij} \, \nabla_k Y^k  
\end{equation}\index{York's decomposition} \index{Helmholtz decomposition}
is the \emph{conformal Killing form} of the vector field $Y^i$. It can be obtained
as the Lie derivative w.r.t. $Y^i$ of the unit-determinant part $g^{-1/3} g_{ij}$ of the metric:
\begin{equation}
(LY)_{ij} = g^{ 1/3} \, \Lie_Y ( g^{-1/3} g_{ij}) \,.
\end{equation}
We can solve for $Y^a$ the transversality condition
\begin{equation}\label{EquationForSolvingTheDiffeoConstraint}
\nabla_j (LY)^{ij} = \nabla_j (X^{ij} -  {\ts \frac 1 3} \, X \, g^{ij}),
\end{equation}
on the left-hand side of which we have a linear second-order differential
operator acting on $Y^i$ (the Ricci tensor comes from commuting
the covariant derivatives: $[\nabla_j , \nabla^i] Y^j = {R^i}_j Y^j$),
\begin{equation}
\nabla_j (LY)^{ij} = (\Delta_\st{L} Y)^i = \laplacian Y^i + {\ts \frac 1 3} \nabla^i \nabla_j Y^j + {R^i}_j Y^j \,.
\end{equation}
Here the operator $\Delta\st{L}$ is strongly elliptic, as can be seen by studying its
\emph{principal symbol}, that is, the matrix obtained by replacing each derivative
by an arbitrary variable $\partial_i \to z_i$ and taking only the highest
derivatives (second):
\begin{equation}
\sigma_\st{P}(\Delta_\st{L}) = {\delta^i}_j \, z^k z_k +{\ts \frac 1 3} z^i z_j \,.
\end{equation}
If this matrix has positive determinant for any value of $z_i$, the operator is
\emph{elliptic}, and if its eigenvalues are always positive the operator is \emph{strongly elliptic}.
Both conditions are realized: $\det \sigma_\st{P} = \frac 4 3 \, (z^k z_k)^3$
and the eigenvalues are $z^k z_k$ with multiplicity 2 and  $\frac 4 3 \, z^k z_k$.

The operator $\Delta_L$ also has the property of being Hermitian with respect to the natural global
inner product between vectors, as can be seen with two integrations by parts,
\begin{equation}\begin{aligned}
(Z_i | (\Delta_\st{L} Y)^i ) & = \int \d^3 x \, \sqrt g \, Z_i \, \nabla_j (LY)^{ij} 
= - \int \d^3 x \, \sqrt g \,\nabla_j Z_i  (LY)^{ij} \\  
& = \int \d^3 x \, \sqrt g \, Y_i \, \nabla_j (LZ)^{ij} = (Y_i | (\Delta_\st{L} Z)^i ) \,.
\end{aligned}\end{equation}

Now we come to the kernel of $\Delta_\st{L}$, which represents the analogue of
harmonic fields in the Helmholtz decomposition. The equation for the kernel is
\begin{equation}
(\Delta_\st{L} \xi)^i = \laplacian \xi^i + {\ts \frac 1 3} \nabla^i \nabla_j \xi^j + {R^i}_j \xi^j = 0 \,.
\end{equation}
On compact manifolds, or on noncompact manifolds but assuming that $Y^a$ asymptotically
approaches zero sufficiently fast, the above equation is equivalent to the vanishing 
of $(L\xi)^{ij}$,
 \begin{equation} 
(L \xi)^{ij} = \nabla^i \xi^j + \nabla^j \xi^i - {\ts \frac 2 3} \, g^{ij} \, \nabla_k \xi^k = 0 \,.
\end{equation}
This equation identifies \emph{conformal Killing vectors} \index{conformal Killing vector}of the metric $g_{ij}$, namely
vectors that generate infinitesimal diffeomorphisms which leave the metric invariant
up to a conformal transformation,
\begin{equation}
\Lie_\xi g_{ij} =\phi \, g_{ij} \,, \qquad \phi =  {\ts \frac 2 3 } \, \nabla_k \xi^k \,.
\end{equation}
In an asymptotically flat space the conformal Killing vectors will not vanish at infinity
and cannot therefore be ignored. One can, however, relax the boundary conditions 
for the $Y^a$ field, requiring that it approach one of the conformal Killing vectors
of Euclidean space. Then Eq.~(\ref{EquationForSolvingTheDiffeoConstraint}) has
a unique solution. These boundary conditions are useful for defining the total 
momentum of the gravitational field in the asymptotically flat case \cite{Niall-York74}.
In the closed case, the good news is that the conformal Killing vectors are always  by construction
orthogonal (according to the natural global inner product) to the source of Eq.~(\ref{EquationForSolvingTheDiffeoConstraint}), namely the divergence of the
traceless part $ \nabla_j (X^{ij} -  {\ts \frac 1 3} \, X \, g^{ij})$ of $X^{ij}$:
\begin{equation}\begin{aligned}
(\xi_i | \nabla_j (X^{ij} -  {\ts \frac 1 3} \, X \, g^{ij})) &=
- \int \d^3 x \sqrt g \, \nabla_j \xi_i (X^{ij} -  {\ts \frac 1 3} \, X \, g^{ij})\\  
&= \int \d^3 x \sqrt g \, ( \nabla_j \xi_i -  {\ts \frac 1 3} \,  \nabla_k \xi^k \, g_{ij}) X^{ij} \\  
&= {\ts \frac 1 2} \int \d^3 x \sqrt g \, (L \xi)_{ij} X^{ij} = 0 \,.
\end{aligned}\end{equation}\index{York's decomposition} \index{Helmholtz decomposition}

Since the source term in Eq.~(\ref{EquationForSolvingTheDiffeoConstraint}) is orthogonal to 
the kernel of the operator $\Delta_\st{L}$, this operator is invertible in the subspace
to which $ \nabla_j (X^{ij} -  {\ts \frac 1 3} \, X \, g^{ij})$ belongs. 
Equation~(\ref{EquationForSolvingTheDiffeoConstraint})  therefore admits a unique solution
modulo conformal Killing vectors,\index{conformal Killing vector}  the addition of which does not change the 
TT-decomposition of $X^{ij}$ because $X^{ij}_\st{L}$ is insensitive to them.

The three terms in the TT-decomposition are orthogonal to each other:
\begin{equation}\begin{aligned}
(X_\st{TT} | X_\st{L}) &=  -2  (\nabla_j X_\st{TT}^{ij} | \xi_i)  - {\ts \frac 2 3} (g_{ij}X_\st{TT}^{ij} | \nabla_k \xi_k ) = 0 \,, \\  
(X_\st{TT} | X_\st{tr}) &= {\ts \frac 1 3} ( g_{ij} X_\st{TT}^{ij} | X ) = 0 \,,\\  
(X_\st{L} | X_\st{tr}) &=  {\ts \frac 1 3} (g_{ij}(LY)^{ij}  | X  ) =0\,.
\end{aligned}\end{equation}

One could further decompose the longitudinal part in the manner of Helmholtz into a pure-spin one
and a pure-scalar part, but this last decomposition is not conformally covariant.

\noindent{\bf Conformal covariance of the decomposition}

Make the conformal transformation
\begin{equation}
\bar g_{ij} = \phi^4 g_{ij} \,, \qquad \bar g^{ij} = \phi^{-4} g^{ij}\,
\end{equation}
of the metric and assume that the transformation acts on a symmetric tensor $X^{ij}$ as follows:
\begin{equation}\label{Conformal TransformationLaw2-Tensor}
\bar X^{ij} = \phi^{-10} X^{ij} \,.
\end{equation}
We will see a posteriori $\phi^{-10}$ is the only scaling law that leads to conformal covariance for a contravariant symmetric 2-tensor. This can be understood by considering the fact that the
metric momenta $p^{ij}$ have to transform in the opposite way to the metric: 
\begin{equation}
\bar p^{ij} = \phi^{-4} p^{ij} \,.
\end{equation}
But $p^{ij}$ is a tensor density, and to have a proper tensor we have to divide it
by $\sqrt g$, which transforms as $\phi^6$. This explains where the
$\phi^{-10}$ factor comes from. More precisely, the exponent of (\ref{Conformal TransformationLaw2-Tensor})
is fixed by the form of (\ref{TTdecomposition2-tensor}). \index{York's decomposition} \index{Helmholtz decomposition}

Let's now consider the transformation of the TT-part of $X^{ij}$. We first recall its definition by  (\ref{TTdecomposition2-tensor}),
\begin{equation}
X^{ij}_\st{TT} = X^{ij} - {\ts \frac 1 3} \, X \, g^{ij} +  (LY)^{ij} =
\phi^{10} (\bar X^{ij} - {\ts \frac 1 3} \,\bar X \, \bar g^{ij} ) +  (LY)^{ij} \,,
\end{equation}
and we then remember that
\begin{equation}
(LY)^{ij}  = g^{ik} g^{jl} g^{1/3} \Lie_Y(g^{-1/3} g_{kl})
= \phi^{4} \bar g^{ik} \bar g^{jl} \bar g^{1/3} \Lie_Y({\bar g}^{-1/3} \bar g_{kl})  \,.
\end{equation}
Let us next denote by $ (\bar L Y)^{ij}$ the conformal Killing form calculated with $\bar g_{ij}$,
\begin{equation}
X^{ij}_\st{TT} = \phi^{10} (\bar X^{ij} - {\ts \frac 1 3} \,\bar X \, \bar g^{ij} ) + \phi^{4}  (\bar L Y)^{ij} \,.
\end{equation}
After these preparations, we define the transformation of $X^{ij}_\st{TT}$ through
\begin{equation}
\bar X^{ij}_\st{TT} = \phi^{-10} X^{ij}_\st{TT} = (\bar X^{ij} - {\ts \frac 1 3} \,\bar X \, \bar g^{ij} ) + \phi^{-6}  (\bar L Y)^{ij} \,.
\end{equation}
We now show that this tensor is TT
with respect to the transformed metric $\bar g_{ij}$. The tracelessness is trivial
because every traceless tensor w.r.t. $g_{ij}$ is traceless also w.r.t. $\bar g_{ij}$,
and the two terms that comprise $X^{ij}_\st{TT}$ are separately traceless.
The transversality is less obvious. It needs to hold with respect to the transformed
covariant derivative $\bar \nabla_i$, which includes the transformed connection
\begin{equation}
\bar \Gamma^i_{jk} =  \Gamma^i_{jk}  +2 ({\delta^i}_j \partial_k \log \phi +{\delta^i}_k \partial_j \log \phi - g_{jk} g^{il} \partial_d \log \phi  ) \,.
\end{equation}
Let's take the covariant deriative w.r.t. $\bar g_{ij}$ of $\bar X^{ij}_\st{TT}$,
\begin{equation}\begin{aligned}
\bar \nabla_j \bar X^{ij}_\st{TT} & = \bar \nabla_j  (\bar X^{ij} - {\ts \frac 1 3} \,\bar X \, \bar g^{ij} ) + \bar \nabla_j \left[ \phi^{-6}  (\bar L Y)^{ij} \right] \\  
& = \bar \nabla_j \left[ \phi^{-10} \left(  X^{ij} - {\ts \frac 1 3} \, X \,  g^{ij} +  (L Y)^{ij} \right) \right]=\bar \nabla_j  (\phi^{-10} W^{ij} )\\
& =\phi^{-10}  \left( \nabla_j  W^{ij} - 10 \, W^{ij} \,\partial_j \log \phi + \Delta \Gamma^i_{kj} \,  W^{kj} + \Delta  \Gamma^j_{jk} \, W^{ik}\right) \,,
\end{aligned}\end{equation}
where I called $W^{ij} =  X^{ij} - {\ts \frac 1 3} \, X \,  g^{ij} +  (L Y)^{ij}$
and $\Delta \Gamma^i_{jk} = 2 ({\delta^i}_j \nabla_k \log \phi +{\delta^i}_k \nabla_j \log \phi - g_{jk} \nabla^i \log \phi  )$. An explicit calculation shows immediately that 
\begin{equation}\label{CovariantDerivativePhiTotheTenSymmetricTensor}
10 \, W^{ij} \,\partial_j \log \phi - \Delta \Gamma^i_{kj} \,  W^{kj} -\Delta  \Gamma^j_{jk} \, W^{ik}
=2 \, g_{jk} W^{jk} \, \partial^i \log \phi\,,
\end{equation}
but $W^{ij}$ is traceless and the above expression vanishes.  Thus, we have proved that
\begin{equation}
\bar \nabla_j \bar X^{ij}_\st{TT} =  \phi^{-10} \left[ \nabla_j  (  X^{ij} - {\ts \frac 1 3} \,  X \,   g^{ij} ) +   \nabla_j  (LY)^{ij} \right] \,,
\end{equation}
and if $X^{ij}_\st{TT}$ was TT w.r.t. $g_{ij}$ then $ \nabla_j  (  X^{ij} - {\ts \frac 1 3} \,  X \,   g^{ij} ) +   \nabla_j (LY)^{ij} = 0$ and $\bar \nabla_j \bar X^{ij}_\st{TT} =  0$, \e{that is, $\bar X^{ij}_\st{TT} $ is
TT with respect to $\bar g_{ij}$.}

It is easy to see that the above statement implies also its converse because the original
metric can be obtained from the barred one through the inverse conformal transformation
$g_{ij} = \phi_\st{inv}^4 \, \bar g_{ij} = \phi^{-4} \, \bar g_{ij}$. This exists because, by
definition, $\phi \neq 0$, and therefore the whole argument can be used to show that
if $\bar X^{ij}_\st{TT}$ is TT w.r.t. $\bar g_{ij}$ then$X^{ij}_\st{TT}$ is TT w.r.t. $g_{ij}$.

We conclude that, given a symmetric 2-tensor $X^{ij}$ on a manifold $\Sigma$ equipped with 
the metric $g_{ij}$, it can be decomposed as (\ref{TTdecomposition2-tensor}).
On a conformally related manifold $\Sigma$ with the metric  $\bar g_{ij} = \phi^4 g_{ij}$,
the tensor $\bar X^{ij} = \phi^{-10} X^{ij}$ decomposes in the same way:
\begin{equation}
\bar X^{ij} = \bar  X^{ij}_\st{TT} + \bar X^{ij}_\st{L} + \bar X^{ij}_\st{tr} \,,
\end{equation}
where $ \bar  X^{ij}_\st{TT} =  \phi^{-10} X^{ij}_\st{TT}$, $\bar X^{ij}_\st{L} =\phi^{-10}   X^{ij}_\st{L}$ and $\bar X^{ij}_\st{tr} =\phi^{-10}  X^{ij}_\st{tr} $, with the vector $Y^i$ that determines the 
longitudinal part being the same for $\bar X^{ij}_\st{L}$ and $X^{ij}_\st{L}$.
\index{York's decomposition} \index{Helmholtz decomposition}

\subsection{Point sources in ADM gravity}\label{AppendicePointParticle}
\index{point source in GR}

The Einstein-Hilbert action coupled with a parametrized point particle of coordinates $y^\mu(s)$ is
\begin{equation}
S_\st{EH+M} = \int d^4 x \, \sqrt g \, R + m \, \int ds \sqrt{g_{\mu\nu}(r) \frac{d y^\mu}{ds}\frac{d y^\nu}{ds}} \,,
\end{equation}
the momentum conjugate to $y^\mu$ is
\begin{equation}
p_\mu = \frac{\delta \mathcal L_\st{M}}{\delta \frac{d y^\mu}{ds} } = \frac{ m \, g_{\mu\nu}(r) \frac{d y^\nu}{ds}
}{ \sqrt{g_{\mu\nu}(r) \frac{d y^\mu}{ds}\frac{d y^\nu}{ds}}
} \,,
\end{equation}
and it satisfies the primary consrtaint
\begin{equation}
g^{\mu\nu} \, p_\mu \, p_\nu = m^2 \,.
\end{equation}
If we parametrize the particle with respect to coordinate time $s=t$, $y^0 =t$,  $\dot y^0 = 1$, then
%
%
\begin{equation}
S_{EH+M} = \int d^4 x \, \sqrt g \, R + m \, \int dt \sqrt{ N^2 - g_{ij}  \left(\frac{d y^i}{dt} + N^i \right)\left( \frac{d y^j}{dt}+ N^j \right)},
\end{equation}
where $N$ and $N^i$ are, respectively, the lapse and the shift. Then the spatial part of the particle momentum reads\index{point source in GR}
\begin{equation}
p_i = \frac{- m \, g_{ij}\left( \frac{d y^j}{dt}+ N^j \right)}{ \sqrt{ N^2 - g_{ij}  \left(\frac{d y^i}{dt} + N^i \right)\left( \frac{d y^j}{dt}+ N^j \right)}} \,.
\end{equation}
Inverting the above expression
\begin{equation}
g_{ij}  \left(\frac{d y^i}{dt} + N^i \right)\left( \frac{d y^j}{dt}+ N^j \right) =
\frac{g^{ij} \, p_i \, p_j \, N^2}{m^2 + g^{ij} \, p_i \, p_j} \,,
\end{equation}
which in turn implies that the action can be written in the following way
\begin{equation}
\begin{aligned}\label{S_EH+M}
S_\st{ADM+M} = & \int dt  \int d^3 x \Big{\{} p^{ij} \, \dot g_{ij} - N \, \mathcal H - N^i \,\mathcal H_i   \\
&+\delta^{(3)}(x^i- r^i) \left[ p_i \, \dot y^i + N^i \,p_i - N \, \sqrt{g^{ij} \, p_i \, p_j +m^2} \right] \Big{\}} \,,
\end{aligned}
\end{equation}
where $\mathcal H$ and $\mathcal H_i$ are the vacuum ADM constraints. The addition of matter changes the constraints  into 
\begin{equation}
\begin{aligned}
\!\! \mathcal H^\st{ADM+M} &= \frac{1}{\sqrt g} \left( p^{ij} p_{ij} - {\frac 1 2} p^2 \right) -  \sqrt g \, R + \delta^{(3)}(x^i- r^i)  \, \sqrt{g^{ij} \, p_i \, p_j +m^2} \,,
\\
\!\! \mathcal H^\st{ADM+M}_i &= - 2 \, \nabla_j p^j{}_i  - \delta^{(3)}(x^i- r^i)  \, p_i\,.
\end{aligned}
\end{equation}

\subsection{The poles of a spherically-symmetric universe}\label{AppendixBoundaryConditionsAtThePoles}
\index{spherical symmetry}

In this Appendix I will present all the evidence collected so far in favour of the boundary conditions $A_\st{N} = A_\st{S} = m_\st{N} = m_\st{S} =0$ for the twin-shell universe of Sec~\ref{TwinShellSec}.

The general solution~(\ref{VacuumSolutionForMu}) to the spherically-symmetric ADM constraints involves a metric that takes the  form
{\medmuskip=0mu
\thinmuskip=0mu
\thickmuskip=0mu
\begin{equation}\label{MetricSolutionOfADMConstraints}
\d s^2 =  \frac{\sigma (\sigma')^2 \d r^2}{A^2  + \left( {\sfrac 2 3} \langle p \rangle  A - 8 \, m  \right) \sigma^{\frac 3 2}  + 4 \, \sigma^2 - {\sfrac 1 9}  \left( 12 \, \Lambda - \langle p \rangle^2 \right) \sigma^3 } 
+ \sigma \left( \d \theta^2 + \sin^2 \theta \d \phi^2 \right) \,.
\end{equation}}
The component $\sigma$ is allowed to go to zero at the poles $r=r_a$ only if the polynomial~(\ref{MordorPolynomial}), and with it the denominator of the $\d r^2$ term, is positive around $\sigma =0$. Looking at the regions of positivity of $\mathscr P(\sqrt{\sigma}/m)$ in Fig.~\ref{MordorFig1}--\ref{MordorFig4} we see that on the plane $\sigma = 0$ the polynomial is always positive, unless $A=0$, in which case it is zero. It is easy to see that the on-shell curves which end at $A=\sigma=0$ will do so in such a way that the polynomial will stay positive all the time. If $m>0$, this means that the behaviour of $\sigma$ for small $A$'s will have to be $\sigma \xrightarrow[A\to 0]{} \left( \frac{\beta}{8m}\right)^{\frac 2 3} |A|^{\frac 4 3} + \mathcal O (|A|^{{\frac 2 3} +\epsilon})$, where $0 \leq \beta < 1$ (while if $m \leq 0$ there is no constraint on the asymptotics of $\sigma$).

For small values of the areal radius (near the poles), we can ignore the term $- {\sfrac 1 9}  \left( 12 \, \Lambda - \langle p \rangle^2 \right) \sigma^3$ in~(\ref{MetricSolutionOfADMConstraints}), and the three independent curvature invariant \emph{densities} take the form:
{\medmuskip=0mu
\thinmuskip=0mu
\thickmuskip=0mu\begin{equation}
\begin{aligned}
R_1 = \sqrt g  \, R  &= \frac{\sin \theta \, |\sigma'|}{2 \, \sigma^{3/2}} \frac{3 \, A^2 }{\sqrt{ A^2+ {\sfrac 2 3} B \sigma^{3/2}+4 \sigma^2}}  \,,
\\
R_2 = \sqrt g \,R^i{}_j R^j{}_i &= \frac{\sin \theta \, |\sigma'|}{8 \, \sigma ^{9/2}}  \frac{27 A^4+6 A^2 B \sigma ^{3/2} +B^2 \sigma ^3}{\sqrt{9 A^2+\sigma ^{3/2} (6 A \langle p\rangle -72 m)+36 \sigma ^2}} \,,
\\
R_3 = \sqrt g \, R^i{}_j R^j{}_k  R^k{}_i  &= \frac{\sin \theta \, |\sigma'|}{96 \, \sigma ^{15/2}} \frac{297 A^6+135 A^4 B \sigma ^{3/2}+27 A^2 B^2 \sigma ^3 + B^3 \sigma ^{9/2} }{\sqrt{9 A^2 + 6 B \sigma ^{3/2} +36 \sigma ^2}} \,,
\end{aligned}
\end{equation}}
where $B=\left( A \langle p \rangle - 12 m\right)$. If $A \neq 0$, all these quantities diverge as $\sigma \to 0$ and we have a curvature singularity at the the poles.\footnote{It is not hard to convince oneself that there is no way to have the $|\sigma'|$ term at the numerator cancel the divergence of the denominator while $\sigma \to 0$. In fact if $\sigma \sim r^n$, then $|\sigma'|/\sigma^{3/2}$ is finite if $n \leq -2$, but then $\sigma$ diverges as $r\to 0$.}
If $A=0$ and $m <0$ the first curvature invariant is zero, but the other two are still divergent. If $m>0$ and $\sigma \sim \left(  \beta/8m\right)^{\frac 2 3} |A|^{\frac 4 3}+ \mathcal O (|A|^{{\frac 4 3} +\epsilon})$ the three terms diverge like
{\medmuskip=0mu
\thinmuskip=0mu
\thickmuskip=0mu\begin{equation}
R_1   \sim \frac{12 m}{A \sqrt{1-\beta } \beta } \,, ~
R_2 \sim \frac{48 \left(12 -4 \beta +\beta ^2\right) m^3}{A^3 \sqrt{1-\beta } \beta ^3}  \,, ~
R_3  \sim \frac{384 \left(
88 - 60 \beta - 18 \beta^2 + \beta^3
\right) m^5}{A^5 \sqrt{1-\beta } \beta ^5} \,.
\end{equation}}\index{spherical symmetry}
So the metric (\ref{MetricSolutionOfADMConstraints}) always has a curvature singularity at the poles, for any value of  the parameters $A$  and $m$, unless $A=m=0$. This should be a sufficient reason to take $A=m=0$ as our boundary conditions around the poles, however Shape Dynamics is concerned with the conformal geometry of the metric, and this is regular (conformally flat) \index{conformal flatness}  even in presence of a curvature singularity. From the perspective of conformal geometry, what the curvature singularity does is to make the theory lose predictivity: in fact the value of $A = A(t)$ at the poles is not fixed by any dynamical equation, and needs to be specified by hand.

To better understand this loss of predictivity, turn now to the vacuum diffeomorphism constraint, $\nabla_j p^j{}_i =0$.  The solution~(\ref{SolutionVacuumDiffConstraint}) of this constraint is:
{\medmuskip=0mu
\thinmuskip=0mu
\thickmuskip=0mu\begin{equation}
p^j{}_i = \mu \left[ \left( {\sfrac 1 3} \langle p \rangle \, \sigma  + \frac{A}{\sqrt \sigma} \right)  \, \delta^j_r \delta^r_i +    \left( {\sfrac 1 3} \langle p \rangle \, \sigma  - {\sfrac 1 2}  \frac{A}{\sqrt \sigma}  \right)   \left( \delta^j_\theta \delta^\theta_i  + \delta^j_\phi \delta^\phi_i \right) \right] \sin \theta  \,.
\end{equation}}
There is one spherically-symmetric ($X^i = \delta^i{}_r \xi(r)$) conformal killing vector of the $S^3$ metric:
\begin{equation}
\nabla^i X^j + \nabla^j X^i - {\sfrac 2 3} g^{ij} \nabla_k X^k =0  \qquad \Rightarrow \qquad
X^i = c \, \frac{\sqrt{\sigma}}{\mu} \,  \delta^i{}_r \,,
\end{equation}
(in isotropic gauge this is just $X^i = c \, \sin r \,  \delta^i{}_r$). This vector field is well-behaved at the poles, where $\sigma \to 0$. Now take the vector field $Y^i = p^i_j X^j$. Its coordinate expression is
\begin{equation}
Y^i =  c \, \delta^i{}_r \left( {\sfrac 1 3} \langle p \rangle \, \sigma^{\frac 3 2}  + A  \right)   \sin \theta  \,.
\end{equation}
The divergence of $Y^i$ is
\begin{equation}
\nabla_i Y^i = (\nabla_i p^i_j) \xi^j + p^{ij} \nabla_i \xi_j =  (\nabla_i p^i_j)\xi^j + {\sfrac 1 3} p \nabla_k \xi^k 
=  (\nabla_i p^i_j)\xi^j + {\sfrac 1 3} \langle p \rangle  \nabla_k \xi^k  \, \sqrt g \,,
\end{equation}
integrating over a spherical region centred around the origin:
\begin{equation}
\begin{aligned}
&\int_{r\leq R} \nabla_i Y^i \d^3 x 
= \int_{r\leq R}  (\nabla_i p^i_j)\xi^j \d^3 x + c {\sfrac {4\pi} 3} \langle p \rangle \, \sigma^{\frac 3 2} (R)
\\
& \qquad \qquad \shortparallel
\\
&\int_{r= R}  Y^i \d \Sigma_i =
4 \pi c \,\left( {\sfrac 1 3} \langle p \rangle \, \sigma^{\frac 3 2} (R)  + A  \right)  
\end{aligned}
\end{equation}\index{spherical symmetry}
which implies
\begin{equation}
\int_{r\leq R}  (\nabla_i p^i_j)\xi^j \d^3 x  = 4 \pi c \, A \,.
\end{equation}
Now notice that, if the region of integration was the annular region $R_1 \leq r \leq R_2$, the result would have been
\begin{equation}
\int_{R_1 \leq r \leq R_2}  (\nabla_i p^i_j) \xi^j \d^3 x  = 0\,.
\end{equation}
The same holds for any region $\Omega$ which does not include the pole. We conclude that 
\begin{equation}\label{DiffeoConstraintWithDiracDeltaSource}
 (\nabla_i p^i_j) \xi^j  = 4 \pi c \, A \, \delta^{(3)} (\vec r) \,.
\end{equation}
This result is analogue to what one gets when considering the vacuum Poisson equation on $\mathbbm{R}^3$ in polar coordinates:
\begin{equation}
\begin{aligned}\Delta V &= \frac{\partial^2 V}{\partial x^2} + \frac{\partial^2 V}{\partial y^2} +\frac{\partial^2 V}{\partial z^2} \\
&= {\frac {1}{r^{2}}}{\frac {\partial }{\partial r}}\left(r^{2}{\frac {\partial V}{\partial r}}\right)+{\frac {1}{r^{2}\sin \theta }}{\frac {\partial }{\partial \theta }}\left(\sin \theta {\frac {\partial V}{\partial \theta }}\right)+{\frac {1}{r^{2}\sin ^{2}\theta }}{\frac {\partial ^{2} V}{\partial \varphi ^{2}}} = 0 \,,
\end{aligned}
\end{equation}
if $V$ is spherically symmetric, the equation reduces to ${\frac {1}{r^{2}}}{\frac {\partial }{\partial r}}\left(r^{2}{\frac {\partial V}{\partial r}}\right)=0$, which admits the general solution:
\begin{equation}
V = \frac{c_1}{r} + c_2 \,.
\end{equation}
This solution has two integration constants, but they can both be fixed by appropriate boundary conditions: $V \xrightarrow[r\to \infty]{} 0$ implies $c_2 =0$ and regularity at the origin implies $c_1 =0$. If we insist on having $c_1 \neq 0$, we find out that we are not solving the original equation (in vacuum), but an equation with some sources concentrated at the origin:
\begin{equation}
\Delta V = -4 \pi \, c_1 \, \delta^{(3)} (\vec r) \,,
\end{equation}
in fact, using cartesian coordinates:
\begin{equation}
\Delta \left(\frac{c_1}{r} + c_2  \right) = - c_1 \, \vec \partial \cdot \left(\frac{x}{r^3},\frac{y}{r^3},\frac{z}{r^3} \right) 
\end{equation}\index{spherical symmetry}
and integrating over a sphere of radius $R$:
\begin{equation}
-c_1 \int_{r\leq R}   \vec \partial \cdot  \left(\frac{x}{r^3},\frac{y}{r^3},\frac{z}{r^3} \right) d^3 x = - \frac{c_1}{r^2} \int_{r=R} d\Sigma = - 4 \pi \, c_1 \,.
\end{equation}
The reason for this is the fact that the spherical coordinate patch covers all of $\mathbbm R^3$ except the origin, which lies on the border of the coordinate chart. Then the elliptic equation $\nabla V =0$ turns into a boundary-value problem, depending on the boundary conditions we choose to impose at $r=0$ and $r=\infty$. If we choose $c_1 \neq 0$, we have effectively changed the vacuum equation into one with a Dirac-delta source concentrated at the origin. Such an equation still coincides with the vacuum Poisson equation in the spherical coordinate chart, which does not include the origin, but in Cartesian coordinates, which cover the origin too, it acquires a source term. Similarly, the solution of the diffeomorphism constraint in spherical coordinates depends on the integration constant $A$, which corresponds, in Cartesian coordinates, to a Dirac-delta source term for the constraint. It is clear now how this ruins the predictivity of the theory: one is free to specify a source term like~(\ref{DiffeoConstraintWithDiracDeltaSource}) as a function of time, and no dynamical equation can fix it for us. One may be interested in this exercise, to model for example some collapsed  matter which has some expansion/contraction, but is concentrated in a small region that we want to approximate as pointlike. However, for the present problem of modelling the gravitational collapse of a distribution of matter, it is clear that we have to require that the effective value of the integration constants $A$ and $m$ at the poles is zero.

\cleardoublepage

\phantomsection

\addcontentsline{toc}{section}{\indexname}
\printindex

\addcontentsline{toc}{section}{\listfigurename}
\listoffigures

\newpage

\addcontentsline{toc}{section}{References}

\providecommand{\href}[2]{#2}\begingroup\raggedright\endgroup

\end{document}